\pdfoutput=1
\documentclass[11pt,twoside,a4paper,cmspaper,final,collab]{cms-tdr}
\def\svnVersion{cecdb9c}\def\svnDate{2021/10/22}\def\cmsCernNoTag{CERN-EP-2021-135}\def\cmsCernDate{\today}\def\cmsMessage{Published in Physical Review D as \href{http://dx.doi.org/10.1103/PhysRevD.104.092013}{\doi{10.1103/PhysRevD.104.092013}.}}
\begin{document}\cmsNoteHeader{TOP-20-001}

\newcommand{\Mtop}{\ensuremath{m_\PQt}\xspace}
\newcommand{\MW}{\ensuremath{m_{\PW}}\xspace}
\newcommand{\mur}{\ensuremath{\mu_\mathrm{R}}\xspace}
\newcommand{\muf}{\ensuremath{\mu_\mathrm{F}}\xspace}
\newcommand{\hd}{\ensuremath{h_\text{damp}}\xspace}

\newcommand{\pp}{\ensuremath{\Pp\Pp}\xspace}
\newcommand{\lpj}{\ensuremath{\Pe{}\text{/}{}\Pgm{}\text{+jets}}\xspace}

\newcommand{\tqh}{\ensuremath{\PQt_\mathrm{h}}\xspace}
\newcommand{\tql}{\ensuremath{\PQt_\ell}\xspace}
\newcommand{\tqhard}{\ensuremath{\PQt_\text{high}}\xspace}
\newcommand{\tqsoft}{\ensuremath{\PQt_\text{low}}\xspace}
\newcommand{\Jbl}{\ensuremath{\PQb_\ell}\xspace}
\newcommand{\Jbh}{\ensuremath{\PQb_\mathrm{h}}\xspace}
\newcommand{\JWa}{\ensuremath{\mathrm{j}_{\PW1}}\xspace}
\newcommand{\JWb}{\ensuremath{\mathrm{j}_{\PW2}}\xspace}

\newcommand{\Pmass}{\ensuremath{P_\mathrm{h}}\xspace}
\newcommand{\Plep}{\ensuremath{P_\mathrm{\ell}}\xspace}
\newcommand{\HNN}{\ensuremath{H_\mathrm{NN}}\xspace}
\newcommand{\LNN}{\ensuremath{L_\mathrm{NN}}\xspace}

\newcommand{\st}{\ensuremath{S_\mathrm{T}}\xspace}
\newcommand{\mevt}{\ensuremath{m_\text{evt}}\xspace}
\newcommand{\thadpt}{\ensuremath{\pt(\tqh)}\xspace}
\newcommand{\tleppt}{\ensuremath{\pt(\tql)}\xspace}
\newcommand{\thardpt}{\ensuremath{\pt(\tqhard)}\xspace}
\newcommand{\tsoftpt}{\ensuremath{\pt(\tqsoft)}\xspace}
\newcommand{\thady}{\ensuremath{\abs{y(\tqh)}}\xspace}
\newcommand{\tlepy}{\ensuremath{\abs{y(\tql)}}\xspace}
\newcommand{\topy}{\ensuremath{\abs{y(\PQt)}}\xspace}
\newcommand{\topbary}{\ensuremath{\abs{y(\PAQt)}}\xspace}
\newcommand{\dy}{\ensuremath{\Delta\abs{y_{\PQt/\PAQt}}}\xspace}
\newcommand{\ady}{\ensuremath{\abs{ \Delta y_{\PQt/\PAQt}}}\xspace}
\newcommand{\ttphi}{\ensuremath{\Delta\phi_{\PQt/\PAQt}}\xspace}
\newcommand{\ttm}{\ensuremath{m(\ttbar)}\xspace}
\newcommand{\ttpt}{\ensuremath{\pt(\ttbar)}\xspace}
\newcommand{\tty}{\ensuremath{\abs{y(\ttbar)}}\xspace}
\newcommand{\cts}{\ensuremath{\cos(\theta^*)}\xspace}
\newcommand{\ptl}{\ensuremath{\pt(\ell)}\xspace}
\newcommand{\thadptvsthady}{\ensuremath{\thadpt\,\vs\,\thady}\xspace}
\newcommand{\ttmvstty}{\ensuremath{\ttm\,\vs\,\tty}\xspace}
\newcommand{\ttmvscts}{\ensuremath{\ttm\,\vs\,\cts}\xspace}
\newcommand{\ttmvsthadpt}{\ensuremath{\ttm\,\vs\,\thadpt}\xspace}
\newcommand{\ttptvsthadpt}{\ensuremath{\ttpt\,\vs\,\thadpt}\xspace}
\newcommand{\topyvstopbary}{\ensuremath{\topy\,\vs\,\topbary}\xspace}
\newcommand{\ttmvsdy}{\ensuremath{\ttm\,\vs\,\dy}\xspace}
\newcommand{\adyvsttm}{\ensuremath{\ady\,\vs\,\ttm}\xspace}

\newcommand{\AMCATNLO}{\textsc{mg}5\_\text{a}\textsc{mc@nlo}\xspace}
\newcommand{\MATRIX}{\textsc{Matrix}\xspace}
\newcommand{\PUPPI}{\textsc{PUPPI}\xspace}
\newcommand{\TOPPP}{\textsc{Top++}\xspace}
\newcommand{\CUET}{CUETP8M2T4\xspace}

\newcommand{\ptvecprime}{\ensuremath{{\vec p}_{\mathrm{T}}^{\kern2pt\prime\kern1pt\text{miss}}}\xspace}

\ifthenelse{\boolean{cms@external}}{\providecommand{\cmsleft}{upper\xspace}}{\providecommand{\cmsleft}{left\xspace}}
\ifthenelse{\boolean{cms@external}}{\providecommand{\cmsright}{lower\xspace}}{\providecommand{\cmsright}{right\xspace}} 
\ifthenelse{\boolean{cms@external}}{\providecommand{\cmsLeft}{Upper\xspace}}{\providecommand{\cmsLeft}{Left\xspace}}
\ifthenelse{\boolean{cms@external}}{\providecommand{\cmsRight}{Lower\xspace}}{\providecommand{\cmsRight}{Right\xspace}} 

\newcommand{\XSECCAPPA}{The data are shown as points with gray (yellow) bands indicating the statistical (statistical and systematic) uncertainties. The cross sections are compared to the predictions of \POWHEG combined with \PYTHIA (P8) or \HERWIG (H7), the multiparton simulation \AMCATNLO(MG)+\PYTHIA FxFx, and the NNLO QCD calculations obtained with \MATRIX. The ratios of the various predictions to the measured cross sections are shown in the lower panels.
}

\newcommand{\XSECCAPPS}{The data are shown as points with gray (yellow) bands indicating the statistical (statistical and systematic) uncertainties. The cross sections are compared to the predictions of \POWHEG{}+\PYTHIA (P8) for the CP5 and \CUET (T4) tunes, \POWHEG{}+\HERWIG (H7), and the multiparton simulation \AMCATNLO(MG)+\PYTHIA. The ratios of the various predictions to the measured cross sections are shown in the lower panels.
}

\newlength\cmsTabSkip\setlength{\cmsTabSkip}{1ex}

\cmsNoteHeader{TOP-20-001}
\title{Measurement of differential \texorpdfstring{\ttbar}{top quark pair} production cross sections in the full kinematic range using lepton+jets events from proton-proton collisions at \texorpdfstring{$\sqrt{s} = 13\TeV$}{sqrt(s) = 13 TeV}}

\date{\today}

\abstract{
Measurements of differential and double-differential cross sections of top quark pair (\ttbar) production are presented in the lepton+jets channels with a single electron or muon and jets in the final state. The analysis combines for the first time signatures of top quarks with low transverse momentum \pt, where the top quark decay products can be identified as separated jets and isolated leptons, and with high \pt, where the decay products are collimated and overlap. The measurements are based on proton-proton collision data at $\sqrt{s} = 13\TeV$ collected by the CMS experiment at the LHC, corresponding to an integrated luminosity of 137\fbinv. The cross sections are presented at the parton and particle levels, where the latter minimizes extrapolations based on theoretical assumptions. Most of the measured differential cross sections are well described by standard model predictions with the exception of some double-differential distributions. The inclusive \ttbar production cross section is measured to be
$\sigma_\ttbar = 791\pm25\unit{pb}$, which constitutes the most precise measurement in the lepton+jets channel to date.
}

\hypersetup{%
pdfauthor={CMS Collaboration},%
pdftitle={Measurement of differential \texorpdfstring{\ttbar}{top quark pair} production cross sections in the full kinematic range using lepton+jets events from proton-proton collisions at \texorpdfstring{$\sqrt{s} = 13\TeV$}{sqrt(s) = 13 TeV}},%
pdfsubject={CMS},%
pdfkeywords={CMS,  top quark, cross sections}}

\maketitle 

\section{Introduction}
Precision measurements of top quark pair (\ttbar) production are important tests of the standard model (SM) since the top quark plays an exceptional role; it is the heaviest known particle and the only quark that can be observed before hadronization.
A detailed understanding of \ttbar production is important for many searches for beyond-SM phenomena, where it often constitutes a major background. In addition, measurements of differential \ttbar production can contribute significantly to the determination of parton distribution functions (PDFs), as well as the extraction of important SM parameters such as the top quark mass \Mtop, the strong coupling constant \alpS{}~\cite{TOP-18-004}, and the top quark Yukawa coupling~\cite{TOP-17-004, TOP-19-008}.  

At the CERN LHC, measurements of differential cross sections have been performed in various \ttbar decay channels at proton-proton (\pp) center-of-mass energies of 7~\cite{Chatrchyan:2012saa,Aad:2015eia}, 8~\cite{Khachatryan:2015oqa,Aad:2015mbv,Aad:2015hna,Khachatryan:2015fwh,Khachatryan:2149620,Aaboud:2016iot,Sirunyan:2017azo}, and 13~\cite{Aaboud:2016xii,Aaboud:2016syx,TOP-16-007,TOP-16-008,TOP-17-002,Aaboud:2018eqg,Aad:2019ntk,Aad:2019hzw}\TeV. In this paper, measurements of differential \ttbar production cross sections are presented by the CMS Collaboration using \lpj events, \ie, with a single electron or muon and jets in the final state. This analysis is based on an integrated luminosity of 137\fbinv at 13\TeV center-of-mass energy, where 35.9\fbinv were recorded in 2016, 41.5\fbinv in 2017, and 59.7\fbinv in 2018. Since the running conditions and the CMS detector changed during this time period, detector performance and calibration measurements are carried out separately for each year.

We use resolved \ttbar reconstruction techniques similar to those introduced in previous CMS analyses~\cite{TOP-16-008,TOP-17-002}. These are applicable if all the \ttbar decay products can be reconstructed as separated leptons and jets in the detector---typically for top quarks with transverse momenta $\pt < 500$\GeV. These results are extended by adding Lorentz-boosted top quarks with collimated and overlapping decay products. This is the first time that both the resolved and boosted techniques are used in a combined analysis. Techniques for the boosted and resolved reconstructions of \tqh and \tql are developed or improved, where \tqh{} (\tql) represents a top quark decaying into a \PQb quark and a \PW boson with a subsequent hadronic (leptonic) decay of the \PW boson. The differential cross sections are extracted by performing a combined fit to various categories that are defined by the reconstruction methods, the lepton flavors, and the three years of data taking. The combination of the different categories provides constraints on the systematic uncertainties and results in an improved precision with respect to previous measurements.    

The differential cross sections are presented at the parton and particle levels. The parton level represents a \ttbar pair before decay. The cross sections are presented in the full phase space of the top quarks. This means for the parton-level measurements that all effects related to top quark decays, hadronization, and limited detector acceptance are corrected based on theoretical assumptions. The uncertainties due to these extrapolations are reduced in the measurements at the particle level, where the \ttbar pair is defined based on jets and leptons that can be directly observed with the detector. The strong relation between particle- and detector-level objects also results in less bin-to-bin migrations and simplifies the unfolding. More details about the definitions of the parton and particle levels are given in Section~\ref{PSTOP}.

At the parton and particle levels, we measure the differential cross sections as a function of the following variables: \thadpt, \tleppt, and their scalar sum \st; the higher \thardpt and the lower \tsoftpt transverse momenta of the top quarks; the rapidities \thady, \tlepy, and the rapidity differences $\dy = \topy - \topbary$, $\ady = \abs{y(\PQt) - y(\PAQt)}$; and the angle between the top quarks in the transverse plane \ttphi. For the \ttbar system the differential cross sections are measured as a function of \tty, \ttpt, the invariant mass \ttm, and \cts, where $\theta^*$ is the angle between the \PQt and the direction of flight of the \ttbar system calculated in the \ttbar rest frame. The identification of a \PQt or \PAQt is done using the charge of the electron or muon. Double-differential cross sections are measured as a function of combinations of these variables: \thadptvsthady, \ttmvstty, \ttmvscts, \ttmvsthadpt, \ttptvsthadpt, \ttmvsdy, \adyvsttm, and \topyvstopbary. From the sum of cross sections in all bins of a distribution, a measurement of the inclusive \ttbar production cross section is obtained. In addition, at the particle level the cross sections are determined as a function of the additional jet multiplicity and as a function of \thadpt, \ttm, and \ttpt in bins of additional jet multiplicity, where additional jets are those that are not used in the reconstruction of the \ttbar system. Finally, differential cross sections as a function of the scalar \pt sum of additional jets \HT, the invariant mass of the top quarks and all additional jets \mevt, and the \pt of the electrons and muons \ptl are presented.

We begin with an overview of the theoretical calculations and simulations of the detector in Section~\ref{SIM}, followed by a detailed discussion of the parton- and particle-level definitions in Section~\ref{PSTOP}. After a short description of the CMS detector and the reconstruction and identification of the involved physics objects in Sections~\ref{DET} and \ref{PHOBJ}, respectively, the resolved and boosted reconstructions of \tqh and \tql are detailed in Sections~\ref{TTREC}--\ref{HADBOOST}. An overview of the event categorization based on the various reconstruction methods is presented in Section~\ref{EVRECO}, before the methods of background subtraction are explained in Section~\ref{RESBKG} for the resolved reconstruction and in Section~\ref{FITBO} for the boosted reconstruction. The extraction of the cross sections using a fit combining all categories is described in Section~\ref{UNFOLDING}. In Section~\ref{SYS}, detailed information about the systematic uncertainties is given. The results at the parton and particle levels are presented and discussed in Section~\ref{RES}, and a summary of the results is provided in Section~\ref{SUM}. Tabulated results are provided in HEPData~\cite{hepdata}.

\section{Signal and background modeling}
\label{SIM}
The Monte Carlo event generator \POWHEG~\cite{Nason:2004rx,Frixione:2007vw,Alioli:2010xd} (version \textsc{powheg-box-v2}, process: hvq~\cite{Powheghvq}) is used to simulate the production of \ttbar events at next-to-leading-order (NLO) accuracy in quantum chromodynamics (QCD). The \POWHEG output is combined with the parton shower (PS) simulation of \PYTHIA~\cite{Sjostrand:2014zea} (version 8.2) using the underlying event (UE) tunes \CUET~\cite{Skands:2014pea, Khachatryan:2015pea, CMS-PAS-TOP-16-021} for the 2016 simulations and CP5~\cite{CP5} for the 2017 and 2018 simulations. The renormalization $\mur$ and factorization $\muf$ scales are set to the transverse mass $\mT = \sqrt{\smash[b]{\Mtop^2 + \pt^2}}$ of the top quark, where $\Mtop=172.5\GeV$ is used. The PDFs are NNPDF30\_nlo\_as\_0118~\cite{Ball:2014uwa} for 2016 and NNPDF31\_nnlo\_hessian\_pdfas~\cite{Ball:2017nwa} for 2017--2018. The detector response is simulated using \GEANTfour{}~\cite{Agostinelli:2002hh}.  The simulations include multiple \pp interactions per bunch crossing (pileup). The simulated distribution of the number of pileup interactions corresponds to the distribution in data for each year. Finally, the same reconstruction algorithms that are applied to the data are used for the simulated events. This \POWHEG{}+\PYTHIA simulation is taken as the default to obtain all corrections for the extraction of the differential cross sections. 

To estimate systematic uncertainties, several variations in the default simulation are used, including simulations with \Mtop varied by $\pm${}1\GeV and variations of the UE parameters representing the uncertainties in the tunes. In other simulations, the parton shower matching scales $\hd = (1.58^{\:+0.66}_{\:-0.59})\Mtop$ (\CUET) and $(1.38^{\:+0.92}_{\:-0.51})\Mtop$ (CP5) are varied within their uncertainties, which are obtained from the UE tunes. In addition, a simulation with a different color reconnection (CR) model~\cite{CMS-PAS-TOP-16-021} that allows interactions of colored particles from resonance decays with other particles, is used.  Distributions corresponding to variations of the scales \mur, \muf, or the PS scales by factors of 0.5 and 2 are obtained by applying event weights to the default simulation. Event weights are also used for the estimation of PDF uncertainties. 

{\tolerance=6000 The measured differential cross sections are compared to predictions obtained using \POWHEG{}+\HERWIG~\cite{Bahr:2008pv} (version 7.1) with tune CH3~\cite{GEN-19-001}. The program \MGvATNLO~\cite{Alwall:2014hca} (version 2.2.2) (\AMCATNLO) is used to simulate \ttbar events with additional partons. All processes with up to two additional partons are calculated at NLO QCD and combined with the \PYTHIA PS simulation using the FxFx~\cite{Frederix:2012ps} algorithm.\par}

{\tolerance=2000 All \ttbar simulations are normalized to the inclusive \ttbar production cross section of $832^{+40}_{-46}\unit{pb}$ calculated with \TOPPP (version 2.0)~\cite{Czakon:2011xx}. This value is determined with next-to-NLO (NNLO) accuracy, including the resummation of next-to-next-to-leading-logarithmic (NNLL) soft-gluon terms, where $\mur = \muf = \Mtop$. Cross sections are calculated using the PDFs MSTW2008nnlo68cl~\cite{MSTW}, CT10nnlo~\cite{CT10}, and NNPDF23\_nnlo\_FFN\_NF5~\cite{NNPDF23}, and the midpoint of their envelopes is used. The uncertainty is evaluated by varying \mur and \muf and adding in quadrature the envelope of the uncertainties obtained for the various PDFs. Differential cross sections with  NNLO QCD accuracy are obtained with \MATRIX~\cite{MATRIXTT1, MATRIXTT2}, where $\mur = \muf = \frac{1}{2}\Bigl[\sqrt{\smash[b]{\Mtop^2 + \pt^2(\PQt)}} + \sqrt{\smash[b]{\Mtop^2 + \pt^2(\PAQt)}}\, \Bigr]$ and the PDF set NNPDF31\_nnlo\_as\_0118 is used.\par}

The main backgrounds from SM processes are simulated applying the same techniques as used for the simulation of \ttbar production. The \AMCATNLO generator is used for the NLO QCD simulation of \PW boson production in association with jets, $s$-channel single top quark production, and Drell--Yan (DY) production in association with jets. The generator \POWHEG~\cite{Re:2010bp} is used for the simulation of $t$-channel single top quark production and single top quark production associated with a \PW boson ($\PQt\PW$). For the latter, contributions from \ttbar production entering the NLO calculation are removed using the diagram removal scheme~\cite{White_2009}. In all cases, the PS and the UE are described by \PYTHIA. Multijet events, \ie, SM events comprised uniquely of jets produced through the strong interaction, are simulated using \PYTHIA. The \PW boson and DY backgrounds are normalized to their NNLO cross sections calculated with \FEWZ~\cite{Li:2012wna} (version 3.1). The $t$-channel single top quark production is normalized to the NLO calculation obtained from \textsc{Hathor}~\cite{Kant:2014oha} (version 2.1). The production of $\PQt\PW$ is normalized to the NLO calculation~\cite{Kidonakis:2012rm}, and the multijet simulation is normalized to the leading-order calculation obtained with \PYTHIA~\cite{Sjostrand:2007gs}.

\section{Definitions at the parton and particle levels}
\label{PSTOP}

The parton level is represented by a \ttbar pair before it decays. In other words, the top quarks are assumed to be stable and all effects related to their decays are corrected for in the measurement. Information from the decay is only used to identify \tqh and \tql, where a \tql with a \PW boson decay involving a \Pgt lepton, regardless of its decay mode, is excluded. The cross sections are presented in the full phase space of the top quarks. Results of a fixed-order \ttbar calculation as obtained with \MATRIX can be directly compared to the parton-level measurement after scaling by the branching fraction of $(28.77\pm0.32)$\% for \lpj events calculated using measured \PW boson branching fractions~\cite{PDG}.  In calculations that combine matrix elements and PS, the top quark momenta are obtained after the combination, \ie, the PS affects the definition of the parton level.

Particle-level objects are constructed from simulated particles with a mean lifetime greater than 30\unit{ps}, obtained from the predictions of \ttbar event generators before any detector simulation. The particle-level objects are further used to define particle-level top quarks. Detailed studies on particle-level definitions can be found in Ref.~\cite{pseudotop}. The exact definitions are summarized below.

\begin{itemize}
\item All simulated electrons and muons, including those from \Pgt lepton decays but not originating from the decay of a hadron, are corrected for effects of bremsstrahlung by adding the momentum of a photon to that of the closest lepton if their separation is $\Delta R < 0.1$, where $\Delta R = \sqrt{\smash[b]{(\Delta \eta)^2 + (\Delta \phi)^2}}$ with the differences in pseudorapidity $\Delta\eta$ and azimuthal angle $\Delta\phi$ between the directions of the photon and lepton. All photons are considered for the momentum correction. We require leptons to have $\pt > 15\GeV$ and $\abs{\eta} < 2.4$.
\item Jets are clustered using the anti-\kt jet algorithm~\cite{Cacciari:2008gp, Cacciari:2011ma} with a distance parameter of 0.4. All particles with the exception of neutrinos are clustered. Jets with $\pt > 25\GeV$ and $\abs{\eta} < 2.4$ are selected if there is no electron or muon, as defined above, within $\Delta R = 0.4$.
\item Particle-level $\PQb$ jets are defined as those jets that contain a $\PQb$ hadron using the ghost-matching technique~\cite{Cacciari_2008}: as a result of the short lifetime of $\PQb$ hadrons only their decay products are considered for the jet clustering. However, to allow their association with a jet, the $\PQb$ hadrons are also included with their momenta scaled down to a negligible value. This preserves the information of their directions, but removes their impact on the jet clustering.
\item The magnitude of the neutrino four-momentum $p_\Pgn$ from the \tql decay is calculated using all neutrinos, including those stemming from decays of hadrons. The vectorial sum of their transverse momenta is used as a proxy for the missing transverse momentum \ptvecprime. The longitudinal component $p_{z}(\Pgn)$ of  $p_\Pgn$ is calculated using the \PW boson mass constraint $(p_\Pgn + p_\ell)^2 = m_{\PW}^2$, where \ptvecprime is taken as the transverse momentum of the neutrino, $p_\ell$ is the four-momentum of the lepton in the \tql decay, and $\MW = 80.4\GeV$~\cite{PDG}. This results in a quadratic equation for the longitudinal component of the neutrino momentum $p_{z}(\Pgn)$. If no real solution exists, the two components of \ptvecprime are scaled separately to find a single solution under the condition of a minimum modification of $p_\mathrm{T}^{\prime \mathrm{miss}}$. The scaled \ptvecprime, together with the calculated solution for $p_{z}(\Pgn)$, form the neutrino momentum. If two real solutions exist, the invariant masses of each of the neutrino solutions, together with the charged lepton and the $\PQb$ jet in the \tql decay, are calculated, and the solution resulting in a mass closer to \Mtop is selected. The event is rejected if the invariant mass is not between 100 and 240\GeV. This method for calculating the neutrino momentum corresponds to the method used at the detector level. 
\item Candidates used to form a boosted \tql are the \PQb jets and any selected lepton within a cone of $\Delta R = 0.4$. The minimum \pt requirement is increased to 50\GeV for these leptons. The candidate momentum is calculated as the momenta sum of the jet, the lepton, and the calculated neutrino. It is ensured that the lepton momentum is not counted twice if it is a constituent of the jet. Finally, $\pt > 380$\GeV, $\abs{\eta} < 2.4$, and an invariant mass between 100 and 240\GeV are required for the boosted \tql candidate.
\item Candidates for a boosted \tqh are \PQb jets defined exactly as the \PQb jets above but clustered with a distance parameter of 0.8. Those jets with $\pt > 380\GeV$ and $\abs{\eta} < 2.4$ are selected if there is no electron or muon, as defined above, within $\Delta R = 0.8$. In addition, the invariant mass of all constituents $m_\mathrm{jet}$ is required to be greater than 120\GeV.
\end{itemize}

Based on these objects, we construct a pair of particle-level top quarks in the \lpj final state. Events with exactly one electron or muon with $\pt > 30\GeV$ and $\abs{\eta} < 2.4$ are selected. Simulated events with an additional particle-level electron or muon are rejected.

If one candidate for a boosted \tql and at least one candidate for a  boosted \tqh exist that are separated at least by $\Delta R = 1.2$, the boosted \tqh with $m_\mathrm{jet}$ closest to $\Mtop$ is selected and the two form the pair of particle-level top quarks. If there is a boosted \tql but no candidate for a boosted \tqh, the event is rejected. The combination of a boosted \tql and a resolved reconstructed \tqh is, in analogy to the detector-level reconstruction (cf.~Section~\ref{EVRECO}), not considered.

If there is no boosted \tql, we find the permutation of jets that minimizes the quantity
\ifthenelse{\boolean{cms@external}}
{
\begin{multline}
[m(p_\Pgn + p_{\ell} + p_{\Jbl}) - \Mtop]^2 + [m(p_{\JWa} + p_{\JWb}) - \MW]^2\\ + [m(p_{\JWa} + p_{\JWb} + p_{\Jbh}) - \Mtop]^2,
\label{PSTOPE1}
\end{multline}
} 
{ 
\begin{equation}
[m(p_\Pgn + p_{\ell} + p_{\Jbl}) - \Mtop]^2 + [m(p_{\JWa} + p_{\JWb}) - \MW]^2 + [m(p_{\JWa} + p_{\JWb} + p_{\Jbh}) - \Mtop]^2,
\label{PSTOPE1}
\end{equation}
}
where $p_{\mathrm{j}_{\PW1}}$ and $p_{\mathrm{j}_{\PW2}}$ are the four-momenta of two light-flavor jet candidates, considered as the decay products of the hadronically decaying \PW boson; and $p_{\PQb_{\ell}}$ and $p_{\PQb_{\mathrm{h}}}$ are the four-momenta of two \PQb jet candidates.  All jets with $\pt > 25\GeV$ and $\abs{\eta} < 2.4$ are considered. At least four jets are required, of which at least two must be $\PQb$ jets. The remaining jets with $\pt > 30\GeV$ and $\abs{\eta} < 2.4$ are defined as additional jets. The best permutation is only accepted if the reconstructed \tqh invariant mass satisfies $100 <  m(\tqh) < 240$\GeV.

Alternatively, we also evaluate the possibility of a selection with a boosted \tqh by minimizing 
\begin{equation}
[m(p_\Pgn + p_{\ell} + p_{\Jbl}) - \Mtop]^2  + [m_\mathrm{jet} - \Mtop]^2,
\label{PSTOPE2}
\end{equation}
where separations of at least $\Delta R = 1.2$ between the \tqh and both the $\ell$ and the $\Jbl$ are required. 
If both reconstruction methods for the resolved and the boosted \tqh are successful, we select the reconstruction for which  $m(\tqh)$ is closer to \Mtop.

Events with a hadronically and a leptonically decaying particle-level top quark are not required to be \lpj events at the parton level. Especially, \Pgt{}+jets events with a leptonically decaying \Pgt lepton can contribute, and also \ttbar dilepton events with additional jets can be identified as \lpj events at the particle level if one lepton fails to pass the selection.

\begin{figure}[tbp]
\centering
\includegraphics[width=0.45\textwidth]{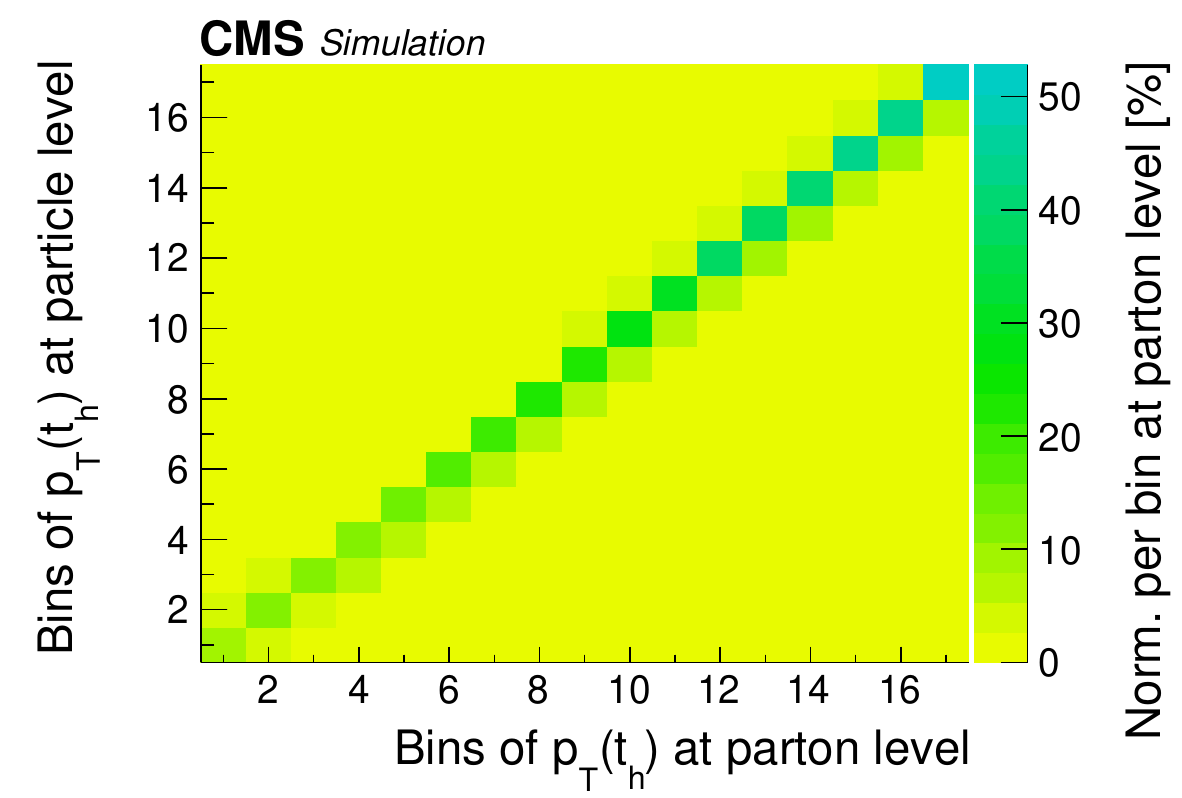}
\includegraphics[width=0.45\textwidth]{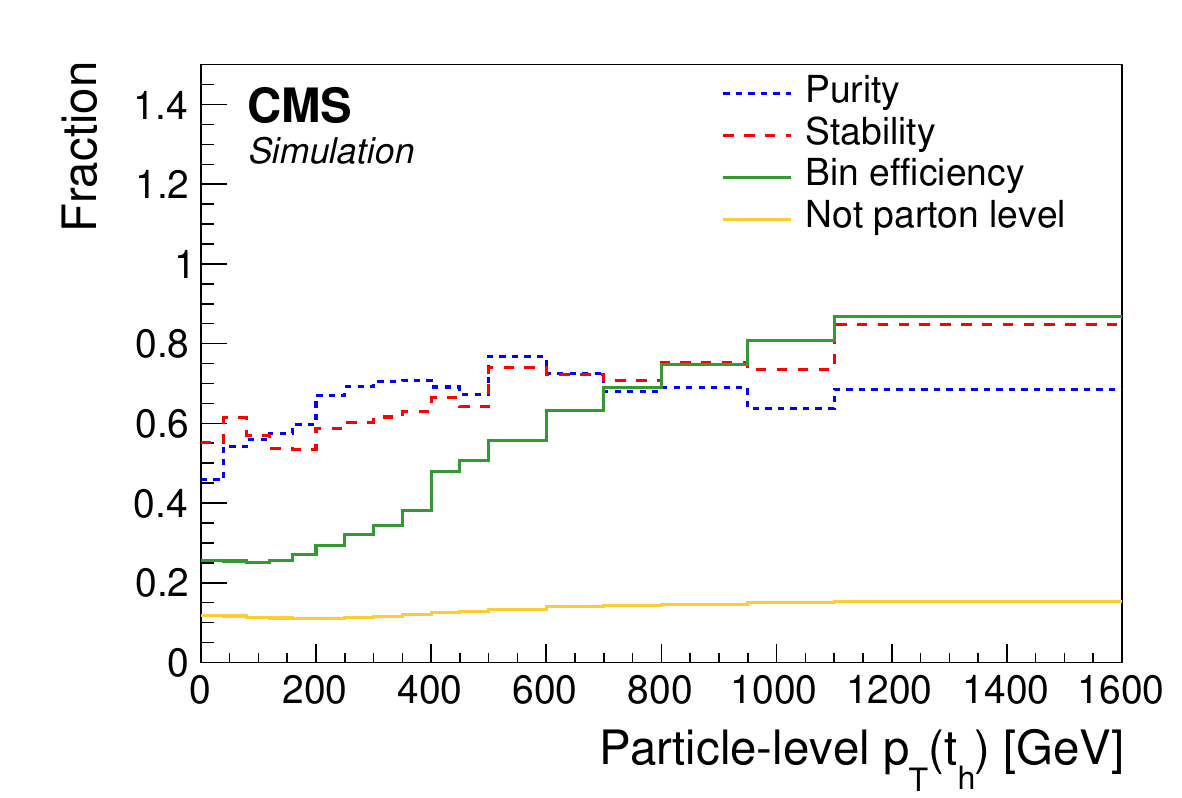}
\caption{Comparison between the \thadpt distributions at the particle and parton levels, extracted from the \POWHEG{}+\PYTHIA simulation. \cmsLeft: the percentage of \pt-bin migration between the particle and parton levels, shown using the color scale to the right of the plot. The bin boundaries are those shown in the \cmsright panel. Each column is normalized such that the sum of its entries corresponds to the fraction of particle-level events in this bin at the parton level in the full phase space. \cmsRight: fraction of parton-level top quarks in the same \pt bin at the particle level (purity), fraction of particle-level top quarks in the same \pt bin at the parton level (stability), ratio of the number of particle- to parton-level top quarks (bin efficiency), and the fraction of particle-level events that are not signal events at the parton level.
}
 \label{PSTOPF1}
\end{figure}

As an example, the comparison between the \thadpt distributions at the particle and parton levels is shown in Fig.~\ref{PSTOPF1}. About 60\% of the \tqh are in the same \pt bins at the parton and particle levels. At high \pt a particle-level \tqh exists for up to 80\% of the parton-level \tqh, while only about 12\% of the particle-level events do not fulfill the definition of parton-level events. This overall good correspondence between the particle and parton levels ensures that the observables are sensitive to the underlying physics of \ttbar production.

\section{The CMS detector}

\label{DET}
The central feature of the CMS detector is a superconducting solenoid of 6\unit{m} internal diameter, providing a magnetic field of 3.8\unit{T}. Within the solenoid volume are a silicon pixel and strip tracker, a lead tungstate crystal electromagnetic calorimeter (ECAL), and a brass and scintillator hadron calorimeter (HCAL), each composed of a barrel and two endcap sections. Forward calorimeters extend the $\eta$ coverage provided by the barrel and endcap detectors. Muons are measured in gas-ionization detectors embedded in the steel flux-return yoke outside the solenoid. A more detailed description of the CMS detector, together with a definition of the coordinate system and relevant kinematic variables, can be found in Ref.~\cite{Chatrchyan:2008zzk}.

Events of interest are selected using a two-tiered trigger system. The first level (L1), composed of custom hardware processors, uses information from the calorimeters and muon detectors to select events at a rate of around 100\unit{kHz} within a fixed latency of about 4\mus~\cite{Sirunyan:2020zal}. The second level, known as the high-level trigger (HLT), consists of a farm of processors running a version of the full event reconstruction software optimized for fast processing, and reduces the event rate to around 1\unit{kHz} before data storage~\cite{Khachatryan:2016bia}. For this measurement events are selected using single electron and muon triggers with \pt thresholds below 34\GeV for isolated leptons and of 50\GeV for nonisolated leptons.

The particle-flow (PF) event algorithm~\cite{PF} aims to reconstruct and identify each individual particle with an optimized combination of information from the various elements of the CMS detector. The energy of muons is obtained from the curvature of the corresponding track. The energy of electrons is determined from a combination of the electron momentum at the primary interaction vertex as determined by the tracker, the energy of the corresponding ECAL cluster, and the energy sum of all bremsstrahlung photons spatially compatible with originating from the electron track. The energy of photons is directly obtained from the ECAL measurement. The energy of charged hadrons is determined from a combination of their momentum measured in the tracker and the matching ECAL and HCAL energy deposits, corrected for zero-suppression effects and for the response function of the calorimeters to hadronic showers. Finally, the energy of neutral hadrons is obtained from the corresponding corrected ECAL and HCAL energy.

\section{Physics object reconstruction}
\label{PHOBJ}
The measurements presented in this paper depend on the reconstruction and identification of electrons, muons, jets, and missing transverse momentum \ptvecmiss associated with neutrinos. Electrons and muons are selected if they are compatible with originating from the primary vertex, which, among the reconstructed \pp interaction vertices, is the one with the largest value of summed physics-object $\pt^2$. The physics objects are the jets, clustered using the anti-\kt jet-finding algorithm~\cite{Cacciari:2008gp,Cacciari:2011ma} with the tracks assigned to the primary vertex as inputs, and the associated missing transverse momentum, taken as the negative vectorial \ptvec sum of those jets.

Isolated electrons~\cite{CMSelectrons} and muons~\cite{Sirunyan:2018} with $\pt > 30$\GeV and $\abs{\eta} < 2.4$ are selected. In 2018 the minimum \pt of electrons was raised to 34\GeV due to increased trigger thresholds. Several quality criteria including isolation and compatibility with the primary vertex are required. The electron and muon reconstruction and selection efficiencies are measured in the data using the ``tag-and-probe'' technique~\cite{TNP}. Depending on \pt and $\eta$, the overall reconstruction and selection efficiency is 50--80\% for electrons and 75--85\% for muons. Nonisolated electrons and muons are required to have $\pt > 50$\GeV due to the higher \pt thresholds for the lepton triggers without an isolation requirement. Apart from the \pt and isolation requirements, the nonisolated electrons and muons have to fulfill the same selection criteria as the isolated leptons. Their reconstruction and trigger efficiencies are measured using \ttbar events in the \Pe{}\Pgm decay channel. The products of efficiencies for the nonisolated electrons and muons are about 80\% and 90\%, respectively. The trigger for nonisolated electrons also requires a jet with $\pt > 165\GeV$. The efficiency of this jet requirement reaches almost 100\% for events with a boosted \tqh candidate as selected in this analysis.   

Jets are clustered from PF candidates using the anti-\kt jet algorithm with a distance parameter of 0.4 (AK4 jets) implemented in the \FASTJET package~\cite{Cacciari:2011ma}. Charged PF candidates originating from a pileup interaction vertex are excluded. The total energy of the jets is corrected for energy depositions from pileup. In addition, \pt- and $\eta$-dependent corrections are applied to correct for detector response effects~\cite{JET}. If an isolated lepton with $\pt > 15\GeV$ within $\Delta R = 0.4$ around a jet exists, the jet is assumed to represent the isolated lepton and is removed from further consideration to prevent counting the lepton momentum twice. The AK4 jets are considered for analysis if they fulfill the kinematic requirements $\pt > 30$\GeV and $\abs{\eta} < 2.4$.

For the identification of $\PQb$ jets, the DeepCSV algorithm~\cite{BTV-16-002} is used. It is based on an artificial neural network (NN) that provides a discriminant to distinguish between \PQb and other flavored jets using features of secondary vertices. An AK4 jet is identified as a $\PQb$ jet if the associated value of the discriminant exceeds a threshold criterion. Two different selection criteria are used: a tight one with an efficiency of about 70--75\% and rejection probabilities of about 88\% for $\PQc$ jets and about 99\% for other jets, and a loose one with corresponding values of 85--90, 55, and 90\%. 

Boosted \tqh candidates are identified using anti-\kt jets with a distance parameter of 0.8 (AK8 jets),  $\pt > 400$\GeV, $\abs{\eta} < 2.4$, and a jet mass $m_\mathrm{jet} > 120$\GeV. In contrast to the AK4 jets, the AK8 jets utilize the \PUPPI~\cite{PUPPI} algorithm for pileup mitigation, and the \pt of all PF constituents are rescaled according to their \PUPPI weights before clustering. Energy corrections are applied to account for detector response effects~\cite{JET}. If an isolated lepton with $\pt > 15\GeV$ within $\Delta R = 0.8$ of a jet exists, that jet is removed from further consideration.

The vector \ptvecmiss is calculated as the negative vectorial \ptvec sum of all PF candidates in the event. Jet energy corrections are also propagated to improve the measurement of \ptvecmiss.

\section{Reconstruction of the resolved \texorpdfstring{\ttbar}{top quark pair} system}
\label{TTREC}
For the resolved reconstruction exactly one isolated electron or muon and at least four AK4 jets are required. If at least two of the jets are identified as \PQb jets by the tight criterion, the event is categorized as ``2t''. If there is one jet passing the tight and another jet passing the loose \PQb tagging criteria, the event is classified as ``1t1l''. About 75\% of the resolved reconstructed events fall into the 2t and the rest into the 1t1l category.

The reconstruction of the \ttbar system in the resolved case follows closely the methods used in previous CMS analyses~\cite{TOP-16-008,TOP-17-002}. The goal is the correct identification of detector-level objects as top quark decay products. For the optimization of the reconstruction algorithm, the following definitions of correct reconstructions based on generator-level information are used. For the particle-level measurements, the jets used to construct the top quarks are spatially matched to detector-level jets within $\Delta R = 0.4$. For the parton-level measurements, the two \PQb quarks from top quark decays and the two light quarks from the \PW boson decay are obtained from the event record of the simulation and spatially matched to detector-level jets with the highest \pt within $\Delta R = 0.4$.        

All possible permutations of assigning detector-level jets to the corresponding \ttbar decay products are tested and a likelihood that each permutation is correct is evaluated based on top quark and \PW boson mass constraints. Only the two jets with the highest $\PQb$ identification probabilities are tested as \Jbl and \Jbh candidates. In each event, the permutation with the highest likelihood is selected.

For all tested permutations, $p_\Pgn$ is calculated using the \PW boson mass constraint $(p_\Pgn + p_\ell)^2 = m_{\PW}^2$, where \ptvecmiss is taken as the transverse momentum of the neutrino. This results in a quadratic equation for the longitudinal component of the neutrino momentum $p_{z}(\Pgn)$. If no real solution exists, which can happen even in signal events due to the finite \ptvecmiss resolution, the $x$ and $y$ components of \ptvecmiss are scaled separately to find a single solution under the condition of a minimum modification of \ptmiss. The scaled \ptvecmiss, together with the calculated solution for $p_{z}(\Pgn)$, form the neutrino momentum. If there are two solutions to the quadratic equation, the invariant mass is calculated using $m(\tql)^2 = (p_\Pgn + p_\ell + p_{\Jbl})^2$ for both $p_\Pgn$ values and the solution closer to \Mtop is selected. For each permutation, the $p_\Pgn$ calculation yields a value of $m(\tql)$, which is used to optimize the selection of \Jbl, as explained below.

\begin{figure}[tbp]
\centering
\includegraphics[width=0.45\textwidth]{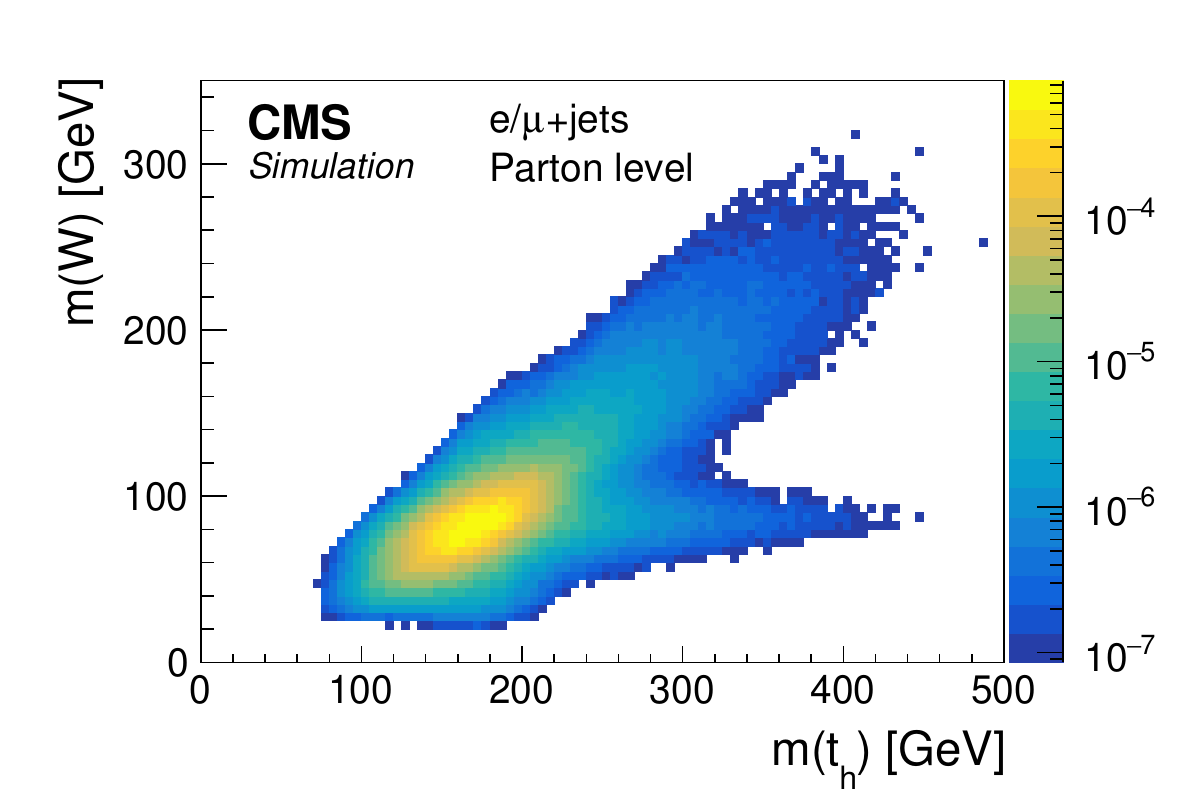}
\includegraphics[width=0.45\textwidth]{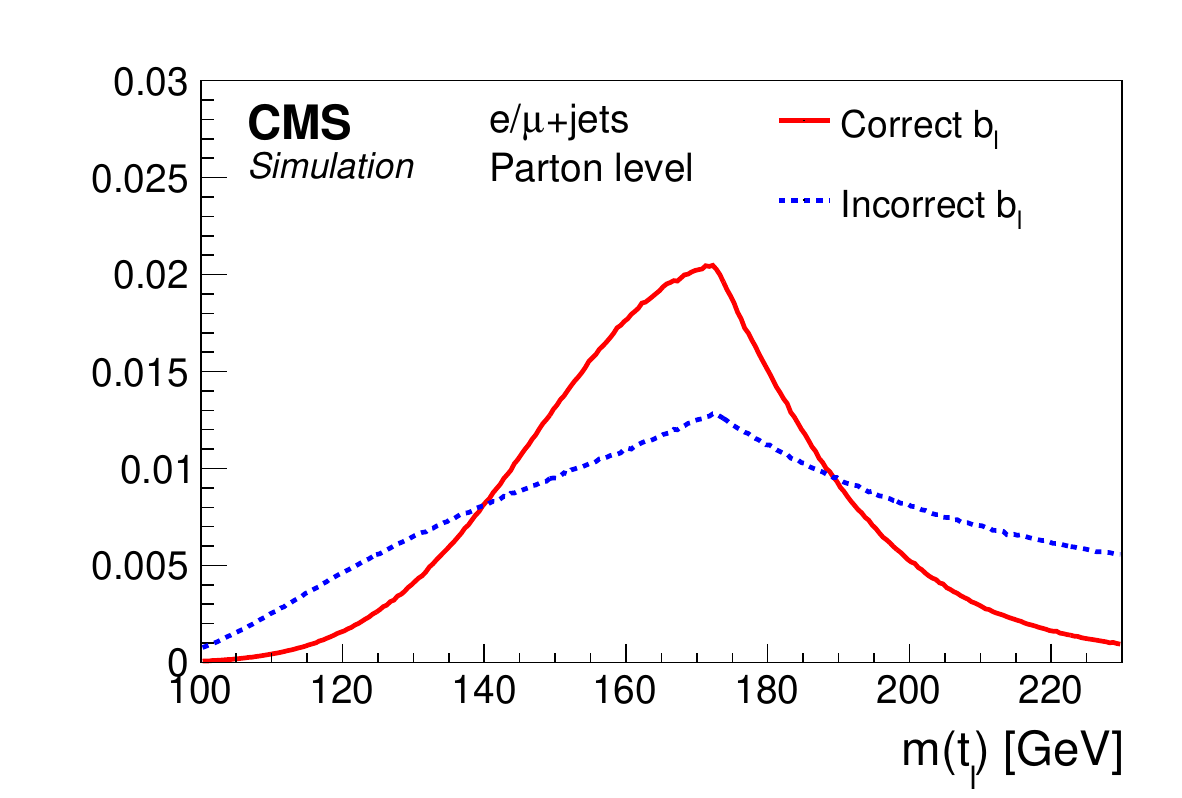}
\caption{Normalized two-dimensional mass distribution of the correctly reconstructed hadronically decaying \PW bosons  and the correctly reconstructed \tqh candidate (\cmsleft). Normalized distributions of the reconstructed $m(\tql)$ for correctly (solid red curve) and incorrectly (dashed blue curve) selected \Jbl (\cmsright). The distributions are taken from the \POWHEG{}+\PYTHIA{} \ttbar simulation for the parton-level measurement.}
\label{TTRECF2}
\end{figure}

The information on $m(\tql)$ and the mass constraints on the hadronically decaying top quark are combined in a likelihood function $\lambda$, given by
\begin{equation}
-\log[\lambda] = -\log[\Pmass(m_2, m_3)] -\log[\Plep(m(\tql))], \label{TTRECEQ1}
\end{equation}
where \Pmass is the two-dimensional probability density found from simulation of the invariant masses of the hadronically decaying \PW bosons and top quarks that are correctly reconstructed. The value of $\lambda$ is maximized to select the best permutation of jets. The probability density \Pmass is calculated as a function of the invariant mass $m_2$ of the two jets tested as the \PW boson decay products, and the invariant mass $m_3$ of the three jets tested as the decay products of the \tqh. The distributions for the correct jets, taken from the \POWHEG{}+\PYTHIA simulation and normalized to unit area, are shown in Fig.~\ref{TTRECF2}\,(\cmsleft). This part of the likelihood function is sensitive to the correct reconstruction of the \tqh. For the 2t (1t1l) category, $-\!\log[\Pmass(m_2, m_3)] < 11\:(9)$ is required. This selection rejects about 50\% of the multijet, \PW boson, and single top quark backgrounds. For higher values of $-\!\log[\Pmass(m_2, m_3)]$, almost all \ttbar events are reconstructed incorrectly and are then considered for a  boosted reconstruction.

The probability density \Plep, also found from simulation, describes the distribution of $m(\tql)$ for a correctly selected \Jbl. In Fig.~\ref{TTRECF2}\,(\cmsright) the normalized distributions of \Plep for correctly selected \Jbl and for other jets are shown. Permutations with $m(\tql) < 100$\GeV or $m(\tql) > 230$\GeV are rejected since they are very unlikely to originate from a correct \Jbl association. This part of the likelihood function is sensitive to the correct reconstruction of the \tql.

\begin{figure*}[tbp]
\centering
 \includegraphics[width=0.42\textwidth]{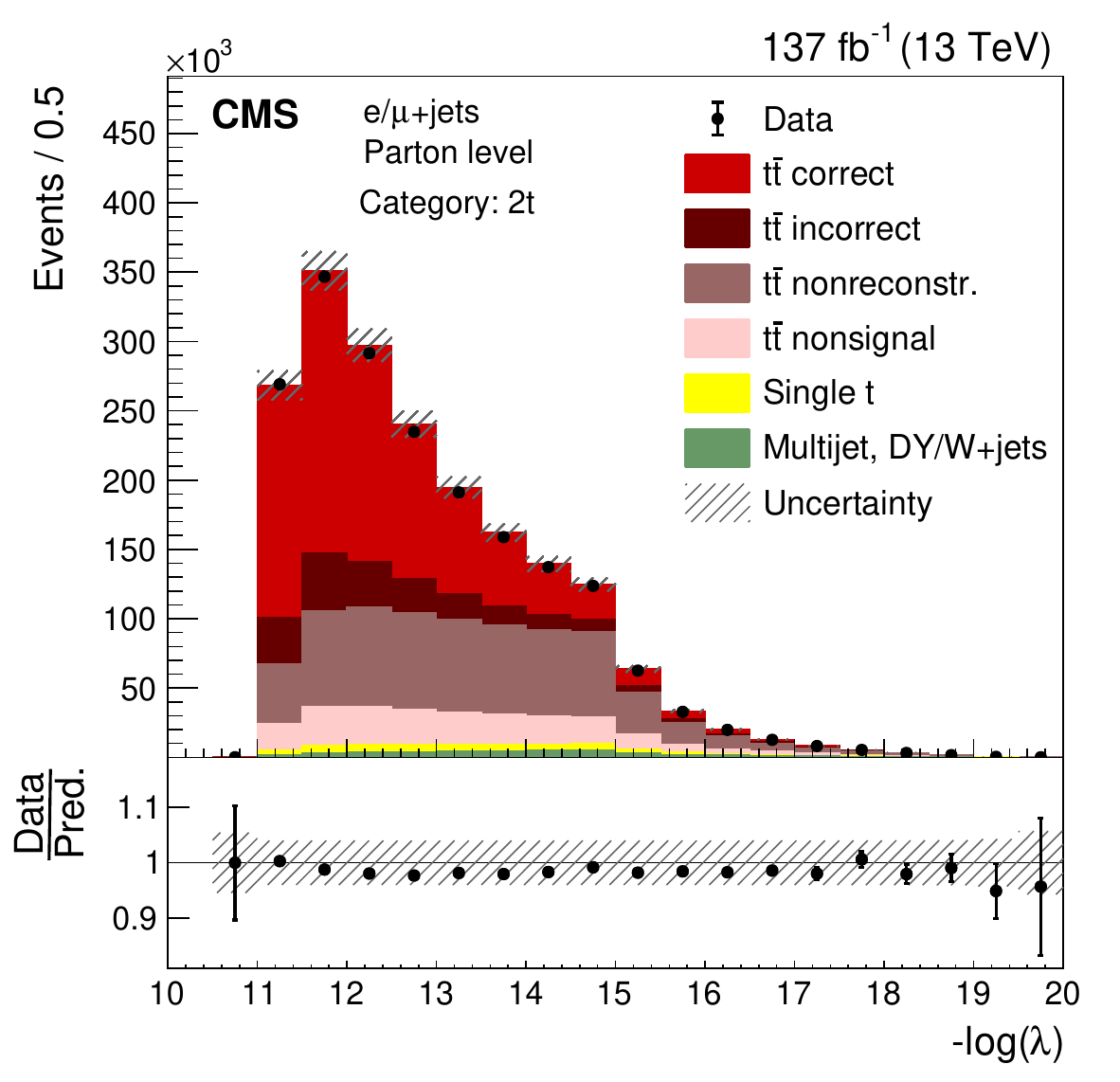}
 \includegraphics[width=0.42\textwidth]{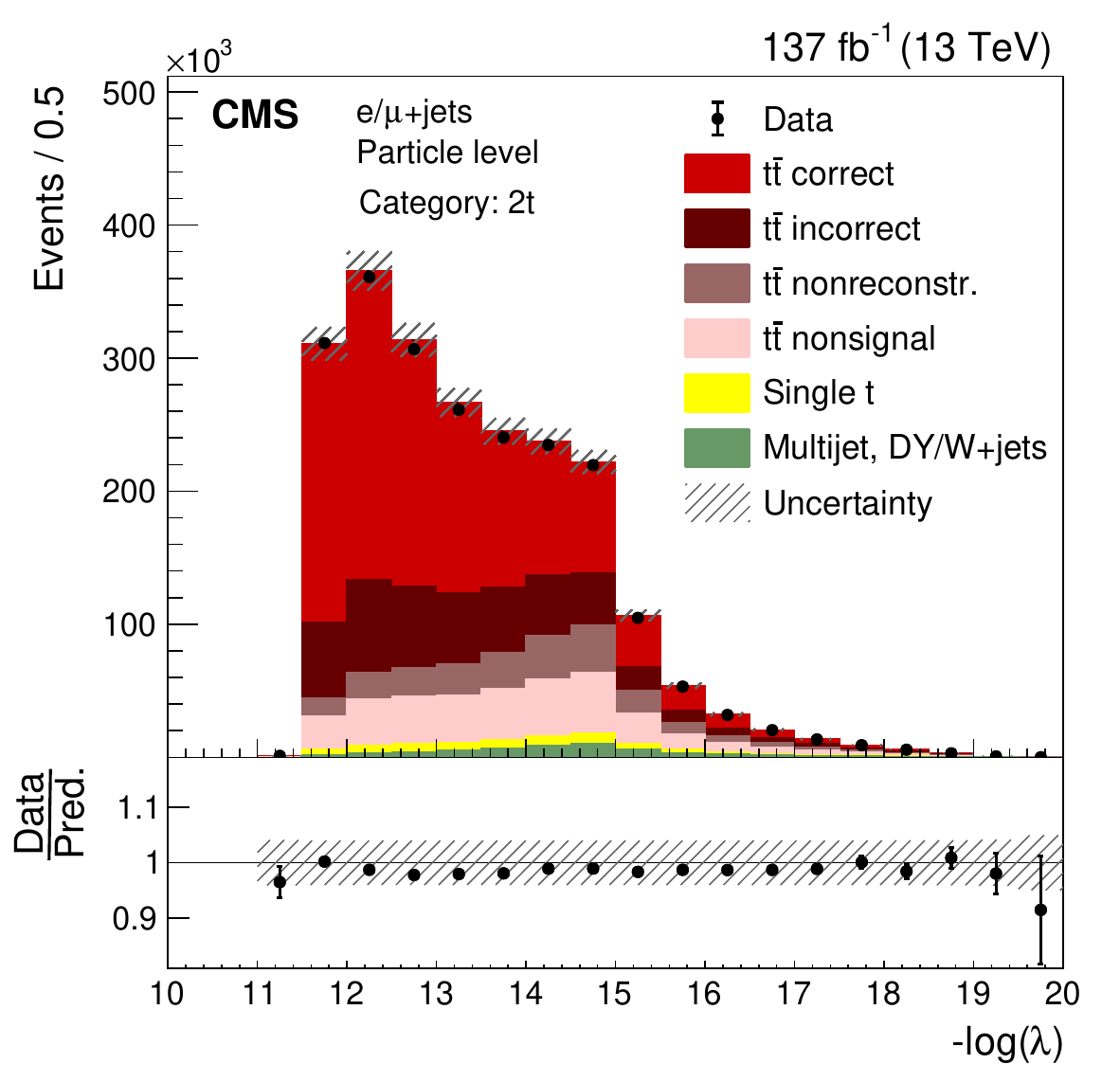}
 \includegraphics[width=0.42\textwidth]{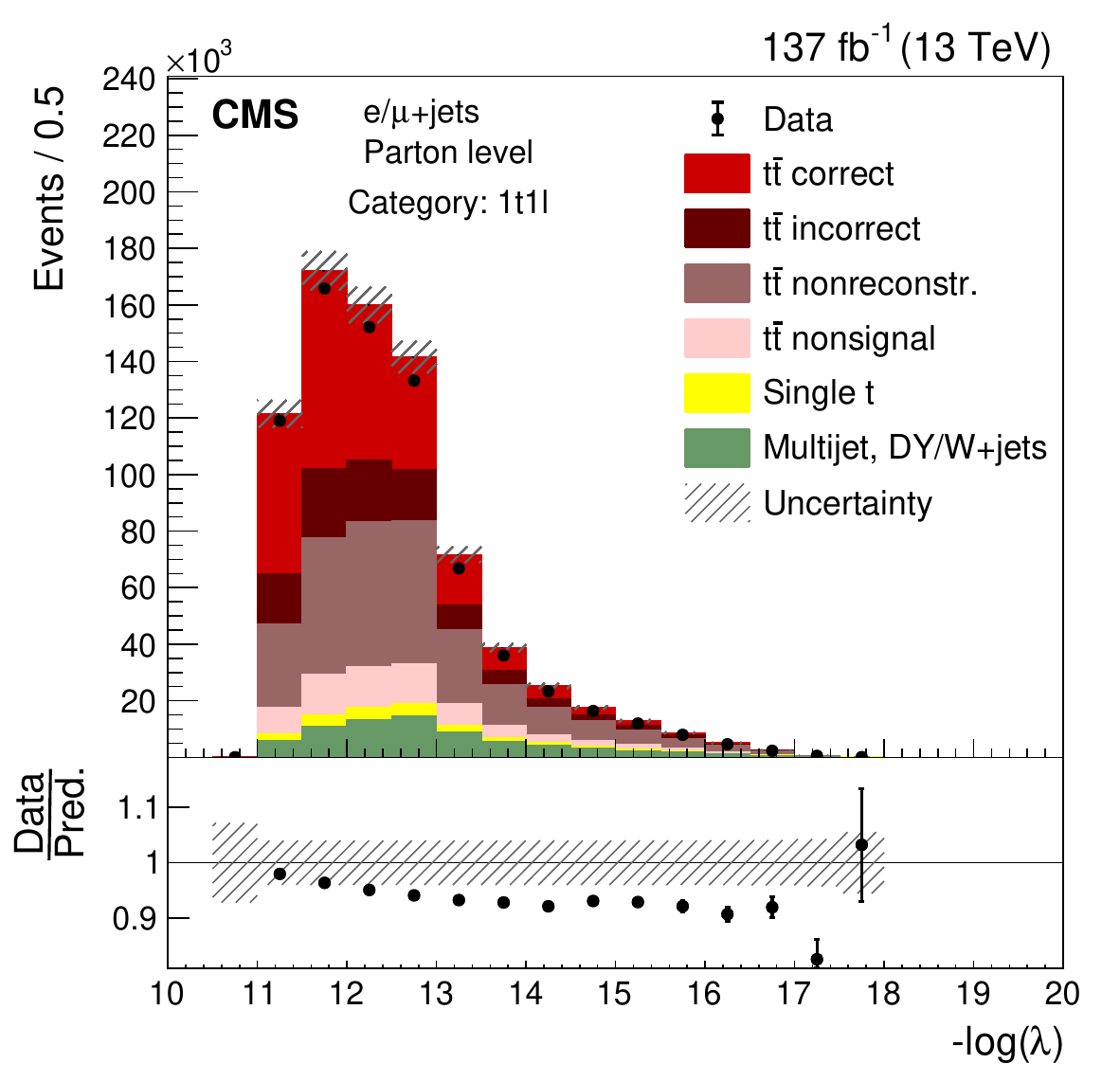}
 \includegraphics[width=0.42\textwidth]{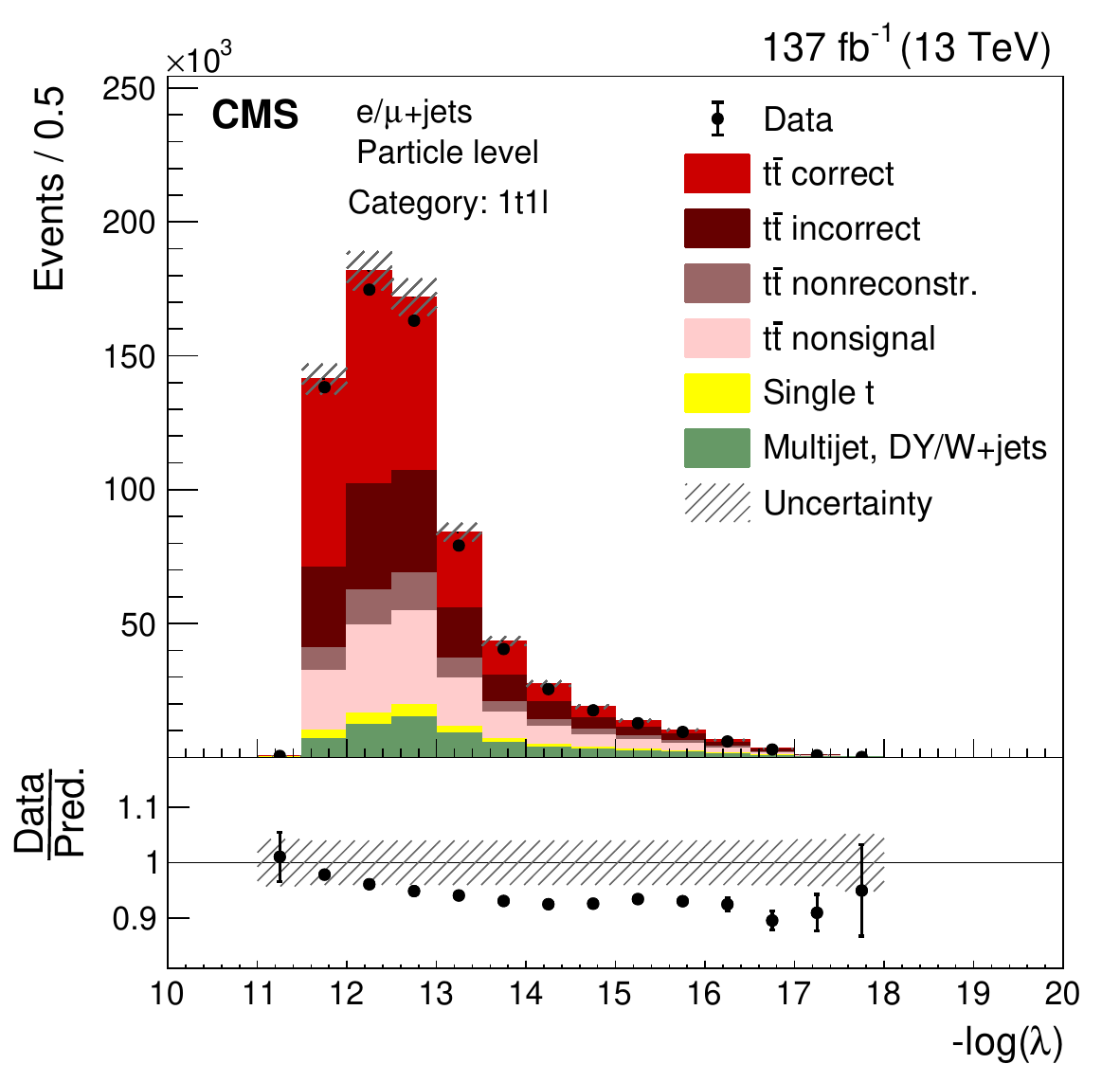}
\caption{Distributions of the negative log-likelihood for the selected best permutation in the 2t (upper) and 1t1l (lower) categories. The comparisons of data (points) and predictions (colored histograms) are shown for the (left) parton- and (right) particle-level measurements. Events generated with \POWHEG{}+\PYTHIA describe \ttbar production. The contribution of multijet, DY, and \PW boson background events is extracted from the data (cf.~Section~\ref{RESBKG}). Combined systematic (cf.~Section~\ref{SYS}) and statistical uncertainties (hatched area) are shown for the total predicted yields. The vertical bars on the points show the statistical uncertainty. The ratios of data to the sum of the predicted yields are provided in the lower panels.}
\label{TTRECF3}
\end{figure*}

The distributions of $-\!\log(\lambda)$ in data and simulation are compared in Fig.~\ref{TTRECF3} for the 2t and 1t1l categories. Here, the \ttbar simulation is split into different categories based on the success of the reconstruction. The category ``\ttbar correct'' contains the events with all decay products correctly identified, ``\ttbar incorrect'' are incorrectly reconstructed but all decay products are available, and ``\ttbar nonreconstructable'' are events with at least one missing decay product caused by detector inefficiencies or acceptance losses. Finally, ``\ttbar nonsignal'' events are \ttbar background events, \ie, having no parton- or particle-level \ttbar pair in the desired decay channel and phase space. 

\section{Identification and reconstruction of boosted leptonically decaying top quarks}
\label{LEPBOOST}
For $\pt(\tql) > 400$\GeV the separation in $\eta$ and $\phi$ between the charged lepton and the \PQb jet becomes increasingly small and the isolation cone for the lepton of $\Delta R = 0.4$ starts overlapping with the constituents of the \PQb jets. In this case, the leptons do not fulfill the isolation criterion. Therefore, nonisolated leptons are used for the reconstruction of the boosted \tql. We look for a loosely \PQb-tagged AK4 jet within $\Delta R = 0.8$ around such a lepton. If the lepton is a constituent of the jet, its momentum is subtracted from the jet momentum to avoid double counting. Afterwards, $p_\Pgn$ is calculated using the \MW mass constraint as in the resolved reconstruction described in Section~\ref{TTREC}. The momentum of the boosted $\tql$ is calculated as the momentum sum of the lepton, \PQb jet, and neutrino. If $\pt(\tql) > 400$\GeV, we consider this as a boosted \tql candidate. The transverse and longitudinal momentum resolution is about 10\% in the selected \pt range.

The reconstructed candidates might be jets containing a misreconstructed lepton, a lepton from a hadron decay, or from a leptonically decaying \PW boson produced within the jet. Based on the following variables, we use an NN to discriminate between a signal \tql and background:
 the invariant mass of the lepton plus \PQb jet system $m(\ell,\:\PQb\text{ jet})$,
the ratios $m_{\PQb\text{ jet}}/m(\ell,\:\PQb\text{ jet})$, where $m_{\PQb\text{ jet}}$ is the invariant mass of the \PQb jet,
$\pt(\ell)/\pt(\ell+\PQb\text{ jet})$,
$I_\mathrm{far}/I_0$, and
$I_\mathrm{near}/I_0$, with the isolation variables $I = \sum_i \pt(o_i) \Delta R^{q}(\ell,\:o_i)$ and $q = -2, 0, 2$ for $I_\mathrm{near}$, $I_0$, and $I_\mathrm{far}$, respectively, where $o_i$ includes all charged, neutral, and photon-like PF objects within $\Delta R = 0.4$ around the lepton. 

Separate NNs are used for $400 < \pt(\tql) < 650$\GeV and $\pt(\tql) > 650$\GeV. The muon and electron channels are combined. The NNs consist of a five-node input layer, three fully connected hidden layers with 20, 10, and 5 nodes, and a single output node. The activation function is hyperbolic tangent in all layers. The logistic loss function is minimized using stochastic gradient descent with the Adam algorithm~\cite{adam}. Training samples of 2000 \ttbar and 2000 multijet, DY, and \PW boson background events are used.

\begin{figure}[tbp]
\begin{center}
 \includegraphics[width=0.42\textwidth]{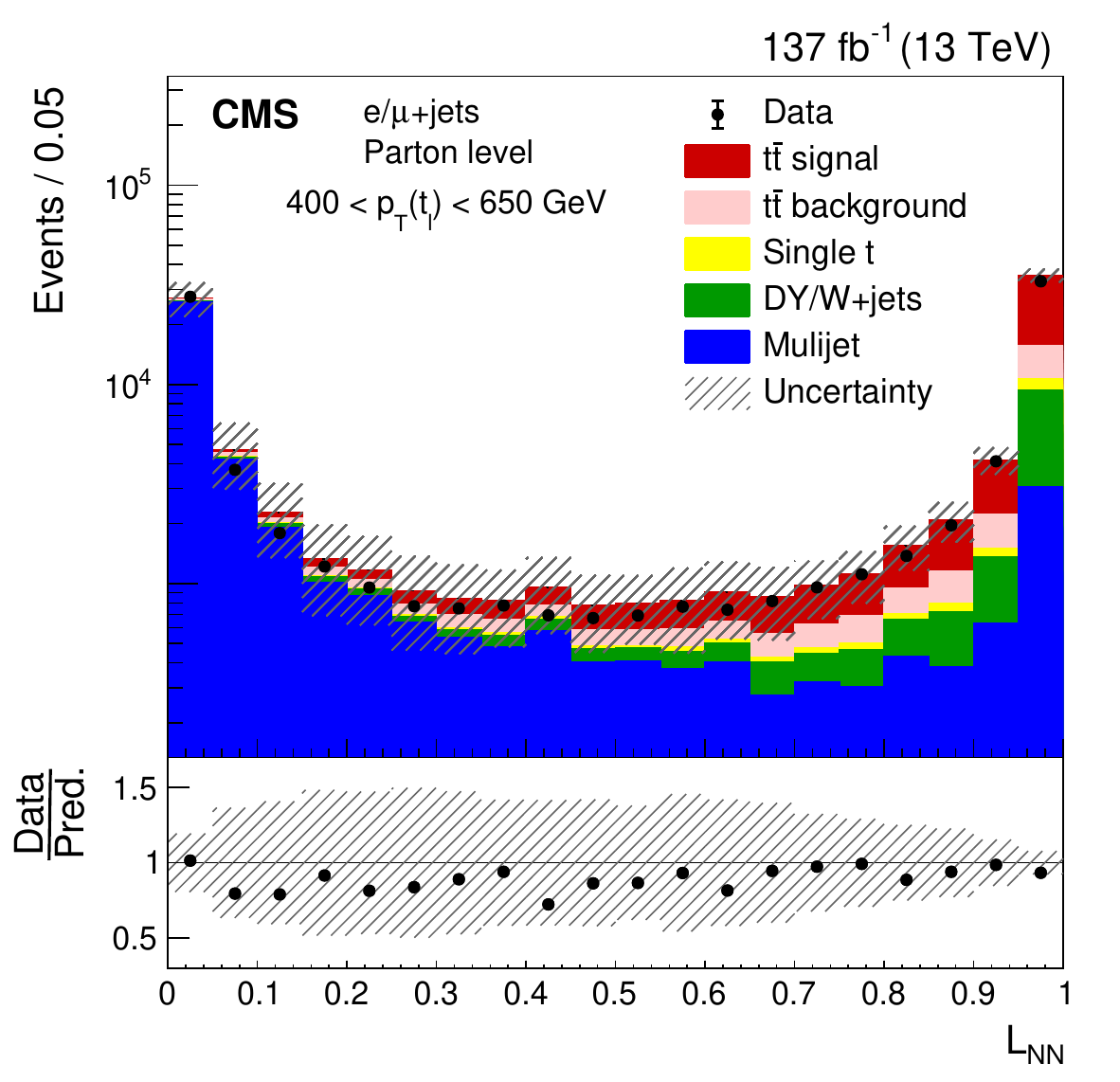}
 \includegraphics[width=0.42\textwidth]{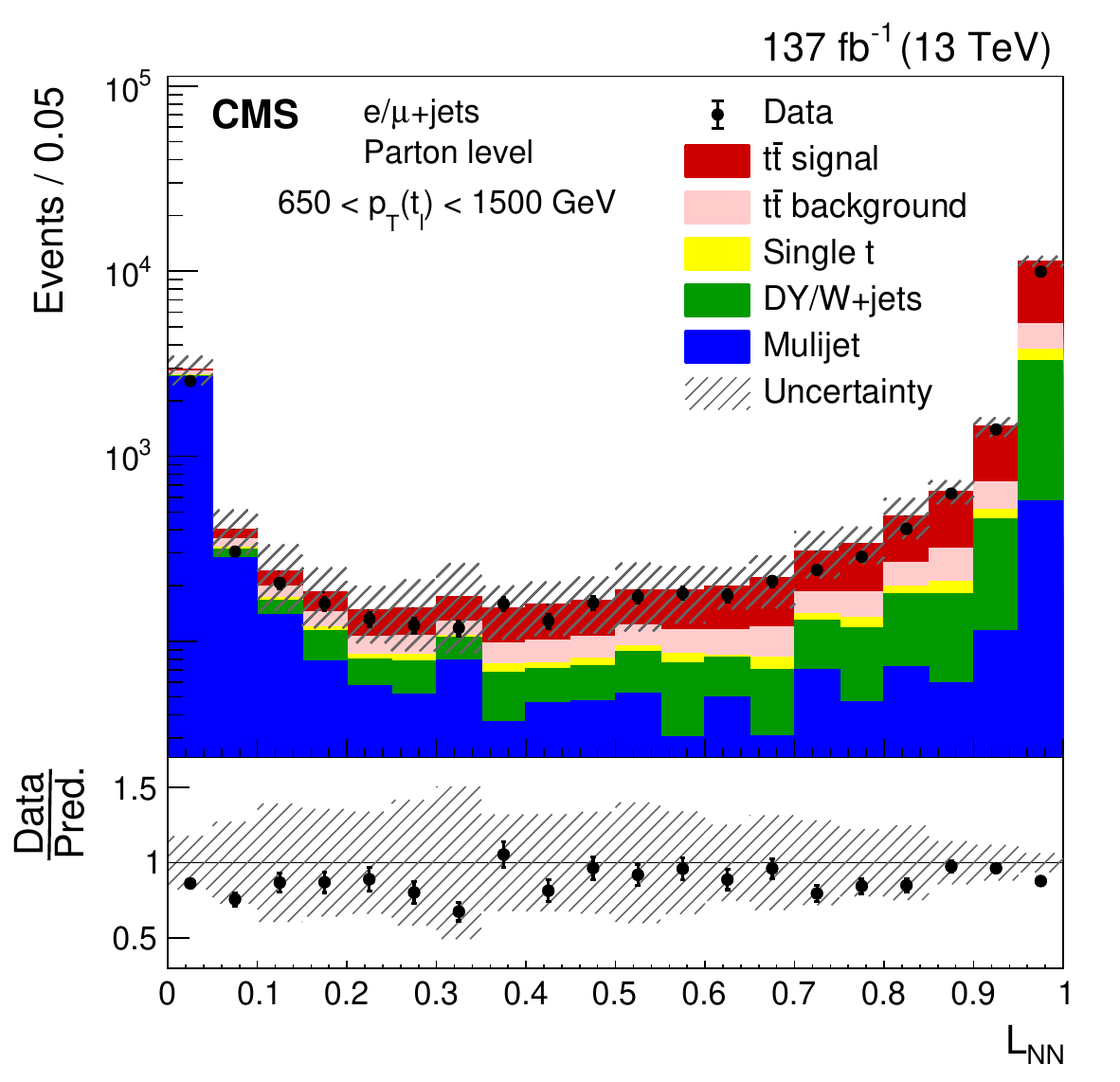}
\end{center}
 \caption{Distributions of the output discriminant \LNN used in the boosted \tql identification for the low-\tleppt (\cmsleft) and high-\tleppt (\cmsright) regions. The data (points) and predictions from simulation (colored histograms) are shown. The hatched area gives the combined statistical and systematic uncertainties in the prediction. The vertical bars on the points show the statistical uncertainty in the data. The ratio of the data to the sum of the individual predictions is displayed in the lower panels.}
 \label{fig:LEPBOOST_3}
\end{figure}

The distribution of the output variable \LNN of the NN is shown in Fig.~\ref{fig:LEPBOOST_3}. The largest uncertainties in the distribution of \LNN are those from the PS and the UE tune, since these affect the isolation variables. The discriminant \LNN is efficient in reducing the multijet background, whereas \PW bosons decaying into leptons and produced within a jet are more difficult to distinguish from the boosted \tql candidate. We select \tql candidates with $\LNN > 0.7$. The efficiency of this selection has been studied using \ttbar events in the $e\mu$ decay channel with one isolated and one nonisolated lepton and has a signal efficiency of about 90\% over the entire \pt range.  

\section{Identification and reconstruction of boosted hadronically decaying top quarks}
\label{HADBOOST}

All selected AK8 jets are considered as \tqh candidates. The pileup subtraction based on individual jet constituents, as performed with the \PUPPI algorithm, results in a significant improvement in the reconstruction of the jet substructure used for the identification of the boosted \tqh. To discriminate the candidates containing decay products of a \tqh from other AK8 jets, several properties are combined using an NN. Most of these quantities are calculated after boosting the jet constituents into their center-of-mass system and clustering them with the anti-\kt jet algorithm with a distance parameter of 0.5 to obtain subjets. Since these subjets are clustered in the jet center-of-mass frame rather than in the laboratory frame, the energy and the angle between the objects are used in the jet definition instead of \pt and $\Delta R$, which are invariant under longitudinal boosts. This boost into the center-of-mass system separates the top quark decay products and helps to identify the typical pattern of the three jets. The following NN input variables are used: 

\begin{itemize}
 \item the invariant masses of all combinations of two subjets are calculated, and the three highest values are used;
 \item the number of combinations of two subjets with invariant masses exceeding 40\GeV;
 \item the invariant masses of all combinations of three subjets are calculated, and the two highest values are used;
 \item the ratio of the highest invariant mass of three subjets to the invariant mass of all constituents;
 \item {\tolerance=8000 the ratios of $N$-jettiness~\cite{PhysRevLett.105.092002} $\tau_{2}/\tau_{1}$, $\tau_{3}/\tau_{2}$, $\tau_{4}/\tau_{3}$, and $\tau_{5}/\tau_{4}$ with $\tau_{N} = \sum_{k} \min(q_{1} p_{k},\:q_{2} p_{k},\:... ,\:q_{N}p_{k})$, where $q_{i}$ with $1 \le i \le N$ are the momenta of the $N$ leading subjets and $p_{k}$ are the momenta of all constituents in the jet rest frame;\par}
 \item the energies of the four most-energetic subjets;
 \item the value of $\abs{(\vec{p}_1 \times \vec{p}_2)\cdot \vec{p}_3}$, where $\vec{p}_1$, $\vec{p}_2$, and $\vec{p}_3$ are the three-momenta of the most-energetic subjets normalized to unity;
 \item the sphericity~\cite{PhysRevD.1.1416} $s = \frac{3}{2}(\lambda_2 + \lambda_3)$ of all subjets, with $\lambda_2$ and $\lambda_3$ the second- and third-highest eigenvalues of the tensor
\begin{equation}
S^{\alpha \beta} = \frac{\sum_i p_i^\alpha p_i^{\beta} / \abs{p_i}} {\sum_i \abs{p_i}},
\end{equation}
 where $p_i$ are the momenta of the subjets in the jet rest frame and $\alpha, \beta$ are the spatial indices;
 \item the three highest-momentum subjets are boosted back to the laboratory frame and their momentum fractions relative to the AK8 jet are calculated.
\end{itemize}
There are always at least three subjets in the \tqh candidates. Variables relying on having four subjets are set to zero in the rare cases when only three subjets are found. 

We use separate NNs for each of four different \thadpt regions: 400--500, 500--700, 700--1000, and ${>}1000$\GeV. The NNs consist of a 21-node input layer, four fully connected hidden layers with 63, 42, 42, and 21 nodes, and a single output node. We found that adding more nodes or layers does not improve the discriminating performance of the NN. The activation function is hyperbolic tangent in all layers. The logistic loss function is minimized using stochastic gradient descent with the Adam algorithm. Training samples of 100\,000 \ttbar signal and 100\,000 \ttbar background jets are used. The output discriminant is referred to as \HNN.  

\begin{figure}[tbp]
\centering
 \includegraphics[width=0.5\textwidth]{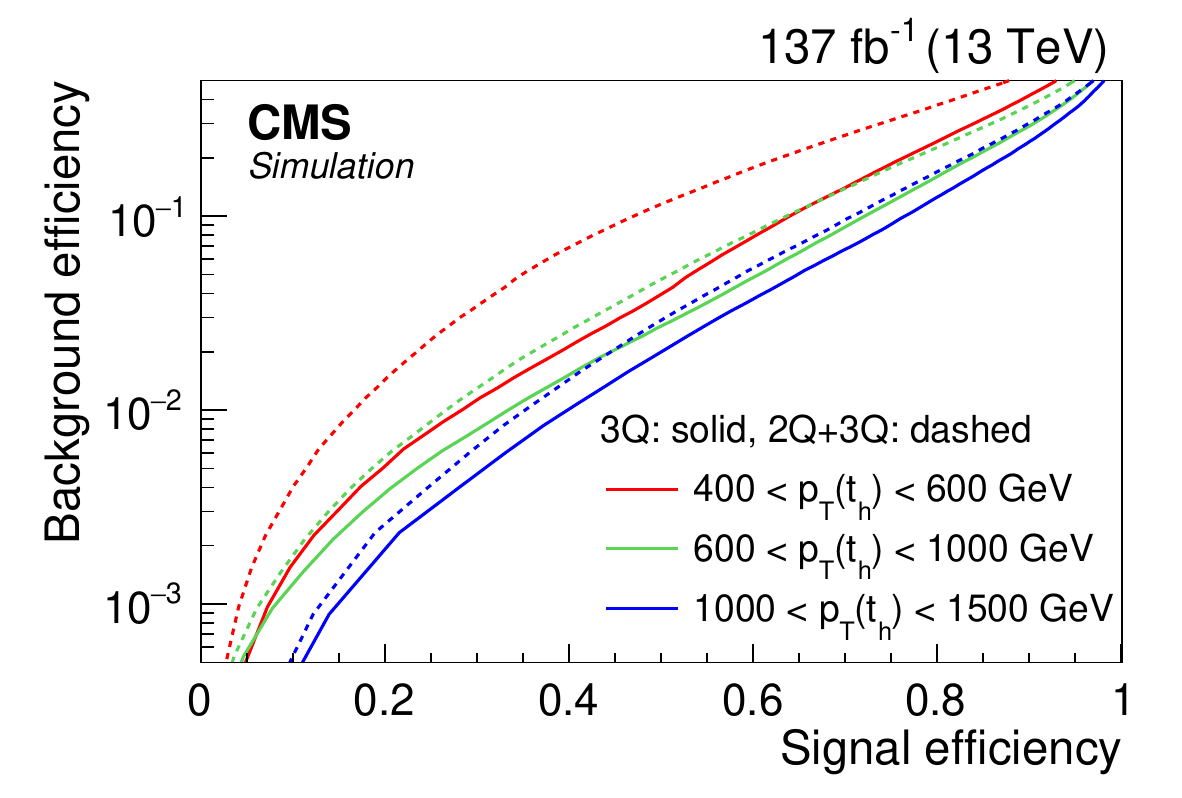}
 \caption{The selection efficiency for background jets as a function of the signal selection efficiency in three \thadpt ranges for 3Q (solid curves) and 2Q+3Q (dashed curves) jets from simulation. An efficiency of 100\% corresponds to the preselection of $\abs{\eta} < 2.4$ and $m_\mathrm{jet} > 120$\GeV.}
 \label{fig:HADBOOST1}
\end{figure}

We consider as signal two types of top quark candidates, labeled 2Q and 3Q, where two and three of the quarks from the \tqh decay are within $\Delta R = 0.8$, respectively. The 2Q candidates represent a significant contribution for $400<\pt<600$\GeV. These are, in general, more difficult to distinguish from background, but they do not provide a significantly worse momentum measurement of the top quarks than 3Q candidates. In Fig.~\ref{fig:HADBOOST1} the performance of the NN is demonstrated, showing the background \vs signal selection efficiency for three different \thadpt ranges.

The definition of 2Q and 3Q candidates is based on the spatial matching of quarks to a jet, which is ambiguous. However, this definition is useful since the properties of two or three comparable hard structures within a jet allow us to identify a boosted \tqh using the NN. We test the effect of the ambiguity by repeating the cross section measurement with $\Delta R = 0.7$ and $0.9$ in the definition. Although this changes the number of events considered as signal and background, in combination with the altered response matrices, very similar cross sections are obtained and their differences are negligible compared with other uncertainties.  
 
\section{Event reconstruction and categorization}
\label{EVRECO}

Depending on the event content, the various reconstruction algorithms are used, and the events are categorized into exclusive categories. The first attempt is the resolved reconstruction described in Section~\ref{TTREC}. If an event passes the required selection criteria of the 2t category, no further reconstruction is tried. The same is true for 1t1l, but in these events no boosted \tqh candidate must exist. If a boosted \tqh candidate exists, the boosted reconstruction methods are used on the event. 

If the event is not categorized as 2t or 1t1l, but an isolated lepton, at least one \PQb jet fulfilling the tight criterion, and at least one boosted \tqh candidate are found, the event falls into the category BHRL (boosted \tqh, resolved \tql). The \tql is reconstructed using the lepton and \PQb jet for which $\Delta R(\ell,\:\PQb\text{ jet})$ is minimal. The neutrino momentum is determined using the \MW mass constraint introduced in Section~\ref{TTREC}. If such a \tql can be reconstructed, all the \tqh candidates with $\Delta R(\tqh,\:\ell) > 1.2$ and $\Delta R(\tqh,\:\PQb\text{ jet}) > 1.2$, are considered. The event is counted once for each available \tqh candidate. Later, a fit of the \HNN distribution is performed to measure the signal yields.

If no isolated lepton exists but at least one boosted \tql candidate with $\LNN > 0.7$ is found, the boosted \tql candidate with the highest value of \LNN is used. The event is counted once for each boosted \tqh candidate with $\Delta R(\tql,\:\tqh) > 1.2$. We refer to this category as BHBL (boosted \tqh, boosted \tql). After combining with the BHRL category, the signal fraction is extracted from a fit of the \HNN distribution. The fitted signal yields corresponds to the number of signal events, since the definition of the boosted \tqh signal does not allow for the presence of more than one signal candidate per event.

The category consisting of boosted \tql and resolved \tqh candidates is not used, since the fraction of these events is small. In addition, we find a low fraction of correctly reconstructed \ttbar events in this category. Therefore, we consider these events as nonreconstructable. 

\begin{figure*}[tbp]
\centering
 \includegraphics[width=0.6\textwidth]{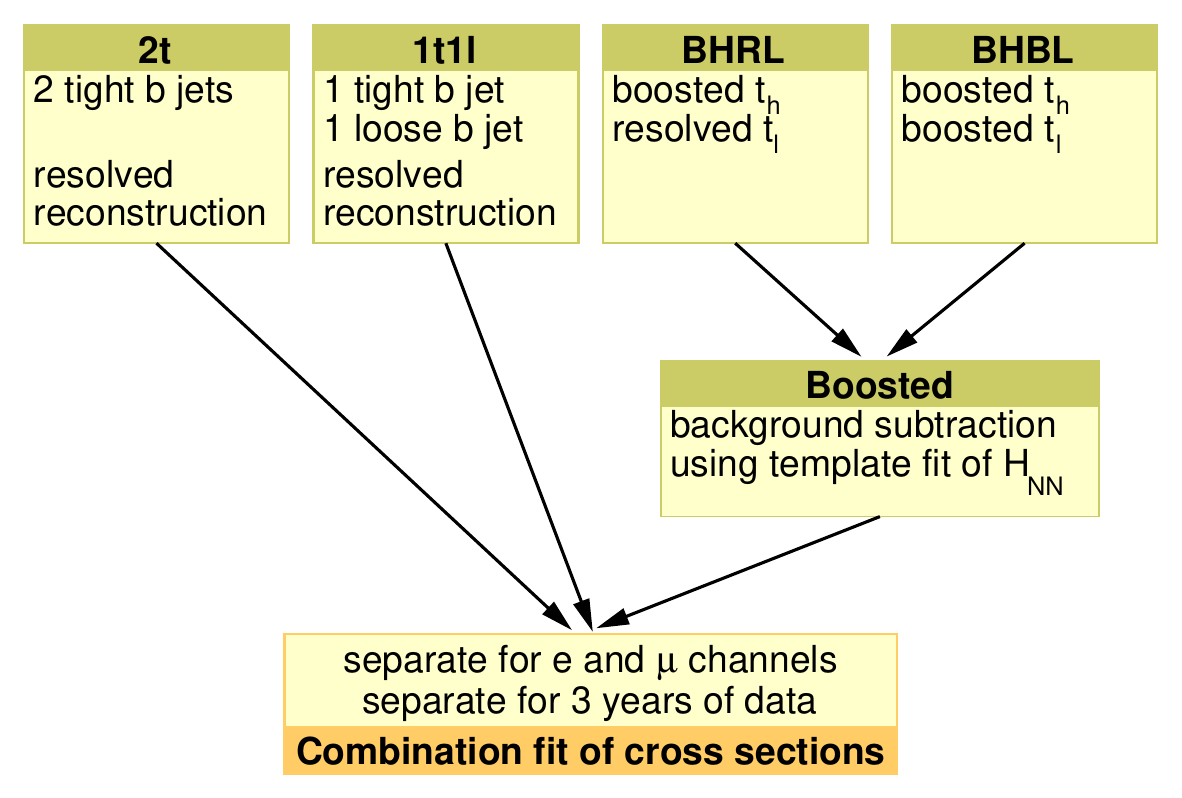}
 \caption{Schematic overview of the categories and how they are used in the analysis.}
 \label{fig:OVERVIEW}
\end{figure*}

\begin{figure*}[tbp]
\centering
 \includegraphics[width=0.4\textwidth]{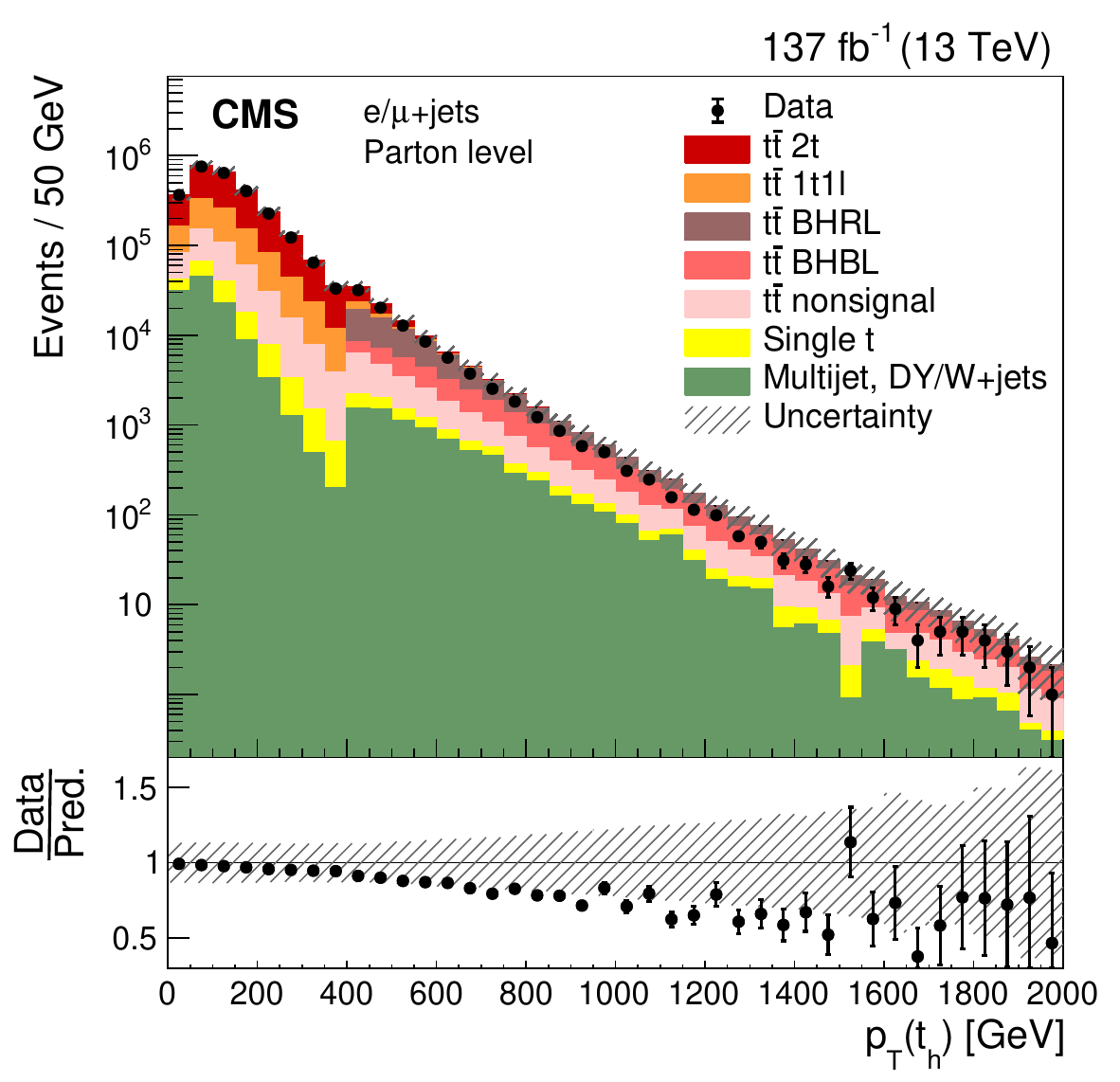}
 \includegraphics[width=0.4\textwidth]{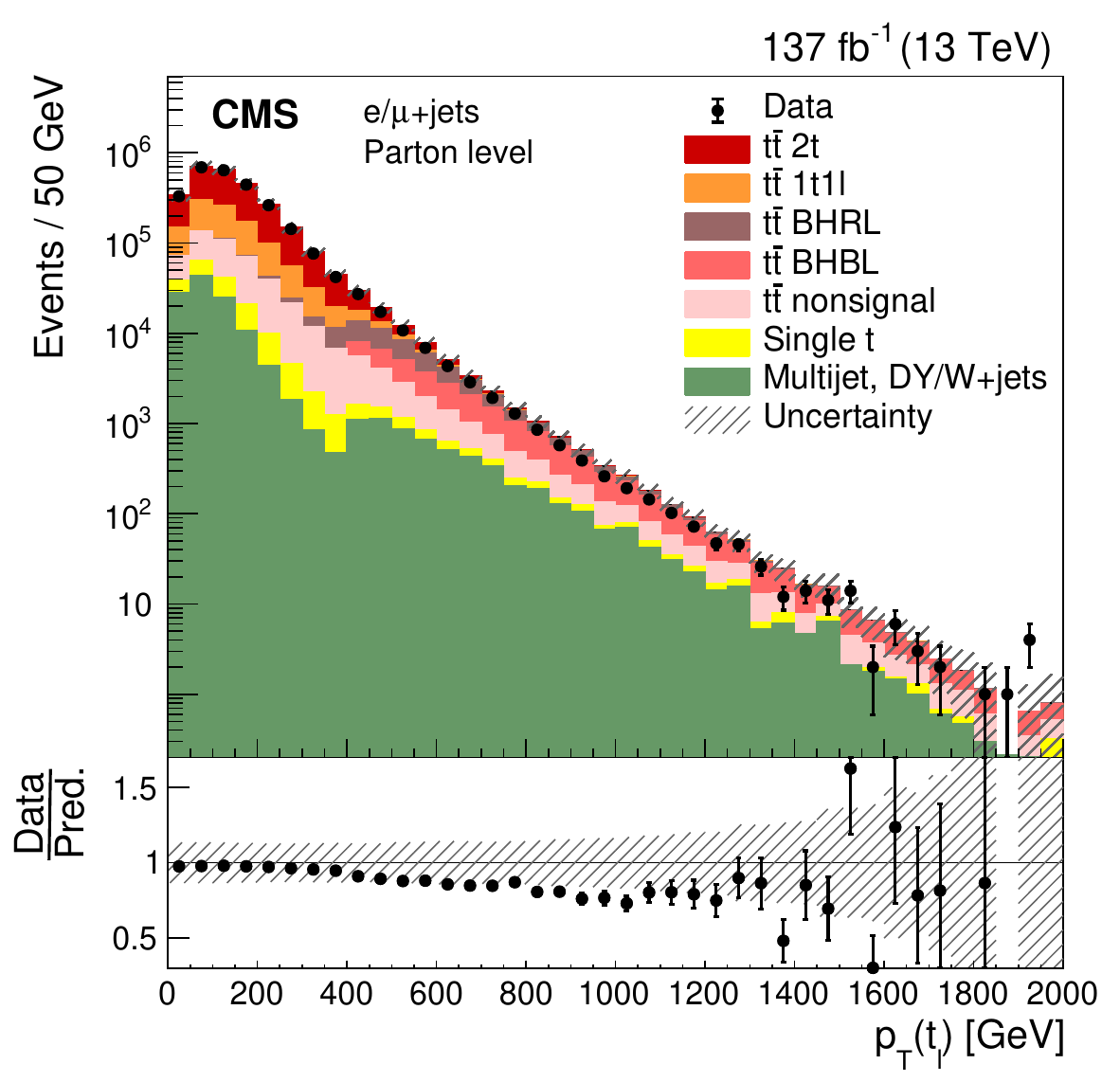}
 \includegraphics[width=0.4\textwidth]{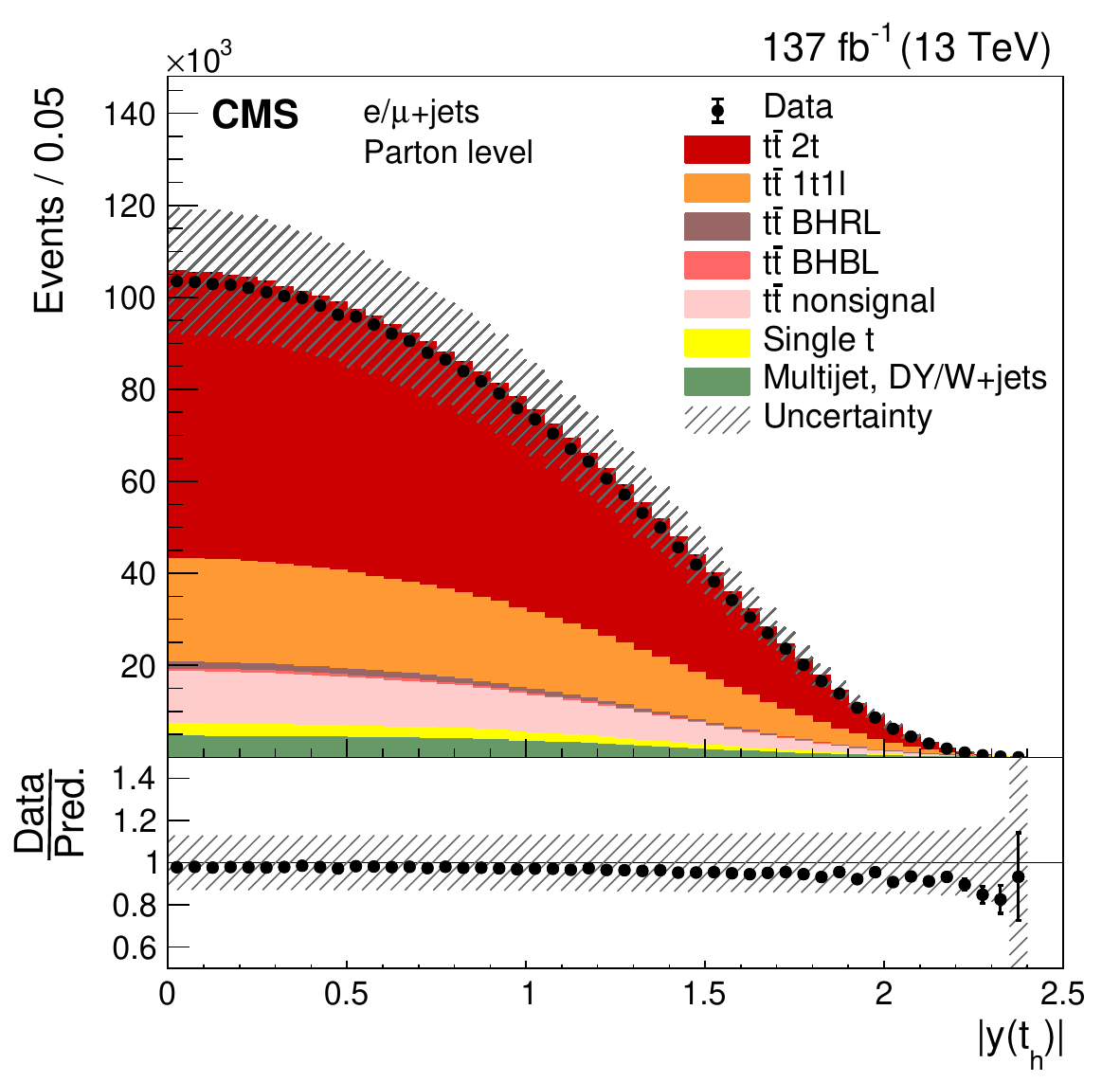}
 \includegraphics[width=0.4\textwidth]{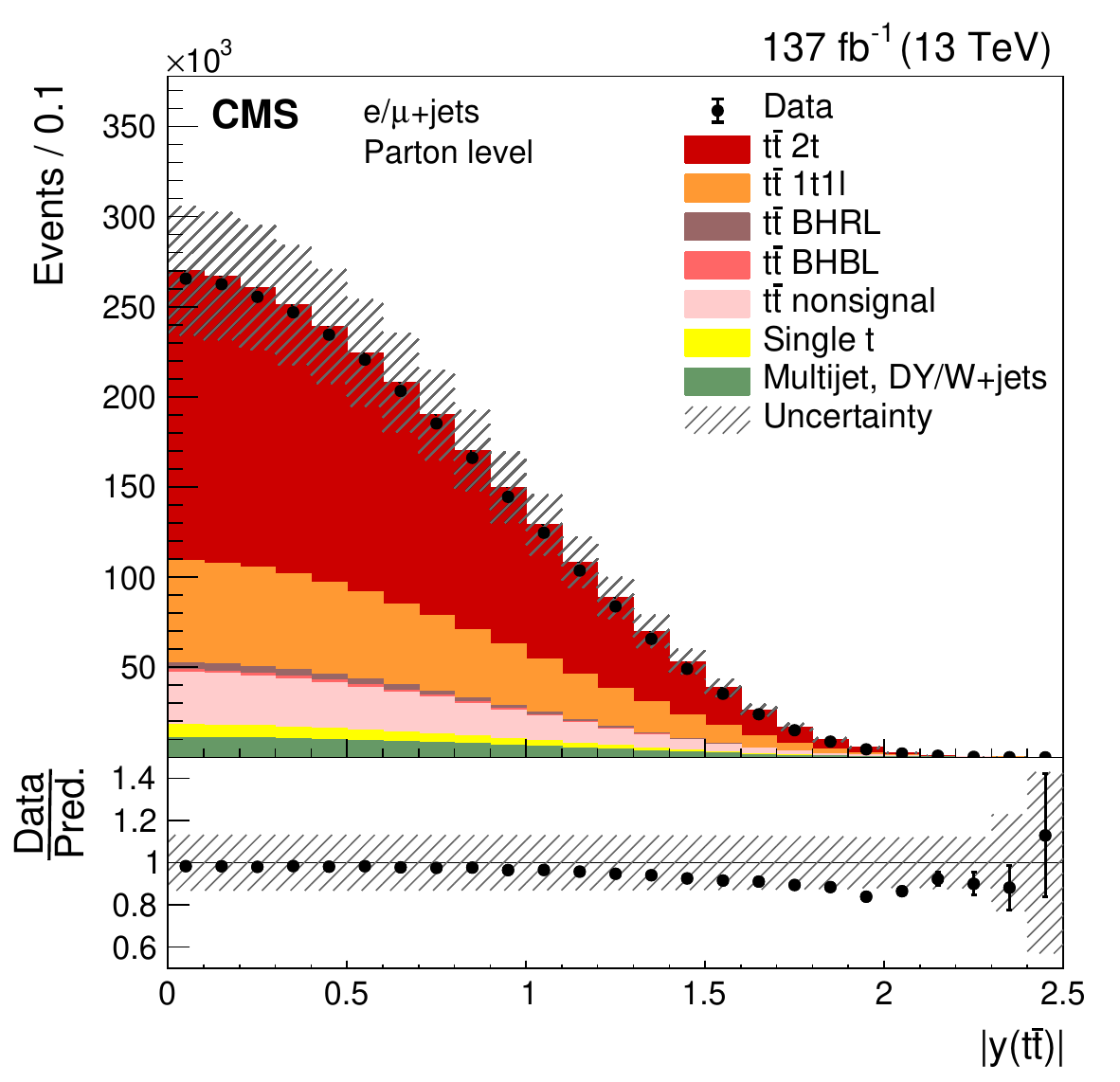}
 \includegraphics[width=0.4\textwidth]{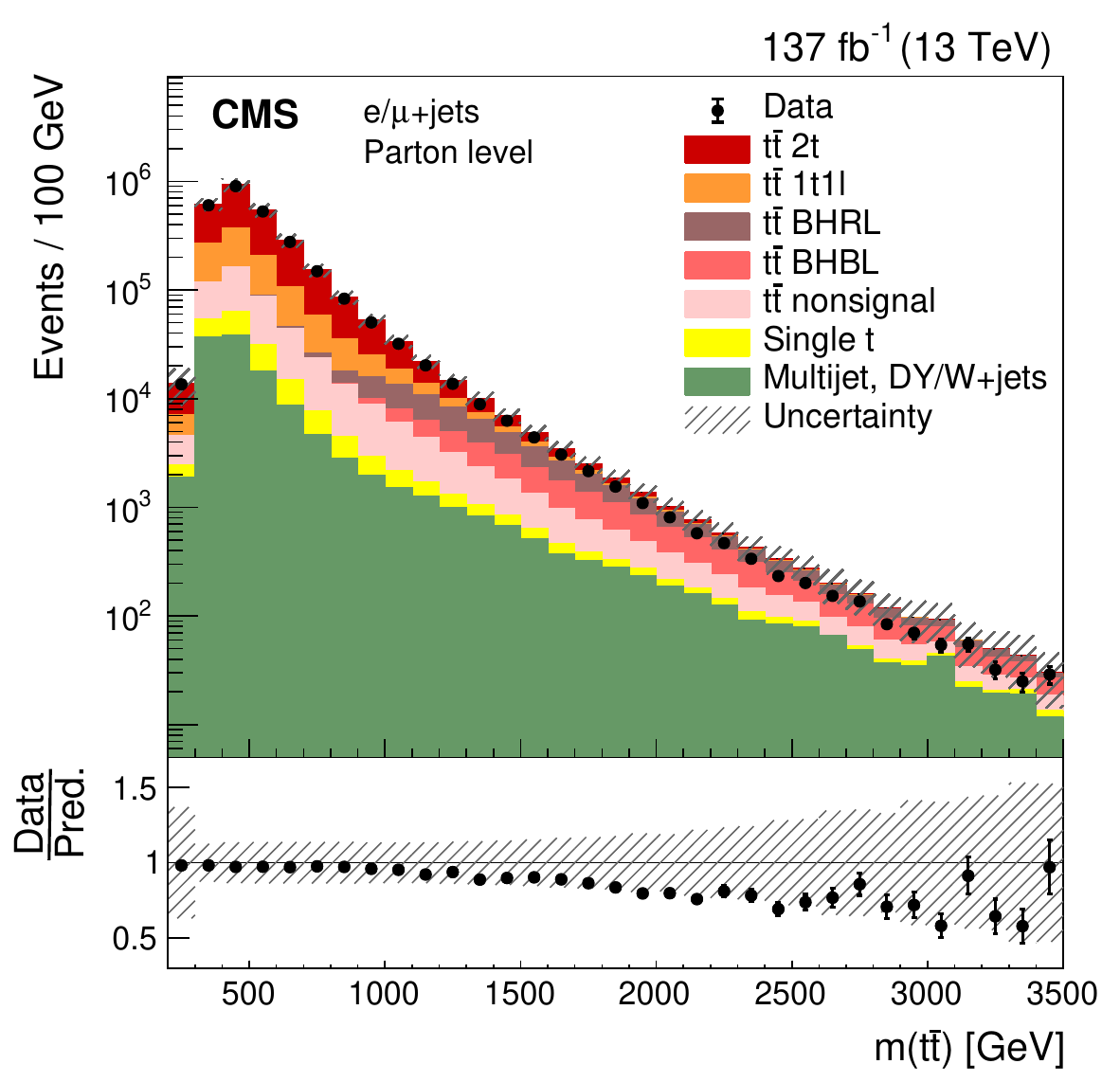}
 \includegraphics[width=0.4\textwidth]{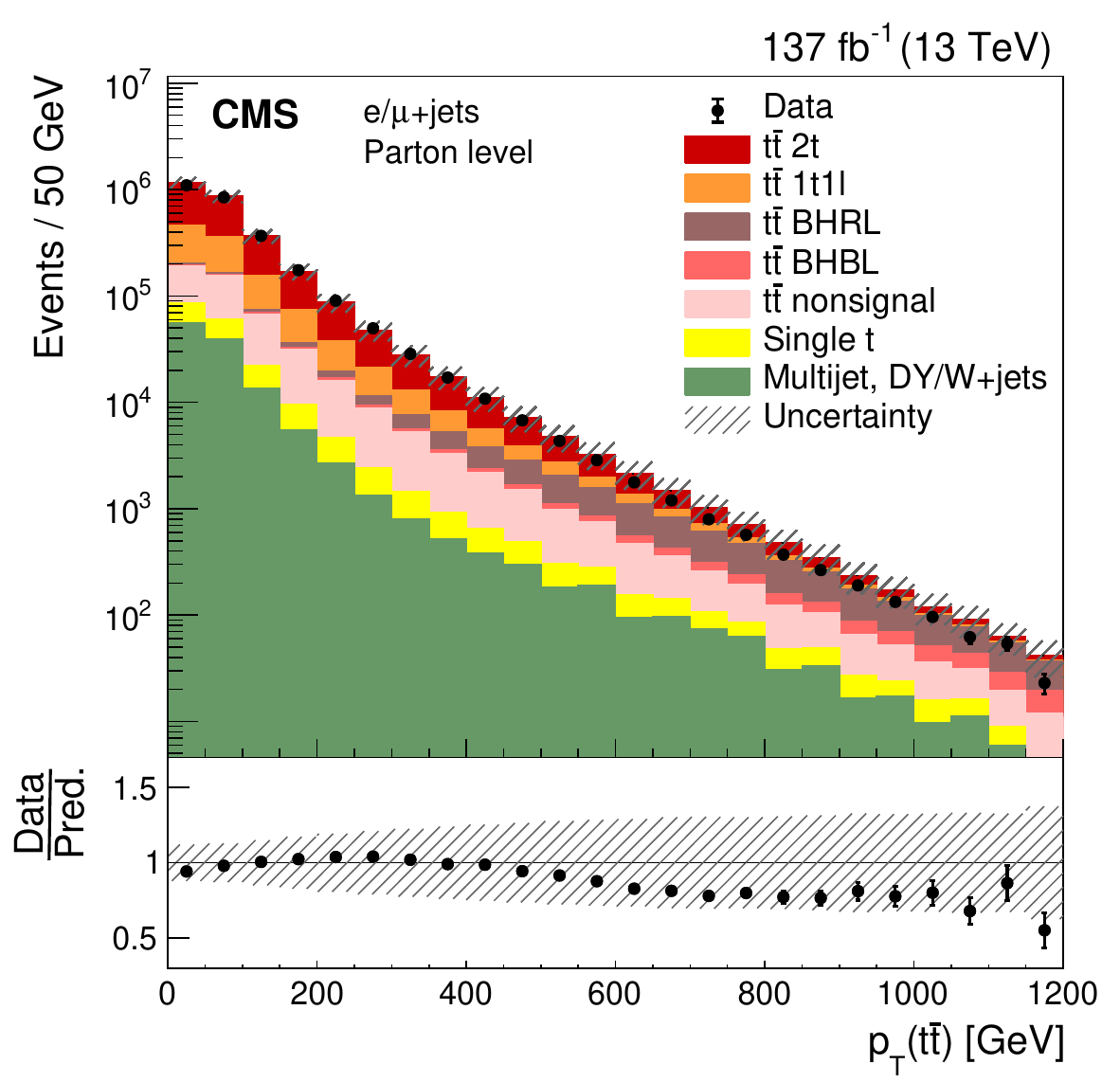}
 \caption{Comparisons of various reconstructed kinematic distributions between data (points) and predictions (colored histograms) obtained for the parton-level measurement. Contributions of the various reconstruction categories are obtained from the \POWHEG{}+\PYTHIA{} \ttbar simulation. The contribution of multijet, DY, and \PW boson background events in the 2t and 1t1l categories are extracted from the data (cf.~Section~\ref{RESBKG}). All other background contributions are taken from simulation. Combined systematic (cf.~Section~\ref{SYS}) and statistical uncertainties (hatched area) are shown for the total predicted yields. The data points are shown with statistical uncertainties. The ratios of data to the sum of the predicted yields are provided in the lower panels.}
 \label{fig:EVRECO1}
\end{figure*}

An overview of the categories and how they are used in the course of the analysis is presented in Fig.~\ref{fig:OVERVIEW}. In Fig.~\ref{fig:EVRECO1} the contributions of the various reconstruction categories to the distributions of several kinematic variables are shown, and the predicted yields are compared to the data. The distributions are given for the parton-level measurements, but look very similar for the particle-level results.

\section{Background subtraction in the resolved categories}
\label{RESBKG}
\begin{figure}[tbp]
\centering
 \includegraphics[width=0.42\textwidth]{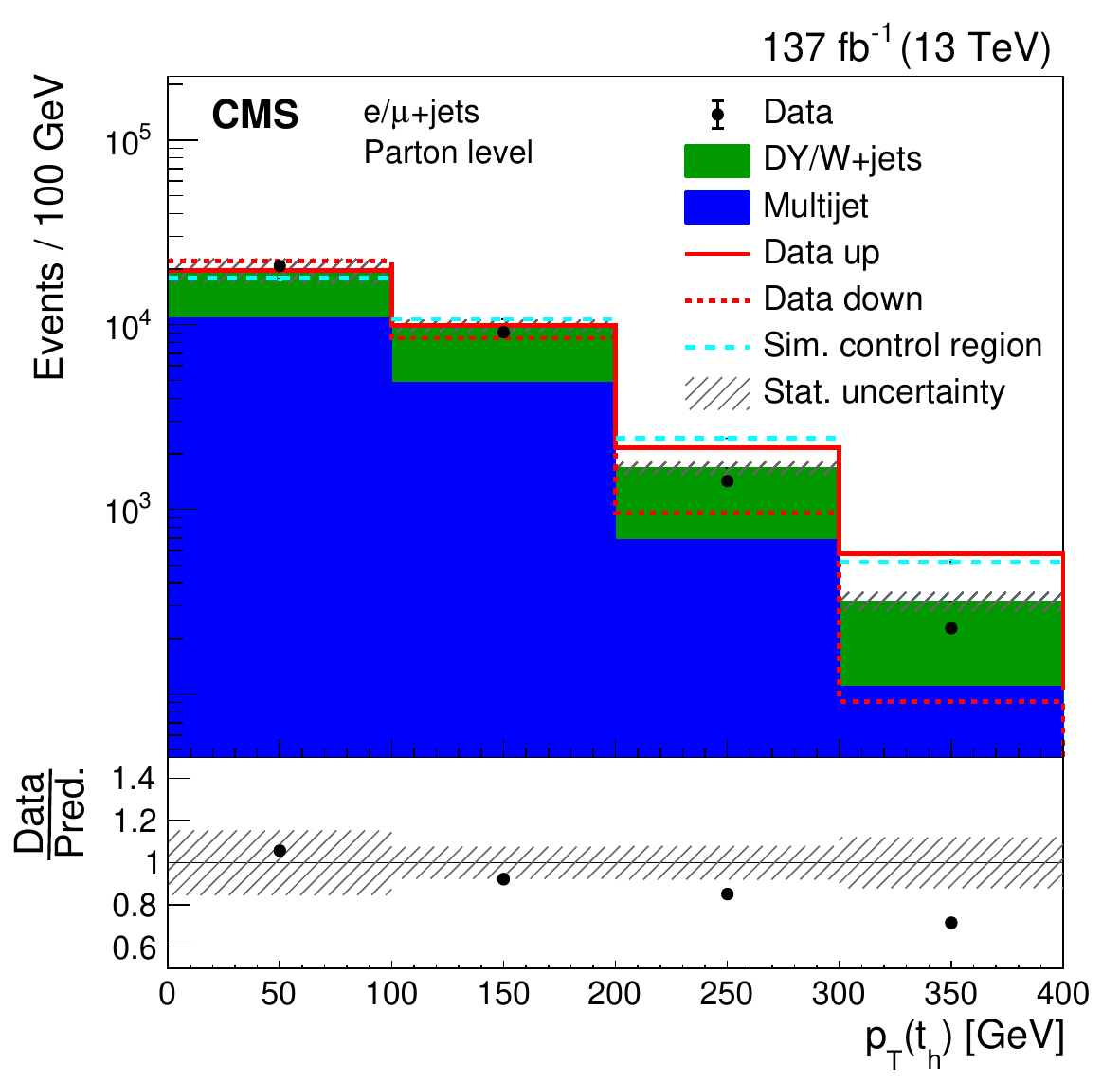}
 \includegraphics[width=0.42\textwidth]{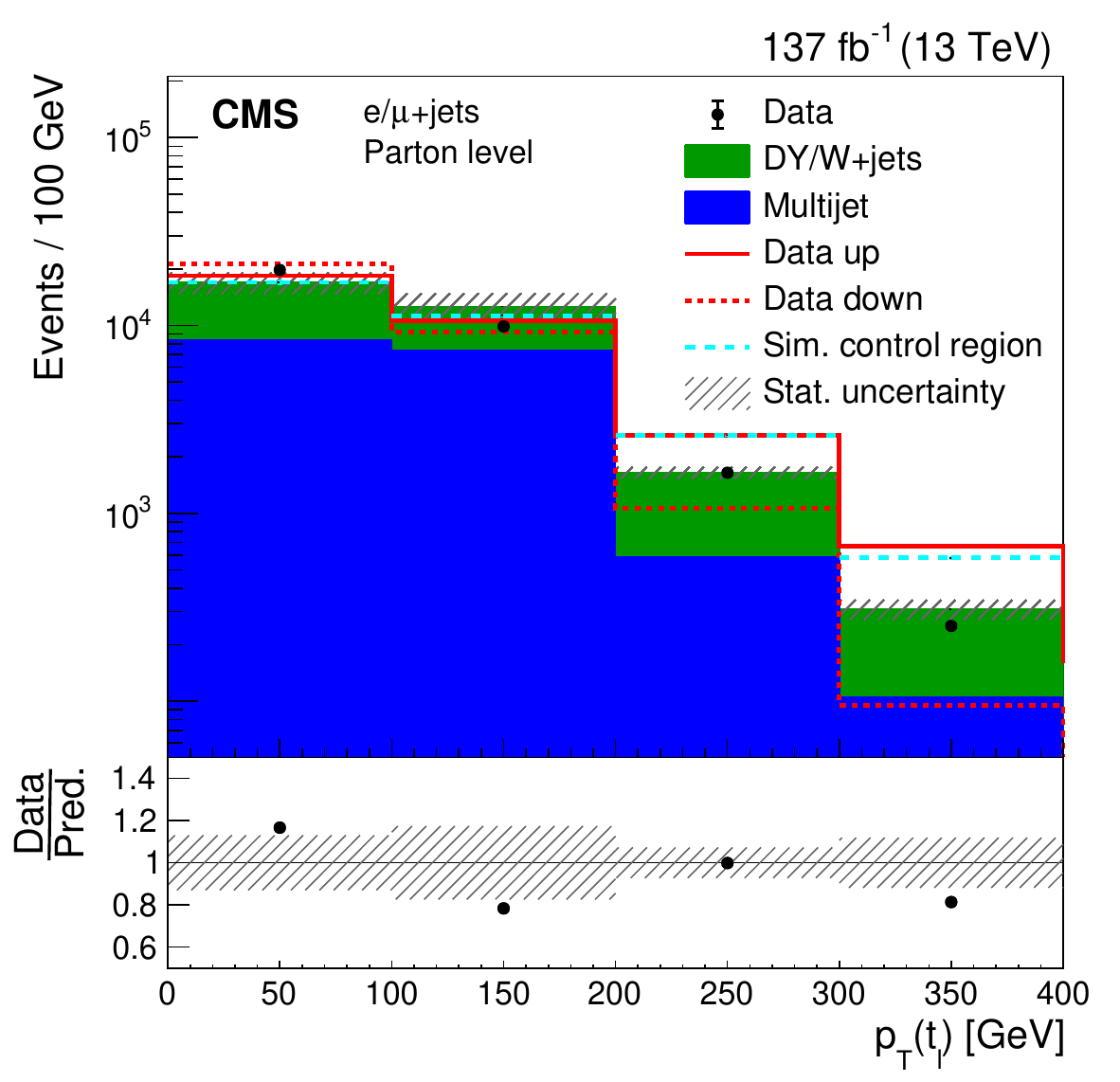}
 \caption{The \thadpt (\cmsleft) and \tleppt (\cmsright) distributions for the multijet and DY/\PW boson backgrounds from data (points) in the control region and from simulation (colored histograms) in the signal region for the 1t1l category.  The red lines show the variations in the control region distribution when shifting the discriminant selection range up (solid lines) and down (dotted lines).  The blue dashed line gives the sum of the multijet and DY/\PW boson predictions in the control region. The hatched band shows the statistical uncertainties in the prediction. The data points are shown with statistical uncertainties. The ratios of data to the predicted yields are provided in the lower panels.}
 \label{fig:RESBKG1}
\end{figure}

 In the resolved categories 2t and 1t1l, the multijet and DY/\PW boson backgrounds, which constitute fractions of 1.5 and 5.5\%, respectively, of the candidate events according to the simulation, are subtracted using combined templates that are obtained from a control region with a reduced \ttbar contribution. In order to obtain the background templates, the resolved reconstruction is performed on events where the highest value of the \PQb tagging discriminant is below the tight \PQb tagging requirements. The exact range is optimized to obtain good agreement between the distributions in the control region and the simulated prediction of the background in the signal region. Since the statistical precision of the background simulation is limited after applying all selection criteria, the agreement can only be assessed using a coarse binning, as shown in Fig.~\ref{fig:RESBKG1}. In this figure, distributions from altered background estimations are also shown. These distributions are obtained by varying up and down the edges of the \PQb tagging discriminant range used in the selection of the control region. The deviations of these distributions from those coming from the default range are used to estimate the systematic uncertainties. The variations in the range are chosen so that the obtained systematic uncertainties are comparable to the differences between the default background estimations and the simulated background predictions in the signal region.

The normalization of this background is obtained from the comparison of data and simulation in the control region. The normalization of the distributions in the control region as well as their shapes are well described by the simulation. After subtracting the predicted yield of \ttbar events, the ratio of the observed and simulated yields is used to scale the predicted event yield in the signal region. The deviation of the scale factor from unity and the statistical uncertainty in the simulation are taken as uncertainties in the normalization of this background. The normalization uncertainty is about 50\%. 

The obtained background predictions with their shape and normalization uncertainties are included in the fits of the cross sections, as described in Section~\ref{UNFOLDING}. Since the statistical uncertainties in the simulation are dominant, the background normalization is considered as uncorrelated among the three years of data taking. However, since the same method is used to derive the background distributions, the shape uncertainties are considered fully correlated among the years. 

Another background contribution is single top quark production, which contributes about 2.5\% in the 2t and 1t1l categories. It is subtracted during the cross section fits using SM expectations obtained from the simulation, where the following dominant uncertainties are included: variations of $\mur$ and $\muf$, jet energy scales, and \PQb tagging efficiency. The variations of $\mur$ and $\muf$ are treated independently from the corresponding variations of the \ttbar simulation, whereas the experimental uncertainties are fully correlated among all processes.

\section{Background subtraction in the boosted categories}
\label{FITBO}

Events with a boosted \tqh, \ie, the categories BHRL and BHBL, are combined for the background subtraction. Template fits based on the \HNN distribution are used to extract the yields of 2Q and 3Q events. Since the shape of the \HNN distribution depends on $\pt(\tqh)$, the template fit is performed using the following bin boundaries in $\pt(\tqh)$: 400, 450, 500, 550, 600, 700, 800, 900, 1000, and 7500\GeV. For each bin of a variable under consideration, we fit nine \pt bins. Because of kinematic restrictions, not all \pt bins are populated for each measured bin of the variable under consideration. In addition, we split the templates into three pseudorapidity regions: $\abs{\eta(\tqh)} < 0.8$, $0.8 < \abs{\eta(\tqh)} < 1.6$, and $1.6 < \abs{\eta(\tqh)} < 2.4$, corresponding to the barrel, barrel-endcap transition, and endcap regions, respectively. For each fitted bin the fractions in these three regions $f_i$ are determined in data and the templates are composed as $f_1 \boldsymbol{T}_1 + f_2 \boldsymbol{T}_2 + f_3 \boldsymbol{T}_3$, where the $\boldsymbol{T}_i$ represent the templates in the $\abs{\eta(\tqh)}$ regions. This takes into account the small $\abs{\eta(\tqh)}$ dependence of the templates without fitting in separate bins of $\abs{\eta(\tqh)}$. We verified  in the simulation that the templates do not depend on any other of the measured variables. Therefore, the templates do not have to be adapted as a function of these variables.   

The \ttbar templates are taken from the simulation after applying the full event selections. The yields of 2Q and 3Q events are fitted separately  using different templates for the two contributions. The \ttbar background templates contain all other boosted \tqh candidates that do not belong in the 2Q and 3Q categories, \ie, no or only a single quark from a \ttbar decay points towards the candidate. The normalizations of these three templates are free parameters in the fit. 

The single top quark background templates are taken from the simulation and normalized to their SM expectation. A Gaussian prior is used to represent the dominant normalization uncertainties in the $\mur$ and $\muf$ variations, jet energy scales, and \PQb tagging efficiency. This corresponds to an overall normalization uncertainty of about 30\%.

The common background templates for multijet and DY/\PW boson production are determined from a control region in data using events with at least one jet with $\pt > 400$\GeV and $\abs{\eta} < 2.4$. In addition, events with an isolated electron or muon with $\pt > 15$\GeV are vetoed, and there must be no boosted \tql candidate with $\LNN > 0.7$ in the event. The contribution of all-hadronic \ttbar events is suppressed by discarding events with more than one \tqh candidate with $\HNN > 0.4$. 

For the multijet template, we require at least one jet passing the tight \PQb tagging criterion in the event. In such events, we select all \tqh candidates that do not overlap with at least one of the \PQb jets. The selection of these events with a \tqh candidate and a \PQb jet is like the signal selection but without a lepton. 

For the DY/\PW boson template, a quark jet enhanced region is needed. Therefore, we require an isolated photon with $\pt > 35$\GeV and $\abs{\eta} < 2.4$ in the event. The additional requirement, that the photon must be separated from all AK4 jets by $\Delta R(\gamma,\:\textrm{jet}) > 0.5$, increases the fraction of photons not stemming from hadron decays. However, these photons are dominantly produced close to jets, where they are radiated from quarks. This selection enhances the overall fraction of quark jets in the event. In such events, we select all \tqh candidates that do not overlap with the photon to construct the background templates.

To verify that the selected \tqh candidates from the control regions provide accurate background templates, we compare the templates obtained from simulation in the signal region to the templates obtained from simulation in the control region and find these in reasonable agreement. In addition, the simulation is able to describe the data in the control region. 

The multijet and DY/\PW boson backgrounds have slightly different shapes due to different compositions of quark flavors and gluon jets. The yields are constrained to the SM predictions using Gaussian priors with an uncertainty of 50\%. These 50\% uncertainties are meant to guide the fit to a minimum similar to the SM expectation. Since these background shapes are very similar, the obtained fraction of the background components is somewhat arbitrary if the constraints are removed. However, this gives the fit the freedom to cope with backgrounds of different parton compositions and does not affect the extracted signal yields, since the signal is well distinguished from the backgrounds.

Systematic uncertainties that affect the \HNN distribution in the simulation are included in the fit for the corresponding templates. These are the uncertainties in the final-state PS scale, \Mtop, UE tune, CR model, pileup, and energy response and resolution of the \tqh candidates. In addition, the effect of the energy scale uncertainty in the individual subjets is estimated. To assess this, we boost the subjets into the laboratory frame and vary their energies by the uncertainty obtained for PUPPI jets with a distance parameter of 0.4. For the up- and down-varied subjets we recalculate \HNN taking into account the change of the center-of-mass system of the AK8 jet. In this way we modify the substructure of the jet according to the typical energy-dependent uncertainty in the jet energy scale. These are the only sources that affect the substructure of jets and hence the \HNN distribution. A more detailed discussion of these uncertainties is presented in Section~\ref{SYS}.   

\begin{figure*}[tbp]
\centering
 \includegraphics[width=0.42\textwidth]{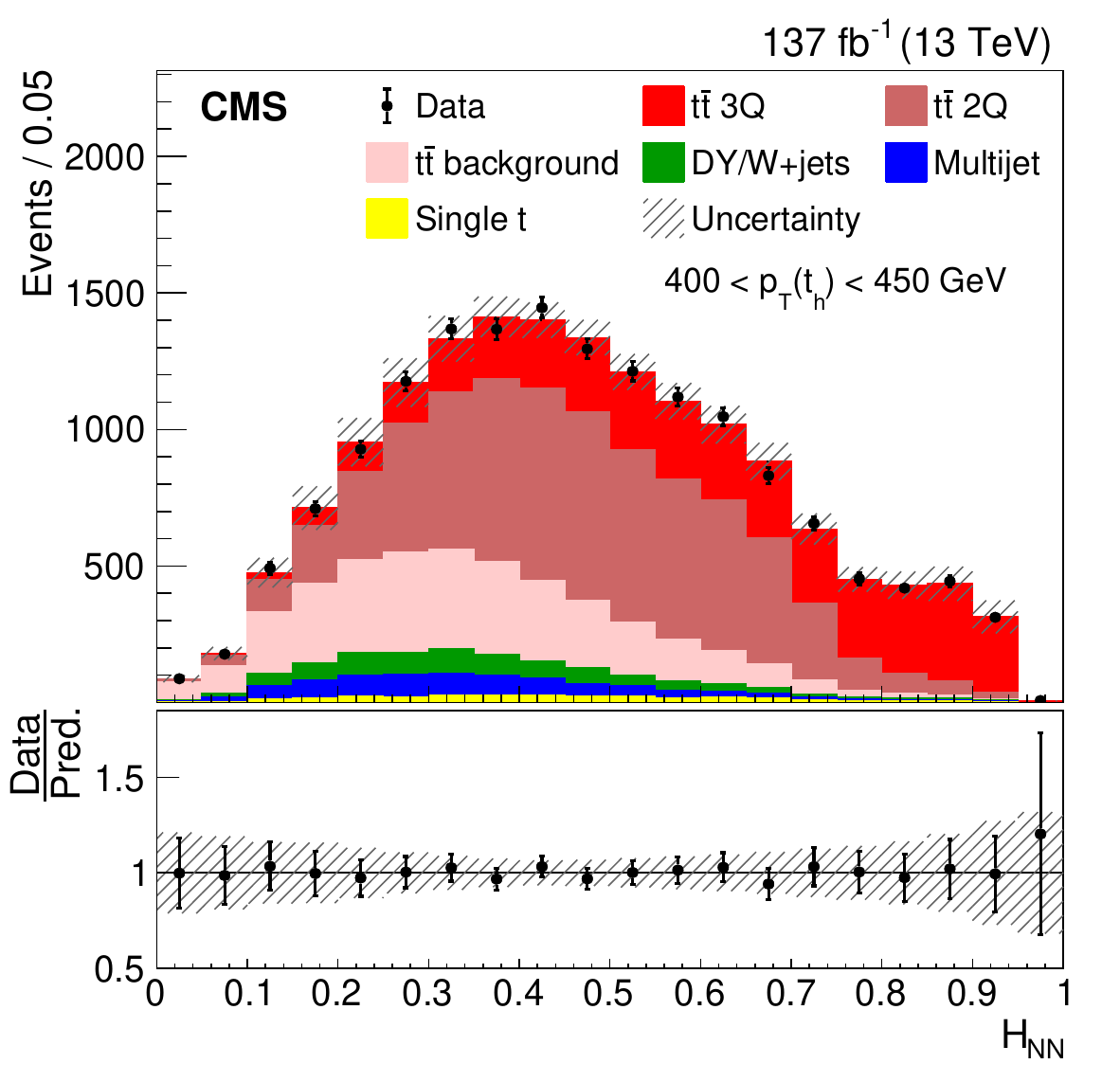}
 \includegraphics[width=0.42\textwidth]{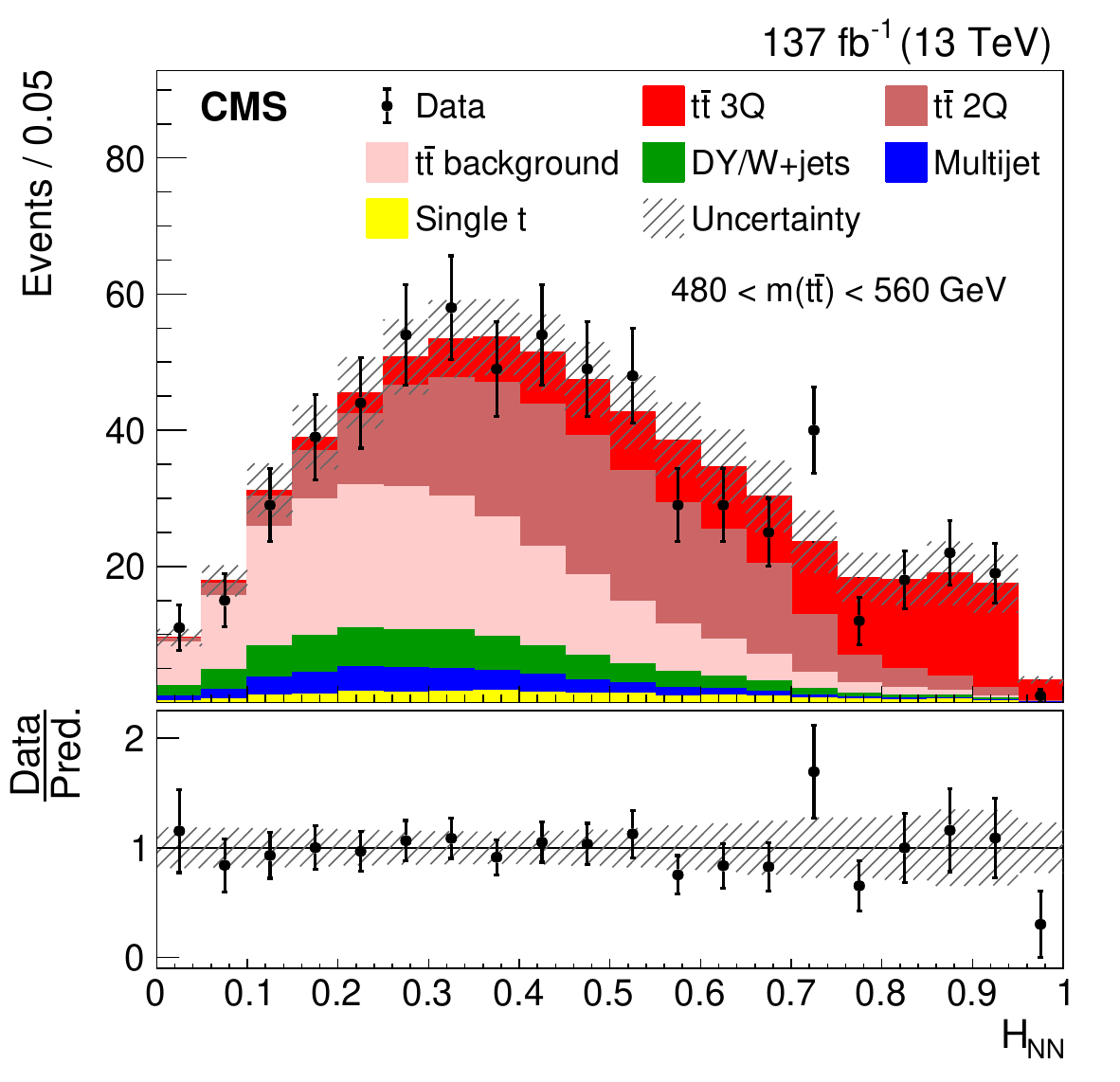}
 \includegraphics[width=0.42\textwidth]{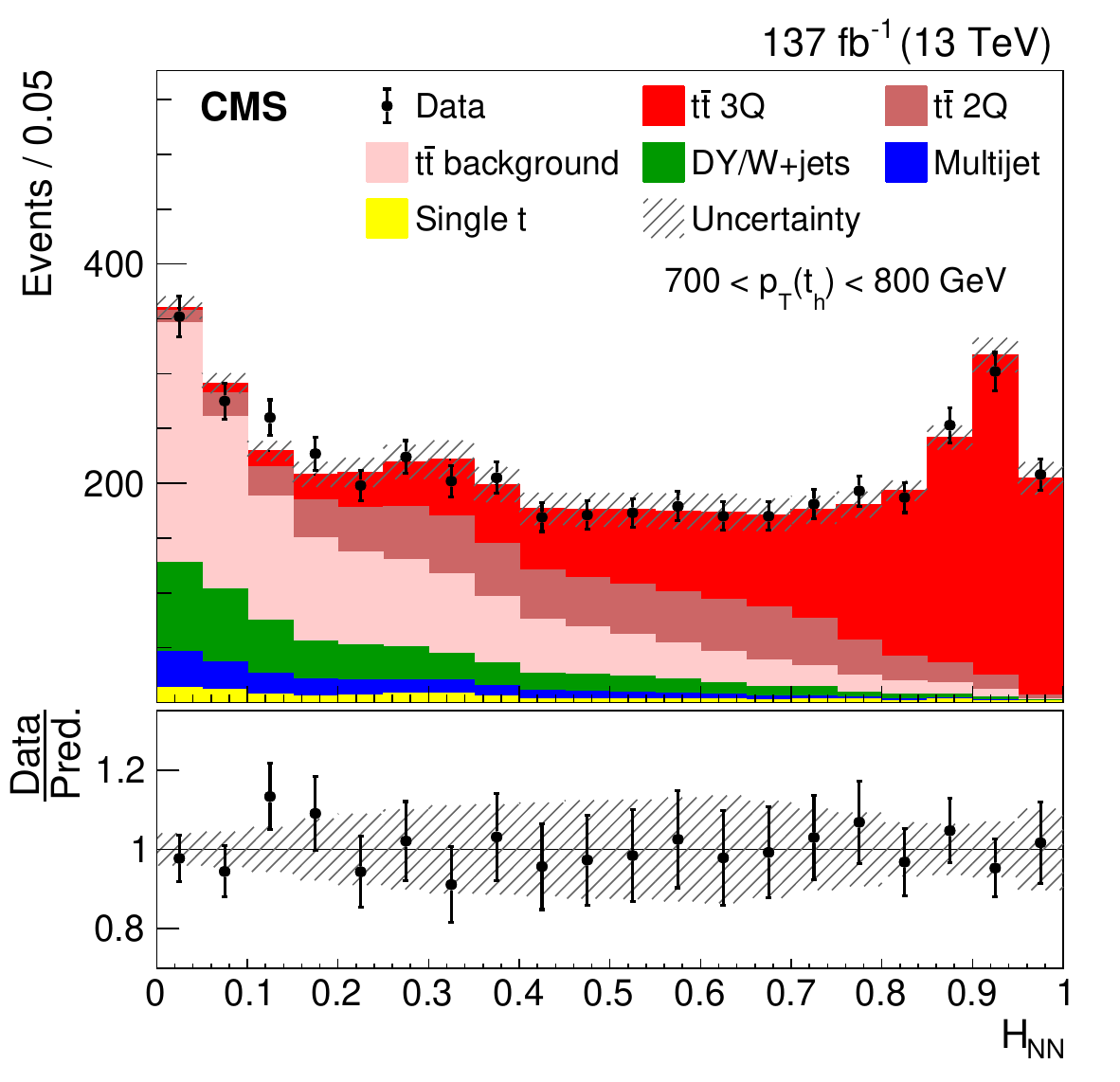}
 \includegraphics[width=0.42\textwidth]{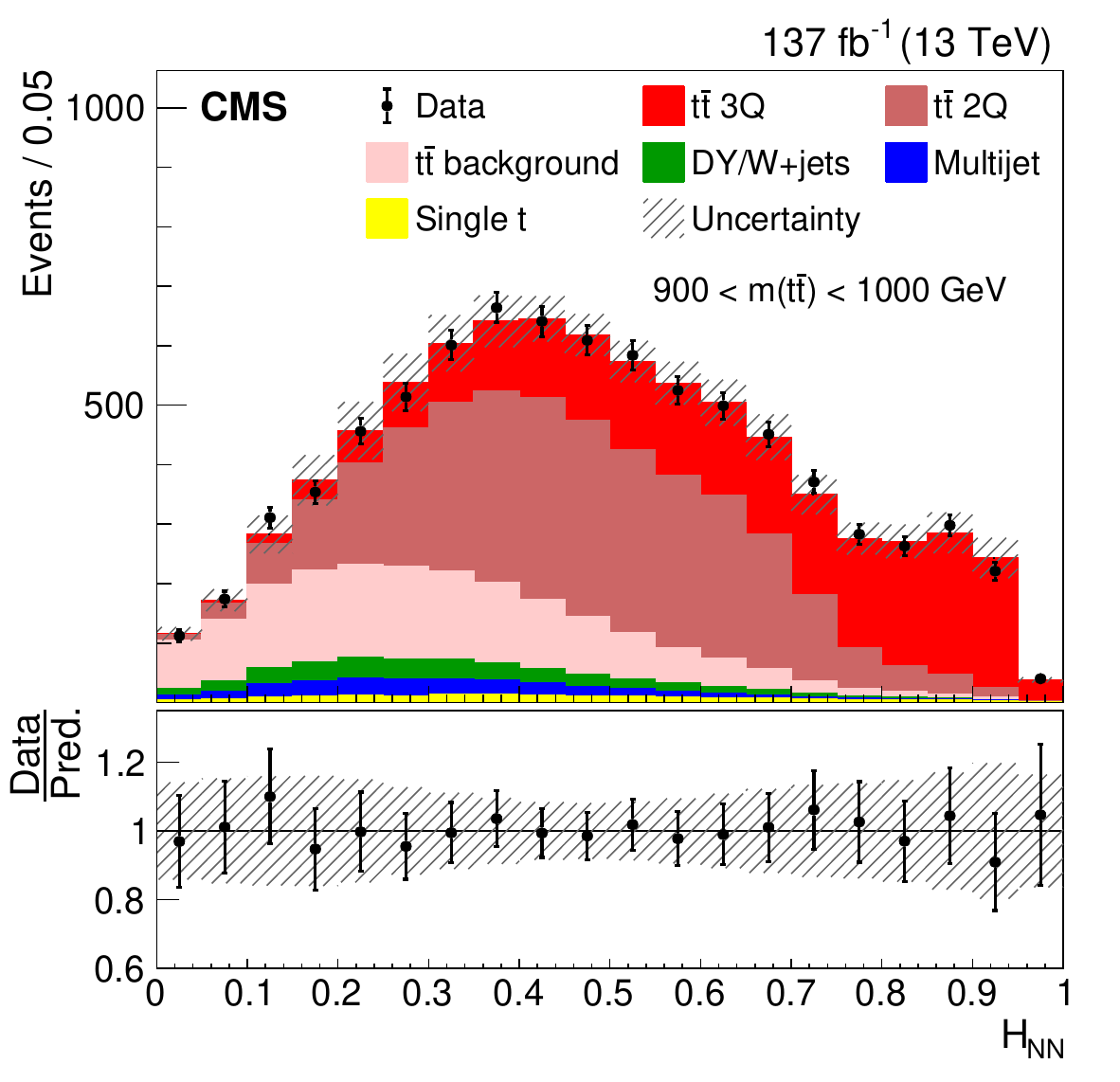}
 \includegraphics[width=0.42\textwidth]{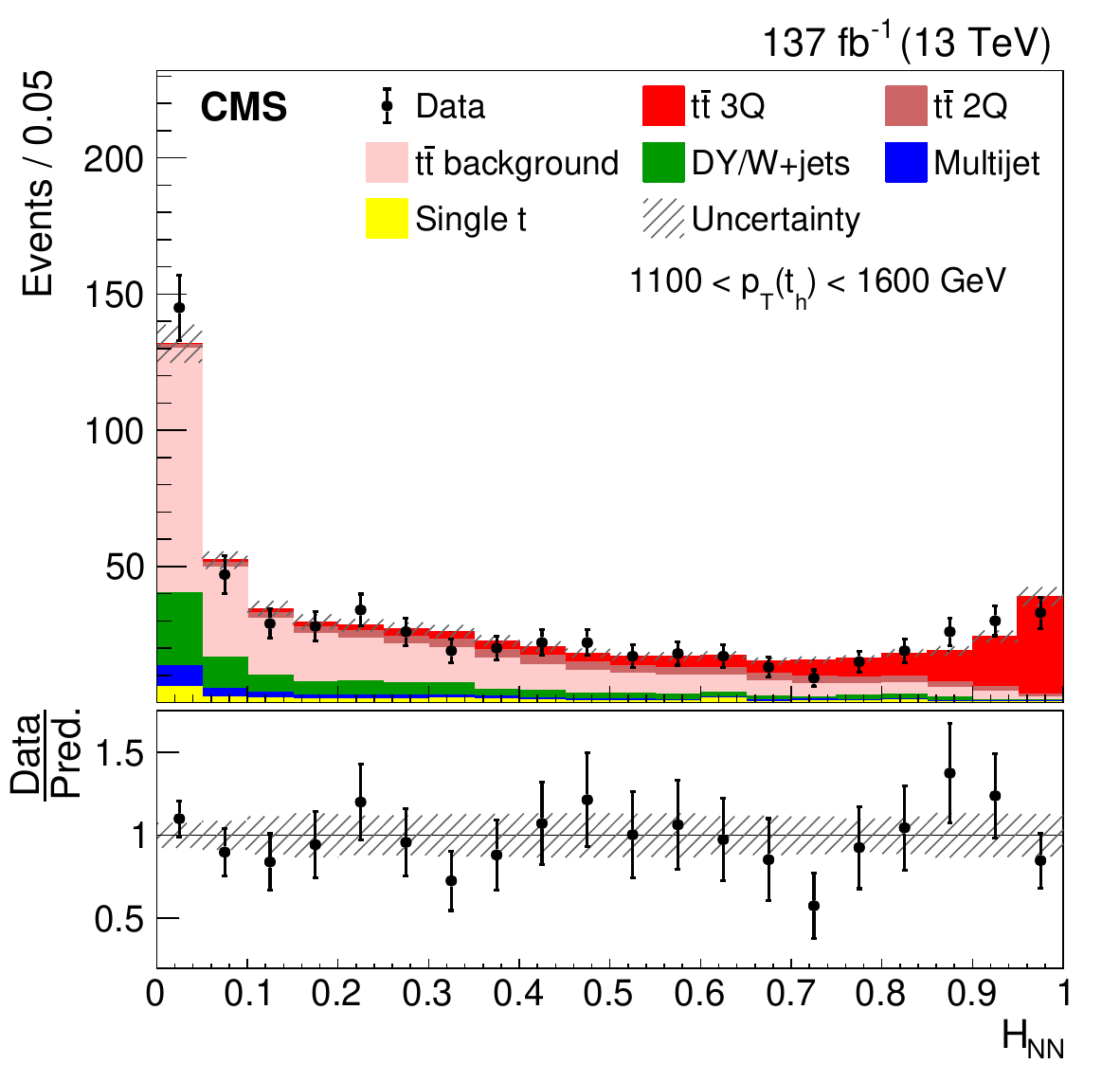}
 \includegraphics[width=0.42\textwidth]{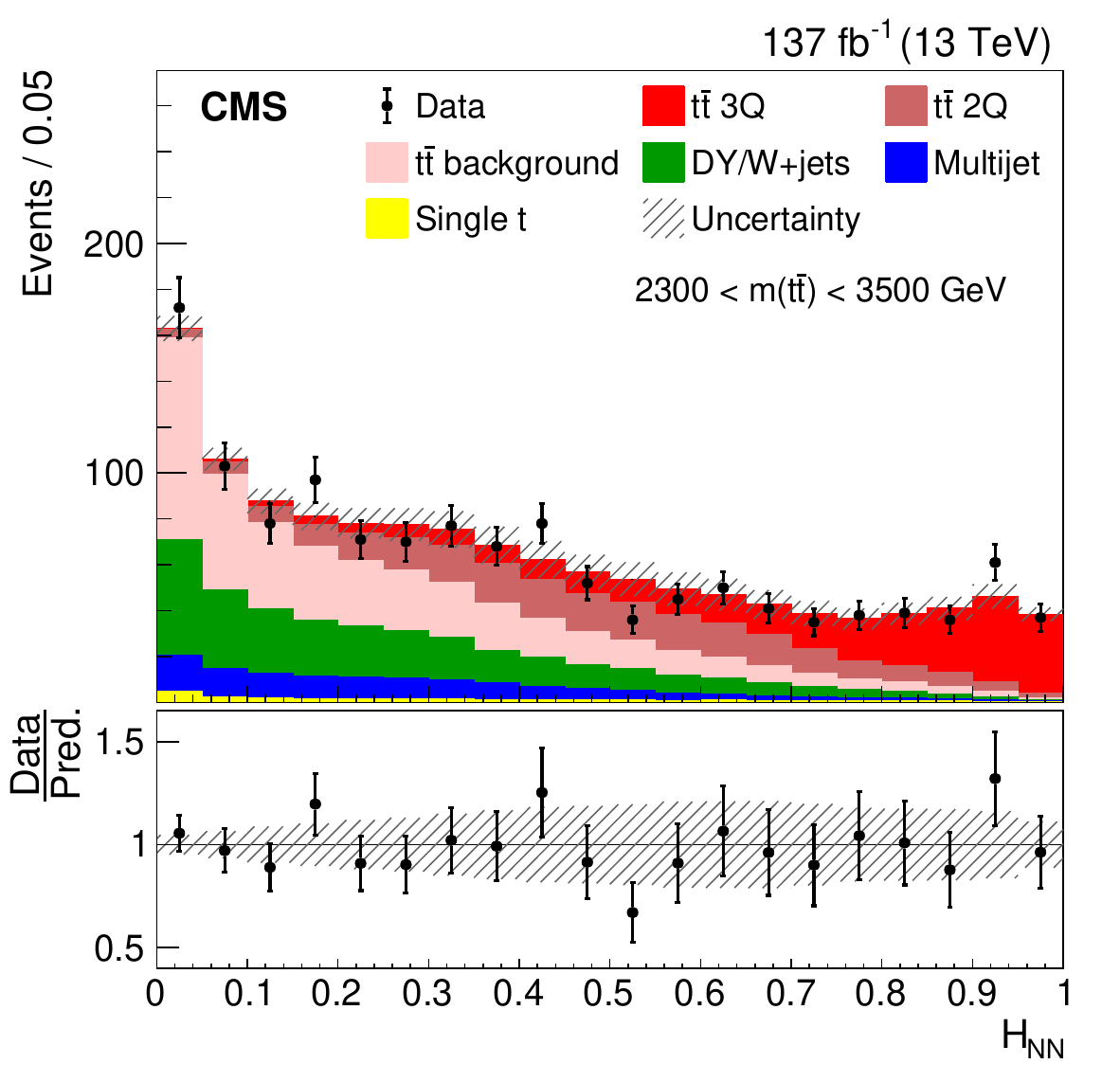}
 \caption{The postfit \HNN distributions in bins of \thadpt (left) and \ttm (right) for the data (points) and simulation (colored histograms). The electron and muon events and all three years of data taking have been combined. The hatched area shows the total uncertainties from the fit.  The vertical bars on the points represent the statistical uncertainty in the data. The ratios of data to the sum of the fitted yields are provided in the lower panels. }
 \label{fig:FITBO1}
\end{figure*}

The various contributions, especially the signal yields for 2Q and 3Q, are obtained by performing a binned maximum likelihood fit simultaneously to all bins of a measurement. The effects of uncertainties in the \HNN templates are parameterized by nuisance parameters and consistently varied in all bins. To reduce statistical fluctuations in the descriptions of the uncertainties, a smoothing algorithm~\cite{LOWESS} is applied to the shape of their relative contributions. The fits are performed separately for the electron and muon channels for each of the three years. As a consistency test, the simulation is fitted, and the extracted event yields are found to be unbiased.  As examples, several postfit \HNN distributions in bins of \thadpt and \ttm are shown in Fig.~\ref{fig:FITBO1}. For the \ttm measurement, the results are summed over the \thadpt bins that are fitted separately for each \ttm bin. However, to obtain the input for the cross section extractions discussed in Section~\ref{UNFOLDING}, we do not use the signal yields with profiled uncertainties. Instead, we perform the fit using the default simulation and repeat it for all relevant up and down variations of the simulation to estimate the systematic uncertainties.

\begin{figure*}[tbp]
\centering
 \includegraphics[width=0.45\textwidth]{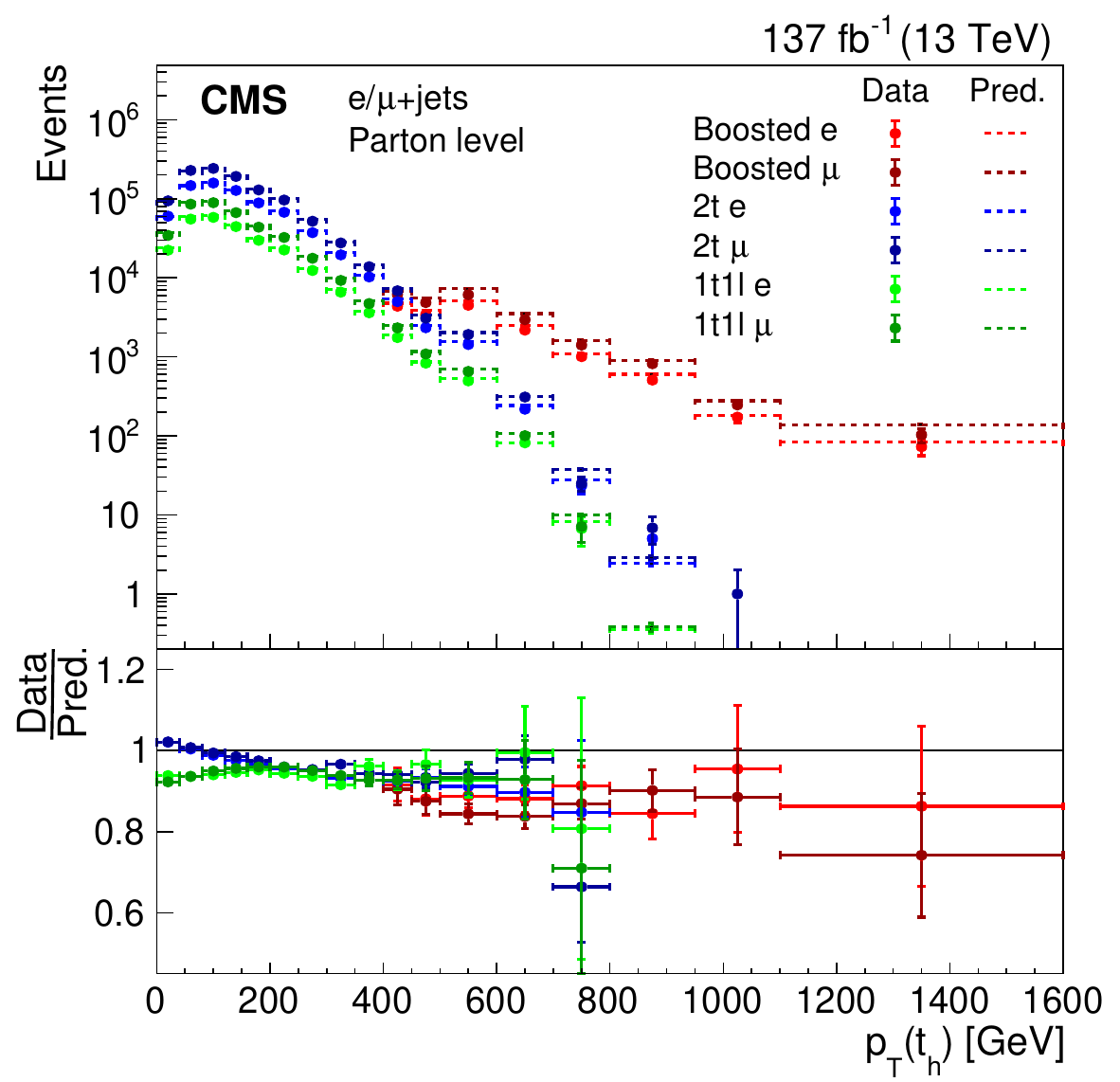}
 \includegraphics[width=0.45\textwidth]{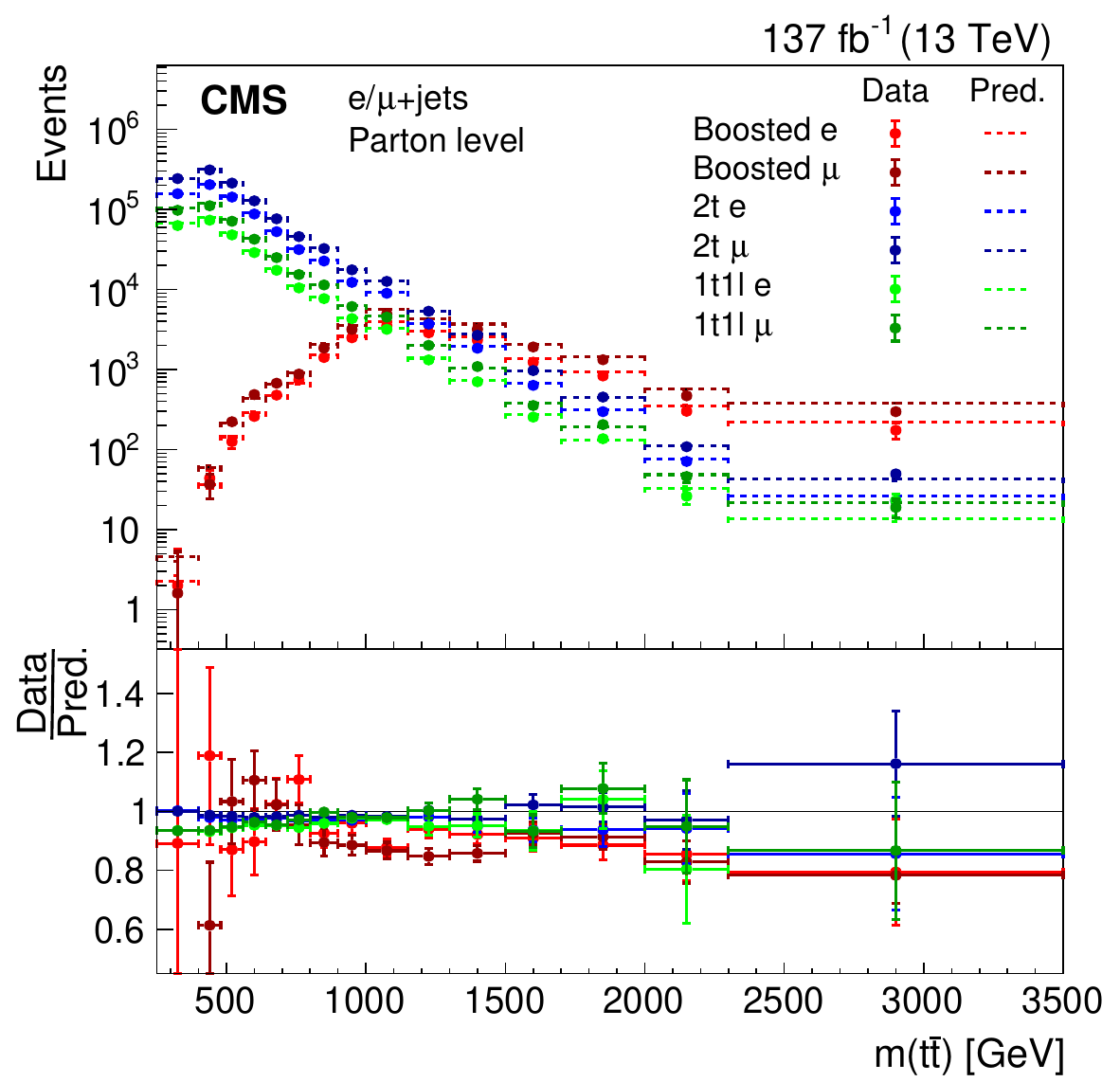}
 \includegraphics[width=0.45\textwidth]{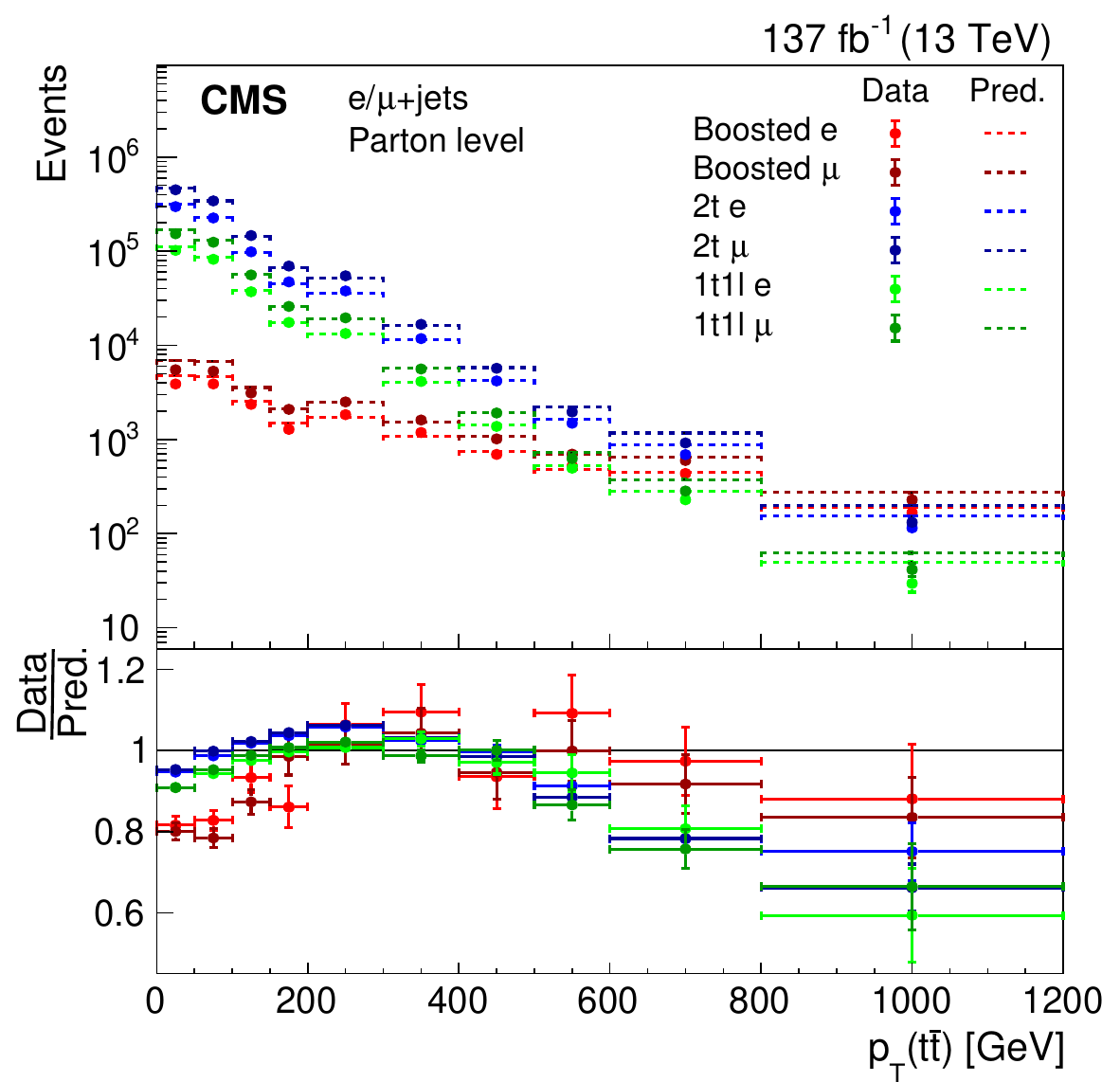}
 \includegraphics[width=0.45\textwidth]{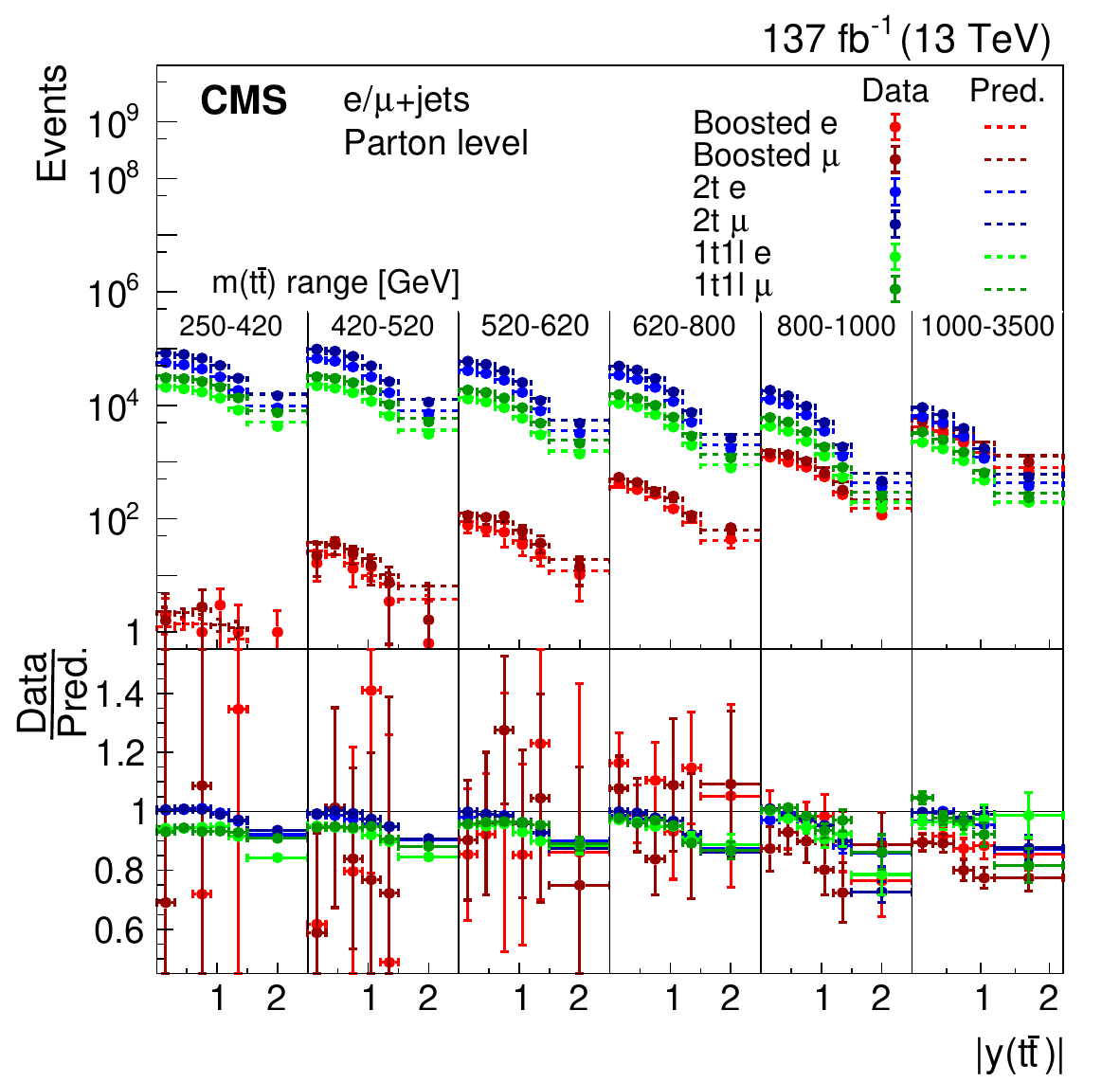}
 \caption{
Distributions of \thadpt (upper left), \ttm (upper right), \ttpt (lower left), and \ttmvstty (lower right) after background subtraction. The points show the data and the dashed lines the predictions for the various events types. The lower panels give the ratio of the data to the predictions.  The vertical bars on the points give the statistical uncertainties, and the horizontal bars the bin widths.}
 \label{fig:FITBO2}
\end{figure*}

To obtain the signal yields as a function of the variable of interest, we add the extracted yields of 2Q and 3Q events in the $\thadpt$ bins. In Fig.~\ref{fig:FITBO2} the extracted yields are compared to the simulation. For comparison these plots also show the yields in the 2t and 1t1l categories that use the resolved reconstruction. The backgrounds in the resolved categories are subtracted using the techniques discussed in Section~\ref{RESBKG}. The ratio of data to the prediction as a function of \thadpt in the upper-left plot of Fig.~\ref{fig:FITBO2} shows a smooth transition between the boosted and resolved reconstructions.  

\section{Extraction of the differential and inclusive cross sections}

\label{UNFOLDING}

As explained in Section~\ref{FITBO}, the BHRL and BHBL categories are combined in the boosted category for the background subtraction fit. The distributions in the 2t, 1t1l, and boosted categories are combined in the fit to extract the differential cross sections. Each of the three categories are measured separately in the electron and muon channels for each of the three years of data taking. This results in 18 categories entering the fit.

\begin{figure*}[tbp]
\centering
 \includegraphics[width=0.45\textwidth]{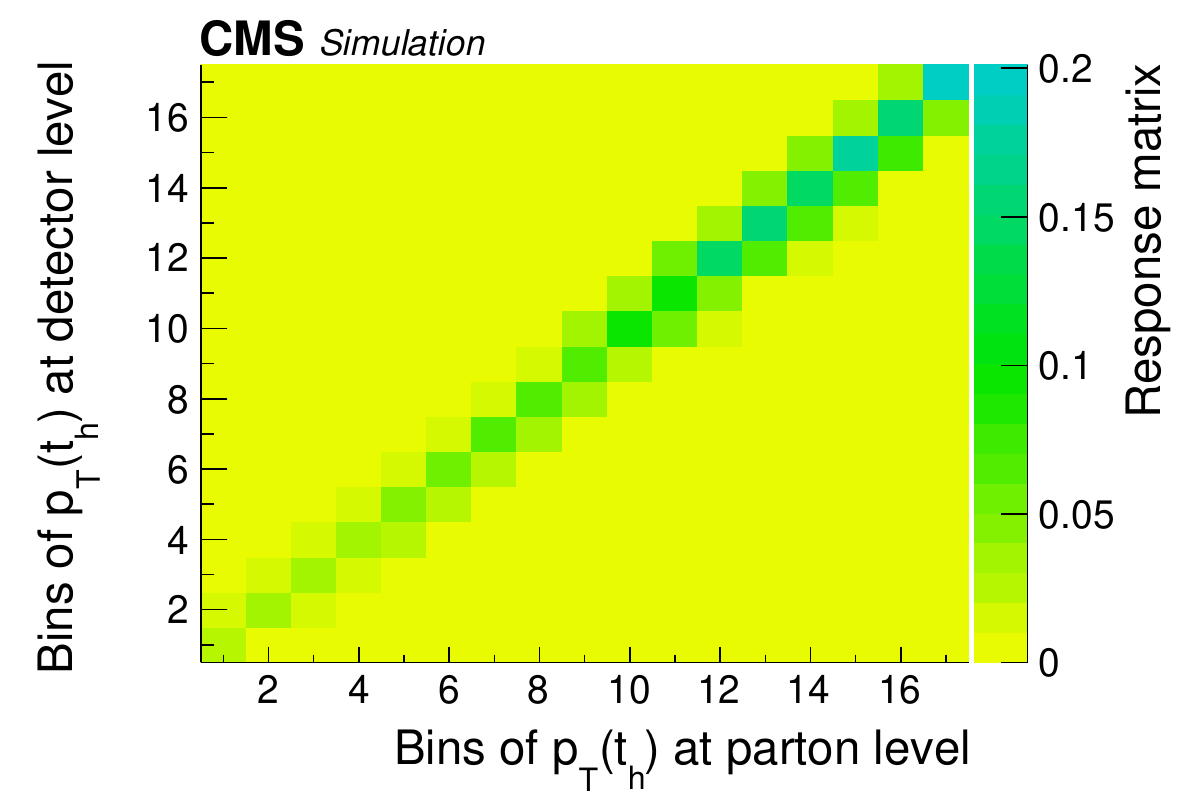}
 \includegraphics[width=0.45\textwidth]{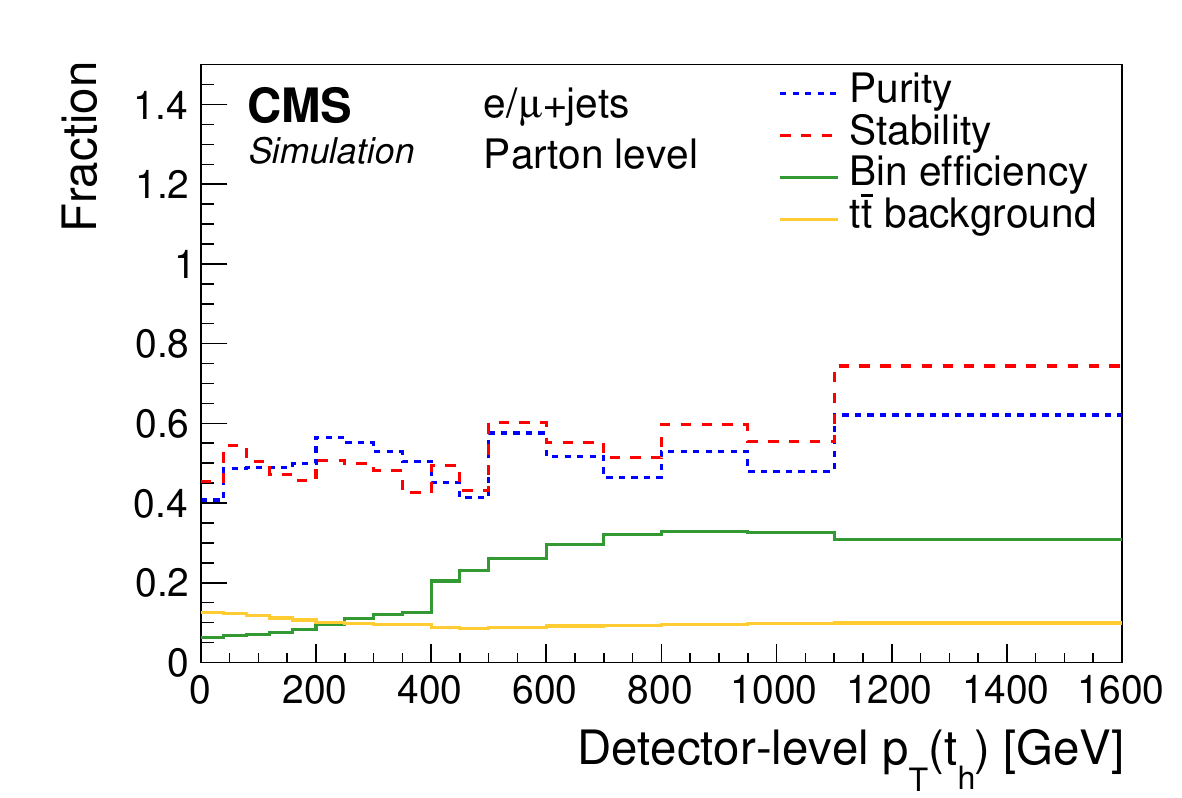}
 \includegraphics[width=0.45\textwidth]{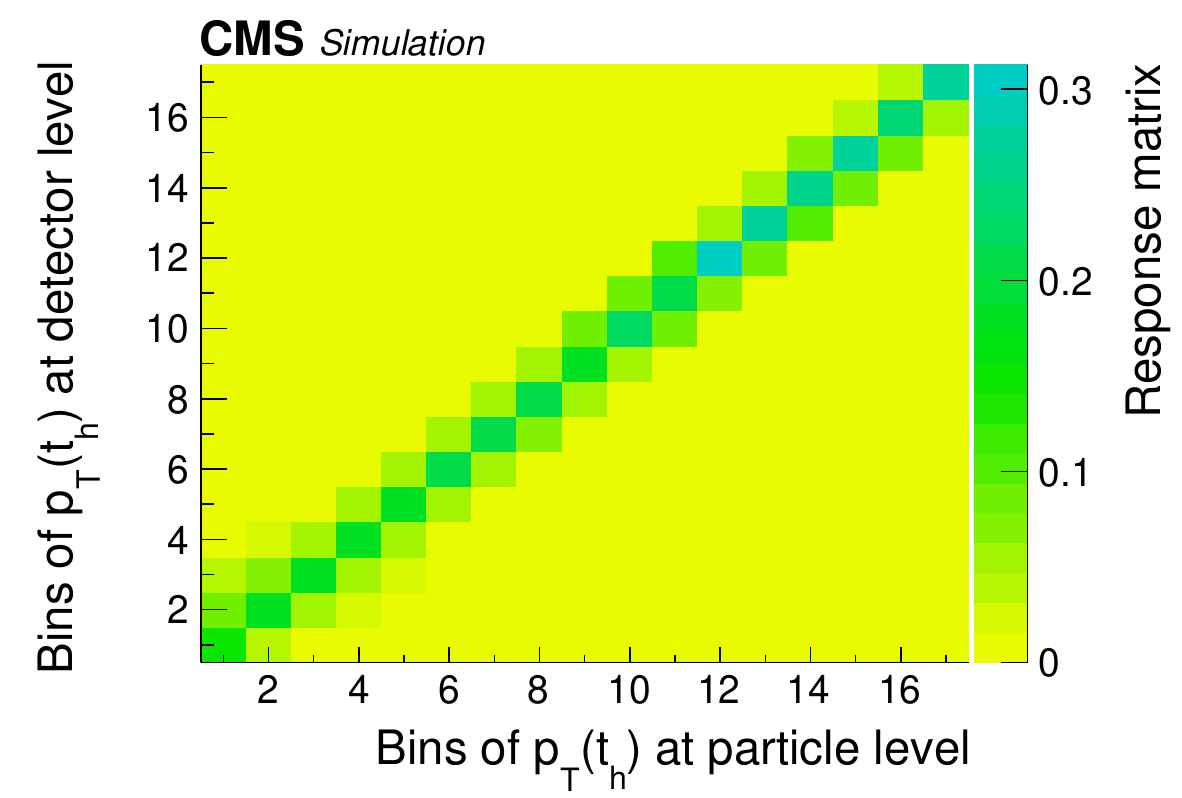}
 \includegraphics[width=0.45\textwidth]{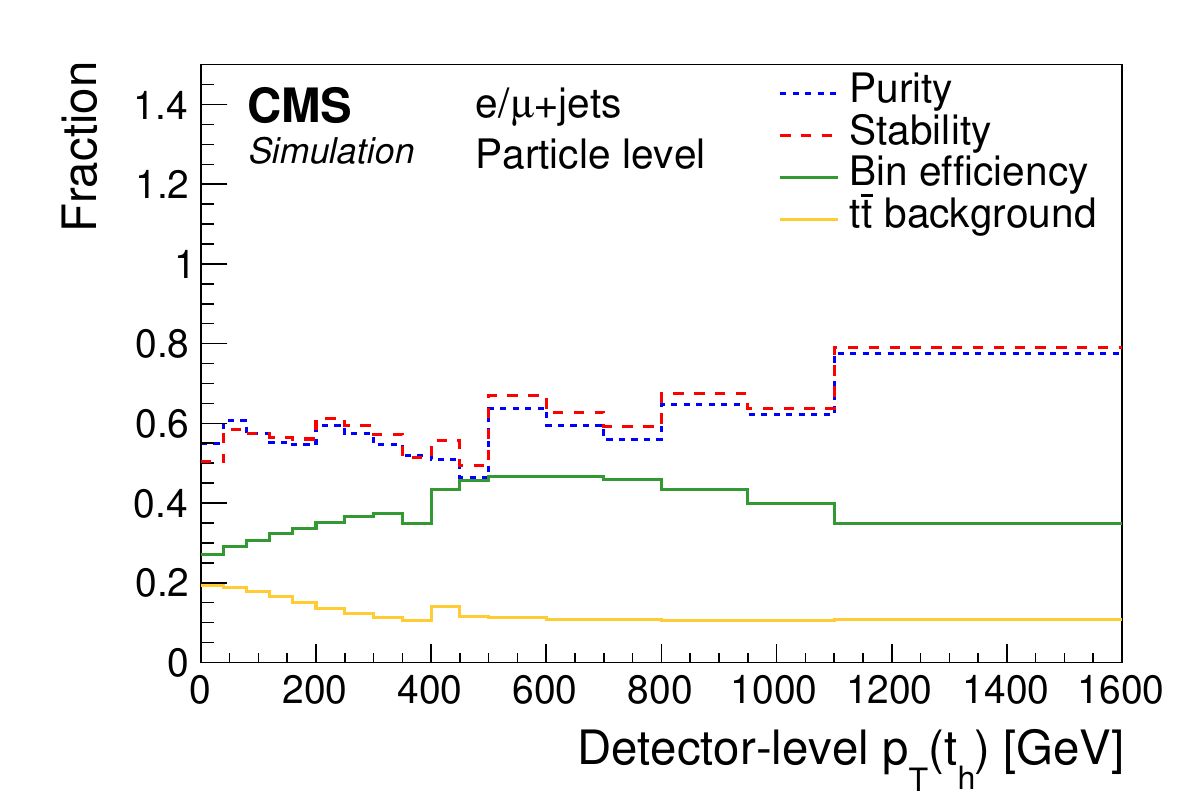}
 \caption{Combined response matrices of all reconstruction categories for the measurements of \thadpt at parton (upper left) and particle (lower left) levels from the \POWHEG{}+\PYTHIA simulation. The purity (fraction of parton-/particle-level events that are reconstructed in the same bin at the detector level), stability (fraction of detector-level events that belong in the same bin at the parton/particle level), the efficiency per bin, and the fraction of \ttbar background for the corresponding parton and particle levels are shown in the right plots.}

 \label{fig:UNFOLDING1}
\end{figure*}

For the extraction of the cross sections, the response matrices $R$ are needed. These map a vector of cross sections $\boldsymbol{\sigma}$ in bins of the measured distribution to the corresponding event yields at the detector level. After adding a vector $\boldsymbol{b}$ giving the non-\ttbar background events in each bin, the prediction for the number of events at detector level
\begin{equation}
\boldsymbol{s} = R \boldsymbol{\sigma} + \boldsymbol{b}
\label{eq:UNFOLDING1}
\end{equation}
 is obtained, which can be compared to the measured event yields. Since non-\ttbar backgrounds are already subtracted in the boosted category, the components of $\boldsymbol{b}$ for that category are all zero. However, the systematic uncertainties in the background subtraction, as discussed in Section~\ref{FITBO}, are taken into account. There are also backgrounds in all categories from \ttbar production. Since these backgrounds scale with the \ttbar cross section, their contributions are encoded in the response matrix $R_{ij}$, whose elements are calculated from the simulated \ttbar events through the equation:
\begin{equation}
R_{ij} = L \sum\limits_m \sum_n \frac{\delta_{ni}(r_{n}+M_{n:})}{M_{n:}}  M_{nm}  \frac{\delta_{mj}}{g_{m}+M_{:m}},
\label{eq:UNFOLFING2}
\end{equation}
where $M_{nm}$ is the two-dimensional distribution of the parton- or particle-level \vs detector-level quantity. The first index $n$ corresponds to a bin at the detector level, the second index $m$ to a bin at the parton/particle level. This distribution is only filled if the quantity can be calculated at both levels. As abbreviations, we define the quantities $M_{i:}$ ($M_{:i}$) as the sum of entries in the $i$-th row (column) of $M$. Events that can be reconstructed but do not contain a defined \ttbar signal are considered as \ttbar background events and enter the distribution $\boldsymbol{r}$. The first ratio in Eq.\,(\ref{eq:UNFOLFING2}) corrects for these \ttbar background events. Events with a defined \ttbar signal at the parton/particle level, but no \ttbar at the detector level, are filled into distribution $\boldsymbol{g}$, entering the second ratio in Eq.\,(\ref{eq:UNFOLFING2}), which represents the losses due to inefficiencies and acceptance. Individual elements of $\boldsymbol{r}$ ($\boldsymbol{g}$) are referred to as $r_i$ ($g_i$). To convert cross sections into event yields, the expression is multiplied by the integrated luminosity $L$. In this analysis, the same binning is used at the parton/particle and detector levels. In Fig.~\ref{fig:UNFOLDING1} we show the response matrices of the $\pt(\tqh)$ measurements obtained from the \POWHEG{}+\PYTHIA simulation, together with their purities (fraction of parton-/particle-level events that are reconstructed in the same bin at the detector level) and stabilities (fraction of detector-level events that belong in the same bin at the parton/particle level). For illustration the combinations of the matrices in the individual categories are shown. The reconstruction performances are very similar for the three years of data taking. 

 Defining $\boldsymbol{o}$ as the observed event yield vector, we calculate the $\chi^2$ of the fit using:
 \ifthenelse{\boolean{cms@external}}
 {
  \begin{multline}
    \chi^2(\boldsymbol{\sigma}, \boldsymbol{\nu}) =  \sum_y \sum_c  \sum_\ell (\boldsymbol{o}_{yc\ell} - \boldsymbol{s}_{yc\ell}(\boldsymbol{\sigma}, \boldsymbol{\nu}))^T\\
     C_{yc\ell}^{-1} (\boldsymbol{o}_{yc\ell} - \boldsymbol{s}_{yc\ell}(\boldsymbol{\sigma}, \boldsymbol{\nu})) + \boldsymbol{\nu}^T Q^{-1} \boldsymbol{\nu},
    \label{eq:UNFOLDING3}
    \end{multline}  
 }
 {
\begin{equation}
\chi^2(\boldsymbol{\sigma}, \boldsymbol{\nu}) =  \sum_y \sum_c  \sum_\ell (\boldsymbol{o}_{yc\ell} - \boldsymbol{s}_{yc\ell}(\boldsymbol{\sigma}, \boldsymbol{\nu}))^T C_{yc\ell}^{-1} (\boldsymbol{o}_{yc\ell} - \boldsymbol{s}_{yc\ell}(\boldsymbol{\sigma}, \boldsymbol{\nu})) + \boldsymbol{\nu}^T Q^{-1} \boldsymbol{\nu},
\label{eq:UNFOLDING3}
\end{equation}
 }
where we sum over the  years ($y=$ 2016, 2017, and 2018), the reconstruction categories ($c=$ 2t, 1t1l, and boosted), and the lepton channels ($\ell$ = e, $\mu$). In the resolved categories, the covariance matrix $C$ is a diagonal matrix with the numbers of observed events per bin, while in the boosted category, since the background was already subtracted, the covariance matrix is the diagonal matrix of the squared statistical uncertainties obtained from the background fits, as described in Section~\ref{FITBO}. In each category, only bins with at least four events are used, \ie, very low-content and unpopulated bins are not included in the fit. For such bins a $\chi^2$ fit is not well defined. However, since the combined event yields of all categories are from several hundred to thousands of events per bin, this does not affect the results.

\begin{figure*}[tbp]
\centering
 \includegraphics[width=0.75\textwidth]{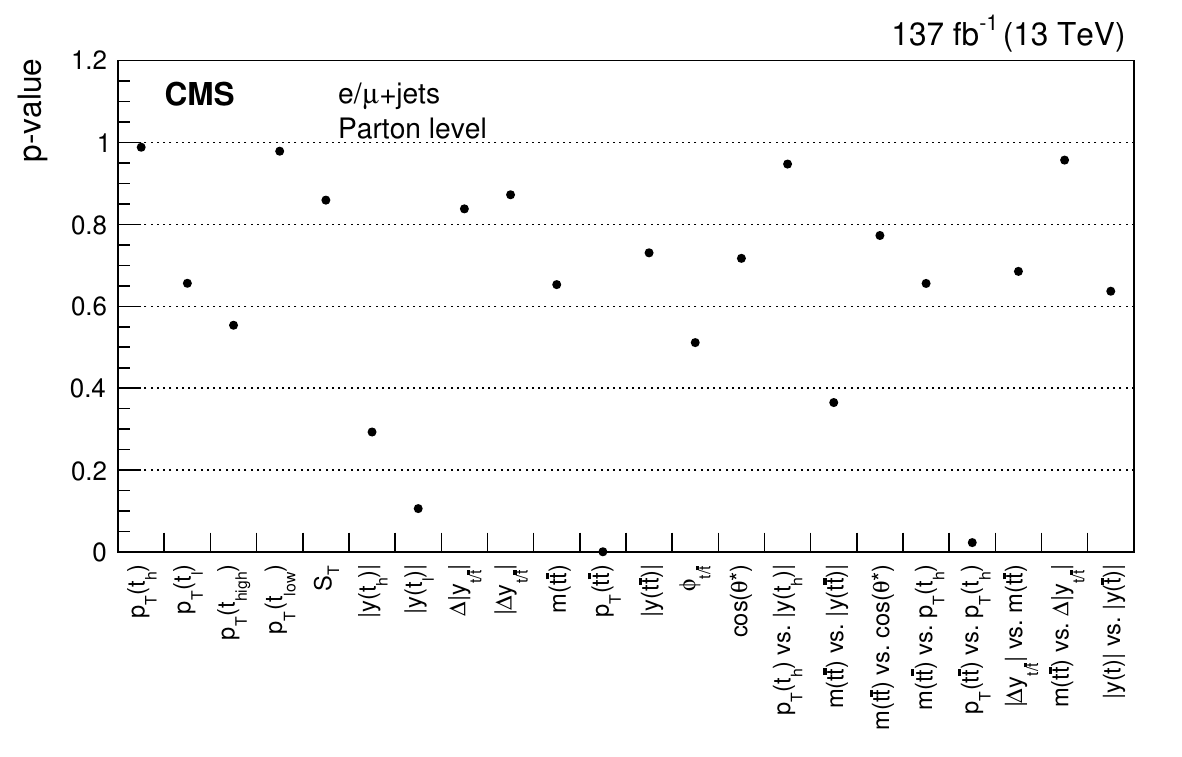}\\
 \includegraphics[width=0.75\textwidth]{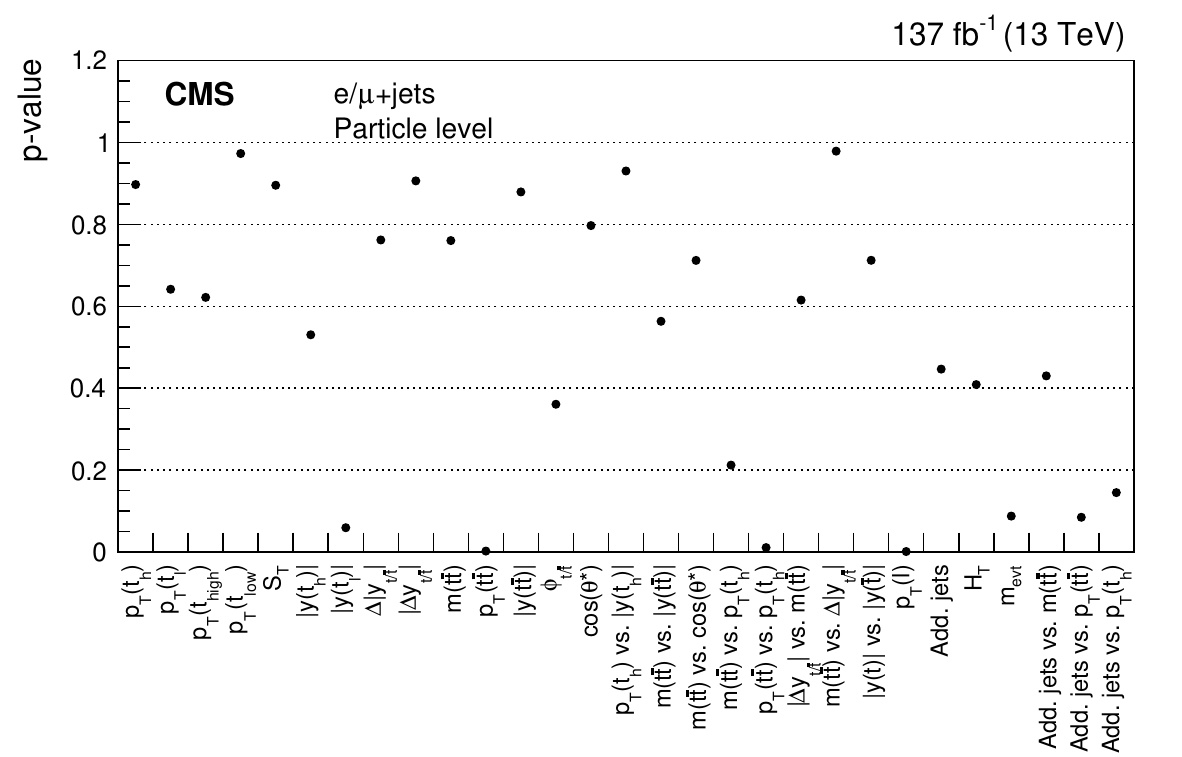}
 \caption{The $p$-values of the $\chi^2$ fits for the listed variables are shown for the parton- (upper) and particle- (lower) level measurements.}
 \label{fig:UNFOLDING2}
\end{figure*}

The $\boldsymbol{\sigma}$ values are free parameters of the fit. The systematic uncertainties affecting $\boldsymbol{s}$ are parameterized as functions of the nuisance parameters $\boldsymbol{\nu}$. These are constrained by the last term in Eq.\,(\ref{eq:UNFOLDING3}), where the matrix $Q$ is the correlation matrix of the nuisance parameters. The correlations of the uncertainty sources among the measurements from different years are important. They are discussed in detail in Section~\ref{SYS}. The goodness of fit is calculated from the minimized $\chi^2$ and the number of degrees of freedom, which is obtained by subtracting the number of bins in $\boldsymbol{\sigma}$ from the used detector-level bins. The corresponding $p$-values, shown in Fig.~\ref{fig:UNFOLDING2}, are reasonable. They follow approximately a uniform distribution between 0 and 1, with a minimum value of 0.2\%. Thus, the differential cross sections are able to describe the data in all categories simultaneously. In all the differential cross sections fits, the nuisance parameter values are not pulled more than two standard deviations. Most of them are either not constrained or only moderately constrained, and their uncertainties are typically reduced by about 30\%.

\section{Systematic uncertainties}
\label{SYS}

In the cross section fits, several sources of systematic uncertainties affect the response matrix. These sources can be split into two main categories: theoretical and experimental uncertainties. The theoretical ones are listed below.

\begin{itemize}
\item The effects of higher-order contributions to the cross section calculations are estimated by varying \mur and \muf separately by a factor of two. Distributions for these variations are obtained using event weights in the \POWHEG{}+\PYTHIA simulation. The two scale variations are included as two separate uncertainties in the fit. Since the \POWHEG calculation is the same for all three years, these uncertainties are considered to be fully correlated.
\item Since the \mur and \muf variations barely account for the differences in the shape of the \pt spectrum from \POWHEG at NLO QCD and from \MATRIX at NNLO QCD precision, an additional uncertainty is introduced, which corresponds to the difference in the \pt spectra of the two calculations. This uncertainty is correlated among measurements in the three years.
\item With the PDF parameterizations NNPDF30\_nlo\_as\_0118 in the \CUET tune of the 2016 simulations and NNPDF31\_nnlo\_hessian\_pdfas in the CP5 tune of the 2017 and 2018 simulations, different PDFs are used. The PDF sets provide 100 variations representing their uncertainty. For each variation a nuisance parameter enters the fit. In addition, a variation of \alpS by 0.002 in the PDF parameterizations is taken into account. The distributions are obtained using the corresponding event weights. Only the uncertainties in the 2017 and 2018 measurements are treated as correlated, whereas the uncertainties in the 2016 measurement, which uses different PDFs, are assumed to be uncorrelated.
\item The uncertainty in the initial-state PS is estimated by varying the shower scale by factors of 0.5 and 2. The corresponding distributions are obtained using event weights. With $\alpS=0.1108$ (0.118) for \CUET (CP5), the values are very similar, and we assume that this uncertainty is fully correlated among the measurements in all years.  
\item The uncertainty in the final-state PS is estimated by varying the shower scale by factors of 0.5 and 2. The corresponding distributions are obtained using event weights. With $\alpS=0.1365$ (0.118) for \CUET (CP5), we assume that this uncertainty is uncorrelated for the two tunes.
\item In \POWHEG, the matching between the matrix element calculation and the PS is controlled by the parameter \hd. The values used with the \CUET and CP5 tunes are $(1.58^{\:+0.66}_{\:-0.59})\Mtop$ and $(1.38^{\:+0.92}_{\:-0.51})\Mtop$, respectively. Since the values are similar, the uncertainty is taken as fully correlated among the measurements in different years. Separate samples produced with the uncertainty variations of \hd are used to obtain these uncertainties.
\item Separate samples produced with $\Mtop = 171.5$\GeV and $\Mtop = 173.5$\GeV are used to conservatively estimate the uncertainty due to the \Mtop measurement. This ${\pm}1\GeV$ variation is fully correlated among the measurements in different years.
\item The uncertainty in the UE modeling is estimated using separate samples that represent the envelope of uncertainties in the tunes. This uncertainty is fully correlated between the measurements in 2017 and 2018, but taken as uncorrelated with the 2016 measurement, where a different tune is used.
\item The fraction of leptonically decaying \PQb hadrons is changed according to the known precision of the corresponding branching fractions~\cite{PDG} using event-based reweighting. This uncertainty is fully correlated among the measurements in different years.       
\item The uncertainty in the CR is assessed using an alternative model where the reconnection of colored particles from resonance decays, which is deactivated by default in \PYTHIA, is activated. The difference between these two is taken as a symmetric uncertainty. We assume that the amount of CR is fully correlated among the measurements in the three years.
\end{itemize}

The following experimental uncertainties are assessed. 
\begin{itemize}
\item The uncertainties in the integrated luminosity are 2.5, 2.3, and 2.5\% for 2016, 2017, and 2018, respectively. Their correlations are between 20 and 30\%~\cite{CMS-LUM-17-003,LUMI17,LUMI18}. These uncertainties and their correlations are directly used in the cross section fits.
\item The uncertainty in the pileup estimation is divided into two sources: one in the inelastic cross section of about 4.5\%~\cite{Aaboud:2016mmw} and another in the instantaneous luminosities. Since the former is dominant and fully correlated among the years, a high correlation of about 85\% is estimated. Response matrices for enhanced and reduced pileup are obtained by applying event weights to alter the distribution of the number of pileup interactions in the simulation.
\item The jet energy scale uncertainty is split into 20 different sources. The combined uncertainties are \pt- and $\eta$-dependent, with a magnitude that varies between 0.3 and 1.8\% for the relevant jets. The correlation among the years is evaluated for each source. The sources affect the AK4 and AK8 jets simultaneously, but in different ways. The differences in the response matrices are obtained by rescaling the jet momenta in the simulation.
\item The uncertainty in the subjet energy scale of boosted \tqh candidates is considered, where subjets with $\pt < 30$\GeV have energy scale uncertainties up to 3\%. This uncertainty has a correlation of 50\% among the years.
\item For AK4 and \PUPPI AK8 jets, separate uncertainties in the energy resolutions are introduced because the different methods of pileup corrections have large effects on the resolutions. The uncertainties are considered uncorrelated among the measurements in different years. The response matrices for different jet resolutions are obtained by rescaling the jet resolution in the simulation.
\item The dominant source of uncertainty in \ptmiss is the uncertainty in the jet energy calibration. Therefore, \ptmiss is also recalculated whenever the jet momenta are rescaled for uncertainty estimations. An additional contribution to the uncertainty comes from PF particles that do not belong to any of the selected jets. A 50\% correlation of this uncertainty among the years is assumed.
\item Uncertainties in the tagging and mistagging efficiencies of various \PQb tagging requirements have been studied~\cite{BTV-16-002}.
In this analysis, tight and loose \PQb tagging requirements are used. To simulate the variations of the efficiencies, event weights are calculated according to the probability to find the combination of correctly and wrongly identified jets under the assumption of altered efficiencies. We alter separately the efficiencies of the loose and tight requirements, and in addition, the efficiency of passing the loose but not the tight requirement. Each of the three variations introduces a nuisance parameter related to the tagging and mistagging efficiencies. We assume a 50\% correlation among the years.
\item For electrons and muons, uncertainties in the trigger and reconstruction efficiencies are included. Since the same methods for measuring these efficiencies in data are used in all years, the corresponding uncertainties are likely to have a common origin, and we assume a correlation of 50\% for these uncertainties among the years.
\item During the 2016 and 2017 data taking, a gradual shift in the timing of the inputs of the ECAL L1 trigger in the region of $\abs{\eta} > 2.0$ caused a trigger inefficiency. For events containing an electron (a jet) with \pt larger than $\approx$50\GeV ($\approx$100\GeV), in the region $2.5 < \abs{\eta} < 3.0$ the efficiency loss is $\approx$10--20\%, depending on \pt, $\eta$, and time. Correction factors were computed from data and applied to the acceptance evaluated by simulation. The uncertainties in these correction factors are propagated to the cross section measurements.
\end{itemize}

The description of uncertainties estimated from separate simulations (\hd, \Mtop, UE tune, and CR model) suffers from statistical fluctuations despite several tens of millions of simulated events. With a two-dimensional smoothing algorithm~\cite{LOWESS} applied to the relative uncertainties in the response matrices, we recover a meaningful description that is consistent among the different channels.

In addition, we estimate the effect of the limited event count in the simulations by repeating the entire fit 100 times with varied response matrices. From these results we calculate the covariance matrix, which is added to the covariance matrix of the other uncertainties obtained from the fits with the default response matrices. The response matrices are varied randomly taking into account statistical uncertainties in the default \POWHEG{}+\PYTHIA simulation and in the simulations used for the uncertainty estimations in \Mtop, \hd, the UE tune, and the CR model. For the uncertainties estimated based on event weights or rescaling of object momenta, the correlations between the default bin contents and the altered bin contents are taken into account when the random variations are generated. Overall, the uncertainty due to limited event count in the simulations is on the order of other leading systematic uncertainties. It becomes a dominant systematic uncertainty only in a few bins, but is always small compared to the statistical uncertainty in the data.

\begin{figure*}[tbp]
\centering
 \includegraphics[width=0.42\textwidth]{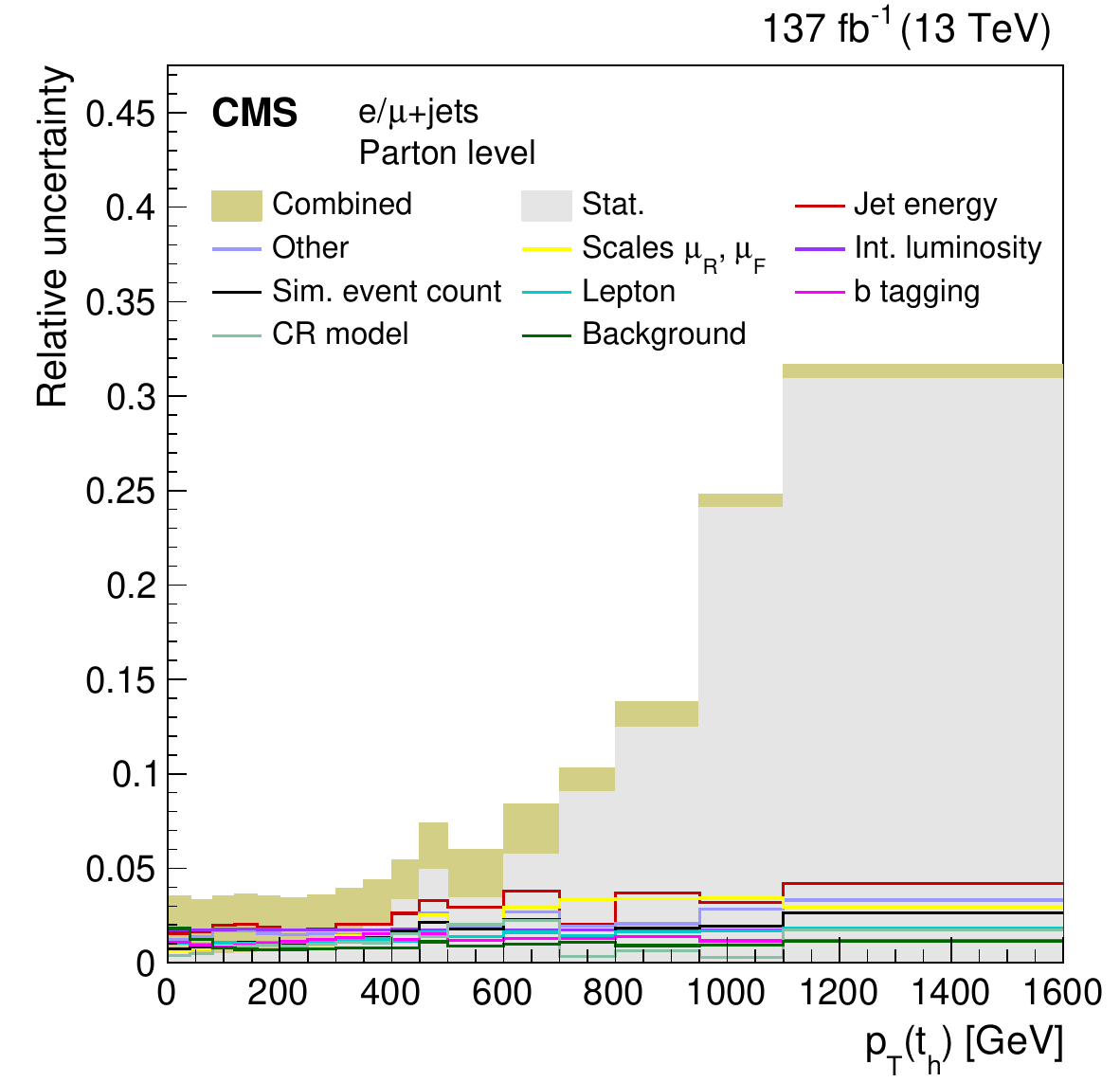}
 \includegraphics[width=0.42\textwidth]{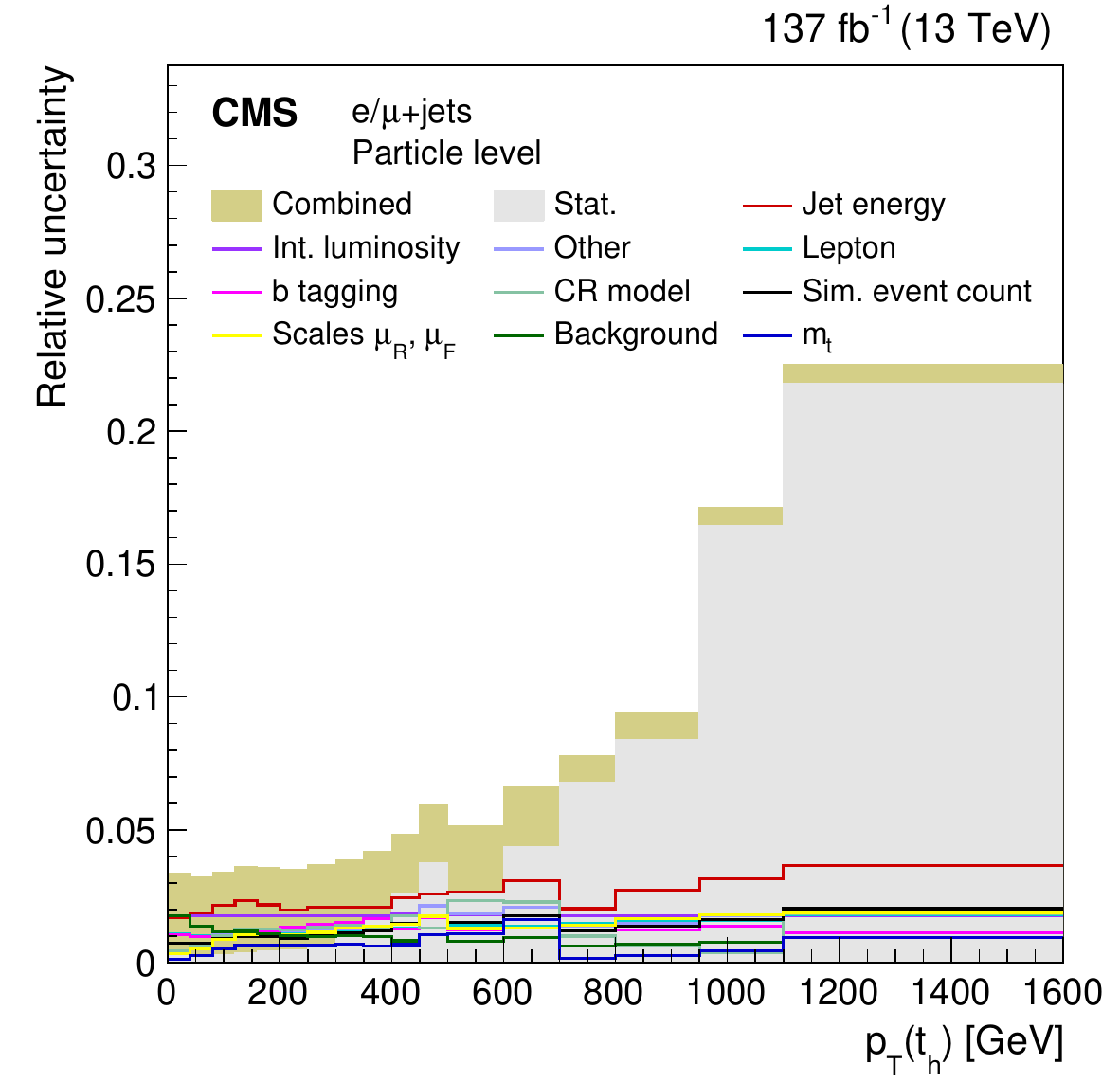}
 \includegraphics[width=0.42\textwidth]{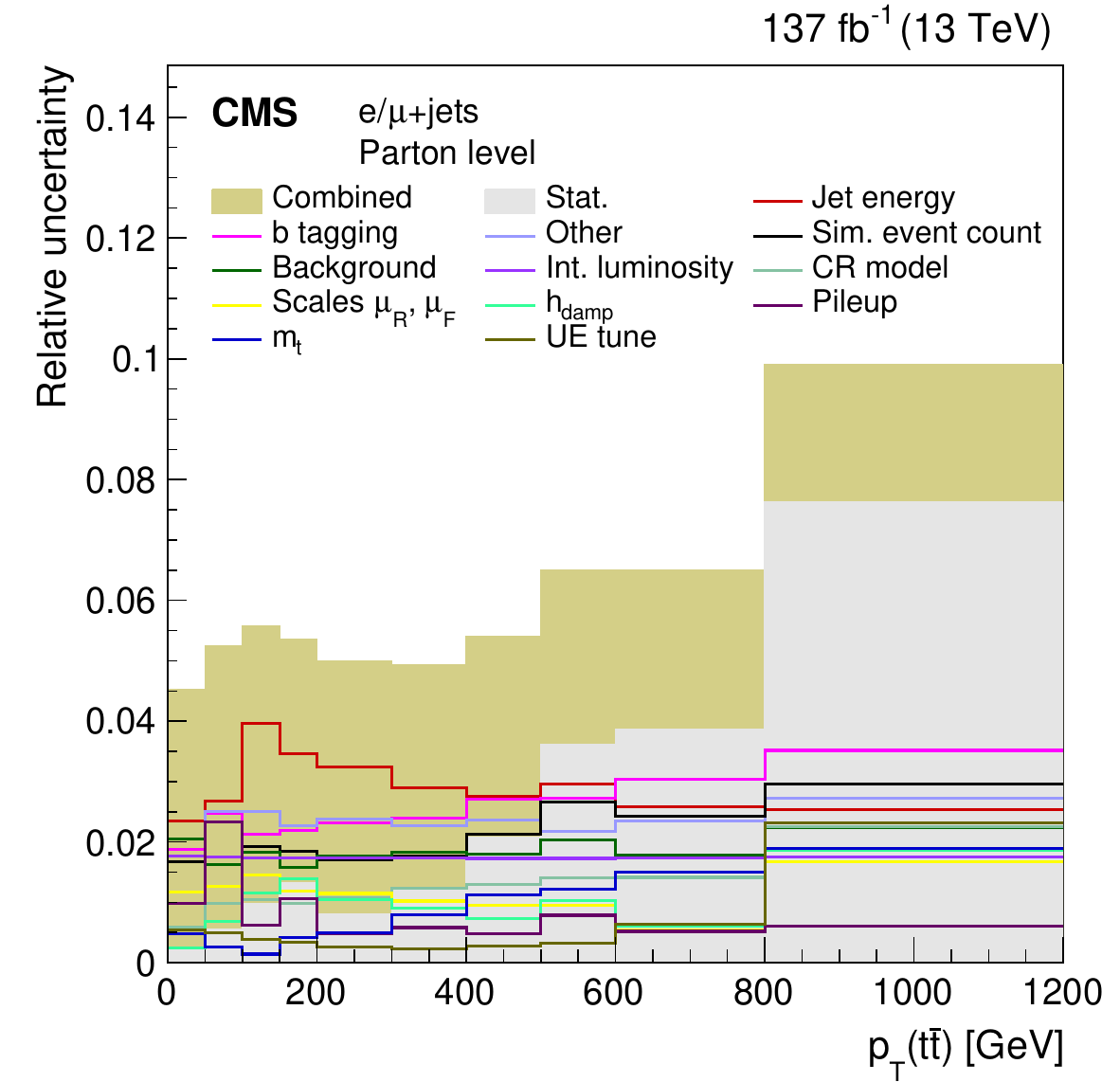}
 \includegraphics[width=0.42\textwidth]{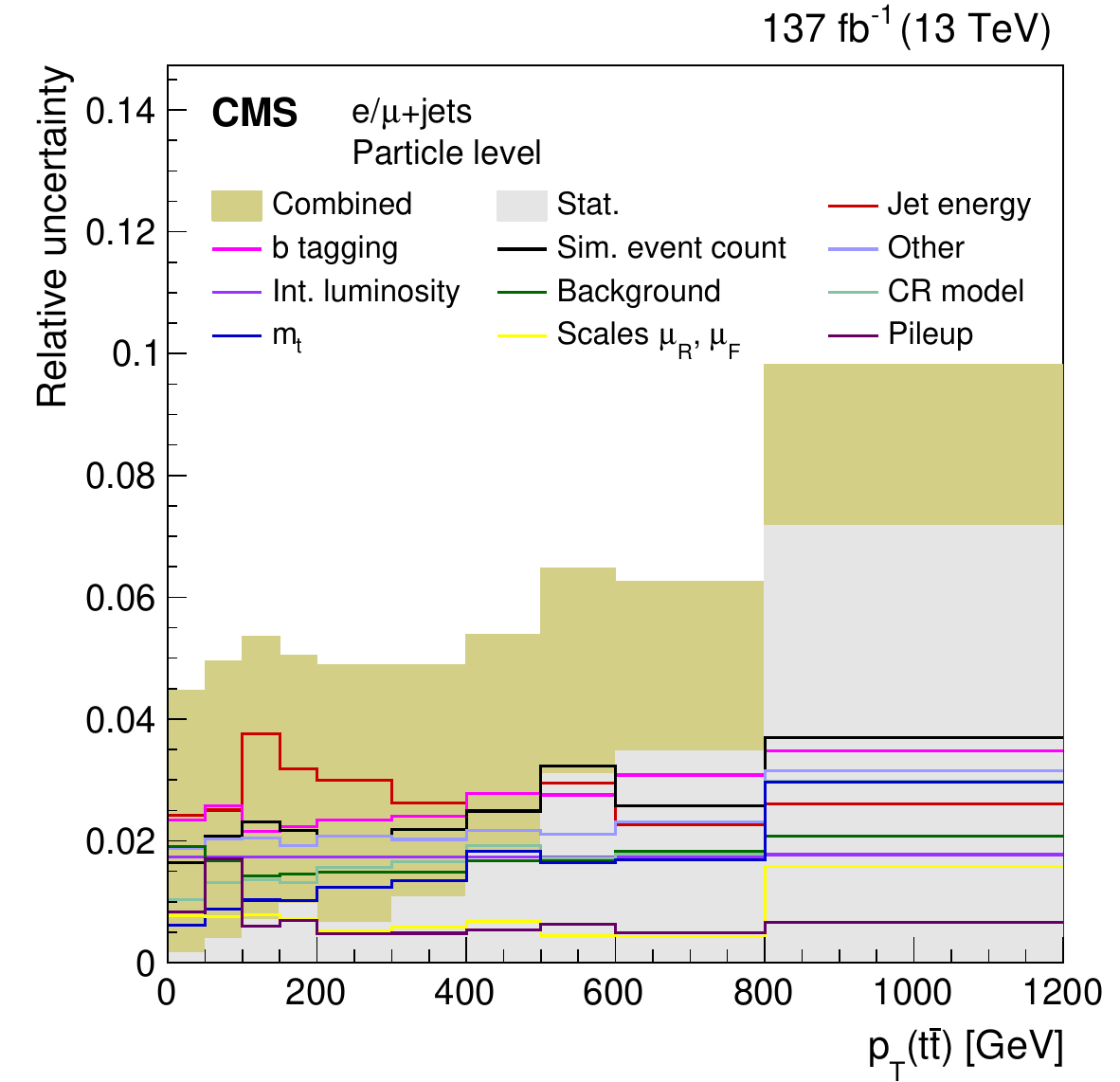}
 \includegraphics[width=0.42\textwidth]{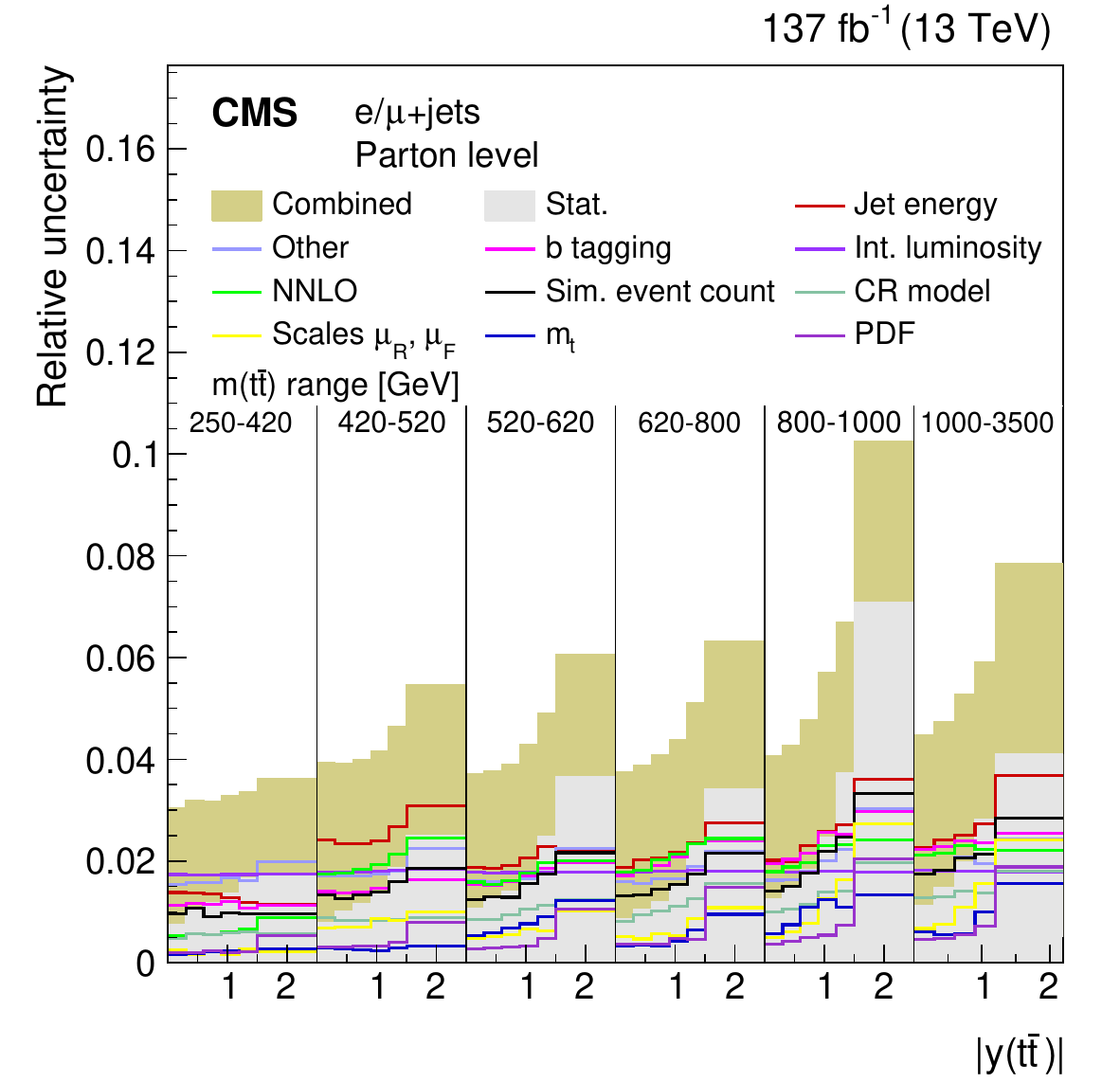}
 \includegraphics[width=0.42\textwidth]{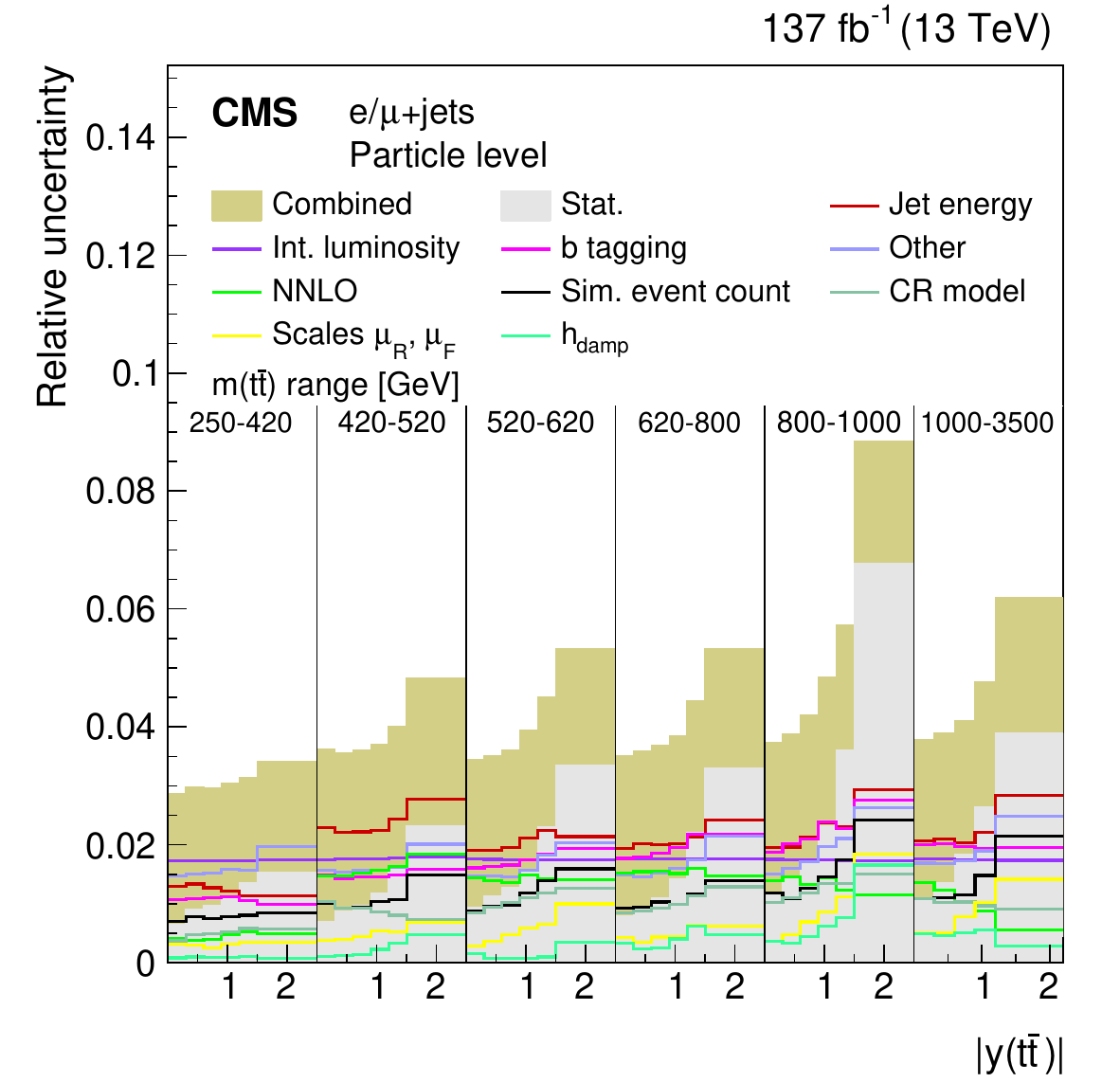}
 \caption{The individual sources of systematic uncertainties in the various parton (left) and particle (right) measurements and their relative contributions to the overall uncertainty. Sources with a maximum uncertainty below 1.5\% are combined in the category ``Other''.}
 \label{fig:SYS1}
\end{figure*}

 In Fig.~\ref{fig:SYS1}, the uncertainties in the measurements of a few differential cross sections are shown, split into individual sources. The statistical uncertainty in the data becomes dominant only in the tails of the distributions, while the bulk is dominated by the systematic uncertainties. The main contributions are from the uncertainty in the jet energy scale and the integrated luminosity. The leading theoretical uncertainties are the variations of \mur, \muf, and the uncertainty that reflects the difference in $\pt(\PQt)$ between the NLO and NNLO calculations.

\section{Results at the parton and particle levels}
\label{RES}

\begin{figure*}
\centering
 \includegraphics[width=0.75\textwidth]{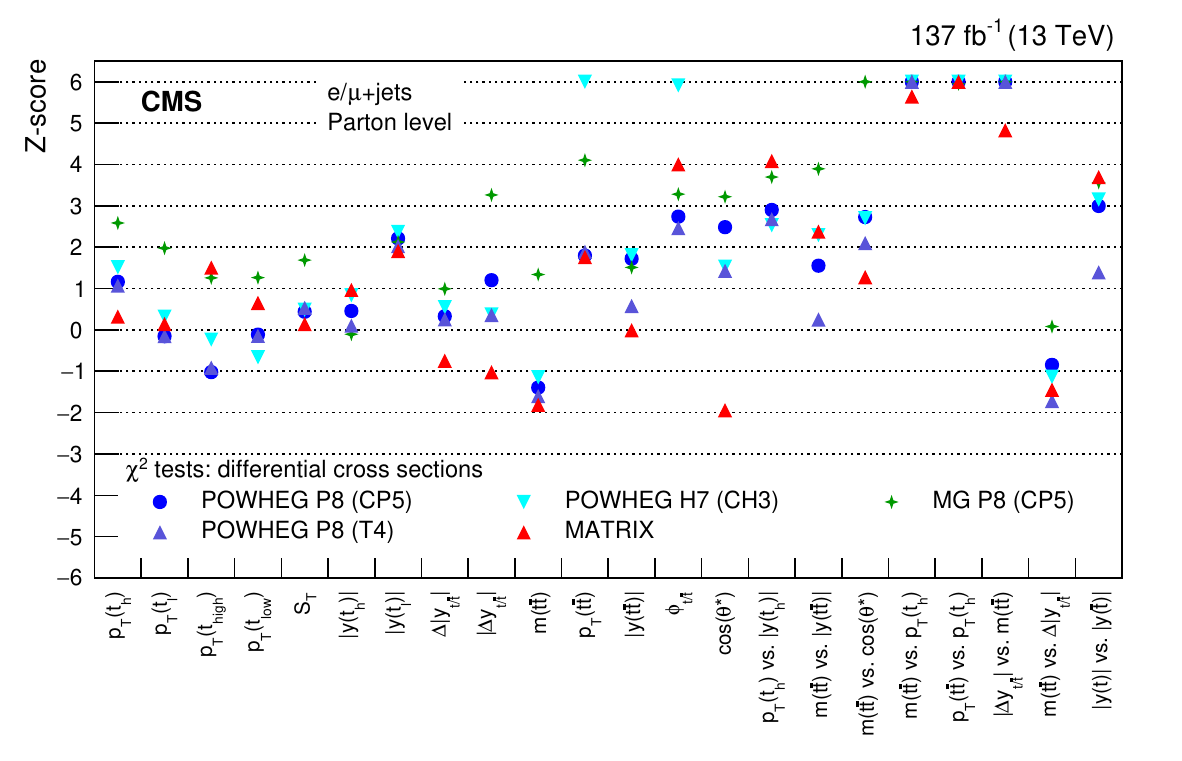}
 \includegraphics[width=0.75\textwidth]{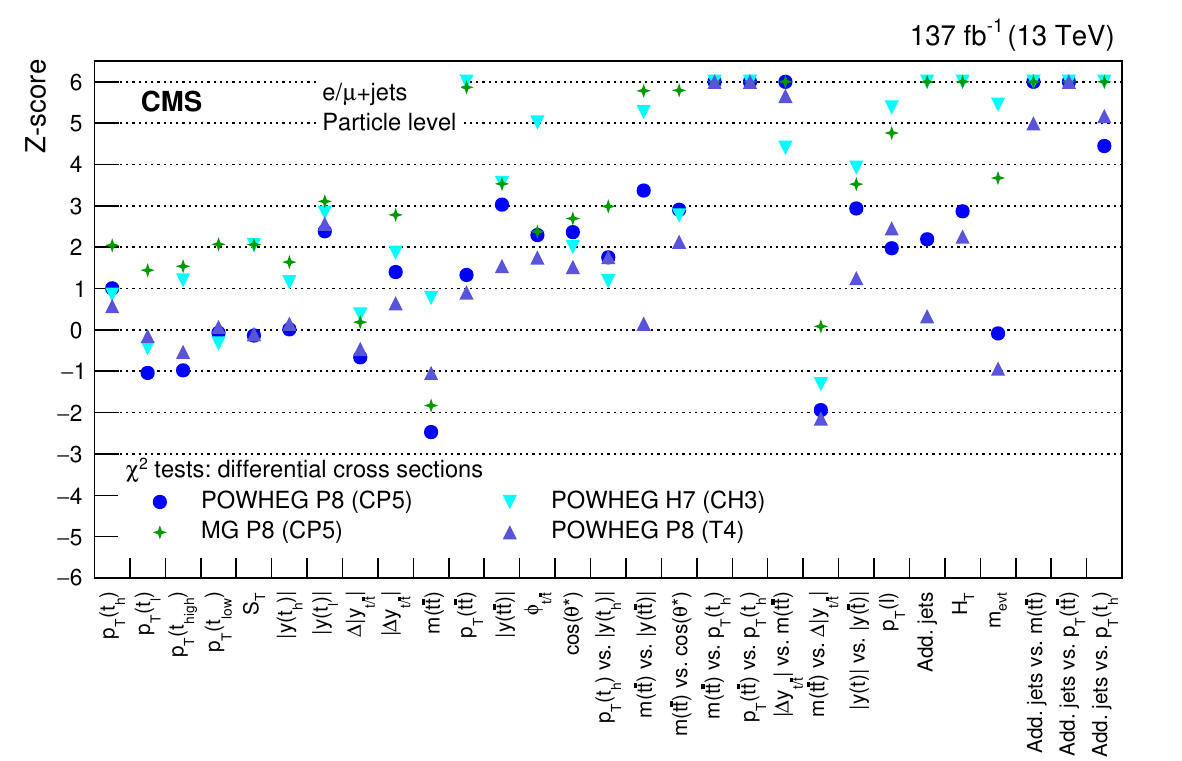}
 \caption{The $Z$-scores from the corresponding $\chi^2$ tests comparing the measured differential cross sections for the shown variables at the parton (upper) and particle (lower) levels to the predictions of \POWHEG{}+\PYTHIA (P8) for the CP5 and \CUET (T4) tunes, \POWHEG{}+\HERWIG (H7), the multiparton simulation \AMCATNLO(MG)+\PYTHIA FxFx, and the NNLO QCD calculations obtained with \MATRIX. The $Z$-scores are truncated at an upper limit of six. The uncertainties in the measurements and the predictions are taken into account in the $\chi^2$ calculation.}
 \label{fig:RES11}
\end{figure*}

\begin{figure*}[tbp]
\centering
 \includegraphics[width=0.42\textwidth]{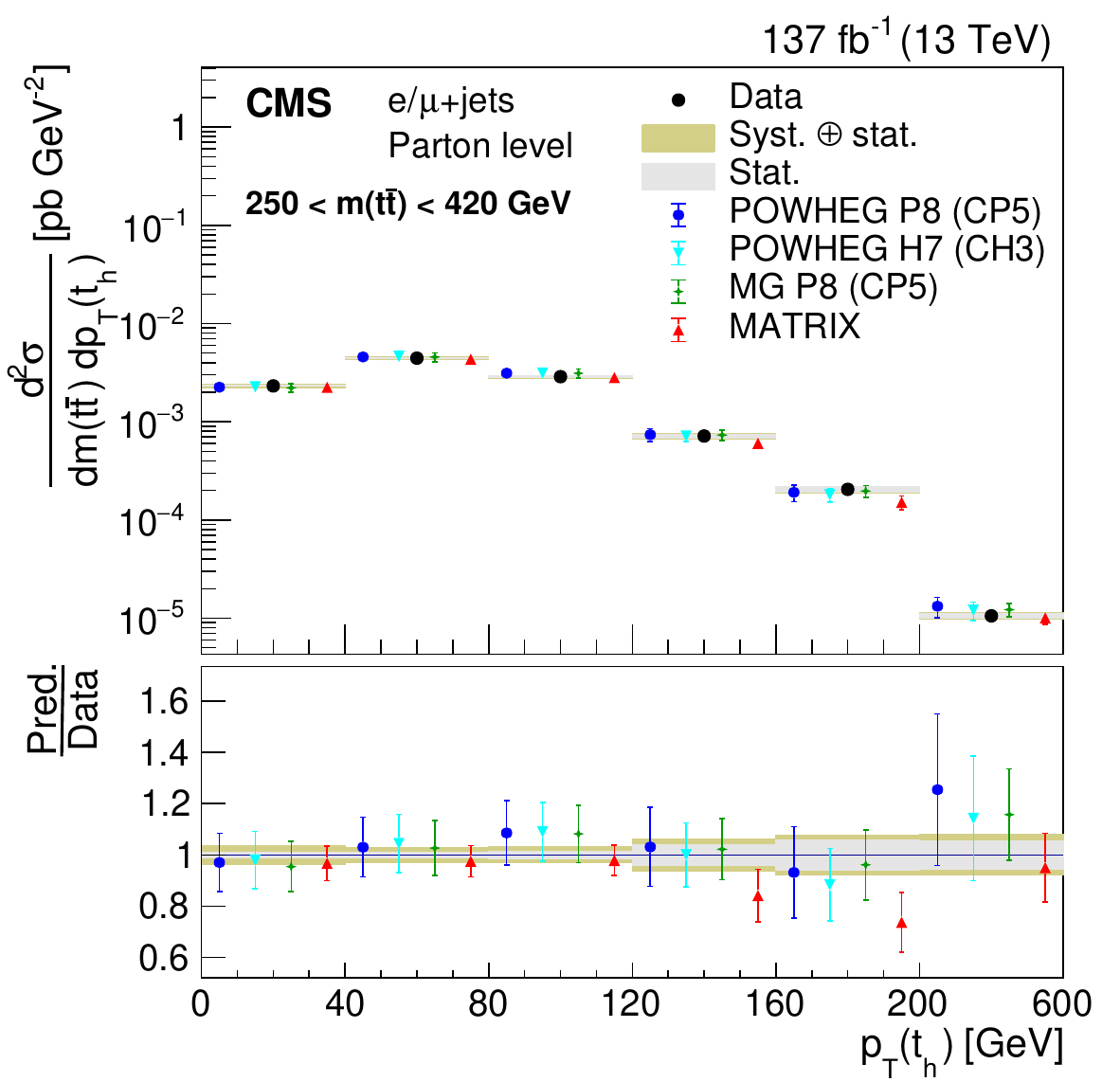}
 \includegraphics[width=0.42\textwidth]{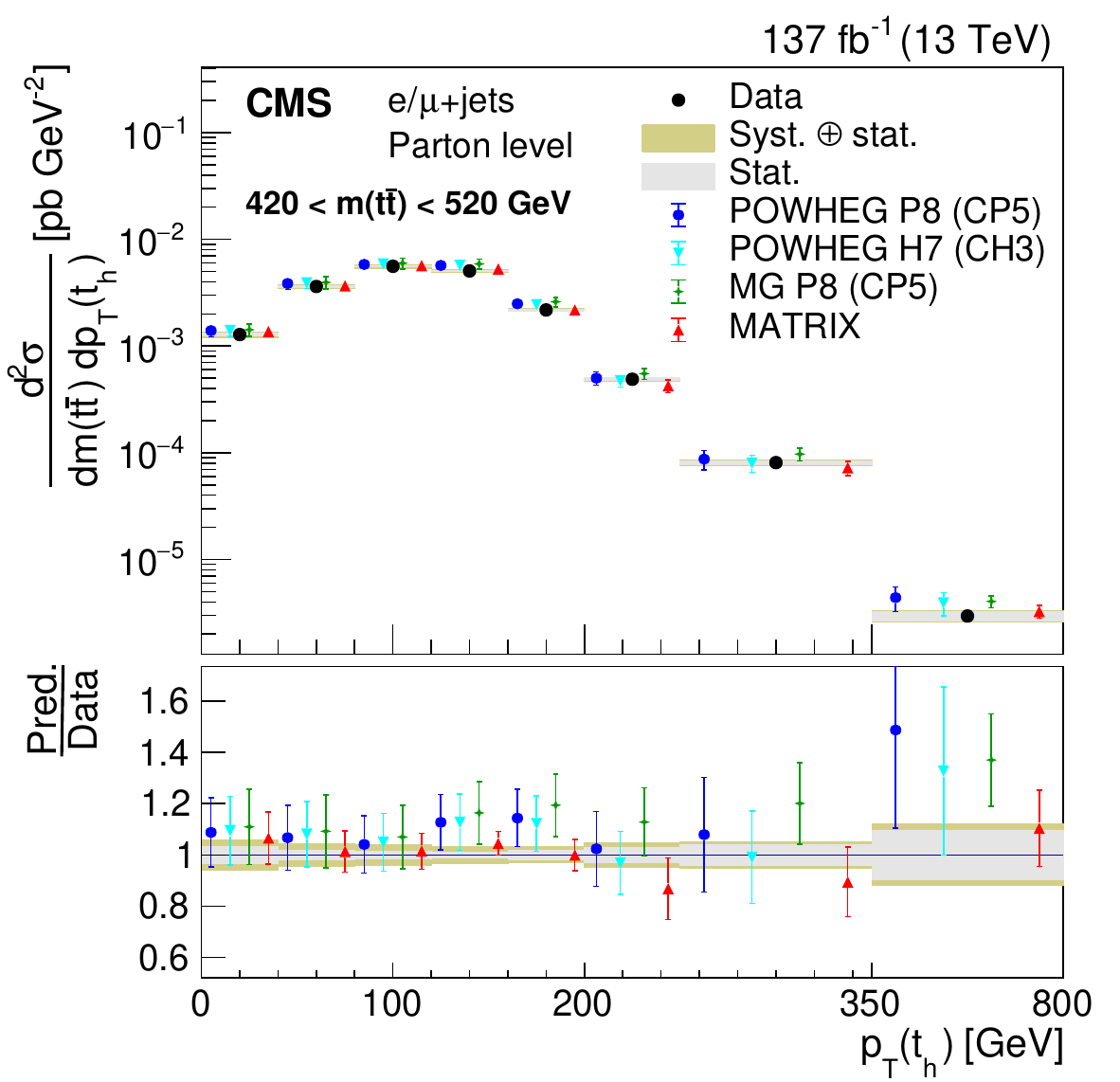}\\
 \includegraphics[width=0.42\textwidth]{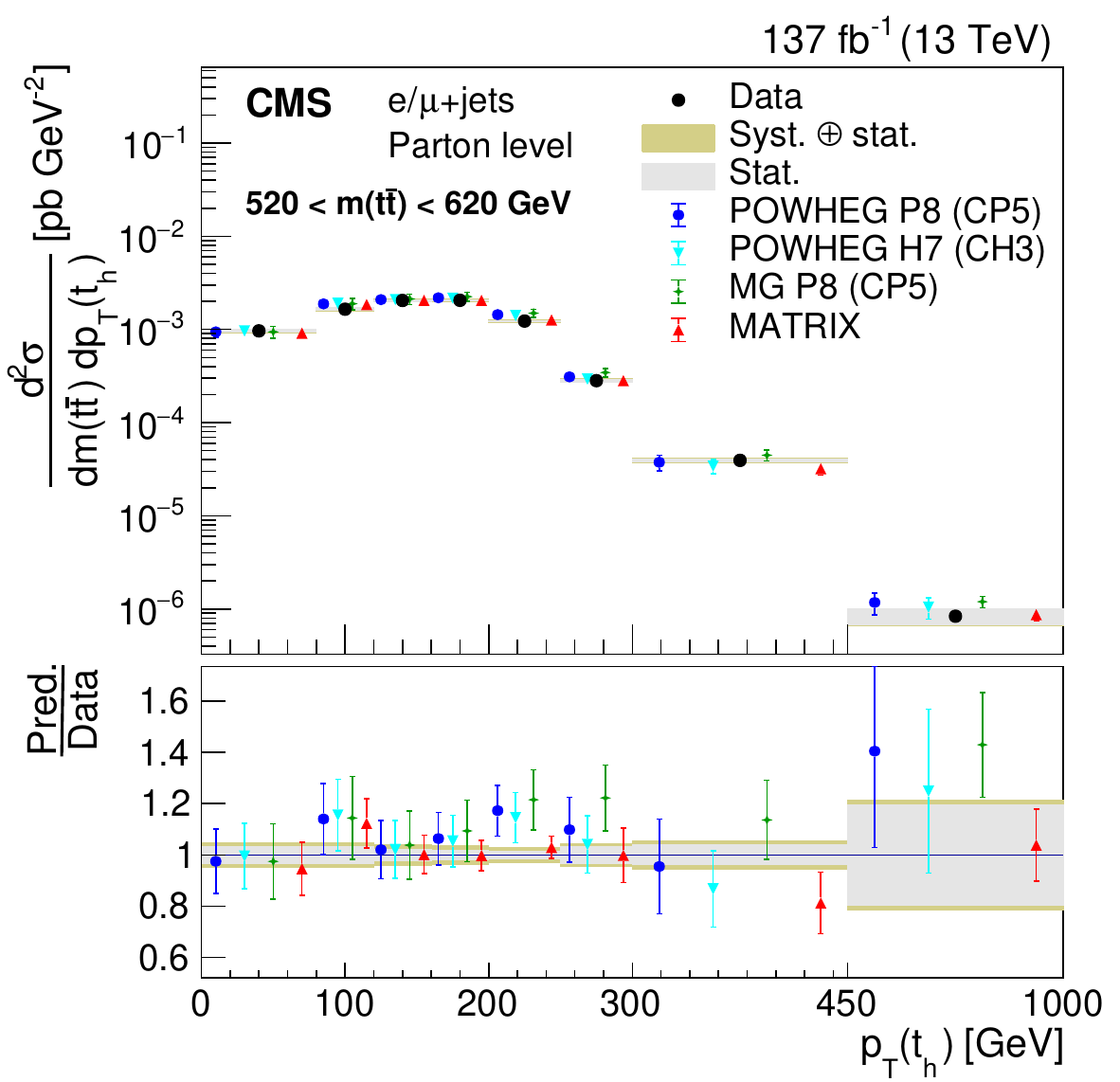}
 \includegraphics[width=0.42\textwidth]{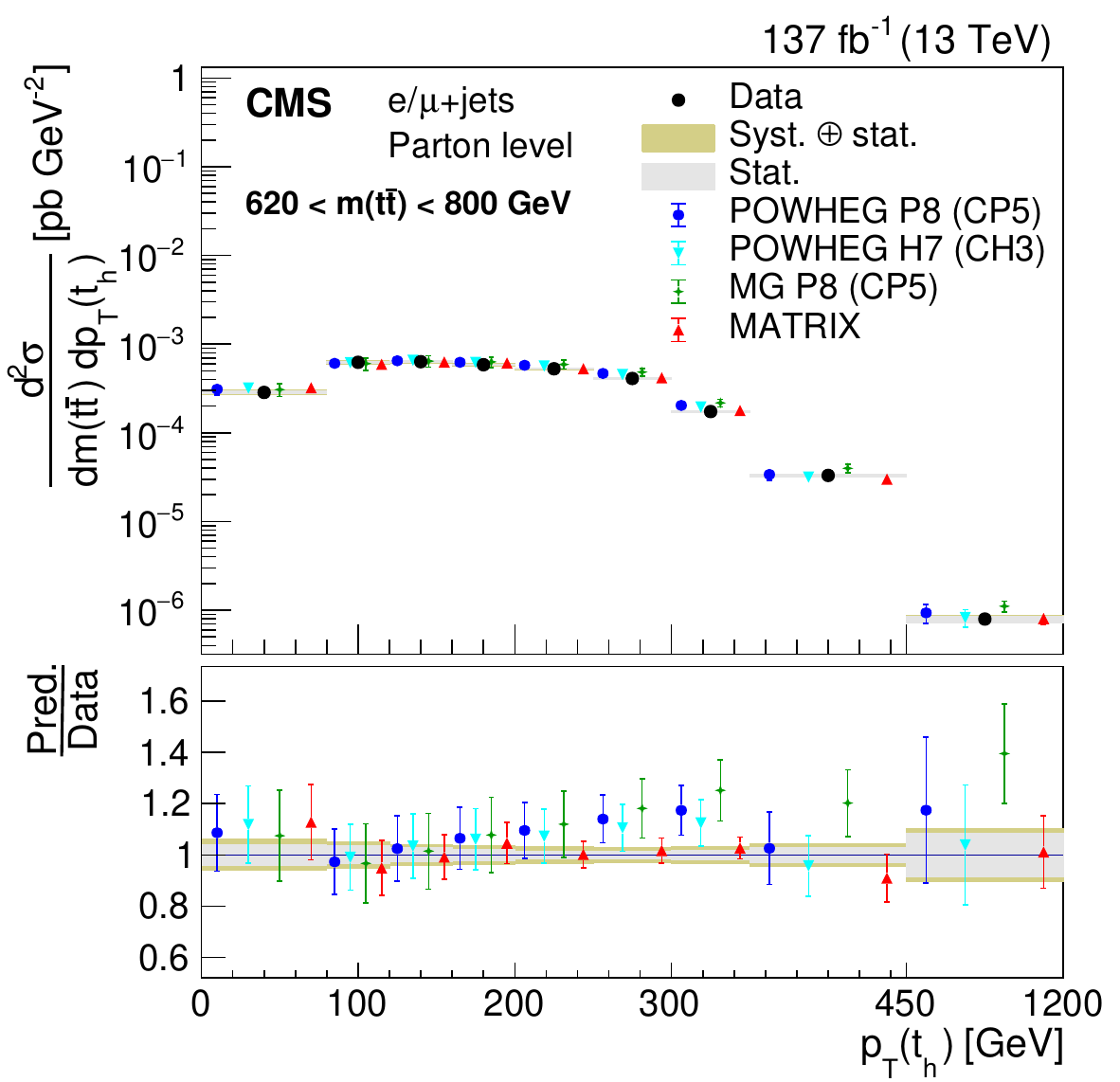}\\
 \includegraphics[width=0.42\textwidth]{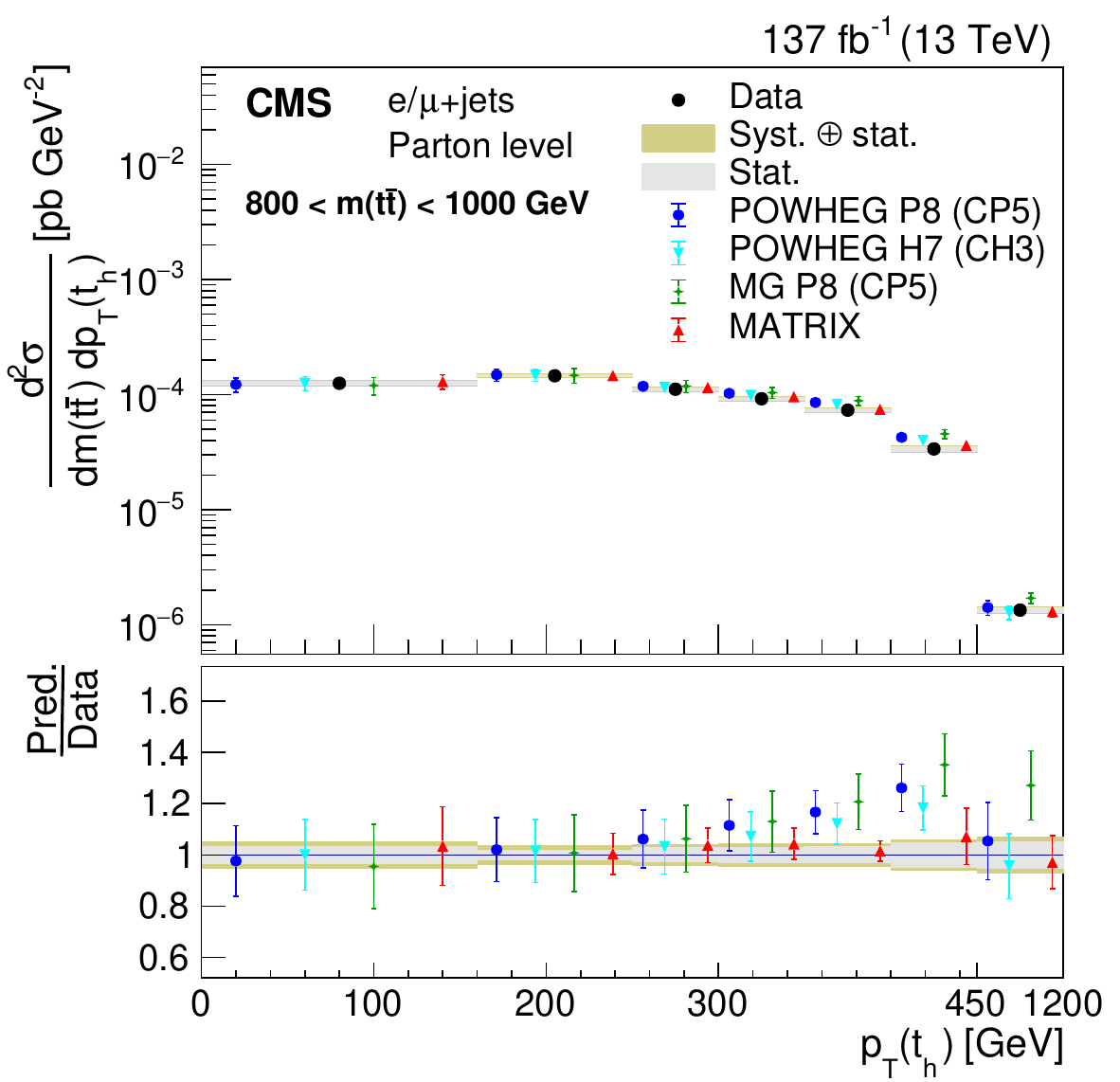}
 \includegraphics[width=0.42\textwidth]{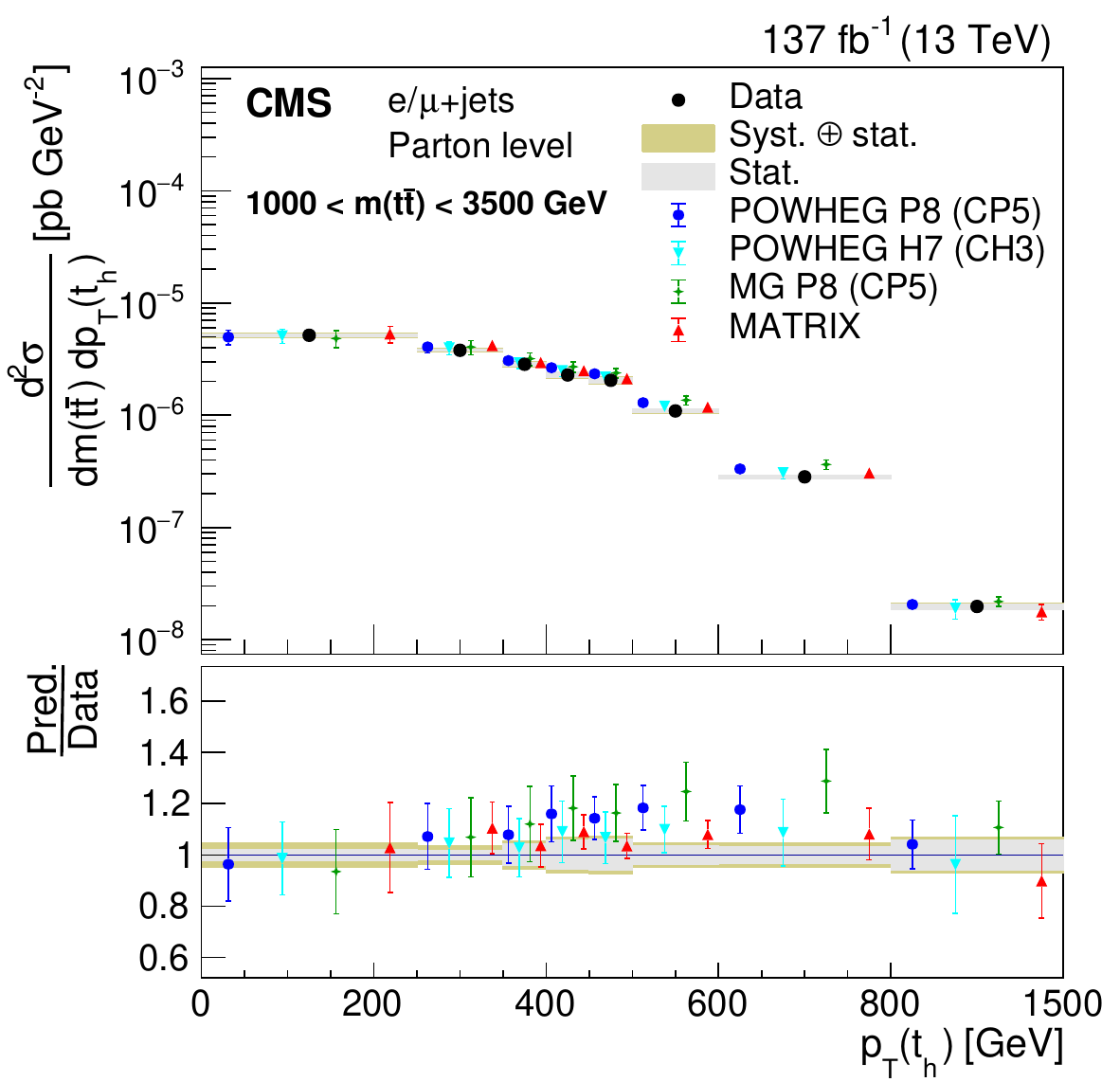}
 \caption{Double-differential cross section at the parton level as a function of \ttmvsthadpt. \XSECCAPPA}
 \label{fig:RES7}
\end{figure*}

\begin{figure*}[tbp]
\centering
 \includegraphics[width=0.42\textwidth]{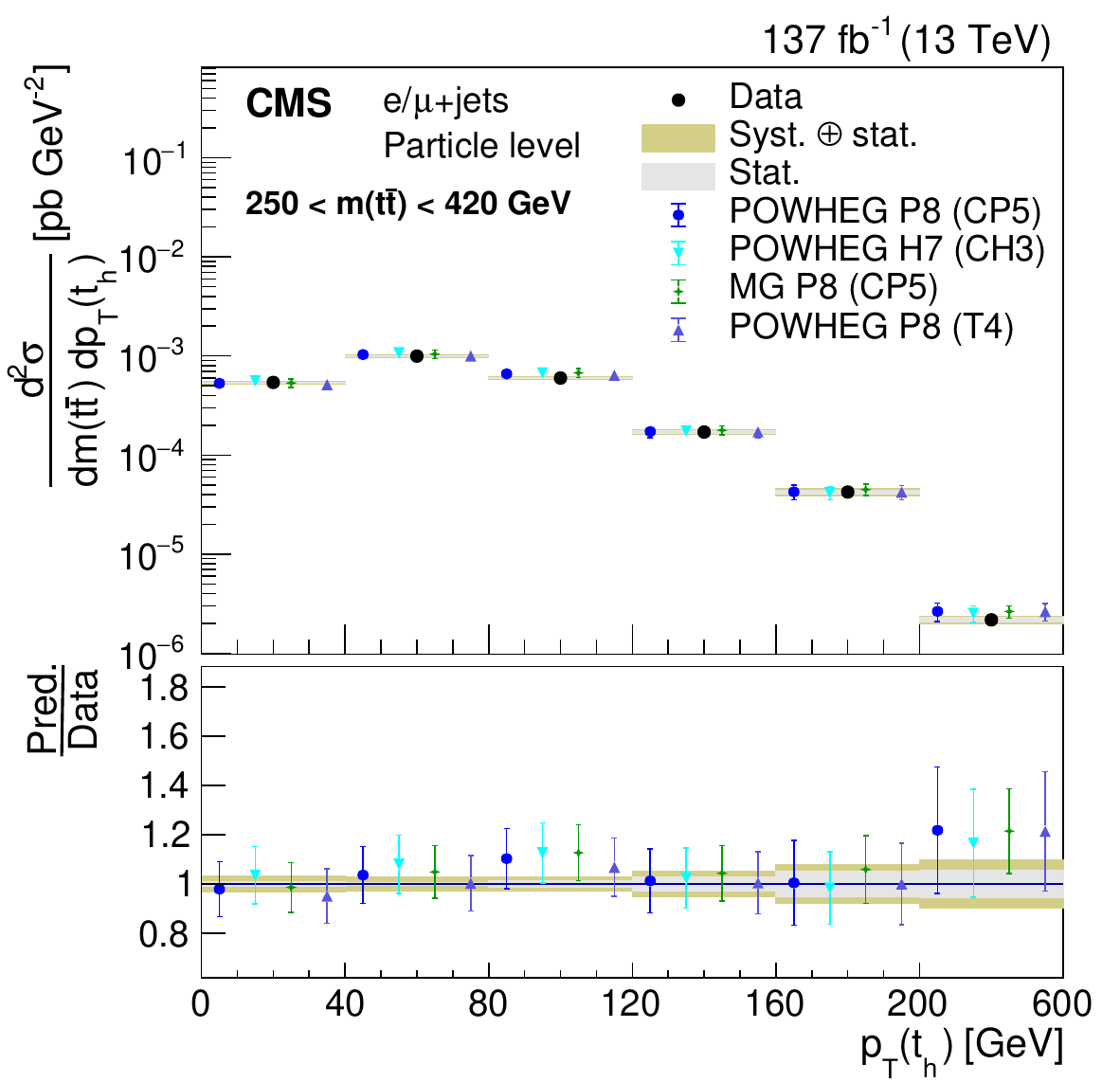}
 \includegraphics[width=0.42\textwidth]{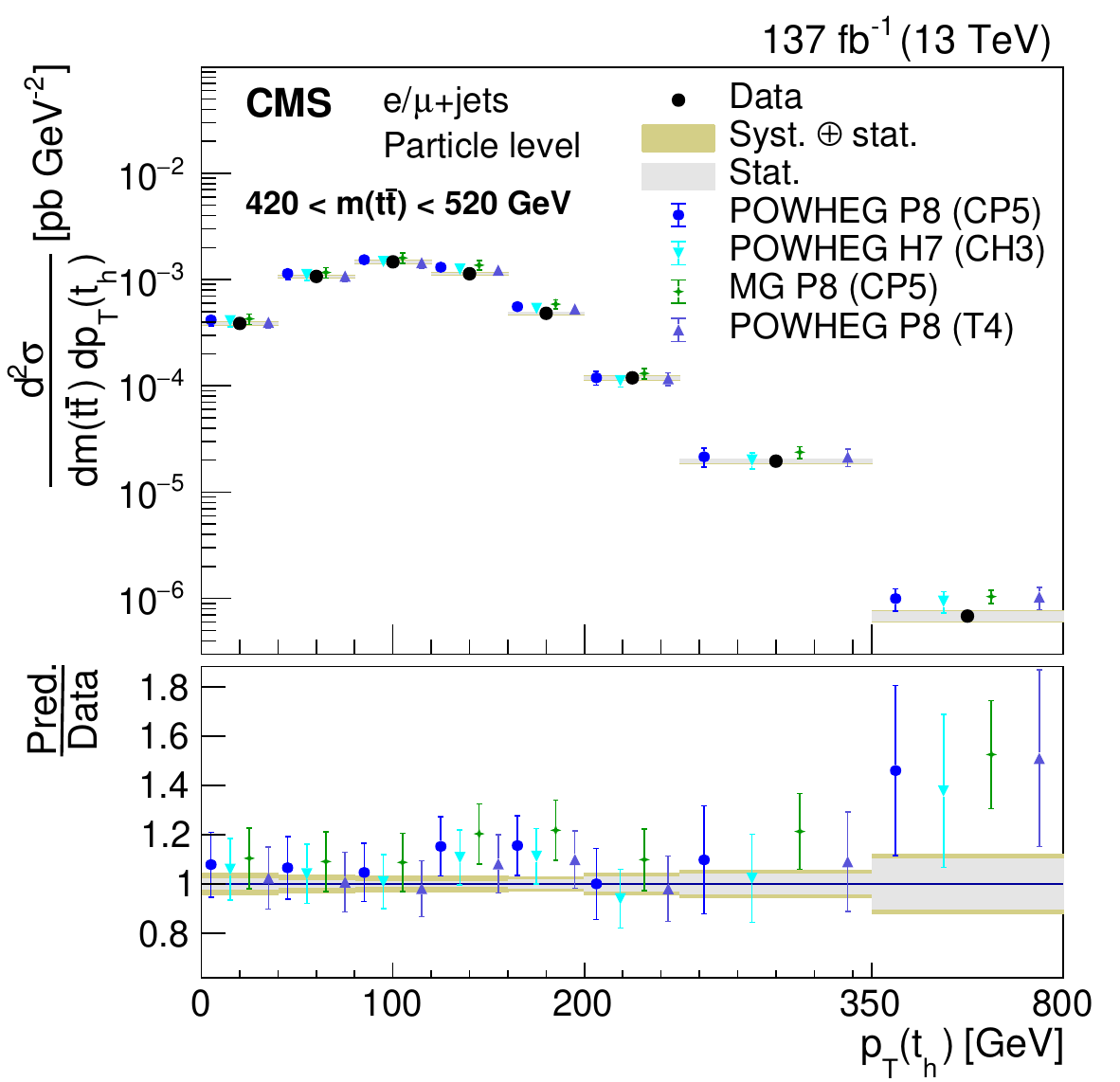}\\
 \includegraphics[width=0.42\textwidth]{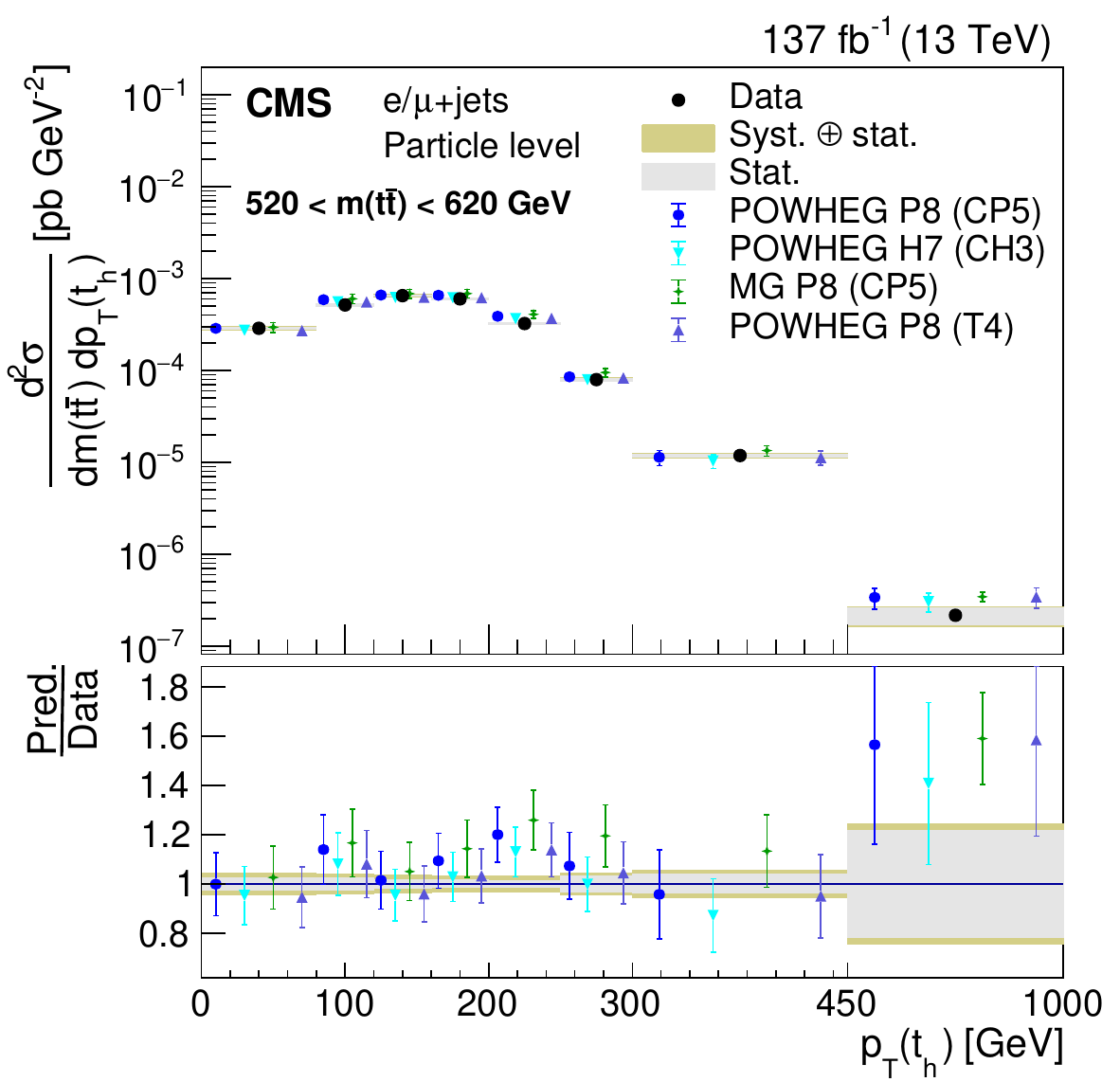}
 \includegraphics[width=0.42\textwidth]{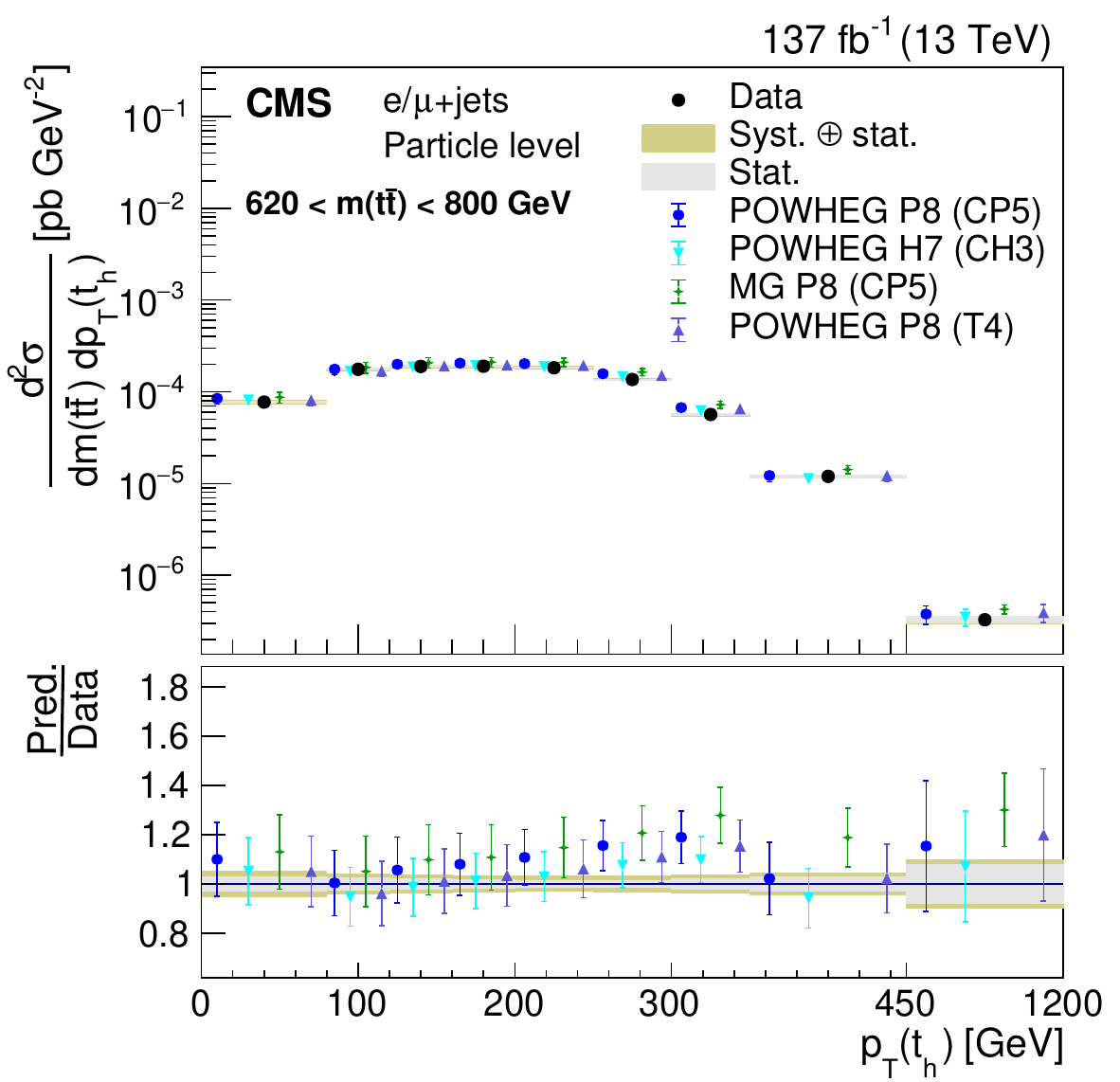}\\
 \includegraphics[width=0.42\textwidth]{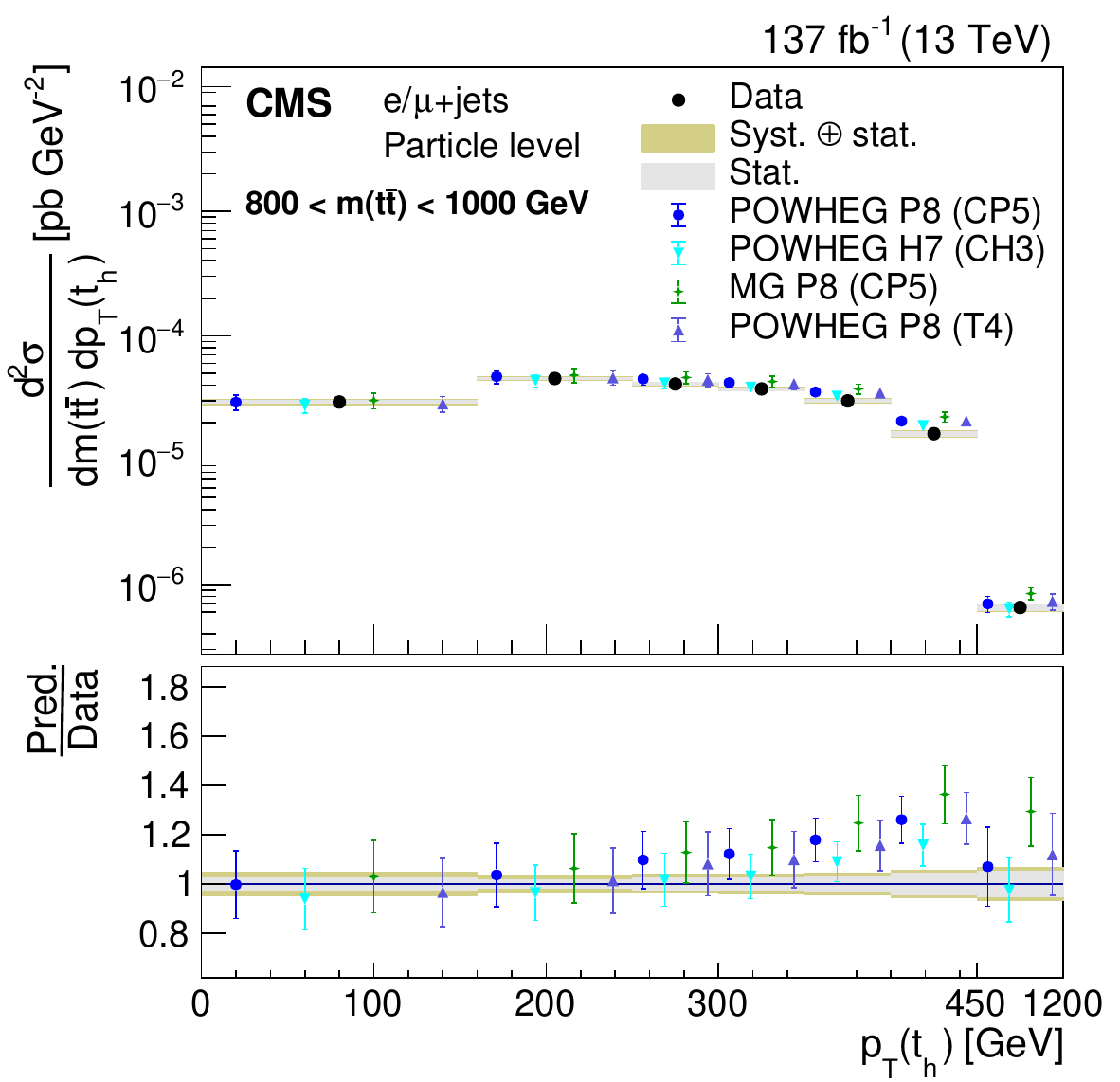}
 \includegraphics[width=0.42\textwidth]{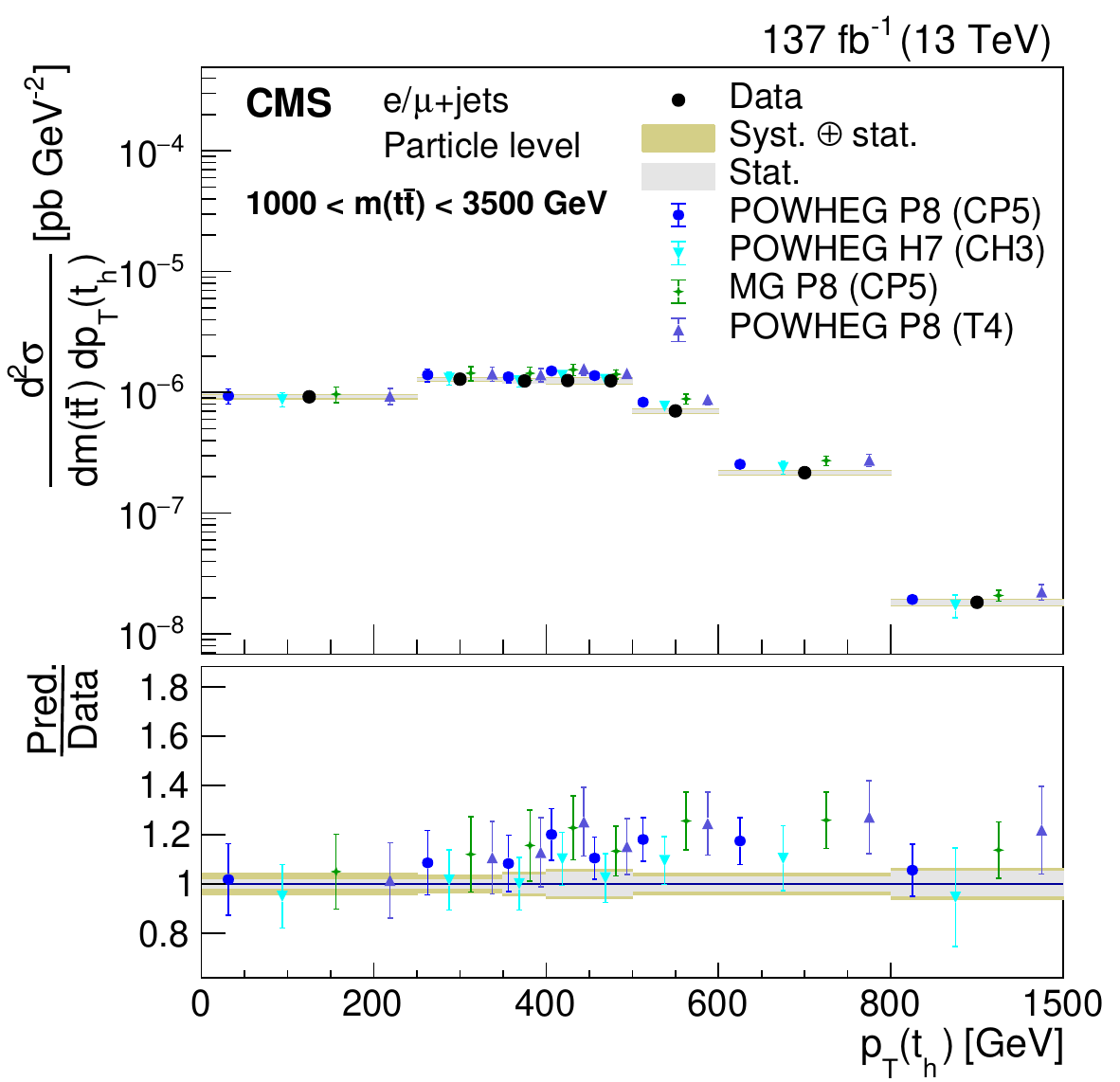}
 \caption{Double-differential cross section at the particle level as a function of \ttmvsthadpt. \XSECCAPPS}
 \label{fig:RESPS7}
\end{figure*}

\begin{figure*}[tbp]
\centering
 \includegraphics[width=0.42\textwidth]{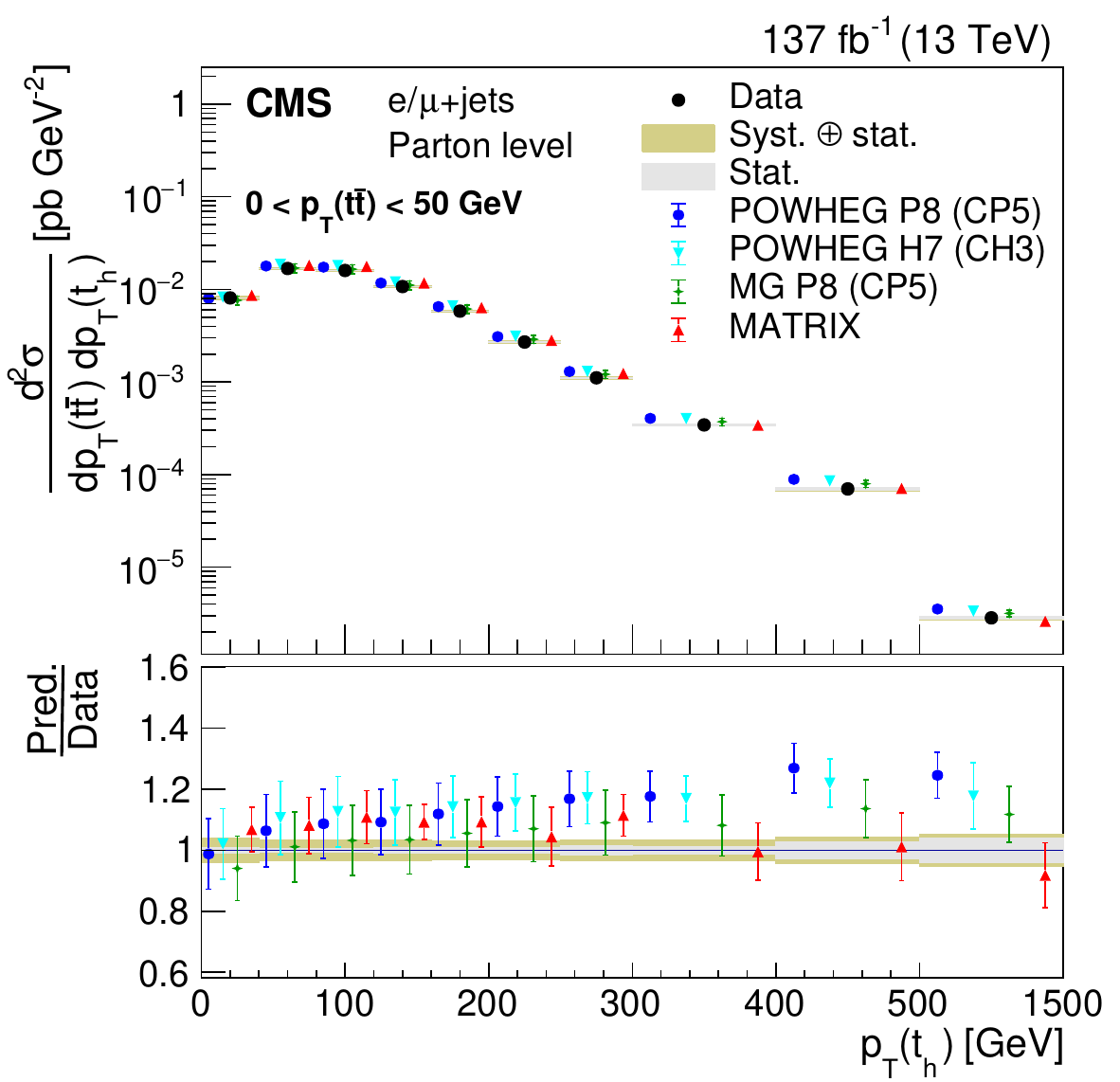}
 \includegraphics[width=0.42\textwidth]{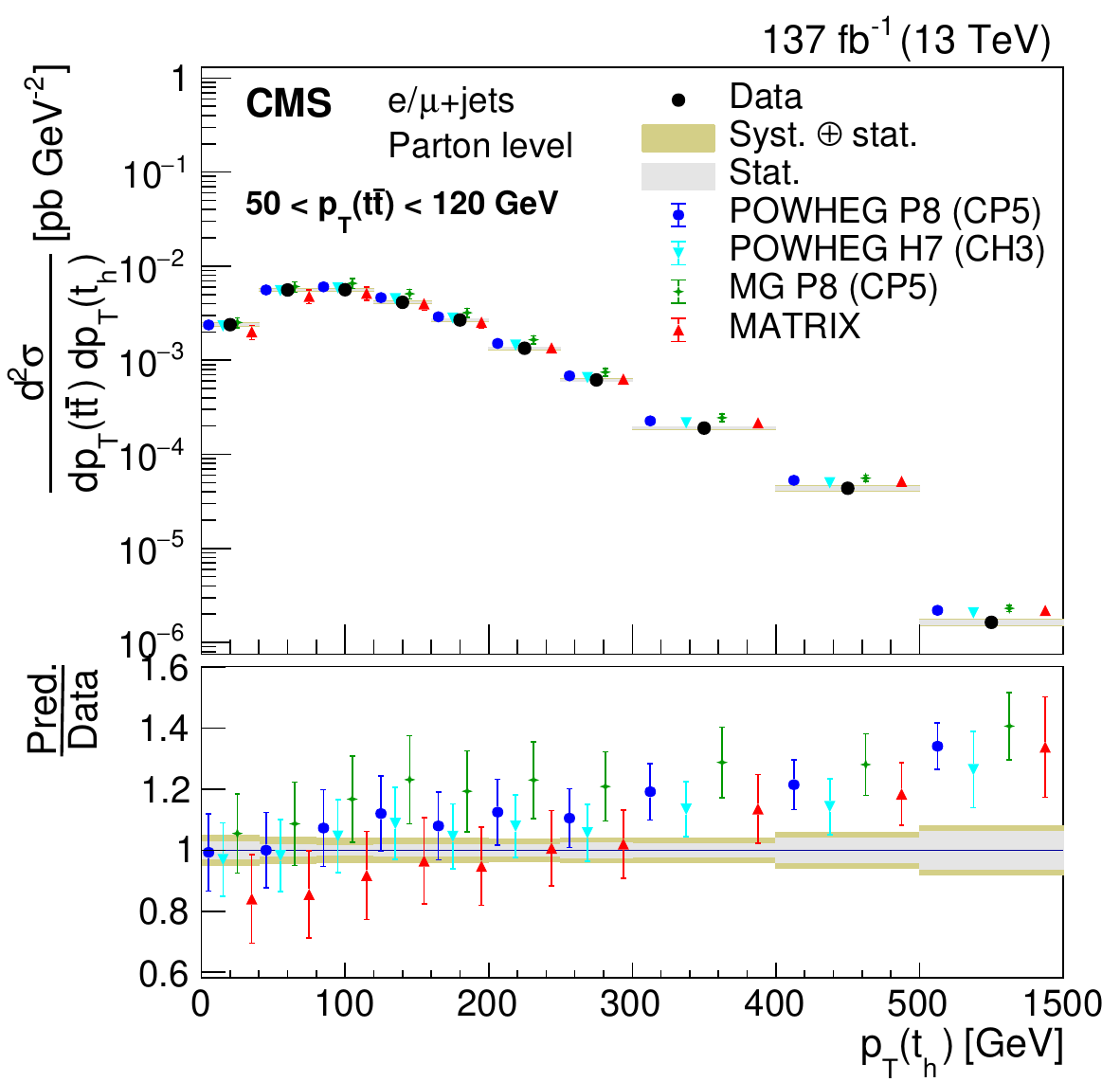}\\
 \includegraphics[width=0.42\textwidth]{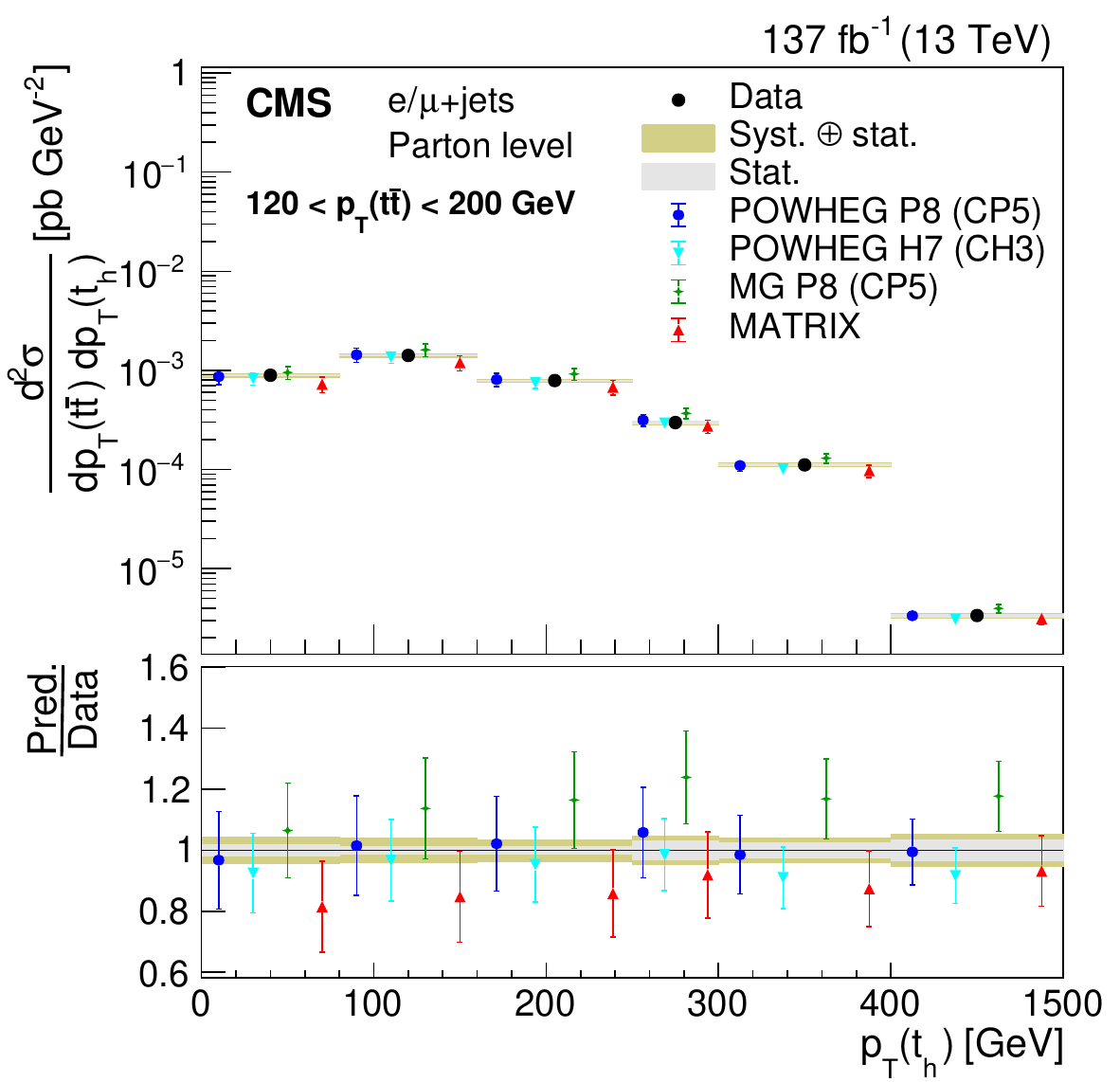}
 \includegraphics[width=0.42\textwidth]{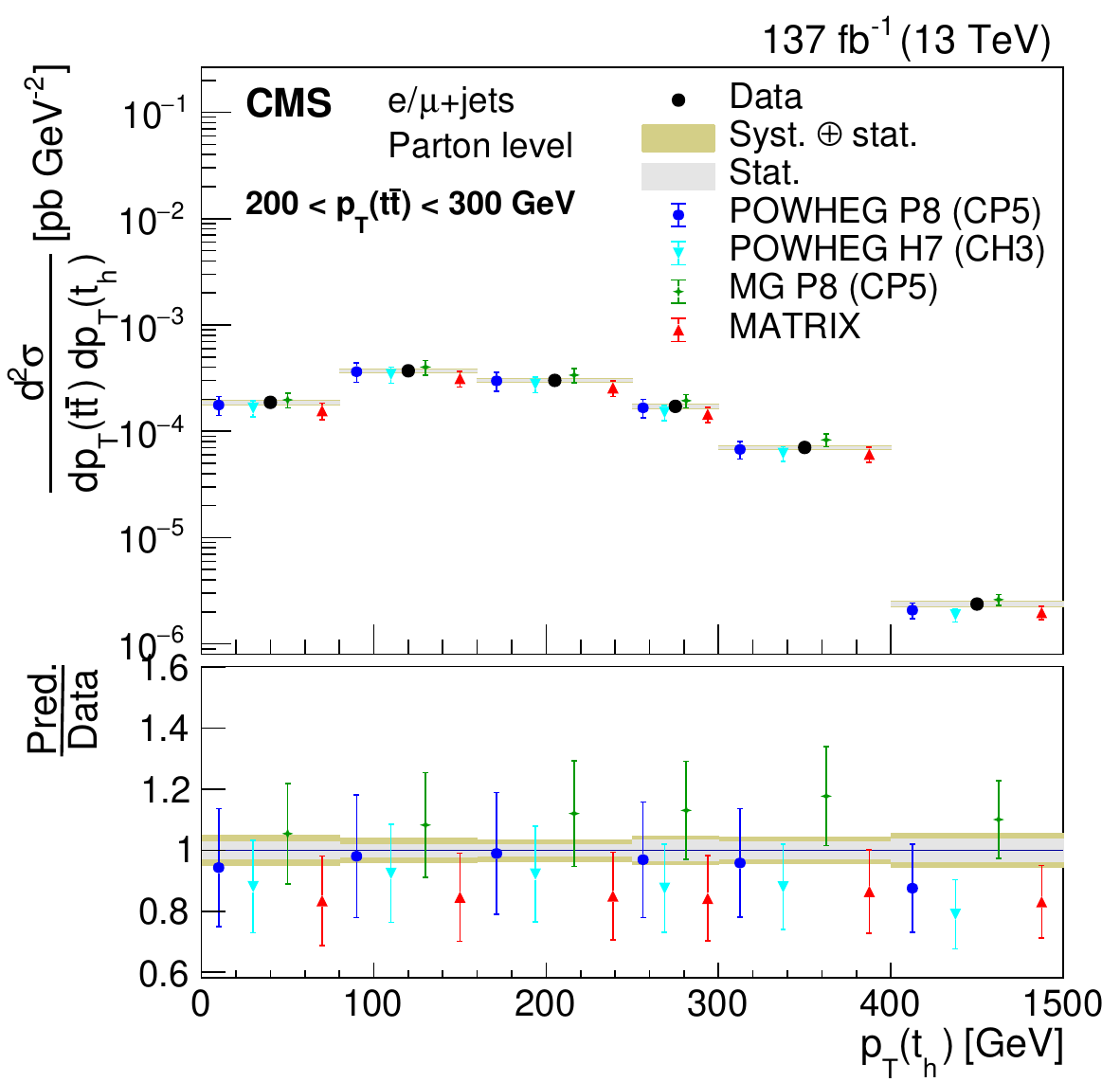}\\
 \includegraphics[width=0.42\textwidth]{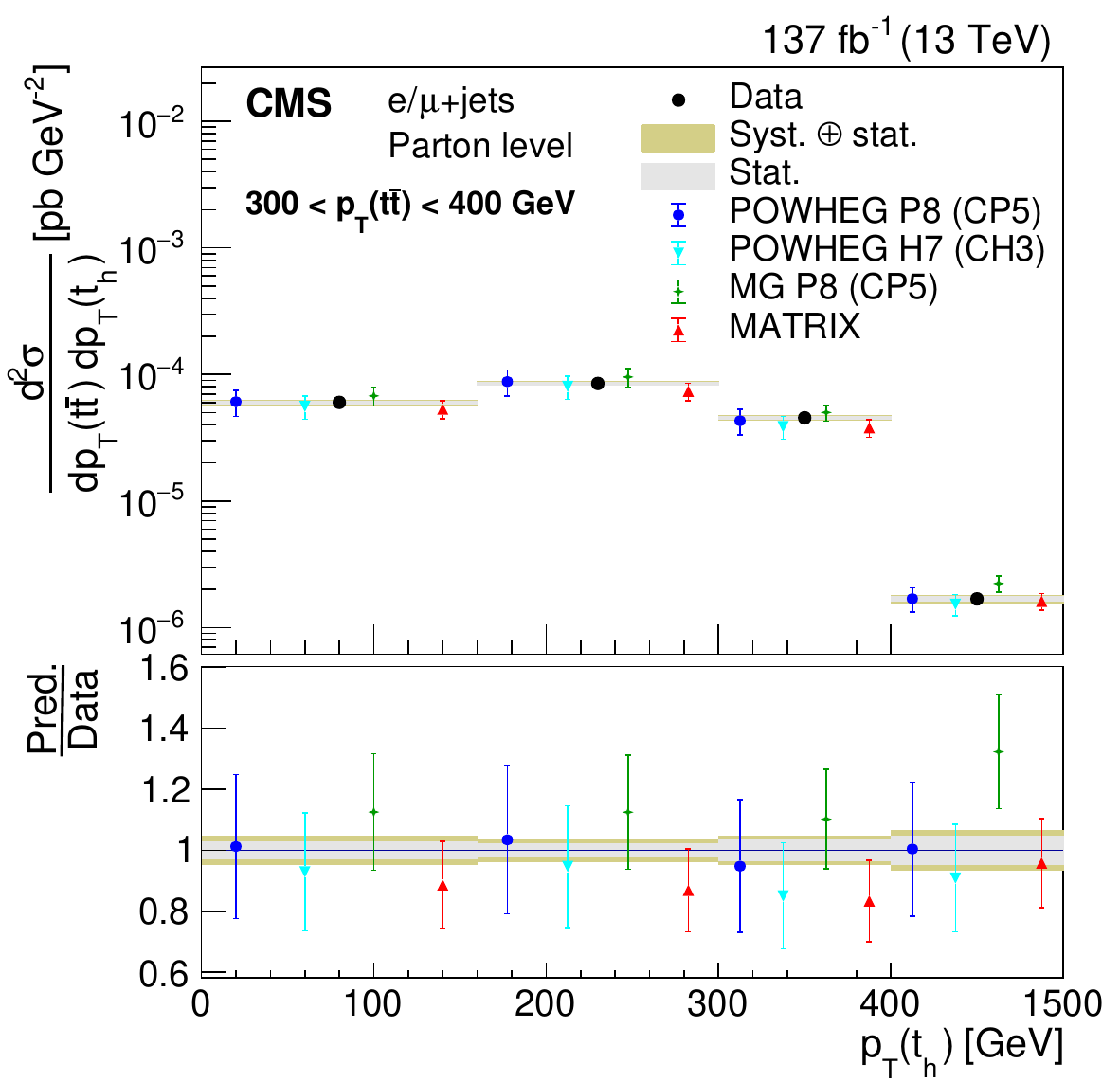}
 \includegraphics[width=0.42\textwidth]{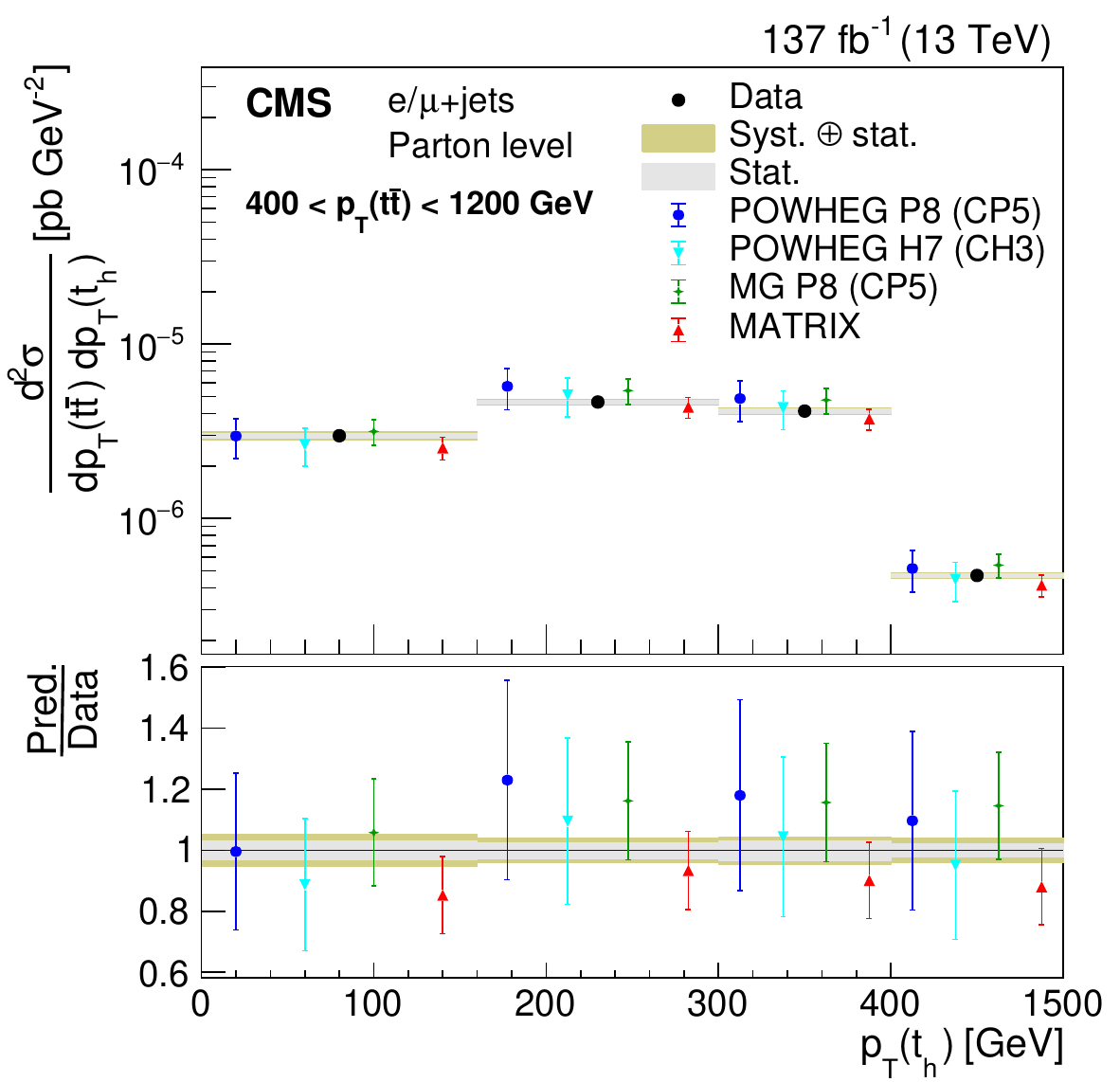}
 \caption{Double-differential cross section at the parton level as a function of \ttptvsthadpt. \XSECCAPPA}
 \label{fig:RES7b}
\end{figure*}

\begin{figure*}[tbp]
\centering
 \includegraphics[width=0.42\textwidth]{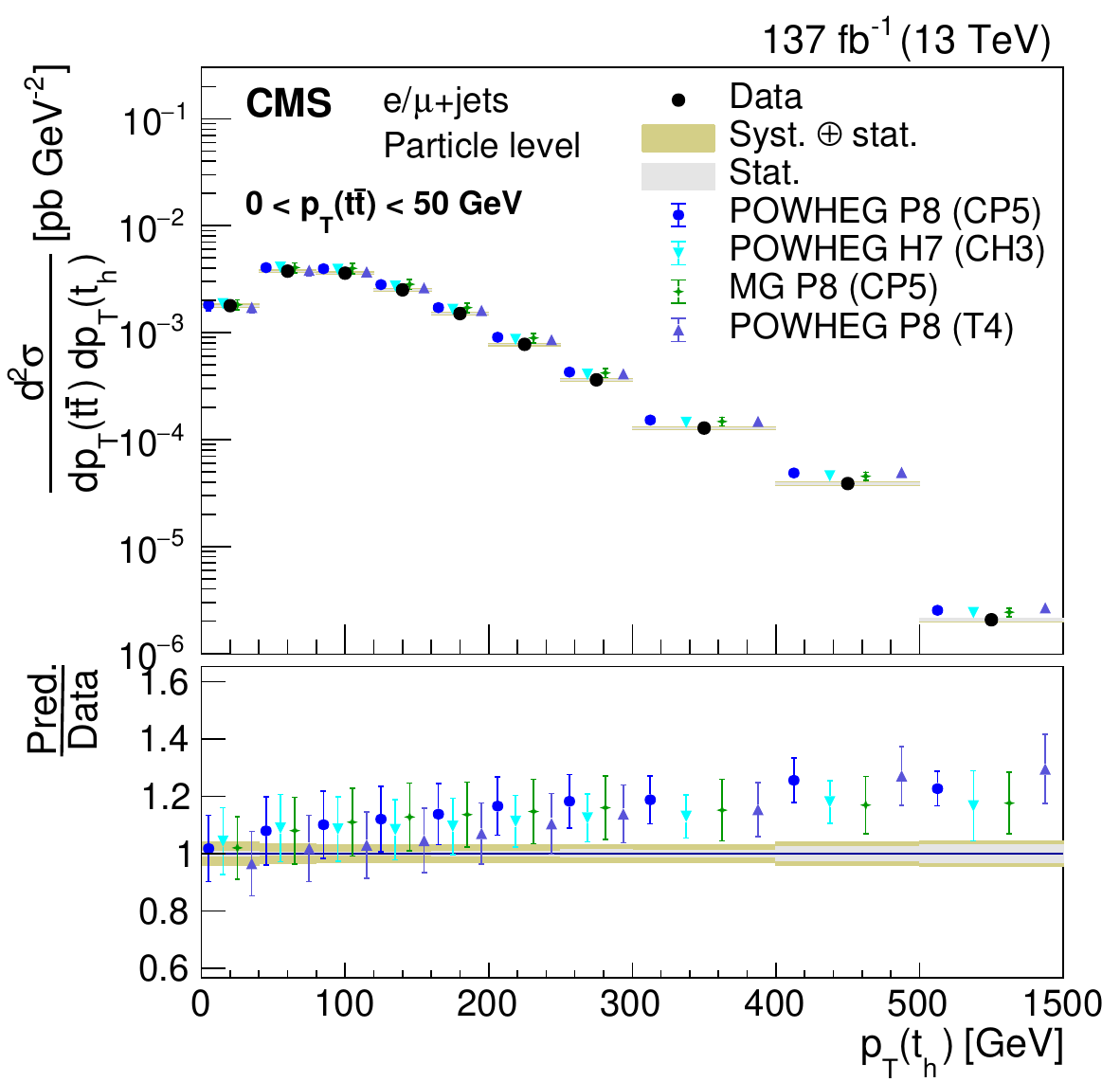}
 \includegraphics[width=0.42\textwidth]{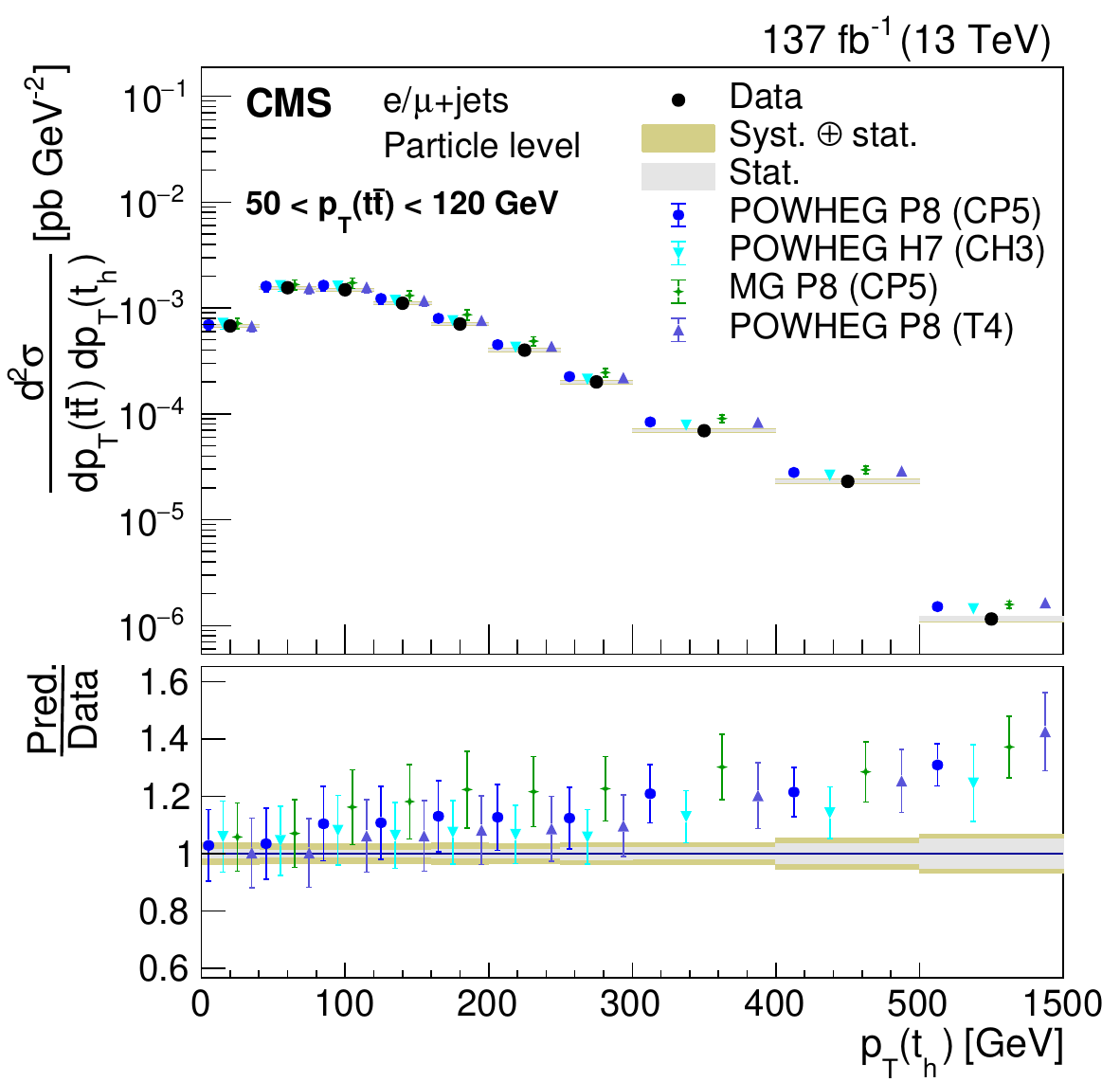}\\
 \includegraphics[width=0.42\textwidth]{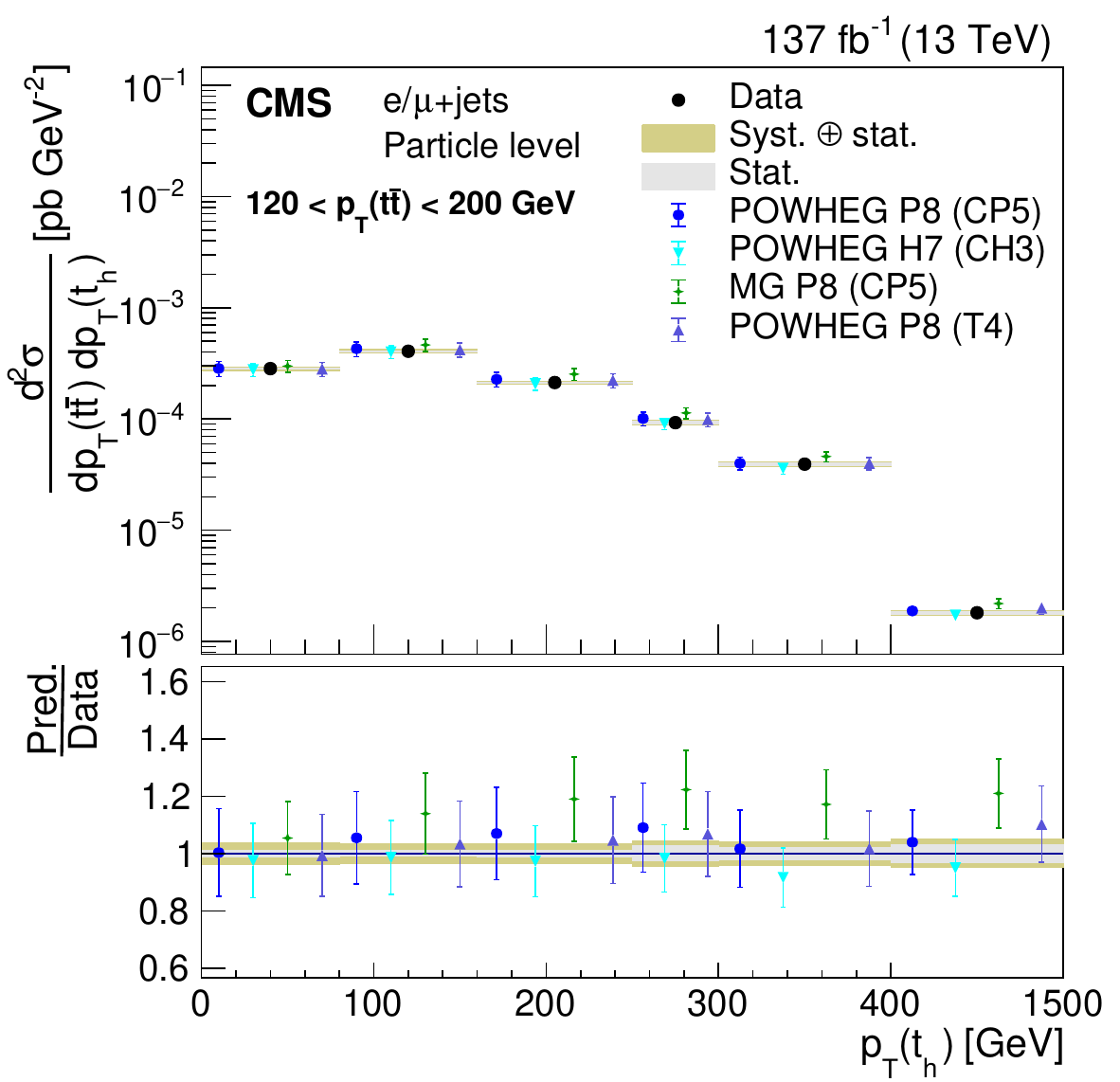}
 \includegraphics[width=0.42\textwidth]{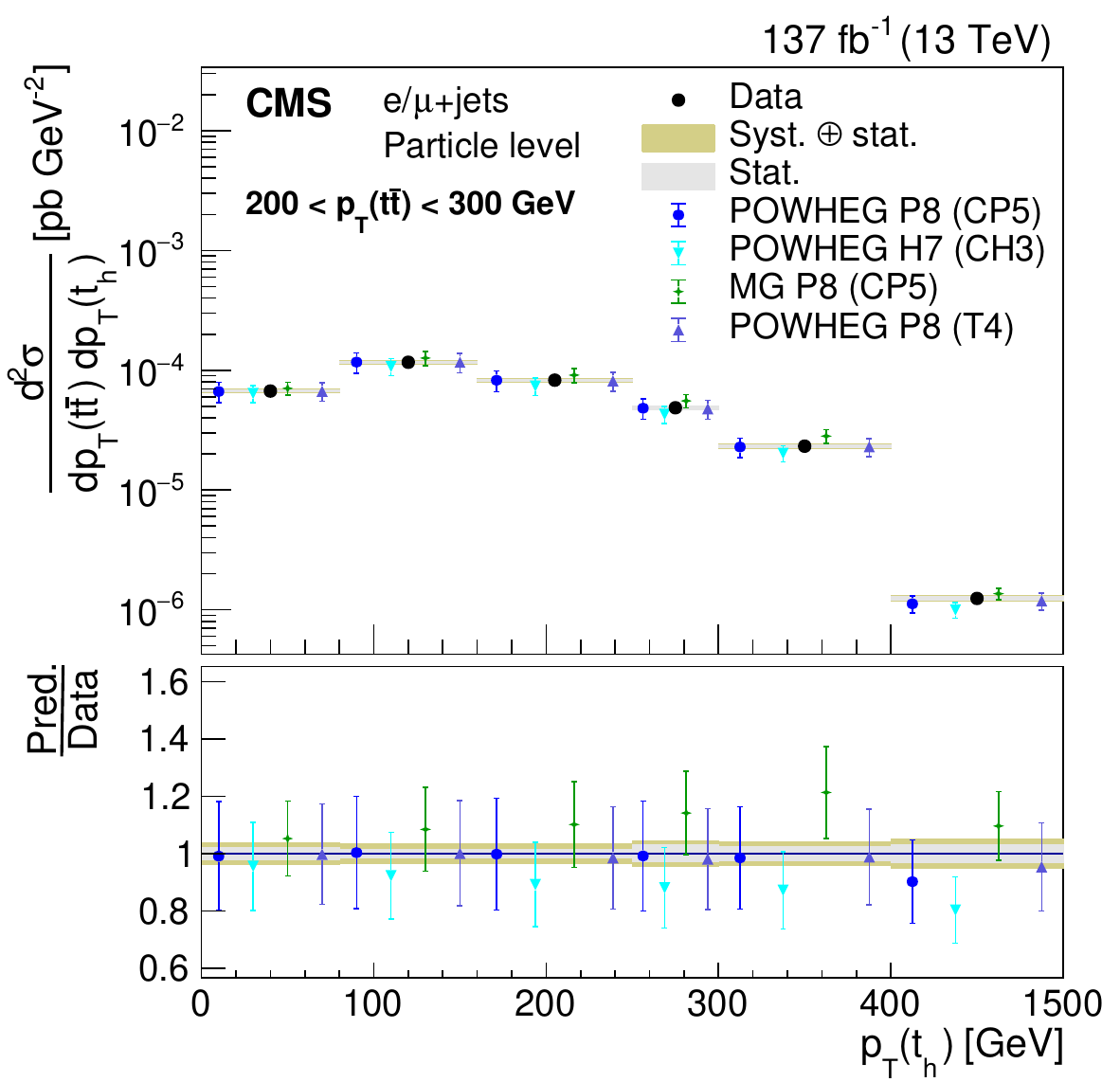}\\
 \includegraphics[width=0.42\textwidth]{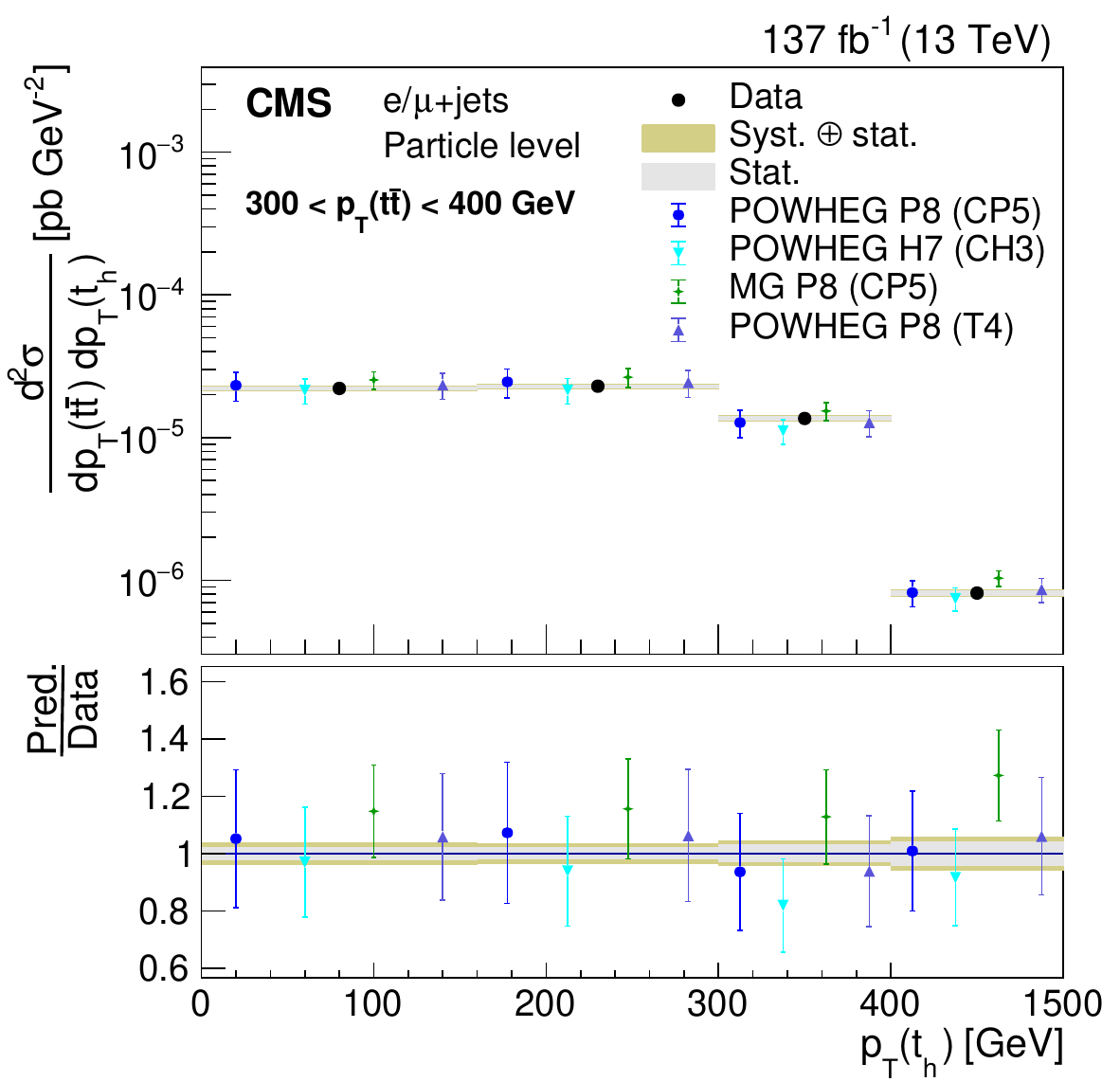}
 \includegraphics[width=0.42\textwidth]{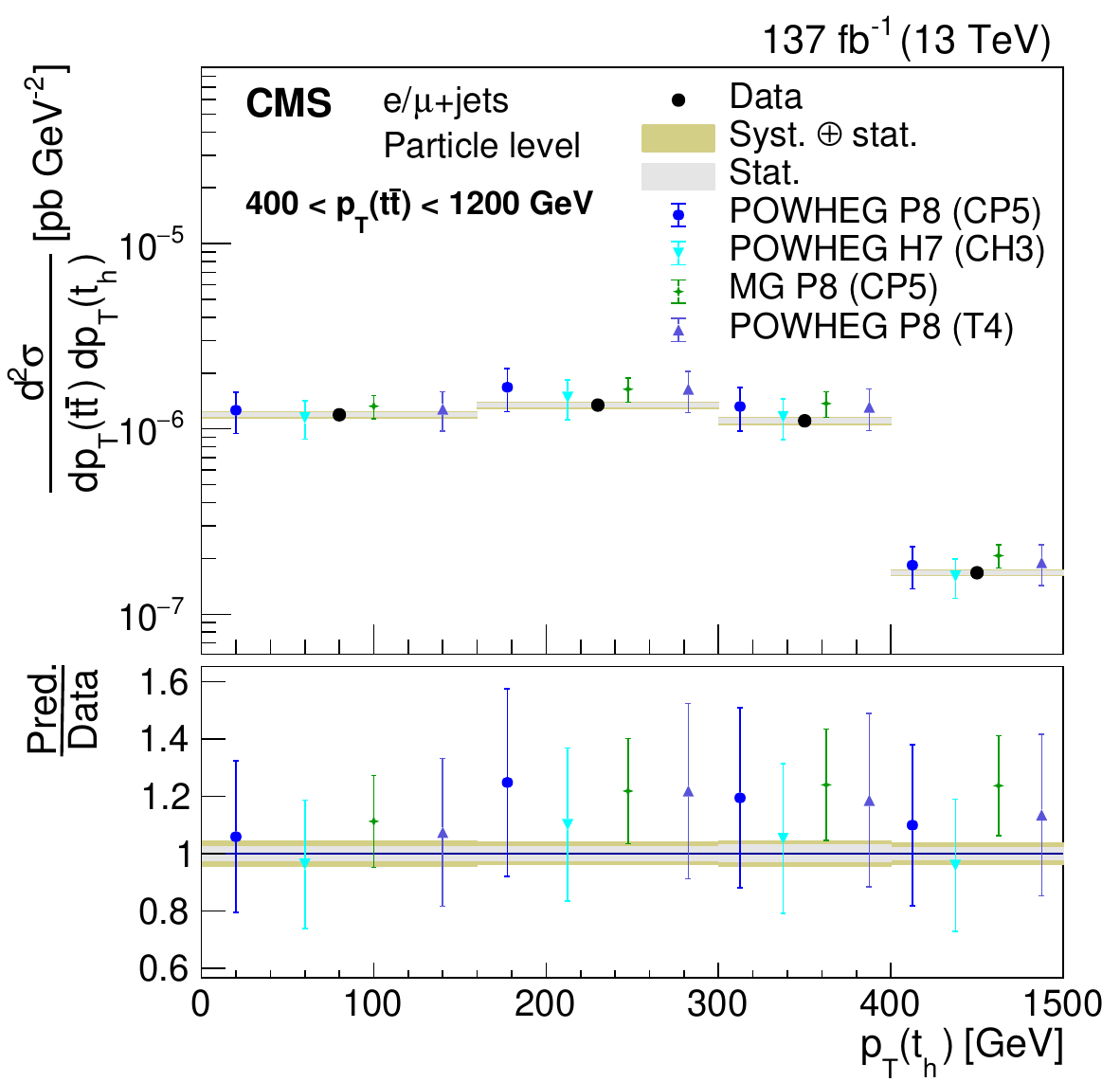}
 \caption{Double-differential cross section at the particle level as a function of \ttptvsthadpt. \XSECCAPPS}
 \label{fig:RESPS7b}
\end{figure*}

\begin{figure*}[tbp]
\centering
 \includegraphics[width=0.42\textwidth]{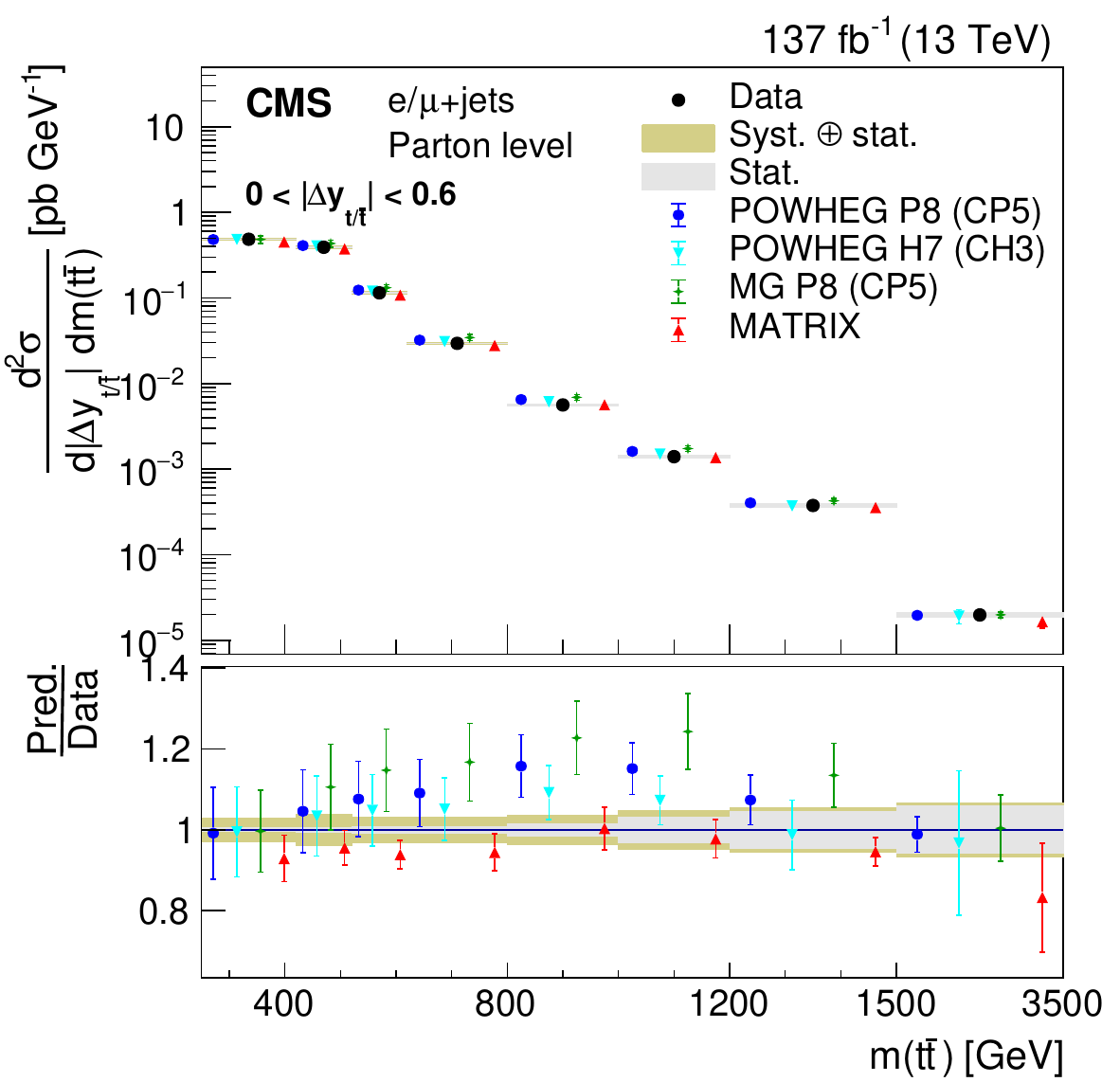}
 \includegraphics[width=0.42\textwidth]{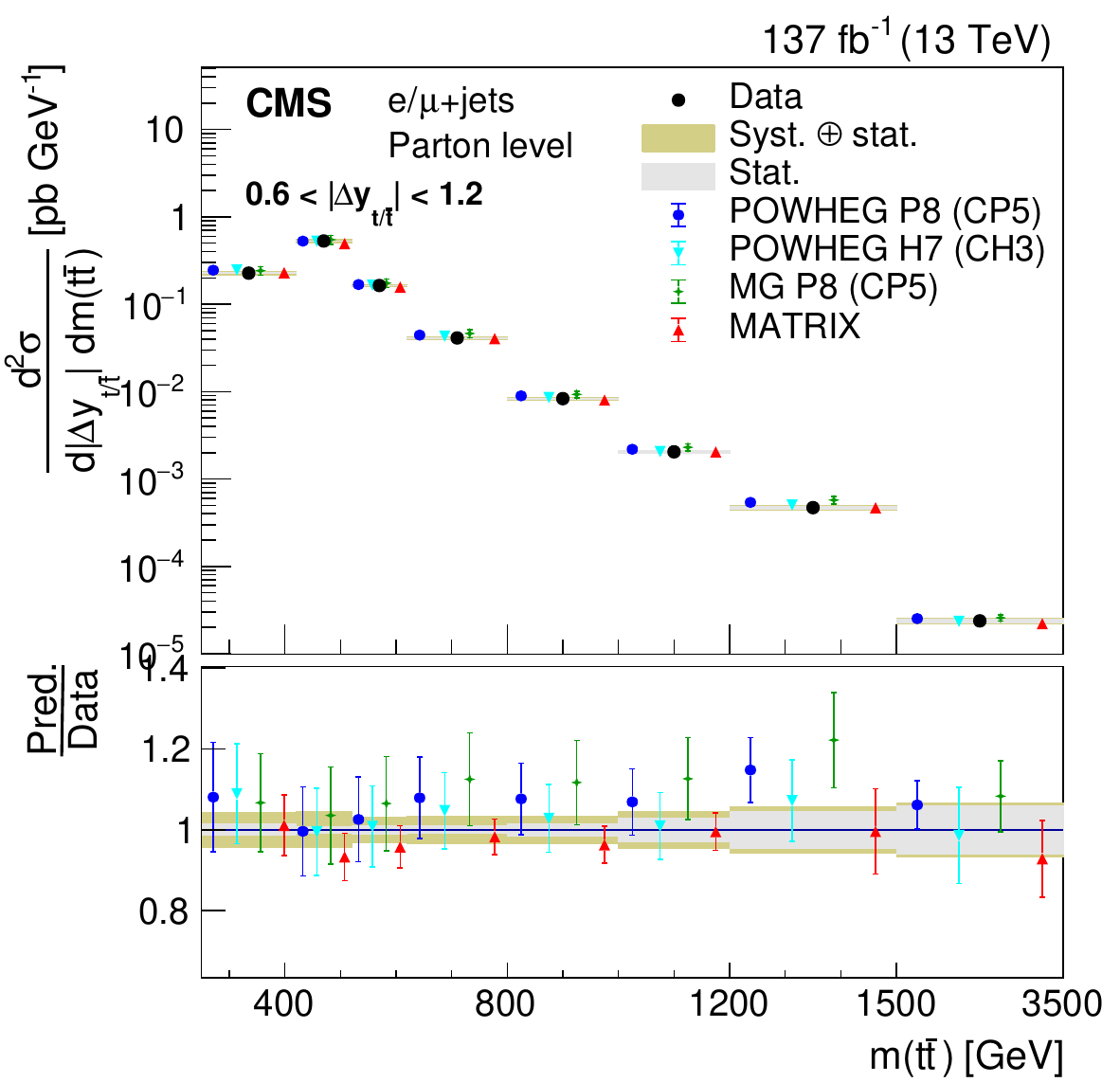}\\
 \includegraphics[width=0.42\textwidth]{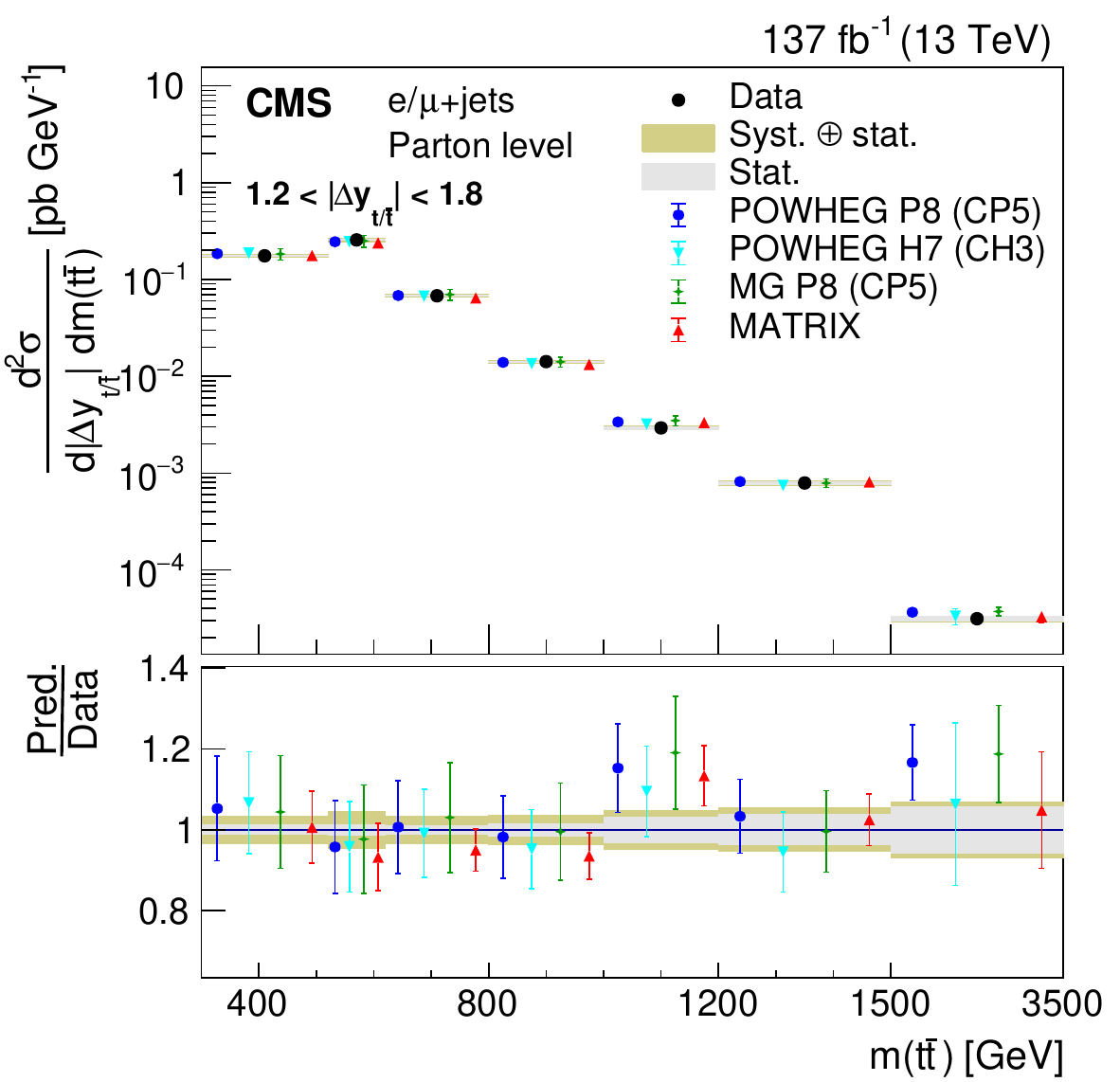}
 \includegraphics[width=0.42\textwidth]{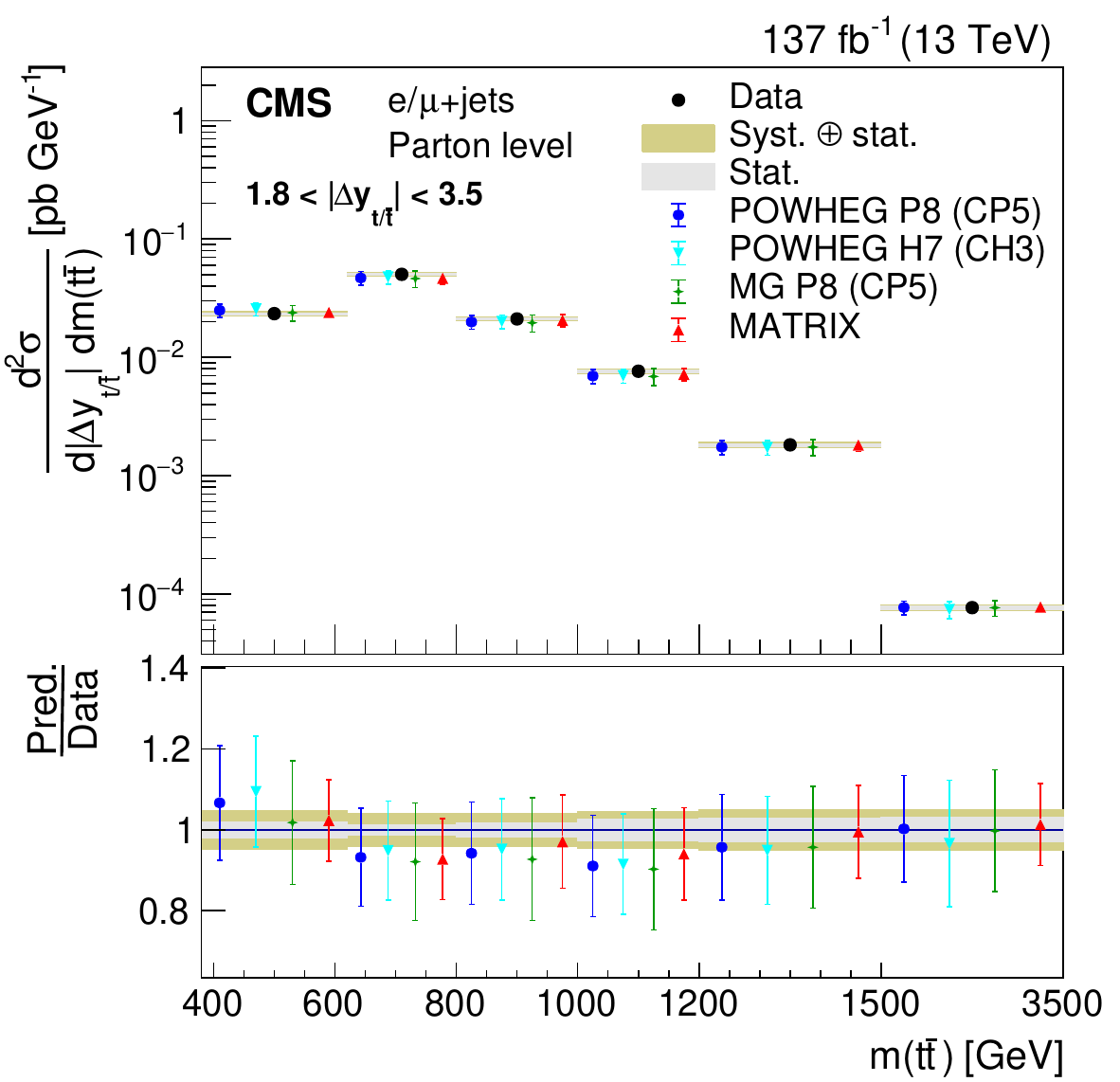}
 \caption{Double-differential cross section at the parton level as a function of \adyvsttm. \XSECCAPPA}
 \label{fig:RES8}
\end{figure*}

\begin{figure*}[tbp]
\centering
 \includegraphics[width=0.42\textwidth]{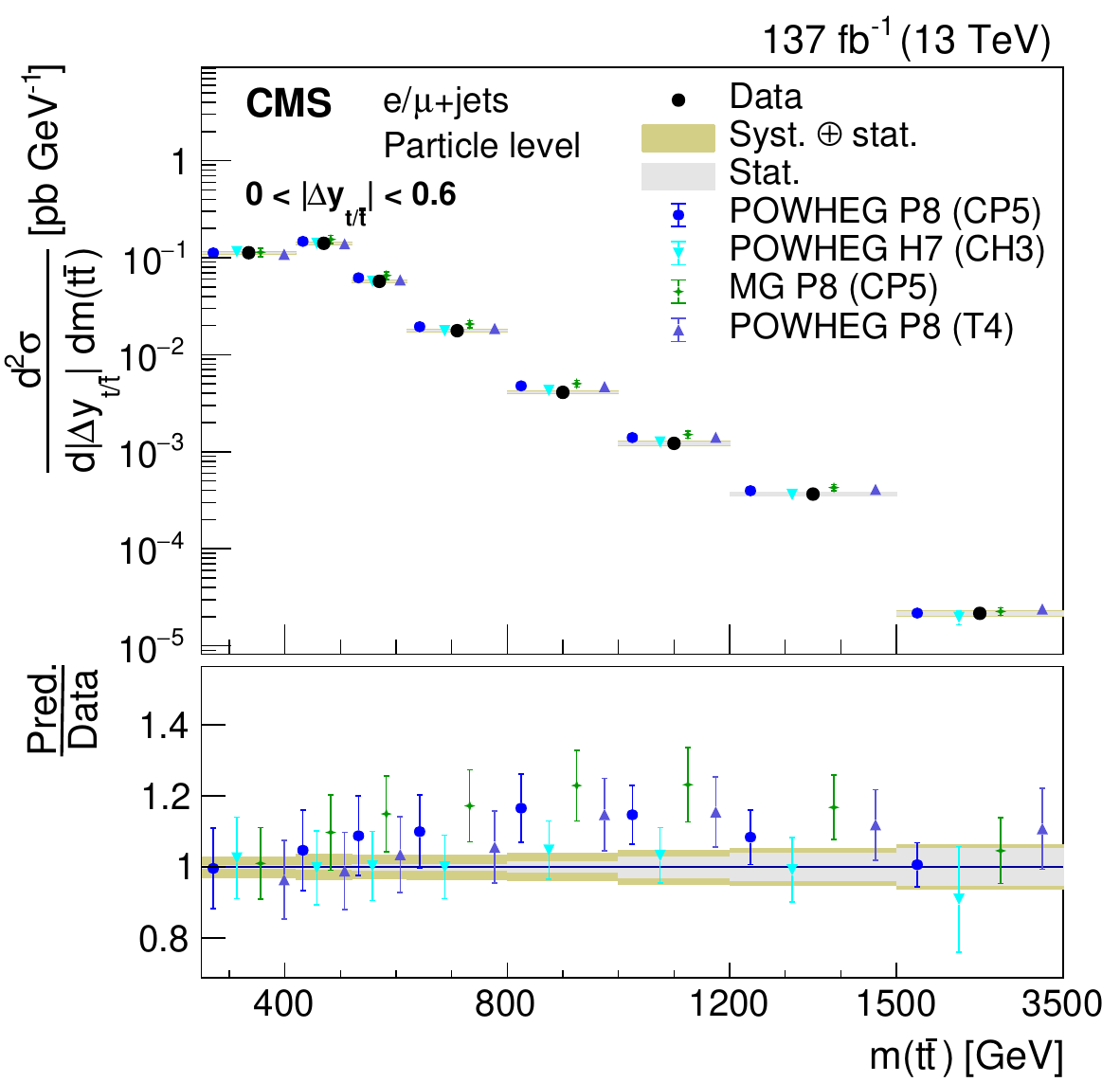}
 \includegraphics[width=0.42\textwidth]{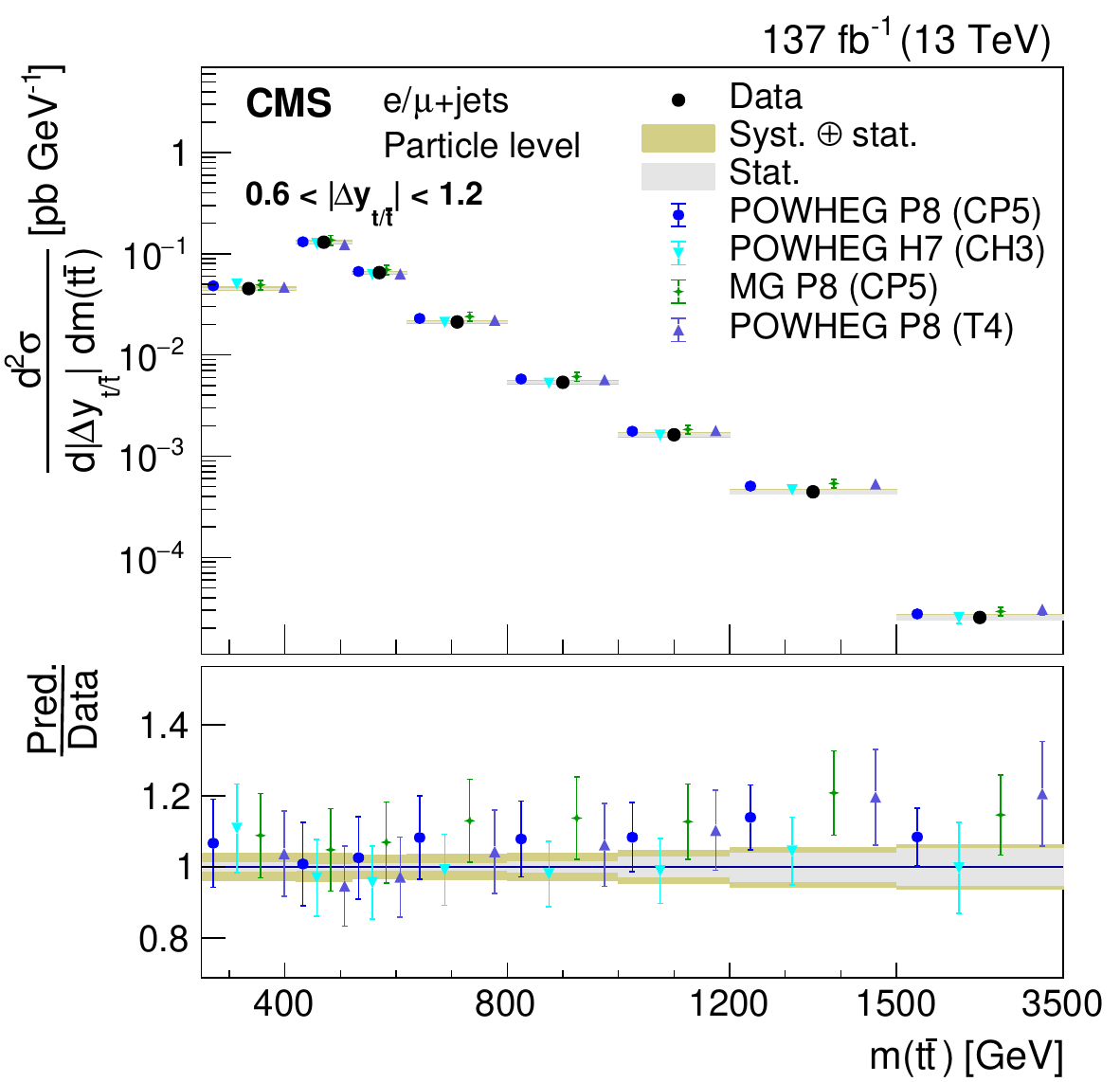}\\
 \includegraphics[width=0.42\textwidth]{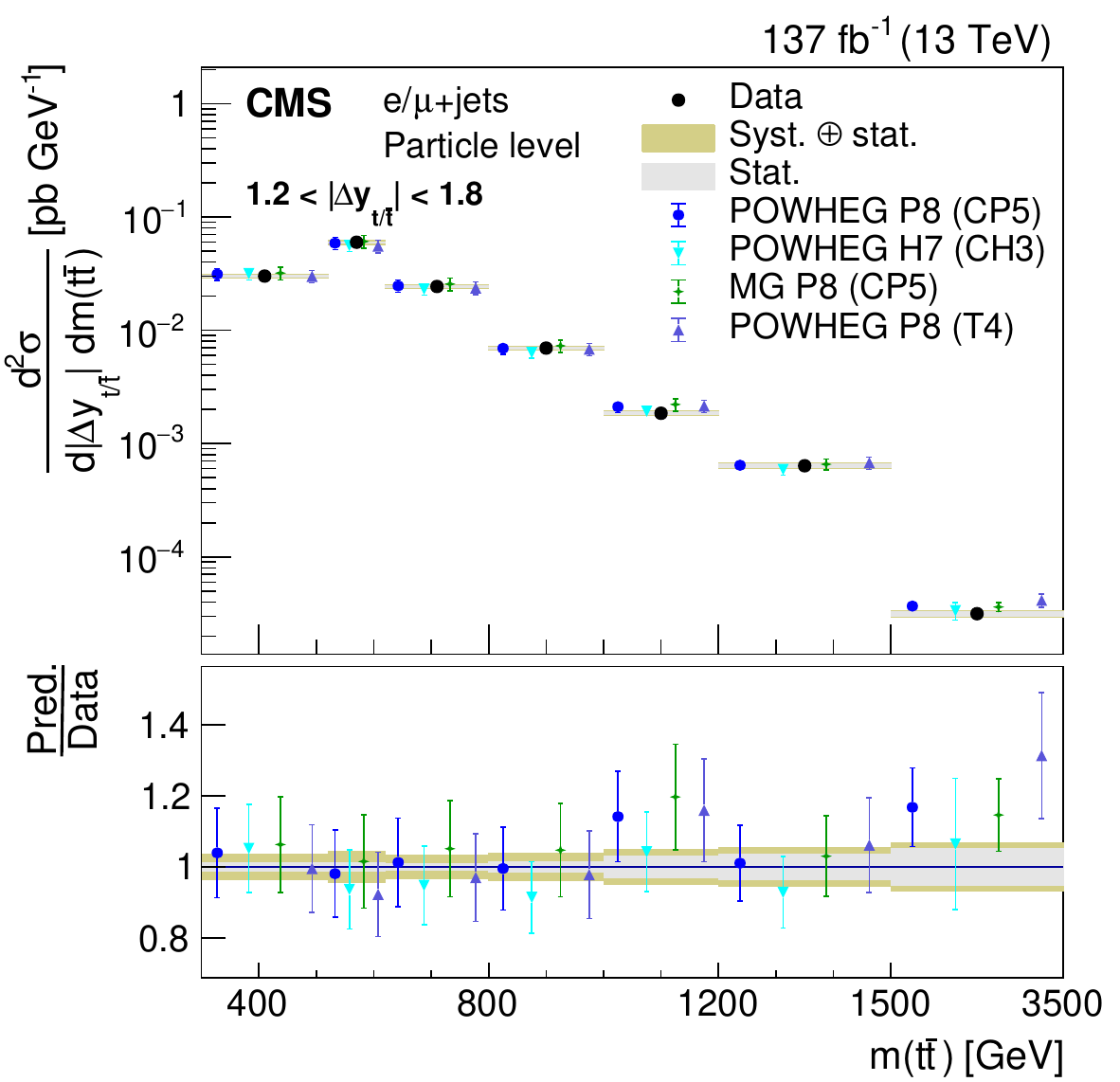}
 \includegraphics[width=0.42\textwidth]{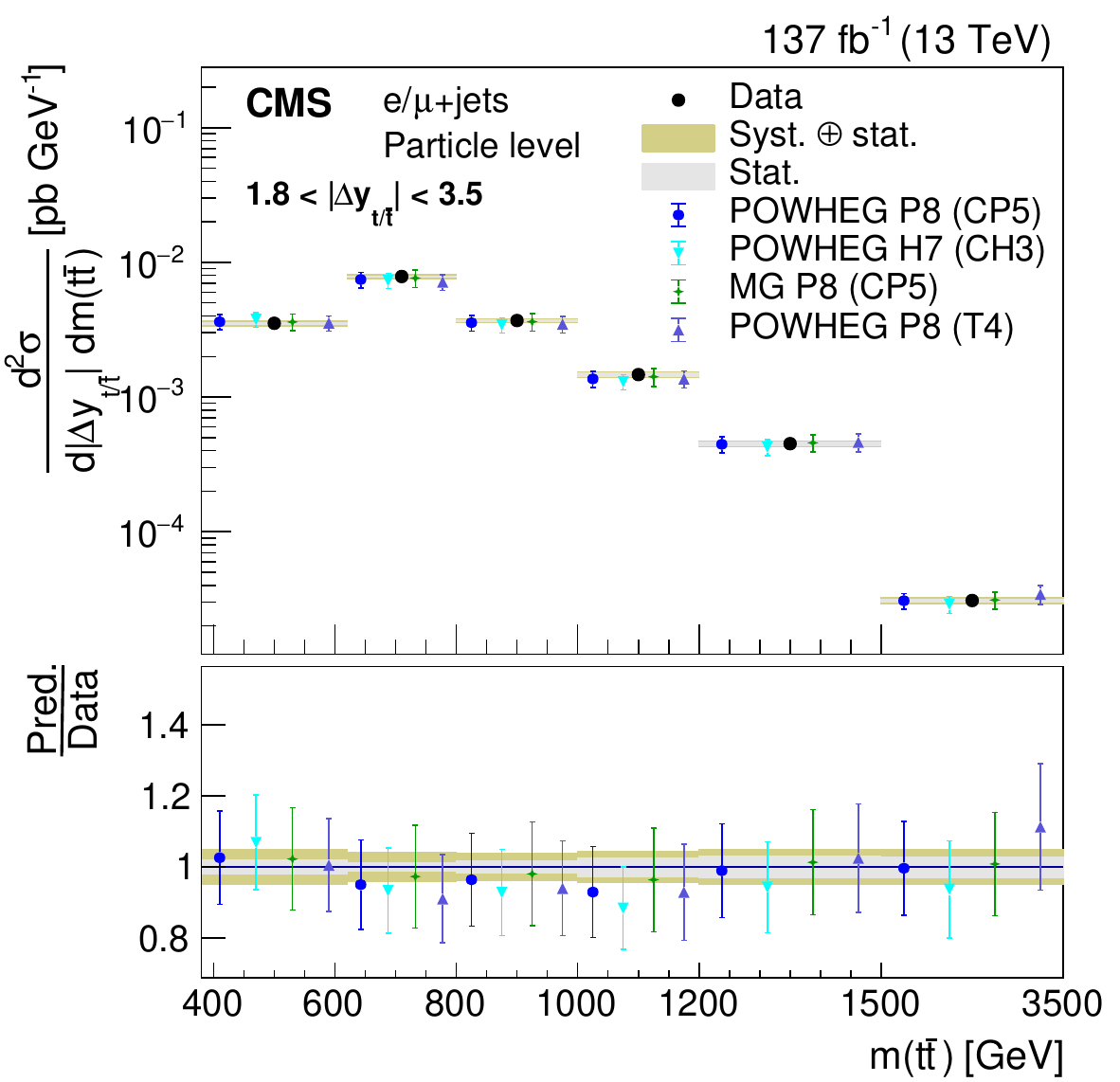}
 \caption{Double-differential cross section at the particle level as a function of \adyvsttm. \XSECCAPPS}
 \label{fig:RESPS8}
\end{figure*}

In Fig.~\ref{fig:RES11} the results of $\chi^2$ tests comparing the measurements with several predictions are shown. To magnify the regions of low and high $p$-values, they are converted into $Z$-scores using the relation 
\begin{equation}
p\text{-value} = 1-\int\limits_{-\infty}^{Z\text{-score}} \text{norm}(x)\,\rd x,
\end{equation}
 where the integrand is the normal distribution with mean 0 and standard deviation 1. For the comparison the covariance matrices of the measurements and the predictions are taken into account. For the \POWHEG{}+\PYTHIA simulations all theoretical uncertainties are considered. For the \MATRIX, \POWHEG{}+\HERWIG, and \AMCATNLO{}+\PYTHIA simulations, we consider the dominant sources of matrix element scales and PDF uncertainties. In addition, the PS scale uncertainties are evaluated for \AMCATNLO{}+\PYTHIA. The PDF uncertainties for \MATRIX are obtained as the envelope of the differences obtained with the CT14~\cite{CT14}, HERA~\cite{HERA}, and MMHT~\cite{MMHT} PDF sets. In all predictions, the individual uncertainty sources are assumed to be correlated among the bins, while the sources themselves are uncorrelated. Most of the distributions are well described by the predictions, and the uncertainty in the NNLO \MATRIX calculation is significantly smaller than in the NLO predictions. For the parton-level measurements, similar $p$-values are obtained for \POWHEG{}+\PYTHIA with the \CUET and CP5 tunes and \POWHEG{}+\HERWIG, while the description by \AMCATNLO{}+\PYTHIA is slightly worse. For the particle-level measurements similar agreements are obtained for \POWHEG{}+\PYTHIA with the \CUET and CP5 tunes. In most cases, the $p$-values of the \AMCATNLO{}+\PYTHIA and \POWHEG{}+\HERWIG predictions are lower. At both levels, the distributions of \ttmvsthadpt, \ttptvsthadpt, and \adyvsttm are not well described by any of the tested predictions. The corresponding distributions are shown in Figs.~\ref{fig:RES7}--\ref{fig:RESPS8}. The one-dimensional distributions of \thadpt, \ttm, and \ttpt are consistent with the predictions at the level of two standard deviations, which calls for further investigations to improve the understanding of the kinematic relations between these variables. In addition, at the particle level the kinematic distributions in bins of jet multiplicity are not well described.        

{\tolerance=8000 The measurements at the parton level are compared to the \POWHEG{}+\PYTHIA (CP5), \POWHEG{}+\HERWIG, \AMCATNLO{}+\PYTHIA, and the \MATRIX predictions. For the particle-level measurements, the \MATRIX prediction is replaced by \POWHEG{}+\PYTHIA (\CUET). In general, when comparing the distributions in data and simulation, very similar trends are observed at both levels.\par}

\begin{figure*}[tbp]
\centering
 \includegraphics[width=0.42\textwidth]{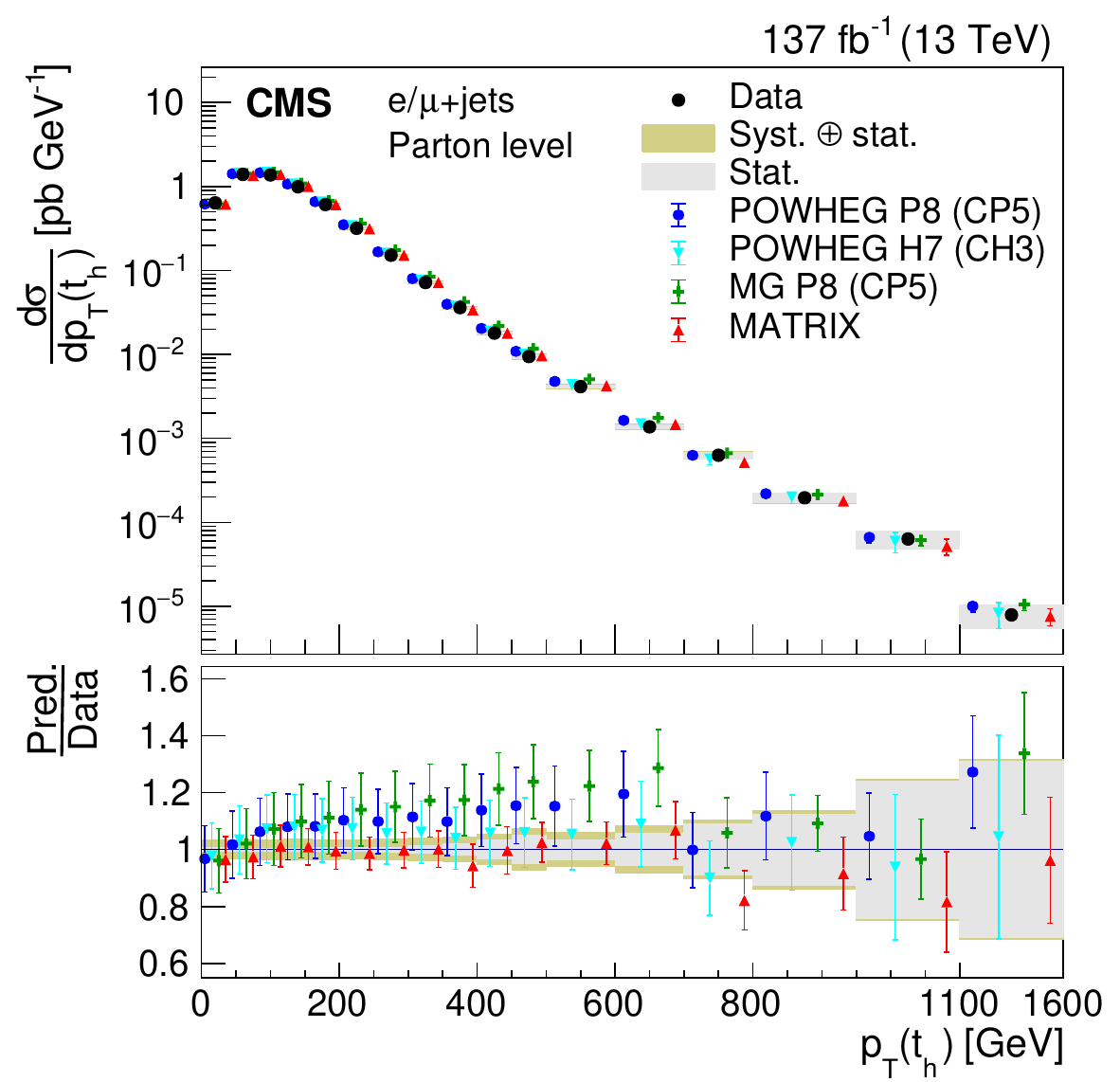}
 \includegraphics[width=0.42\textwidth]{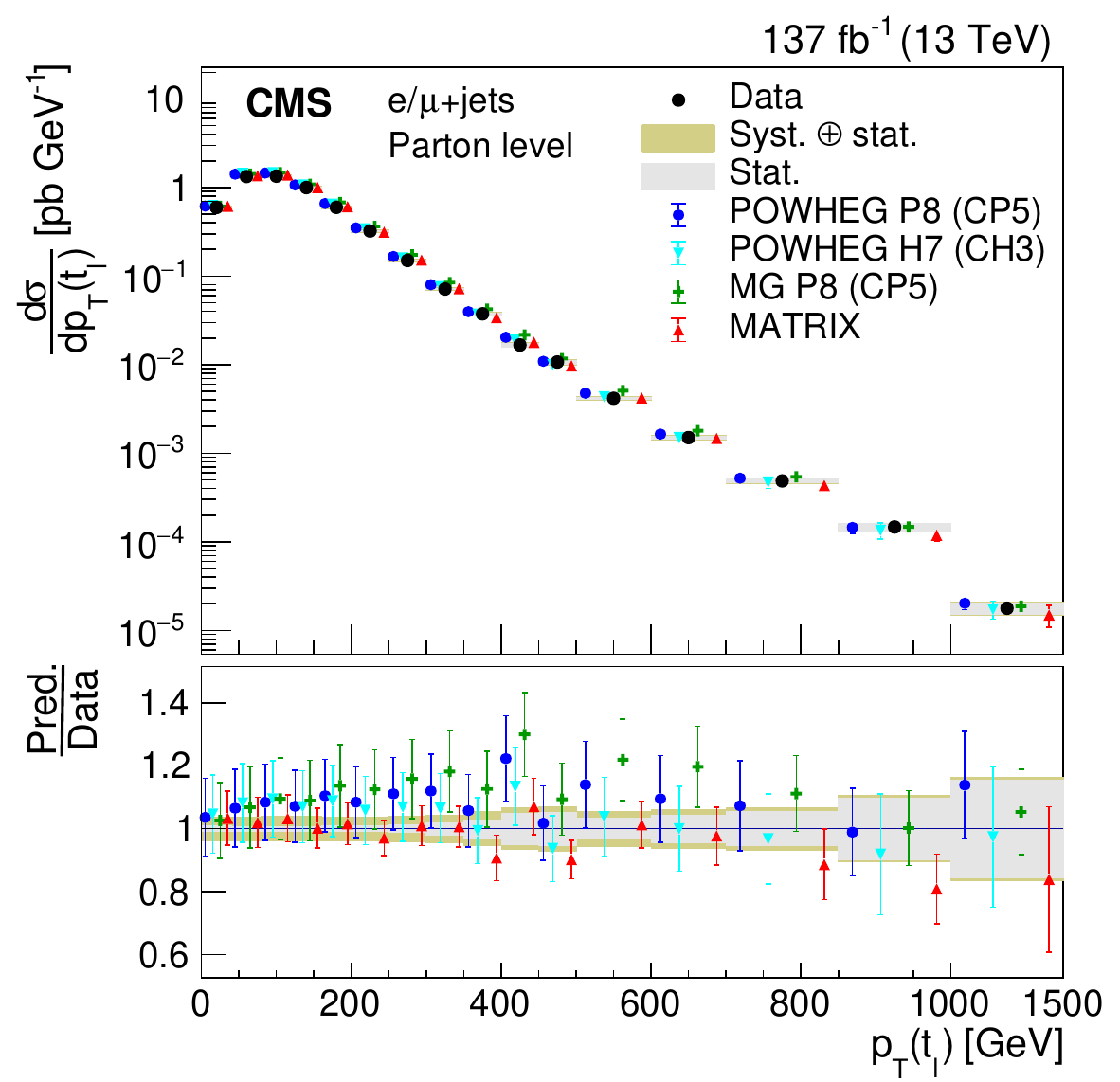}
 \includegraphics[width=0.42\textwidth]{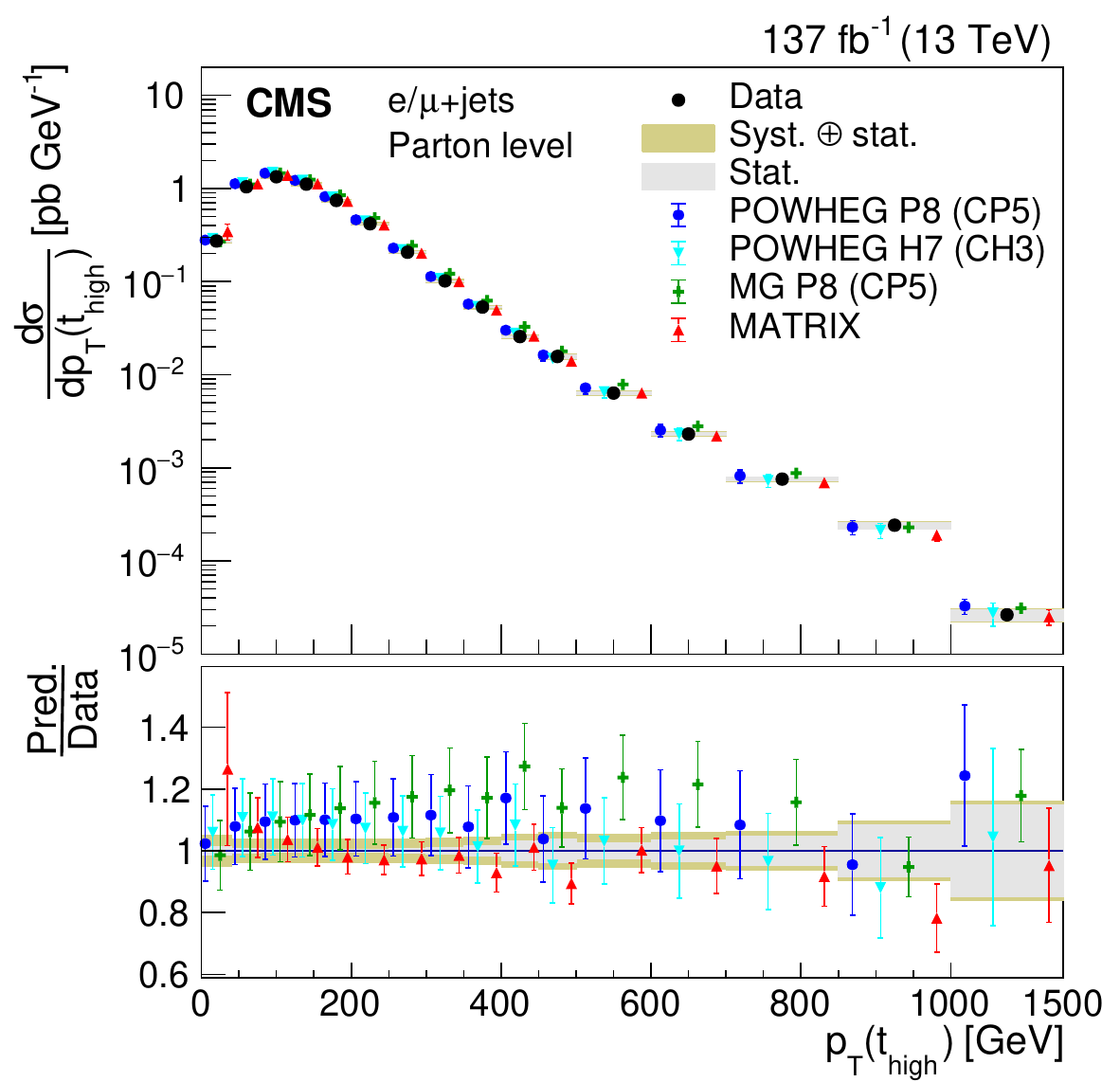}
 \includegraphics[width=0.42\textwidth]{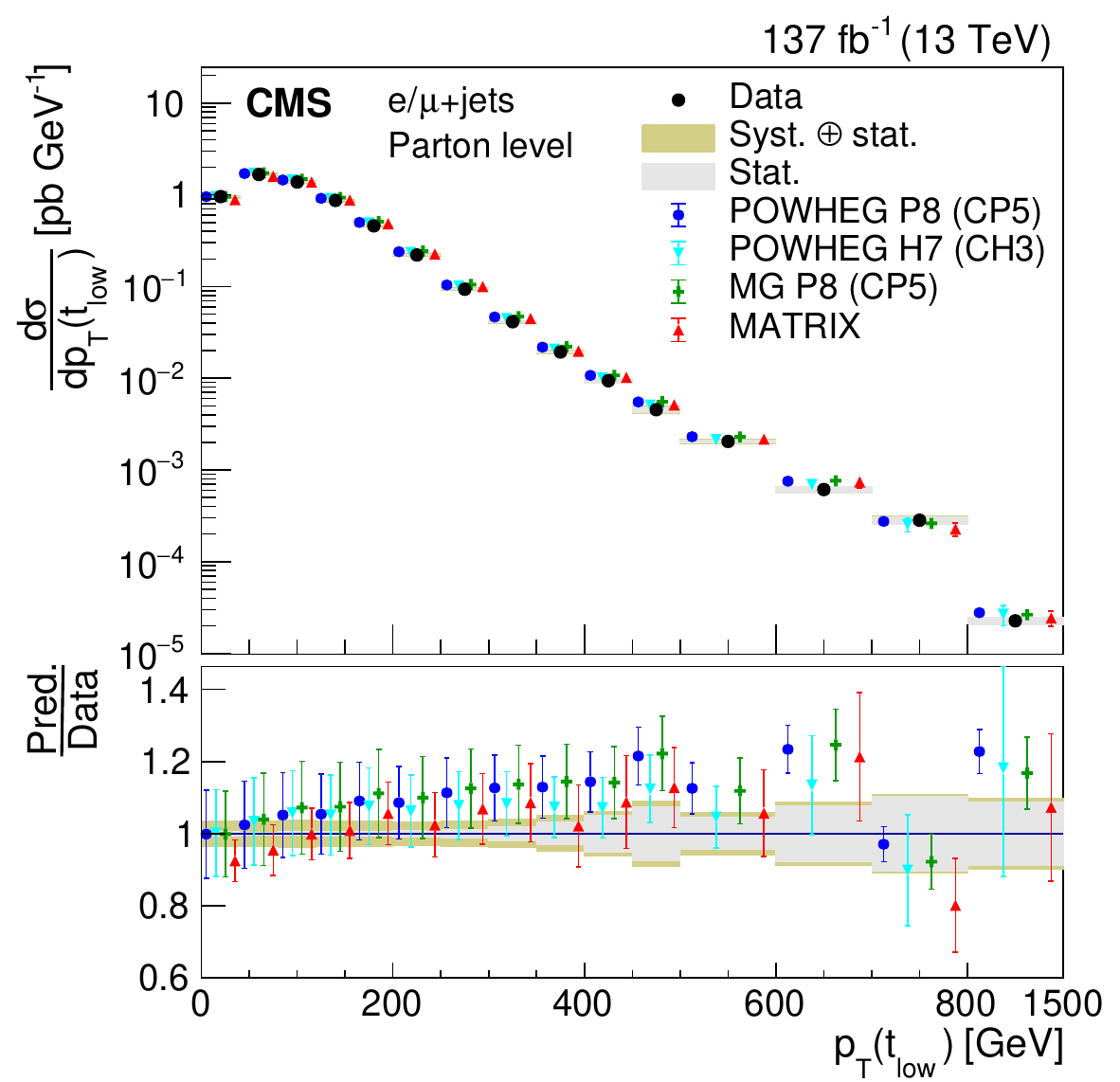}
 \includegraphics[width=0.42\textwidth]{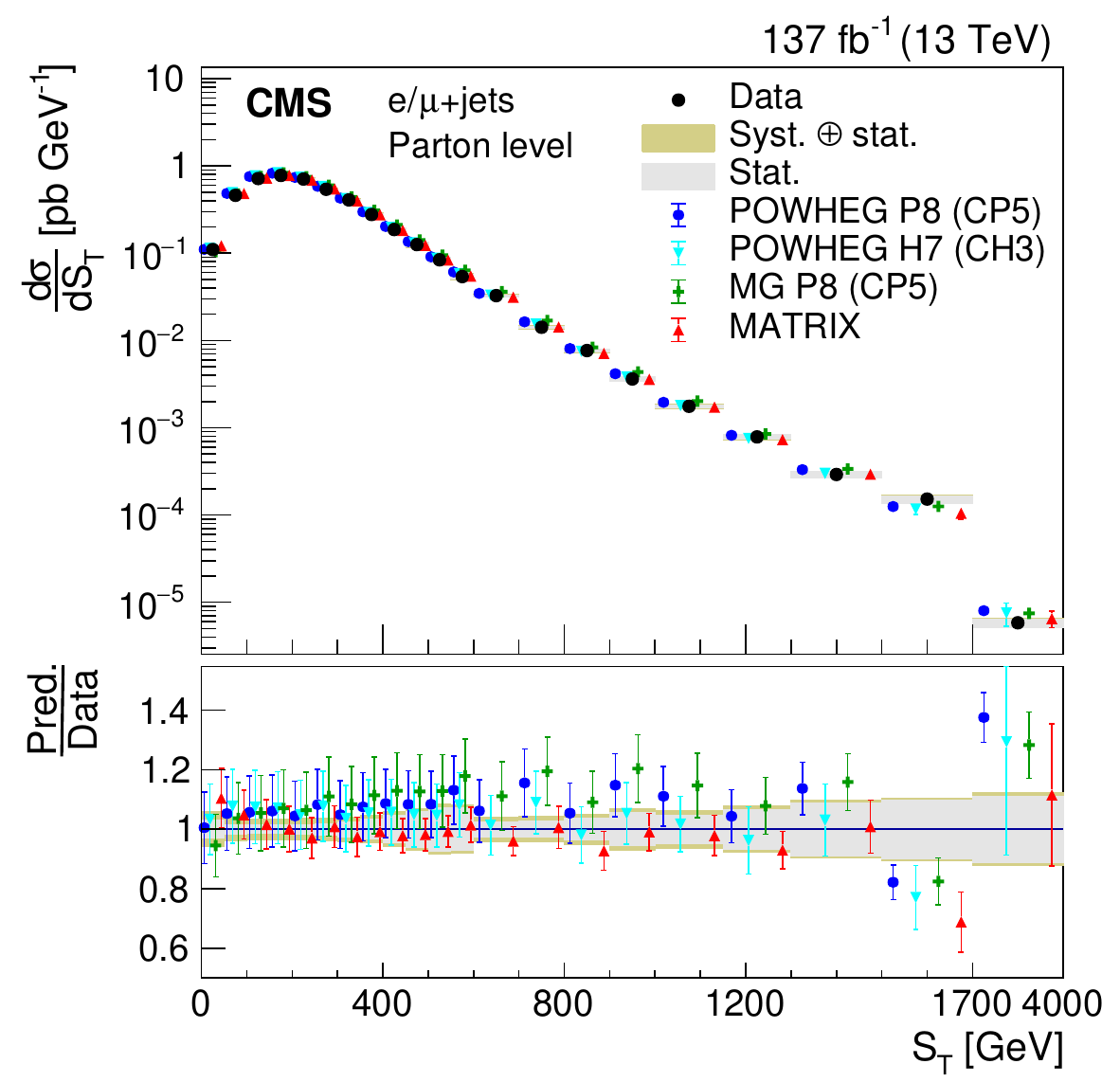}
 \caption{Differential cross sections at the parton level as a function of \thadpt, \tleppt, \thardpt, \tsoftpt, and \st. \XSECCAPPA}
 \label{fig:RES1}
\end{figure*}

The differential cross sections as a function of \thadpt, \tleppt, \thardpt, \tsoftpt, and \st are presented in Figs.~\ref{fig:RES1} and \ref{fig:RESPS1} at the parton and particle levels, respectively. In these and most of the following figures of the differential cross sections, the displayed width of the last bin is reduced. The exact boundaries of these bins are indicated by the axis labels. For better visibility, the horizontal marker positions of the predictions are shifted with respect to the bin centers. 

Compared with the previous CMS measurement~\cite{TOP-17-002} the precision is significantly improved, e.g., for a top quark $\pt < 500$\GeV the uncertainty is reduced by about 50\%. Since it is easier to perform the background subtraction (cf.~Section~\ref{FITBO}) in bins of \thadpt, the measurement as a function of \thadpt is extended towards higher \pt compared to the measurement as a function of \tleppt. The \pt spectra are softer than predicted by the NLO calculations at low \pt. For $\pt > 500$\GeV the predictions overestimate the measured cross sections by about 20\%. However, the NNLO QCD calculation performed with \MATRIX describes the data significantly better, with the exception of \tsoftpt. As discussed in Ref.~\cite{MATRIXTT2}, \thardpt and \tsoftpt cannot be accurately described by fixed-order calculations; resummation effects of the Sudakov type have to be taken into account. At the parton level, the differential cross sections as a function exclusively of \tqh or \tql should be the same, and this is used to check the consistency of the results obtained with the different top quark decays. 

\begin{figure*}[tbp]
\centering
 \includegraphics[width=0.42\textwidth]{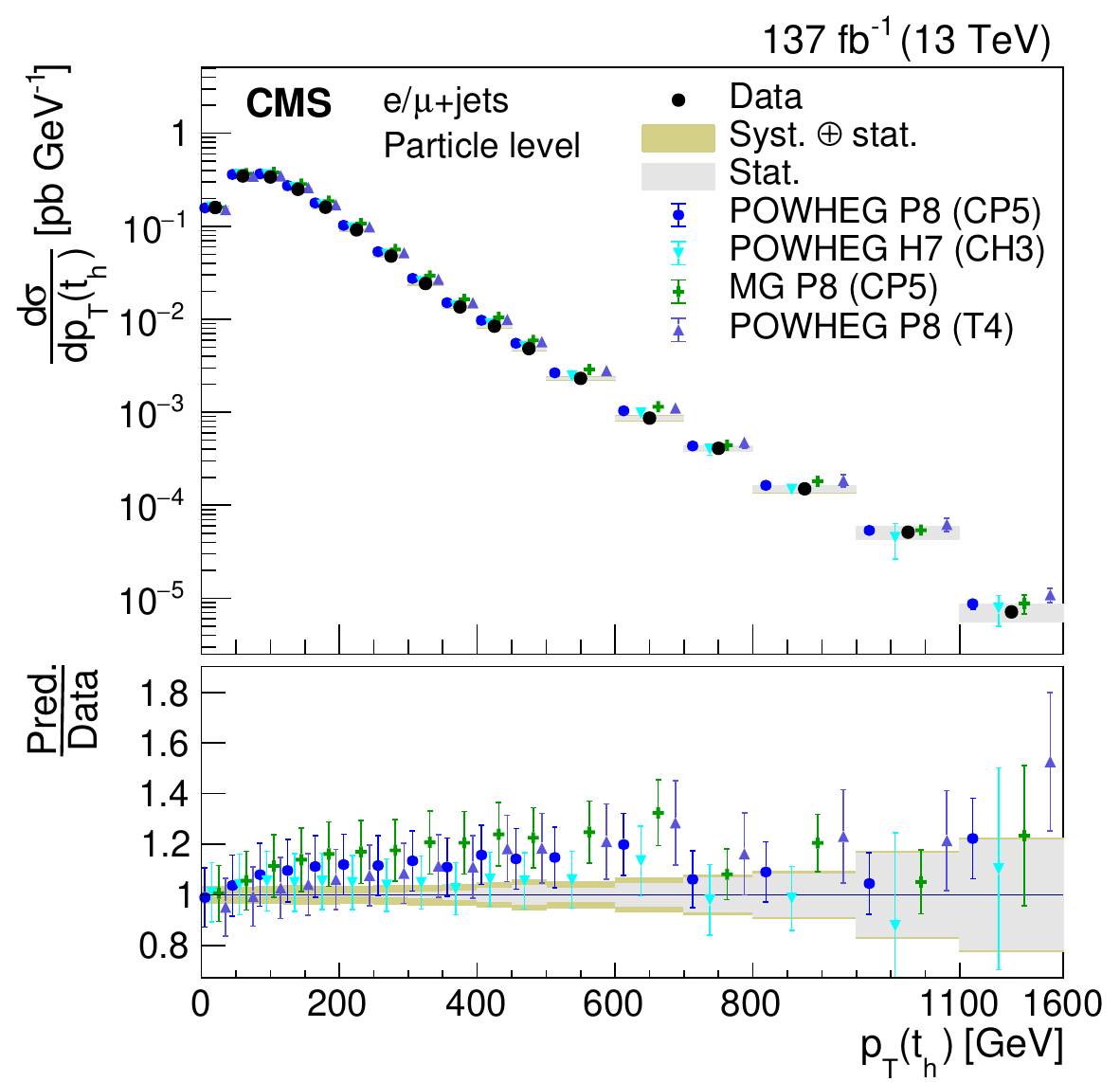}
 \includegraphics[width=0.42\textwidth]{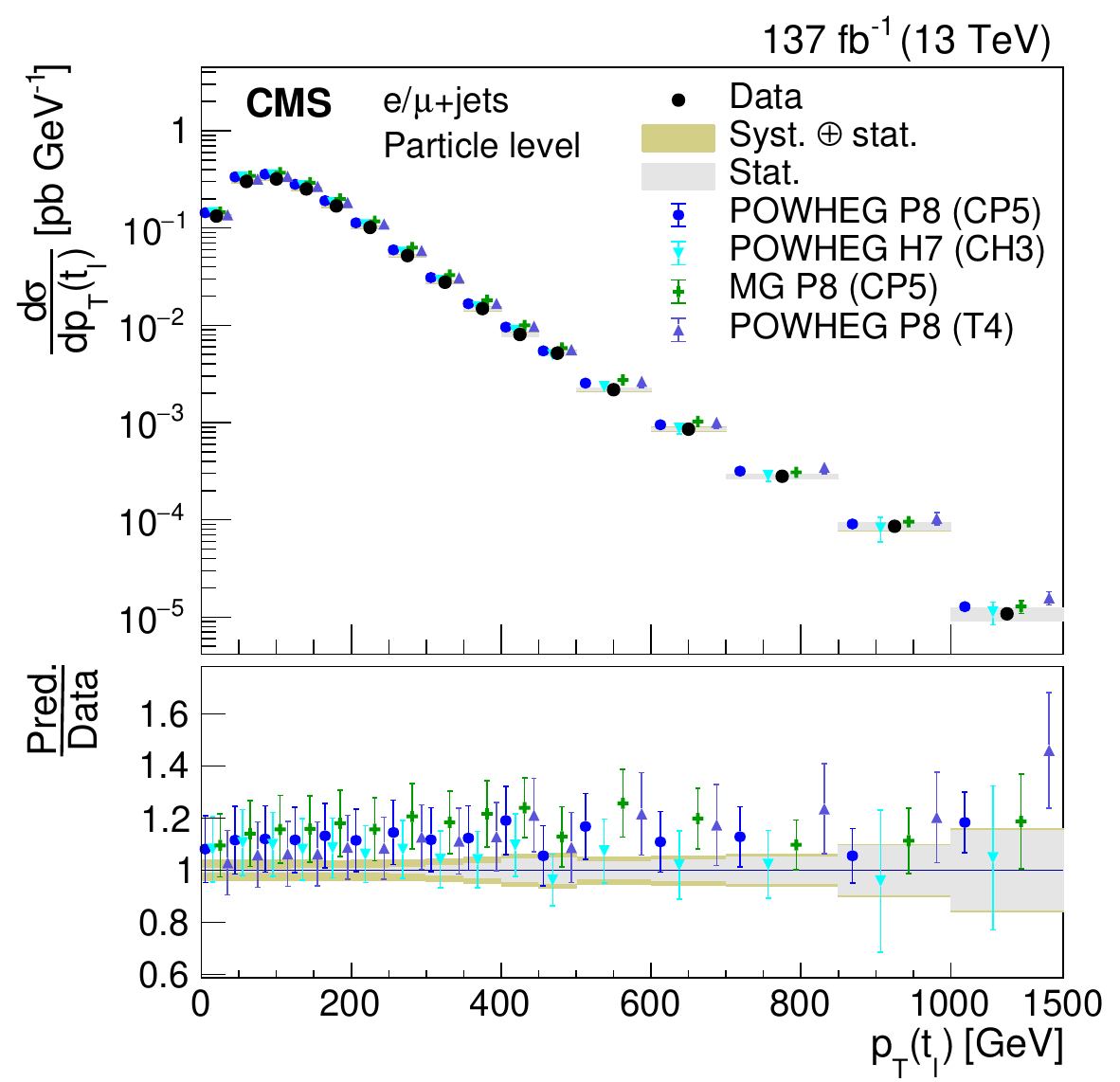}
 \includegraphics[width=0.42\textwidth]{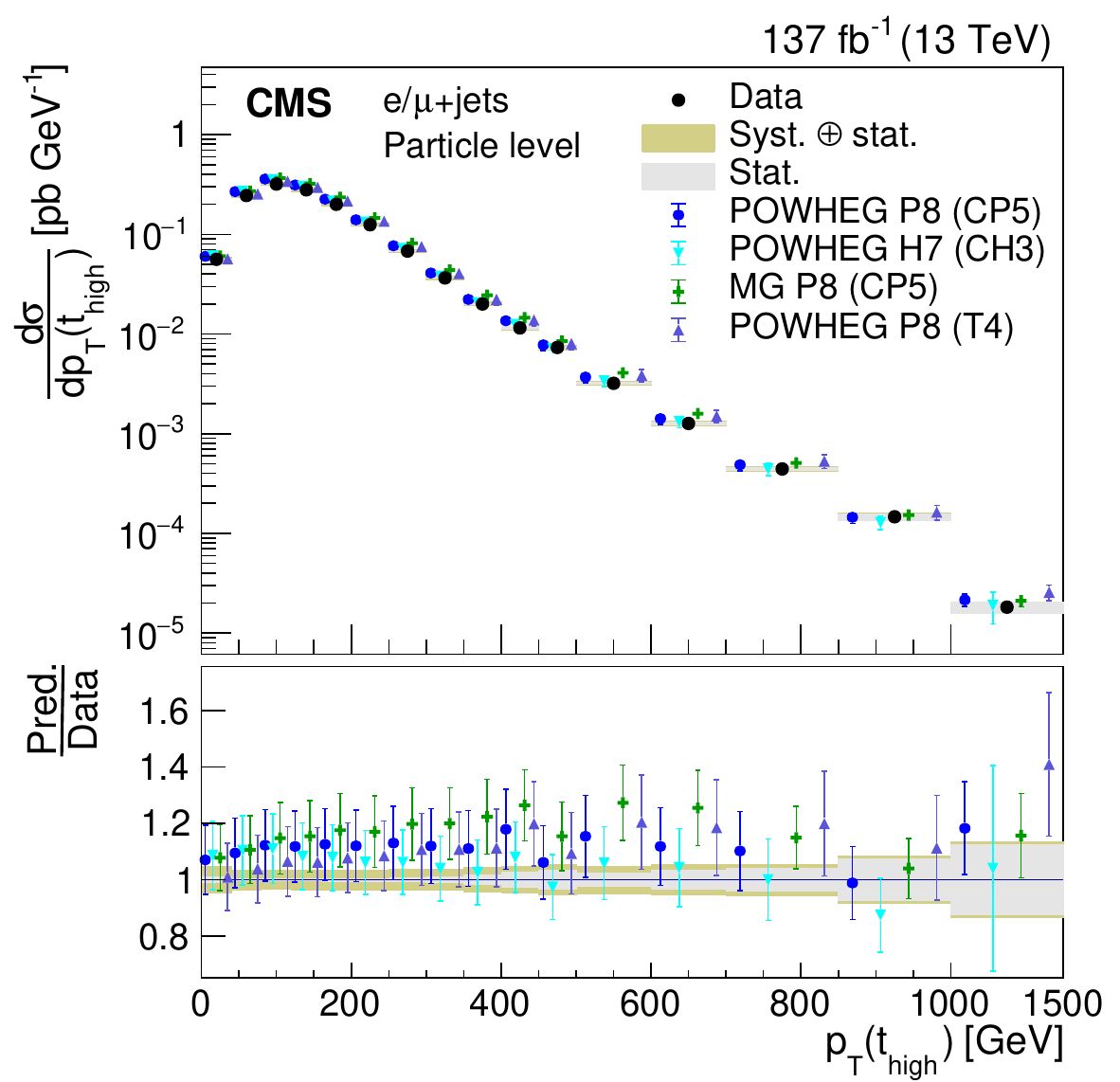}
 \includegraphics[width=0.42\textwidth]{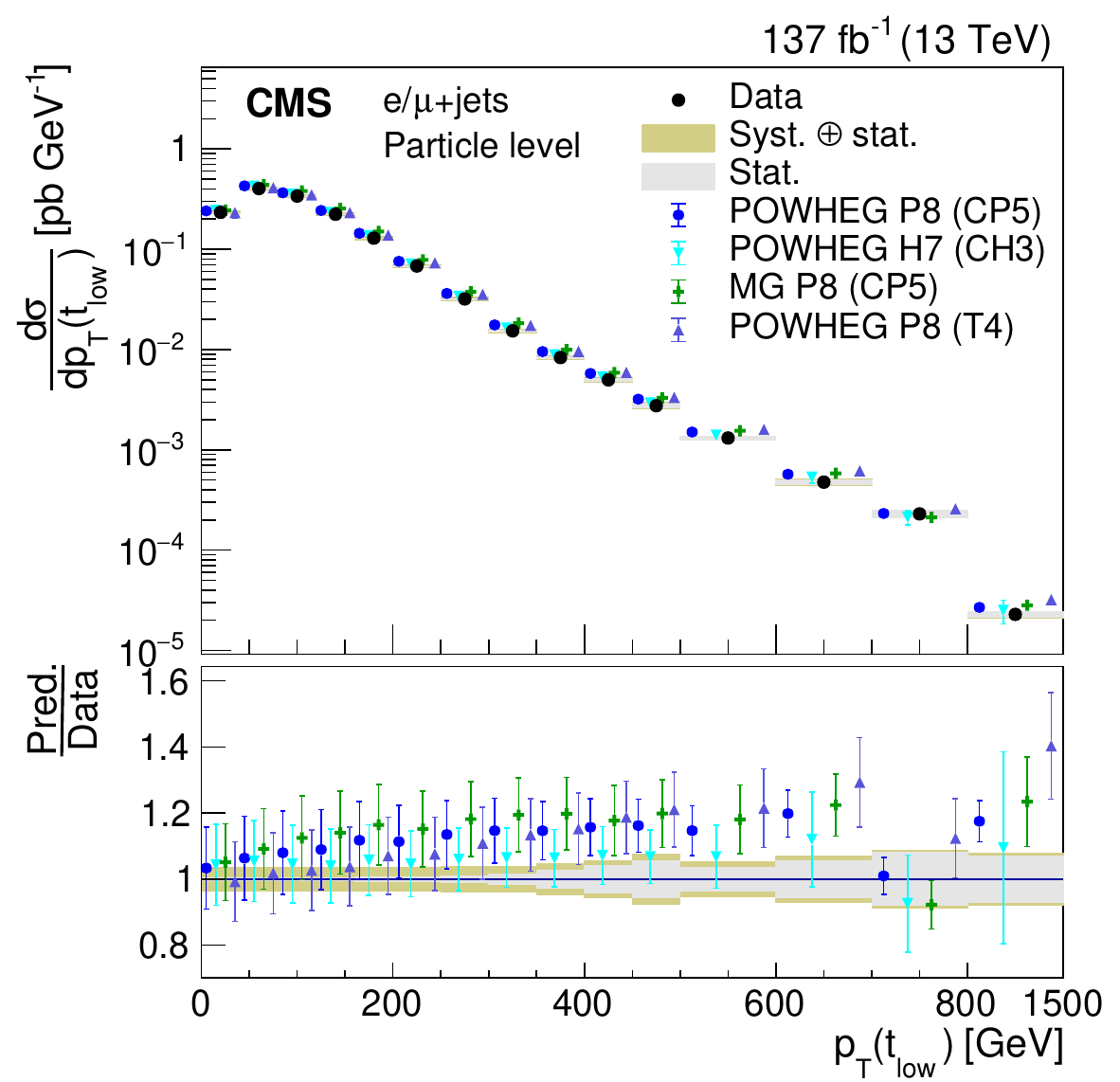}
 \includegraphics[width=0.42\textwidth]{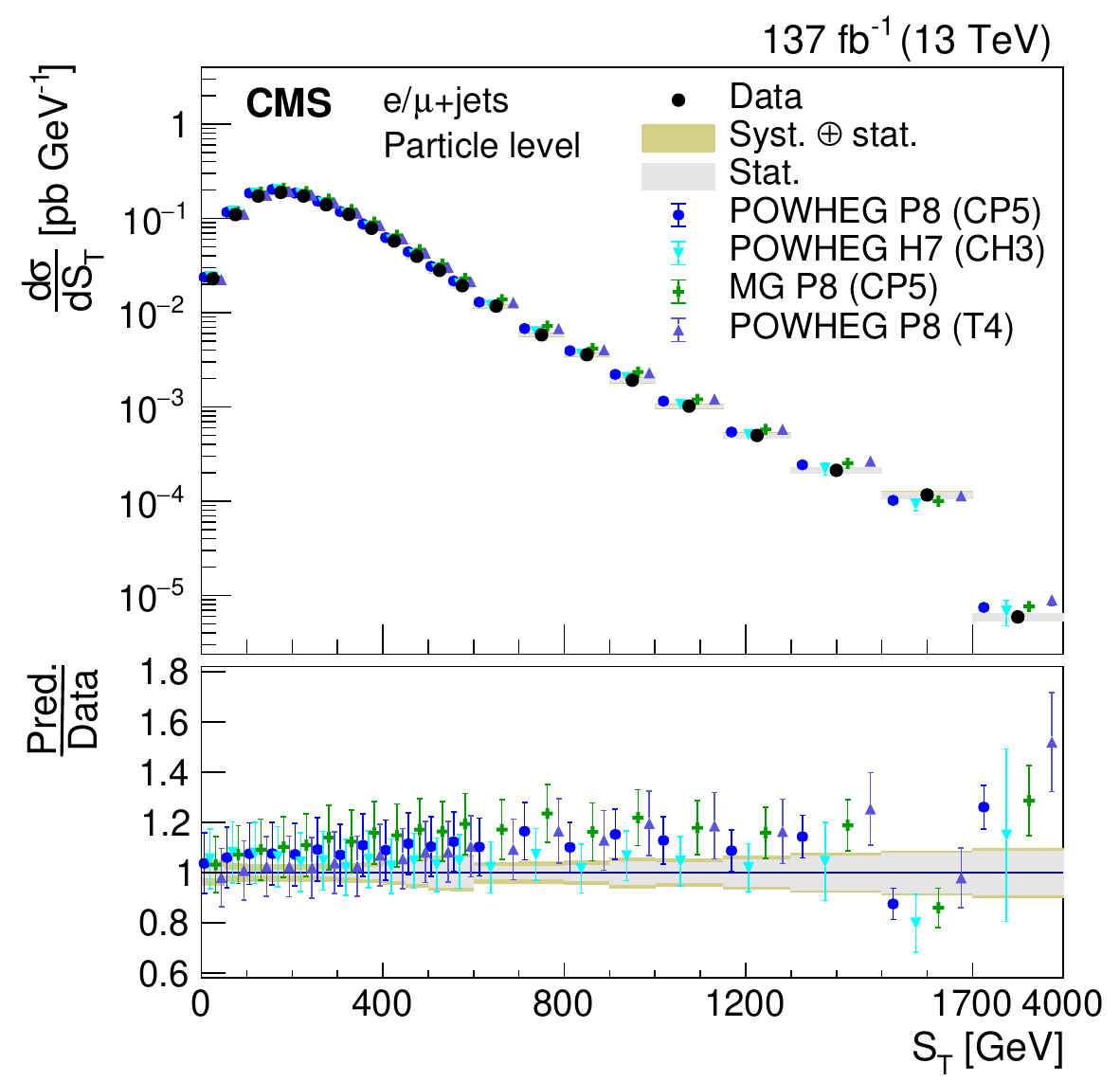}
 \caption{Differential cross sections at the particle level as a function of \thadpt, \tleppt, \thardpt, \tsoftpt, and \st. \XSECCAPPS}
 \label{fig:RESPS1}
\end{figure*}

\begin{figure*}[tbp]
\centering
 \includegraphics[width=0.42\textwidth]{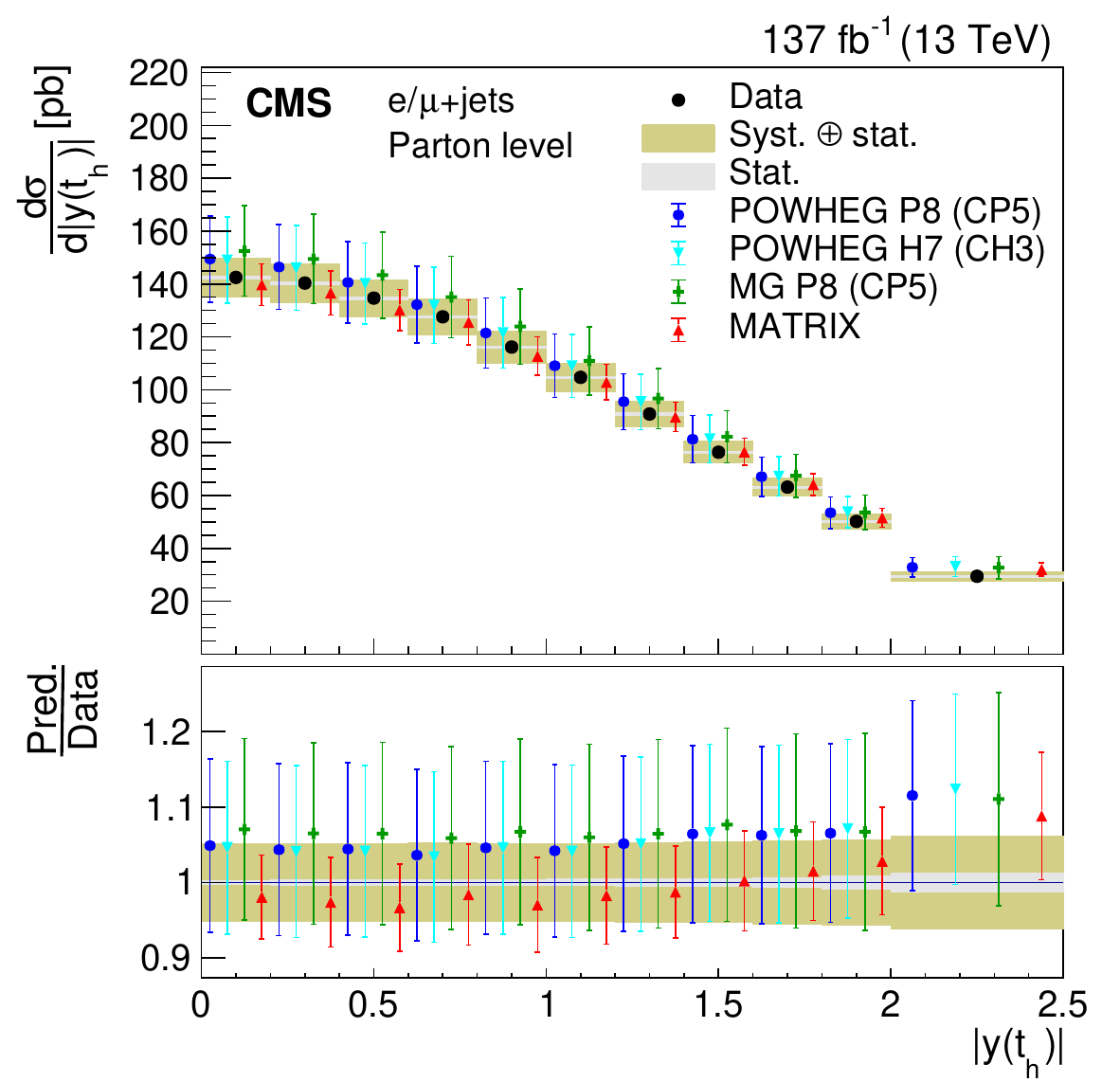}
 \includegraphics[width=0.42\textwidth]{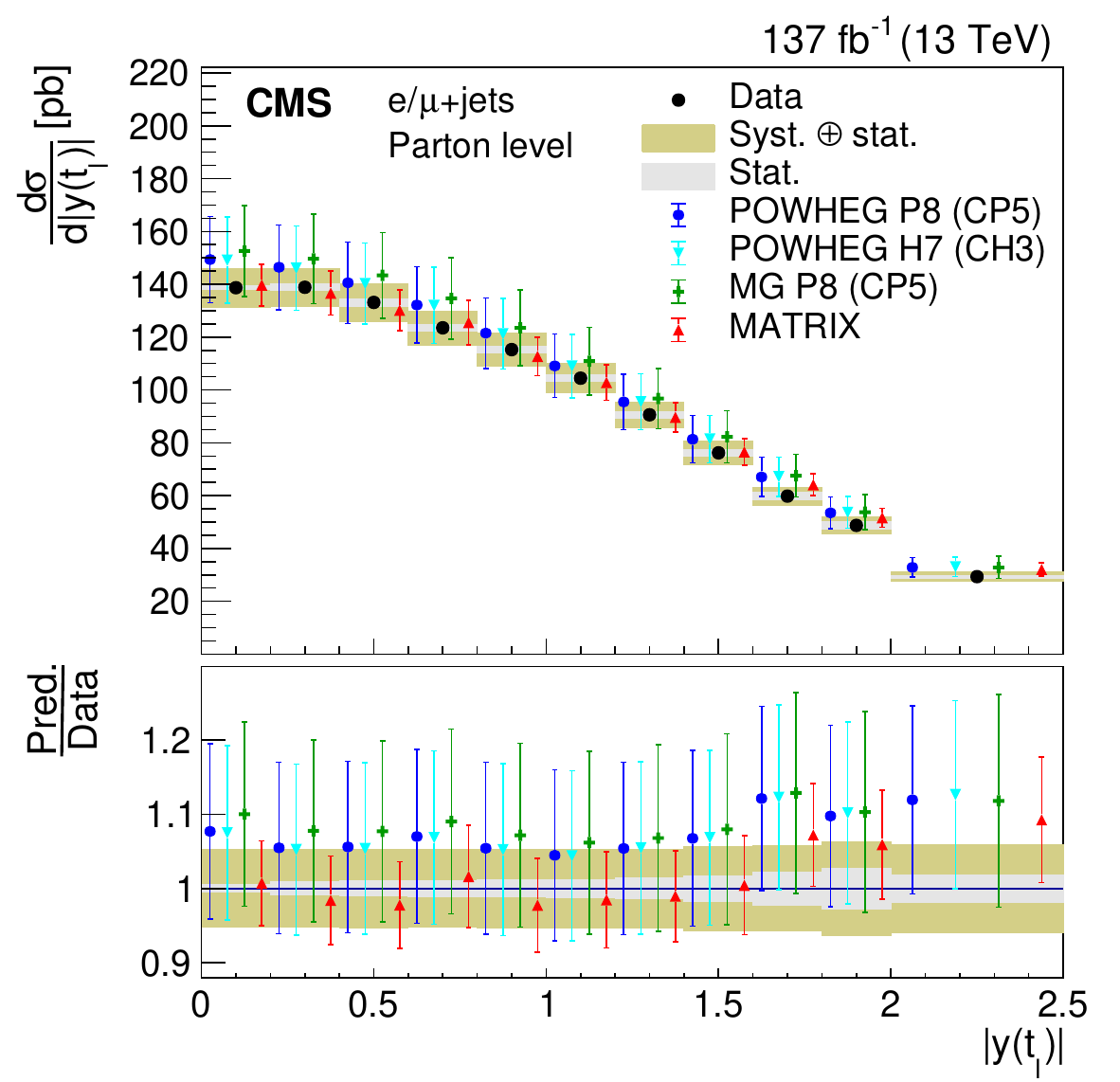}
 \includegraphics[width=0.42\textwidth]{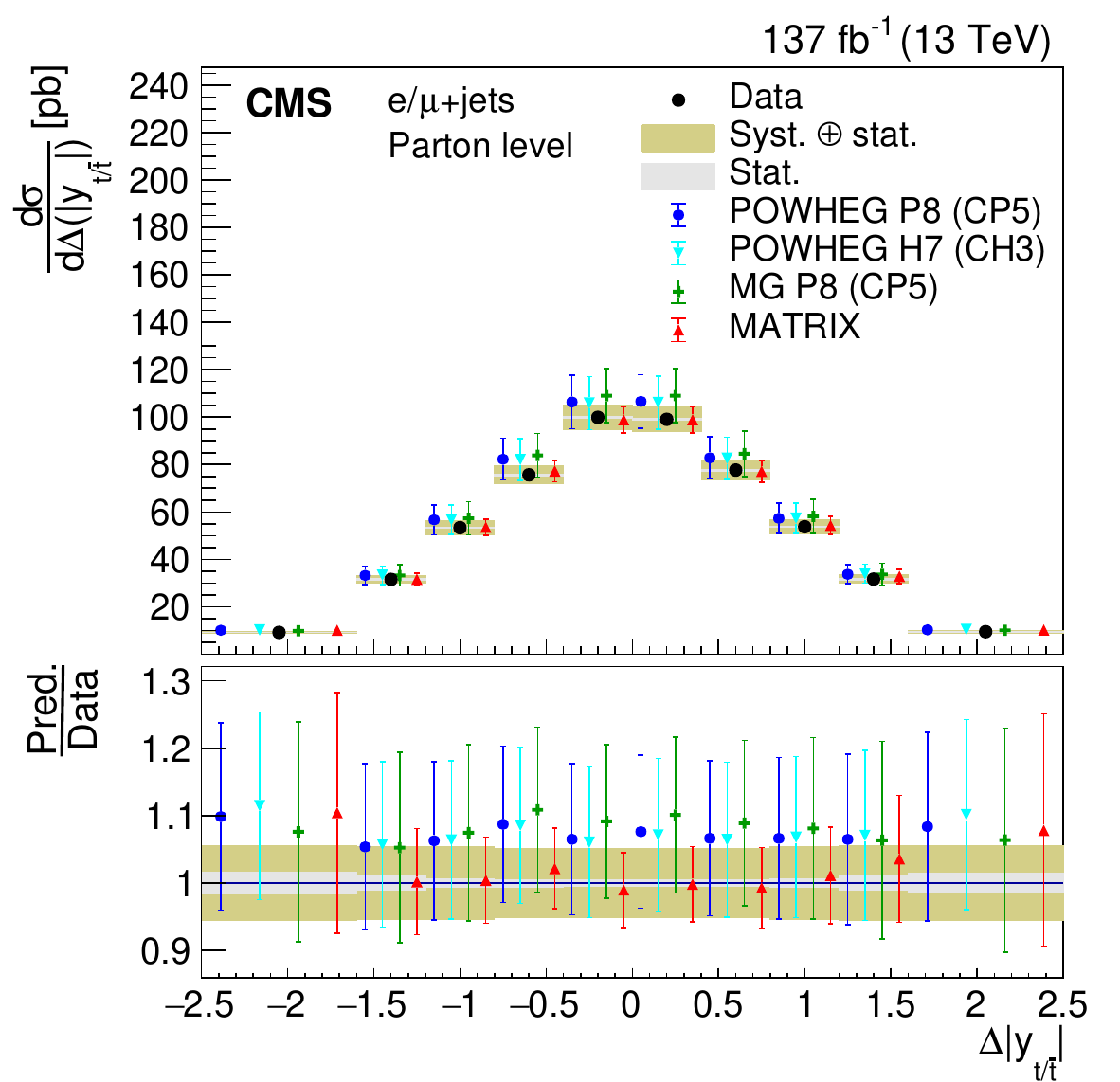}
 \includegraphics[width=0.42\textwidth]{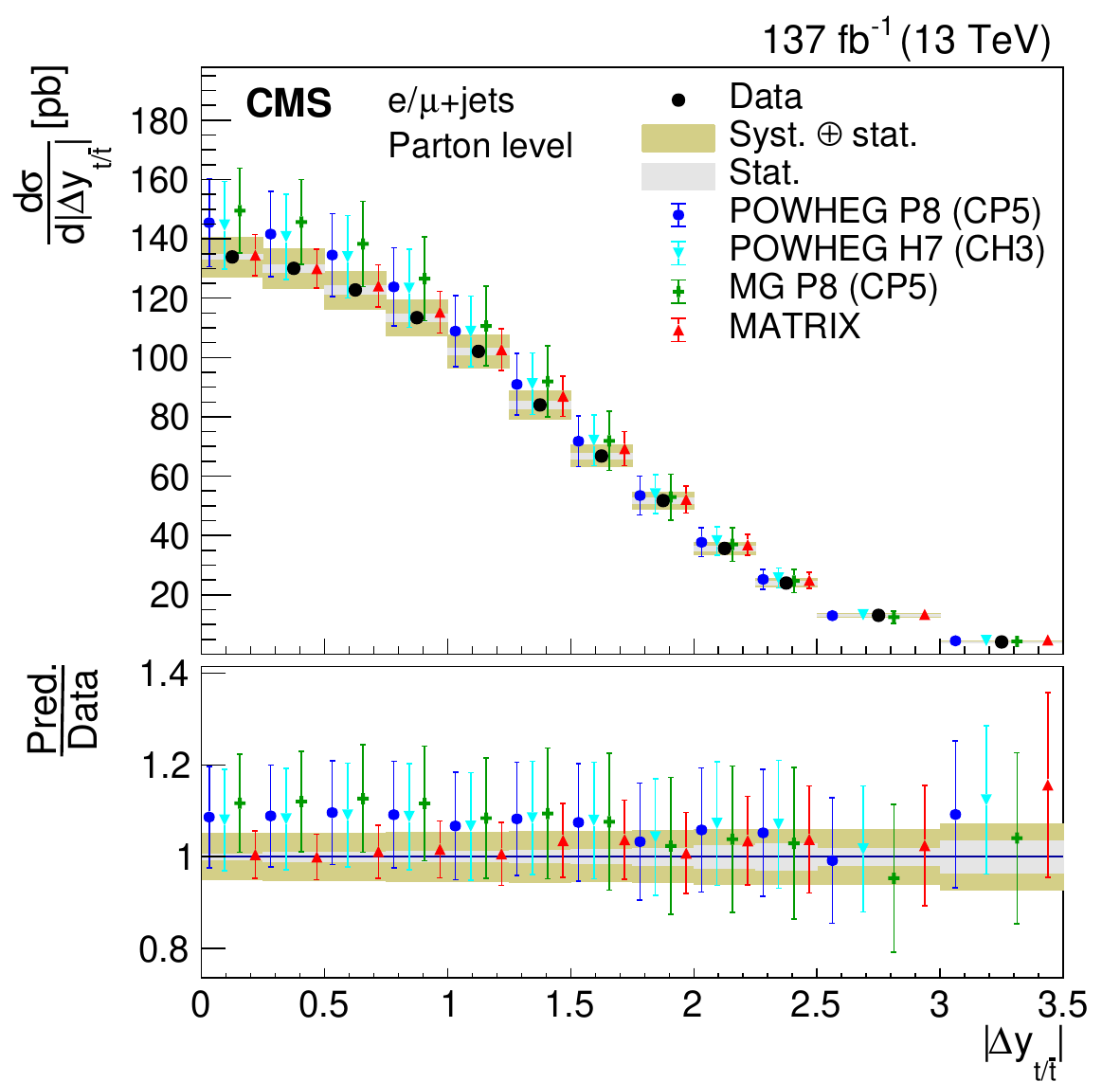}
 \caption{Differential cross sections at the parton level as a function of \thady, \tlepy, and the differences \dy and \ady. \XSECCAPPA}
 \label{fig:RES2}
\end{figure*}

\begin{figure*}[tbp]
\centering
 \includegraphics[width=0.42\textwidth]{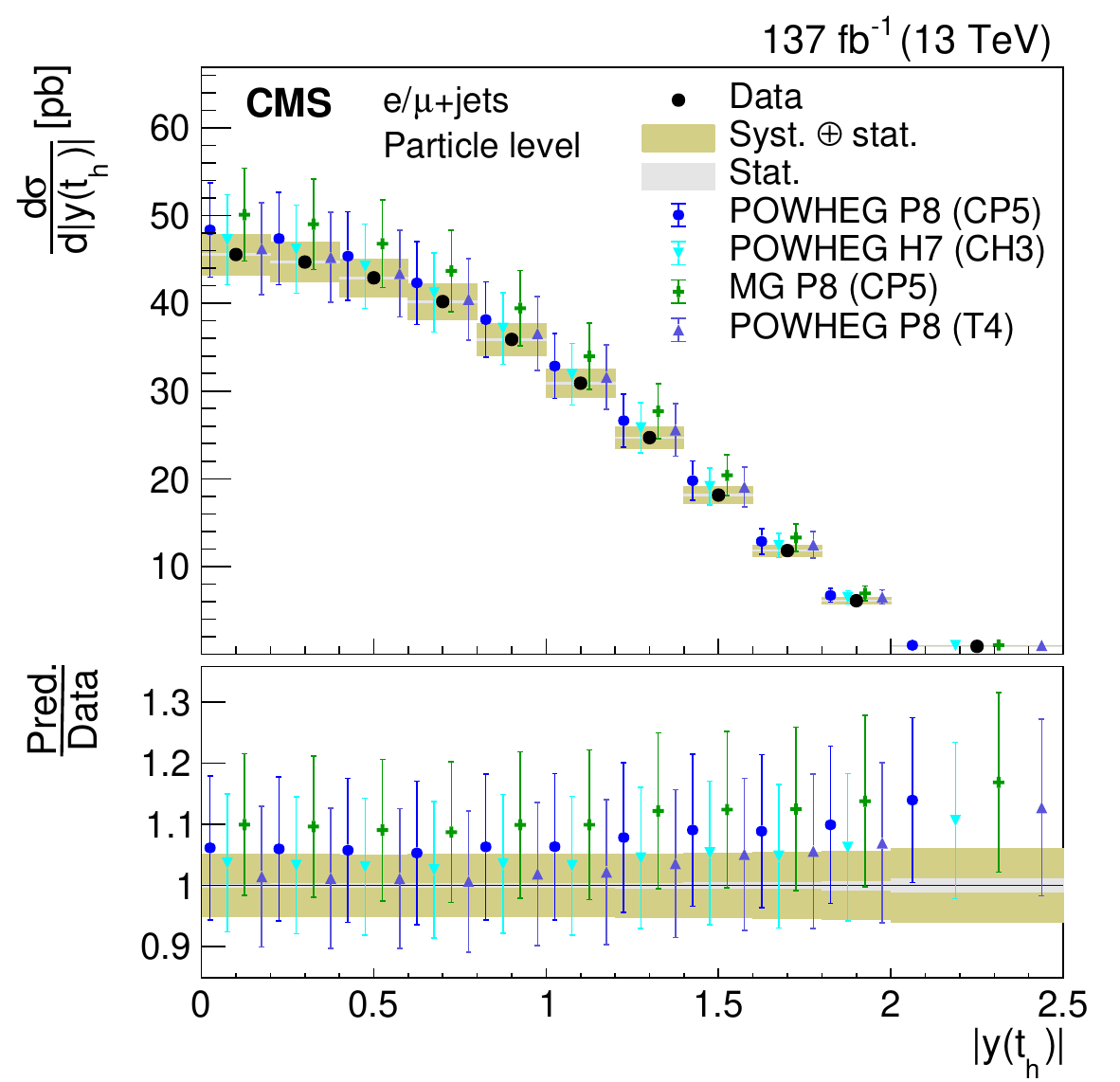}
 \includegraphics[width=0.42\textwidth]{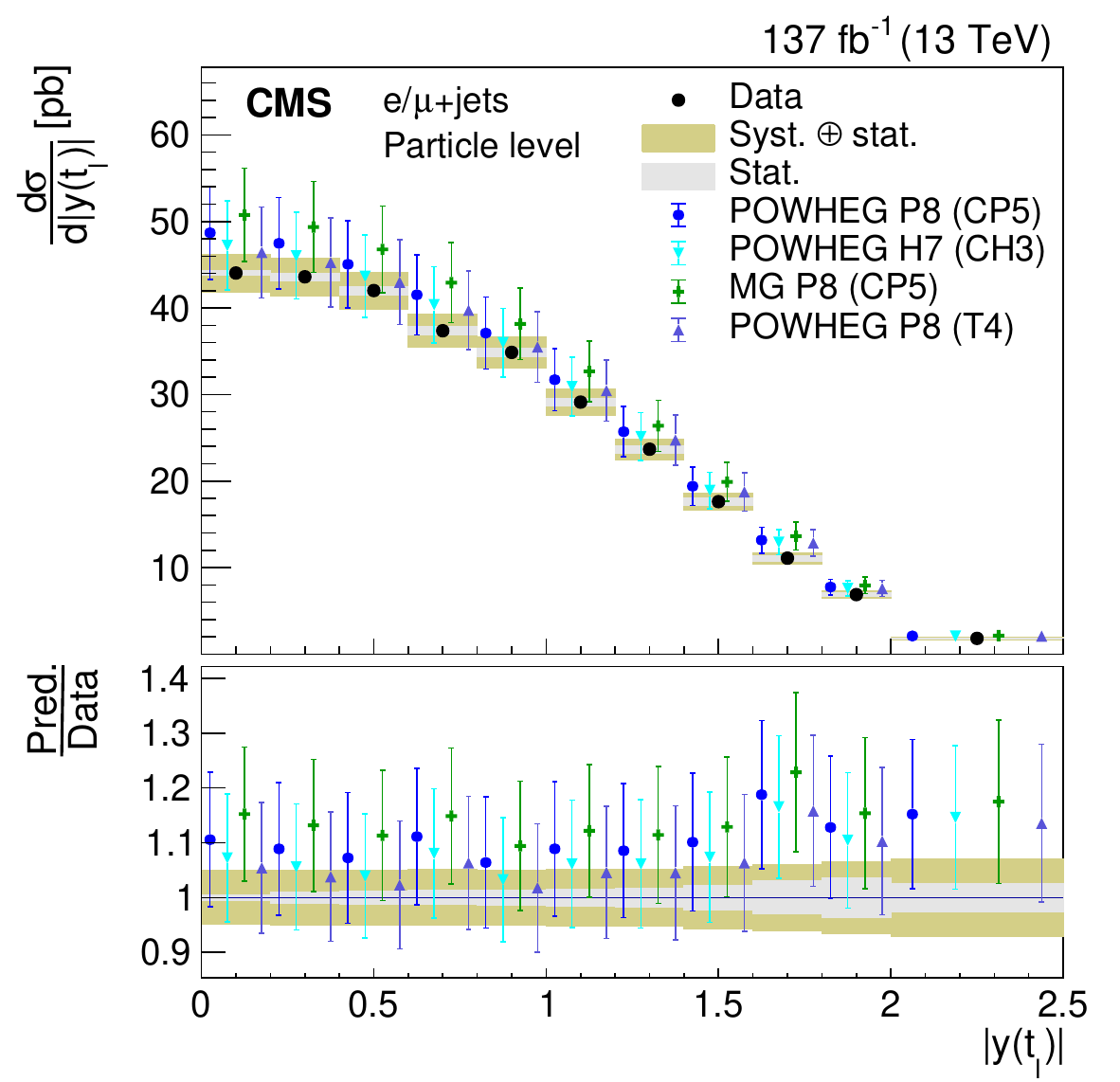}
 \includegraphics[width=0.42\textwidth]{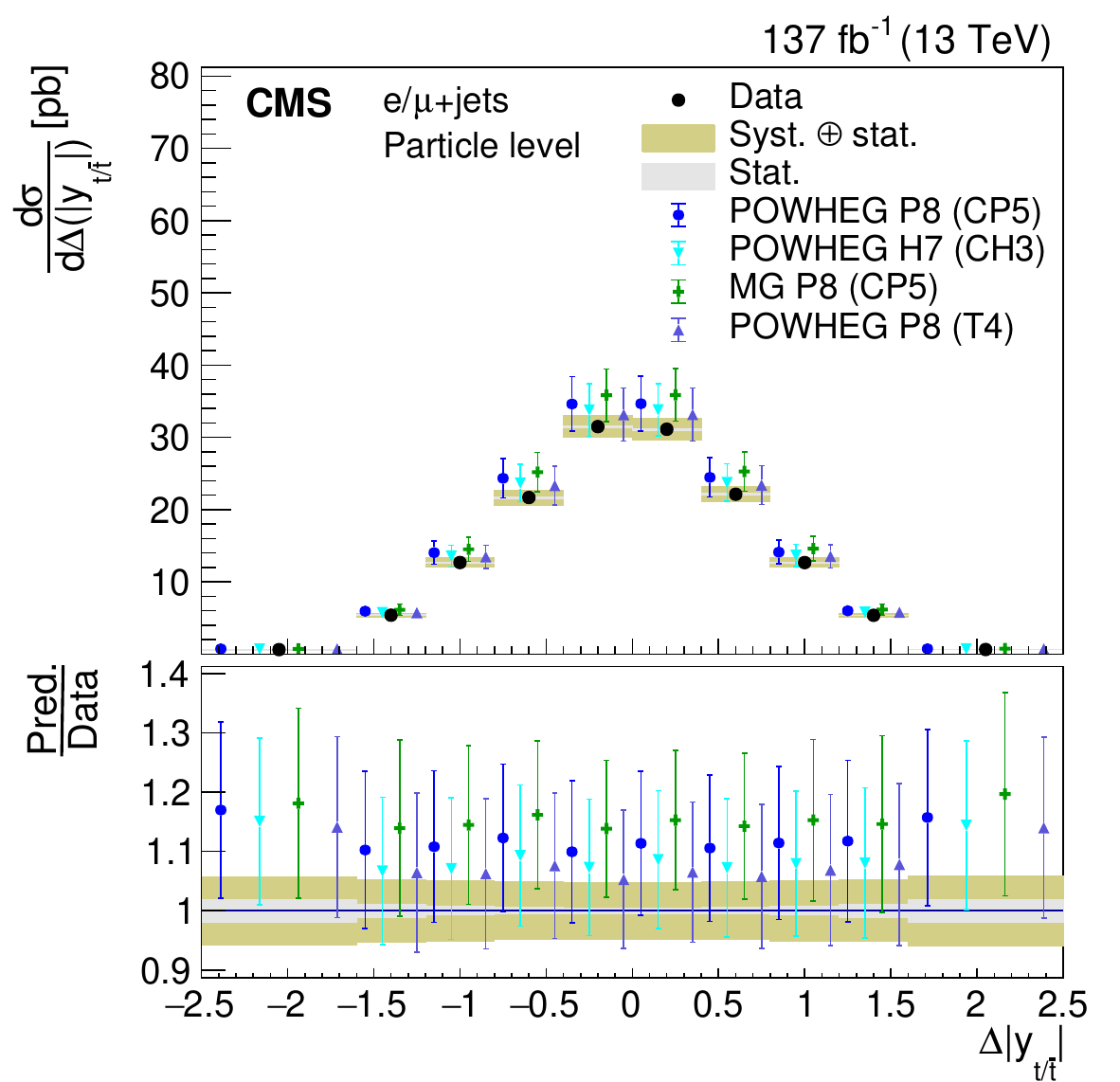}
 \includegraphics[width=0.42\textwidth]{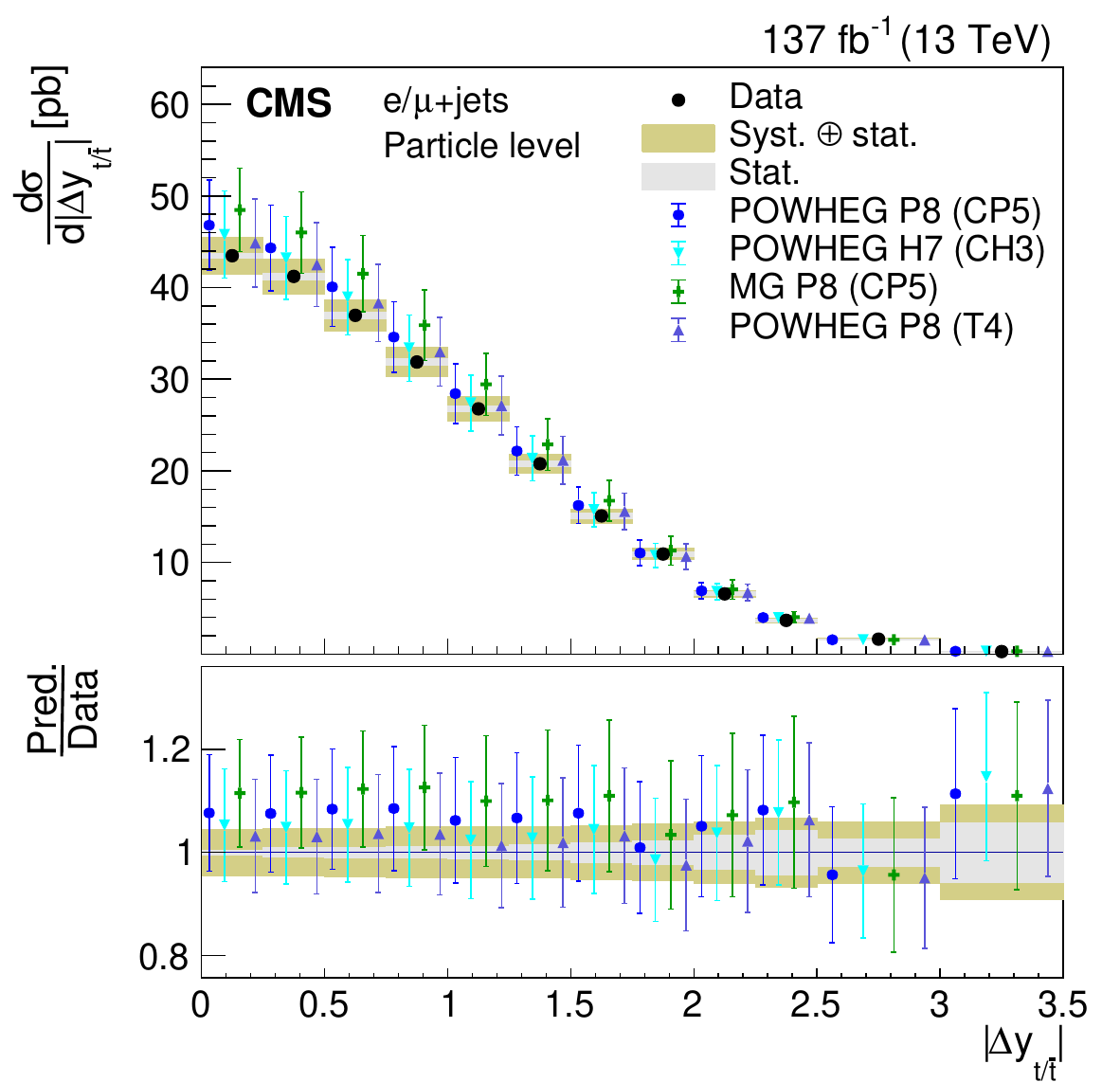}
 \caption{Differential cross sections at the particle level as a function of \thady, \tlepy, and the differences \dy and \ady. \XSECCAPPS}
 \label{fig:RESPS2}
\end{figure*}

Figures~\ref{fig:RES2} and \ref{fig:RESPS2} show the distributions of \thady, \tlepy, and the differences \dy and \ady. They are all well described by the simulations. 

\begin{figure*}[tbp]
\centering
 \includegraphics[width=0.42\textwidth]{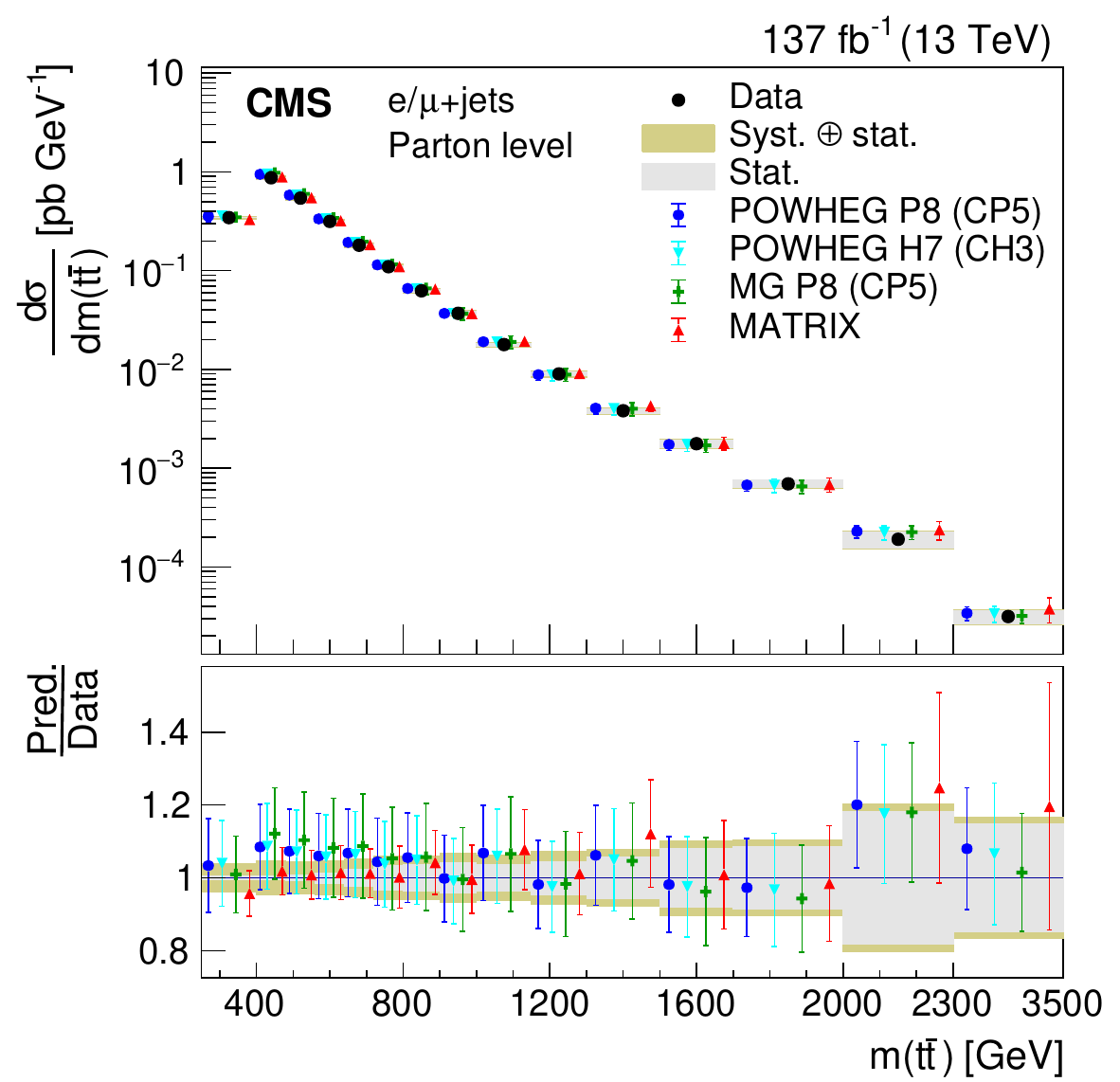}
 \includegraphics[width=0.42\textwidth]{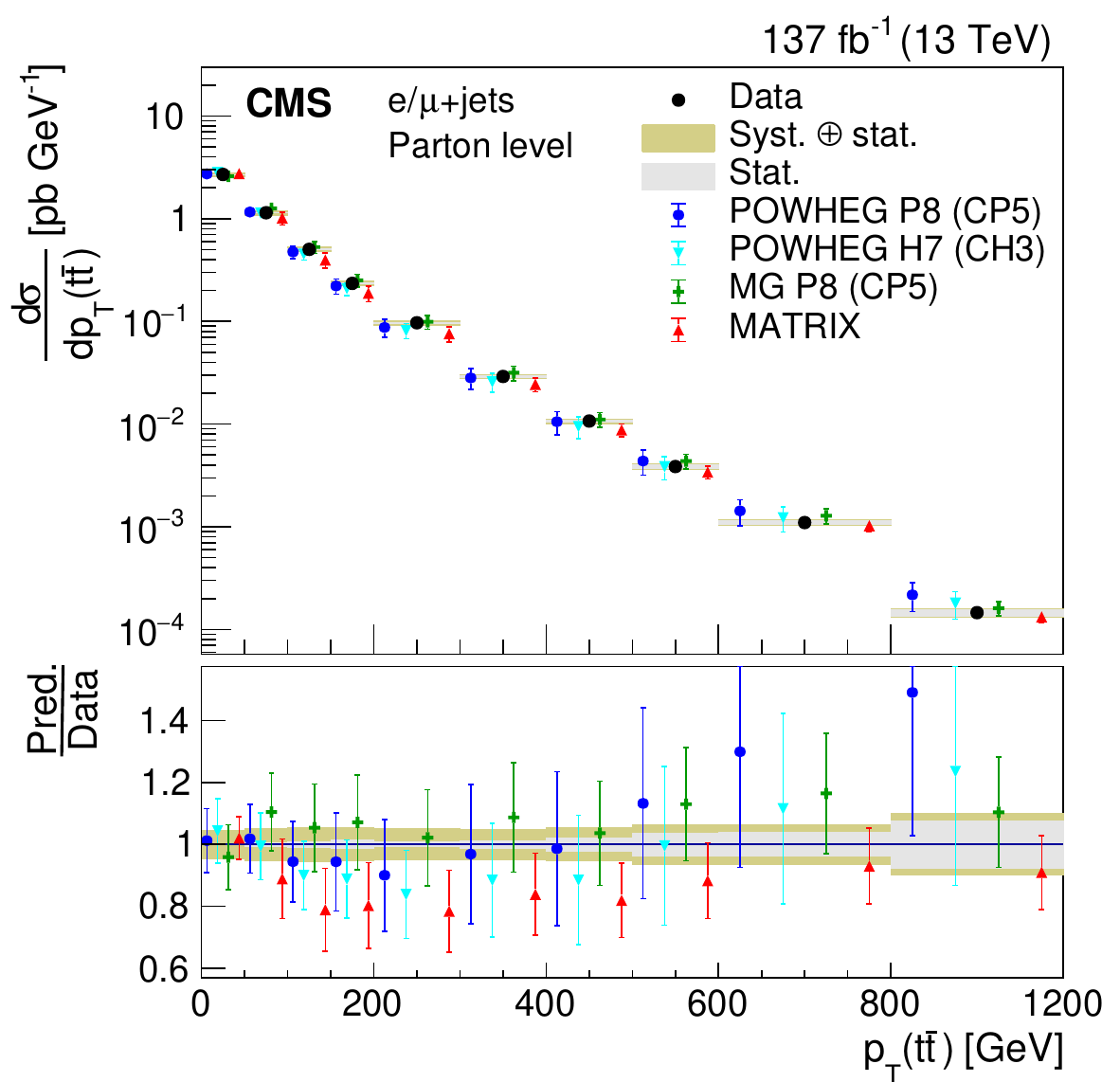}
 \includegraphics[width=0.42\textwidth]{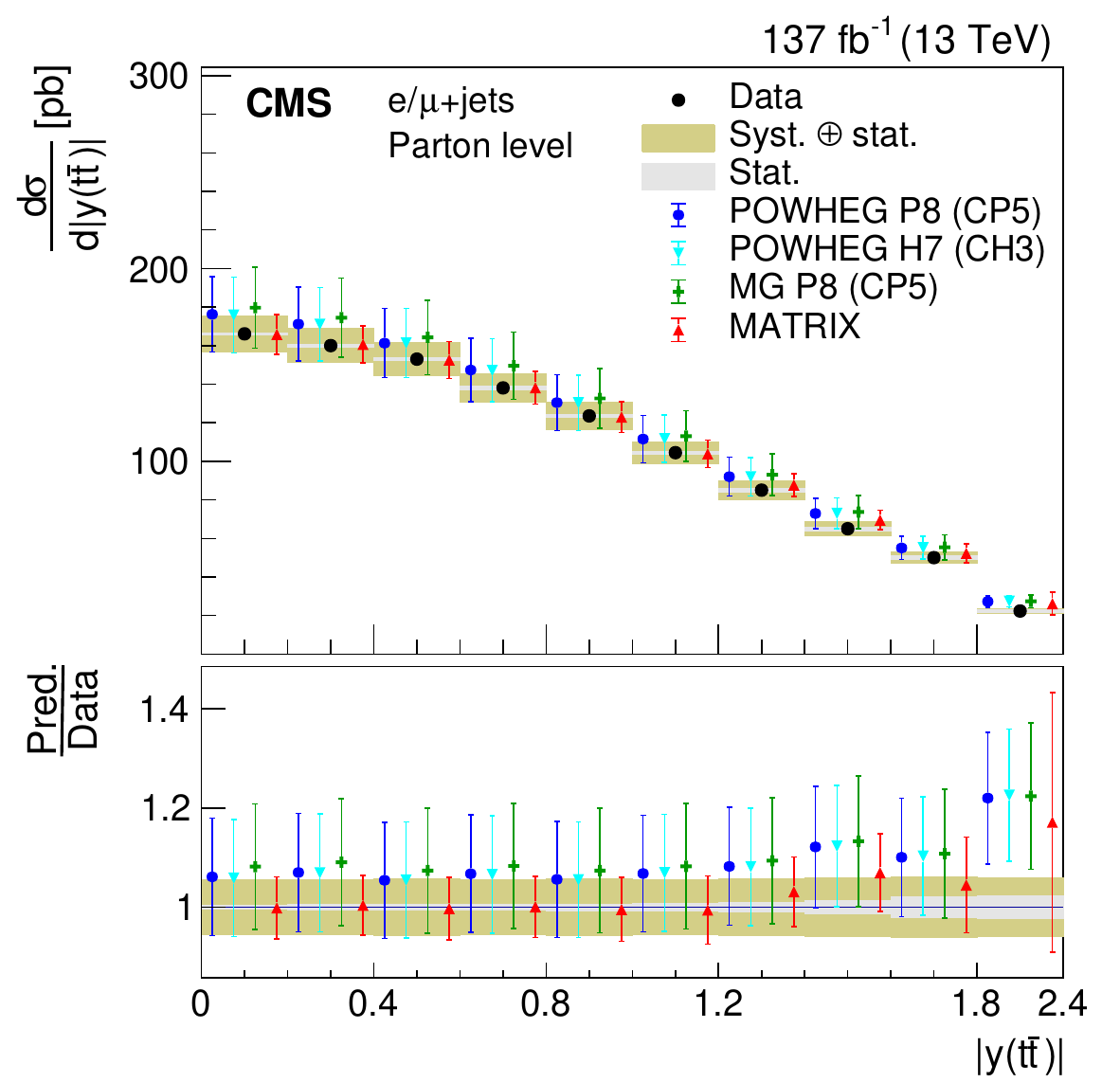}
 \includegraphics[width=0.42\textwidth]{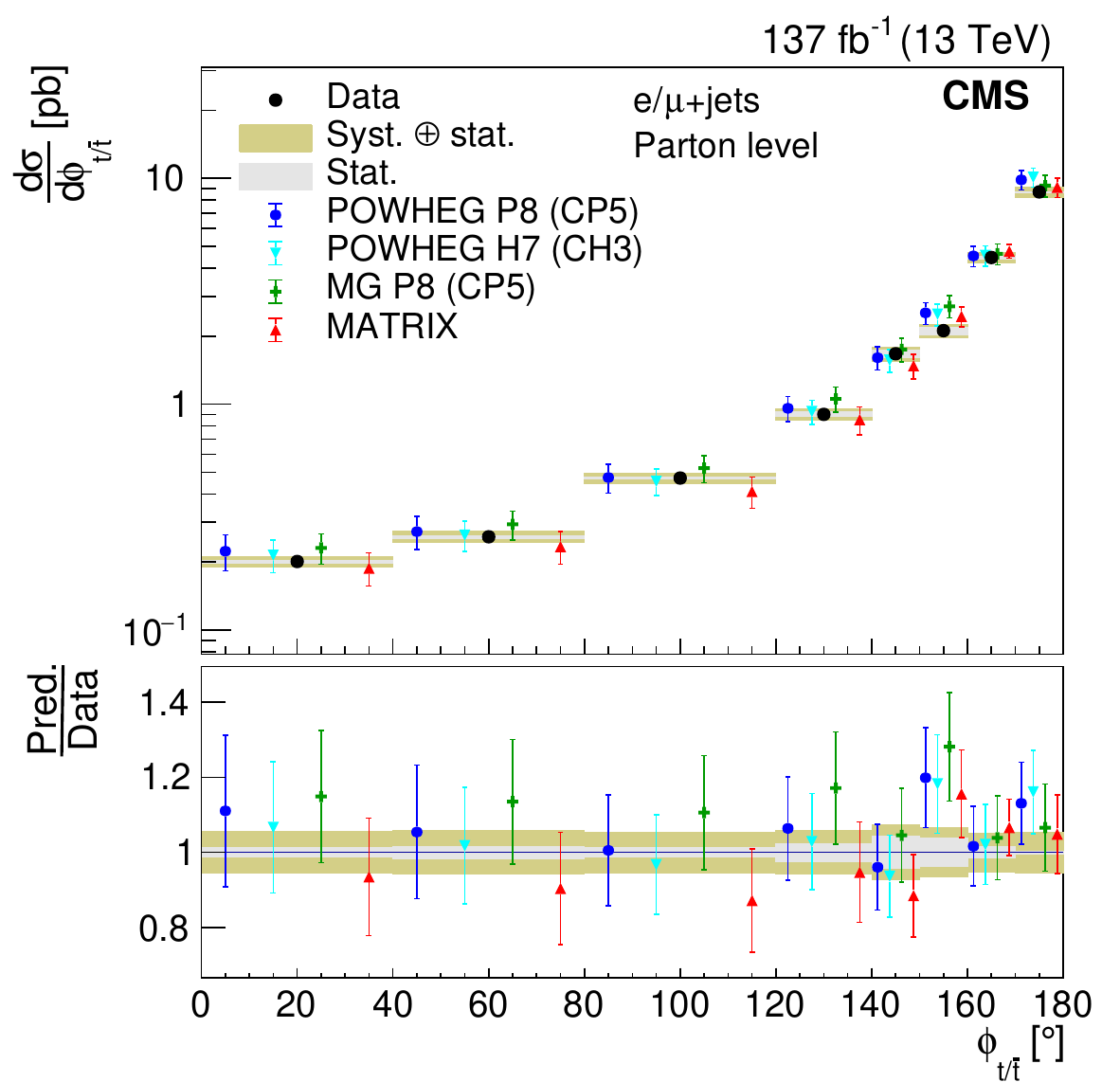}
 \includegraphics[width=0.42\textwidth]{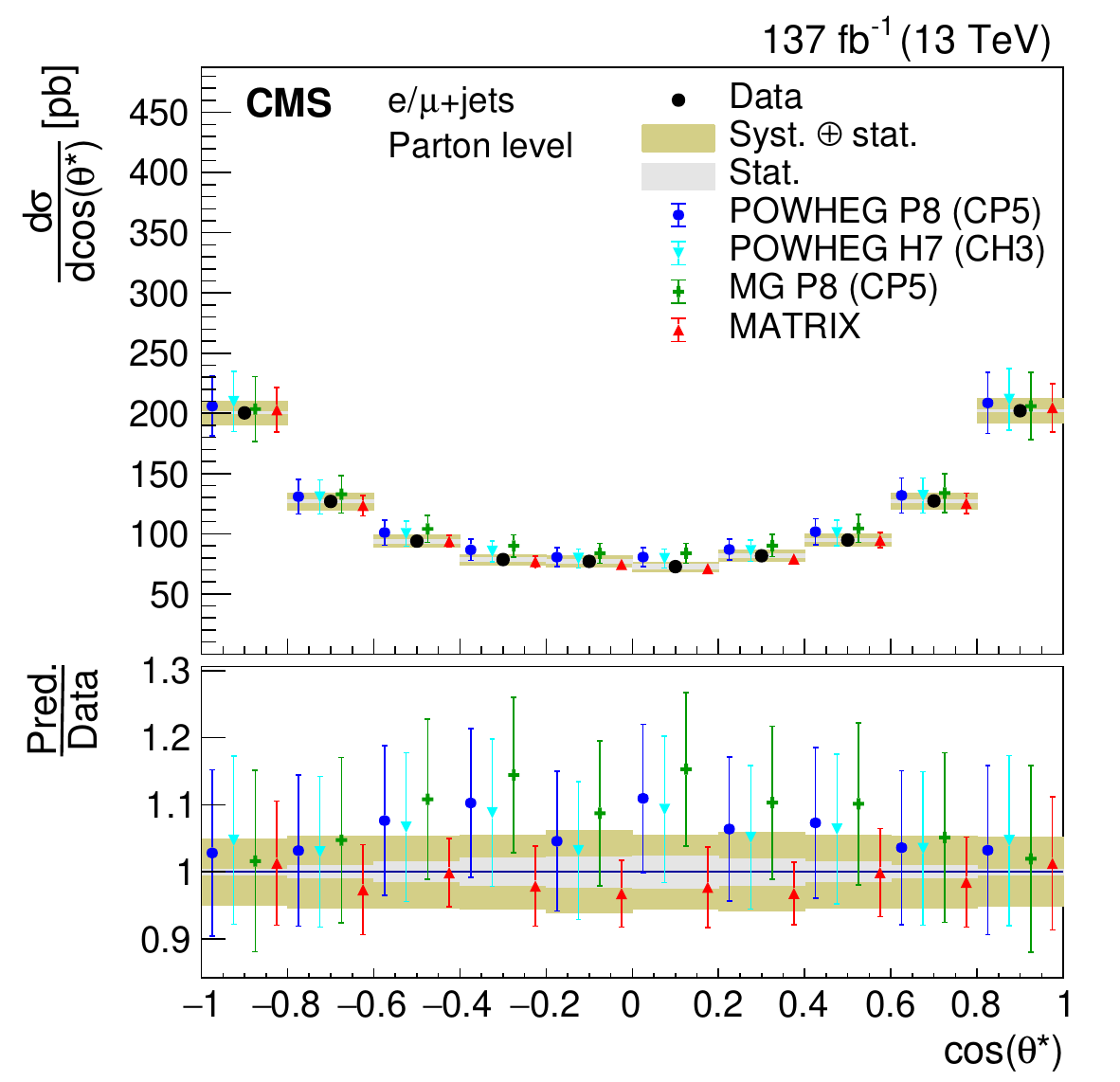}
 \caption{Differential cross sections at the parton level as a function of kinematic variables of the \ttbar system. \XSECCAPPA}
 \label{fig:RES3}
\end{figure*}

\begin{figure*}[tbp]
\centering
 \includegraphics[width=0.42\textwidth]{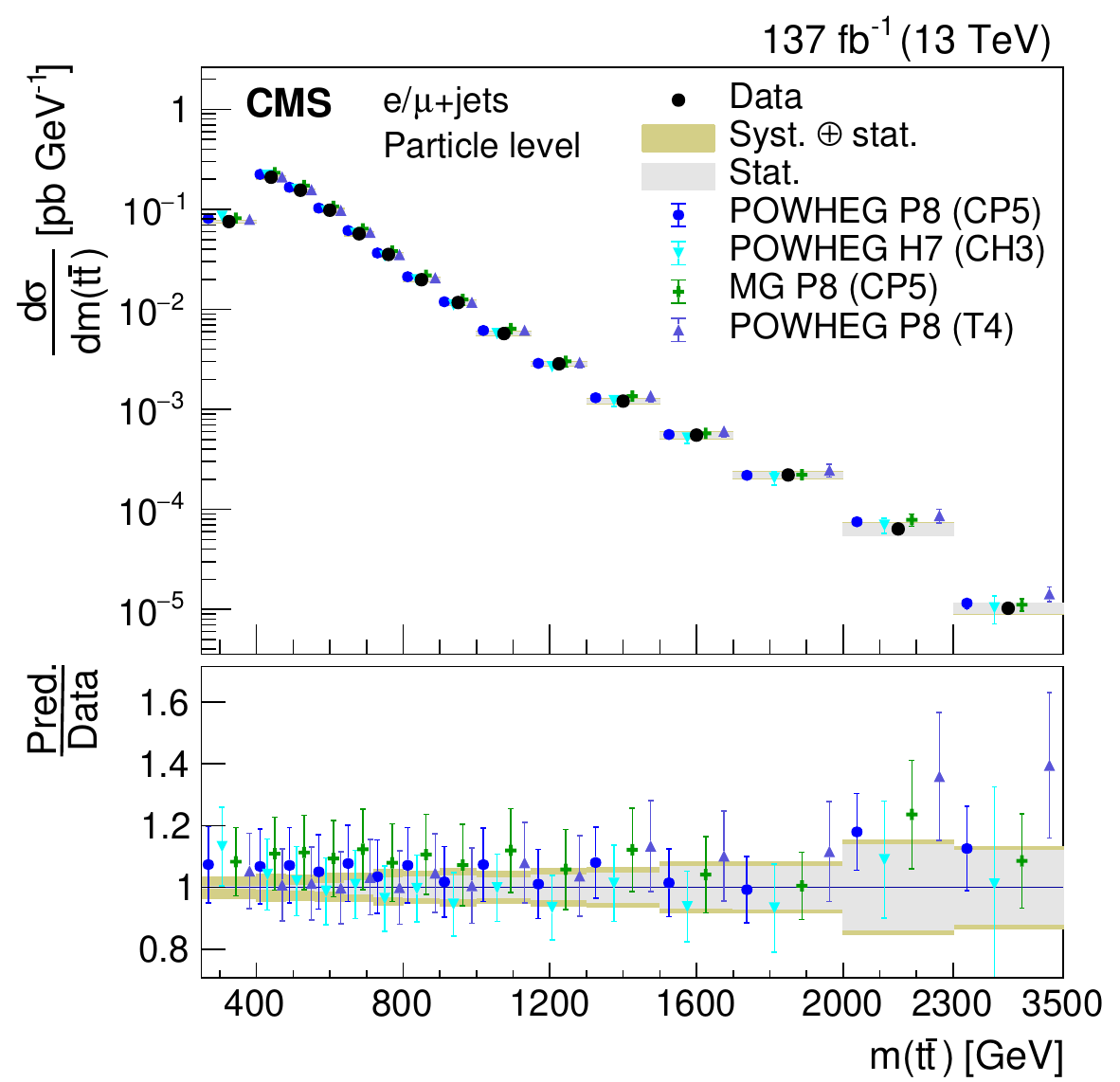}
 \includegraphics[width=0.42\textwidth]{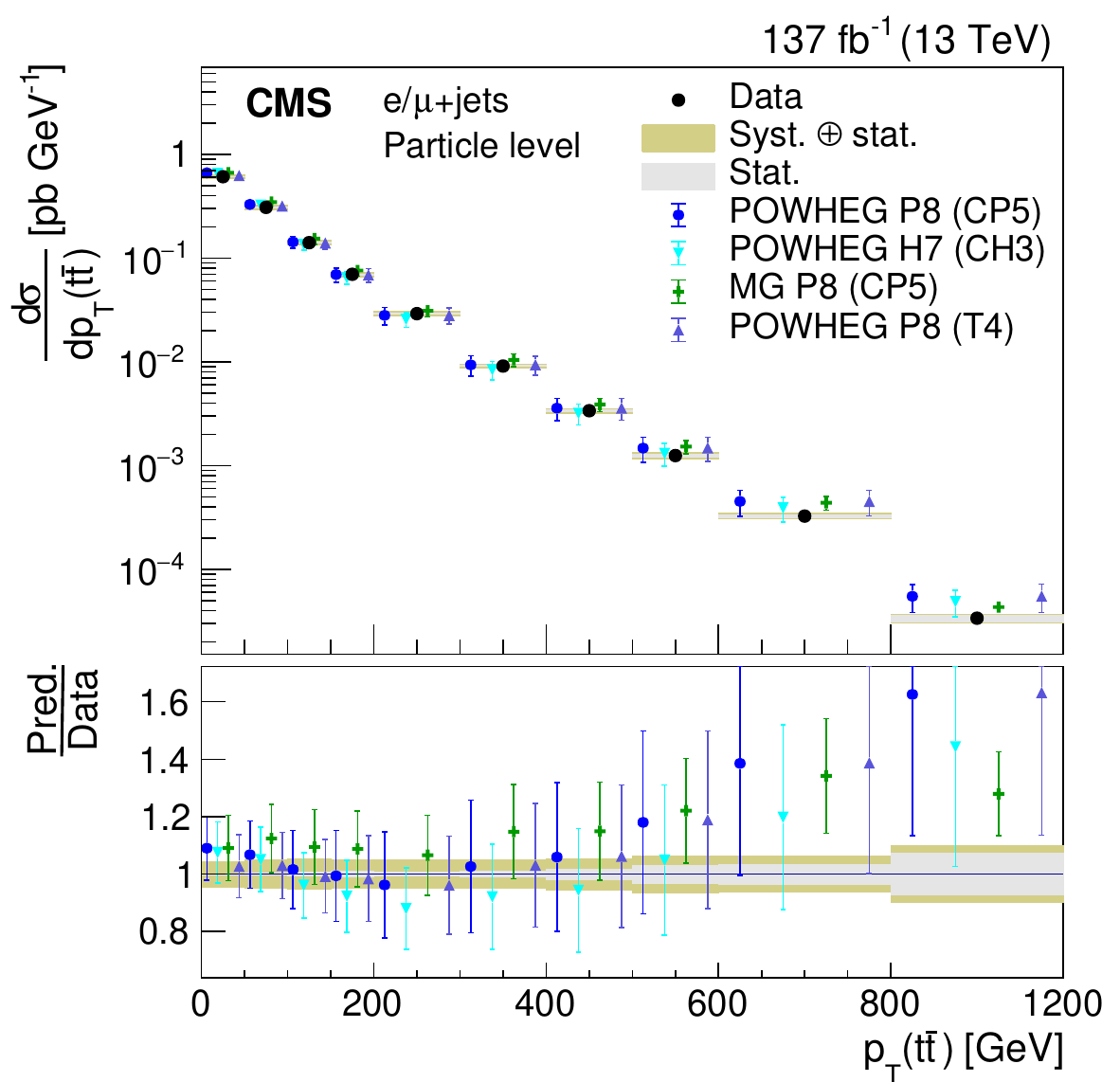}
 \includegraphics[width=0.42\textwidth]{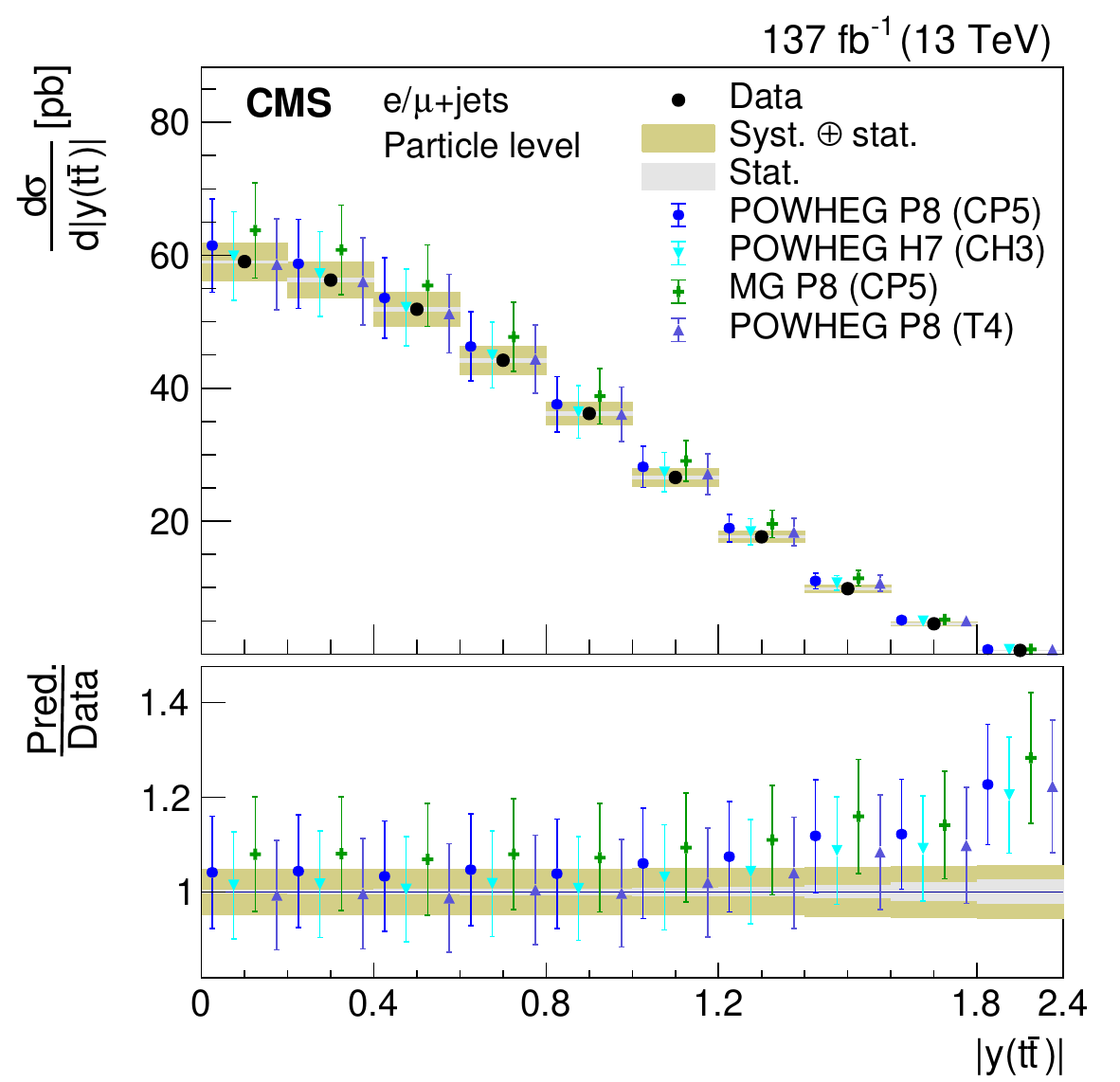}
 \includegraphics[width=0.42\textwidth]{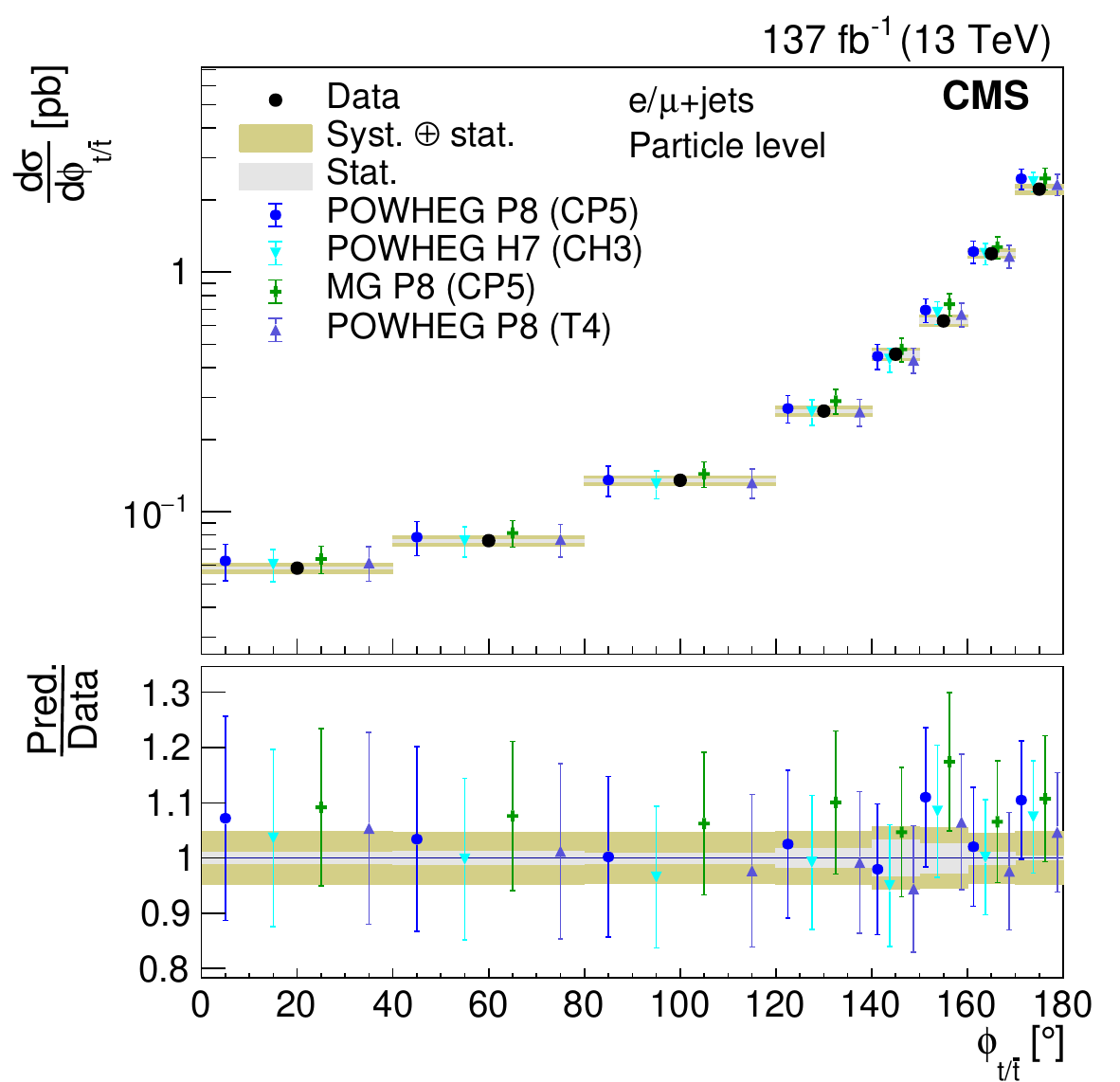}
 \includegraphics[width=0.42\textwidth]{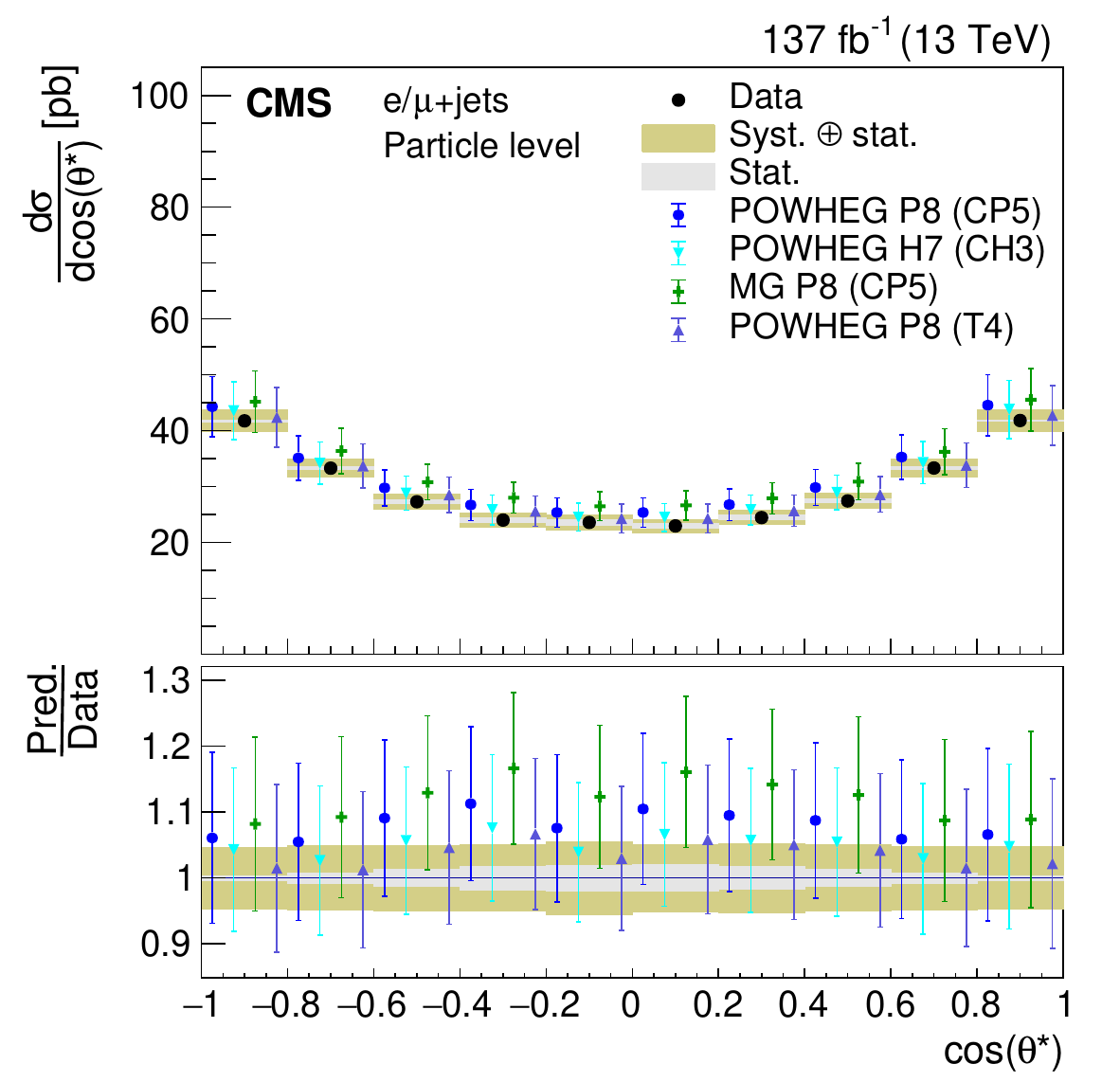}
 \caption{Differential cross sections at the particle level as a function of kinematic variables of the \ttbar system. \XSECCAPPS}
 \label{fig:RESPS3}
\end{figure*}

In Figs.~\ref{fig:RES3} and \ref{fig:RESPS3} the differential cross sections are shown as a function of the \ttbar kinematic variables \ttm, \ttpt, \tty, \ttphi, and \cts. A small modulation is observed in the \ttpt distribution, \ie, the data are underestimated at about 300\GeV and overestimated at very high values. A discrepancy also occurs at high values of \tty, where all simulations predict a higher cross section than observed. Despite these small deviations, there is no significant disagreement between data and predictions. 

\clearpage

\begin{figure*}[tbp]
\centering
 \includegraphics[width=0.42\textwidth]{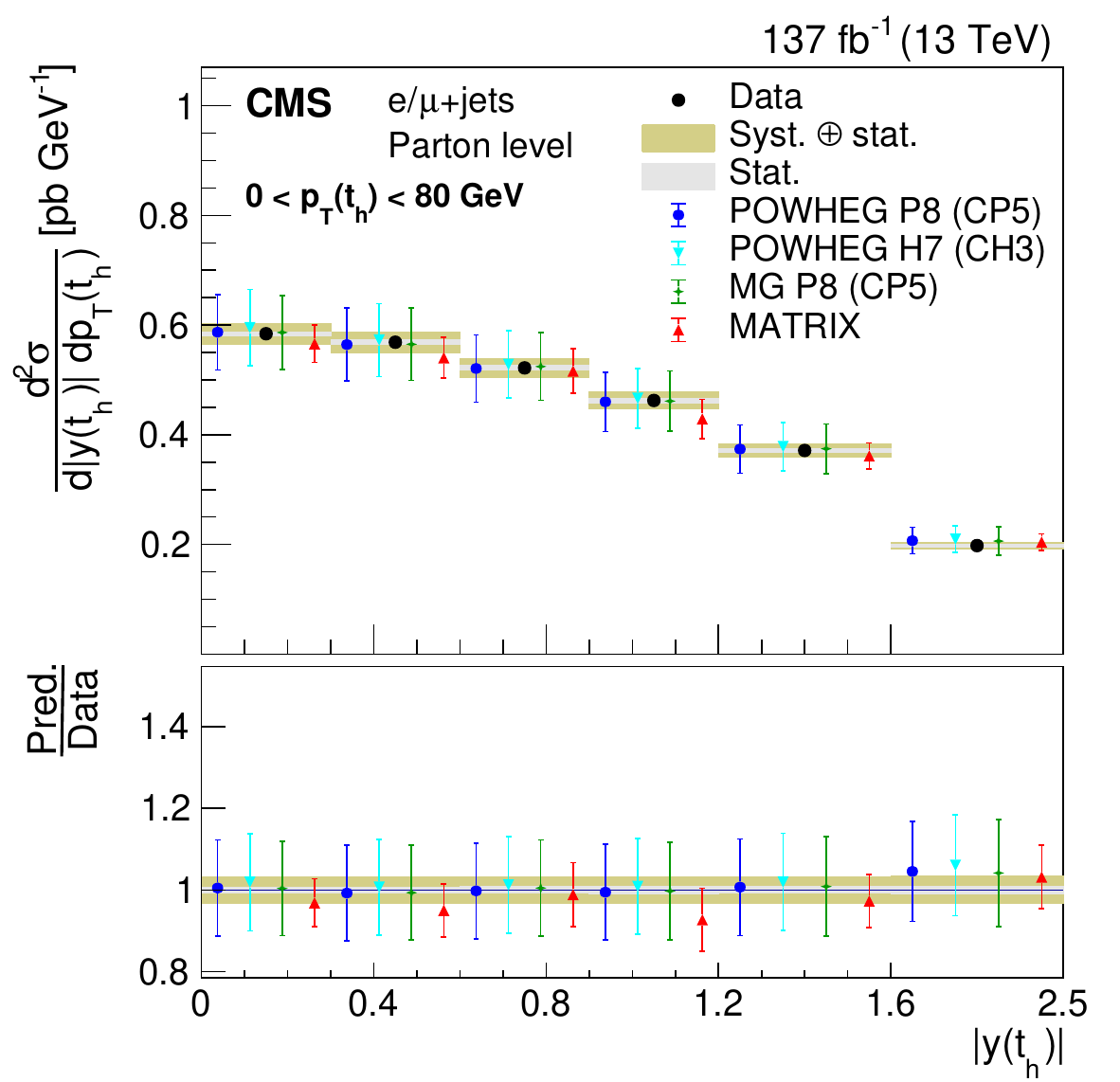}
 \includegraphics[width=0.42\textwidth]{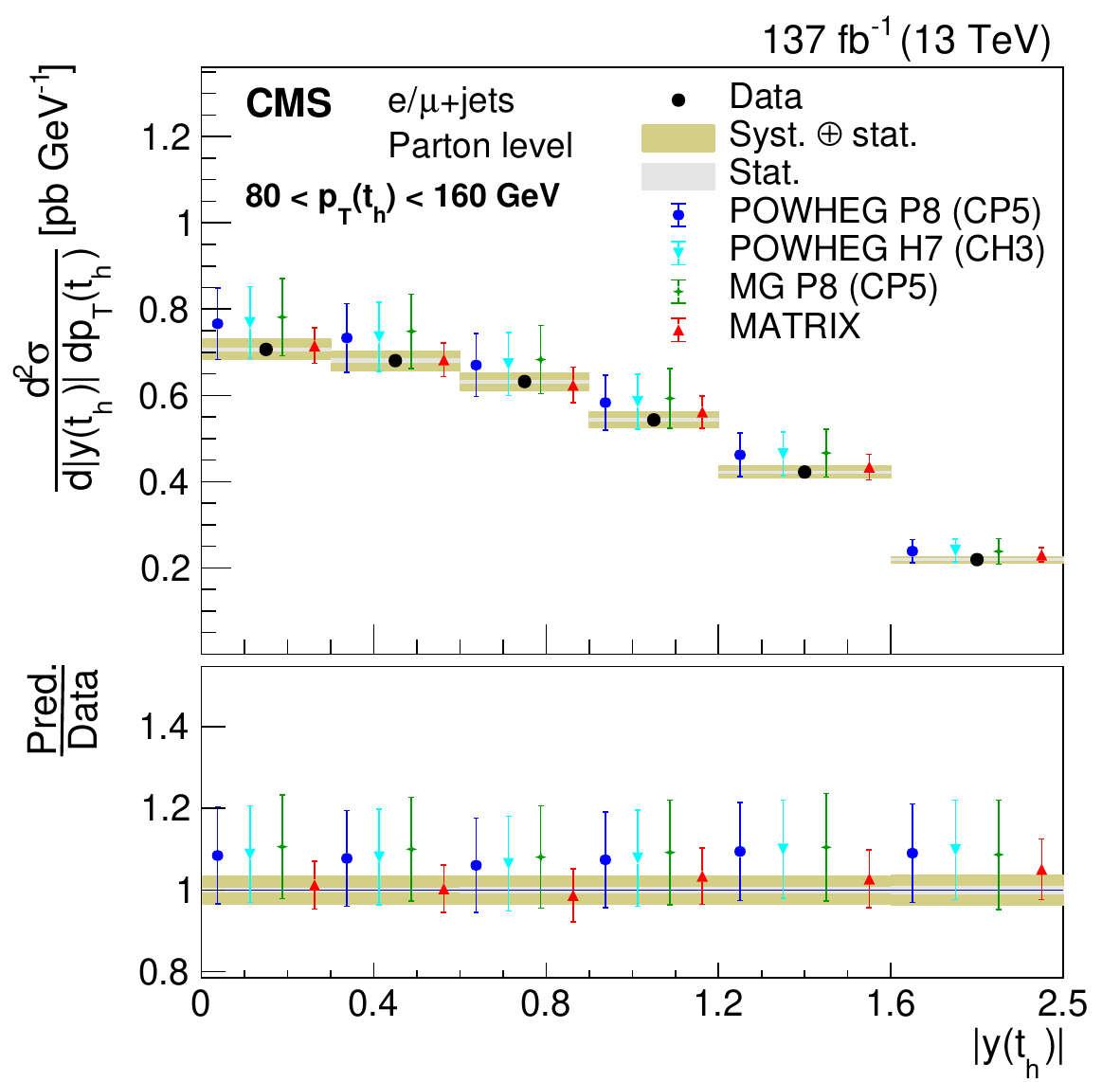}\\
 \includegraphics[width=0.42\textwidth]{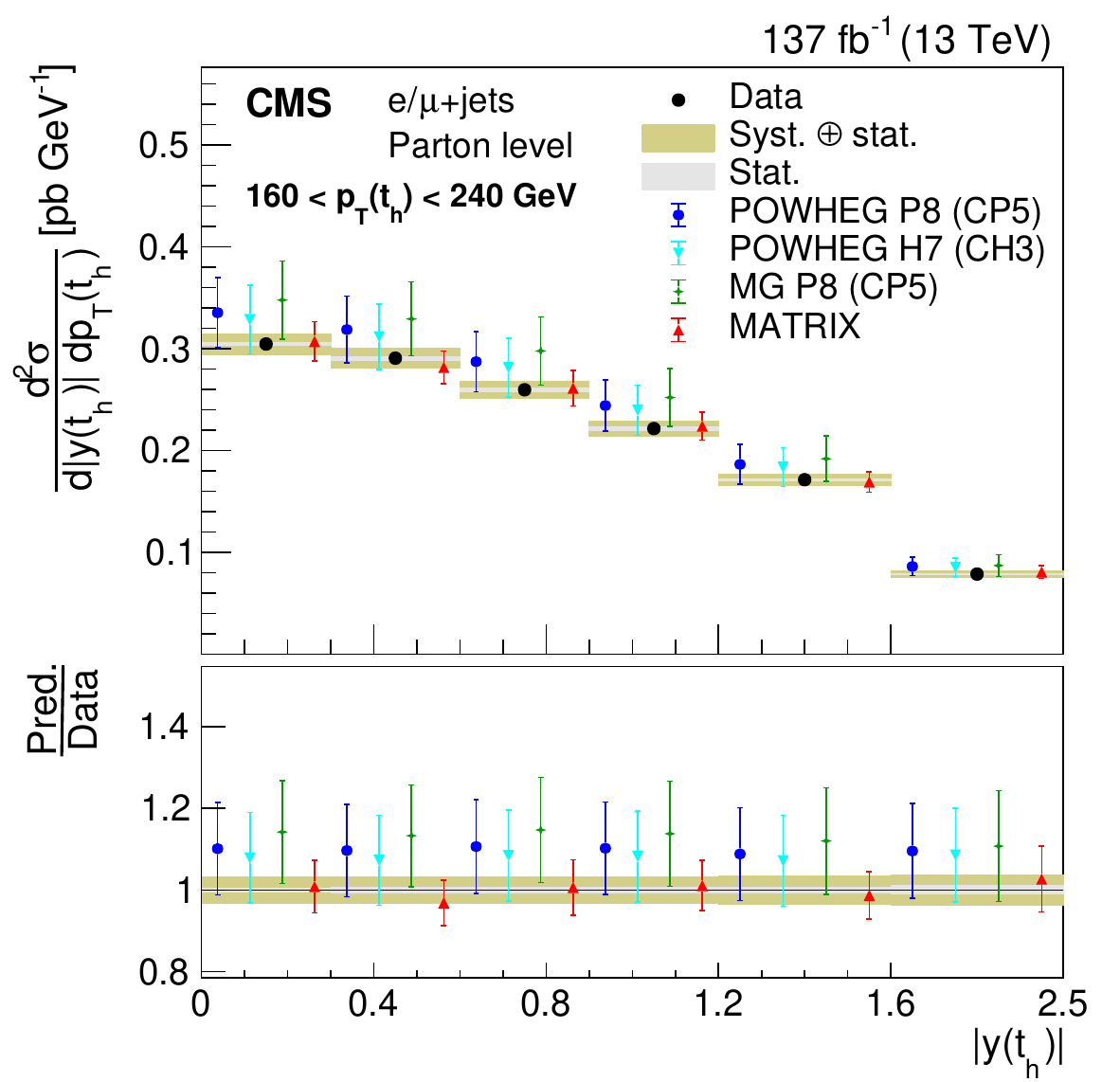}
 \includegraphics[width=0.42\textwidth]{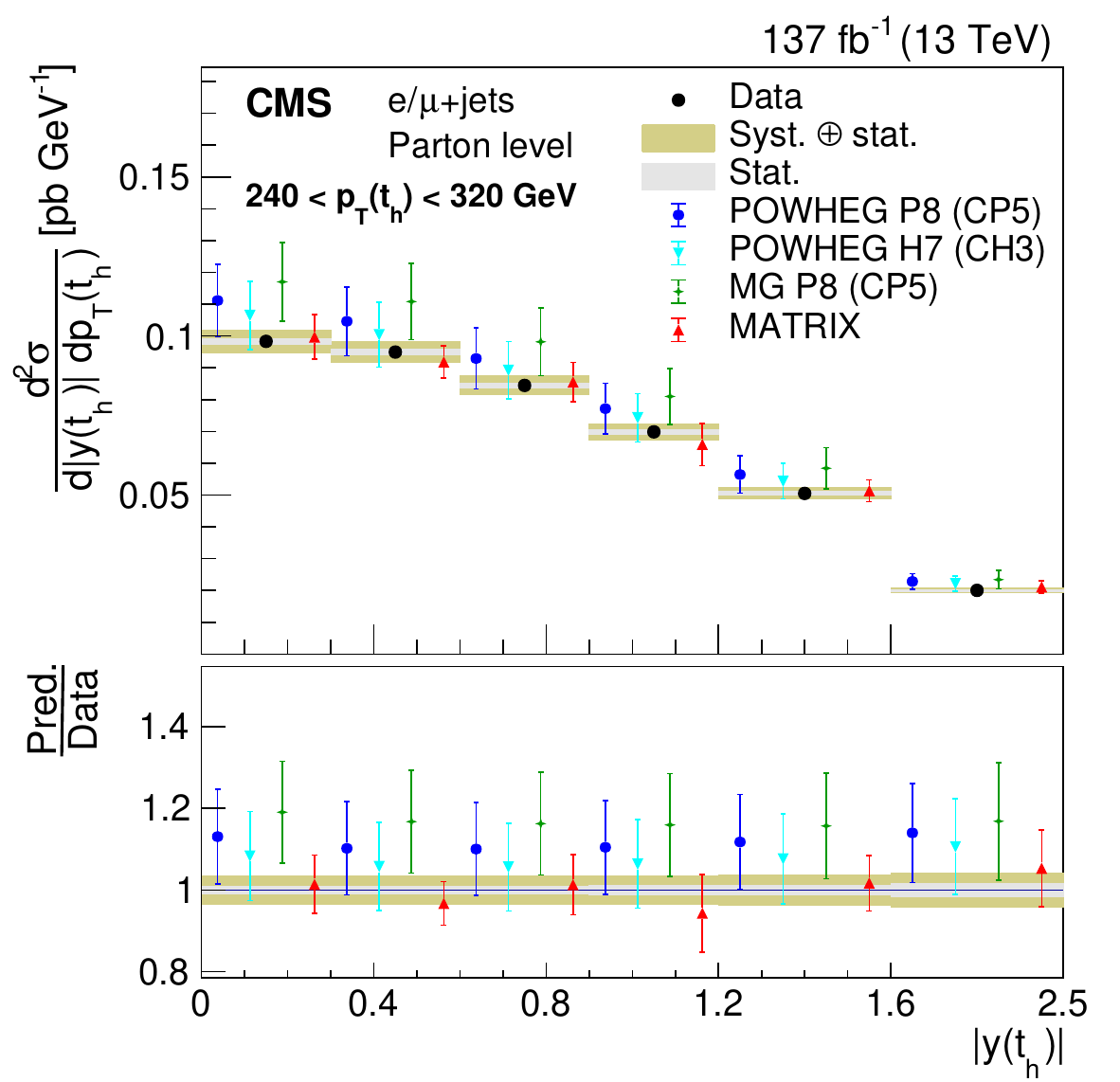}\\
 \includegraphics[width=0.42\textwidth]{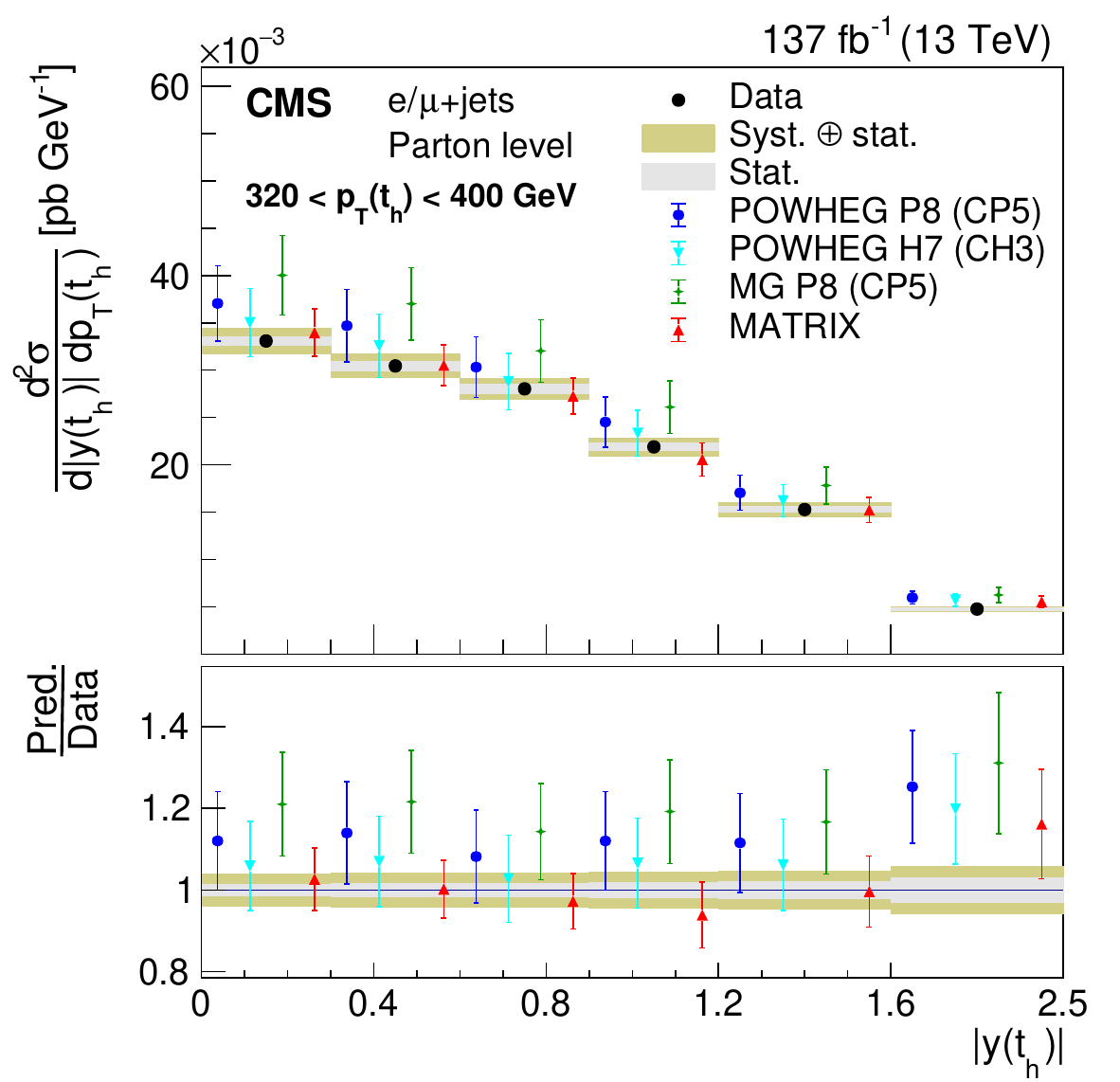}
 \includegraphics[width=0.42\textwidth]{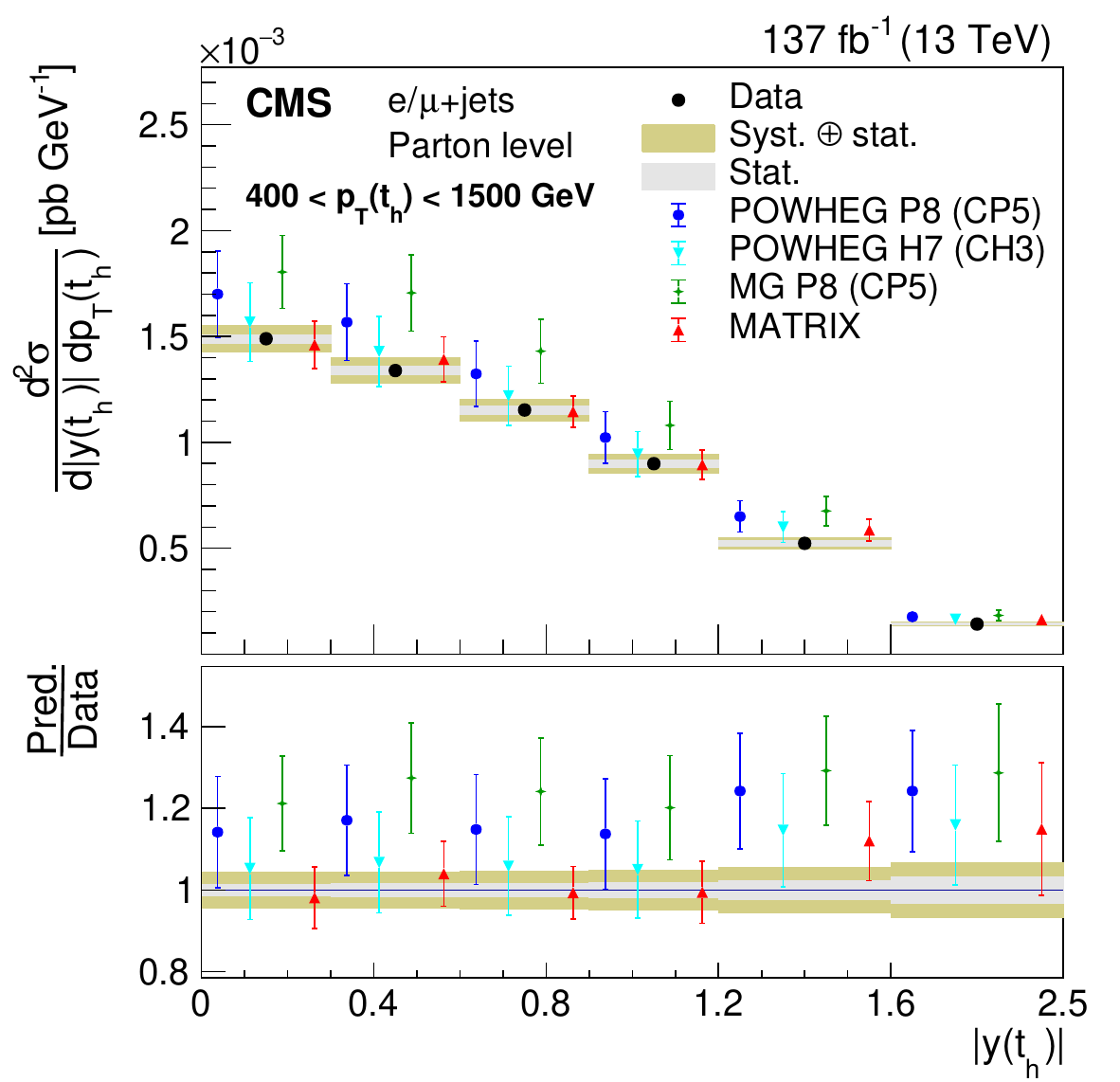}
 \caption{Double-differential cross section at the parton level as a function of \thadptvsthady. \XSECCAPPA}
 \label{fig:RES4}
\end{figure*}

\begin{figure*}[tbp]
\centering
 \includegraphics[width=0.42\textwidth]{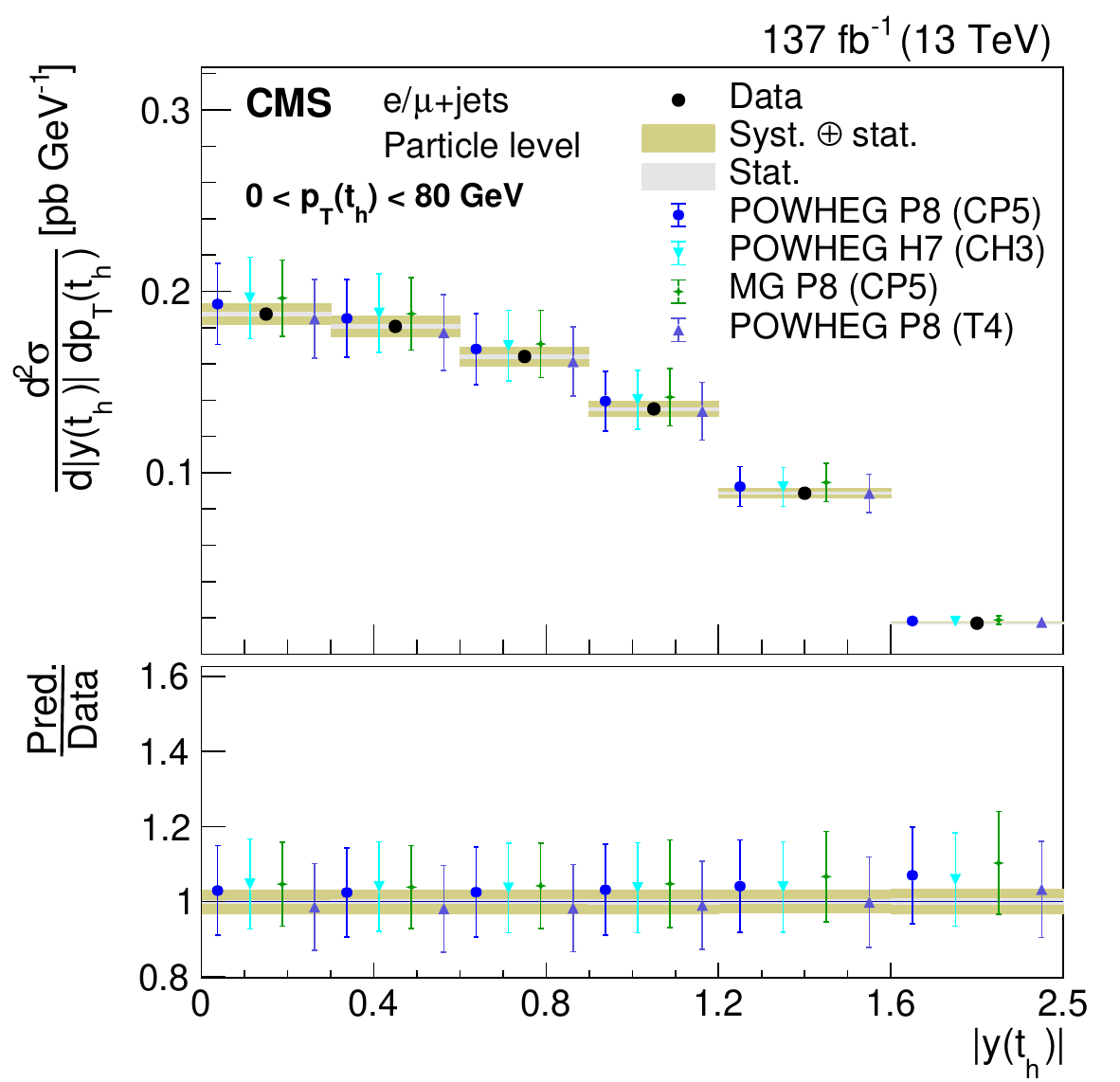}
 \includegraphics[width=0.42\textwidth]{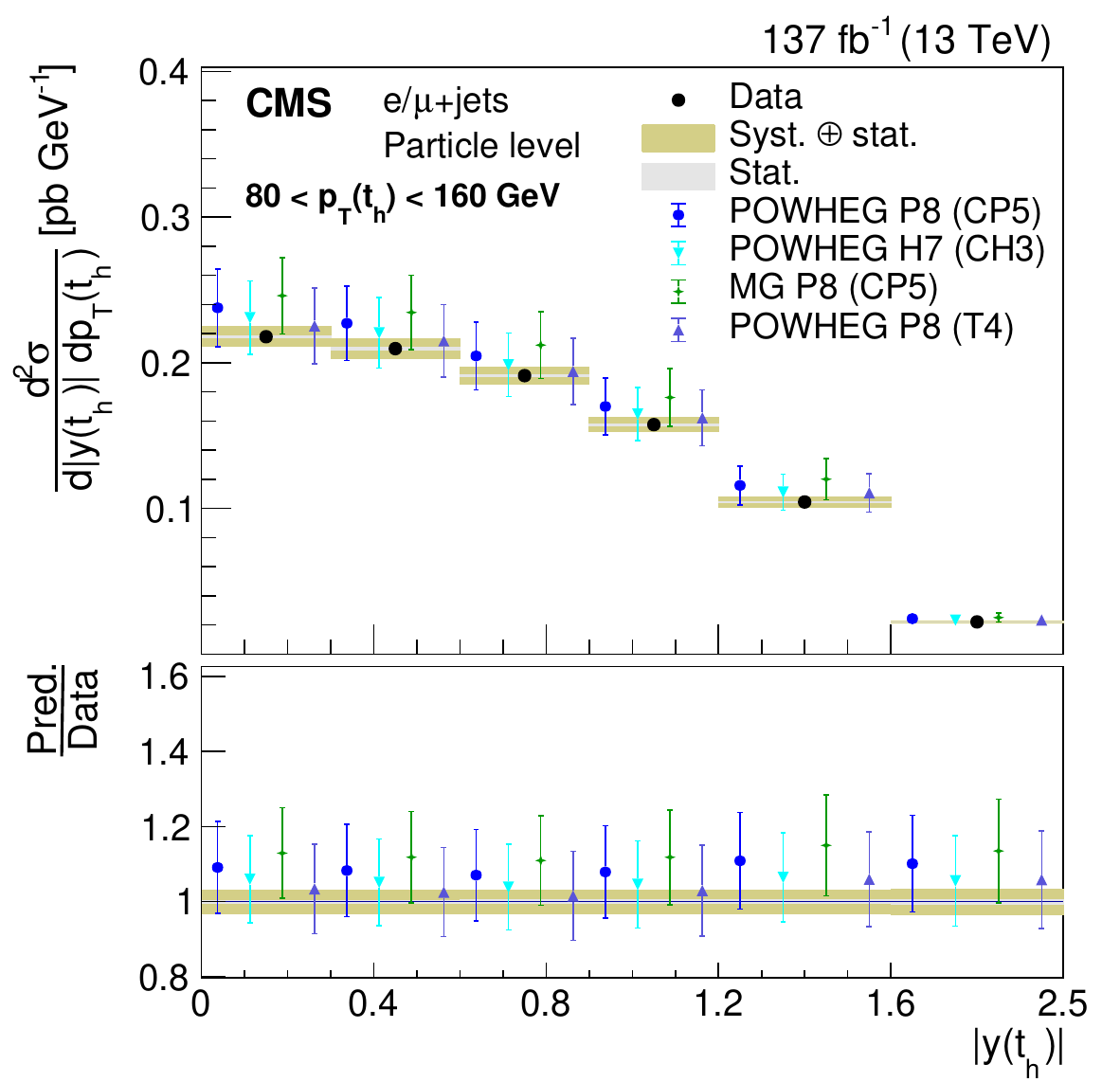}\\
 \includegraphics[width=0.42\textwidth]{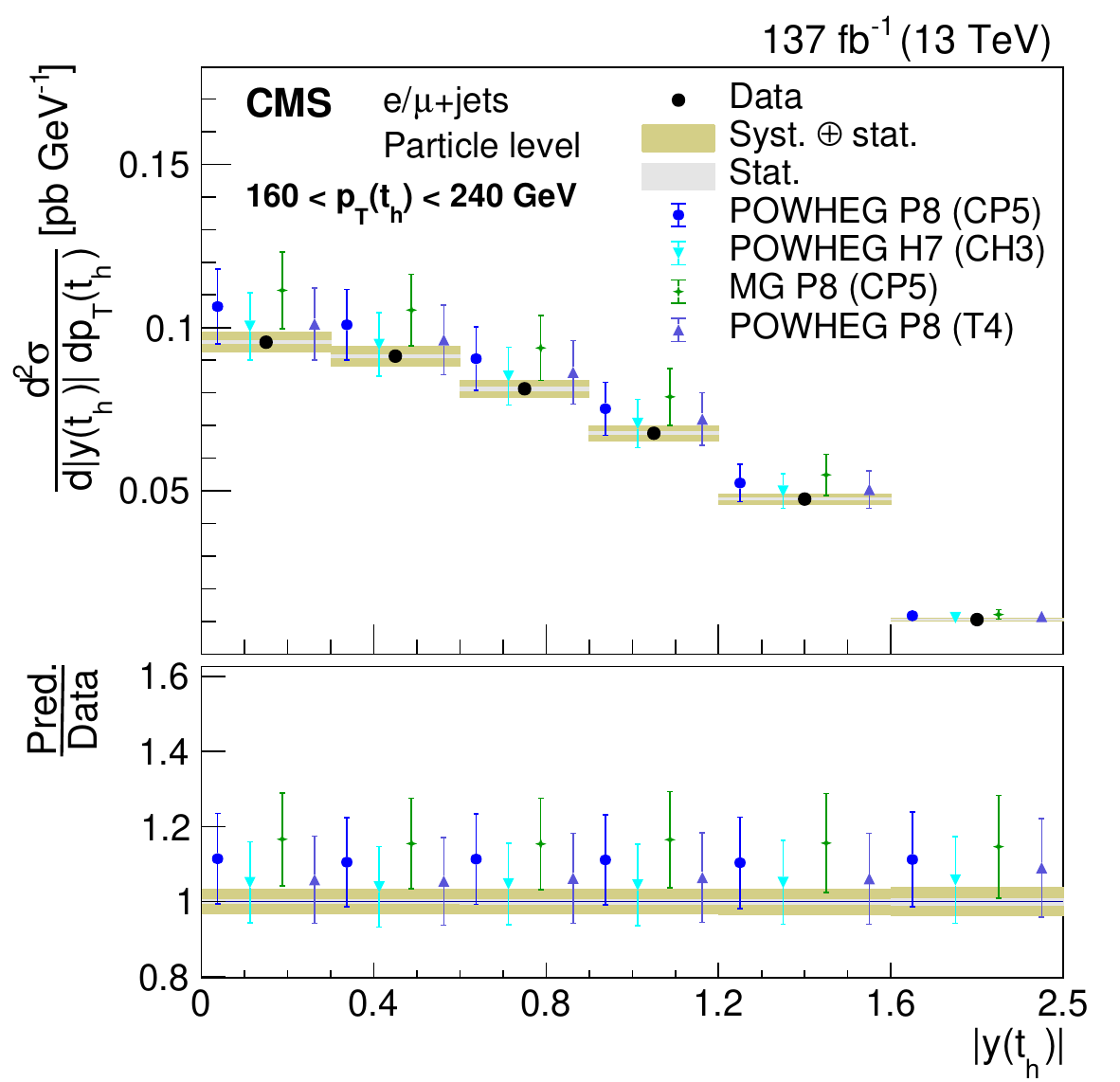}
 \includegraphics[width=0.42\textwidth]{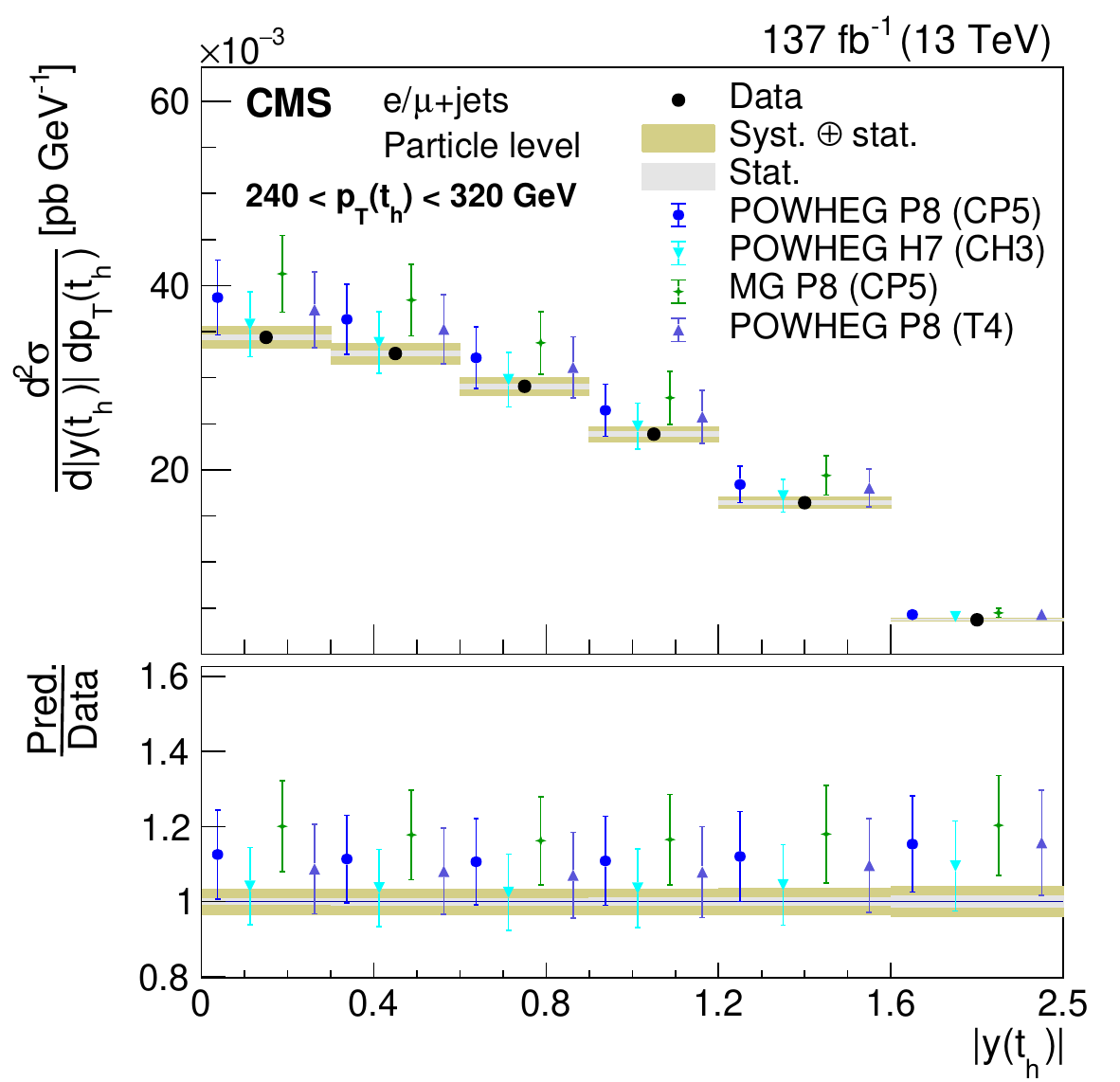}\\
 \includegraphics[width=0.42\textwidth]{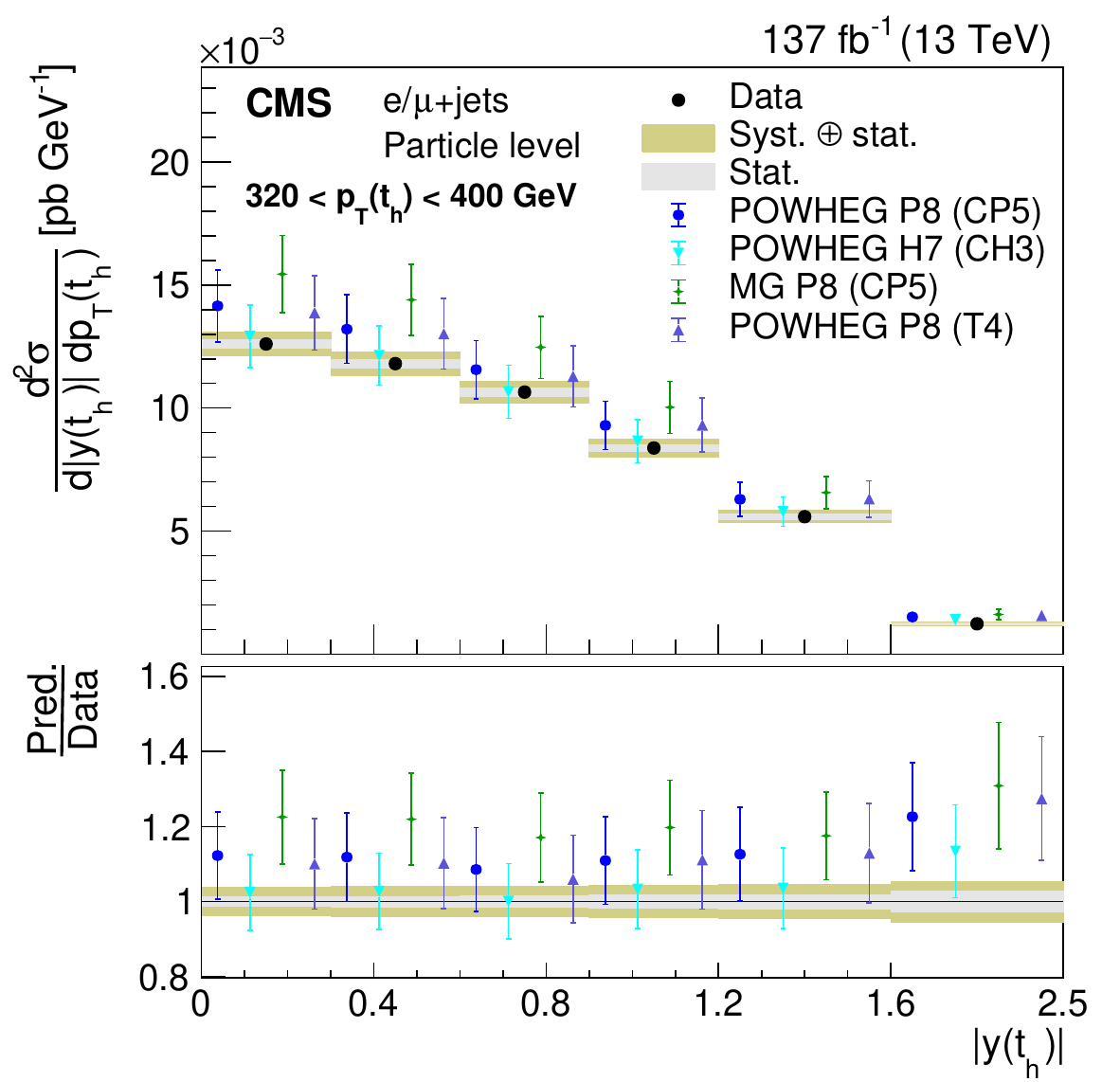}
 \includegraphics[width=0.42\textwidth]{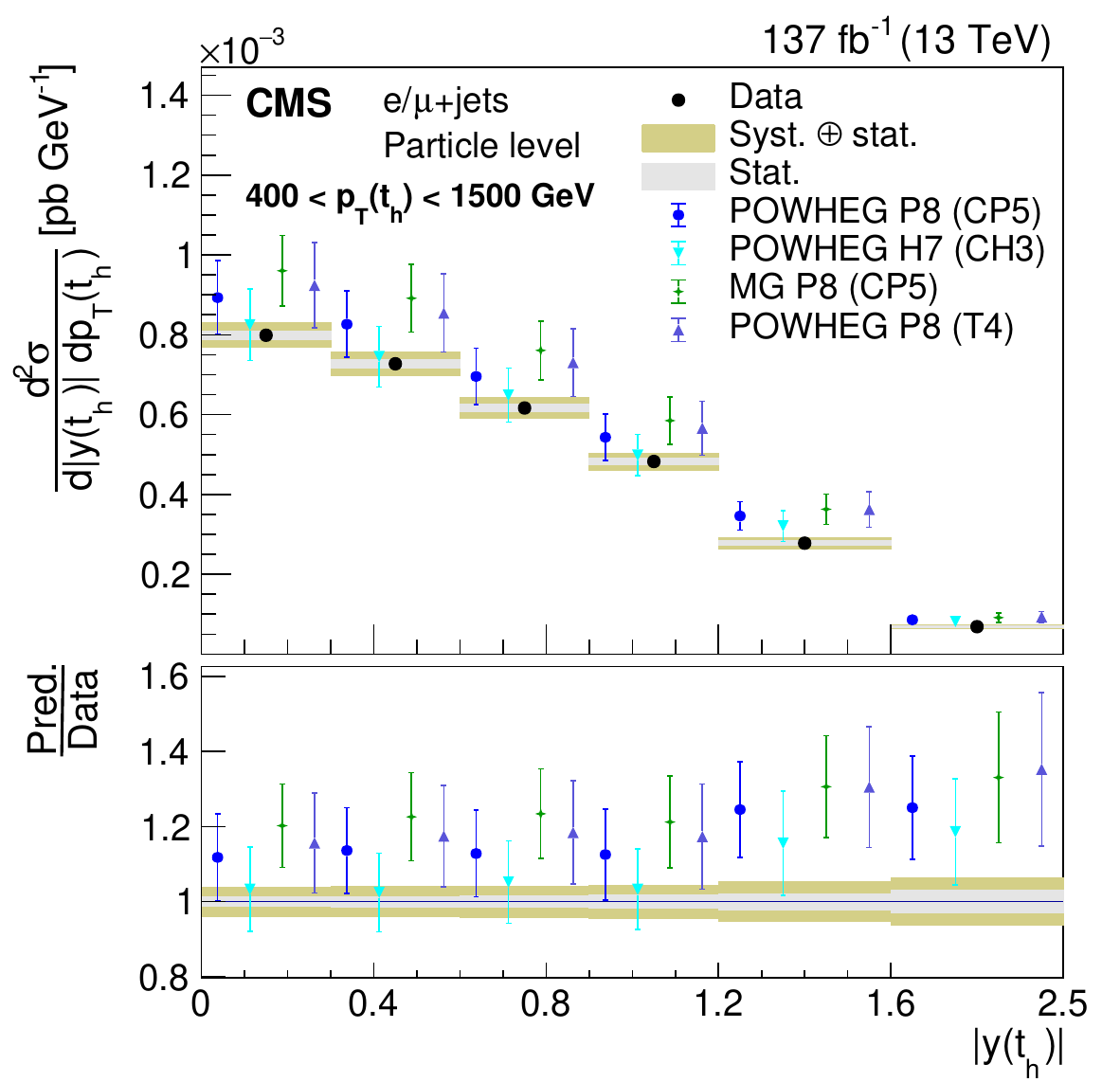}
 \caption{Double-differential cross section at the particle level as a function of \thadptvsthady. \XSECCAPPS}
 \label{fig:RESPS4}
\end{figure*}

\begin{figure*}[tbp]
\centering
 \includegraphics[width=0.42\textwidth]{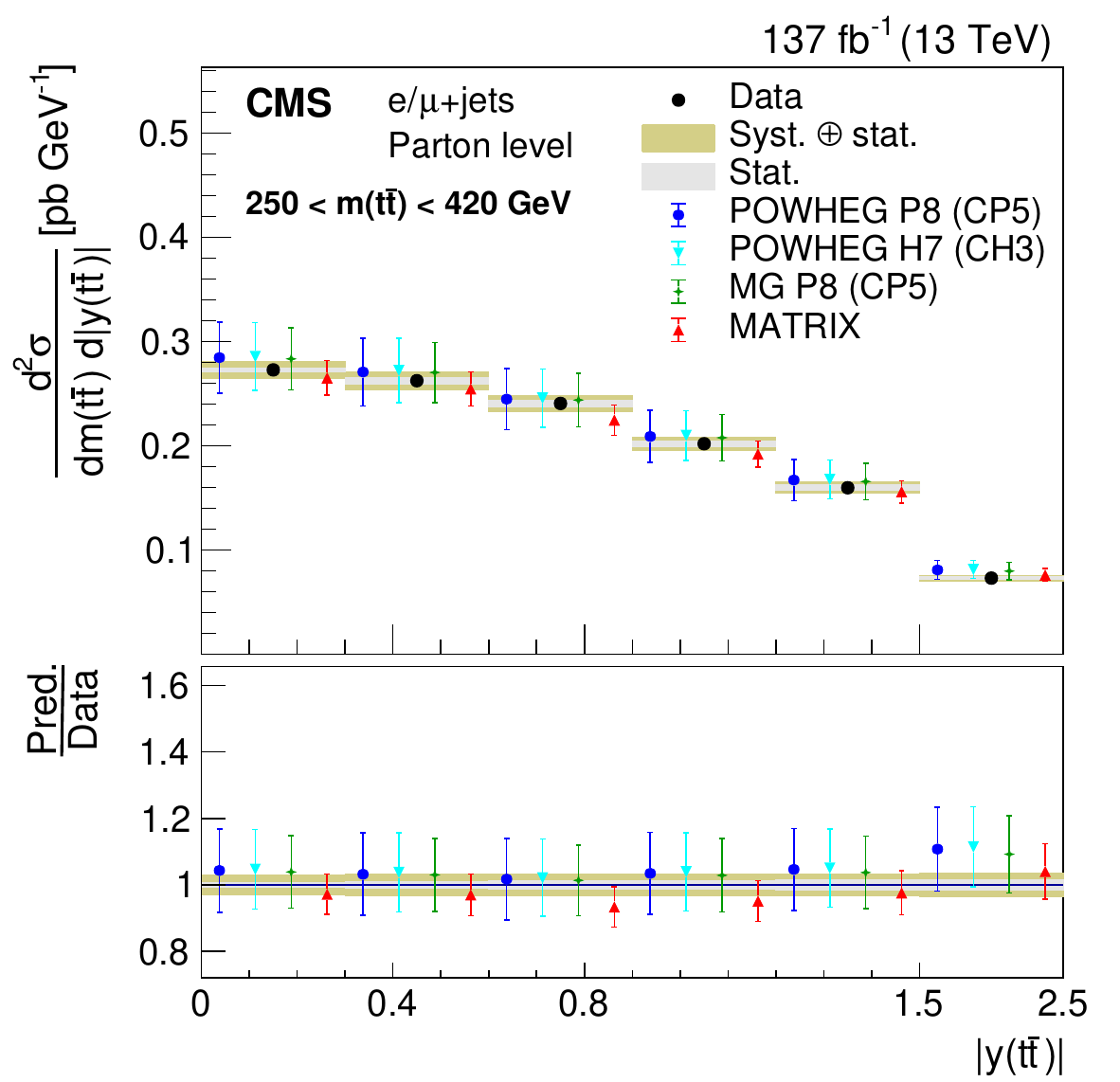}
 \includegraphics[width=0.42\textwidth]{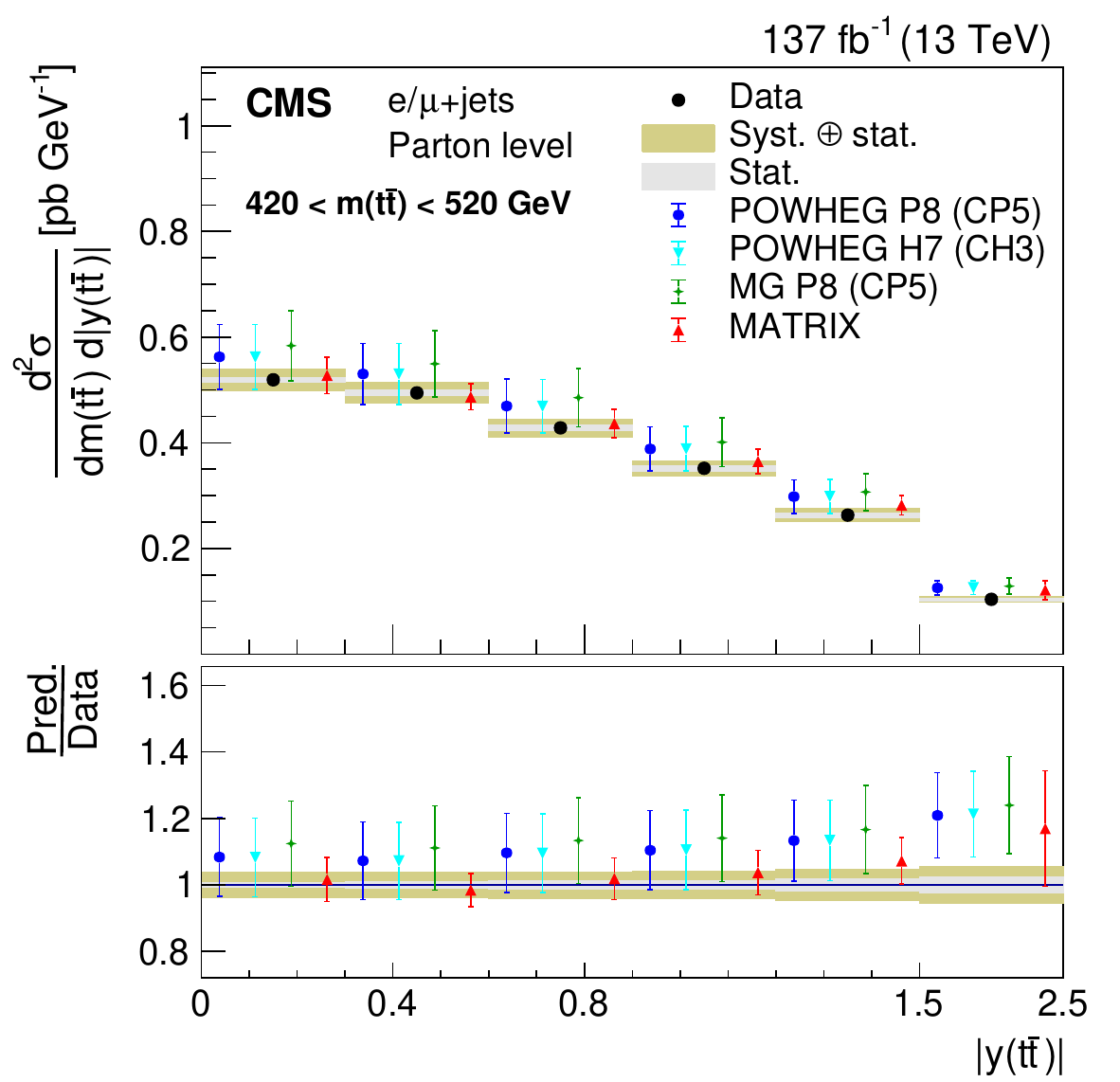}\\
 \includegraphics[width=0.42\textwidth]{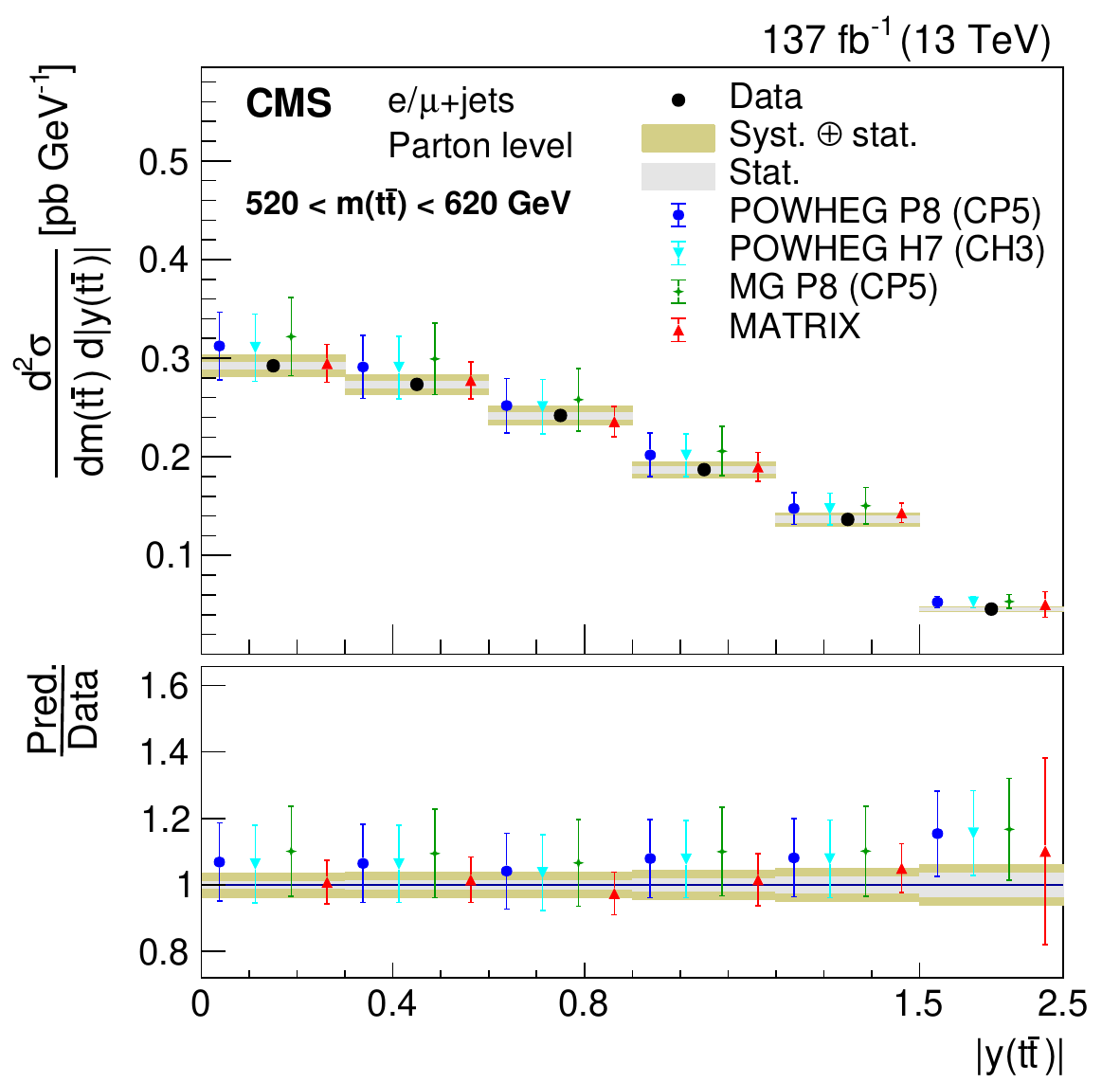}
 \includegraphics[width=0.42\textwidth]{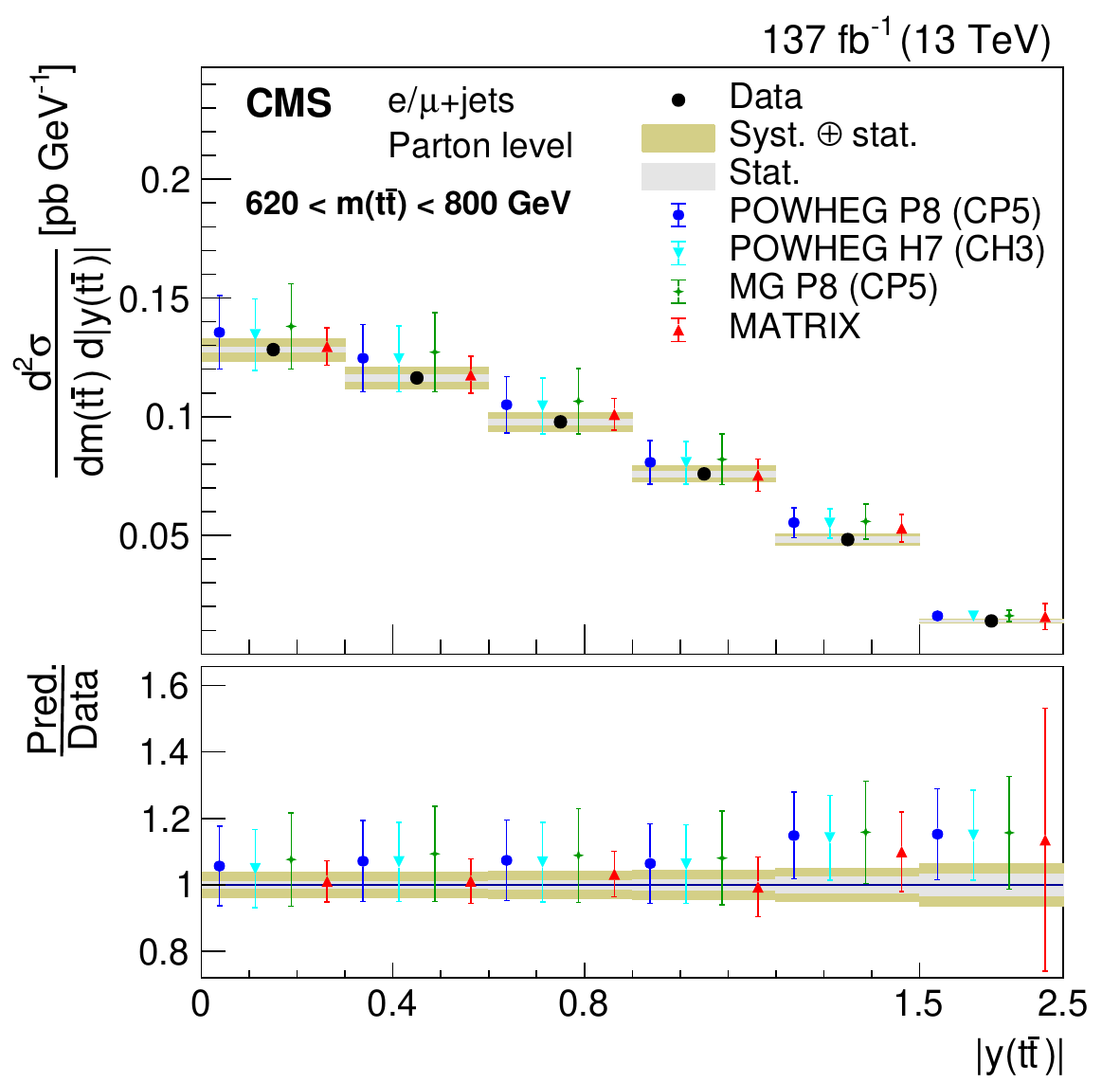}\\
 \includegraphics[width=0.42\textwidth]{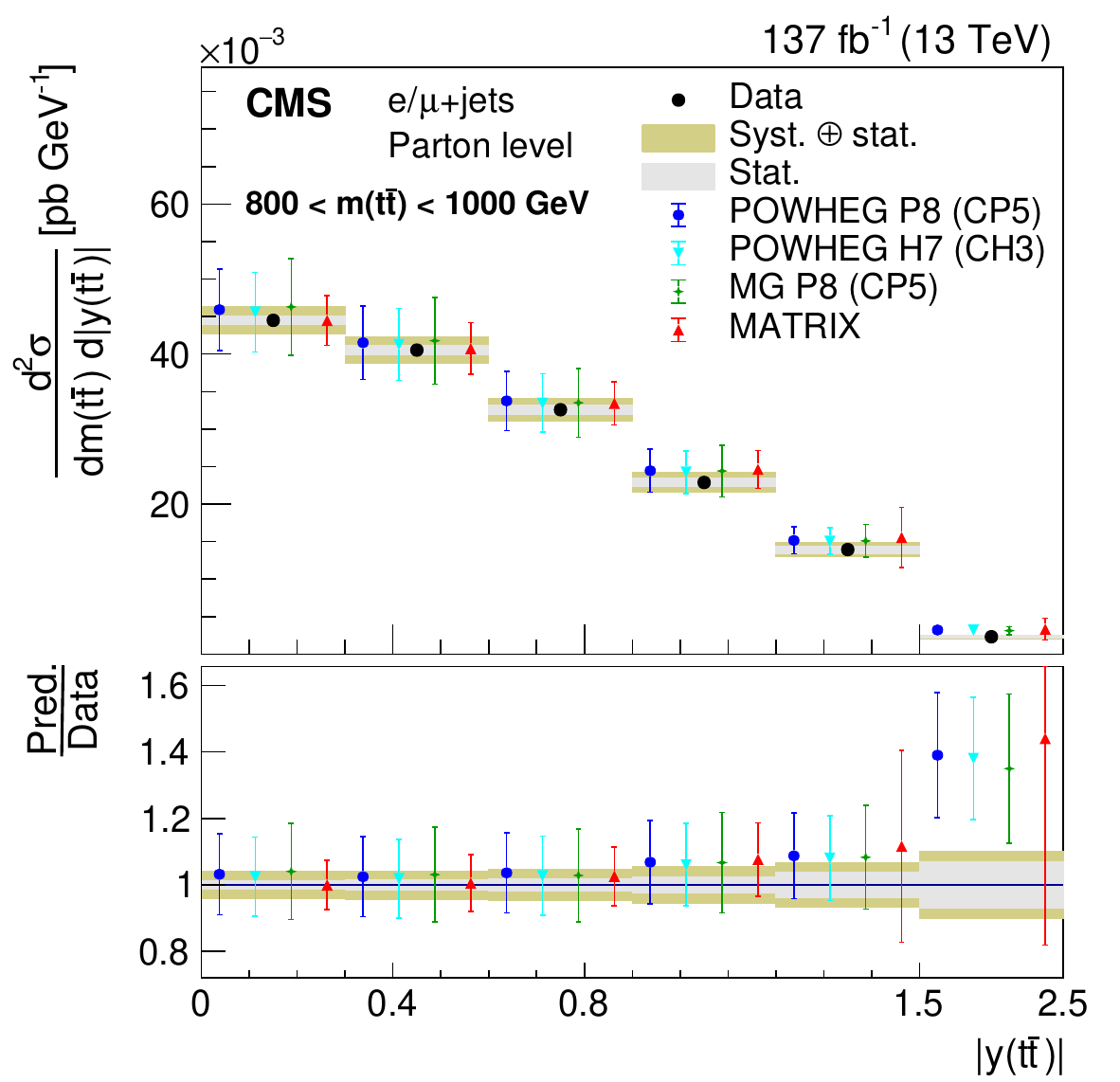}
 \includegraphics[width=0.42\textwidth]{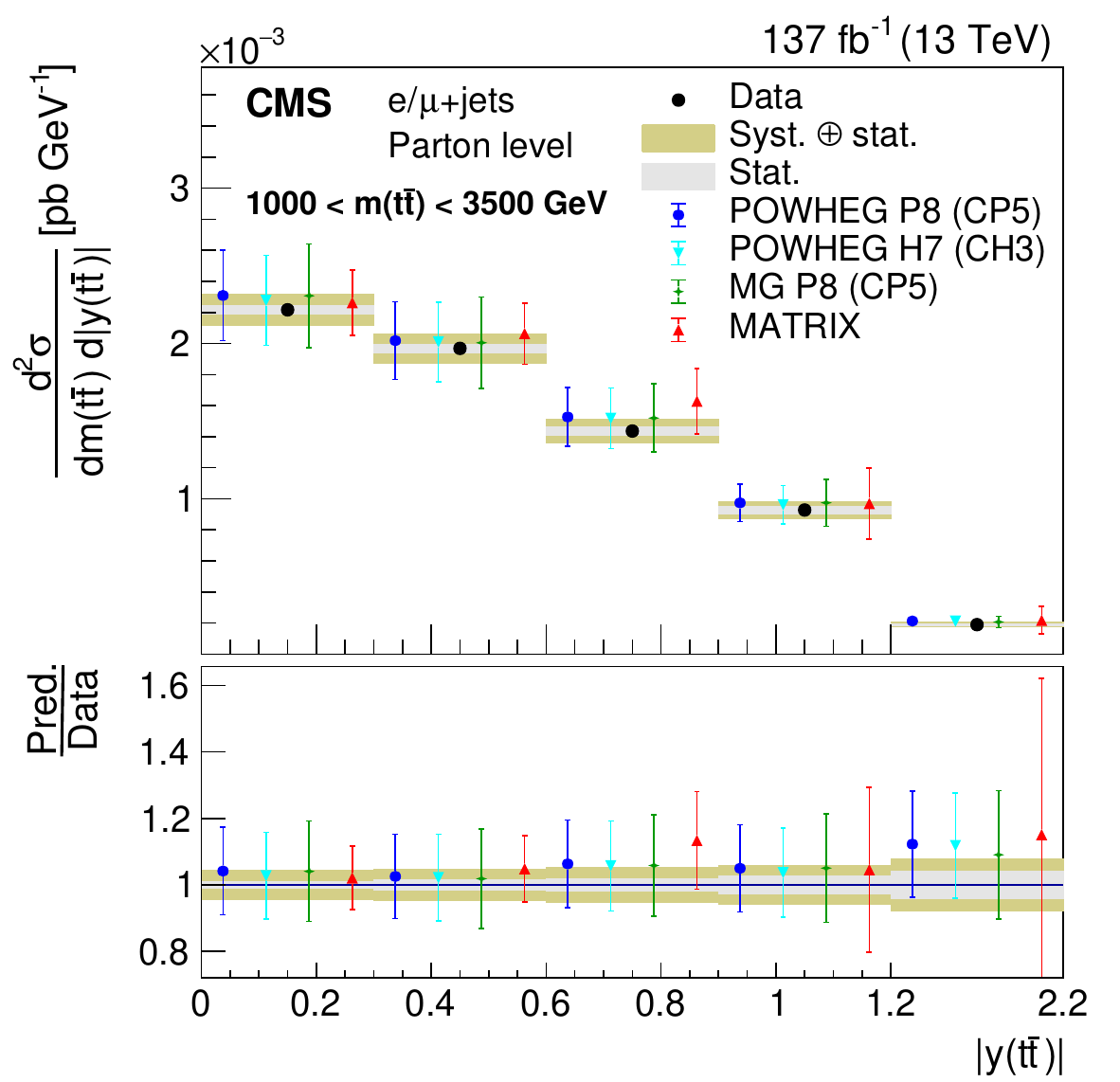}
 \caption{Double-differential cross section at the parton level as a function of \ttmvstty. \XSECCAPPA}
 \label{fig:RES5}
\end{figure*}

\begin{figure*}[tbp]
\centering
 \includegraphics[width=0.42\textwidth]{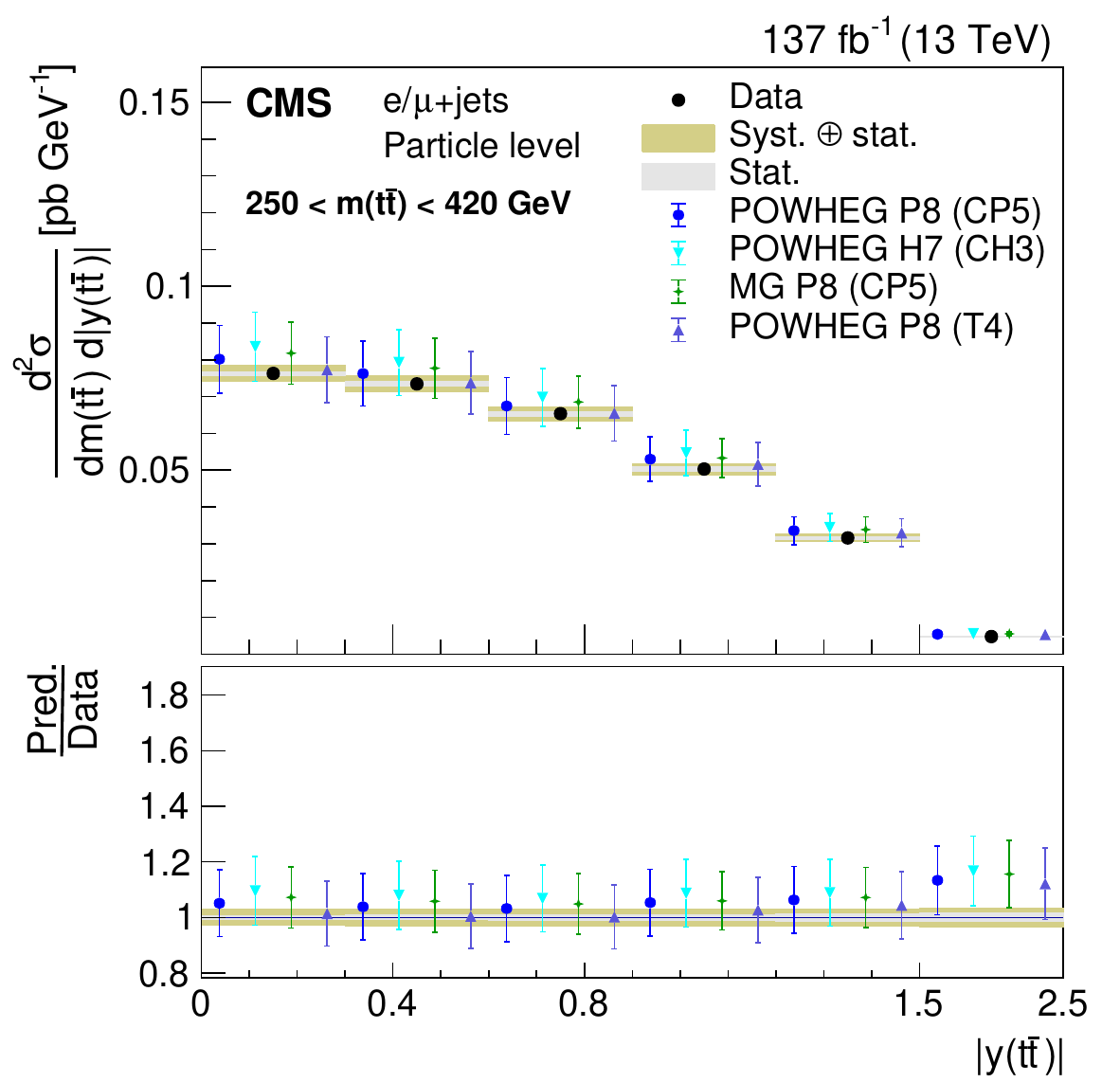}
 \includegraphics[width=0.42\textwidth]{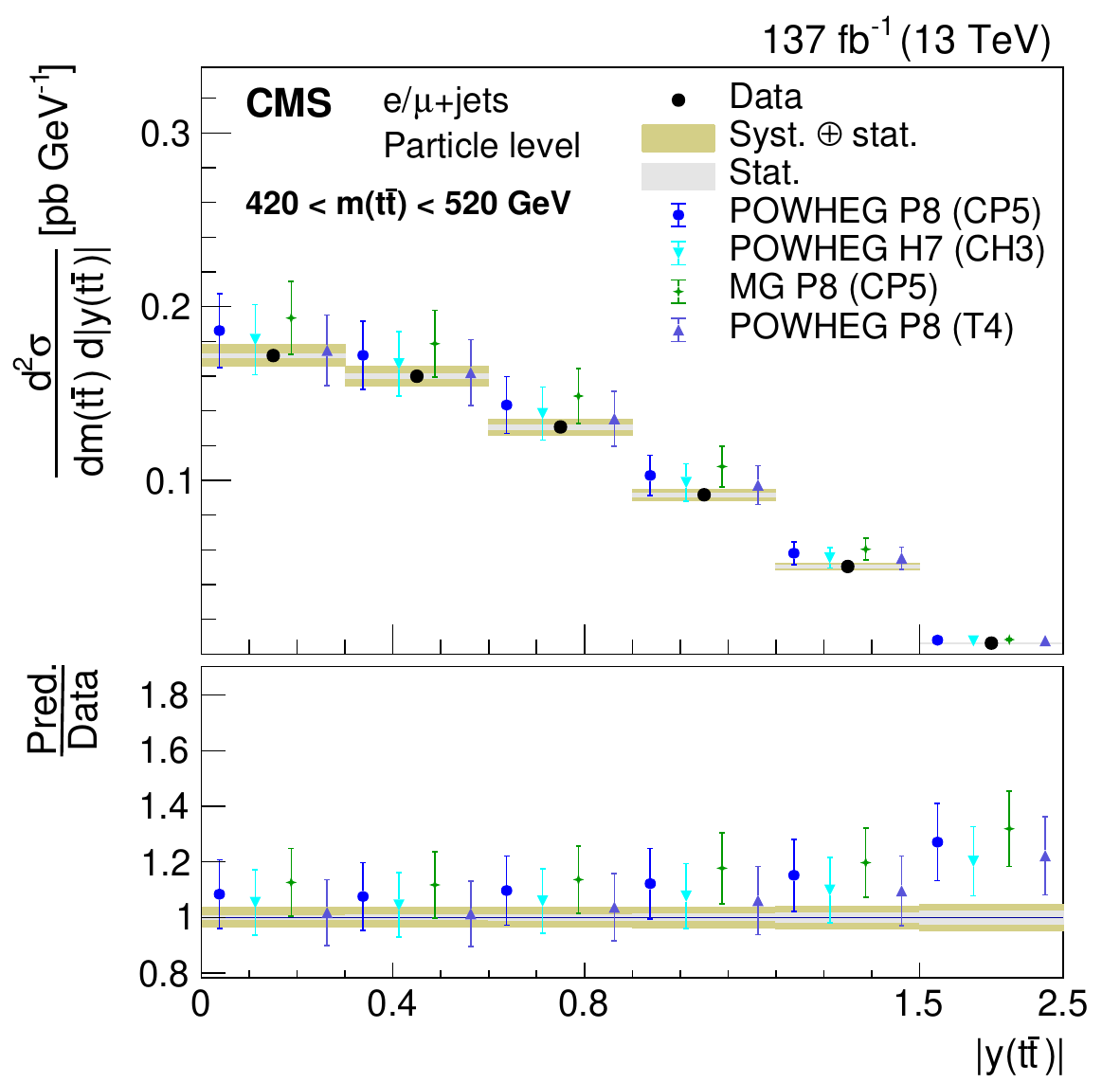}\\
 \includegraphics[width=0.42\textwidth]{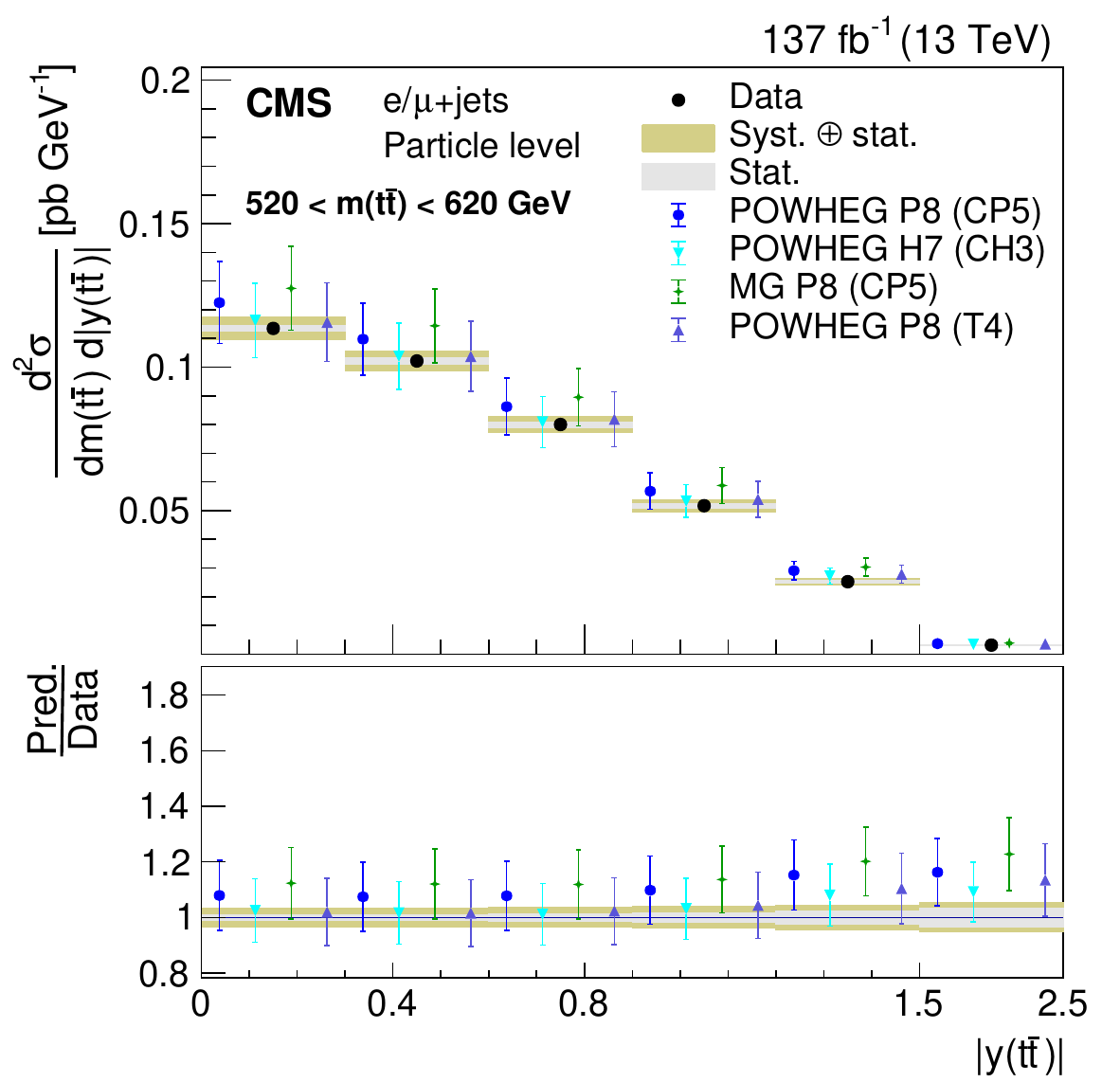}
 \includegraphics[width=0.42\textwidth]{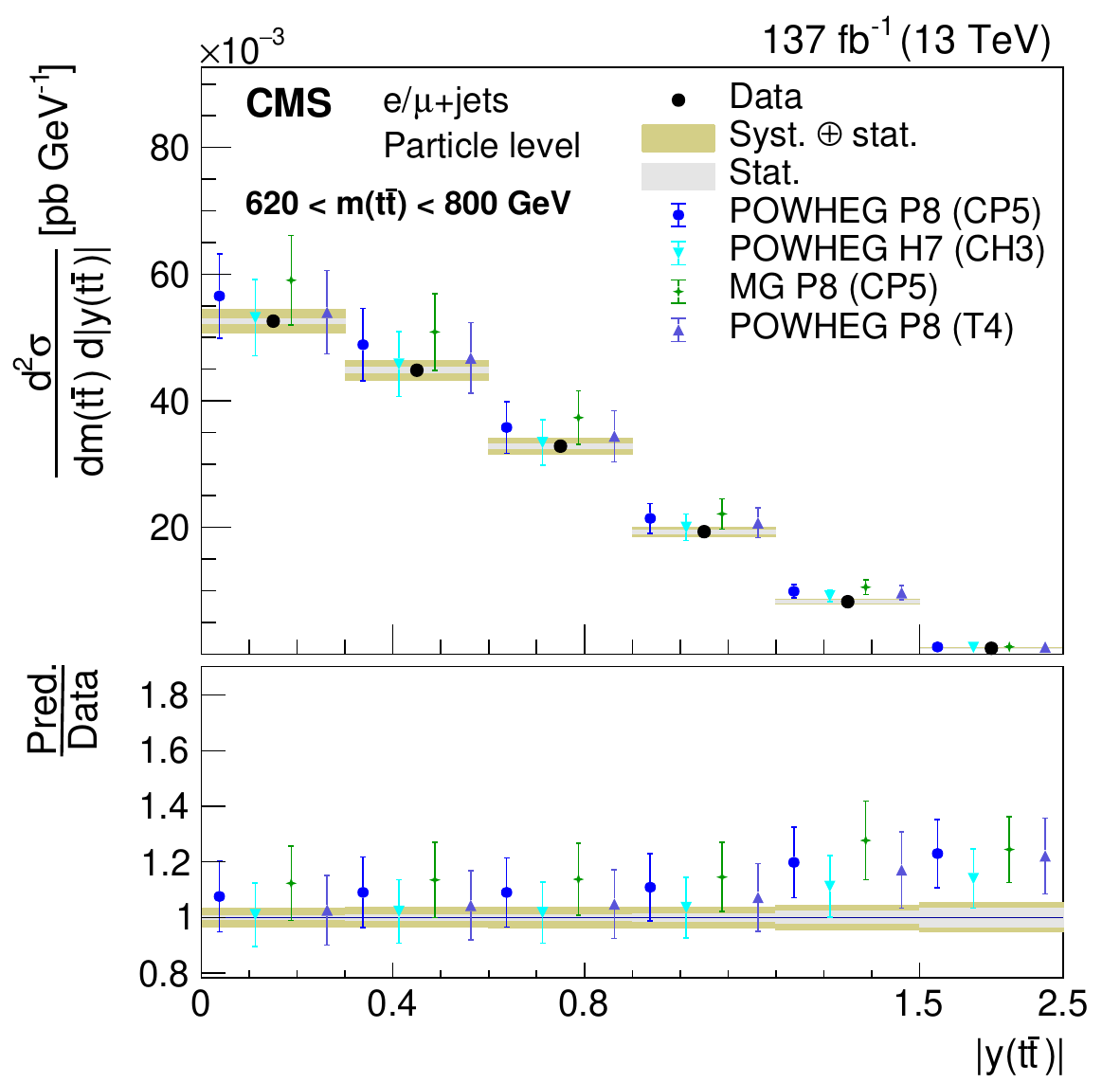}\\
 \includegraphics[width=0.42\textwidth]{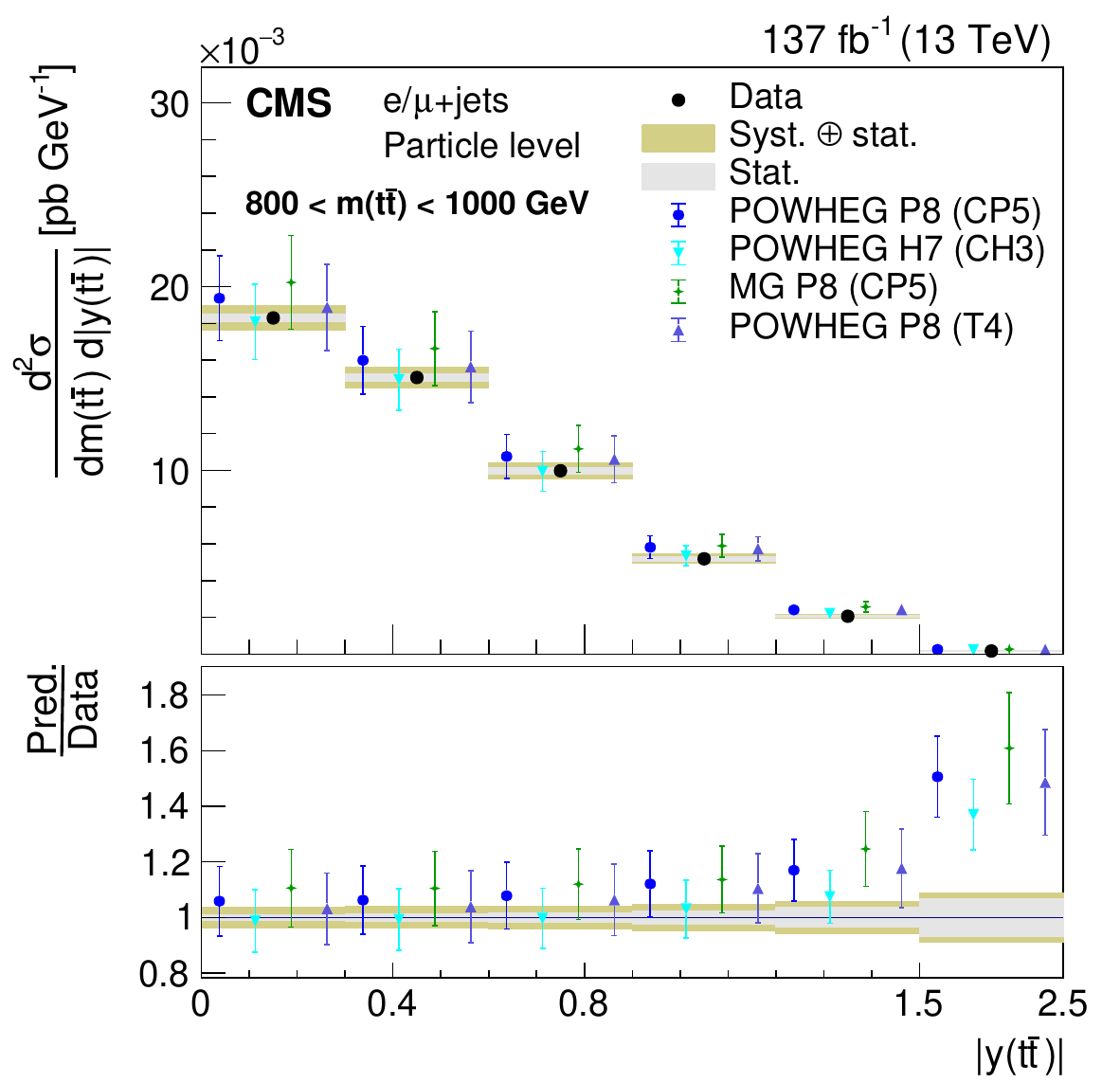}
 \includegraphics[width=0.42\textwidth]{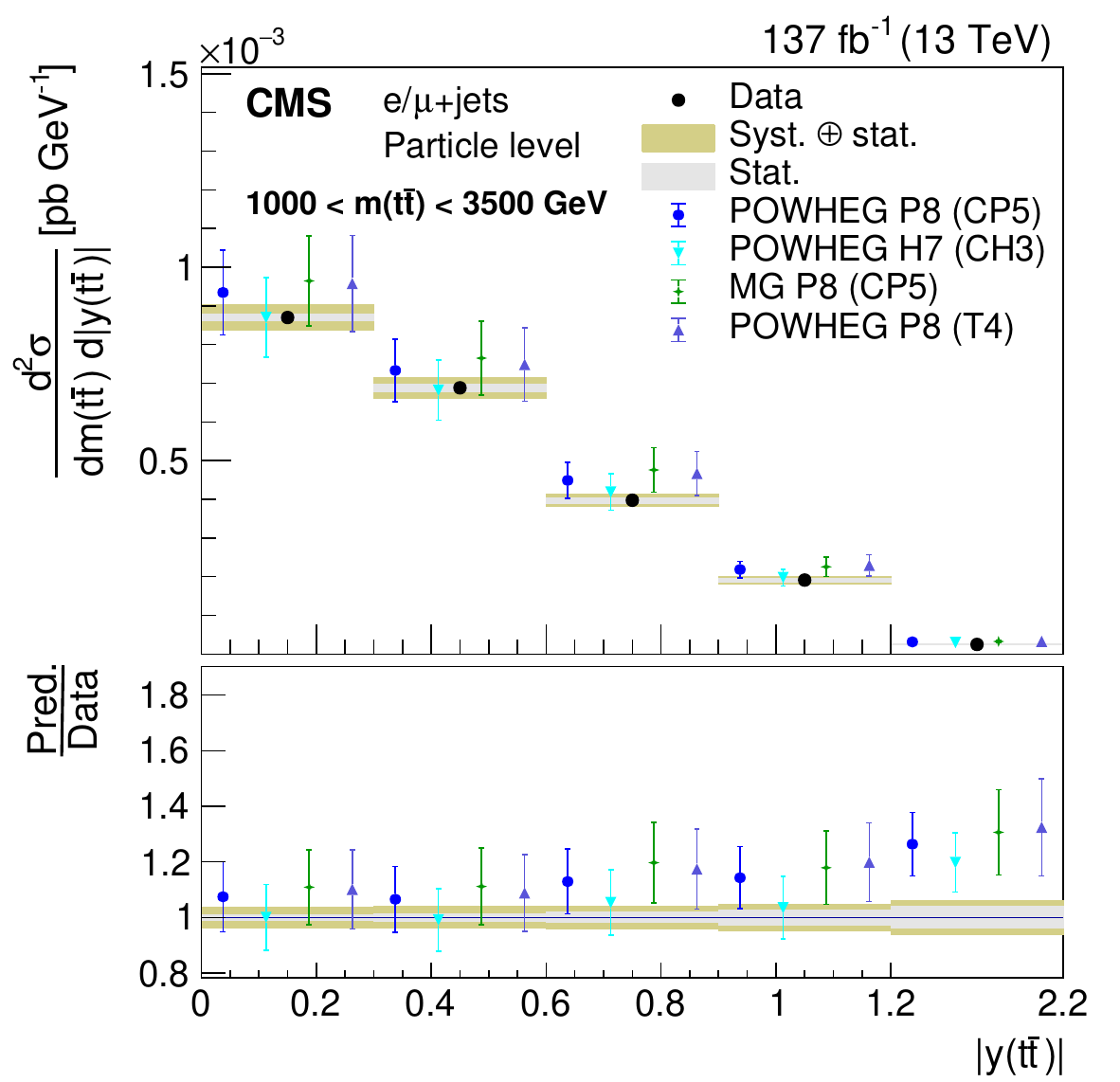}
 \caption{Double-differential cross section at the particle level as a function of \ttmvstty. \XSECCAPPS}
 \label{fig:RESPS5}
\end{figure*}

\begin{figure*}[tbp]
\centering
 \includegraphics[width=0.42\textwidth]{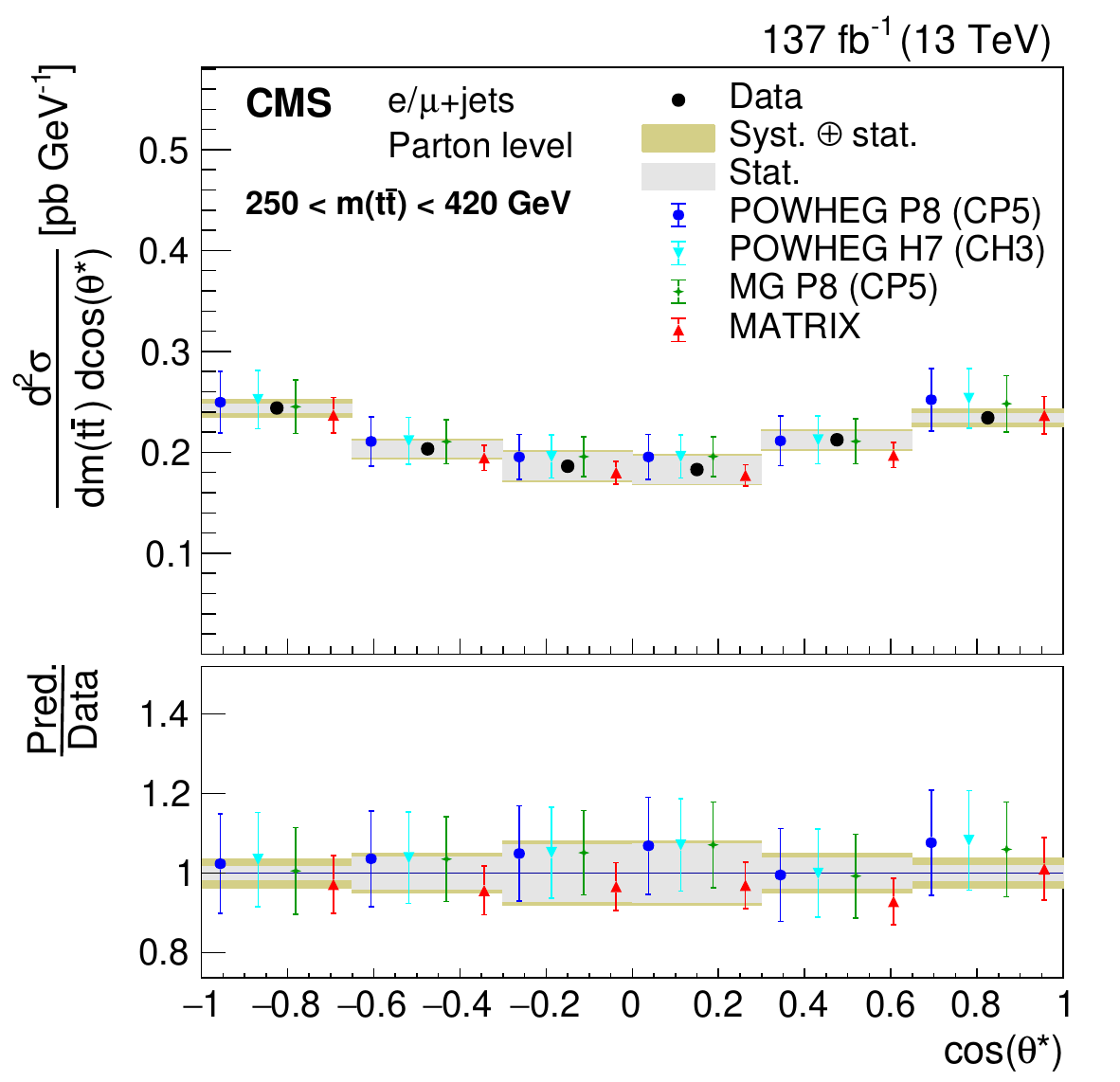}
 \includegraphics[width=0.42\textwidth]{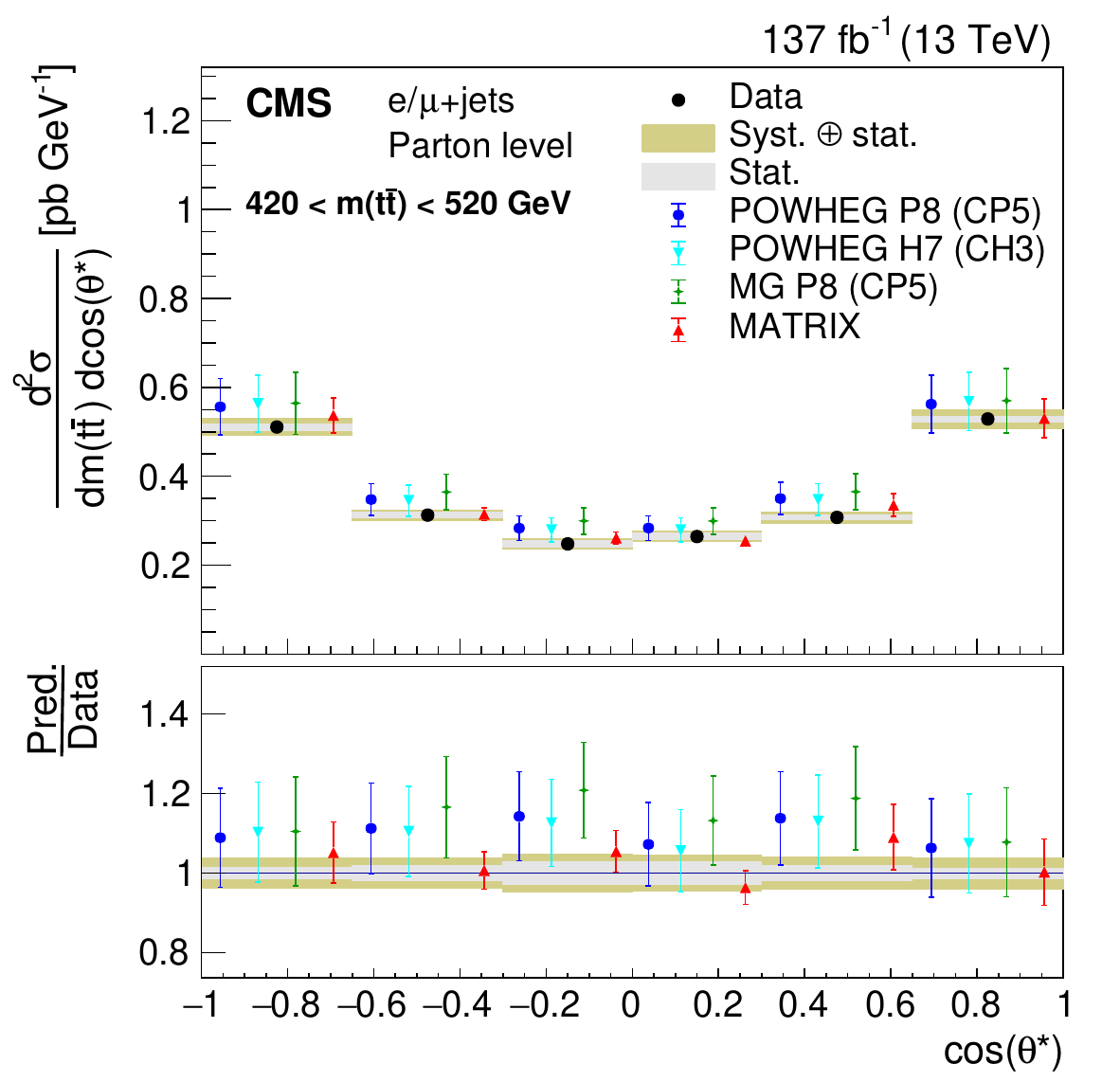}\\
 \includegraphics[width=0.42\textwidth]{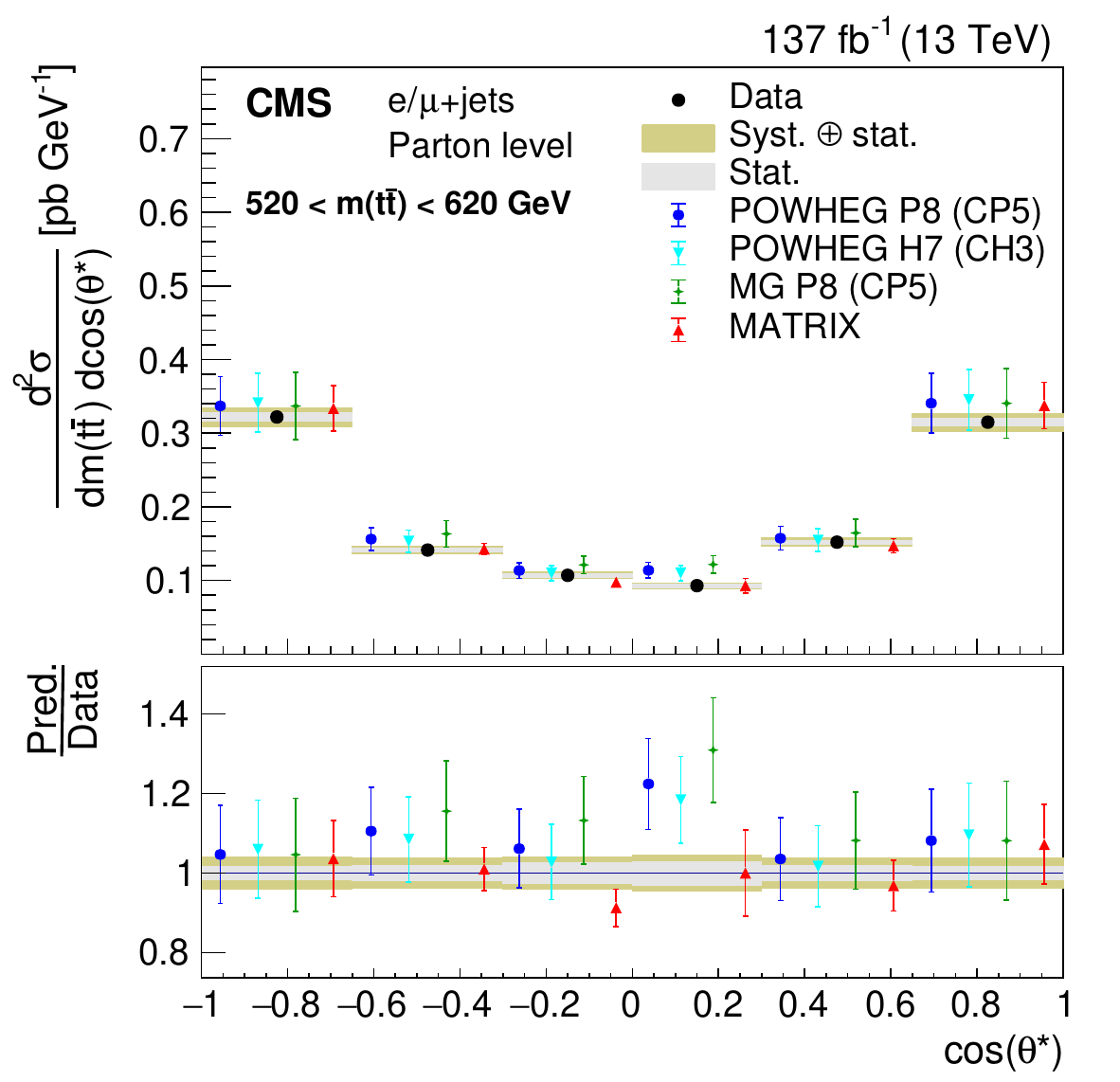}
 \includegraphics[width=0.42\textwidth]{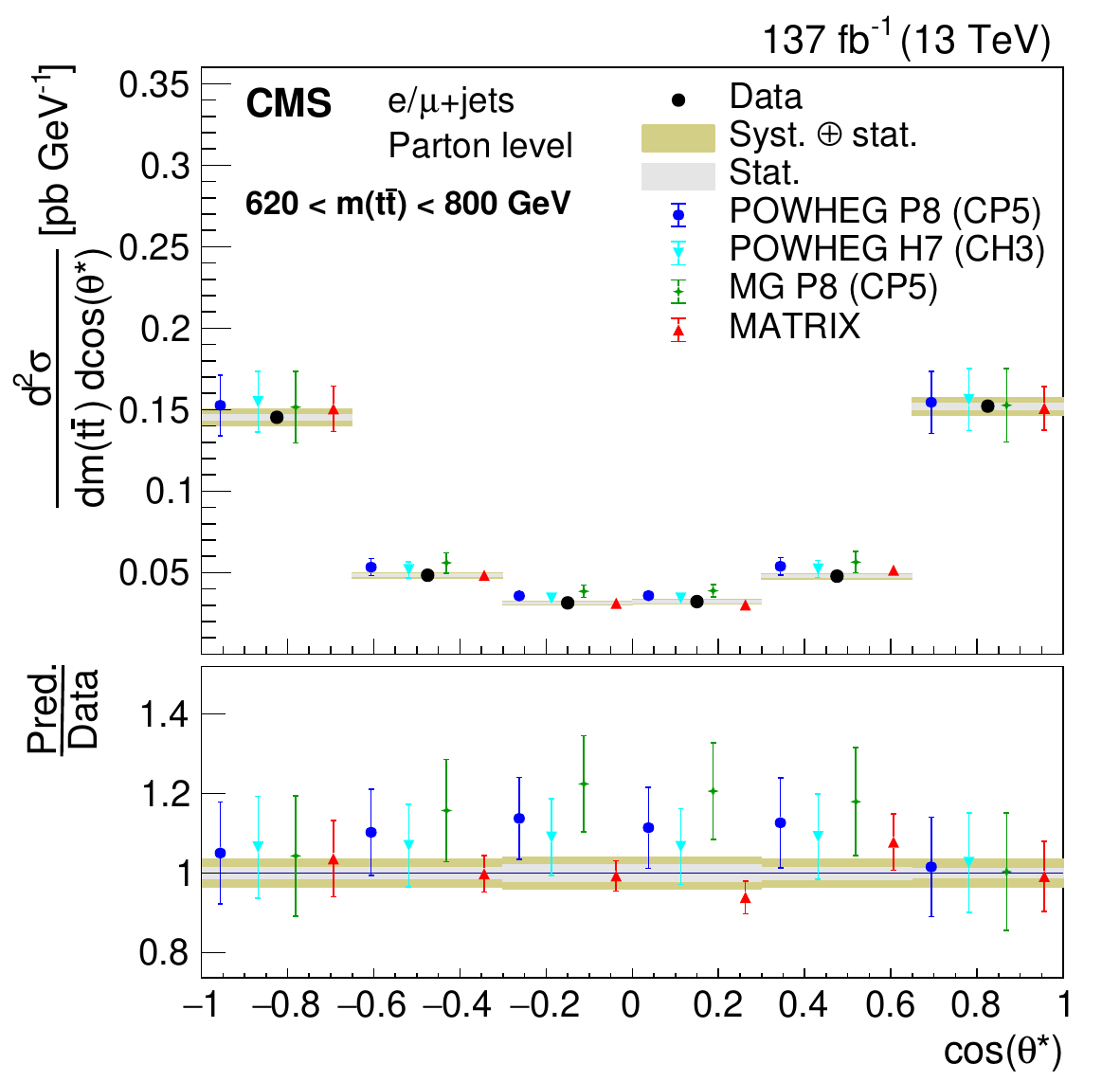}\\
 \includegraphics[width=0.42\textwidth]{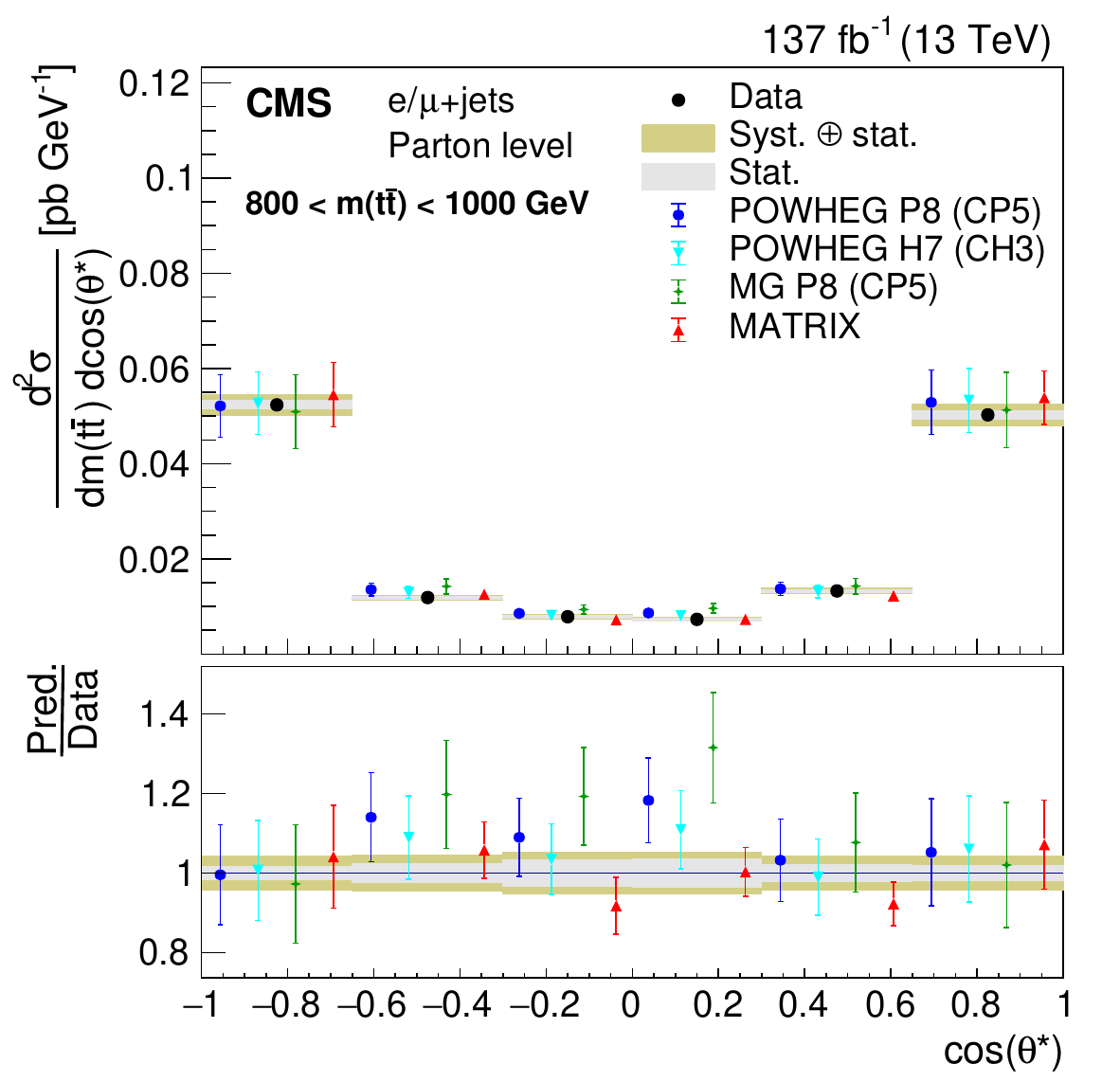}
 \includegraphics[width=0.42\textwidth]{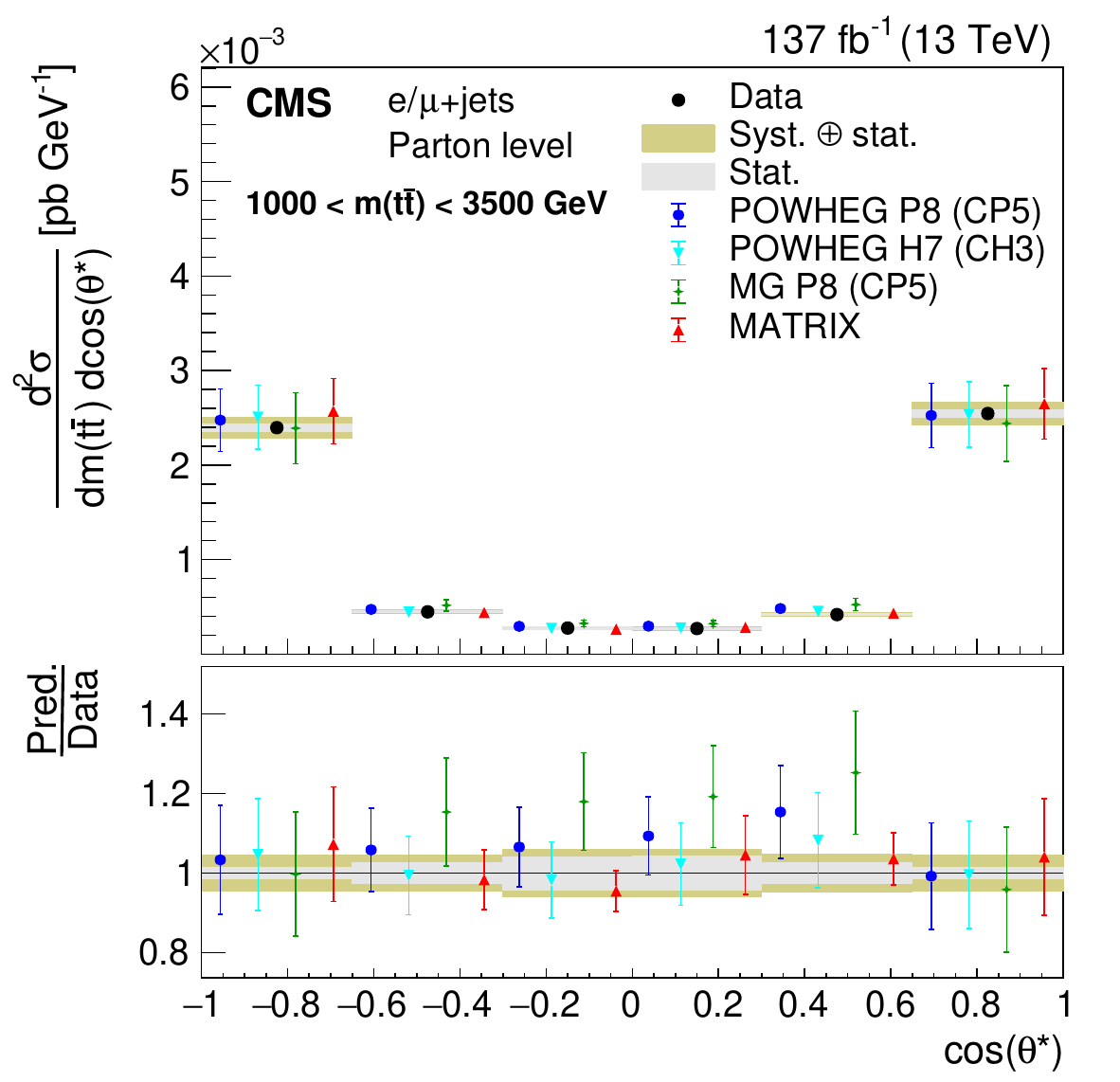}
 \caption{Double-differential cross section at the parton level as a function of \ttmvscts. \XSECCAPPA}
 \label{fig:RES6}
\end{figure*}

\begin{figure*}[tbp]
\centering
 \includegraphics[width=0.42\textwidth]{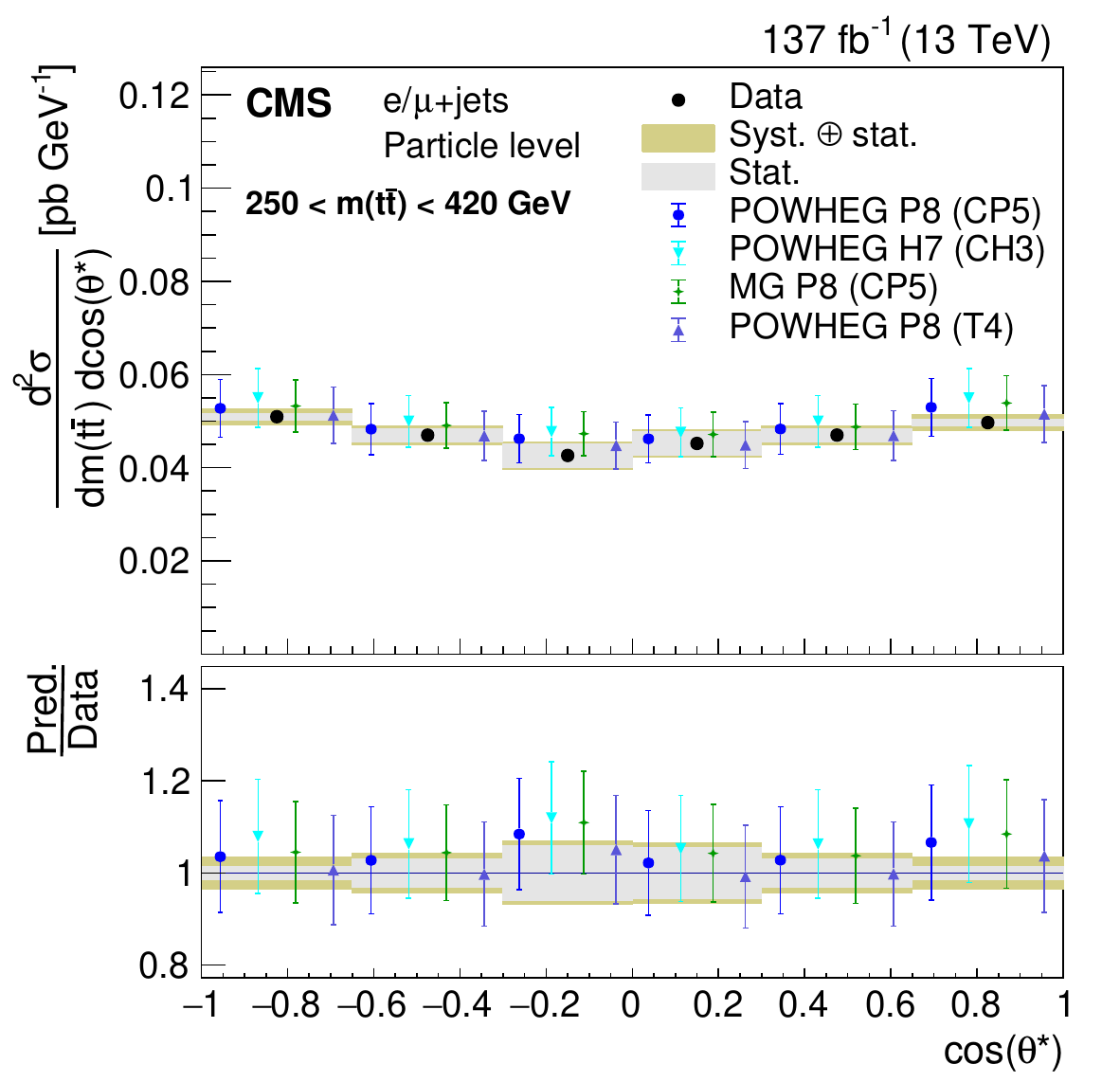}
 \includegraphics[width=0.42\textwidth]{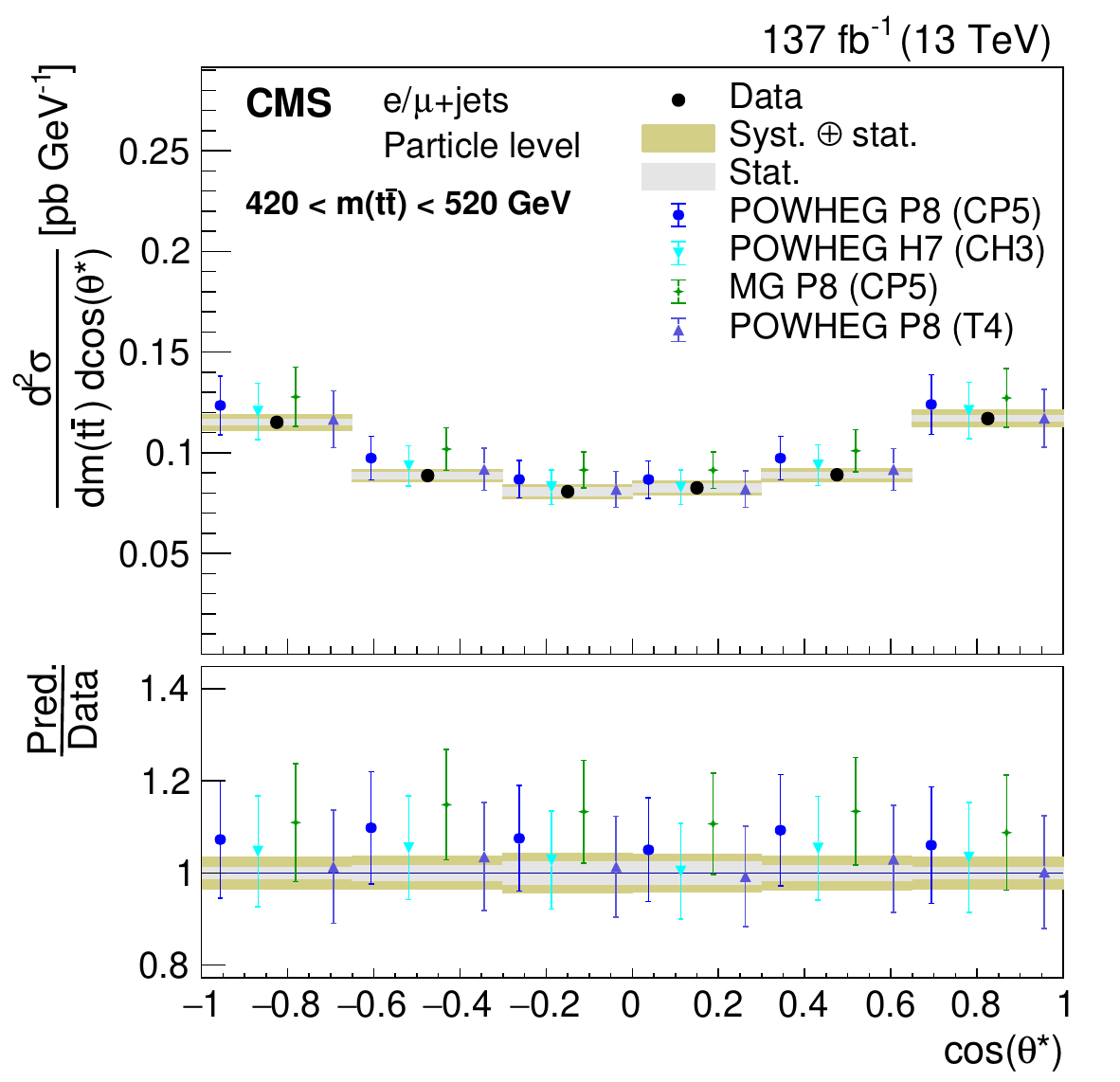}\\
 \includegraphics[width=0.42\textwidth]{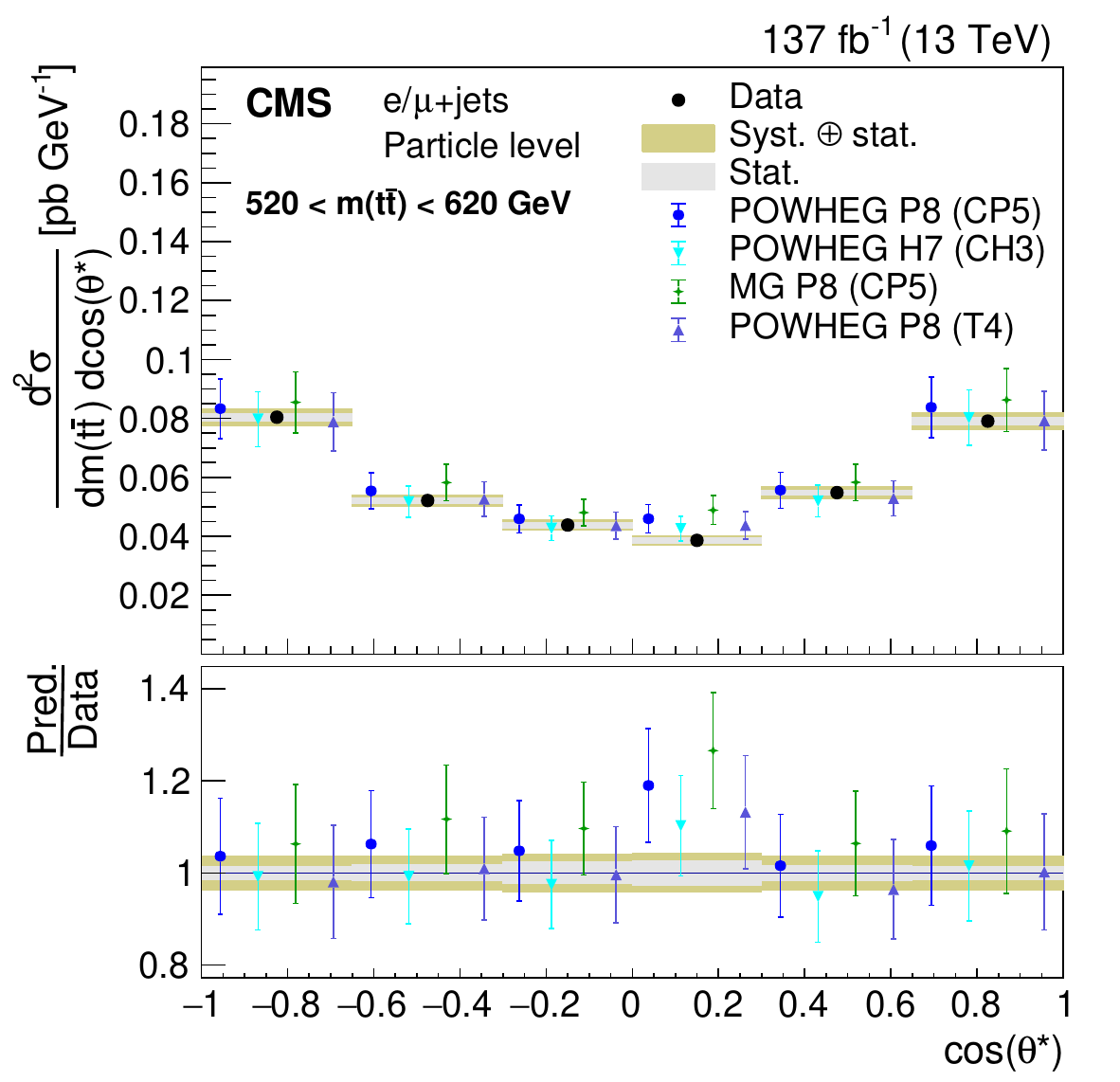}
 \includegraphics[width=0.42\textwidth]{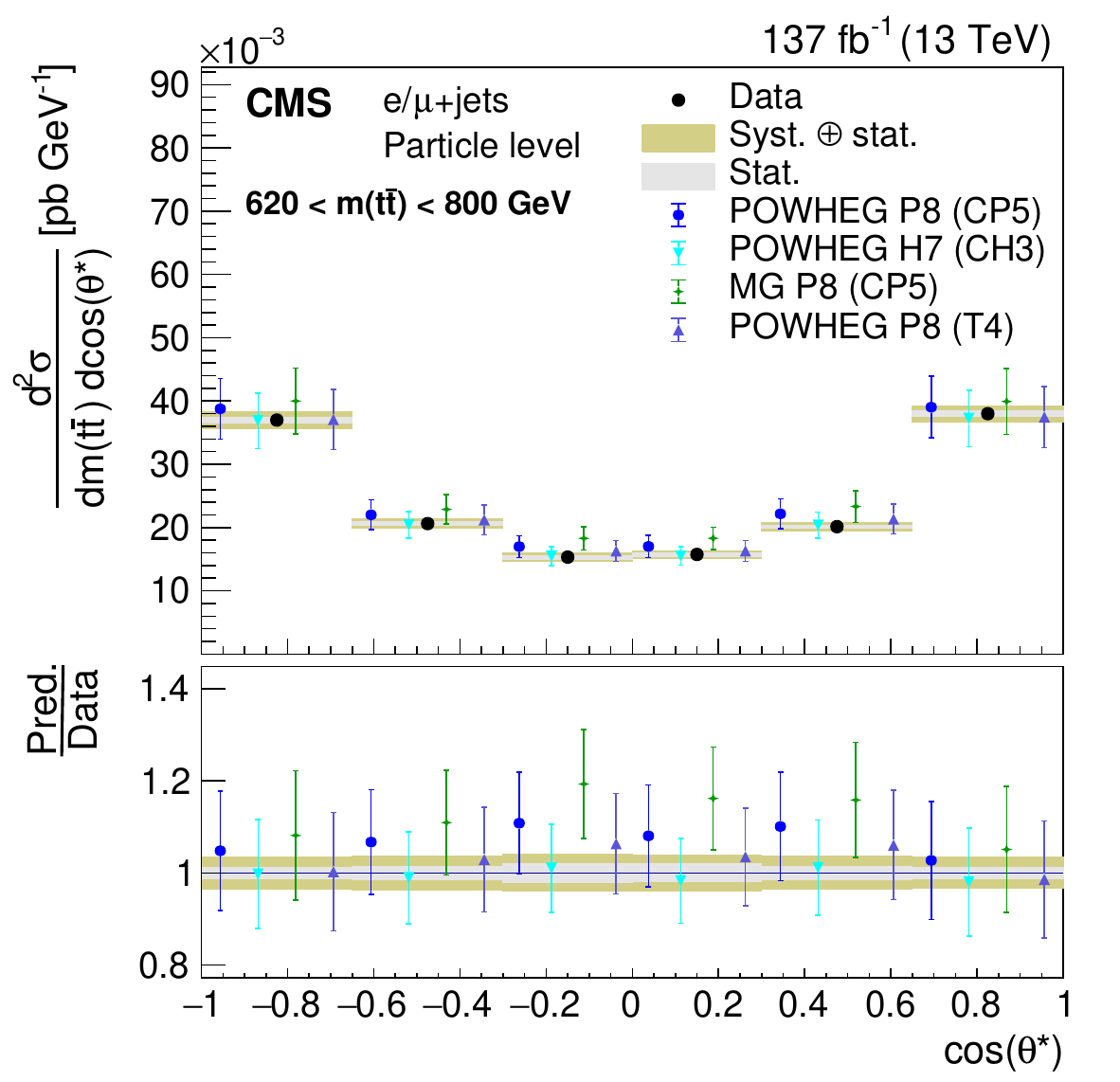}\\
 \includegraphics[width=0.42\textwidth]{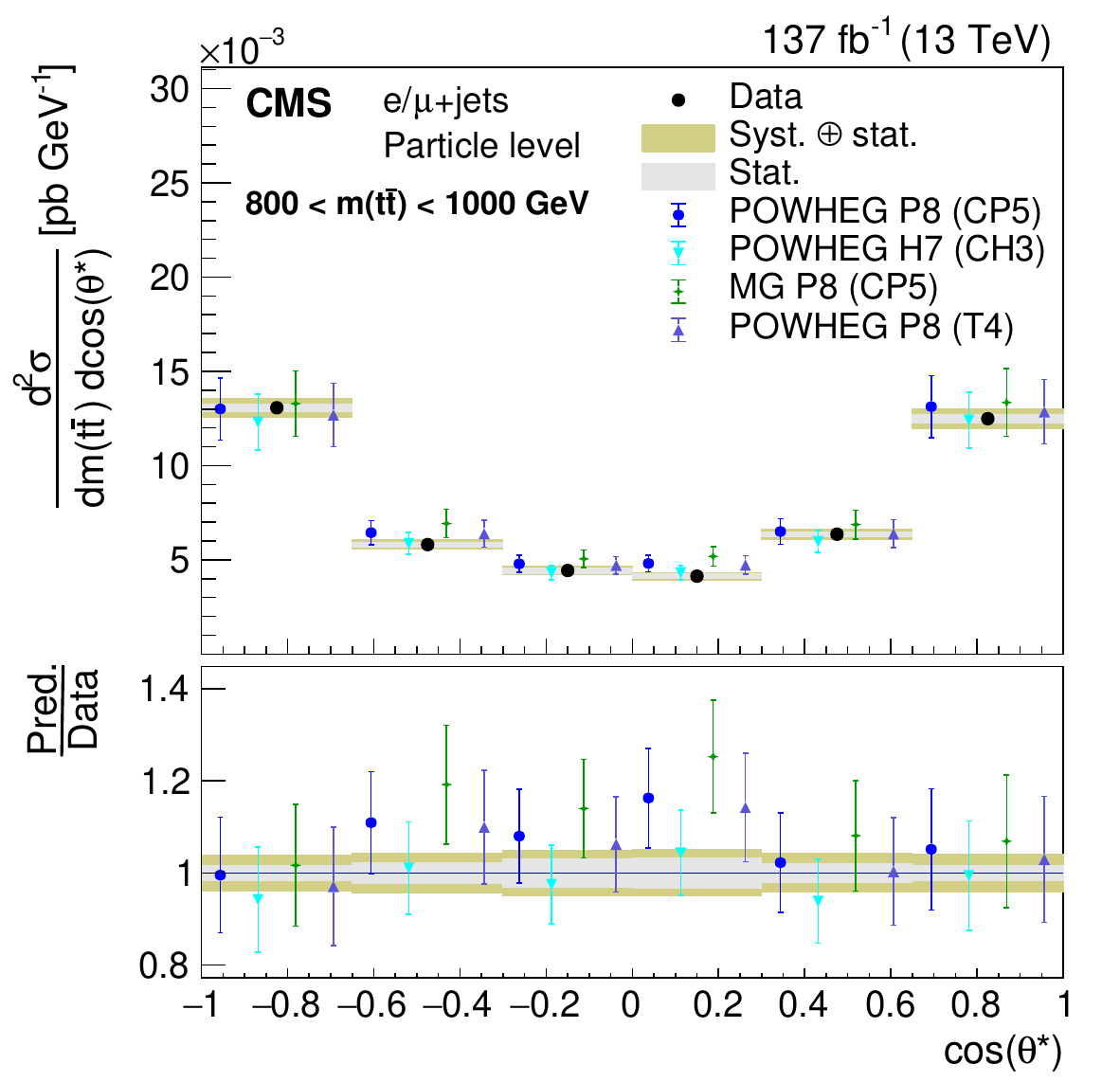}
 \includegraphics[width=0.42\textwidth]{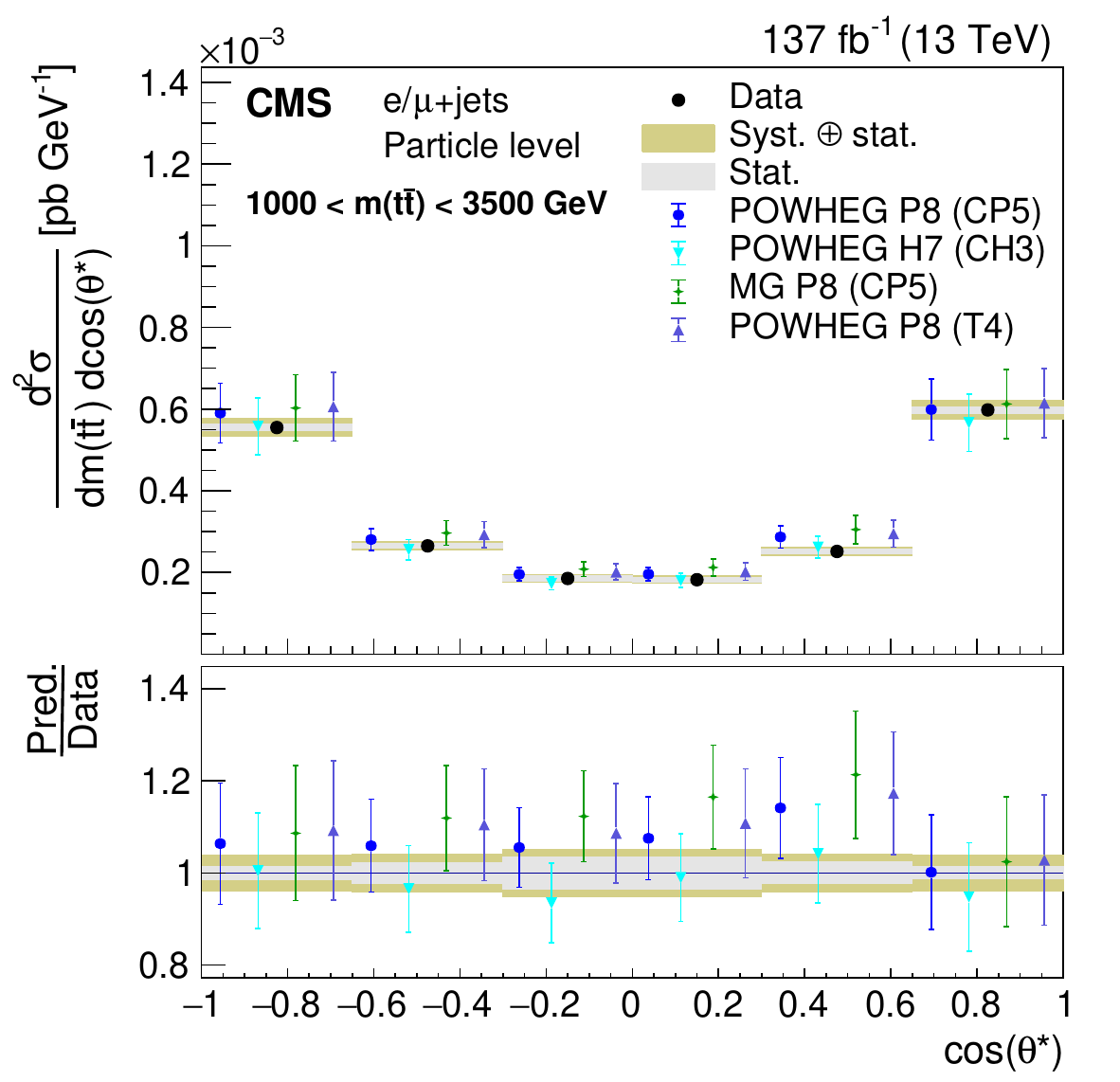}
 \caption{Double-differential cross section at the particle level as a function of \ttmvscts. \XSECCAPPS}
 \label{fig:RESPS6}
\end{figure*}

\begin{figure*}[tbp]
\centering
 \includegraphics[width=0.42\textwidth]{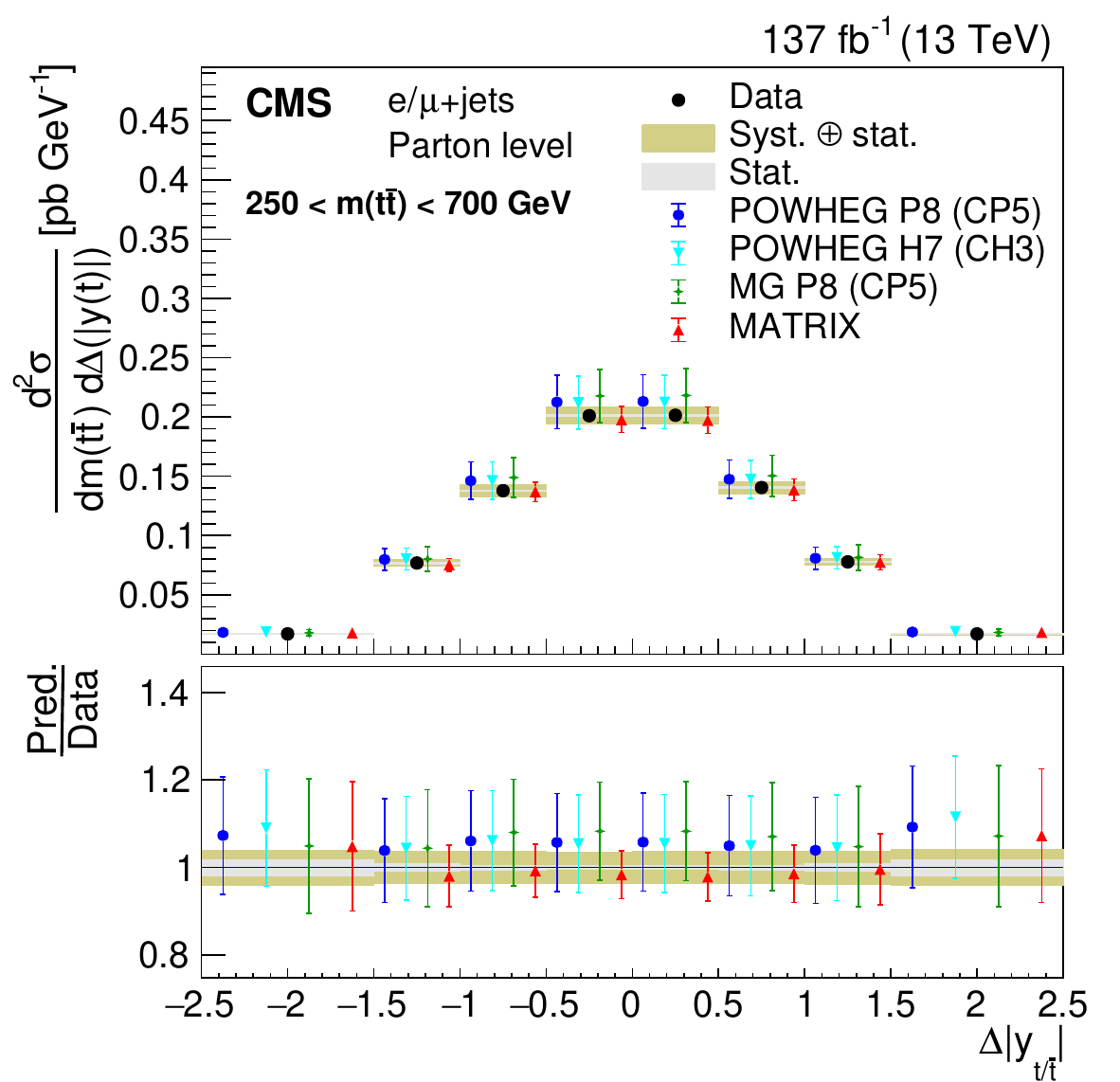}
 \includegraphics[width=0.42\textwidth]{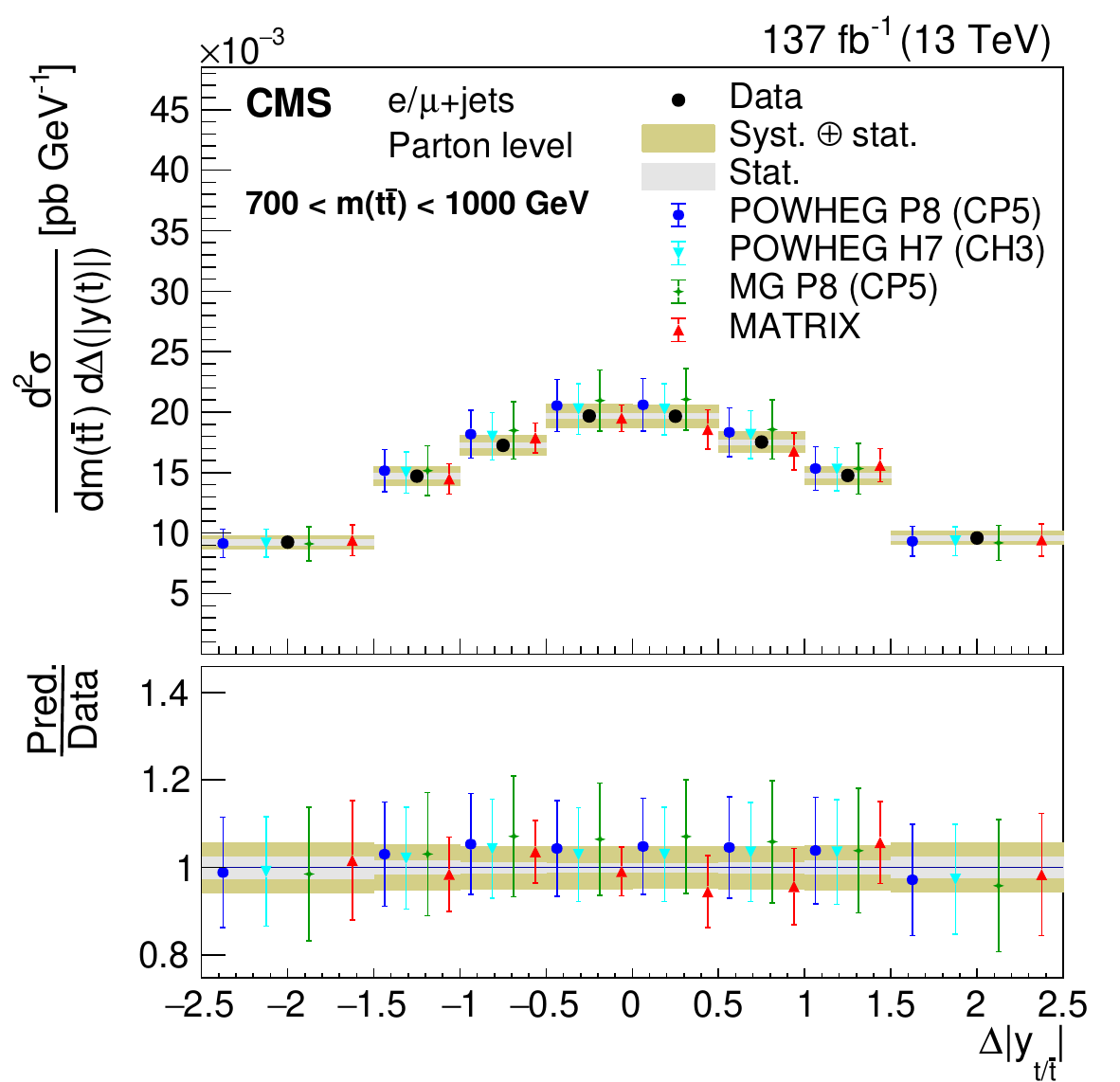}\\
 \includegraphics[width=0.42\textwidth]{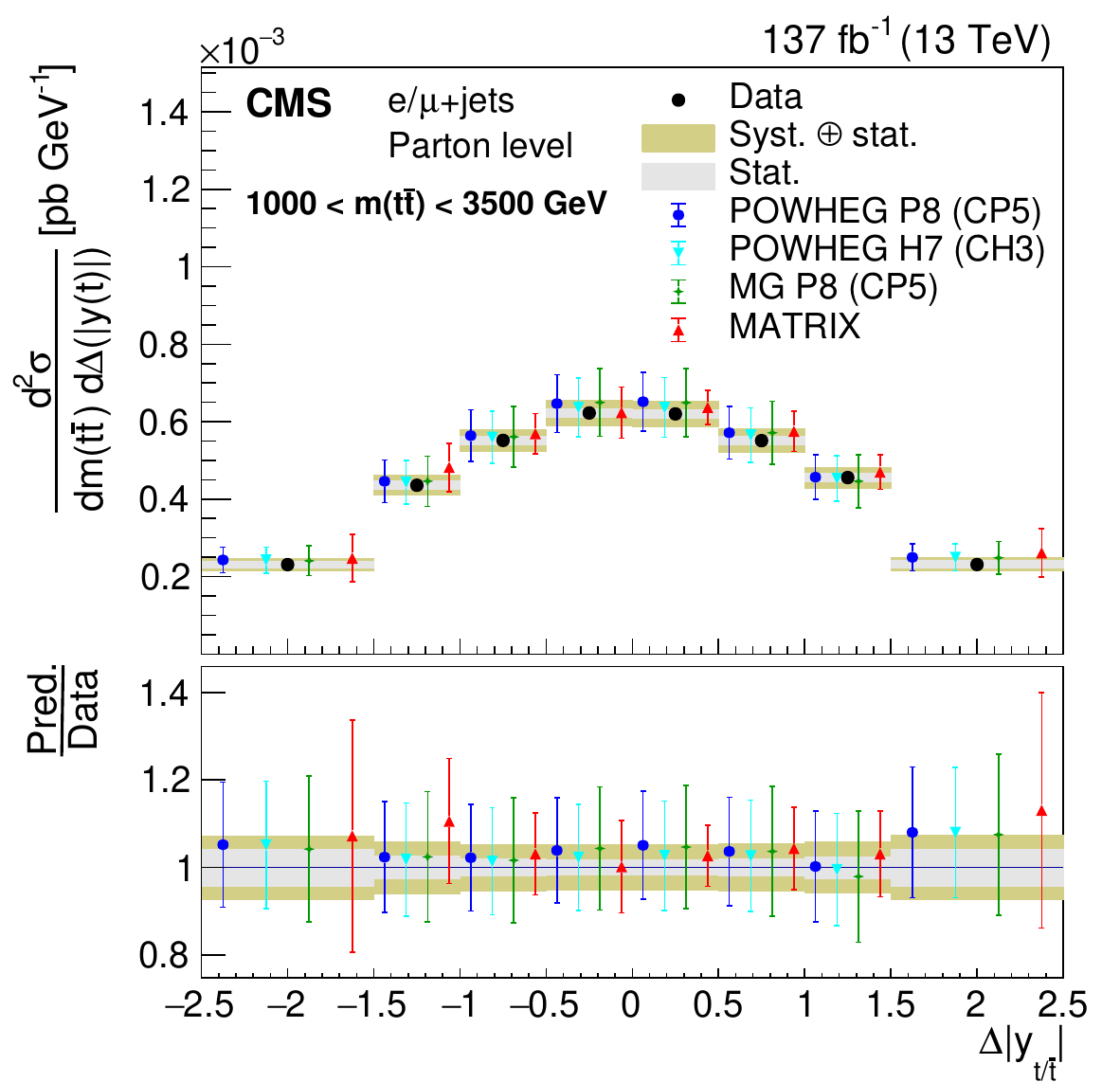}
 \caption{Double-differential cross section at the parton level as a function of \ttmvsdy. \XSECCAPPA}
 \label{fig:RES9}
\end{figure*}

\begin{figure*}[tbp]
\centering
 \includegraphics[width=0.42\textwidth]{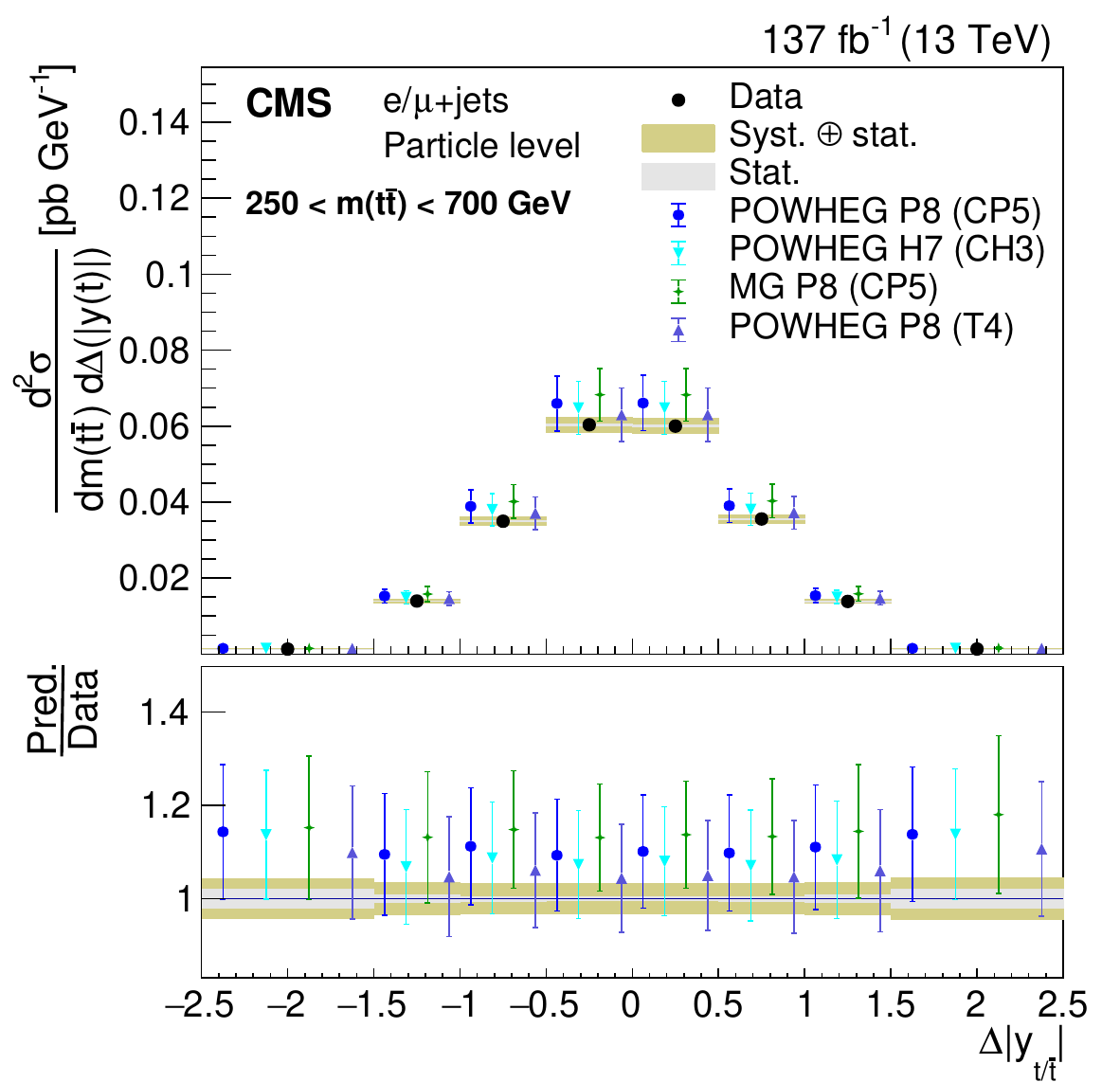}
 \includegraphics[width=0.42\textwidth]{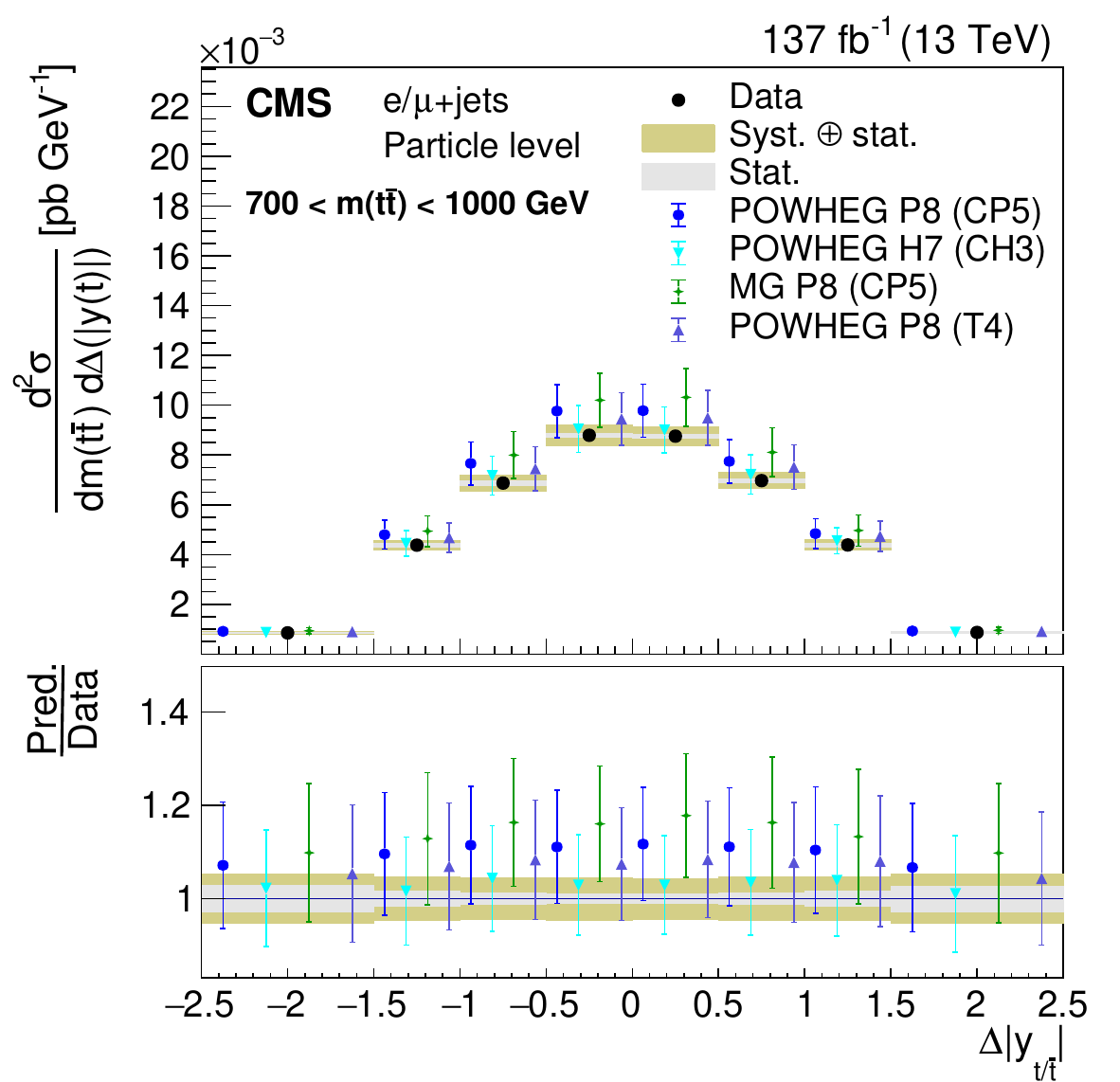}\\
 \includegraphics[width=0.42\textwidth]{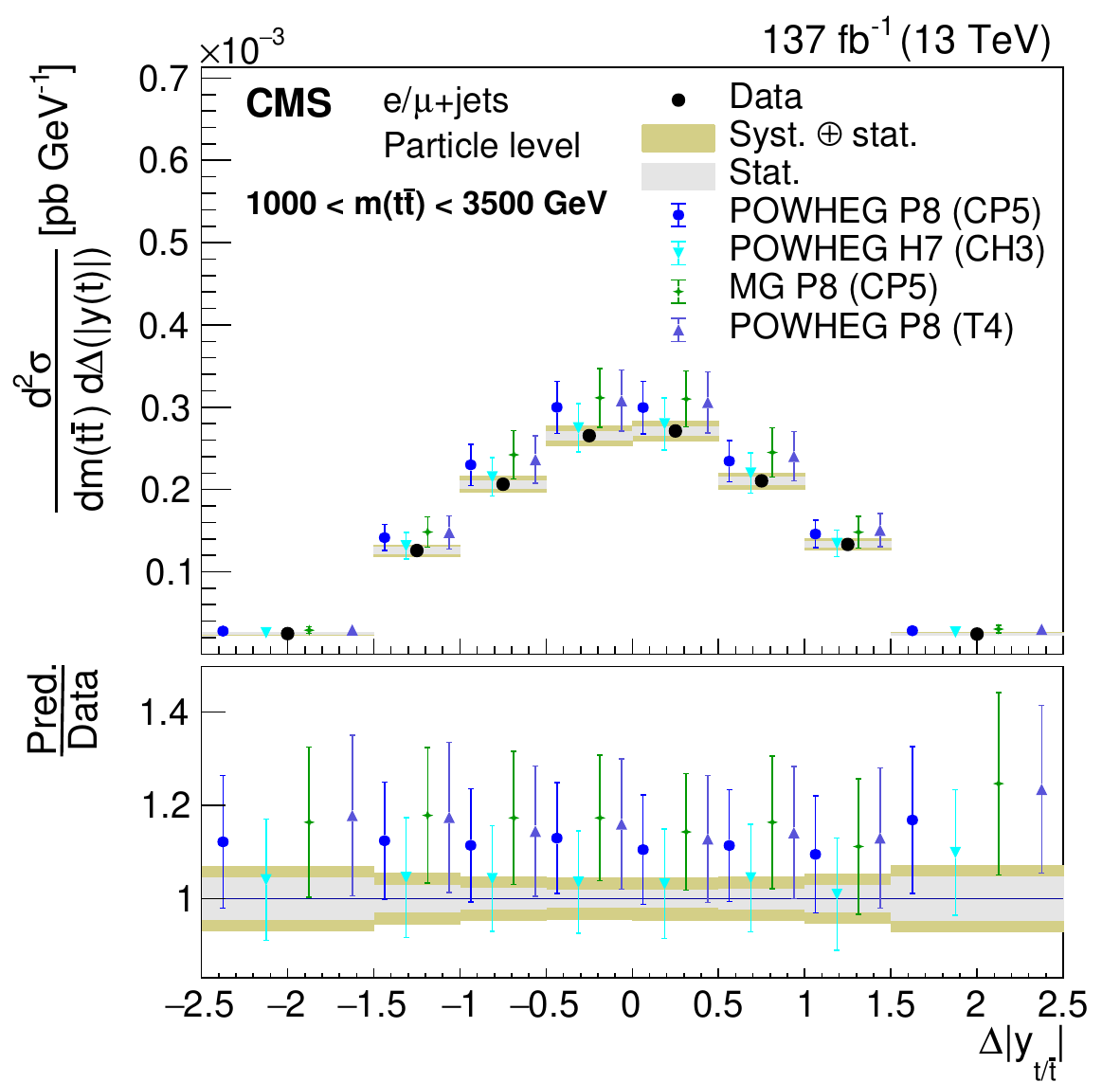}
 \caption{Double-differential cross section at the particle level as a function of \ttmvsdy. \XSECCAPPS}
 \label{fig:RESPS9}
\end{figure*}

\begin{figure*}[tbp]
\centering
 \includegraphics[width=0.42\textwidth]{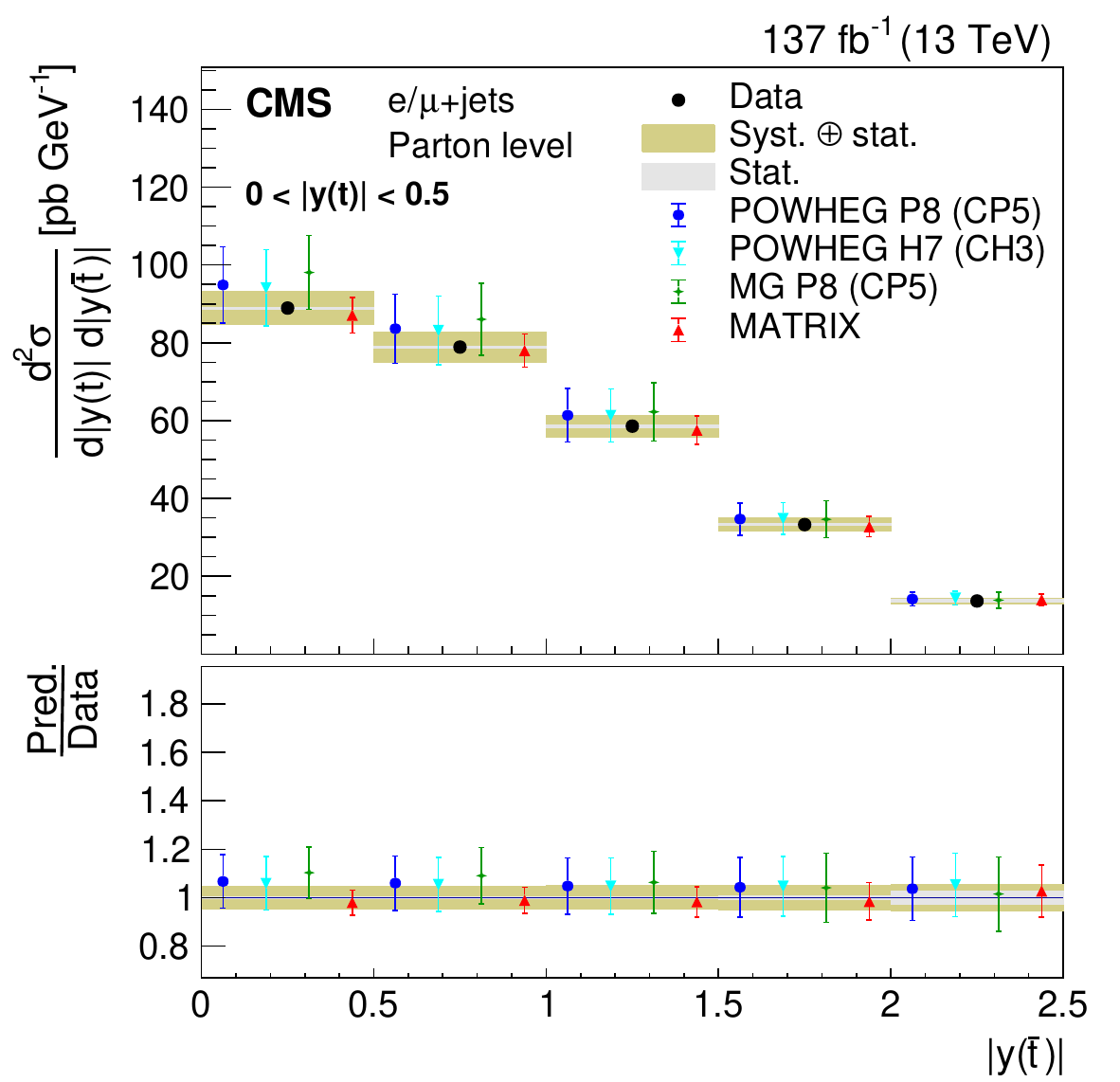}
 \includegraphics[width=0.42\textwidth]{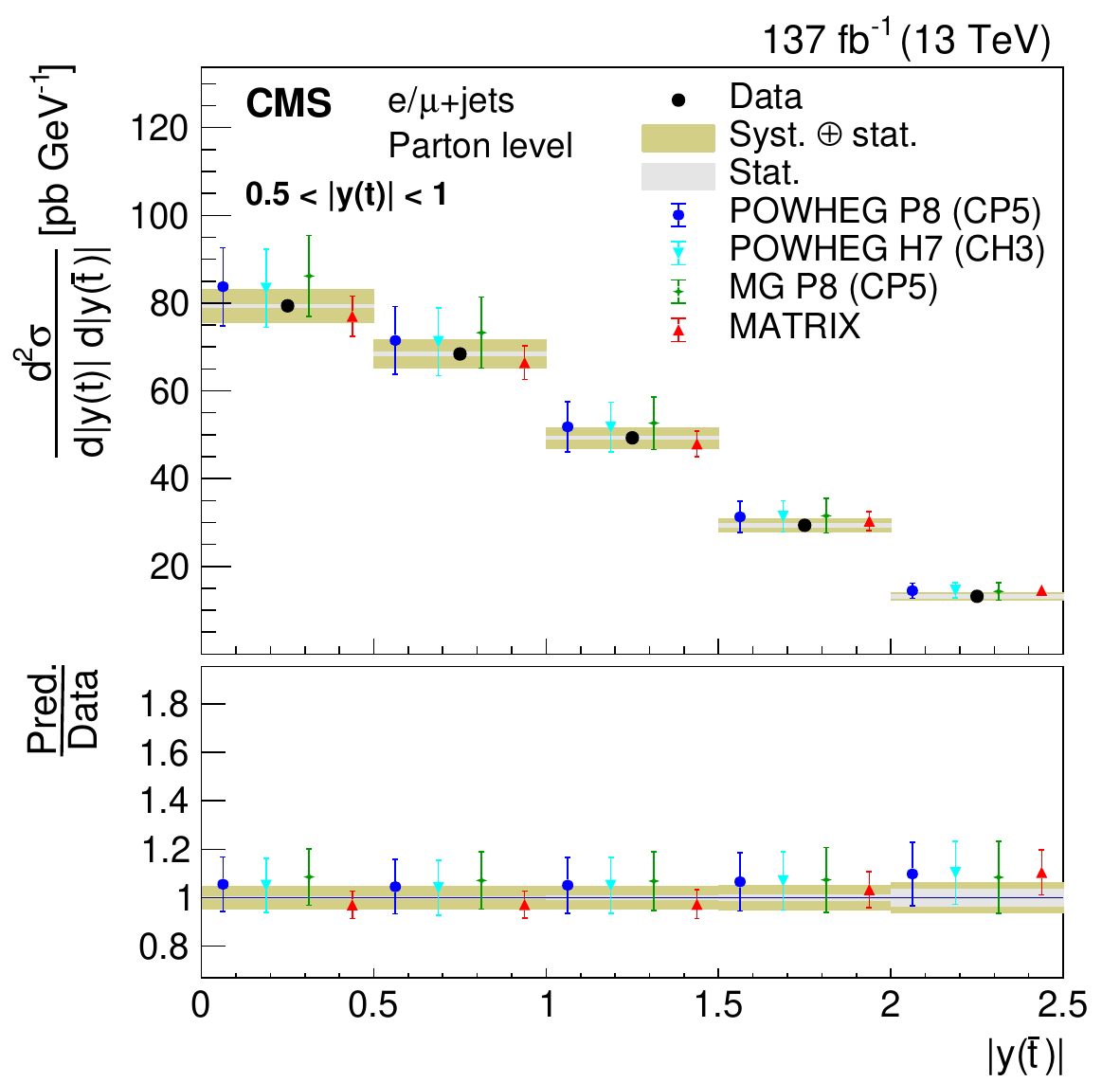}\\
 \includegraphics[width=0.42\textwidth]{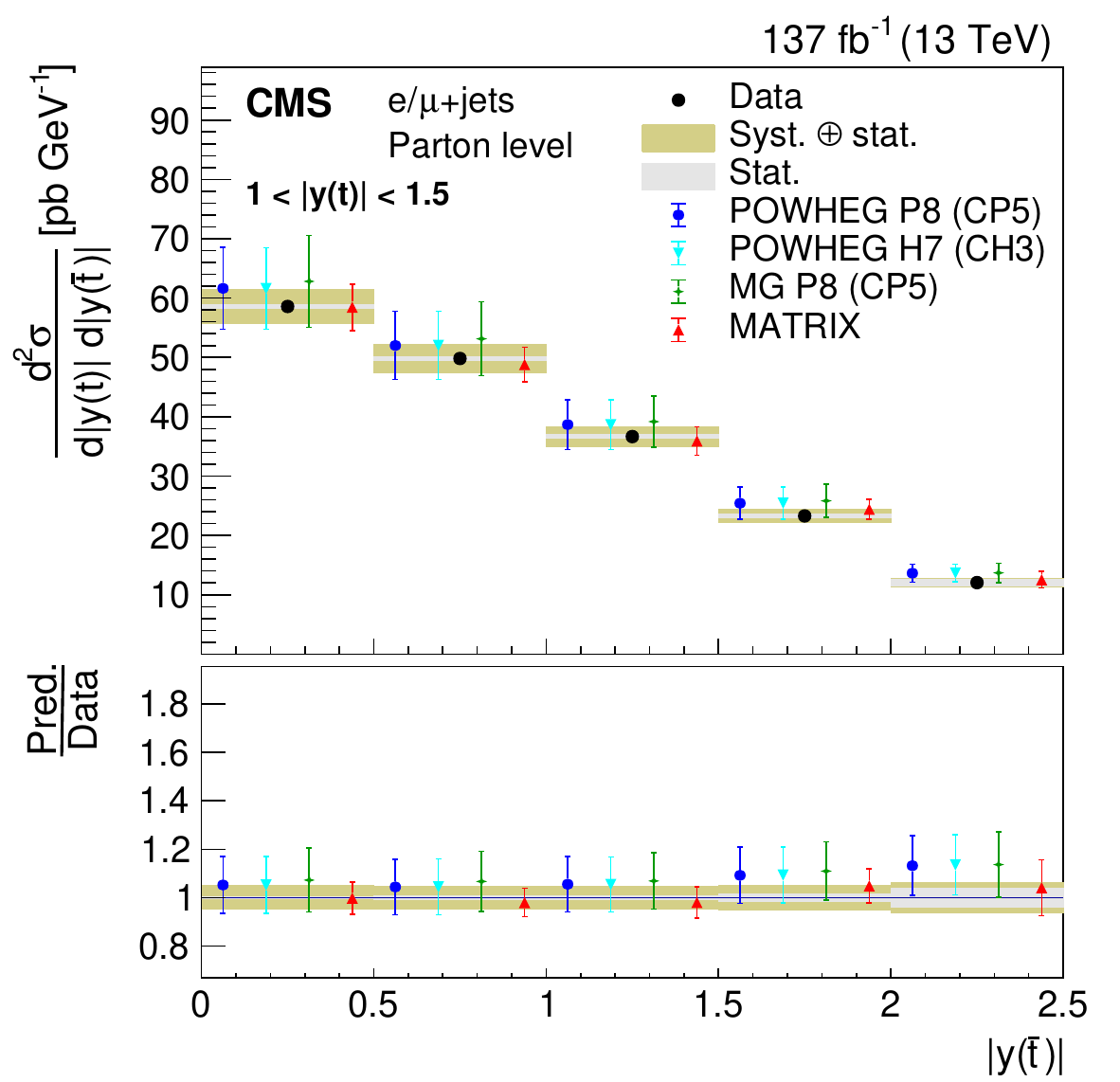}
 \includegraphics[width=0.42\textwidth]{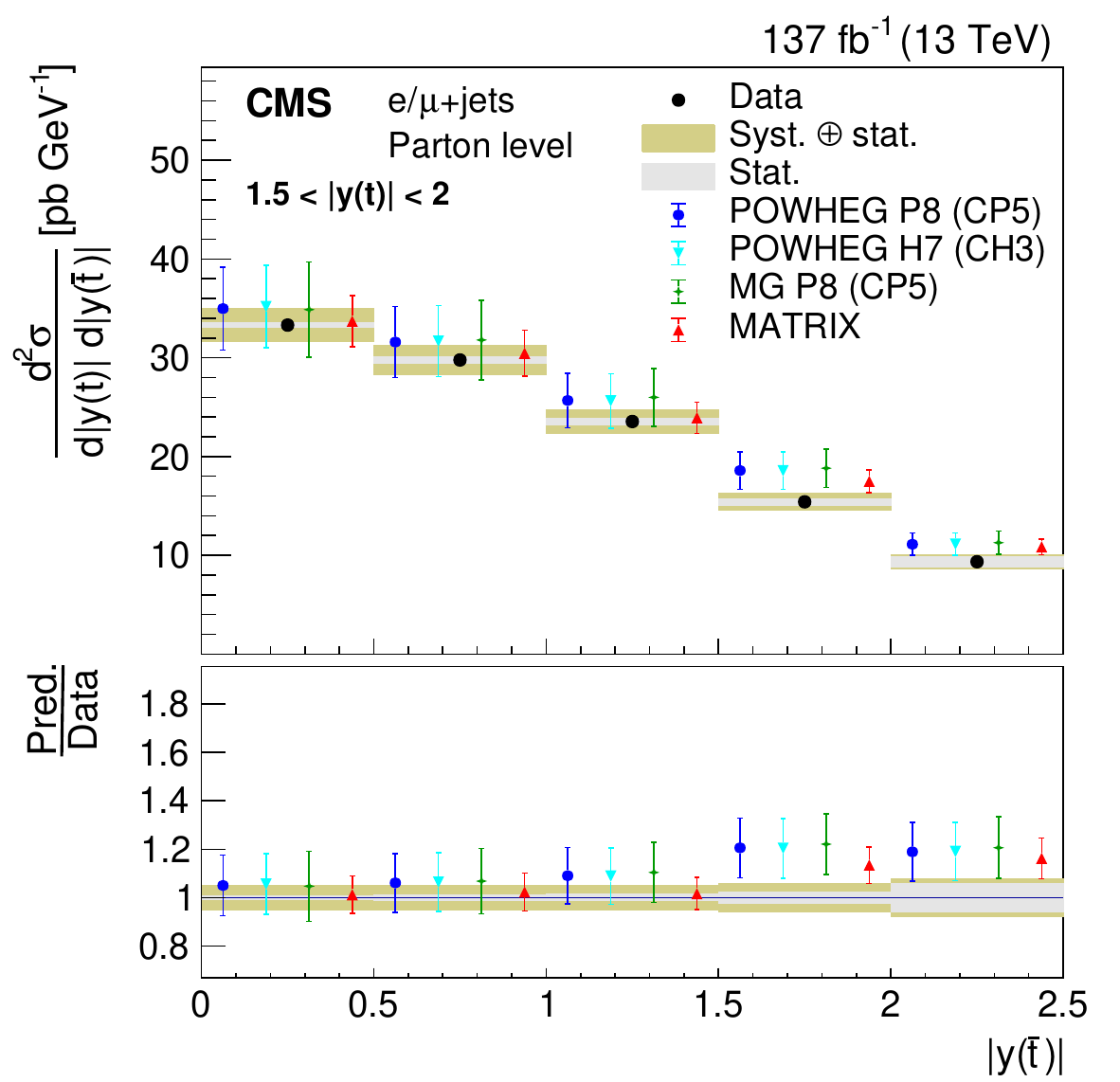}\\
 \includegraphics[width=0.42\textwidth]{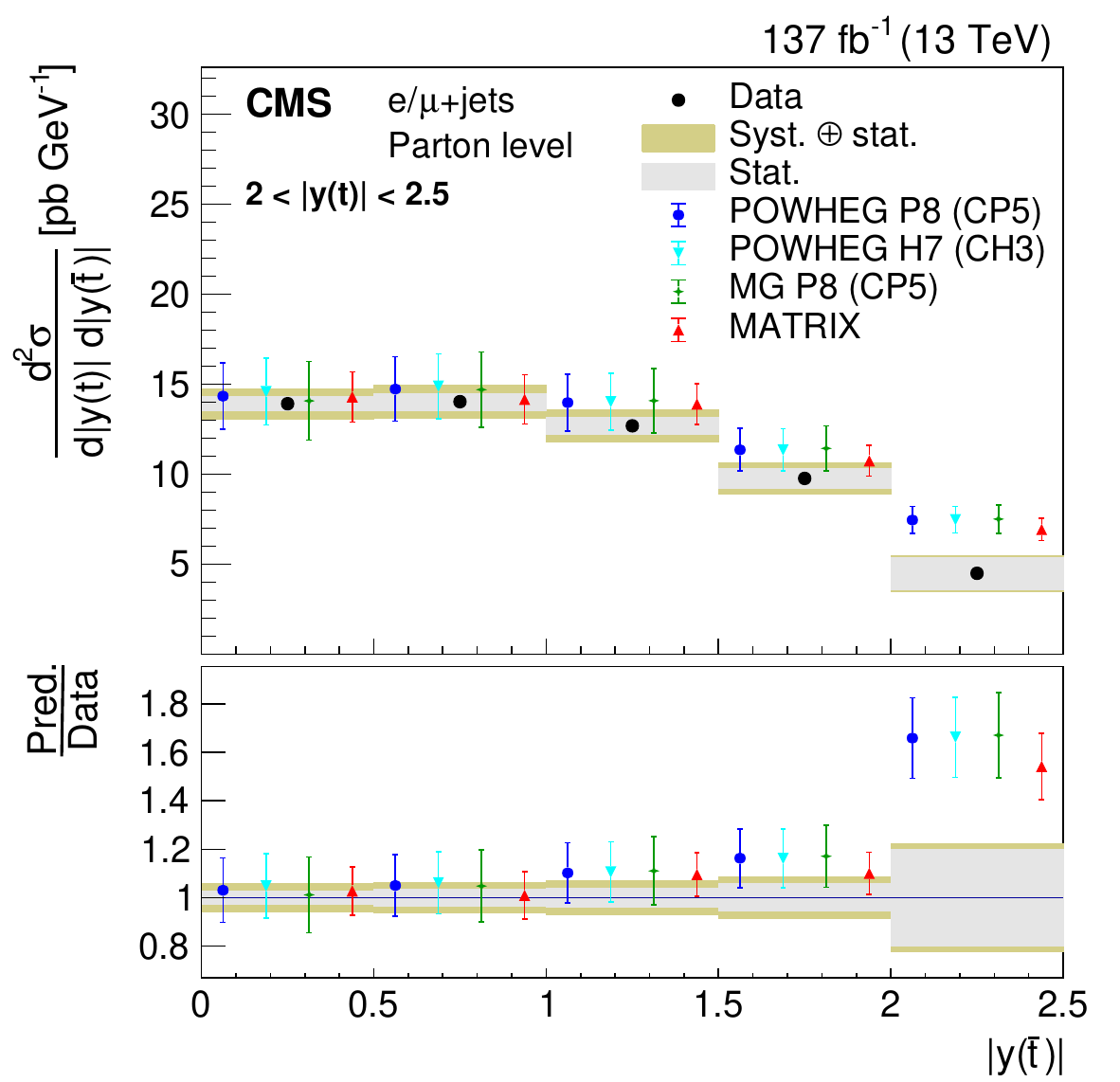}
 \includegraphics[width=0.42\textwidth]{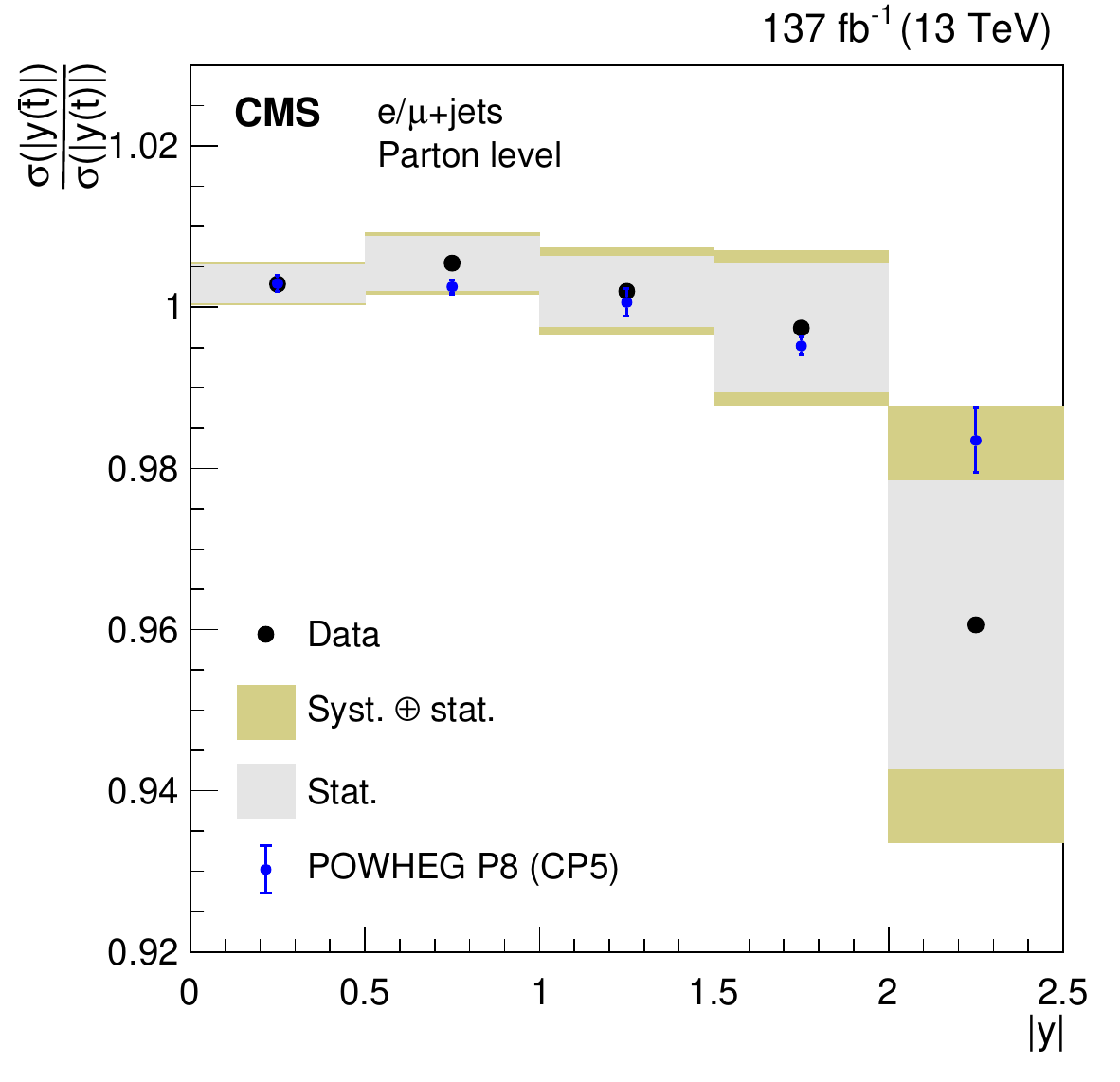}
 \caption{Double-differential cross section at the parton level as a function of \topyvstopbary. \XSECCAPPA Lower right: ratio of $\abs{y(\PAQt)} / \abs{y(\PQt)}$.}
 \label{fig:RES10}
\end{figure*}

\begin{figure*}[tbp]
\centering
 \includegraphics[width=0.42\textwidth]{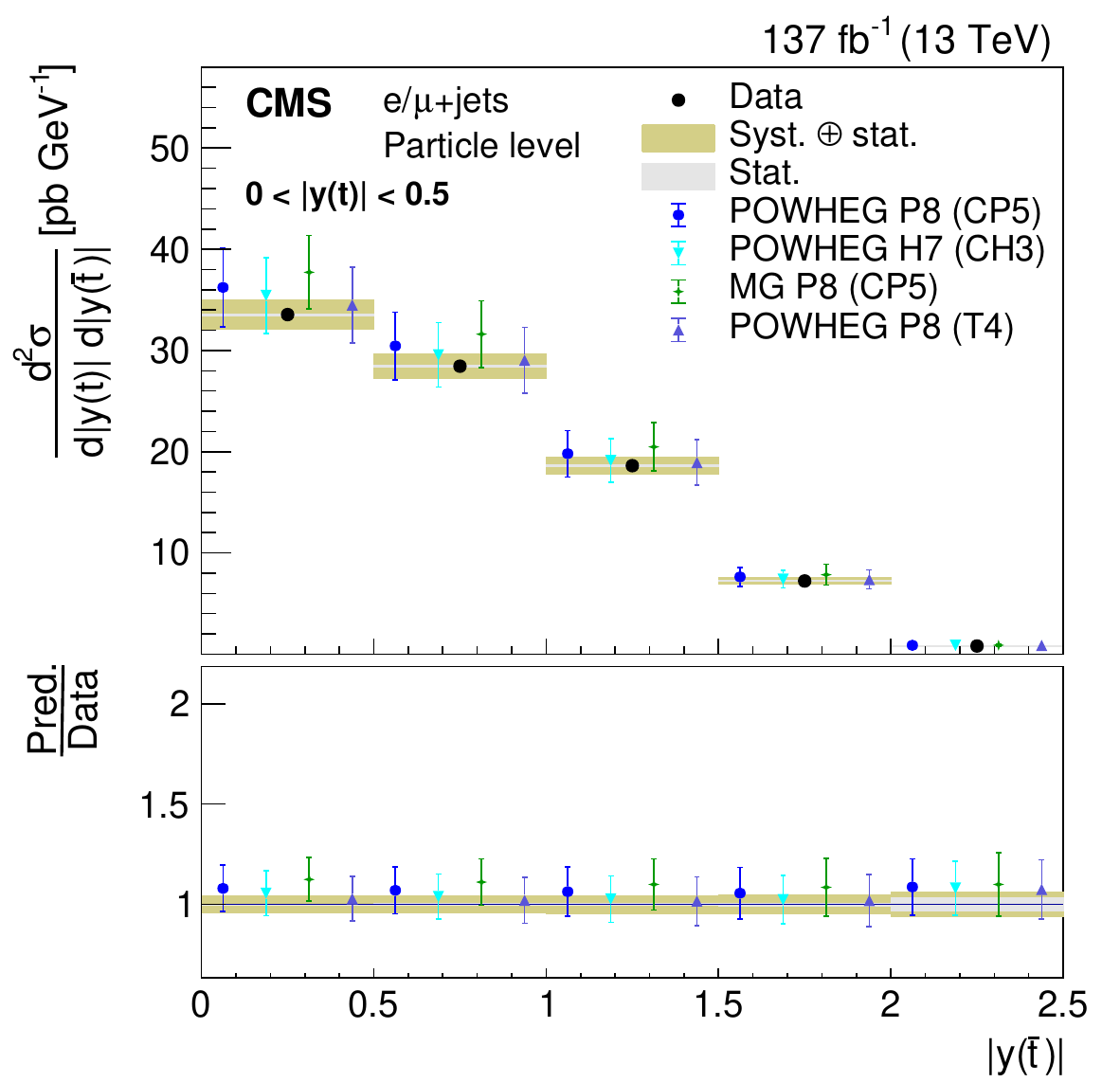}
 \includegraphics[width=0.42\textwidth]{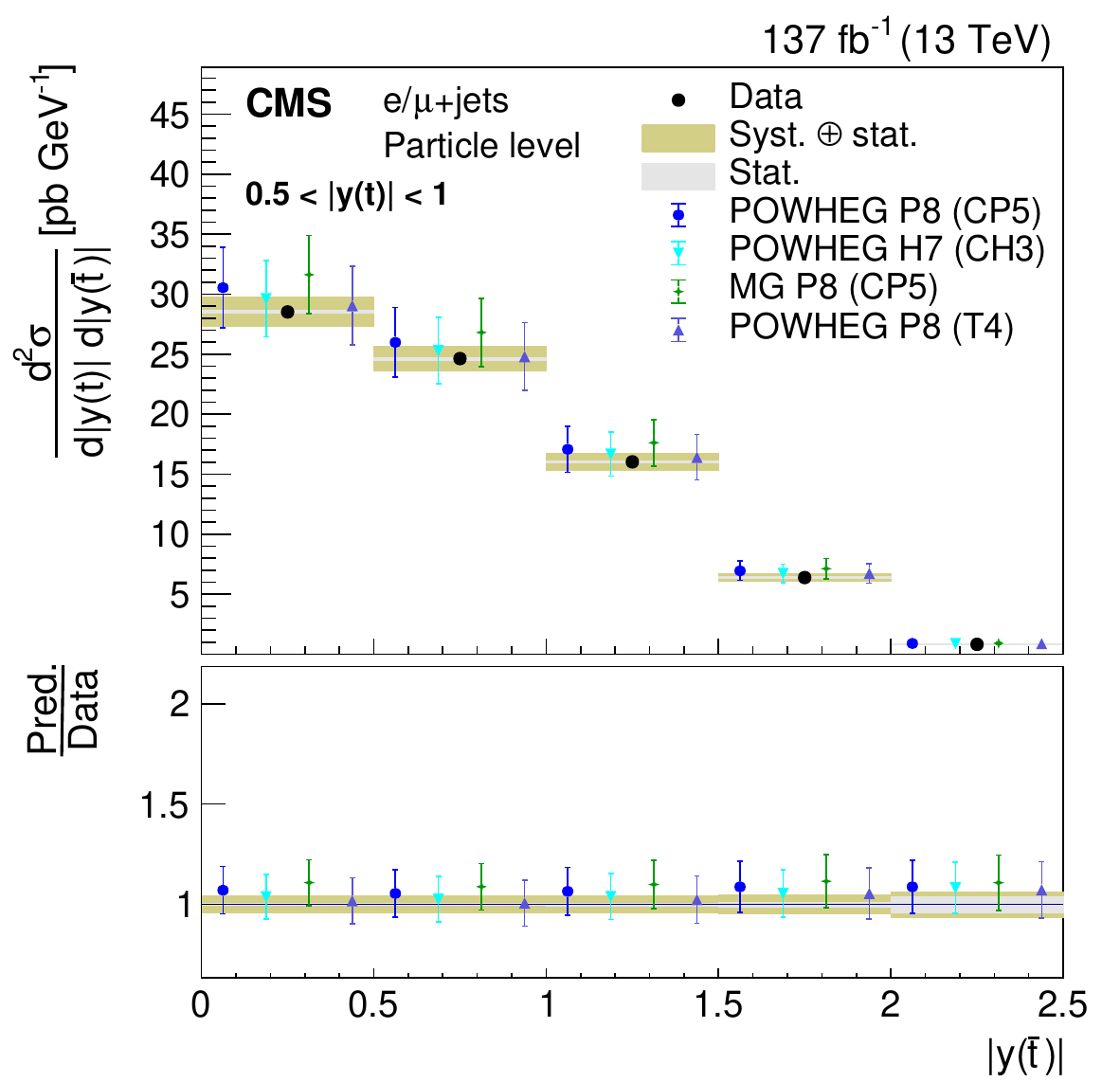}\\
 \includegraphics[width=0.42\textwidth]{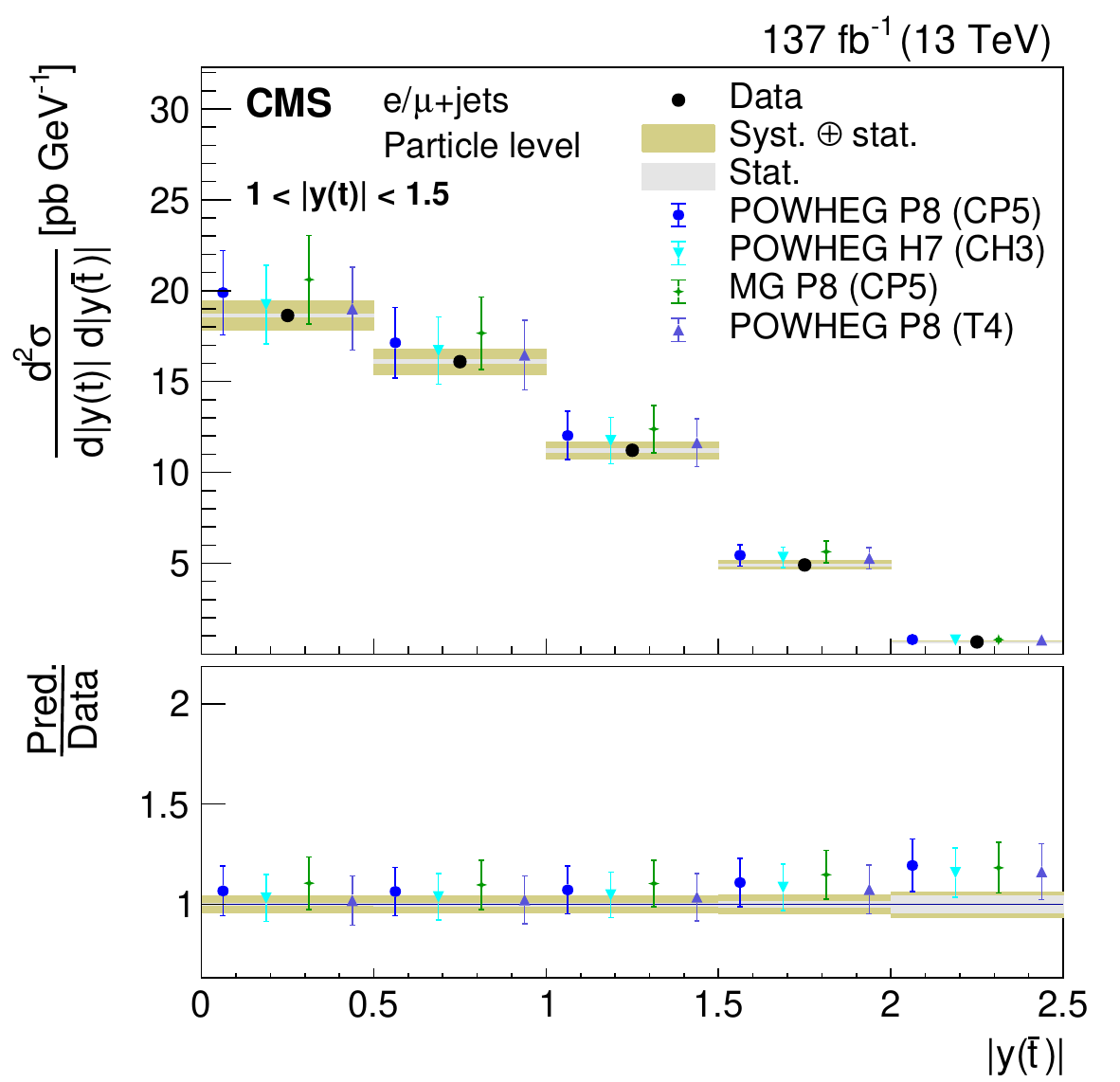}
 \includegraphics[width=0.42\textwidth]{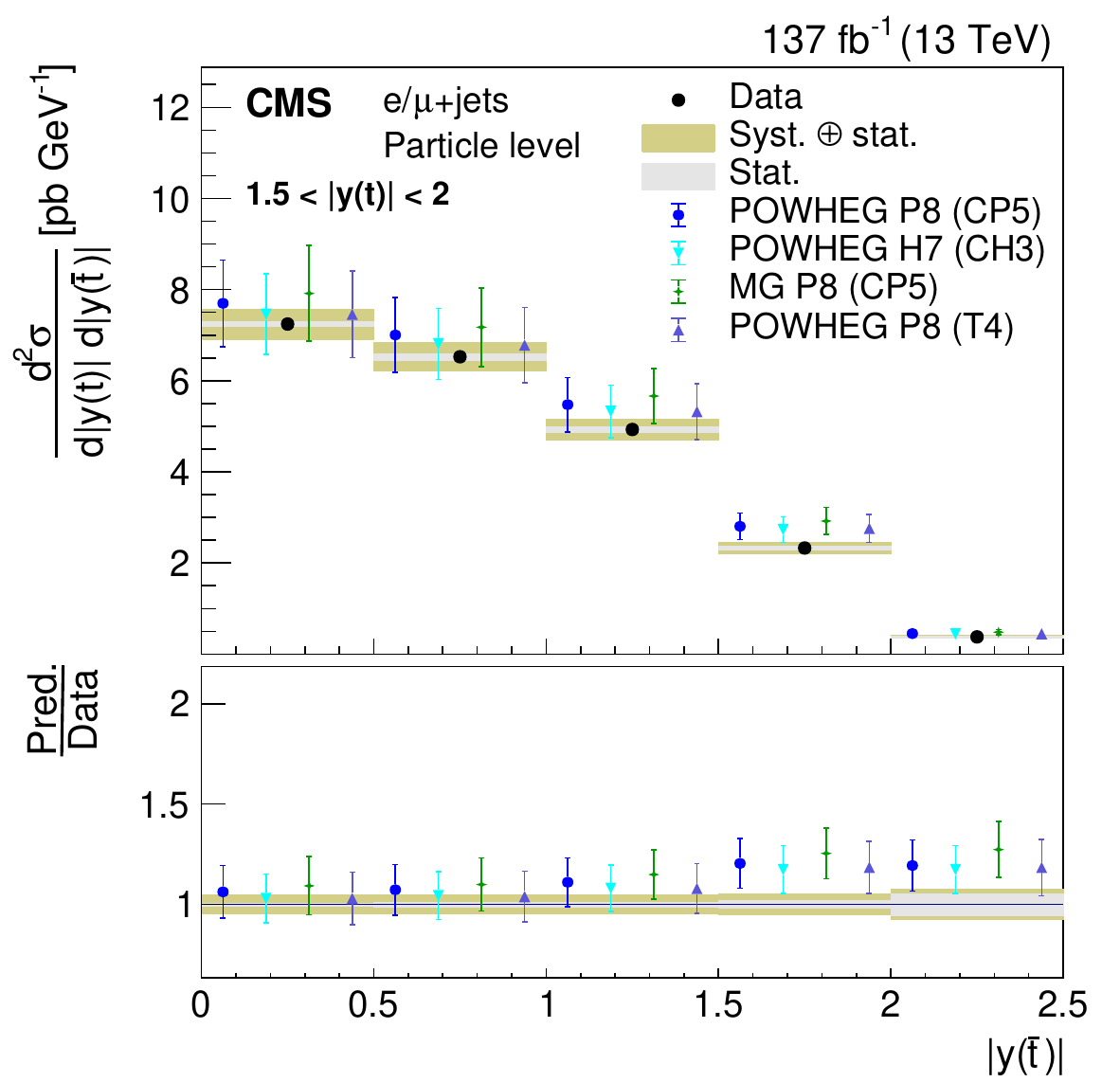}\\
 \includegraphics[width=0.42\textwidth]{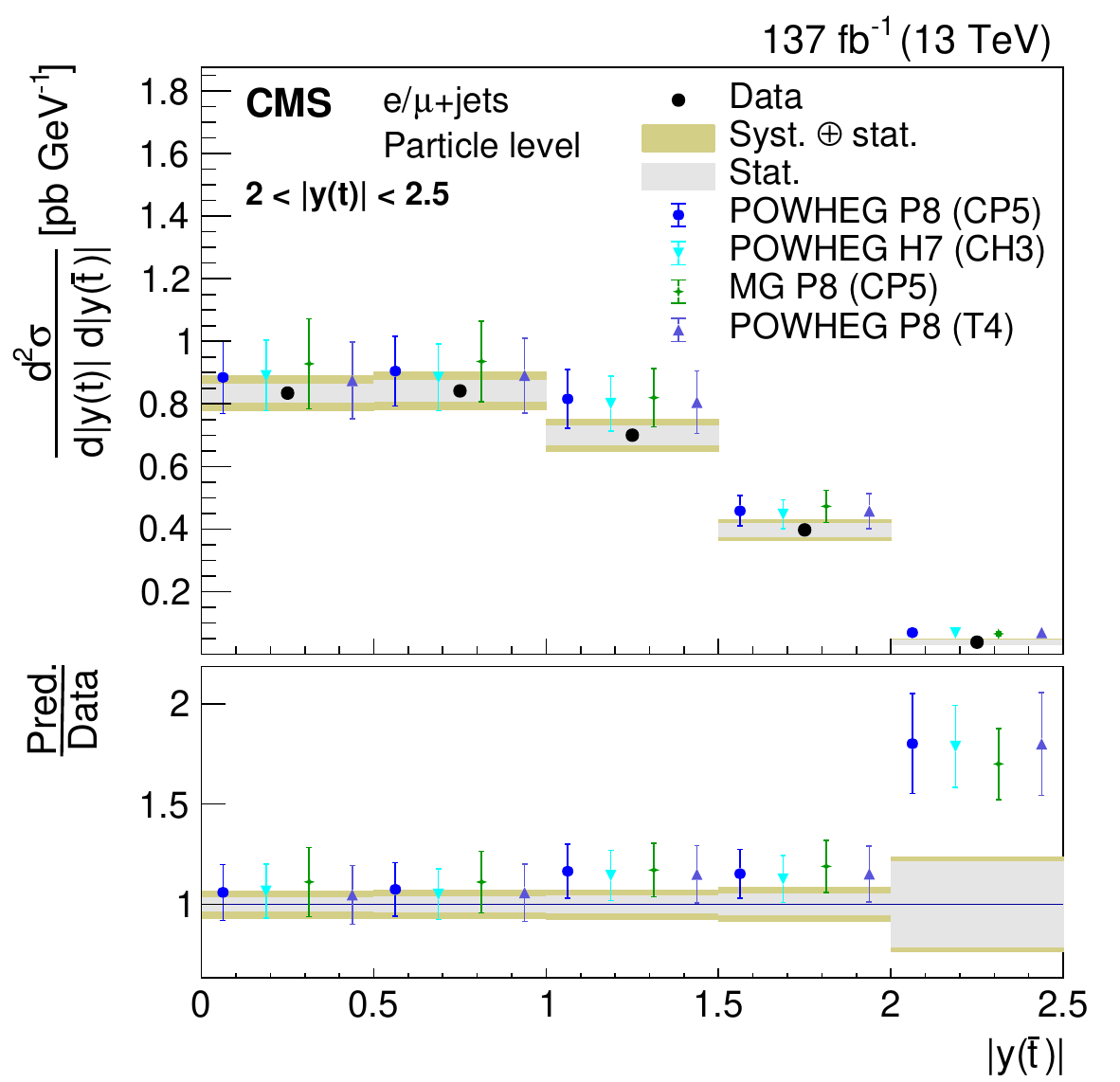}
 \includegraphics[width=0.42\textwidth]{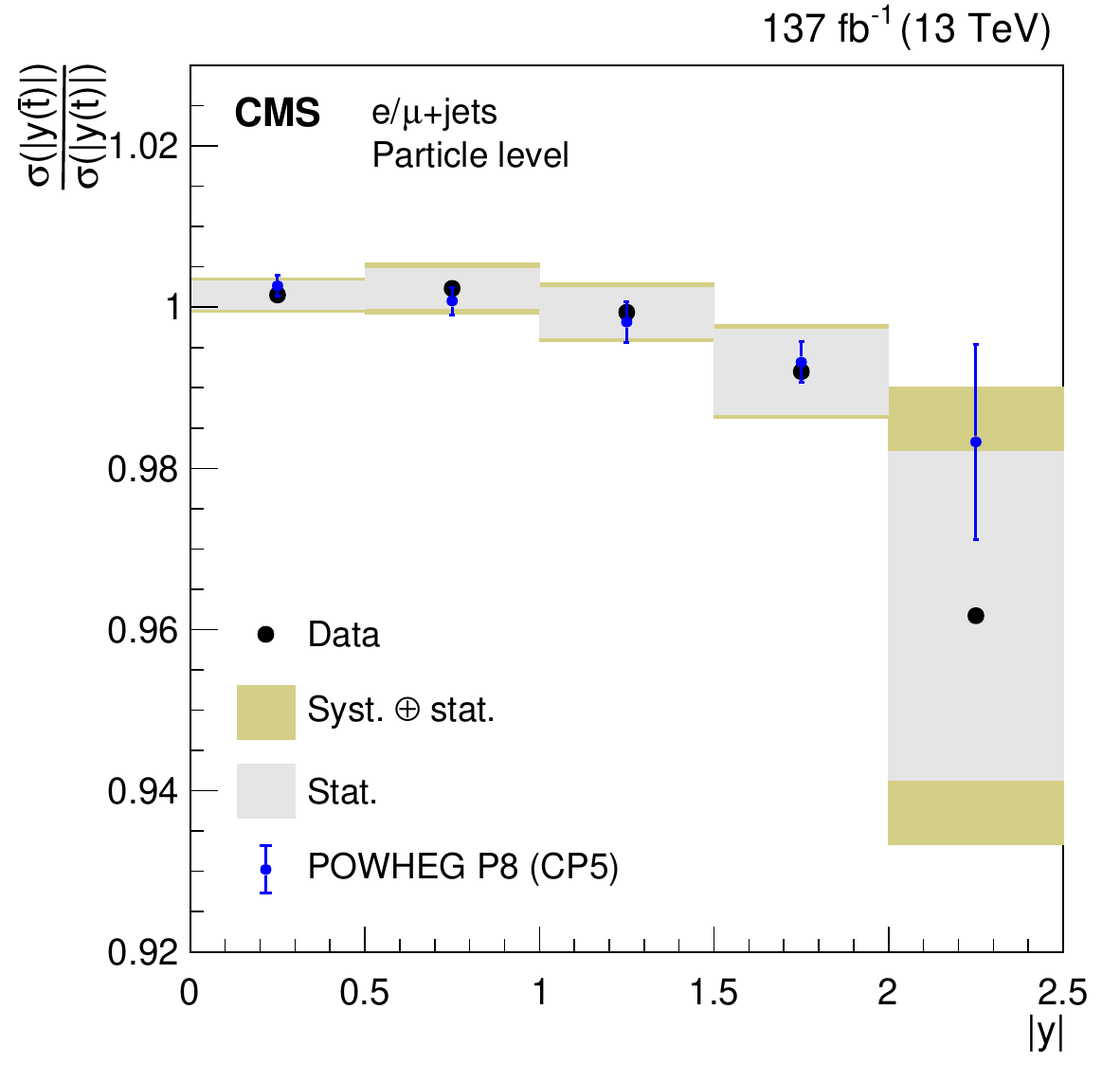}
 \caption{Double-differential cross section at the particle level as a function of \topyvstopbary. \XSECCAPPS Lower right: ratio of $\abs{y(\PAQt)} / \abs{y(\PQt)}$}
 \label{fig:RESPS10}
\end{figure*}

{\tolerance=1000 The double-differential cross sections are shown in Figs.~\ref{fig:RES4}--\ref{fig:RESPS10}. The distributions of \thadptvsthady shown in Figs.~\ref{fig:RES4} and \ref{fig:RESPS4} are well described by most of the predictions. The NLO calculations overestimate the cross sections in all \thady bins in the high-\thadpt region. The corresponding distributions for \tql are not shown, since the information they add is marginal. However, we measure consistent distributions of \tqh and \tql at the parton level.\par}

As mentioned earlier, a higher cross section is predicted than seen in the data at high values of \tty. This effect persists in different regions of \ttm, as shown in the \ttmvstty measurements in Figs.~\ref{fig:RES5} and \ref{fig:RESPS5} .

In Figs.~\ref{fig:RES6} and \ref{fig:RESPS6} the measurements of \ttmvscts are shown. These double-differential cross sections are well described by the simulations. 

The measurements of \ttmvsthadpt, shown in Figs.~\ref{fig:RES7} and \ref{fig:RESPS7}, are not well described by any of the predictions, as indicated by the $\chi^2$ tests given in Fig.~\ref{fig:RES11}. While the measured \pt spectra are in agreement or even harder than predicted at low \ttm, the spectra are softer at high \ttm. Such an opposite trend in the different \ttm regions, along with the bin-by-bin correlations of uncertainties, results in low $p$-values. The mismodeling is even more apparent in the normalized differential cross sections in Figs.~\ref{fig:RESNORM7} and \ref{fig:RESNORMPS7}. 

The measurements of \ttptvsthadpt are shown in Figs.~\ref{fig:RES7b} and \ref{fig:RESPS7b}. Compared to the various predictions, the softer measured \thadpt spectrum is most evident in the low-\ttpt region, while at high \ttpt the measurements are well described.    

The measurements of \adyvsttm are displayed in Figs.~\ref{fig:RES8} and \ref{fig:RESPS8}. This distribution is known to be sensitive to electroweak corrections, especially to the top quark Yukawa coupling~\cite{TOP-17-004, TOP-19-008}. The NNLO calculation results in an improved description of this distribution.      

In Figs.~\ref{fig:RES9} and \ref{fig:RESPS9} the measurements of \ttmvsdy are presented. They are well described by all predictions. Note that it is problematic to use these measurements of \dy to extract information on the charge asymmetry of \ttbar production in \pp collisions because the acceptance corrections introduce a significant asymmetry in these measurements.

In Figs.~\ref{fig:RES10} and \ref{fig:RESPS10} the double-differential cross sections as a function of \topyvstopbary are shown. They are used to calculate the ratio of cross sections of \PQt and \PAQt as a function of rapidity when the bin-by-bin correlations are taken into account correctly. These ratios are displayed in the lower-right plots of Figs.~\ref{fig:RES10} and \ref{fig:RESPS10}. Differences in the rapidity of \PQt and \PAQt are a direct consequence of the charge asymmetry, where on average $\abs{y(\PQt)}$ is less central than $\abs{y(\PAQt)}$. In contrast to measurements based on \dy, the acceptance correction of this double-differential measurement does not depend on the asymmetry in the simulation. However, no significant difference between $\abs{y(\PQt)}$ and $\abs{y(\PAQt)}$ is observed. The simulation predicts that the main effect of the charge asymmetry is expected at high rapidities, where the measurement is limited by the statistical accuracy and detector acceptance.     
 
\clearpage

\begin{figure*}[tbp]
\centering
 \includegraphics[width=0.75\textwidth]{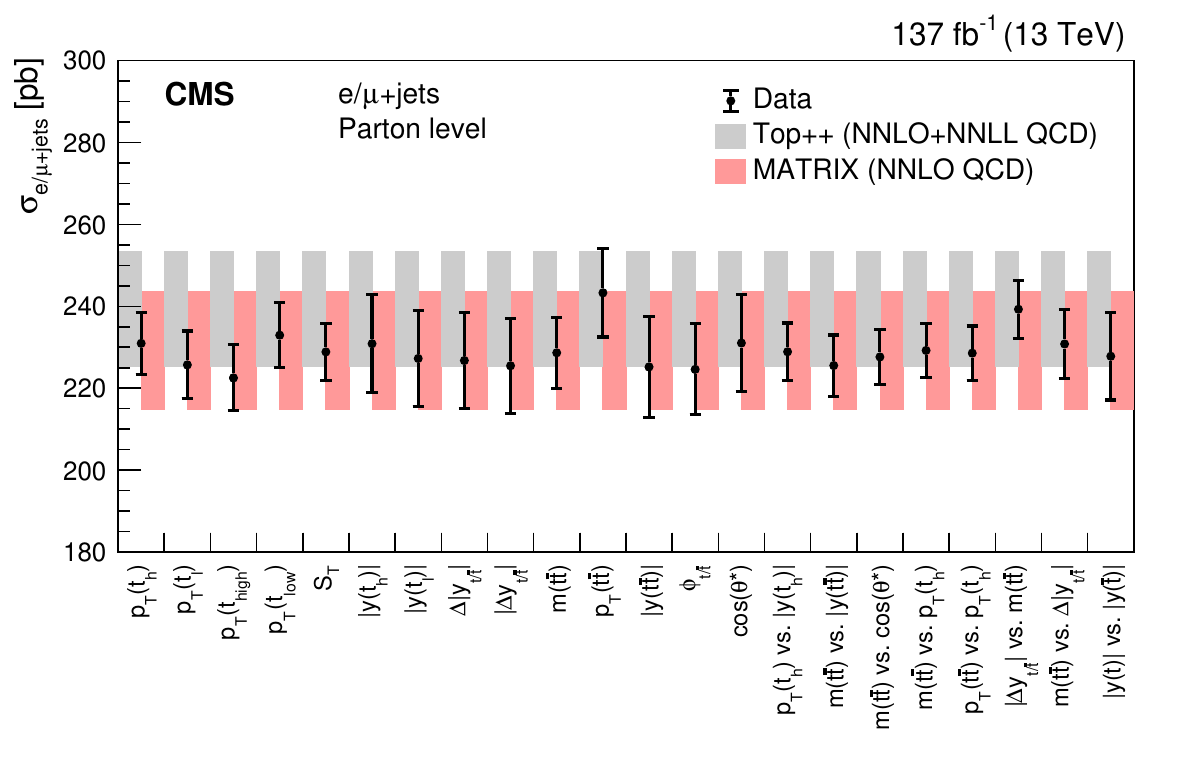}
 \caption{Measurements of the \ttbar production cross sections $\sigma_{\lpj}$ with their total uncertainty obtained as the sum of the cross sections in all bins of a distribution as a function of the kinematic variable used in the determination. The results are compared to the \TOPPP and the \MATRIX predictions with their corresponding uncertainties.}
 \label{fig:RES12}
\end{figure*}

The sum of the cross sections in all bins of a distribution at the parton level corresponds to the \ttbar production cross section in the \lpj channel. The cross sections outside the measured ranges are predicted from the simulation to be negligibly small. Only for the rapidity-related distributions there is a contribution of up to 3\% estimated from the simulation. These corrections are taken into account to determine the different measurements of the total cross section shown in Fig.~\ref{fig:RES12}. 

All cross section values are similar, although it is difficult to judge the degree of agreement because they do not represent independent measurements since several of the observables are strongly correlated. We choose for our final result the cross section obtained from the measurement of \ttmvscts. According to the simulation, this is expected to be the most precise measurement because its response matrix is little affected by systematic uncertainties, and the measurement can most effectively constrain those uncertainties. This expectation is confirmed and we find a value of
\begin{equation}
\sigma_{\lpj} = 227.6\pm 6.8\unit{pb}.
\end{equation}
Only the measurement of \ttmvsthadpt has a marginally smaller uncertainty in data. 
With a branching fraction of $(28.77\pm0.32)$\%~\cite{PDG} for the decay of \ttbar to \lpj, the total \ttbar production cross section becomes
\begin{equation}
\sigma_\ttbar = 791\pm 25\unit{pb}.
\end{equation}
When breaking down the uncertainty into different sources, we find
\begin{equation}
\sigma_\ttbar = 791\pm1\stat\pm21\syst\pm 14\lum\unit{pb},
\end{equation}
where the last uncertainty comes from that in the integrated luminosity.

All individual sources of systematic uncertainty and their values are given in Table~\ref{tab:SYSUNCS}. This result is in good agreement with the SM expectation of $797^{+39}_{-51}\,\mathrm{(scale)} \pm 39\,\mathrm{(PDF)}\unit{pb}$ obtained with \MATRIX, and $832^{+40}_{-46}\unit{pb}$ obtained with \TOPPP~\cite{Czakon:2011xx}, as discussed in Section~\ref{SIM}. 

\begin{table}[tbp]
\centering
\topcaption{The sources of systematic uncertainty and their absolute and relative values in the measurement of $\sigma_\ttbar$.}
\begin{scotch}{lcc}
Source & \multicolumn{2}{c}{Uncertainty}  \\
  &  [pb]  & [\%]  \\\hline
Jet energy & 11  & 1.38 \\
Branching fraction & 8.8  & 1.11 \\
Lepton & 7.8  & 0.98 \\
NNLO & 7.6  & 0.96 \\
\PQb tagging & 7.0  & 0.88 \\
Sim. event count & 6.5  & 0.82 \\
Background & 6.1  & 0.77 \\
CR model & 5.5  & 0.69 \\
Jet energy resolution & 3.4  & 0.43 \\
Scales \mur, \muf & 3.2  & 0.41 \\
Initial-state PS scale & 3.2  & 0.40 \\
Final-state PS scale & 2.7  & 0.34 \\
Subjet energy & 2.4  & 0.31 \\
\PQb mistagging & 2.2  & 0.28 \\
UE tune & 2.2  & 0.27 \\
\Mtop & 2.1  & 0.26 \\
PDF & 1.9  & 0.25 \\
\hd & 1.5  & 0.19 \\
L1 trigger & 0.5  & 0.07 \\
Pileup & 0.4  & 0.05 \\[\cmsTabSkip]
Total syst. & 21  & 2.66 \\
Total stat. & 0.6  & 0.07 \\
Int. luminosity & 14  & 1.75\\
\end{scotch}
\label{tab:SYSUNCS}
\end{table}

\begin{figure*}[tbp]
\centering
 \includegraphics[width=0.75\textwidth]{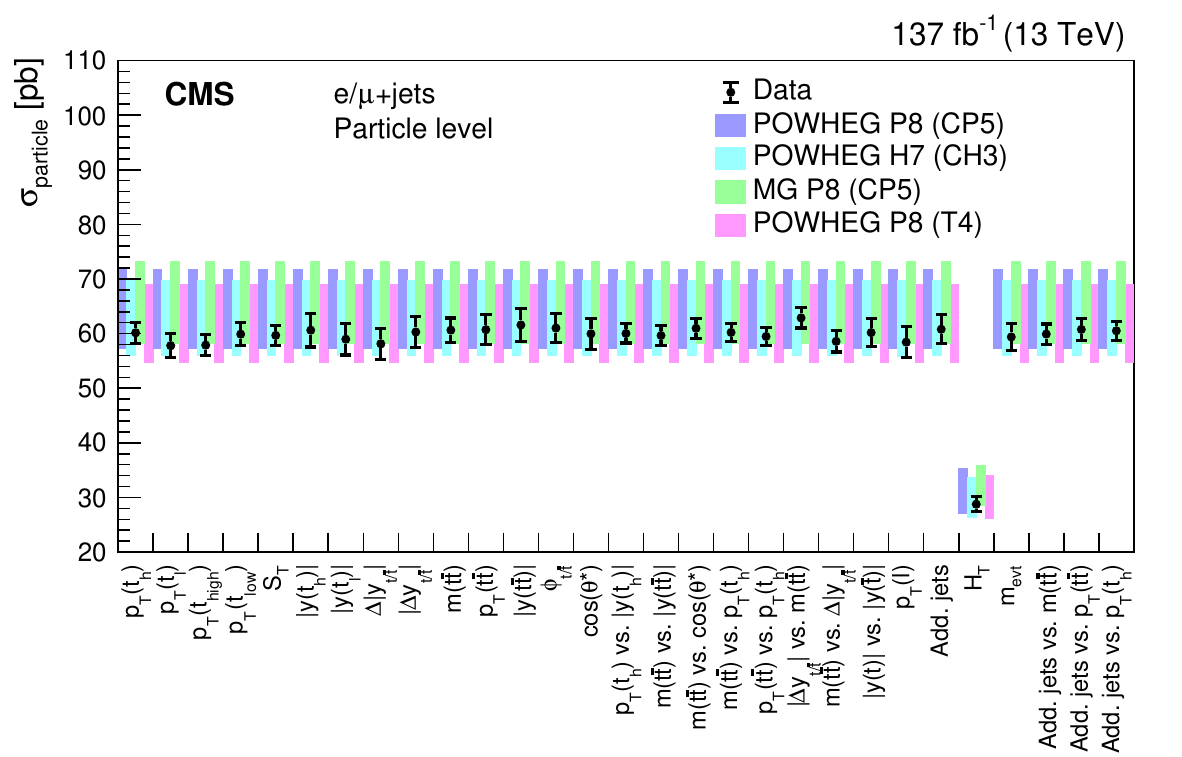}
 \caption{Measurements of the \ttbar production cross sections $\sigma_\mathrm{particle}$ at the particle level and their total uncertainties as a function of the kinematic variable used in the determination. The results are compared to the predictions of \POWHEG{}+\PYTHIA (P8) for the CP5 and \CUET (T4) tunes, \POWHEG{}+\HERWIG (H7), and the multiparton simulation \AMCATNLO(MG)+\PYTHIA.}
 \label{fig:RESPS12}
\end{figure*}

By adding the cross sections in all bins of a distribution at the particle level, the inclusive cross section $\sigma_\mathrm{particle}$ is obtained. The exception is the cross section at the parton level as a function of \HT, where events with zero additional jets do not contribute. This leads to a significantly smaller cross section. As shown in Fig.~\ref{fig:RESPS12}, the $\sigma_\mathrm{particle}$ values are similar for all the other distributions. The simulations predict that the cross section the cross section obtained from the measurement of \ttmvscts to be the most precise. This is confirmed in data, and the measured value is 
\begin{equation}
\sigma_\mathrm{particle} = 61.0\pm 1.8\unit{pb}.
\end{equation}
The individual sources of systematic uncertainty in this measurement and their values are given in Table~\ref{tab:SYSUNCSPS}.

\begin{table}[tbp]
\centering
\topcaption{The sources of systematic uncertainty and their absolute and relative values in the measurement of $\sigma_\mathrm{particle}$.}
\begin{scotch}{lcc}
Source & \multicolumn{2}{c}{Uncertainty}  \\
  &  [pb]  & [\%]  \\\hline
Jet energy & 0.85  & 1.40 \\
\PQb tagging & 0.67  & 1.11 \\
Sim. event count & 0.63  & 1.03 \\
Lepton & 0.59  & 0.96 \\
Background & 0.47  & 0.78 \\
CR model & 0.44  & 0.72 \\
NNLO & 0.32  & 0.52 \\
\Mtop & 0.27  & 0.45 \\
Scales \mur, \muf & 0.27  & 0.45 \\
Jet energy resolution & 0.19  & 0.31 \\
\PQb mistagging & 0.19  & 0.31 \\
UE tune & 0.18  & 0.30 \\
PDF & 0.15  & 0.24 \\
Subjet energy & 0.12  & 0.20 \\
Final-state PS scale & 0.08  & 0.13 \\
\hd & 0.07  & 0.11 \\
Pileup & 0.05  & 0.08 \\
L1 trigger & 0.05  & 0.08 \\
Initial-state PS scale & 0.03  & 0.05 \\[\cmsTabSkip]
Total syst. & 1.46  & 2.39 \\
Total stat. & 0.04  & 0.06 \\
Int. luminosity & 1.06  & 1.74\\
\end{scotch}
\label{tab:SYSUNCSPS}
\end{table}

\clearpage

\begin{figure*}[tbp]
\centering
 \includegraphics[width=0.42\textwidth]{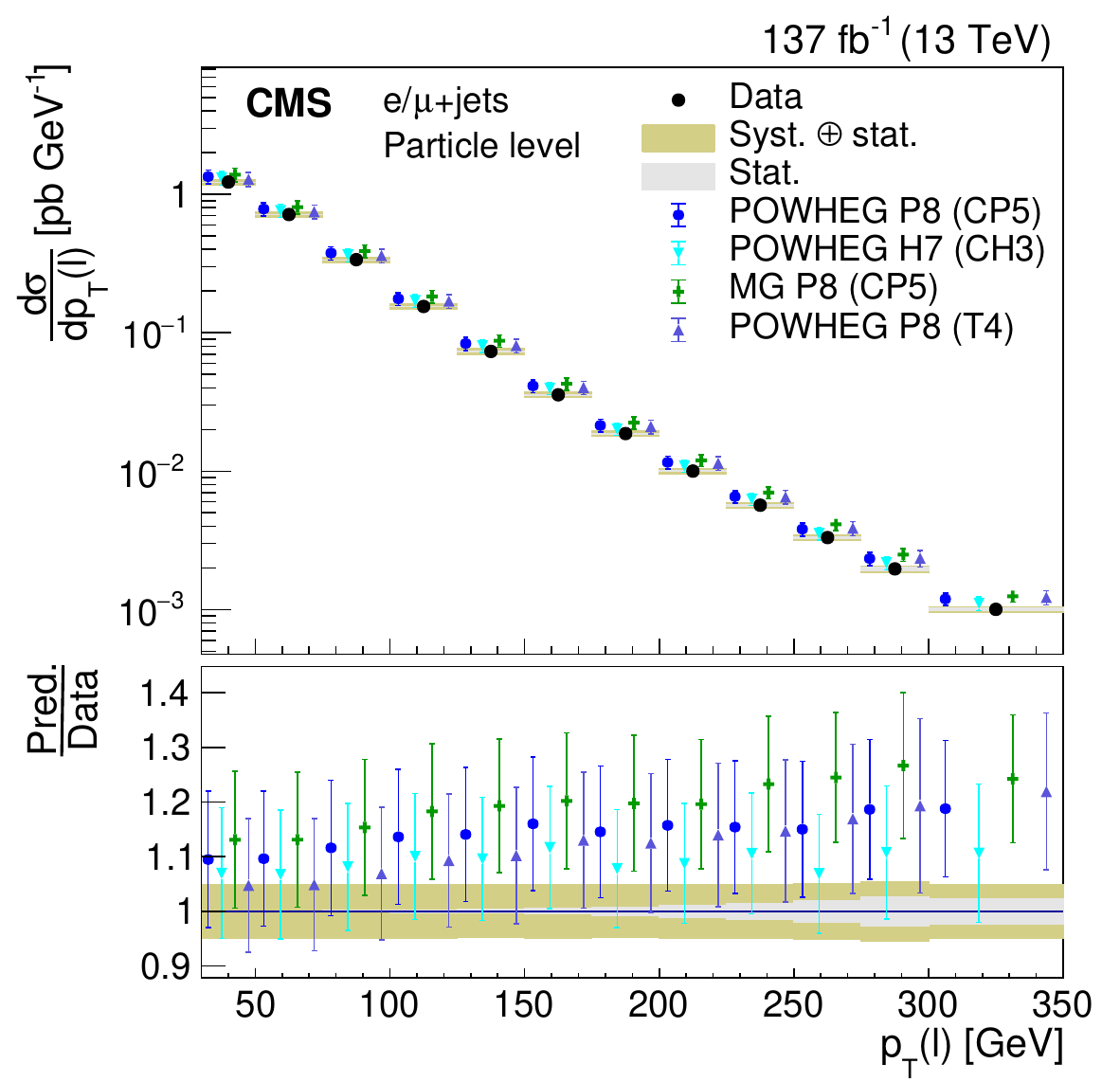}
 \includegraphics[width=0.42\textwidth]{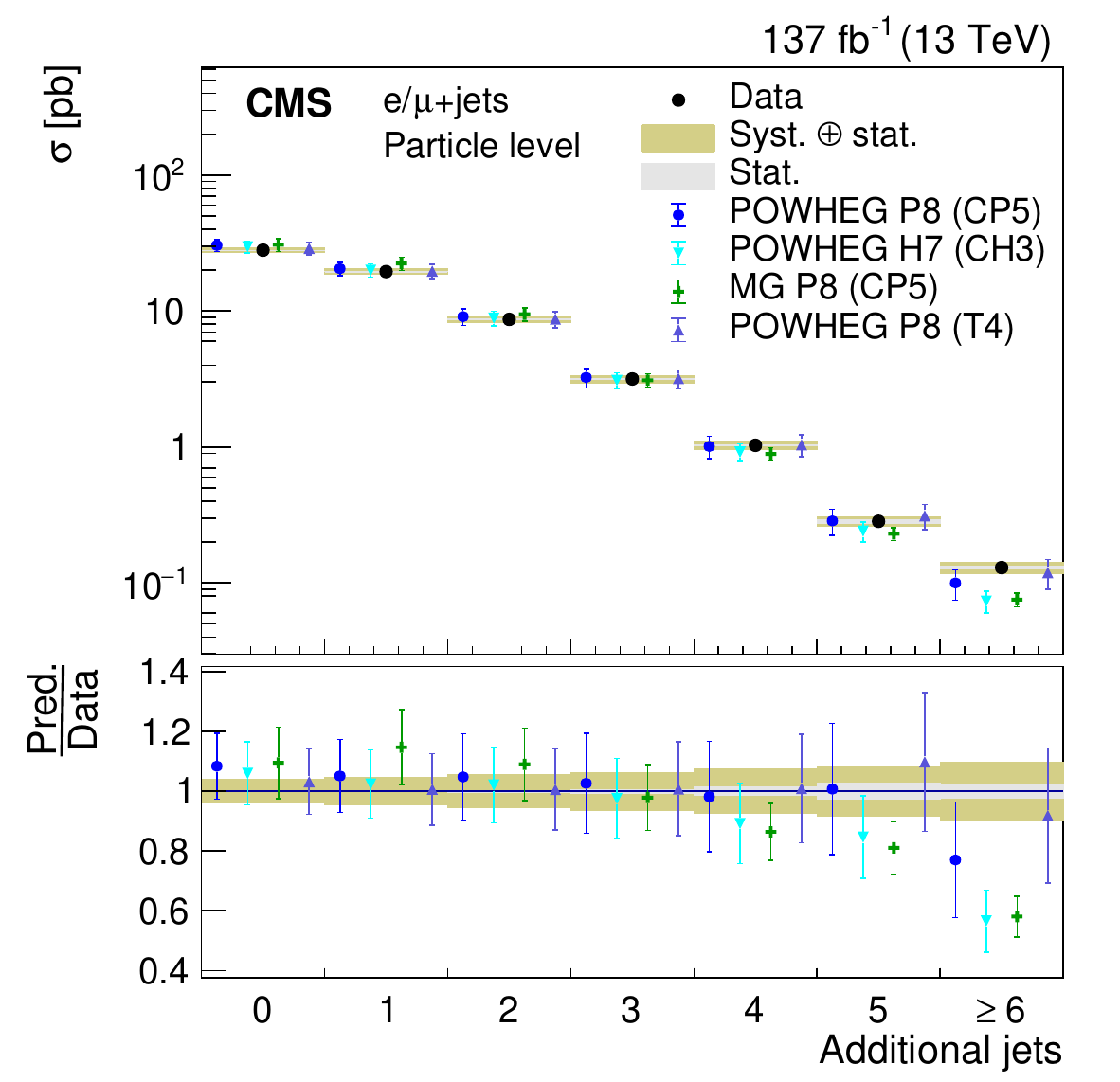}
 \includegraphics[width=0.42\textwidth]{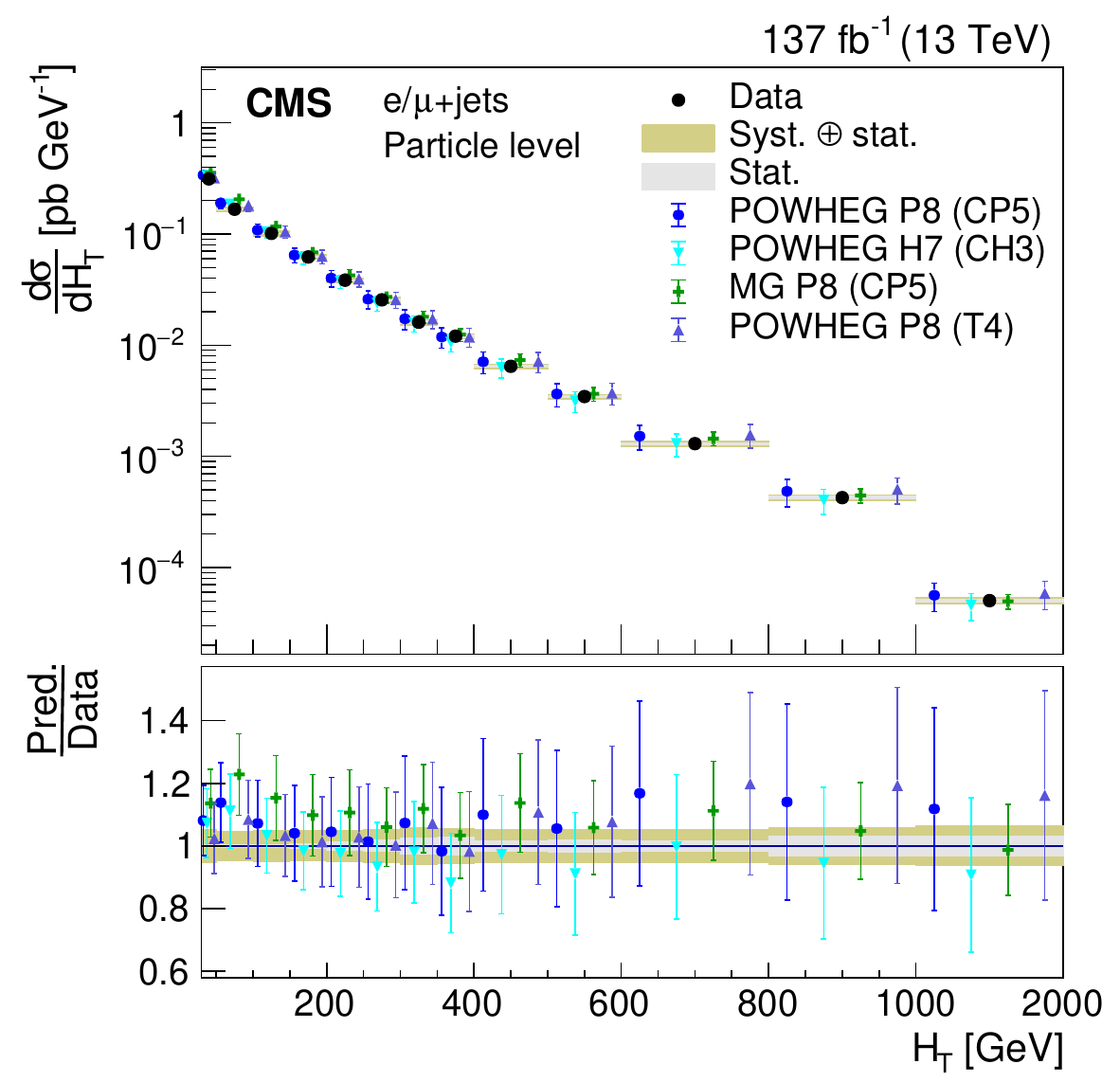}
 \includegraphics[width=0.42\textwidth]{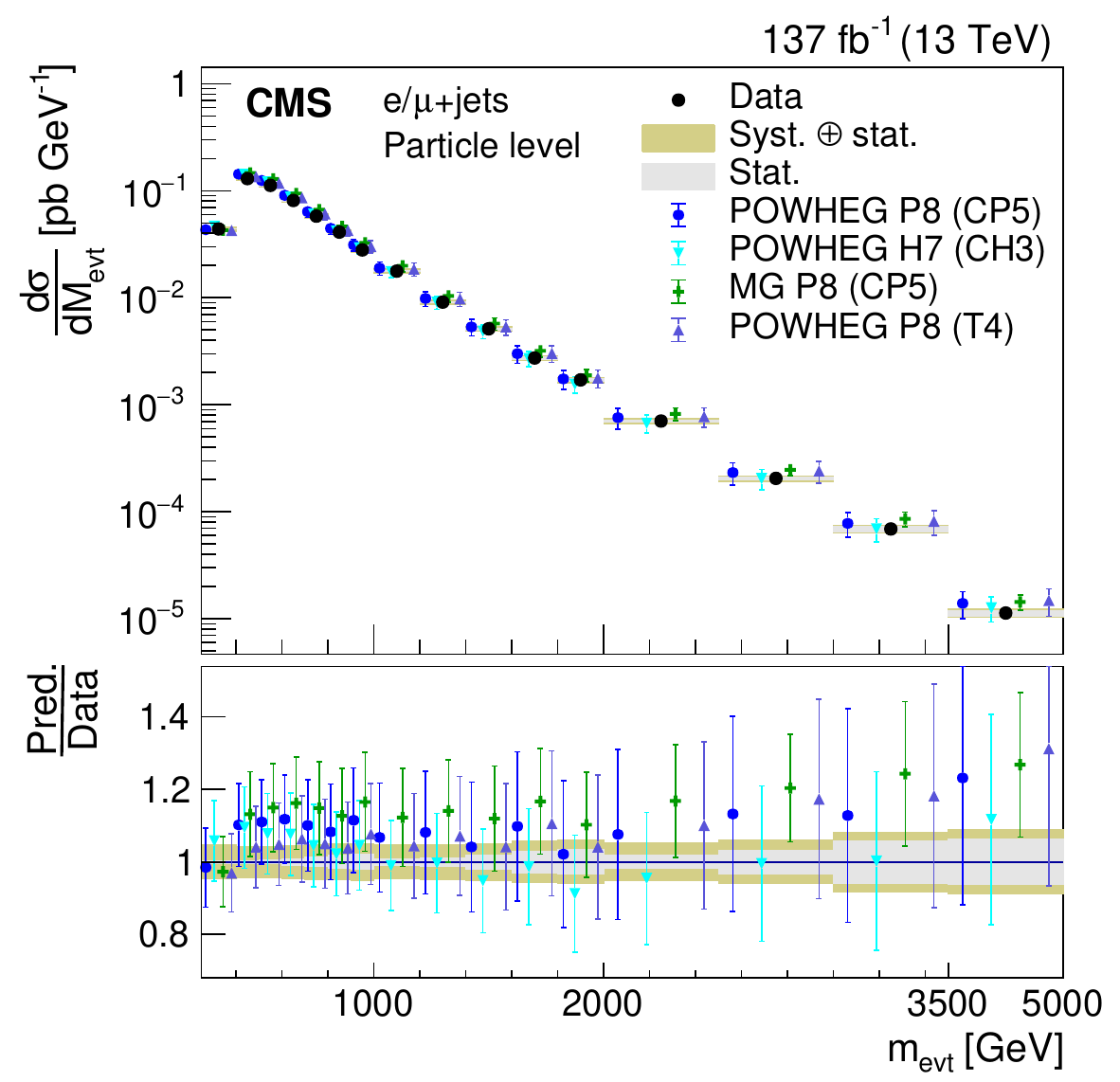}
 \caption{Differential cross sections at the particle level as a function of \ptl, jet multiplicity, \HT, and \mevt. \XSECCAPPS}
 \label{fig:RESPS13}
\end{figure*}

\begin{figure*}[tbp]
\centering
 \includegraphics[width=0.42\textwidth]{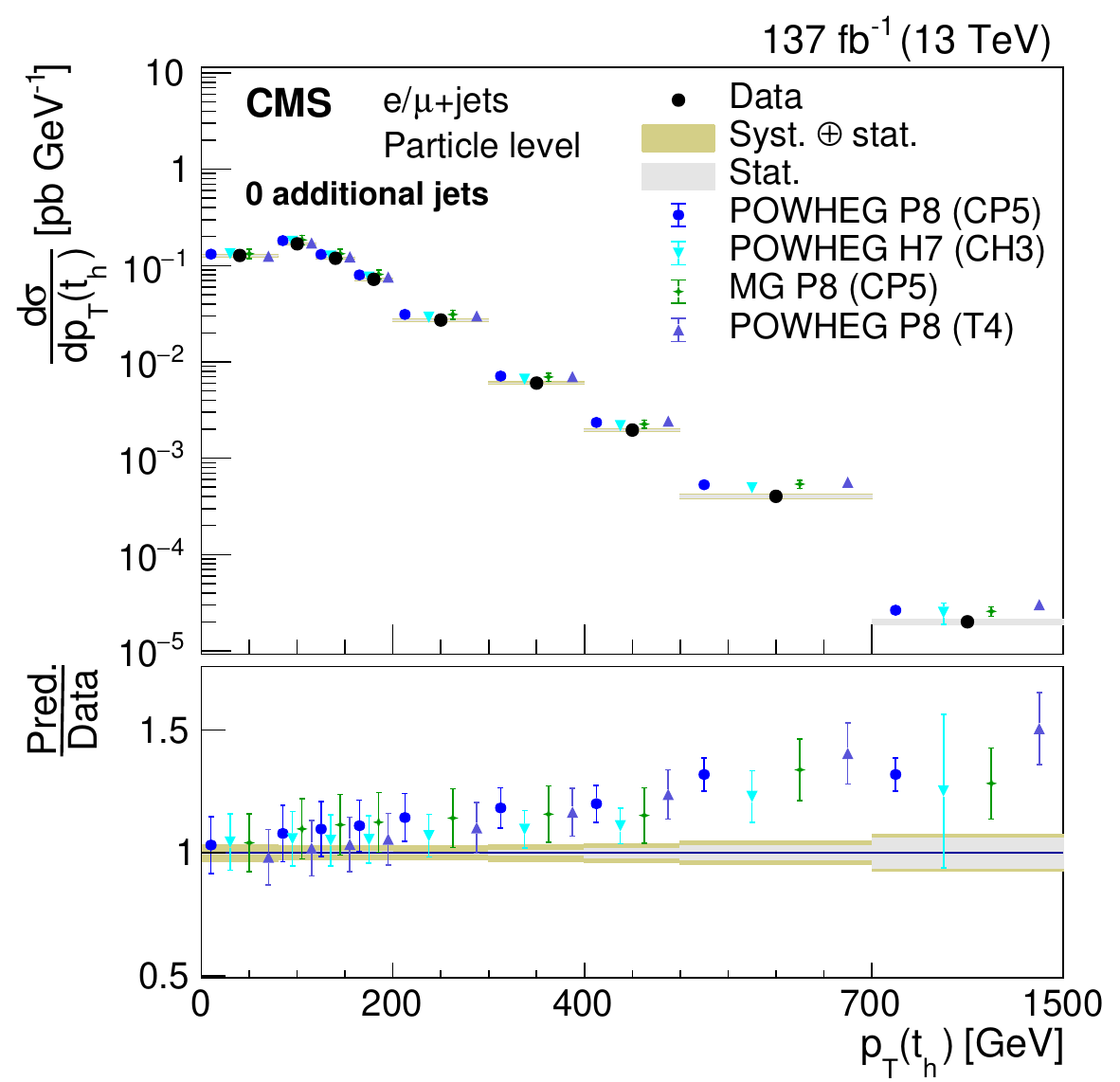}
 \includegraphics[width=0.42\textwidth]{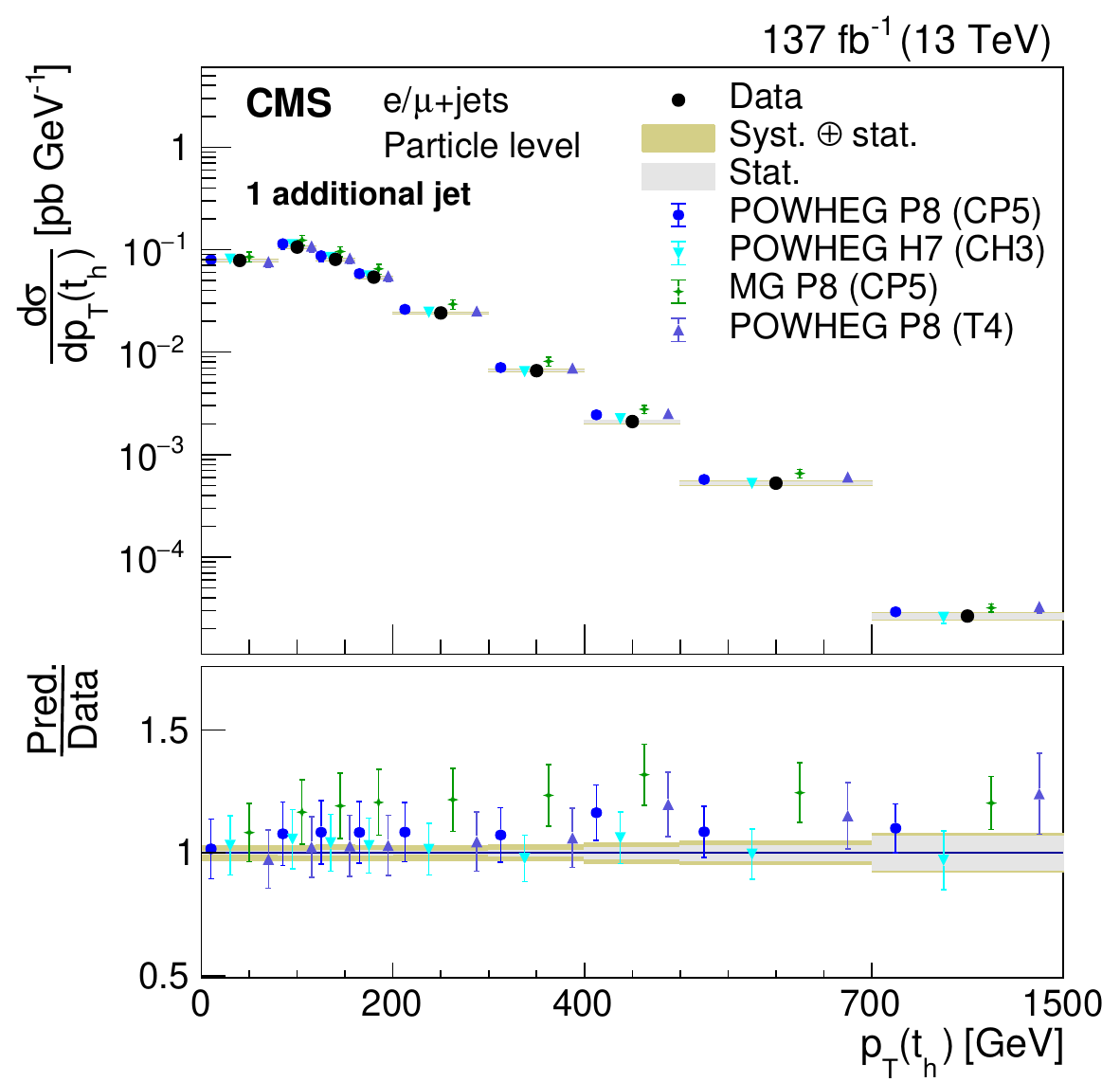}\\
 \includegraphics[width=0.42\textwidth]{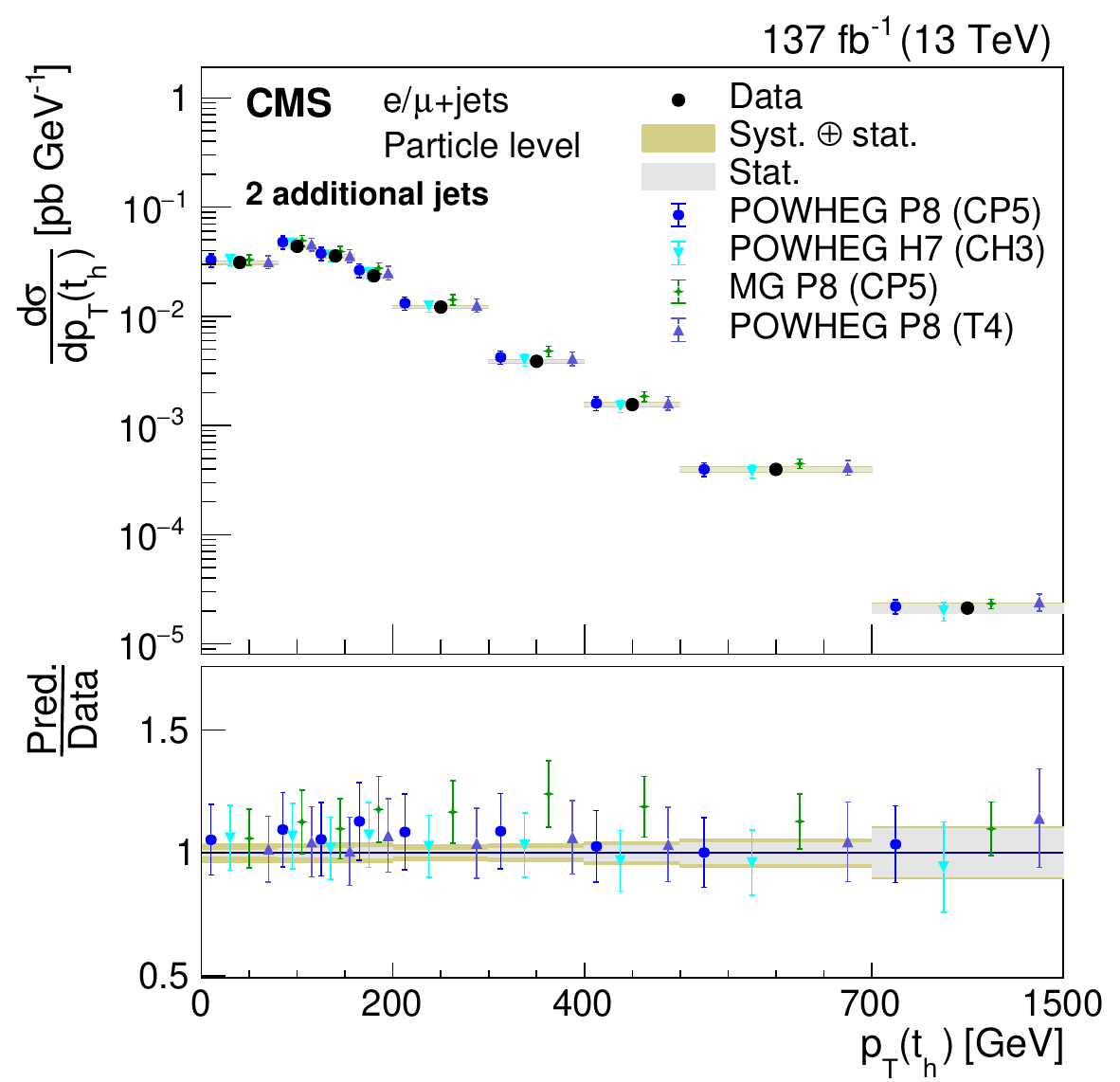}
 \includegraphics[width=0.42\textwidth]{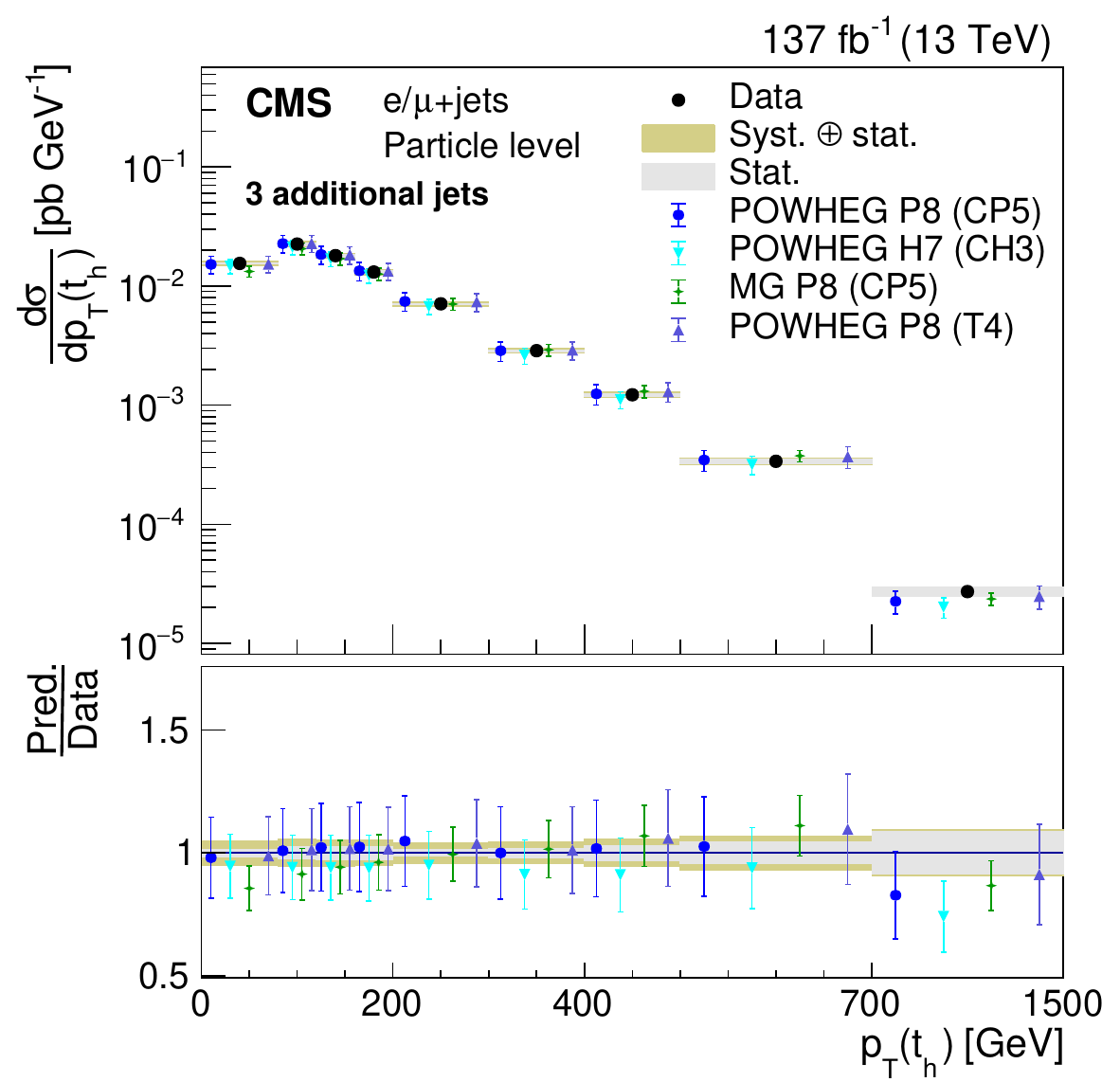}
 \caption{Differential cross section at the particle level as a function of \thadpt in bins of jet multiplicity. \XSECCAPPS}
 \label{fig:RESPS14}
\end{figure*}

\begin{figure*}[tbp]
\centering
 \includegraphics[width=0.42\textwidth]{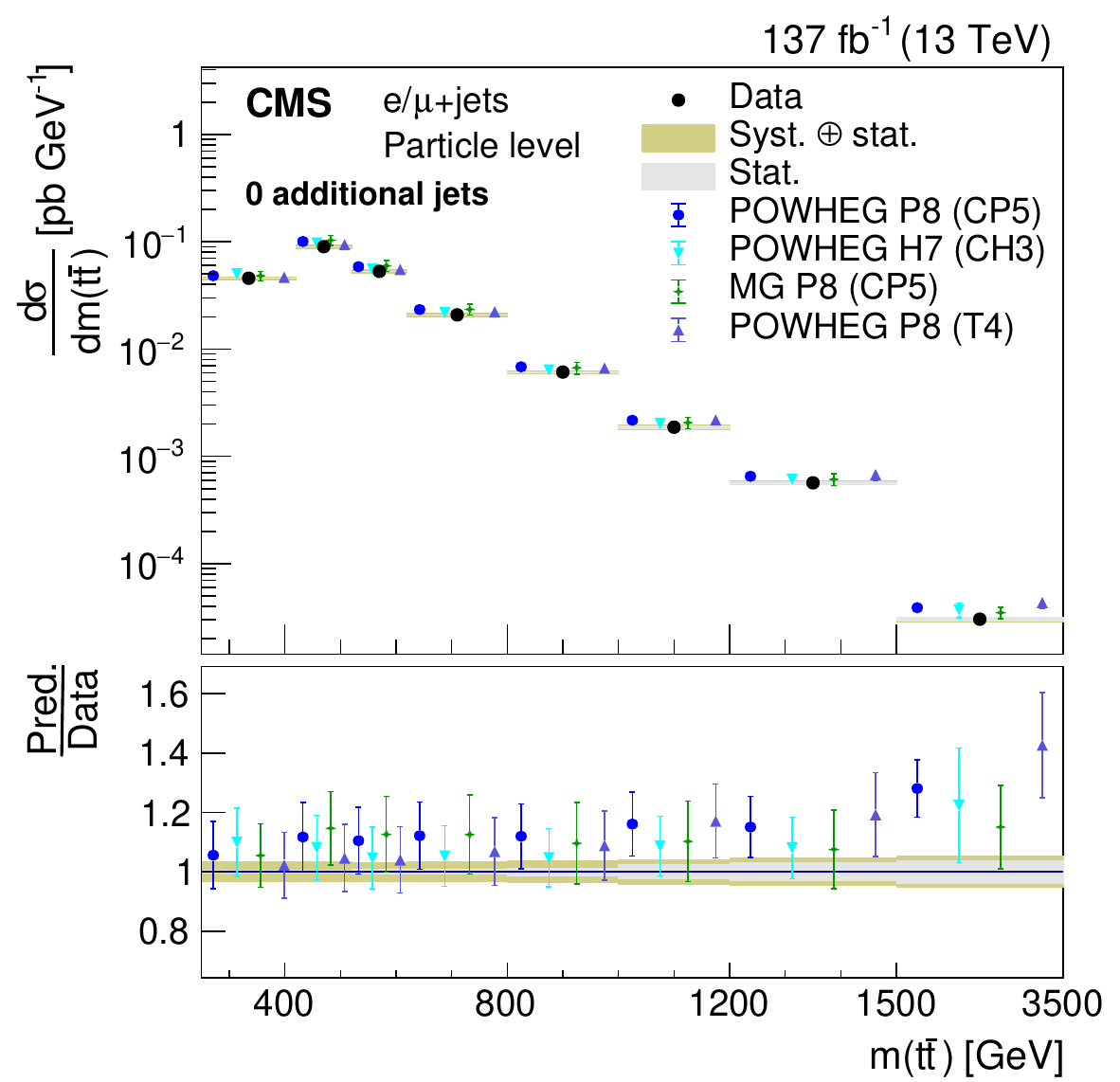}
 \includegraphics[width=0.42\textwidth]{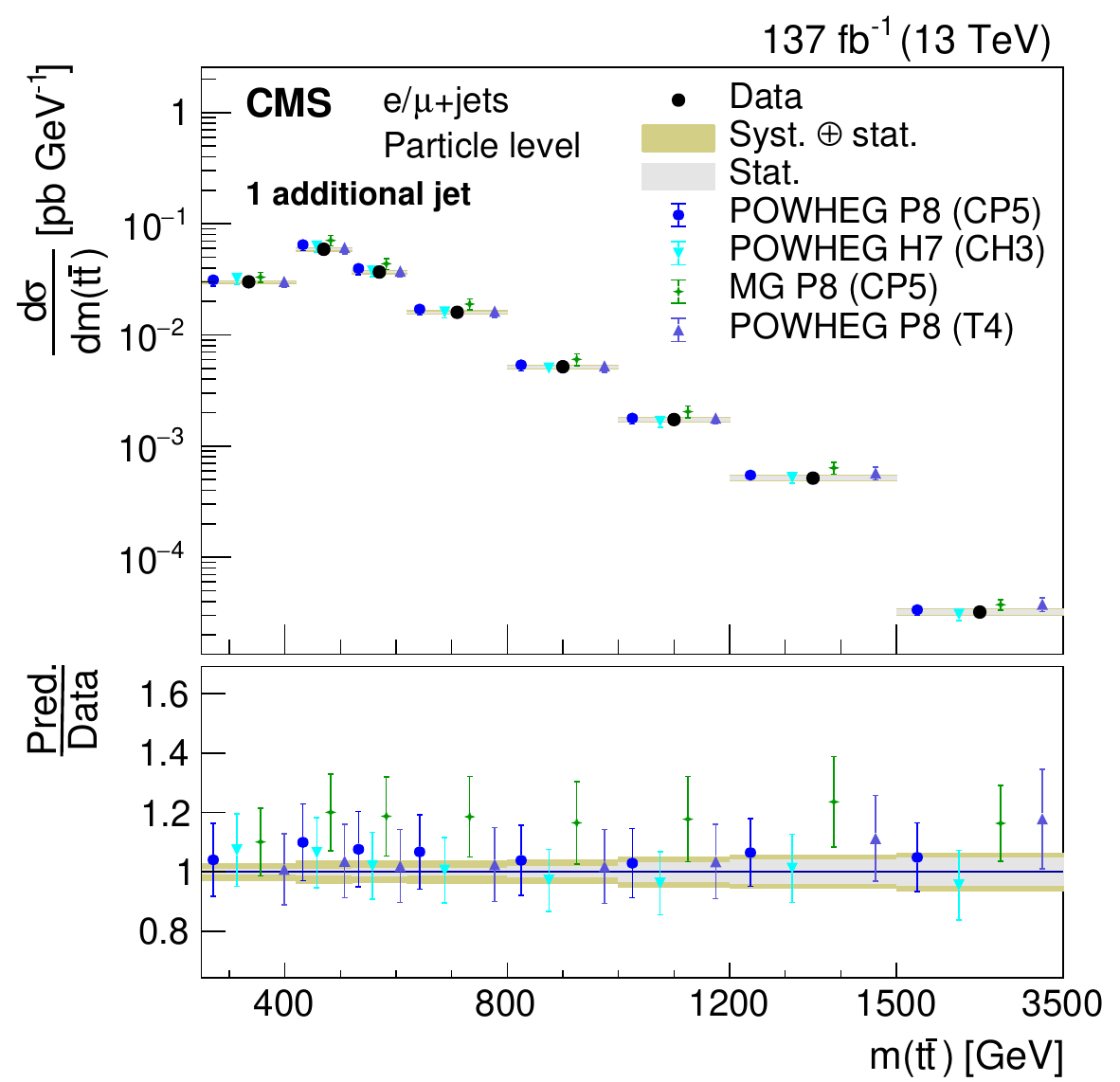}\\
 \includegraphics[width=0.42\textwidth]{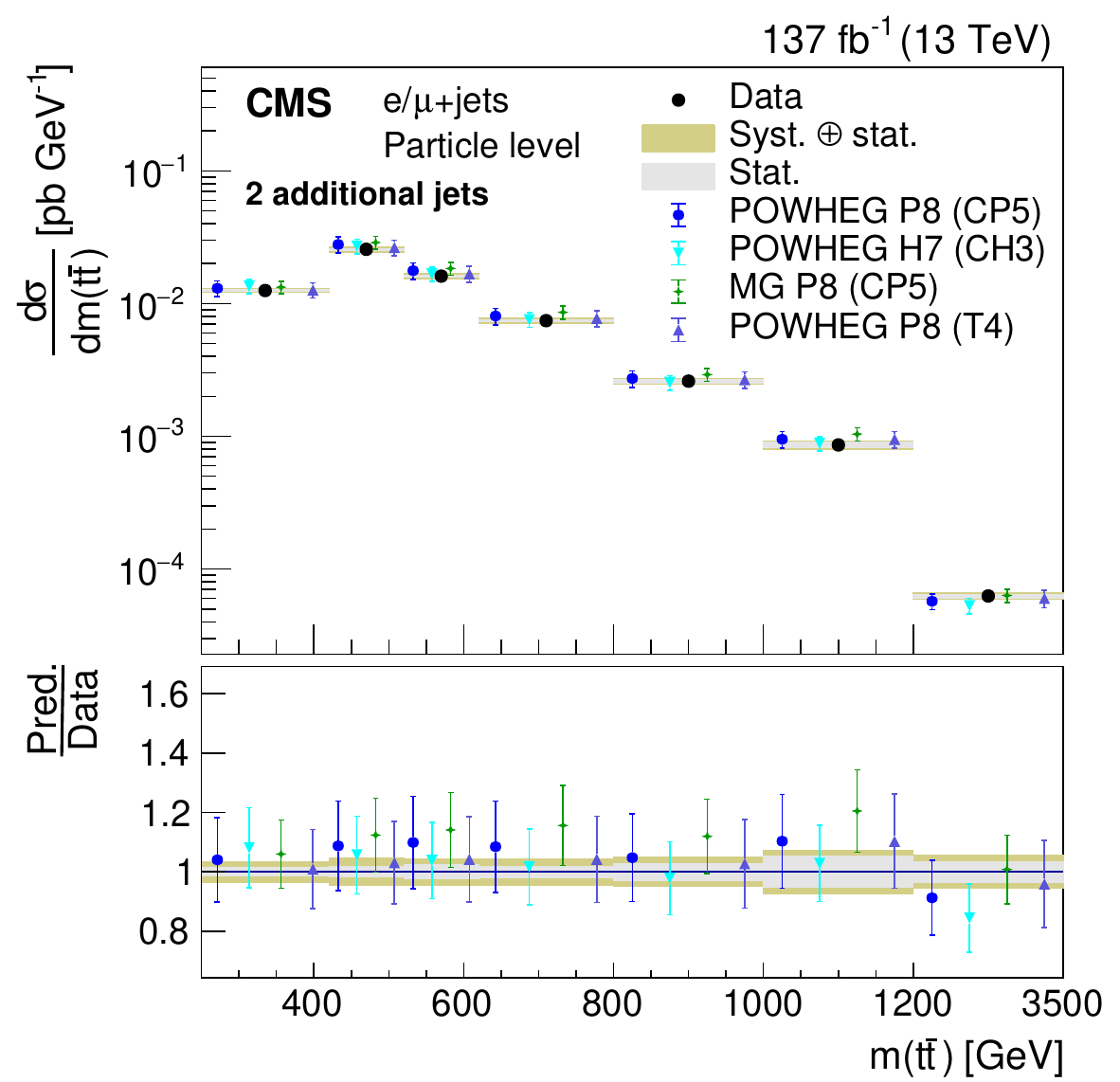}
 \includegraphics[width=0.42\textwidth]{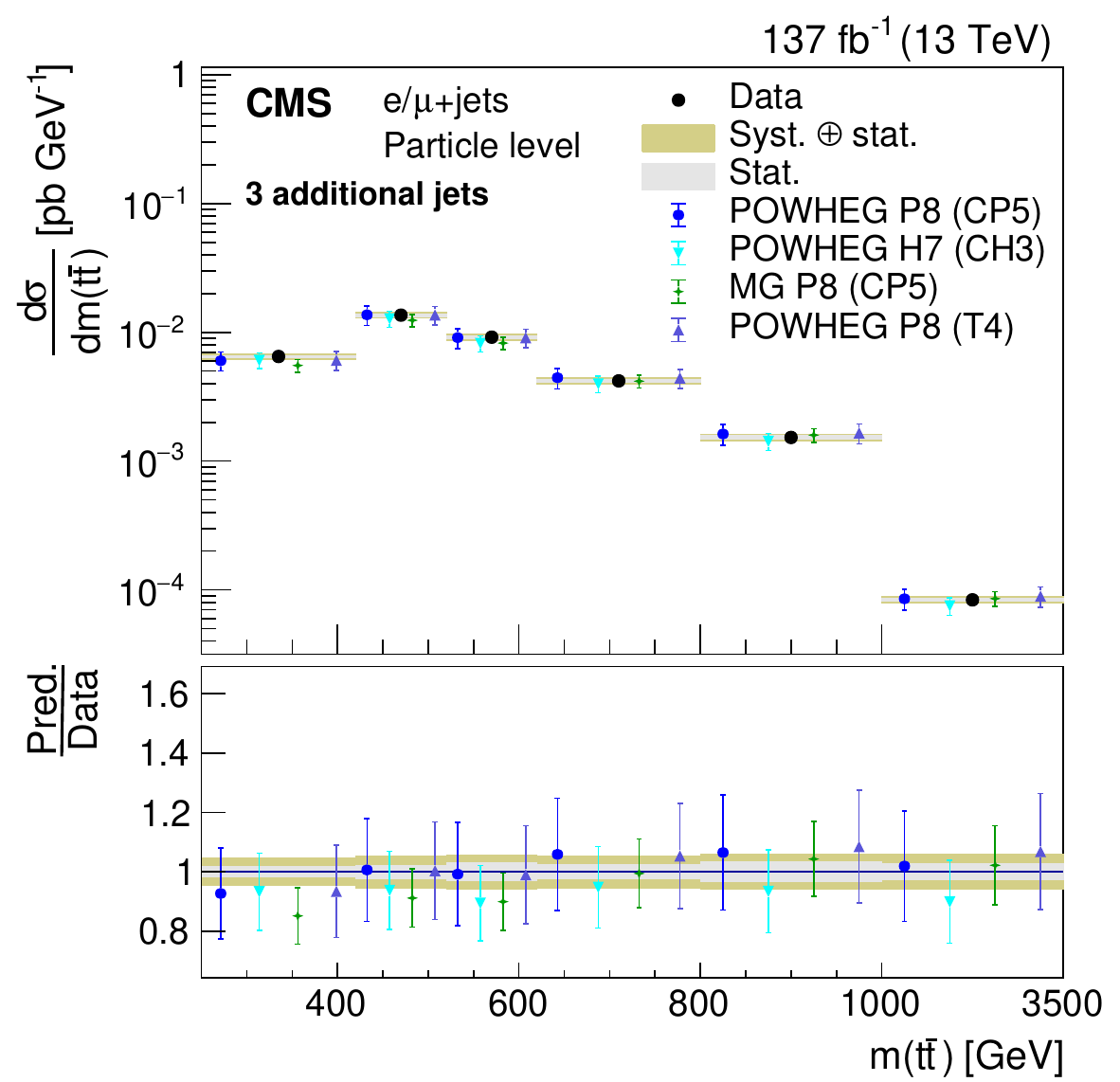}
 \caption{Differential cross section at the particle level as a function of \ttm in bins of jet multiplicity. \XSECCAPPS}
 \label{fig:RESPS15}
\end{figure*}

\begin{figure*}[tbp]
\centering
 \includegraphics[width=0.42\textwidth]{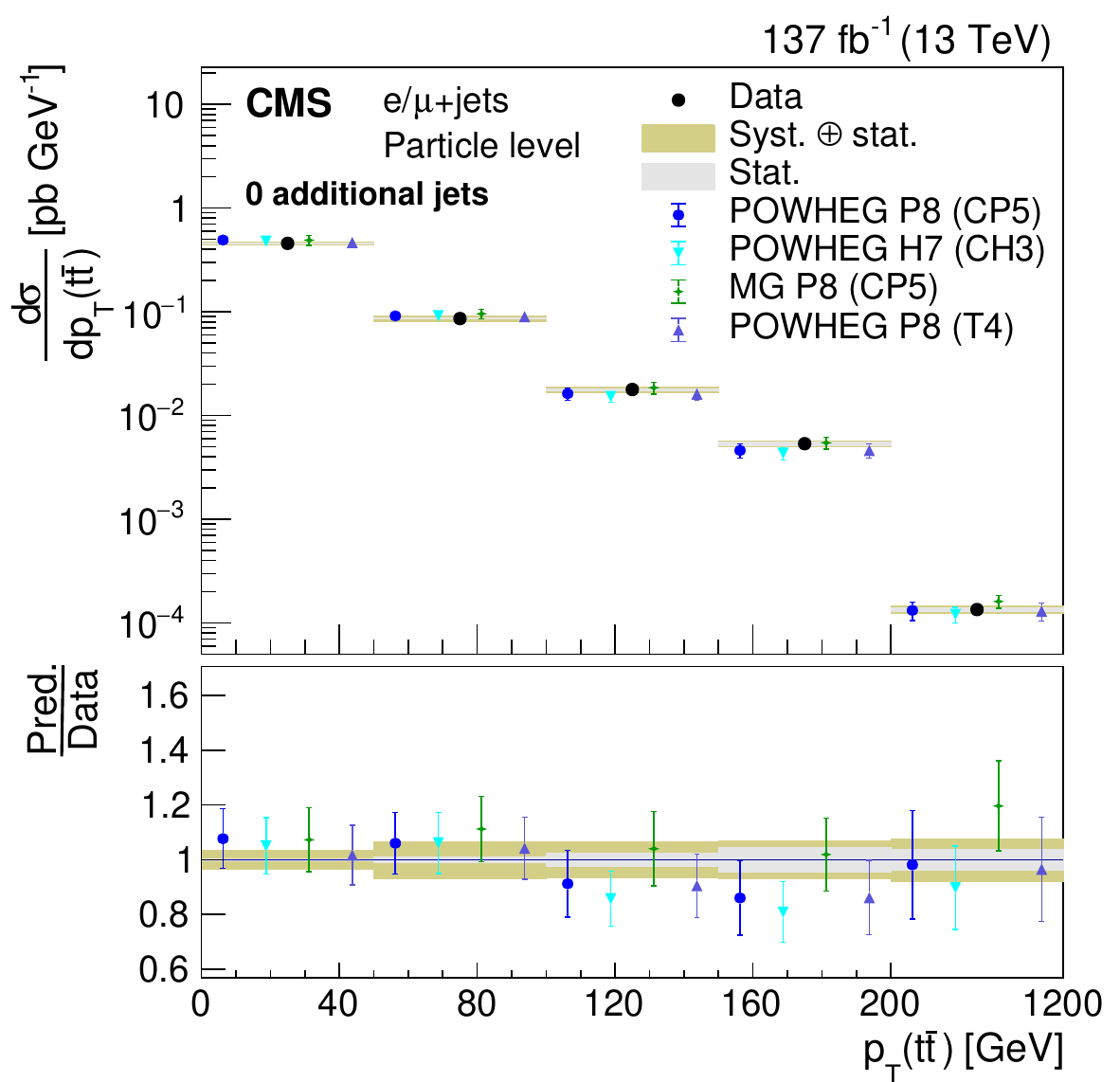}
 \includegraphics[width=0.42\textwidth]{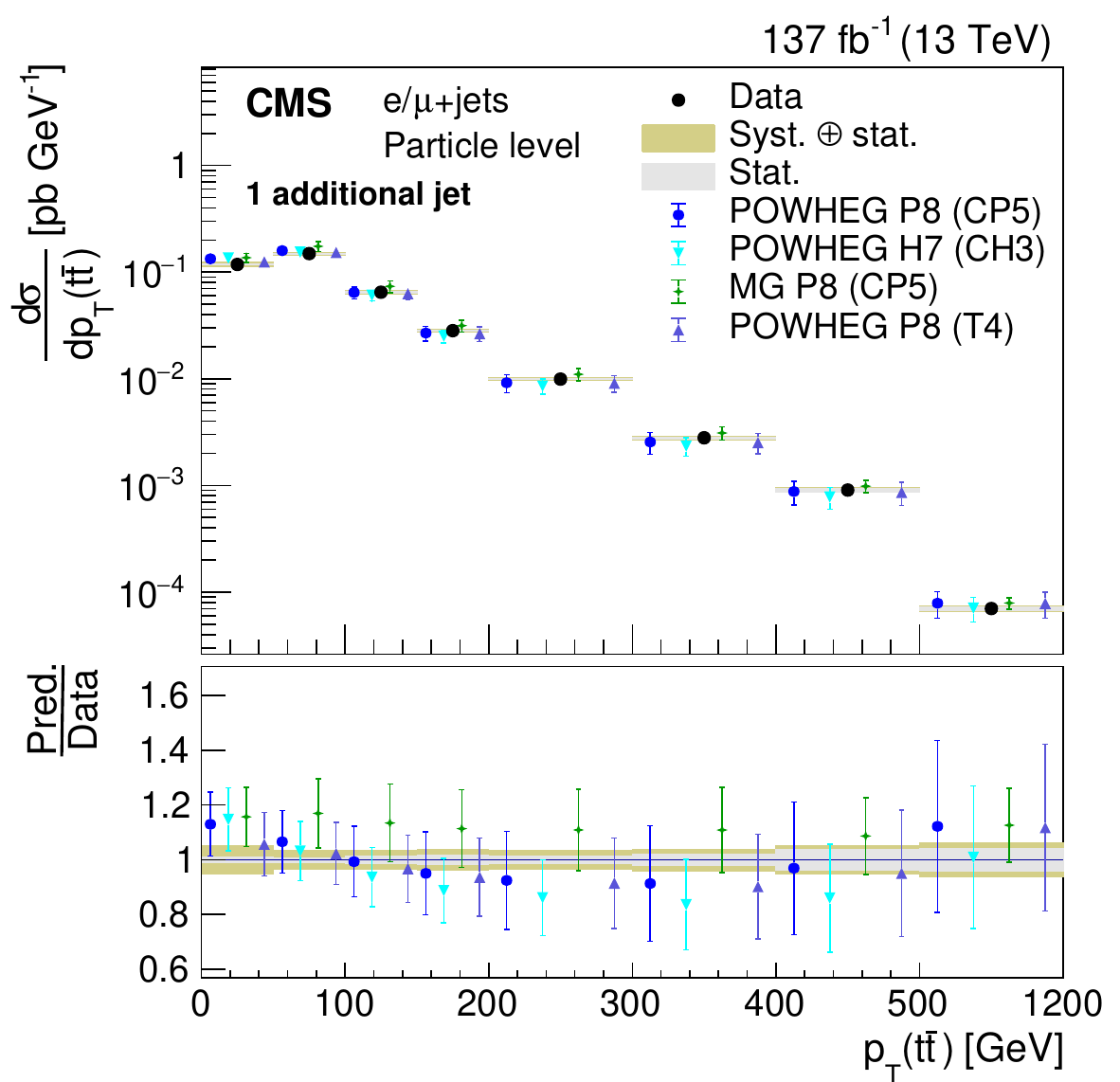}\\
 \includegraphics[width=0.42\textwidth]{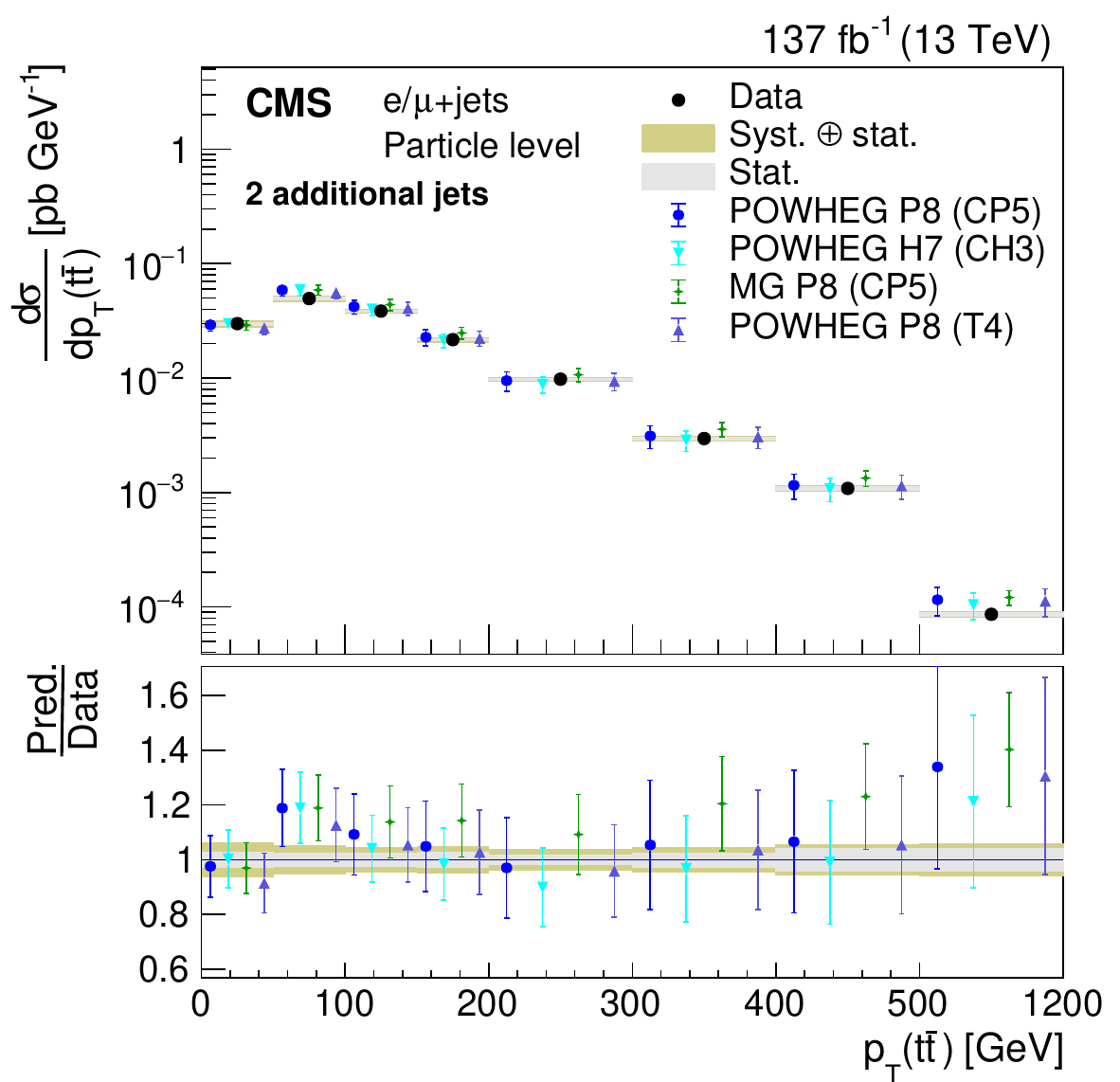}
 \includegraphics[width=0.42\textwidth]{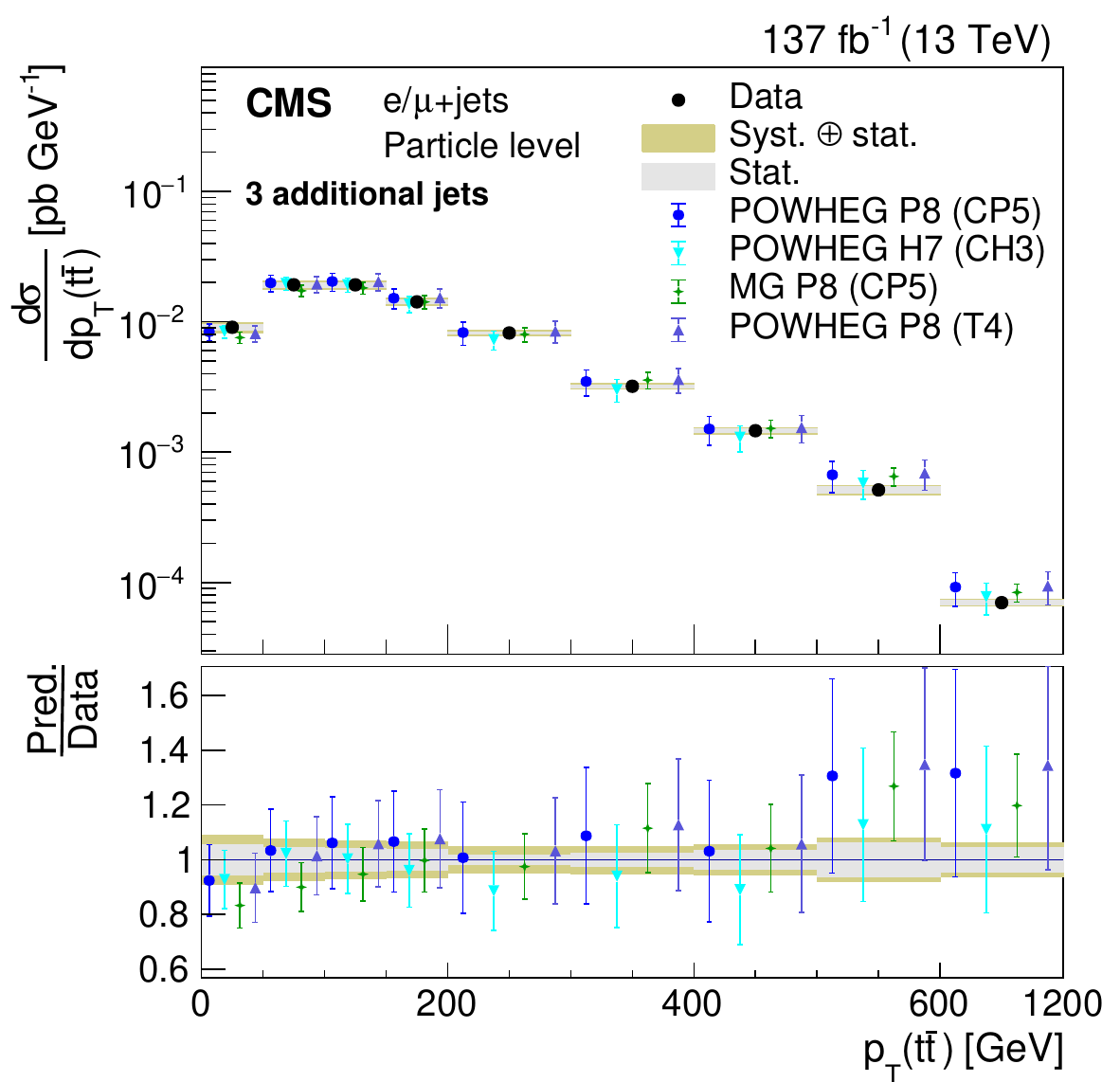}
 \caption{Differential cross section at the particle level as a function of \ttpt in bins of jet multiplicity. \XSECCAPPS}
 \label{fig:RESPS16}
\end{figure*}

In Figs.~\ref{fig:RESPS13}--\ref{fig:RESPS16} the differential cross sections that are only measured at the particle level are shown. Of special note in Fig.~\ref{fig:RESPS13} is the softer \ptl spectrum in data, compared to the prediction, a quantity that is not directly affected by the modeling of jets. Also, the number of additional jets with $\pt > 30$\GeV and $\abs{\eta} < 2.4$ is higher in data compared to most of the predictions. Only \POWHEG{}+\PYTHIA with the \CUET tune makes a correct prediction of the jet multiplicity. The distributions of \HT and \mevt are well described. Figure~\ref{fig:RESPS14} shows that the measured \thadpt spectrum is only softer than the predictions if there is no or, to a smaller extent, one additional jet. For the higher multiplicities the spectrum is better described. Good agreement between data and predictions is observed for the differential cross sections as a function of \ttm and \ttpt measured for different numbers of additional jets, as given in Figs.~\ref{fig:RESPS15} and \ref{fig:RESPS16}, respectively. In Appendix~\ref{NORMXSEC}, all the measured differential cross sections are presented normalized to unity.

\clearpage

\section{Summary}
\label{SUM}

Measurements of differential and double-differential top quark pair (\ttbar) production cross sections have been presented as a function of many kinematic properties of the top quarks and the \ttbar system at the parton and particle levels, where the latter reduces extrapolations based on theoretical assumptions. In addition, the number of additional jets and kinematic variables in bins of jet multiplicities have been measured at the particle level. The data correspond to an integrated luminosity of 137\fbinv recorded by the CMS experiment at the LHC in proton-proton collisions at $\sqrt{s} = 13$\TeV. The \ttbar cross sections are measured in the \lpj channels with a single electron or muon and jets in the final state. For the first time the full spectra of differential cross sections are determined using a combination of resolved and boosted \ttbar topologies, in which the \ttbar decay products can be either identified as separated jets and isolated leptons or as collimated and overlapping jets, respectively. The combination of multiple reconstruction categories provides constraints on the systematic uncertainties and results in a significantly improved precision with respect to previous measurements. For a top quark transverse momentum $\pt < 500$\GeV the uncertainty is reduced by about 50\% compared with the previous CMS measurement~\cite{TOP-17-002}. The dominant sources of systematic uncertainties are the jet energy scale, integrated luminosity, and \ttbar modeling.

Most differential distributions are found to be compatible with the standard model predictions of the event generators \POWHEG{}+\PYTHIA, \POWHEG{}+\HERWIG, and \AMCATNLO{}+\PYTHIA. In addition, the parton-level cross sections are compared to the next-to-next-to-leading-order quantum chromodynamics calculations obtained with \MATRIX that come with a significantly reduced theoretical uncertainty. A softer top quark \pt spectrum is observed compared to most of the next-to-leading-order predictions. Deviations between the predictions and data are observed when the top quark \pt is measured in bins of the \ttbar invariant mass and \pt. The \POWHEG{}+\HERWIG and \AMCATNLO{}+\PYTHIA simulations do not give a good description of the observed jet multiplicities and related observables such as the scalar \pt sum of additional jets. The total \ttbar production cross section is measured to be
\begin{equation}
\sigma_\ttbar = 791\pm 25\unit{pb}.
\end{equation}
When breaking down the uncertainty into different sources, we find
\begin{equation}
\sigma_\ttbar = 791\pm1\stat\pm21\syst\pm 14\lum\unit{pb},
\end{equation}
where the last uncertainty comes from that in the integrated luminosity. The measured value of $\sigma_\ttbar$ is in good agreement with the standard model expectation.

\begin{acknowledgments}

\hyphenation{Bundes-ministerium Forschungs-gemeinschaft Forschungs-zentren Rachada-pisek} We congratulate our colleagues in the CERN accelerator departments for the excellent performance of the LHC and thank the technical and administrative staffs at CERN and at other CMS institutes for their contributions to the success of the CMS effort. In addition, we gratefully acknowledge the computing centers and personnel of the Worldwide LHC Computing Grid and other centers for delivering so effectively the computing infrastructure essential to our analyses. Finally, we acknowledge the enduring support for the construction and operation of the LHC, the CMS detector, and the supporting computing infrastructure provided by the following funding agencies: the Austrian Federal Ministry of Education, Science and Research and the Austrian Science Fund; the Belgian Fonds de la Recherche Scientifique, and Fonds voor Wetenschappelijk Onderzoek; the Brazilian Funding Agencies (CNPq, CAPES, FAPERJ, FAPERGS, and FAPESP); the Bulgarian Ministry of Education and Science; CERN; the Chinese Academy of Sciences, Ministry of Science and Technology, and National Natural Science Foundation of China; the Ministerio de Ciencia Tecnolog\'ia e Innovaci\'on (MINCIENCIAS), Colombia; the Croatian Ministry of Science, Education and Sport, and the Croatian Science Foundation; the Research and Innovation Foundation, Cyprus; the Secretariat for Higher Education, Science, Technology and Innovation, Ecuador; the Ministry of Education and Research, Estonian Research Council via PRG780, PRG803 and PRG445 and European Regional Development Fund, Estonia; the Academy of Finland, Finnish Ministry of Education and Culture, and Helsinki Institute of Physics; the Institut National de Physique Nucl\'eaire et de Physique des Particules~/~CNRS, and Commissariat \`a l'\'Energie Atomique et aux \'Energies Alternatives~/~CEA, France; the Bundesministerium f\"ur Bildung und Forschung, the Deutsche Forschungsgemeinschaft (DFG), under Germany's Excellence Strategy -- EXC 2121 ``Quantum Universe" -- 390833306, and under project number 400140256 - GRK2497, and Helmholtz-Gemeinschaft Deutscher Forschungszentren, Germany; the General Secretariat for Research and Innovation, Greece; the National Research, Development and Innovation Fund, Hungary; the Department of Atomic Energy and the Department of Science and Technology, India; the Institute for Studies in Theoretical Physics and Mathematics, Iran; the Science Foundation, Ireland; the Istituto Nazionale di Fisica Nucleare, Italy; the Ministry of Science, ICT and Future Planning, and National Research Foundation (NRF), Republic of Korea; the Ministry of Education and Science of the Republic of Latvia; the Lithuanian Academy of Sciences; the Ministry of Education, and University of Malaya (Malaysia); the Ministry of Science of Montenegro; the Mexican Funding Agencies (BUAP, CINVESTAV, CONACYT, LNS, SEP, and UASLP-FAI); the Ministry of Business, Innovation and Employment, New Zealand; the Pakistan Atomic Energy Commission; the Ministry of Science and Higher Education and the National Science Center, Poland; the Funda\c{c}\~ao para a Ci\^encia e a Tecnologia, Portugal; JINR, Dubna; the Ministry of Education and Science of the Russian Federation, the Federal Agency of Atomic Energy of the Russian Federation, Russian Academy of Sciences, the Russian Foundation for Basic Research, and the National Research Center ``Kurchatov Institute"; the Ministry of Education, Science and Technological Development of Serbia; the Secretar\'{\i}a de Estado de Investigaci\'on, Desarrollo e Innovaci\'on, Programa Consolider-Ingenio 2010, Plan Estatal de Investigaci\'on Cient\'{\i}fica y T\'ecnica y de Innovaci\'on 2017--2020, research project IDI-2018-000174 del Principado de Asturias, and Fondo Europeo de Desarrollo Regional, Spain; the Ministry of Science, Technology and Research, Sri Lanka; the Swiss Funding Agencies (ETH Board, ETH Zurich, PSI, SNF, UniZH, Canton Zurich, and SER); the Ministry of Science and Technology, Taipei; the Thailand Center of Excellence in Physics, the Institute for the Promotion of Teaching Science and Technology of Thailand, Special Task Force for Activating Research and the National Science and Technology Development Agency of Thailand; the Scientific and Technical Research Council of Turkey, and Turkish Atomic Energy Authority; the National Academy of Sciences of Ukraine; the Science and Technology Facilities Council, UK; the US Department of Energy, and the US National Science Foundation.

Individuals have received support from the Marie-Curie program and the European Research Council and Horizon 2020 Grant, contract Nos.\ 675440, 724704, 752730, 758316, 765710, 824093, and COST Action CA16108 (European Union) the Leventis Foundation; the Alfred P.\ Sloan Foundation; the Alexander von Humboldt Foundation; the Belgian Federal Science Policy Office; the Fonds pour la Formation \`a la Recherche dans l'Industrie et dans l'Agriculture (FRIA-Belgium); the Agentschap voor Innovatie door Wetenschap en Technologie (IWT-Belgium); the F.R.S.-FNRS and FWO (Belgium) under the ``Excellence of Science -- EOS" -- be.h project n.\ 30820817; the Beijing Municipal Science \& Technology Commission, No. Z191100007219010; the Ministry of Education, Youth and Sports (MEYS) of the Czech Republic; the Lend\"ulet (``Momentum") Program and the J\'anos Bolyai Research Scholarship of the Hungarian Academy of Sciences, the New National Excellence Program \'UNKP, the NKFIA research grants 123842, 123959, 124845, 124850, 125105, 128713, 128786, and 129058 (Hungary); the Council of Scientific and Industrial Research, India; the Latvian Council of Science; the National Science Center (Poland), contracts Opus 2014/15/B/ST2/03998 and 2015/19/B/ST2/02861; the National Priorities Research Program by Qatar National Research Fund; the Ministry of Science and Higher Education, project no. 0723-2020-0041 (Russia); the Programa de Excelencia Mar\'{i}a de Maeztu, and the Programa Severo Ochoa del Principado de Asturias; the Stavros Niarchos Foundation (Greece); the Rachadapisek Sompot Fund for Postdoctoral Fellowship, Chulalongkorn University, and the Chulalongkorn Academic into Its 2nd Century Project Advancement Project (Thailand); the Kavli Foundation; the Nvidia Corporation; the SuperMicro Corporation; the Welch Foundation, contract C-1845; and the Weston Havens Foundation (USA).
\end{acknowledgments}

\bibliography{auto_generated}

\clearpage

\numberwithin{table}{section}
\numberwithin{figure}{section}
\appendix

\section{Normalized cross sections at the parton and particle levels}
\label{NORMXSEC}

\begin{figure*}[tbp]
\centering
 \includegraphics[width=0.42\textwidth]{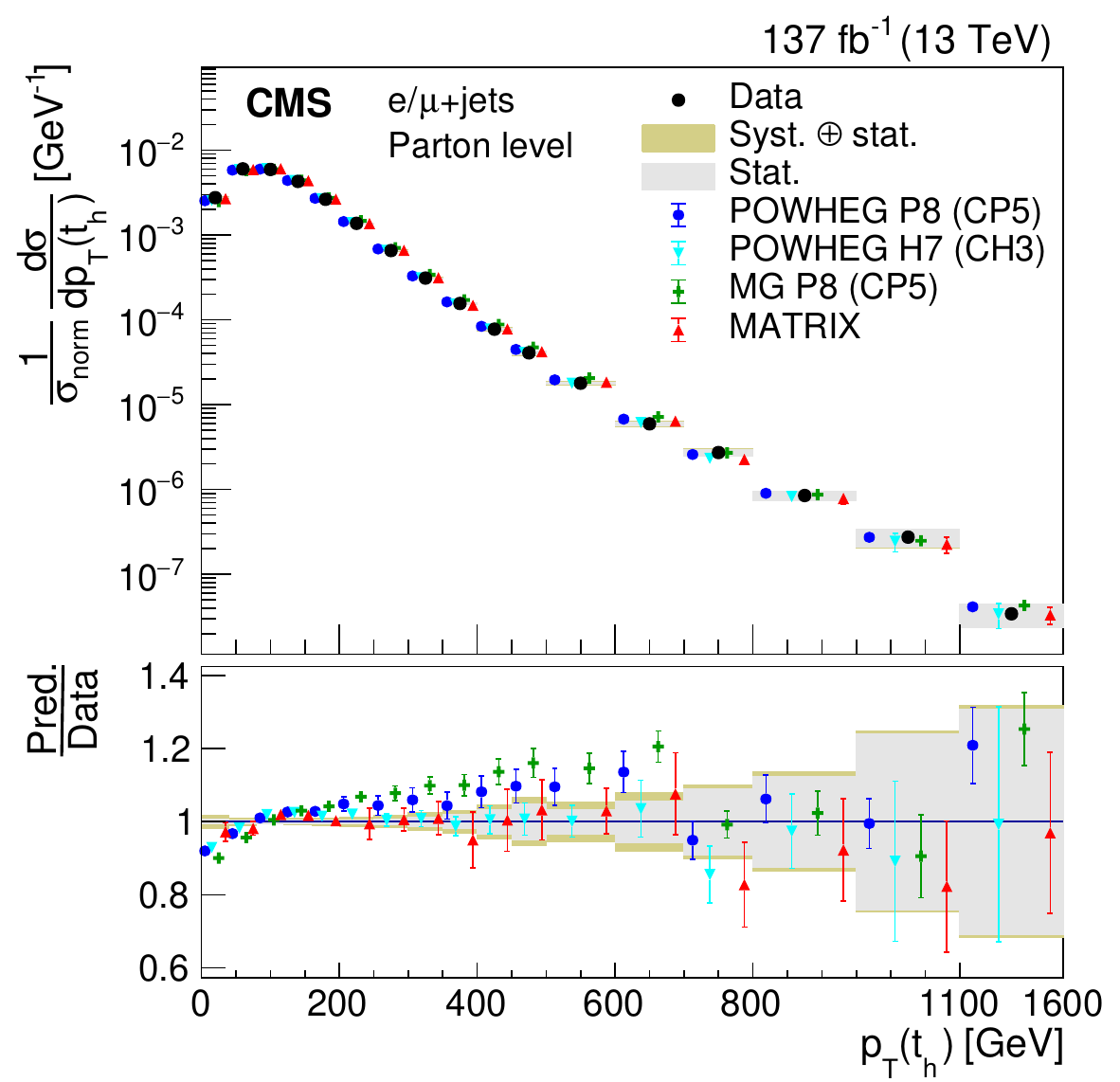}
 \includegraphics[width=0.42\textwidth]{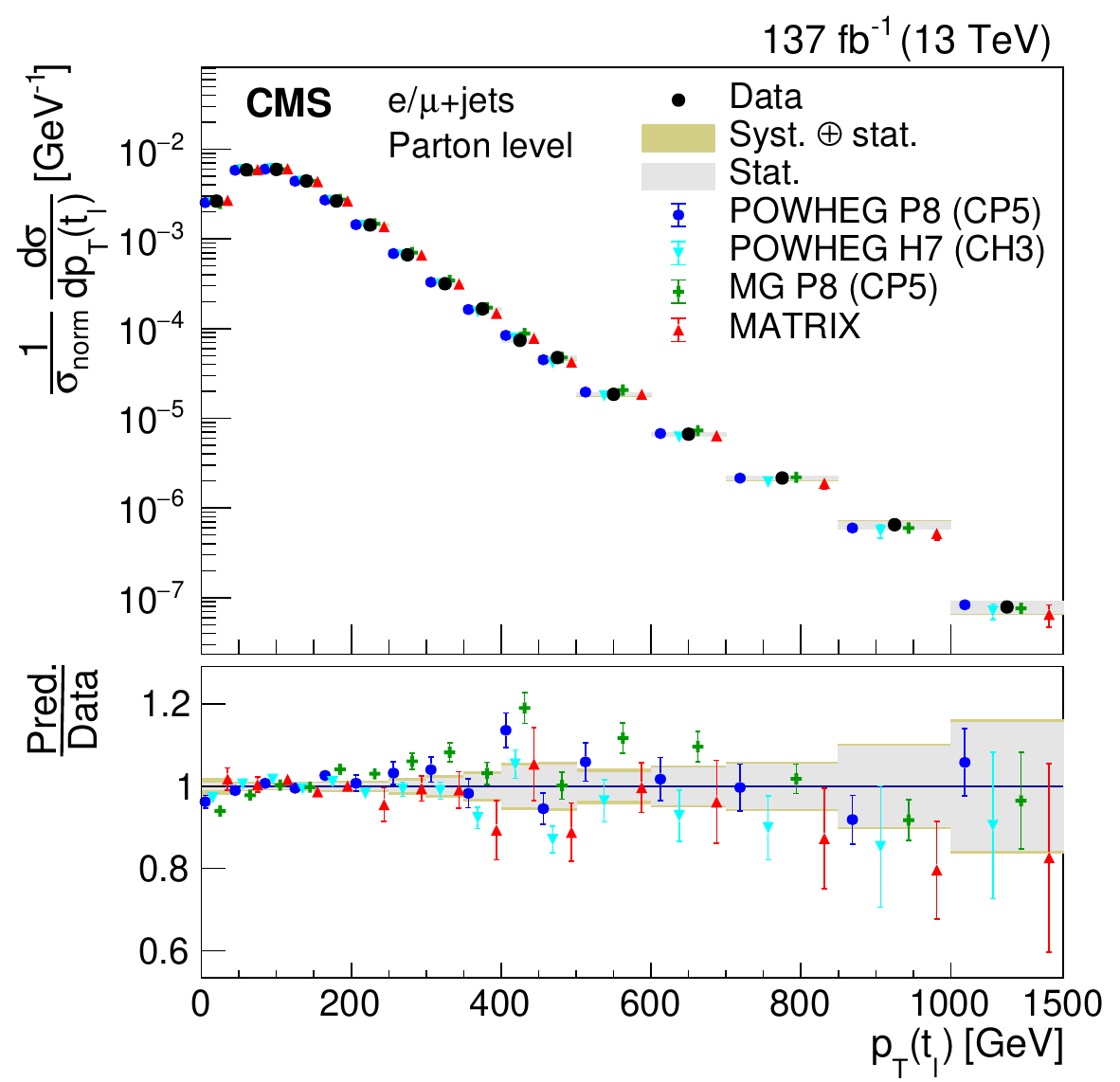}
 \includegraphics[width=0.42\textwidth]{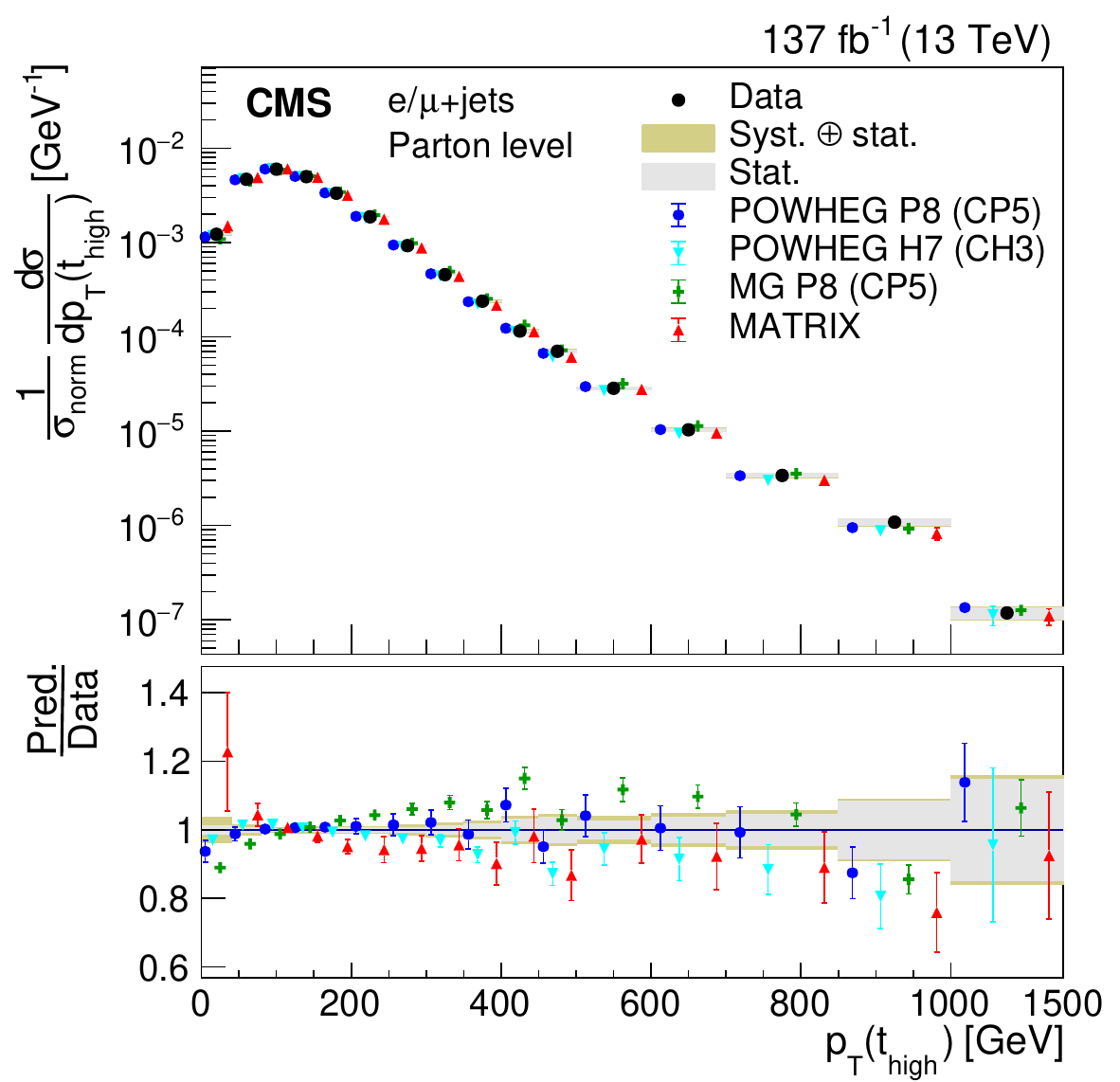}
 \includegraphics[width=0.42\textwidth]{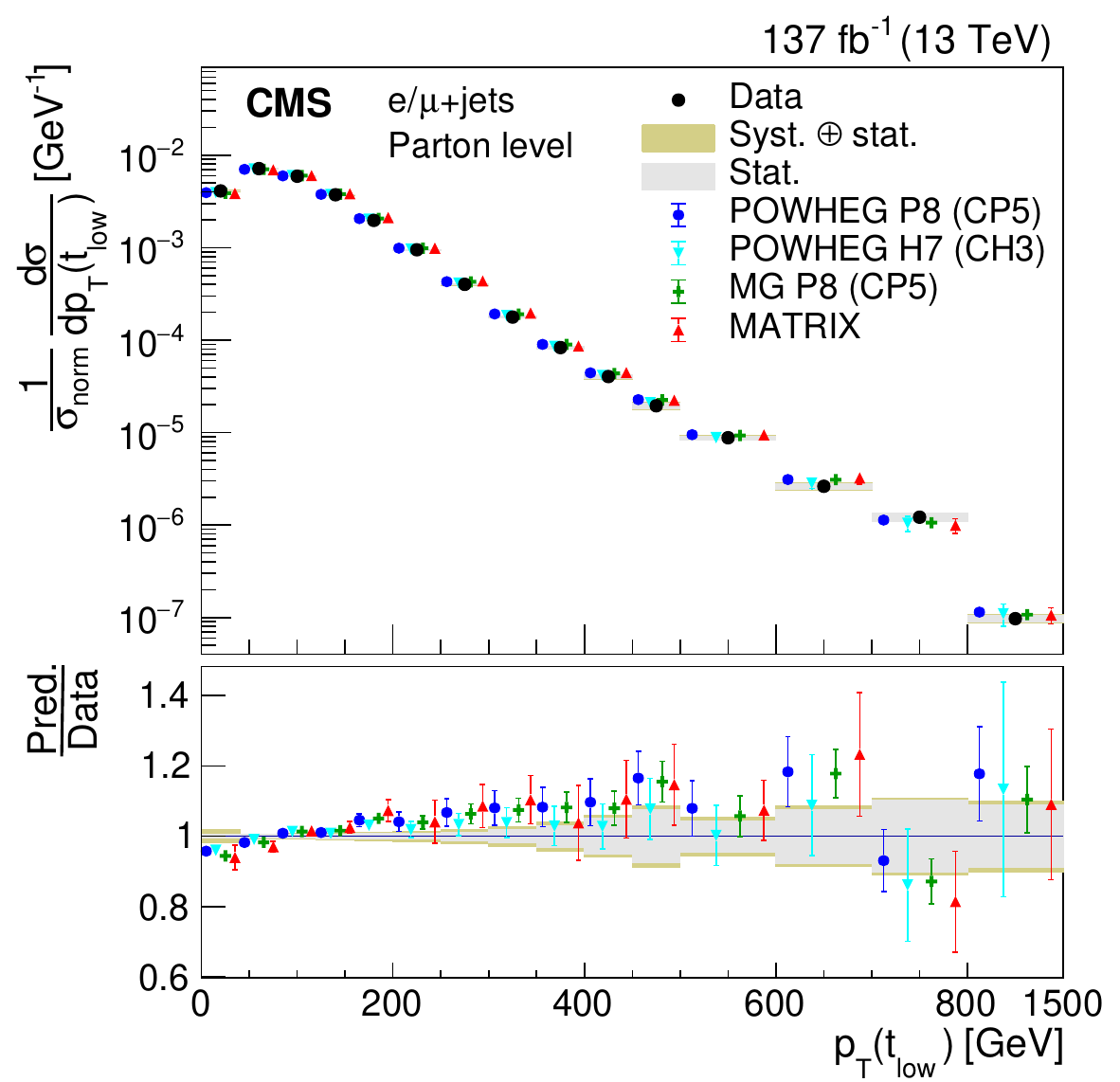}
 \includegraphics[width=0.42\textwidth]{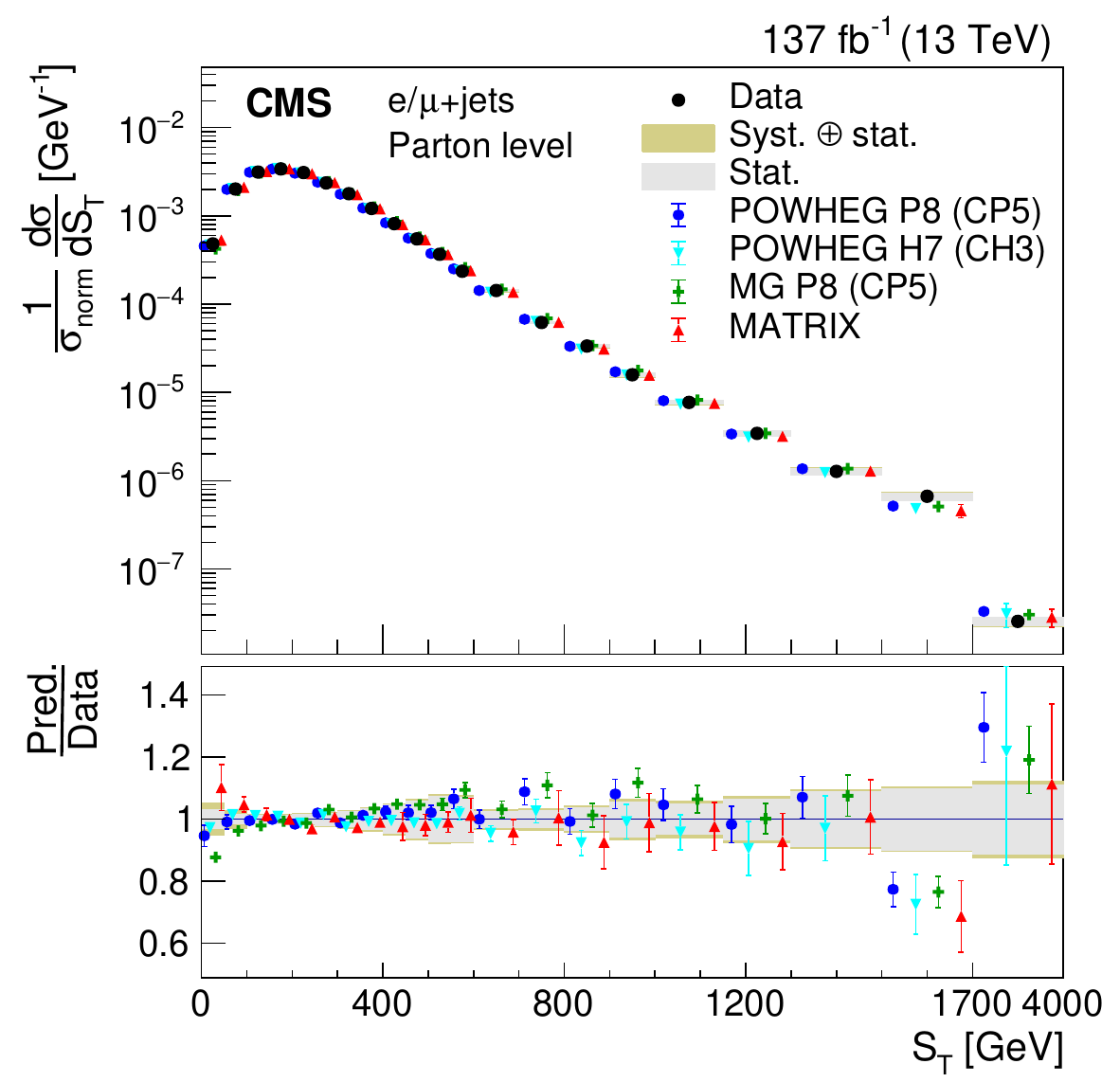}
 \caption{Normalized differential cross sections at the parton level as a function of \thadpt, \tleppt, \thardpt, \tsoftpt, and \st. \XSECCAPPA}
 \label{fig:RESNORM1}
\end{figure*}

\begin{figure*}[tbp]
\centering
 \includegraphics[width=0.42\textwidth]{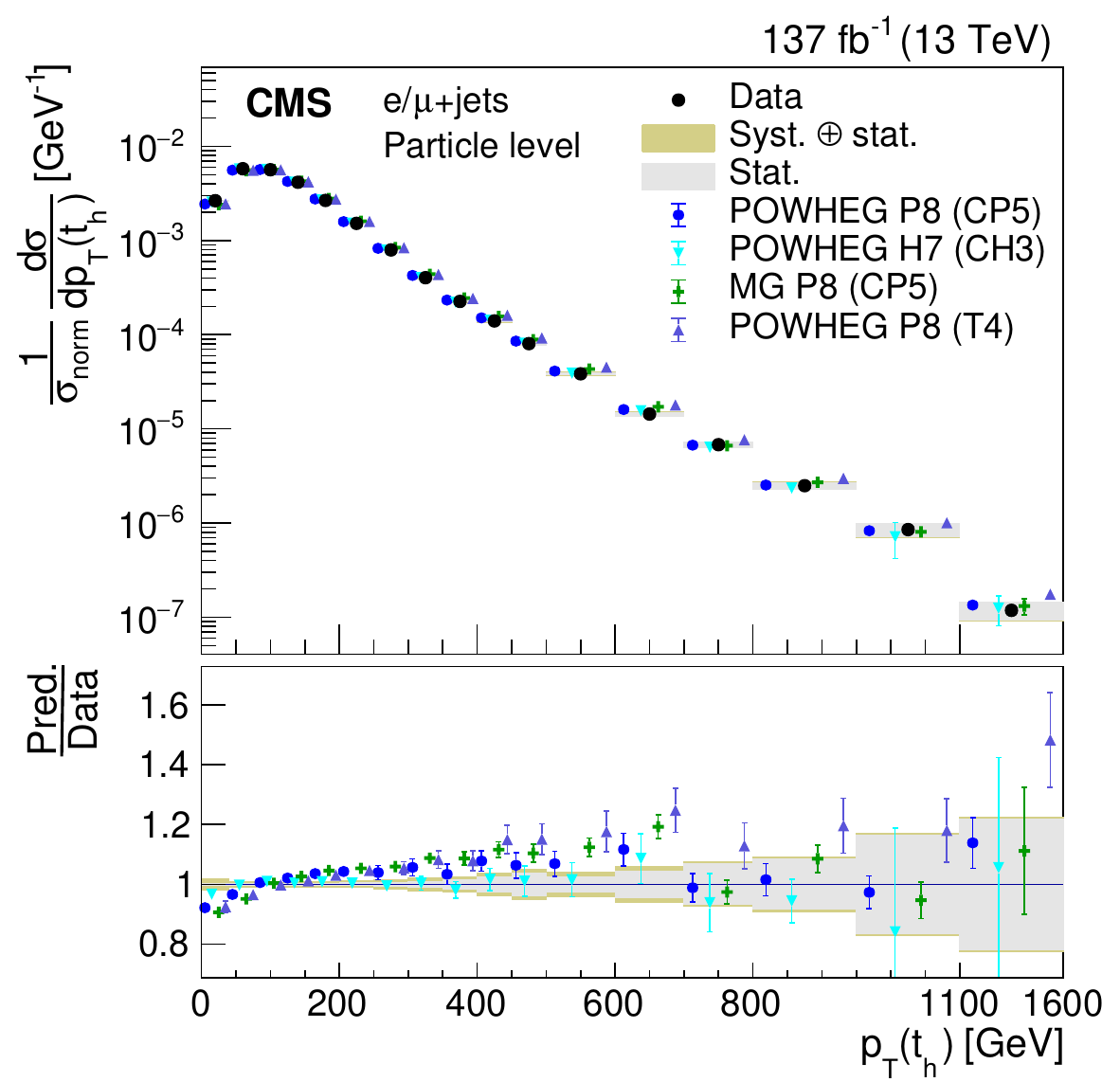}
 \includegraphics[width=0.42\textwidth]{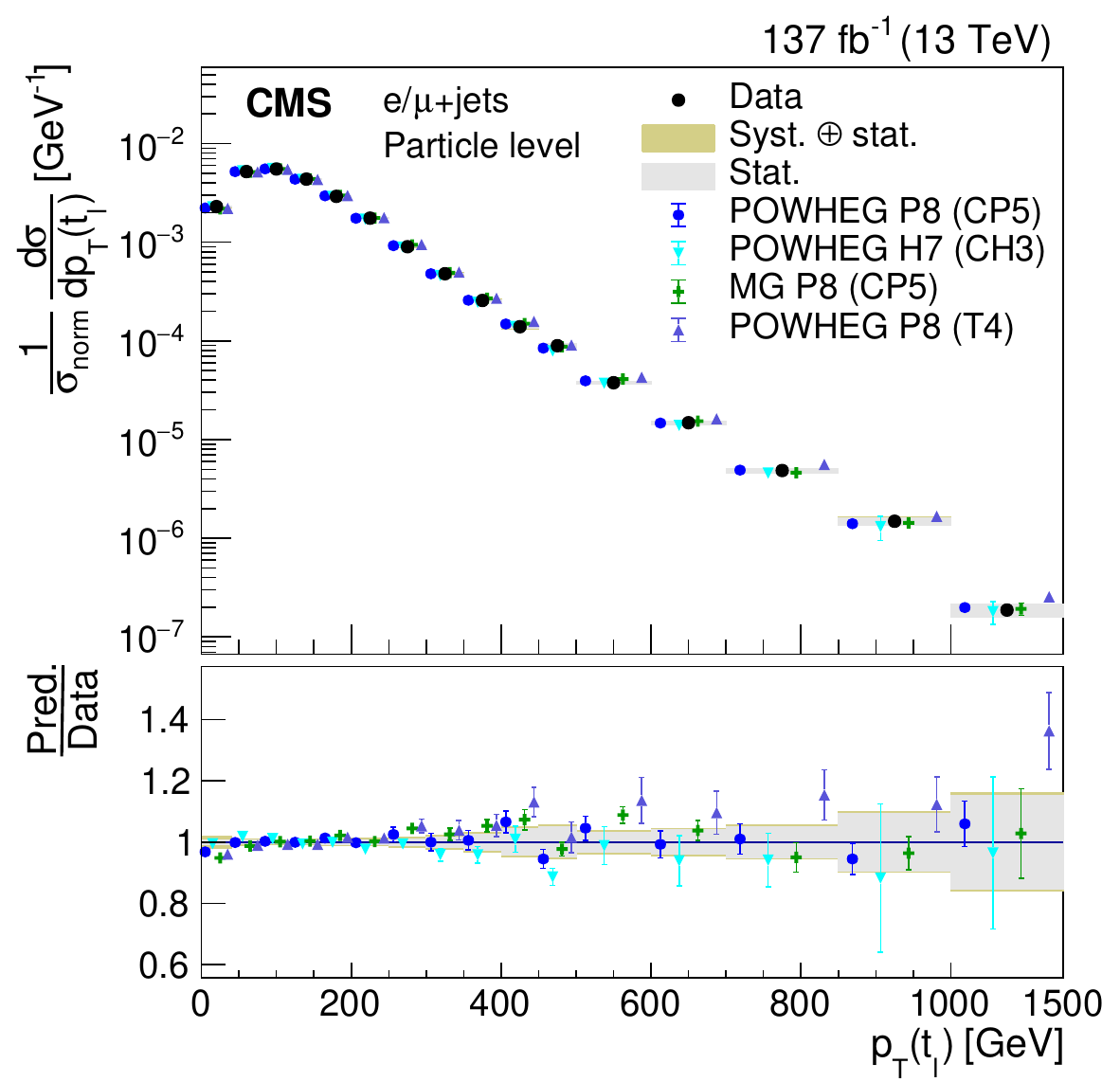}
 \includegraphics[width=0.42\textwidth]{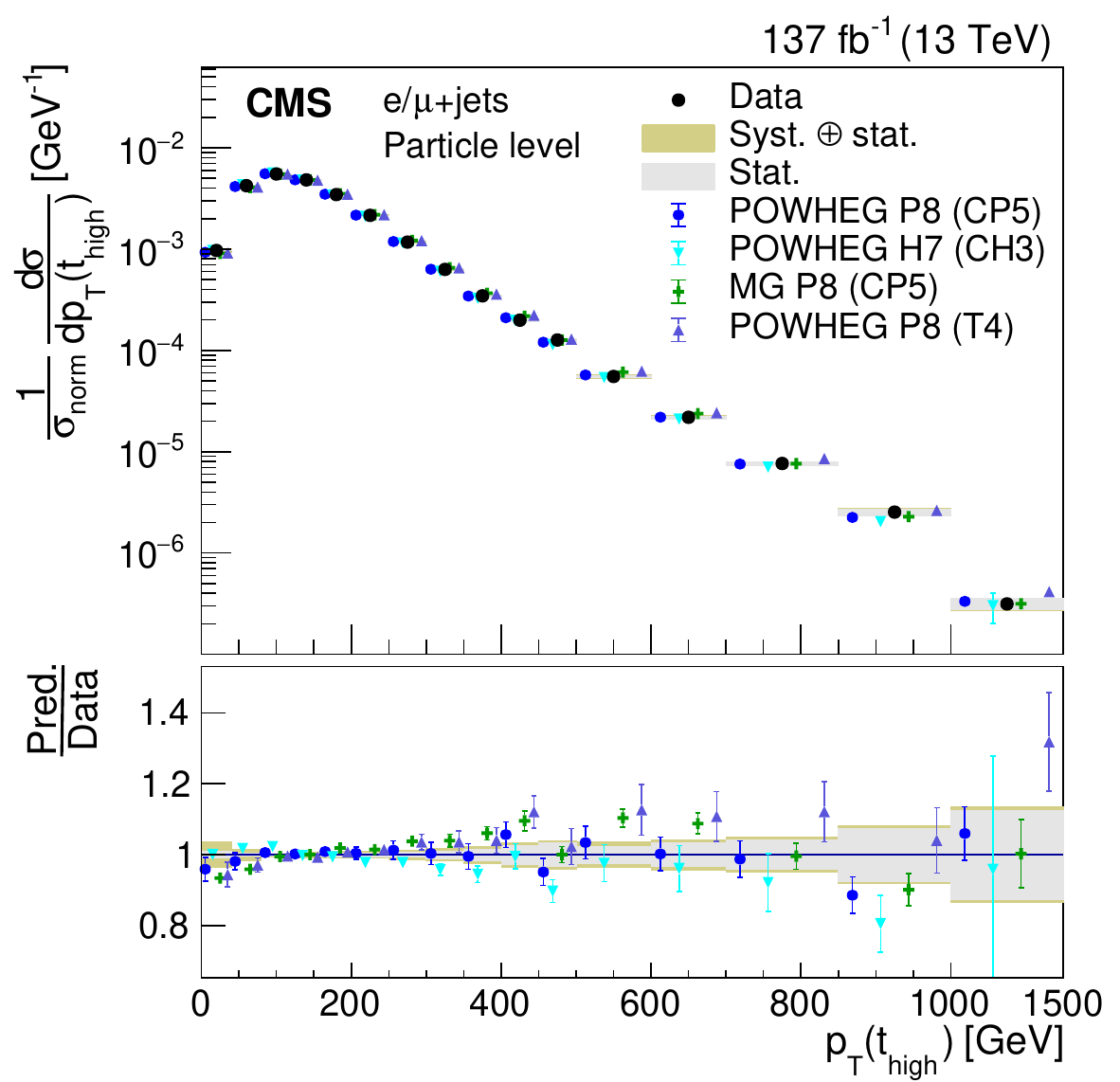}
 \includegraphics[width=0.42\textwidth]{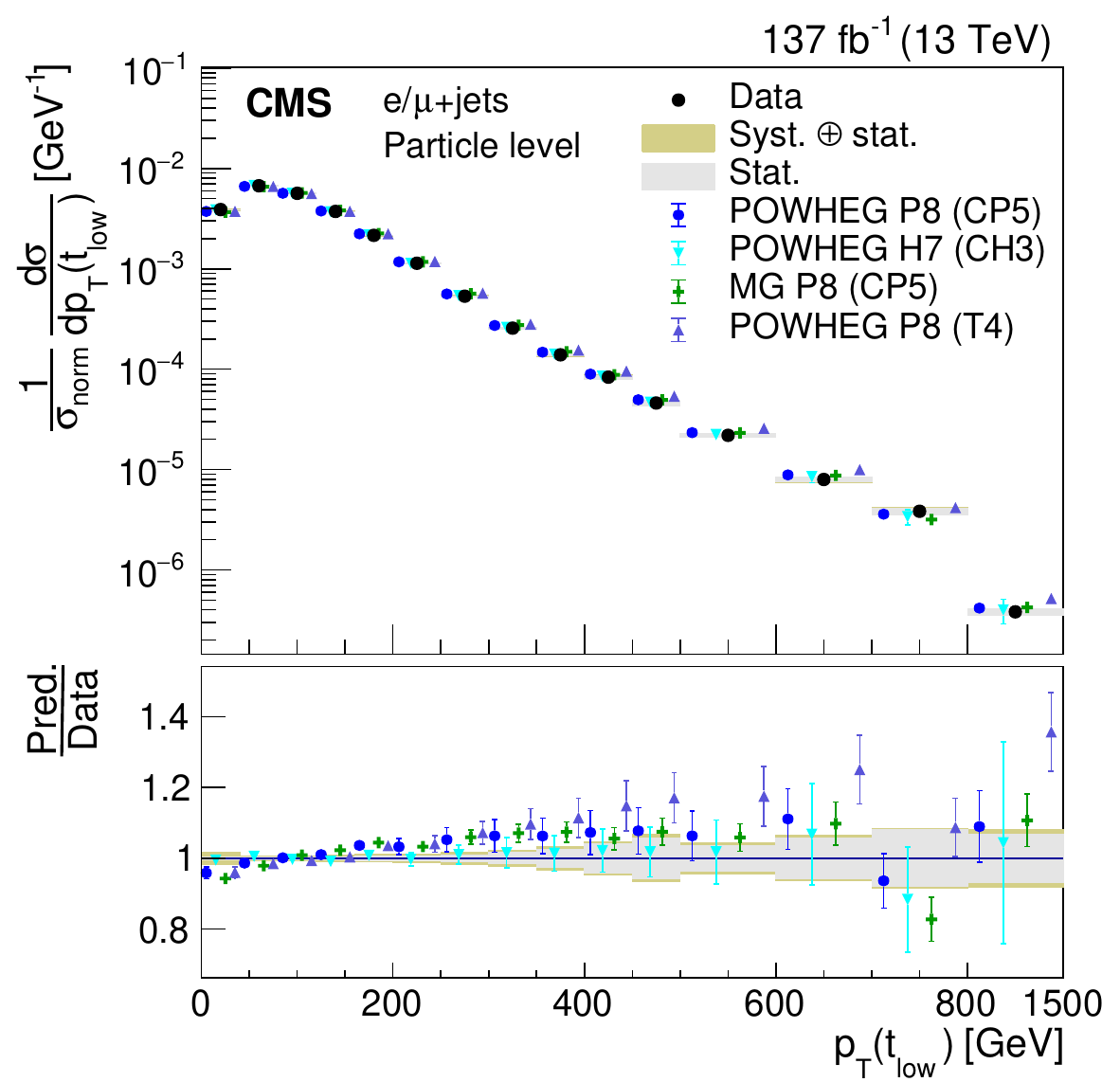}
 \includegraphics[width=0.42\textwidth]{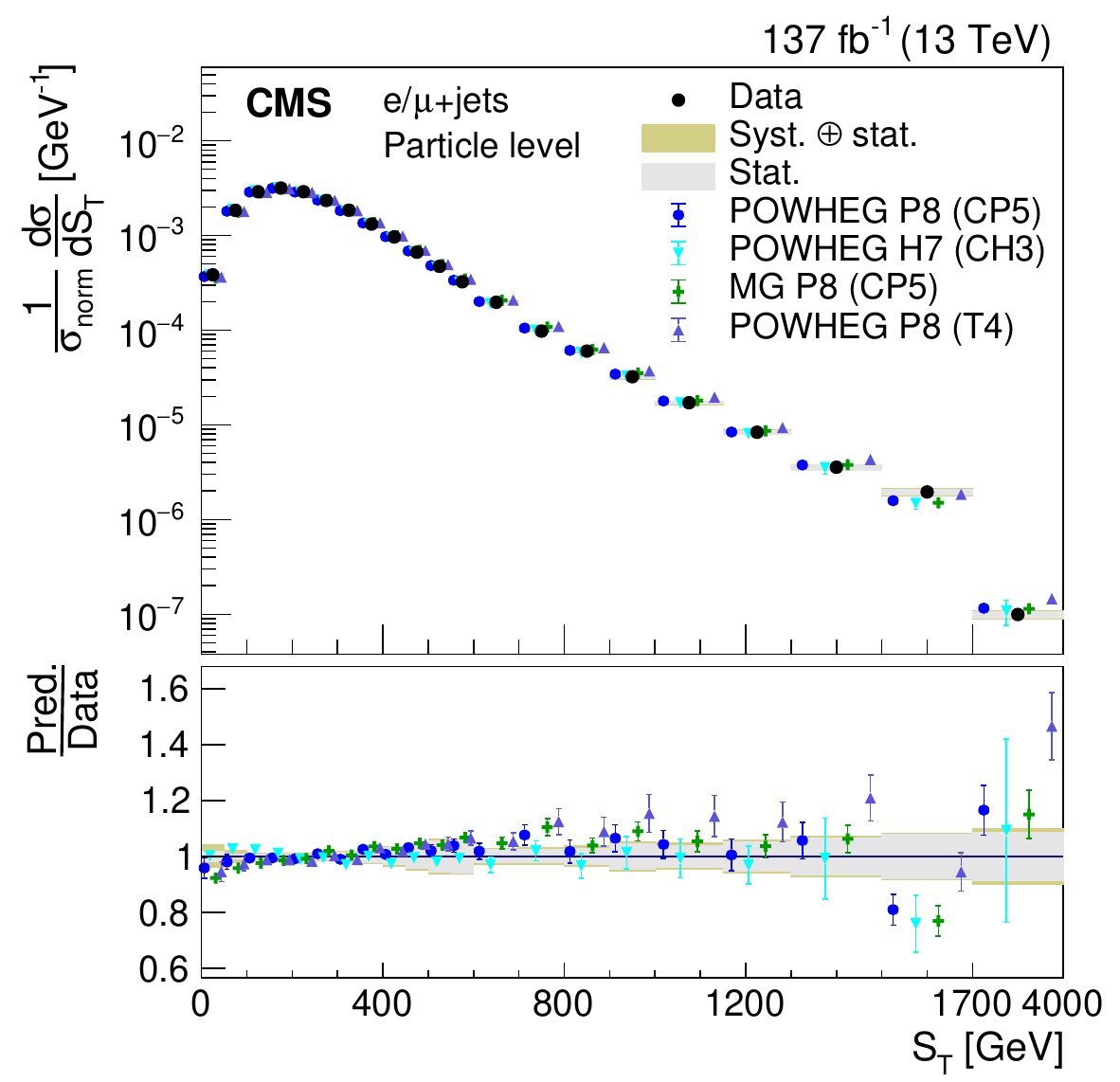}
 \caption{Normalized differential cross sections at the particle level as a function of \thadpt, \tleppt, \thardpt, \tsoftpt, and \st. \XSECCAPPS}
 \label{fig:RESNORMPS1}
\end{figure*}

\begin{figure*}[tbp]
\centering
 \includegraphics[width=0.42\textwidth]{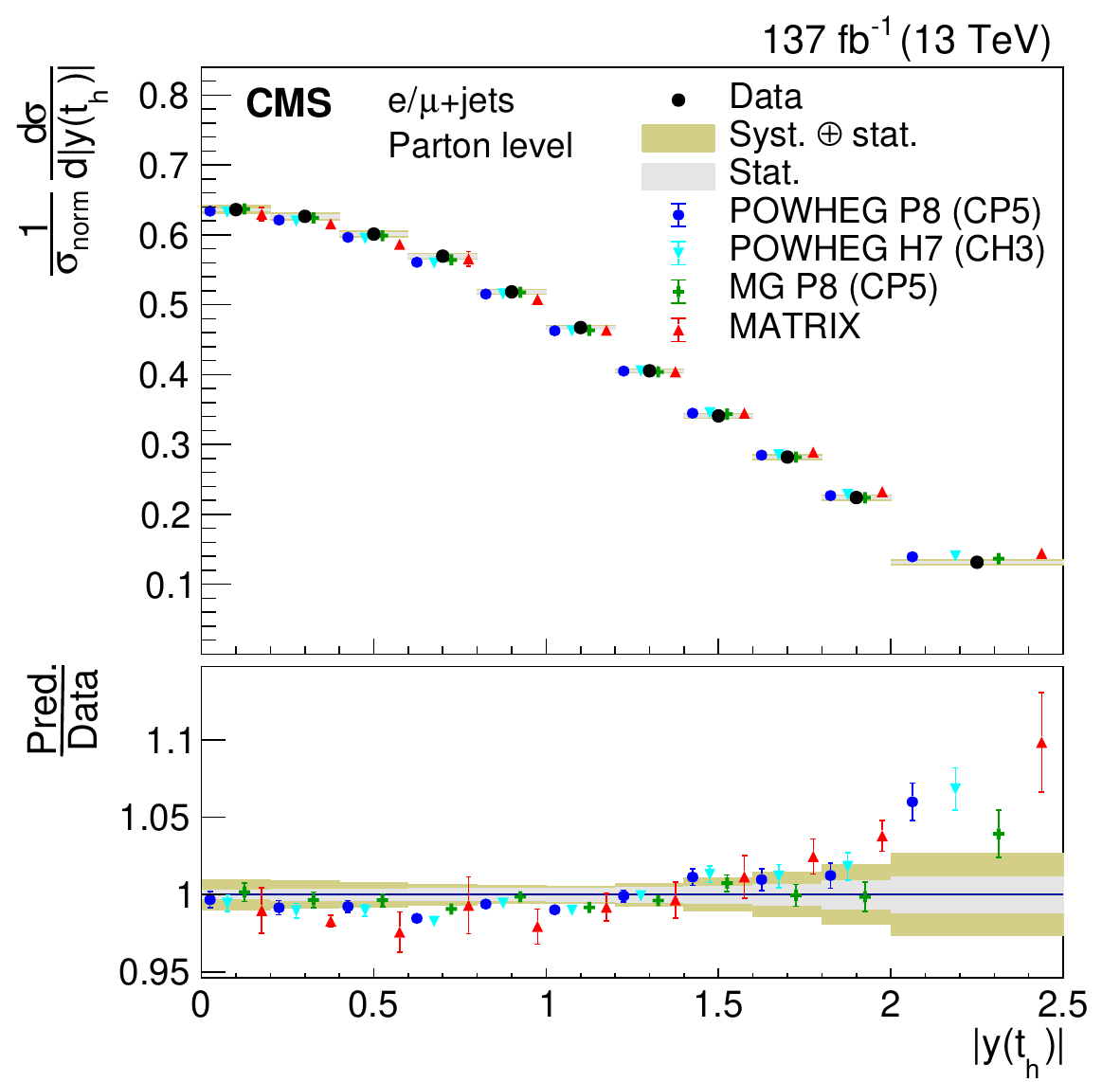}
 \includegraphics[width=0.42\textwidth]{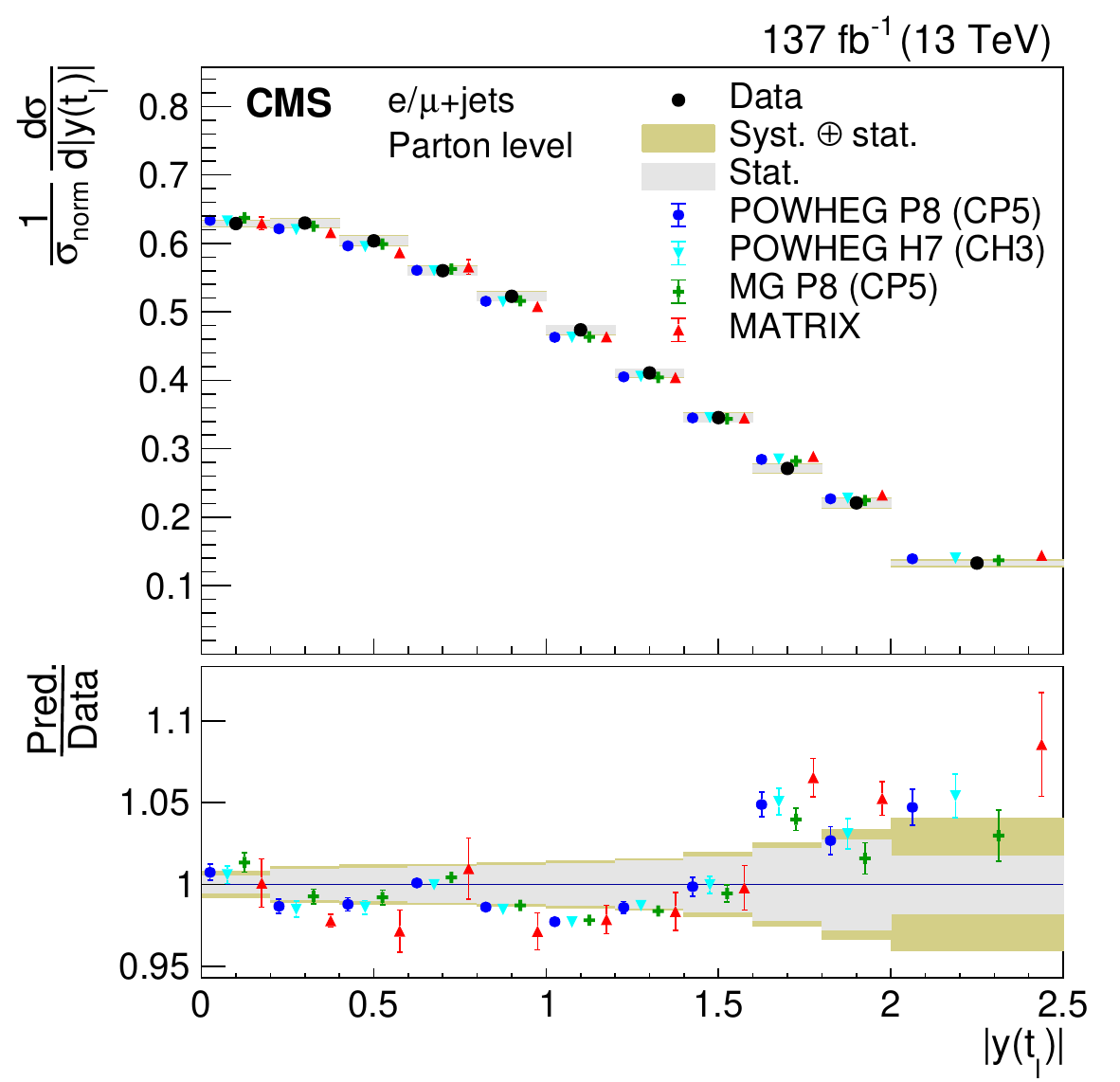}
 \includegraphics[width=0.42\textwidth]{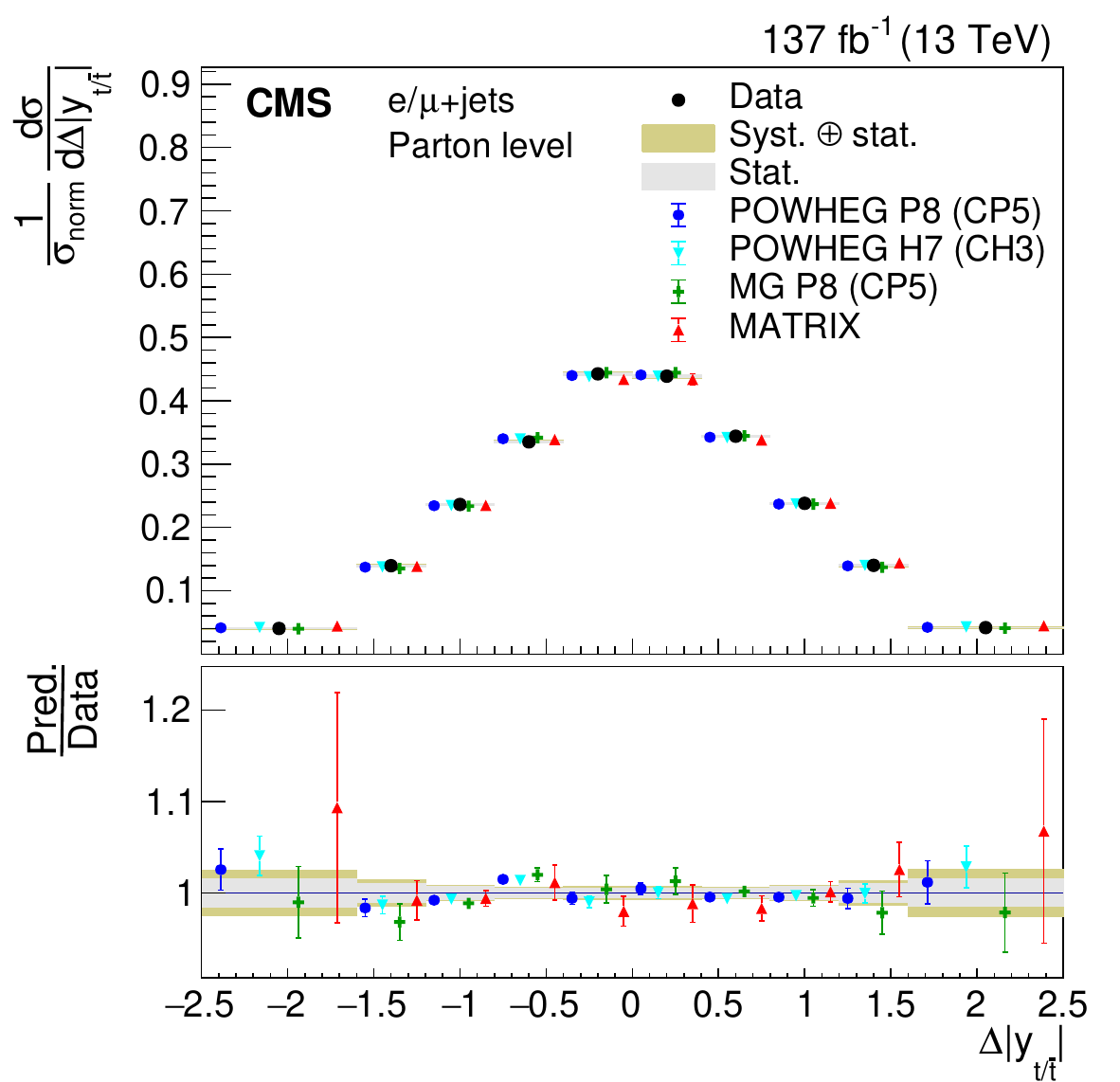}
 \includegraphics[width=0.42\textwidth]{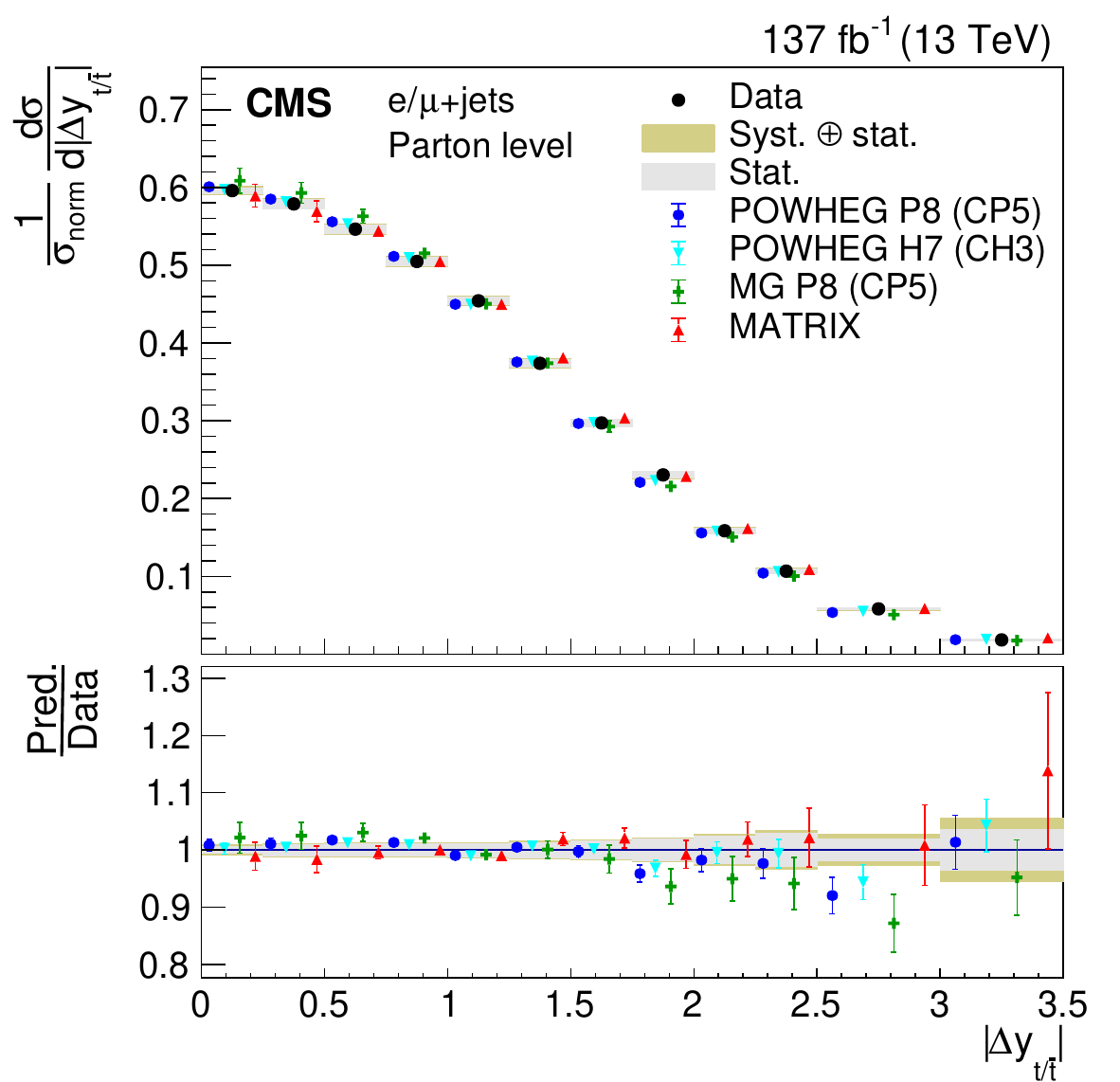}
 \caption{Normalized differential cross sections at the parton level as a function of \thady, \tlepy, and the differences \dy and \ady. \XSECCAPPA}
 \label{fig:RESNORM2}
\end{figure*}

\begin{figure*}[tbp]
\centering
 \includegraphics[width=0.42\textwidth]{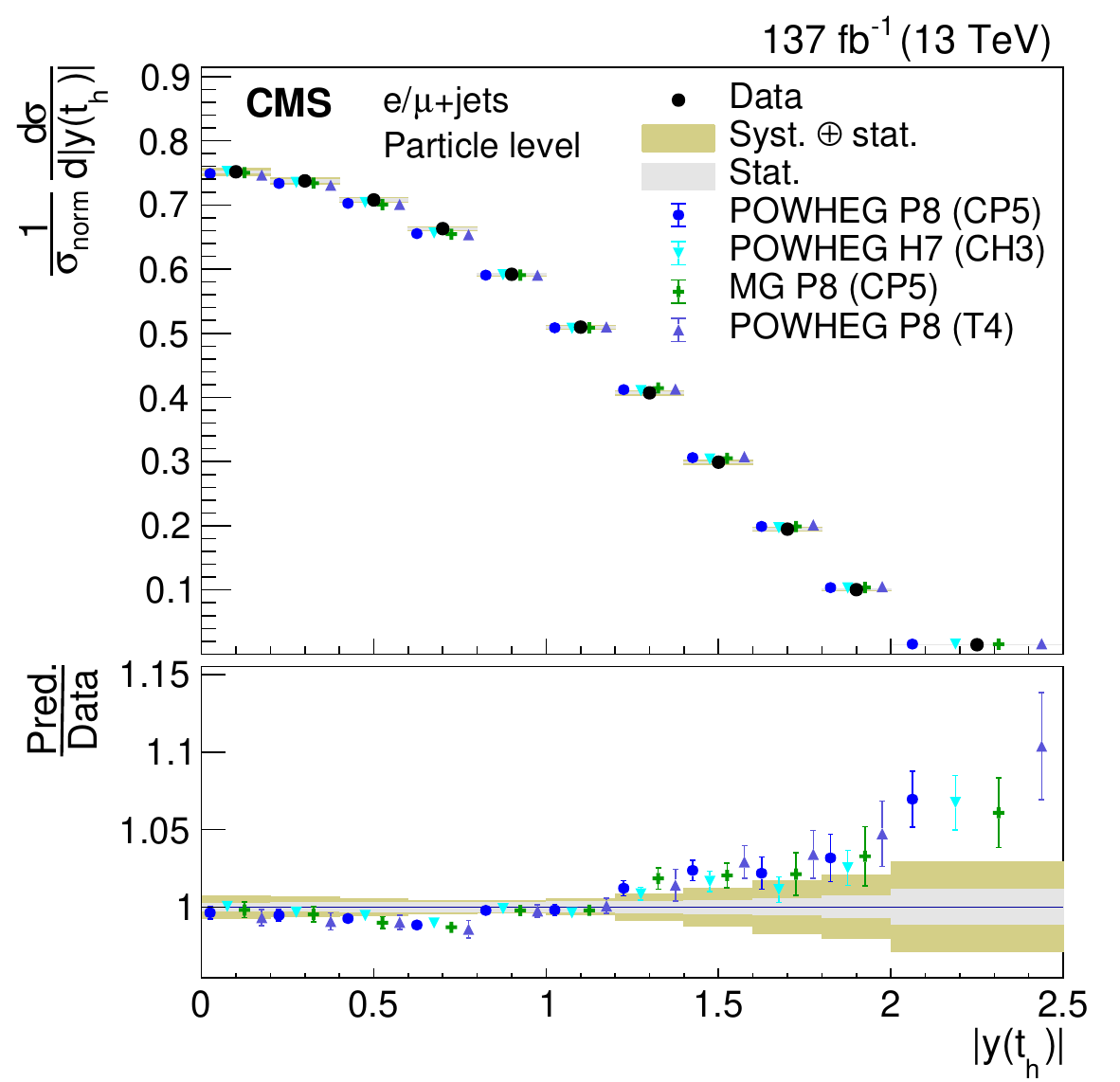}
 \includegraphics[width=0.42\textwidth]{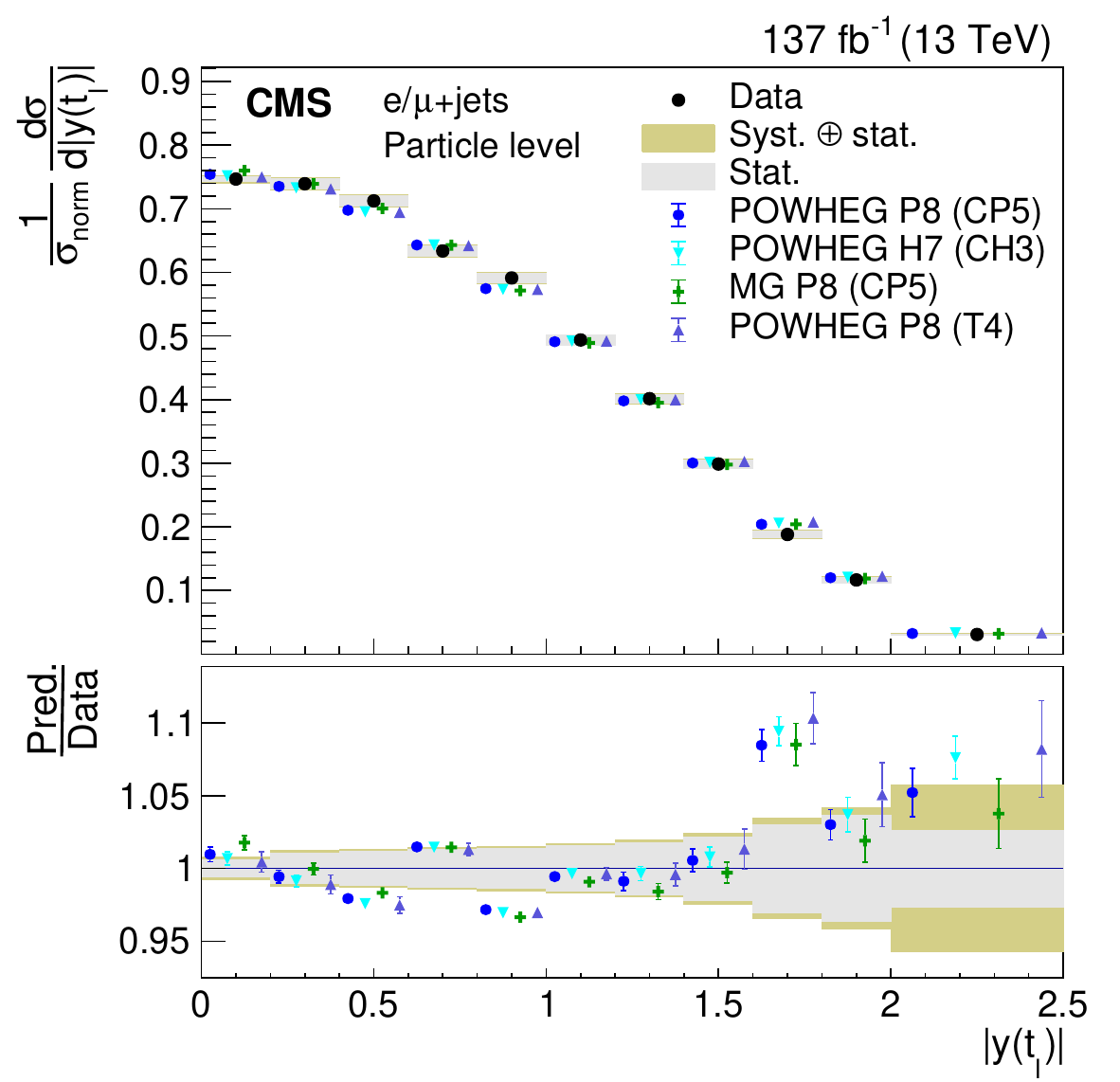}
 \includegraphics[width=0.42\textwidth]{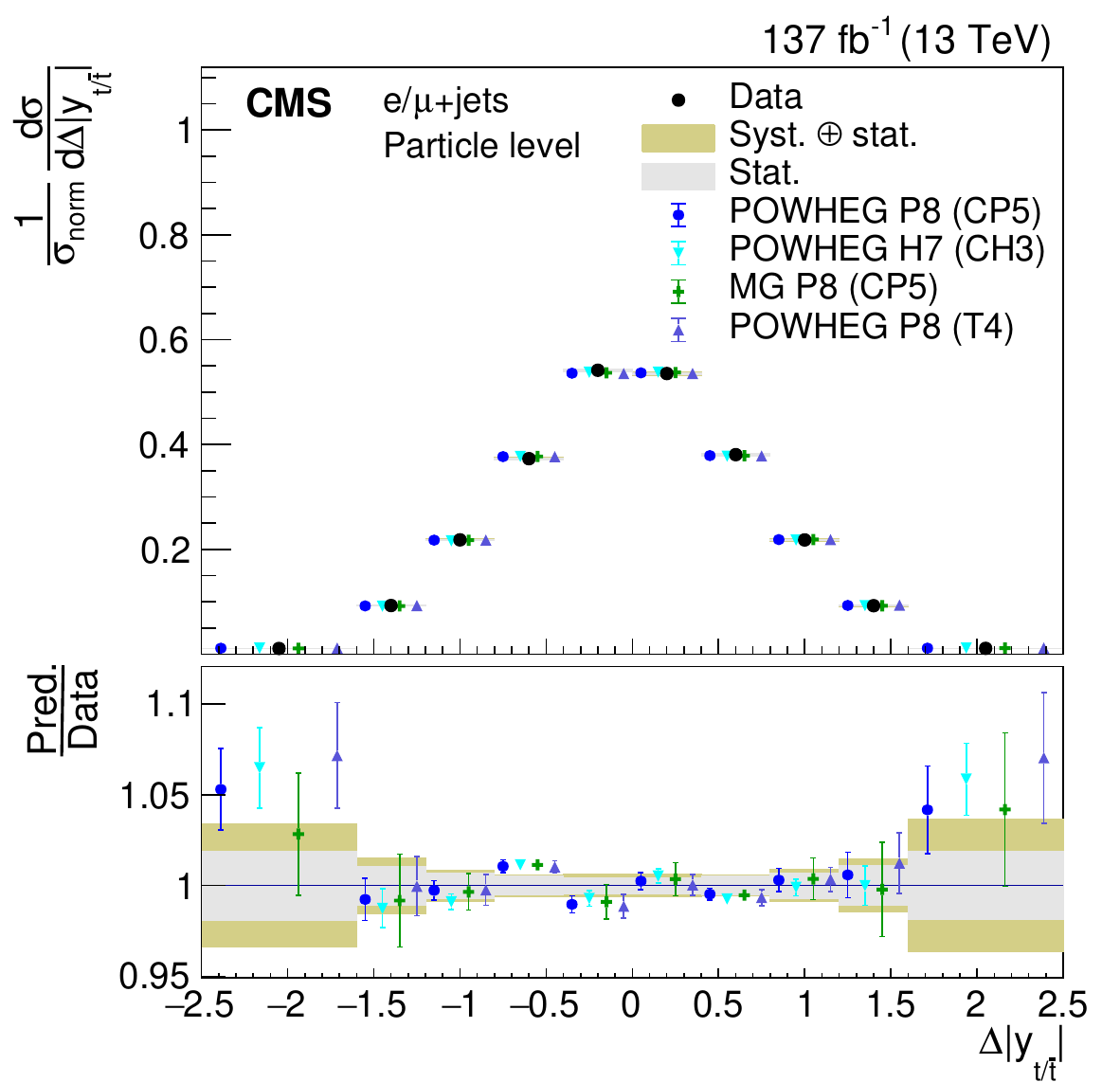}
 \includegraphics[width=0.42\textwidth]{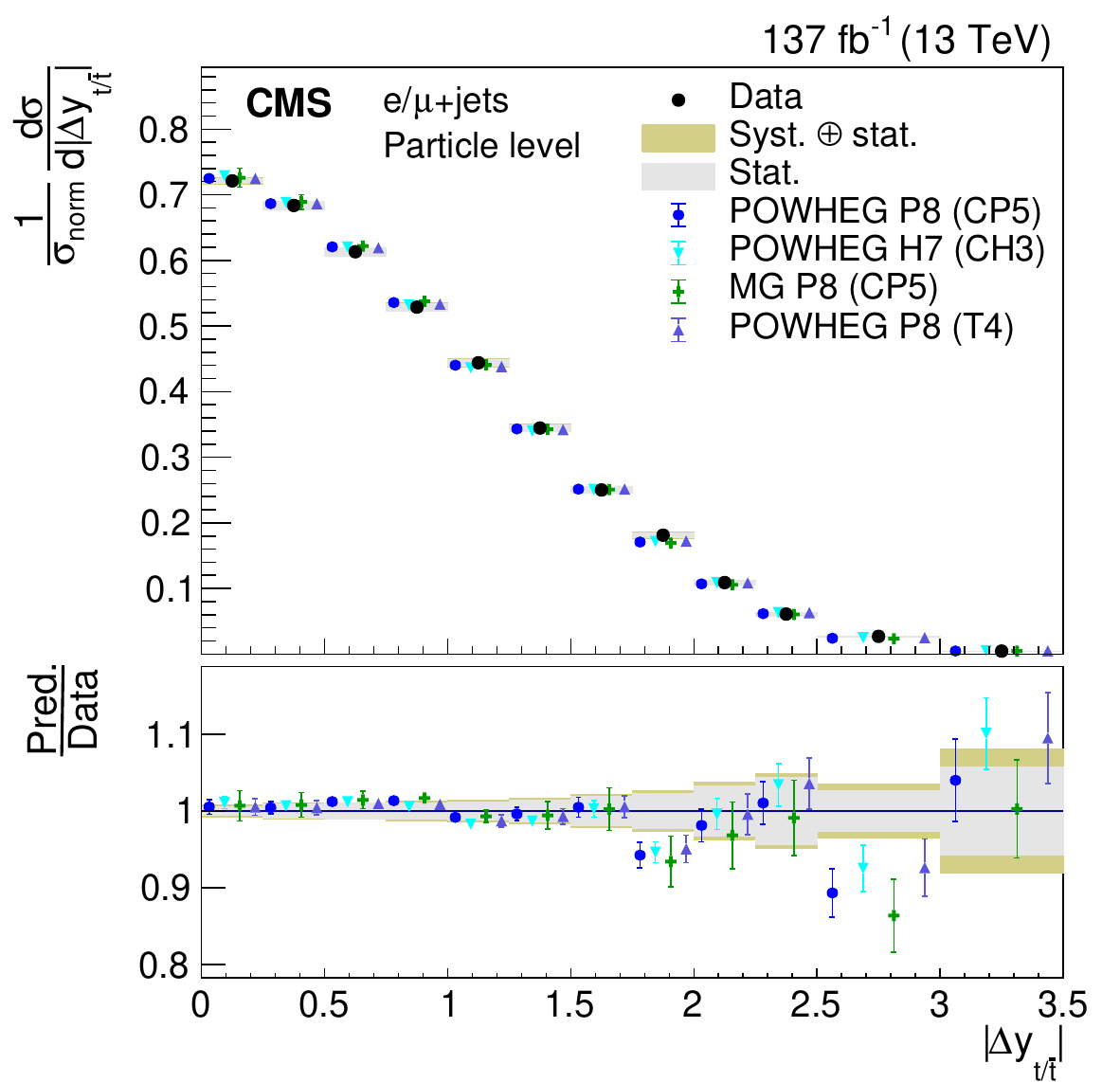}
 \caption{Normalized differential cross sections at the particle level as a function of \thady, \tlepy, and the differences \dy and \ady. \XSECCAPPS}
 \label{fig:RESNORMPS2}
\end{figure*}

\begin{figure*}[tbp]
\centering
 \includegraphics[width=0.42\textwidth]{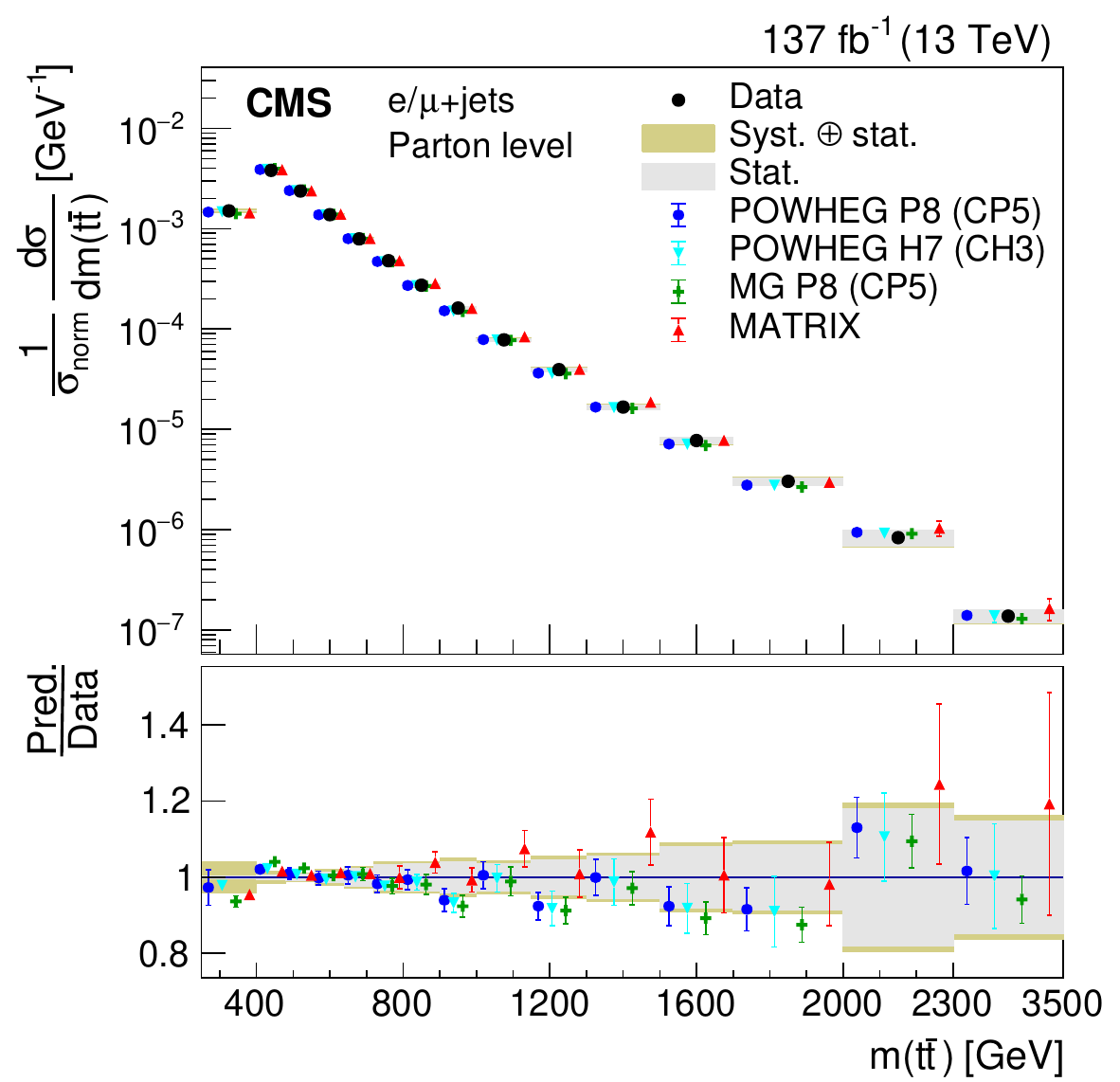}
 \includegraphics[width=0.42\textwidth]{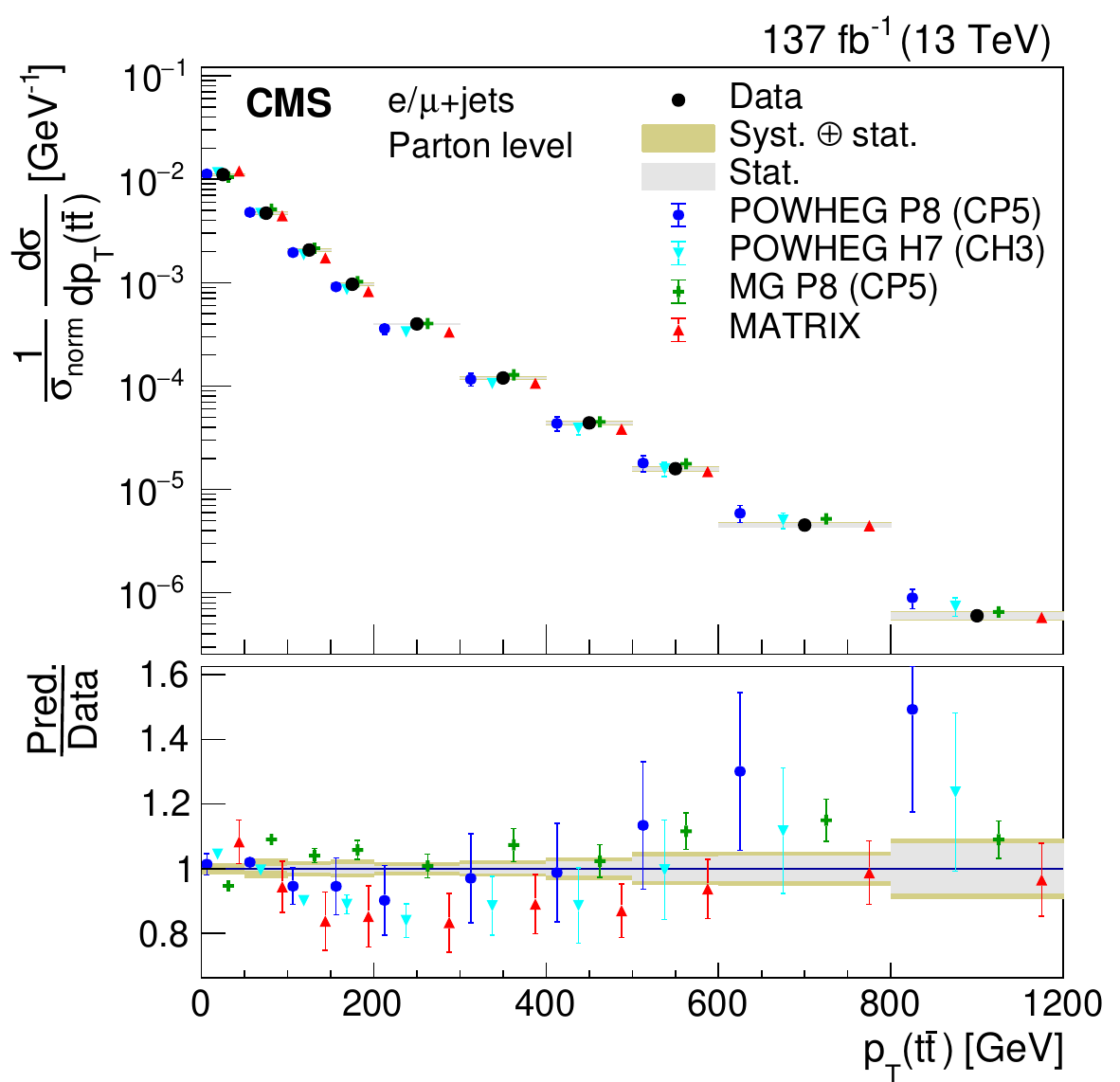}
 \includegraphics[width=0.42\textwidth]{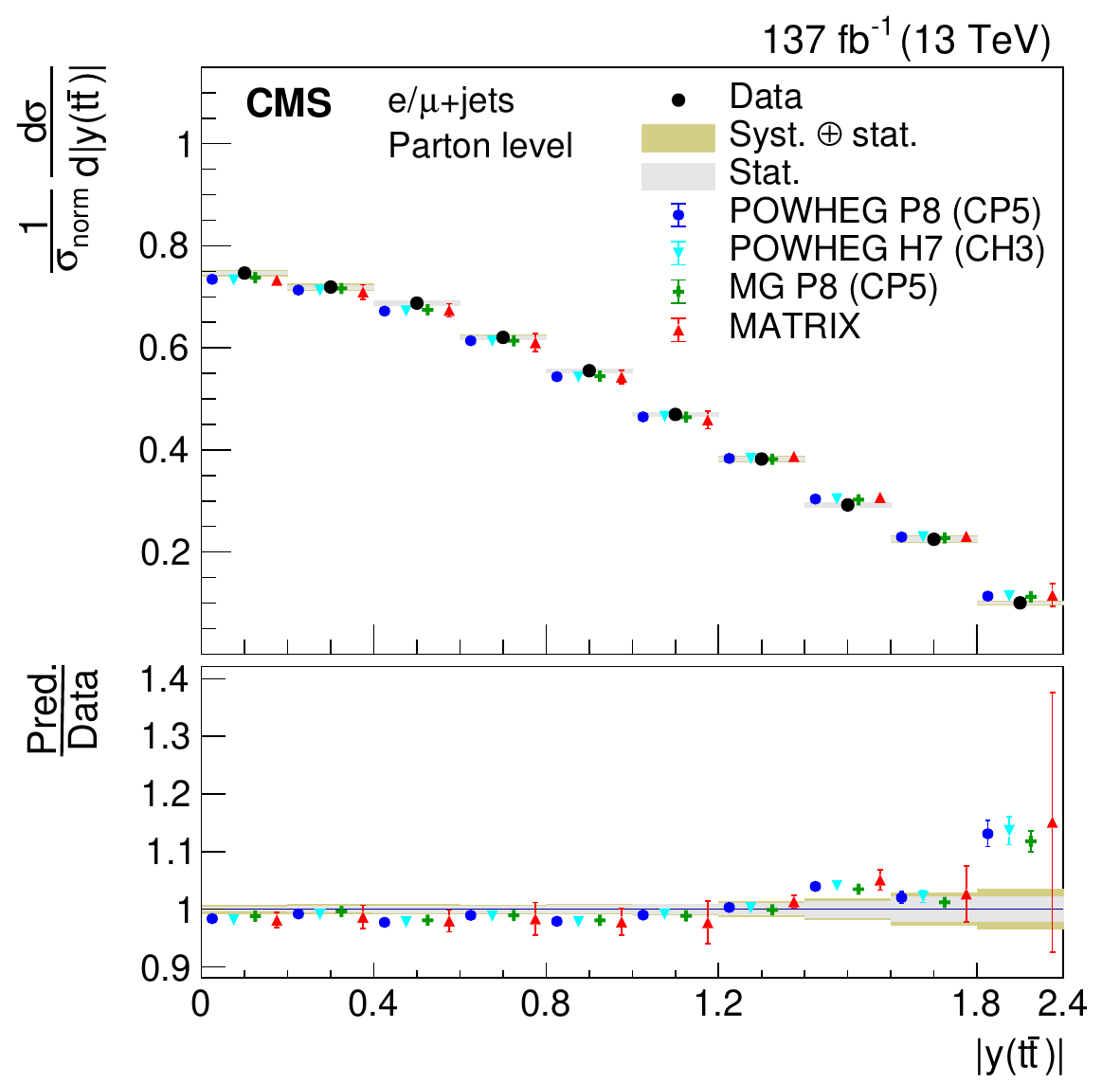}
 \includegraphics[width=0.42\textwidth]{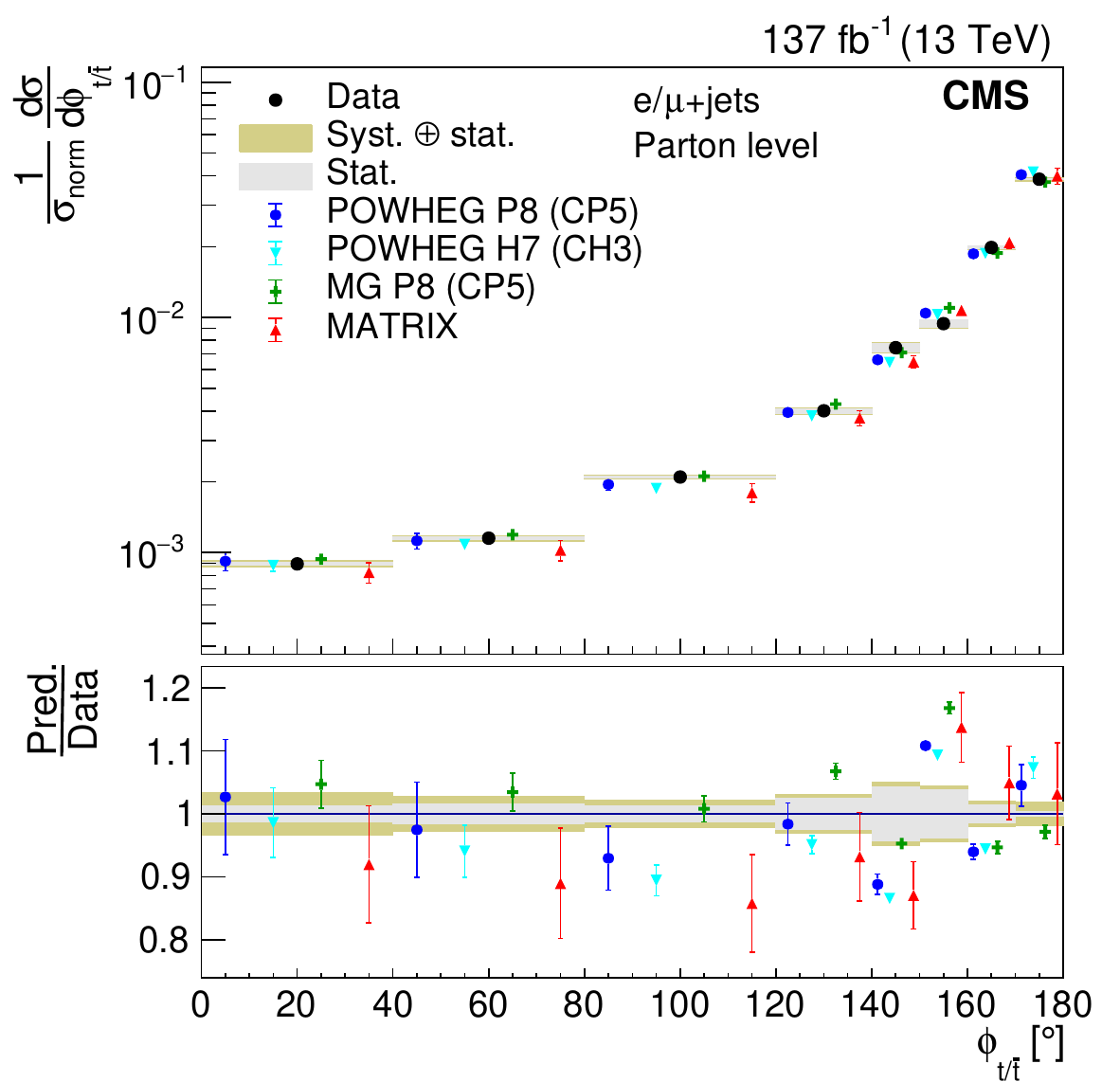}
 \includegraphics[width=0.42\textwidth]{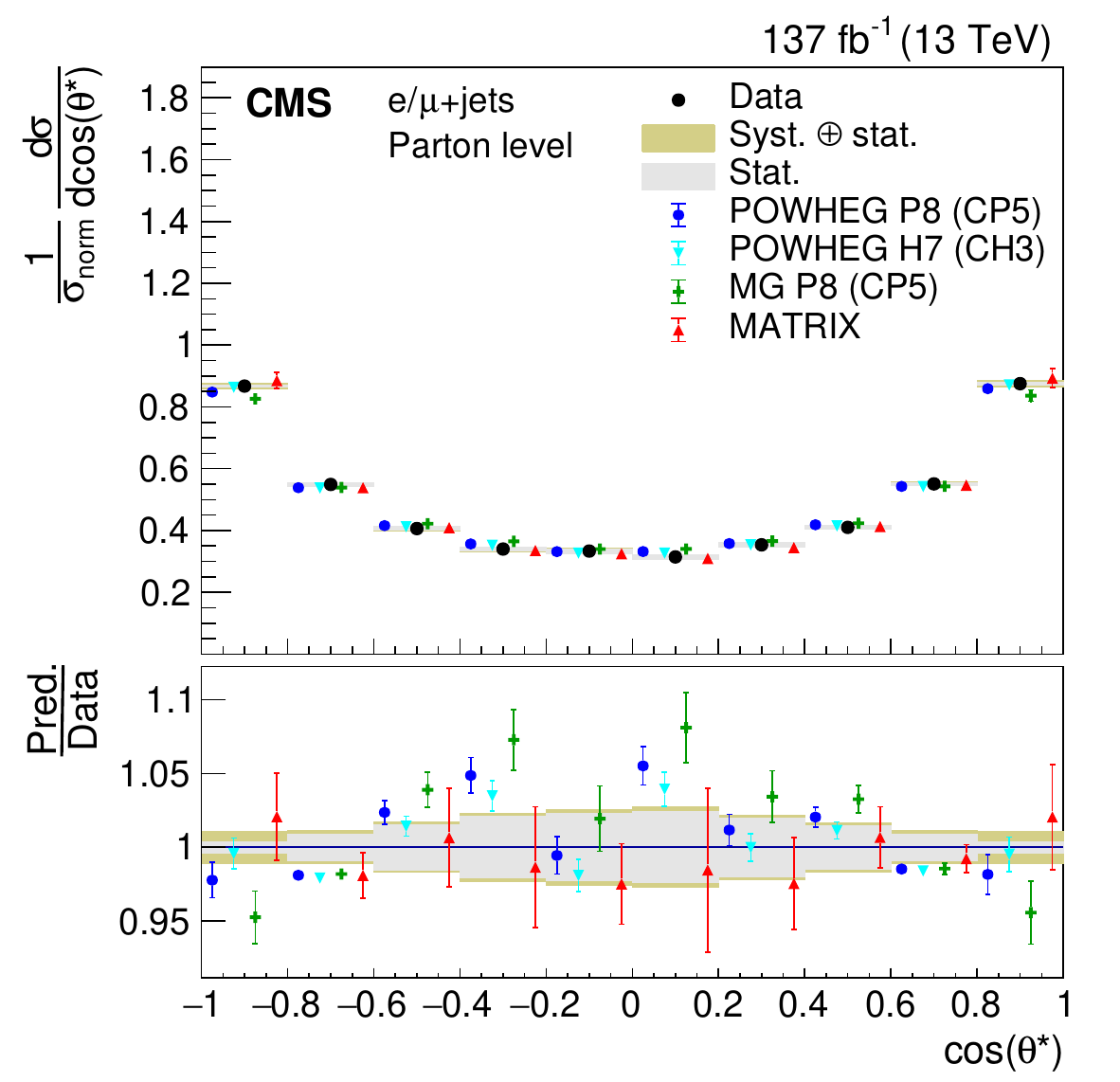}
 \caption{Normalized differential cross sections at the parton level as a function of kinematic variables of the \ttbar system. \XSECCAPPA}
 \label{fig:RESNORM3}
\end{figure*}

\begin{figure*}[tbp]
\centering
 \includegraphics[width=0.42\textwidth]{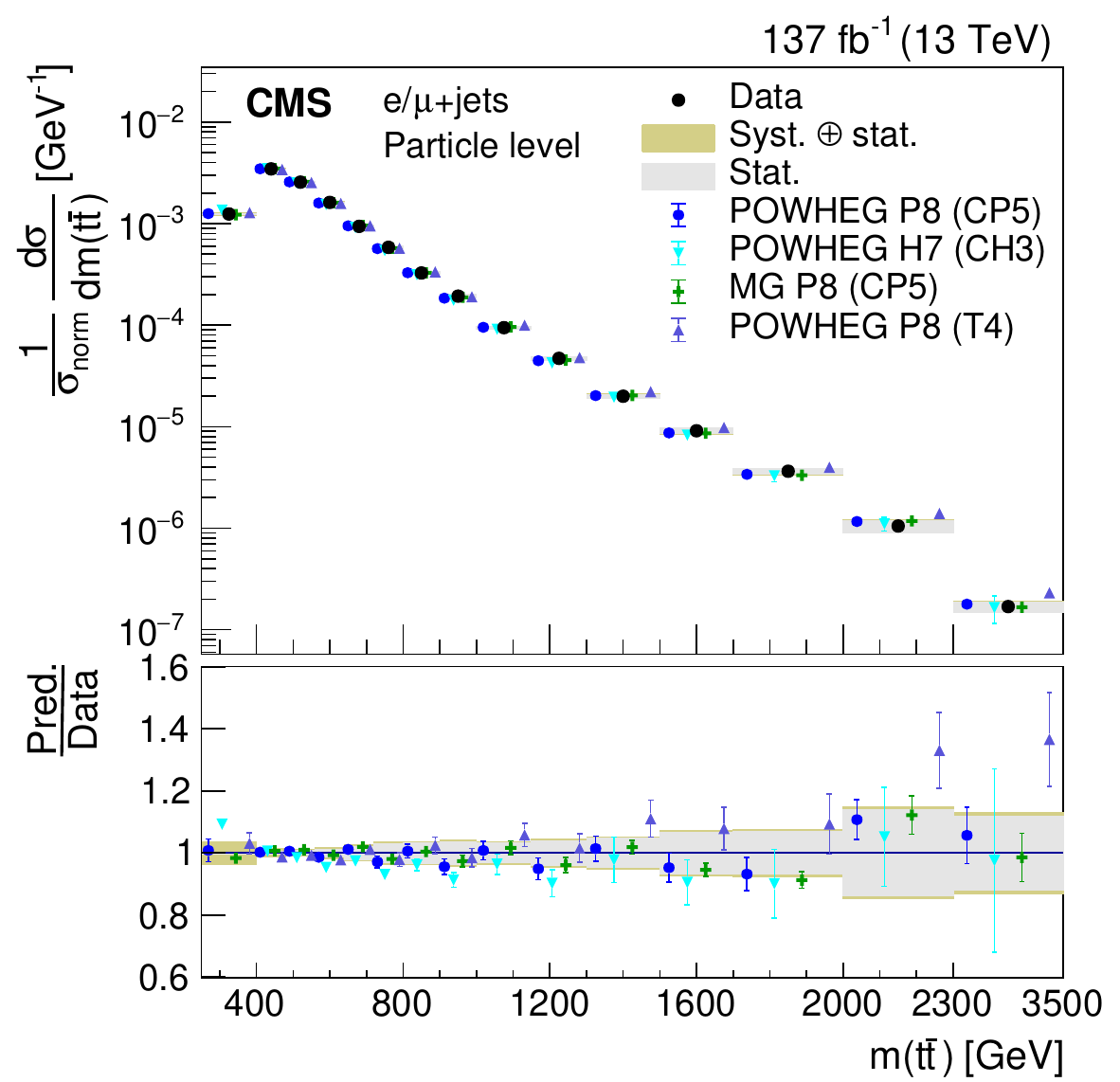}
 \includegraphics[width=0.42\textwidth]{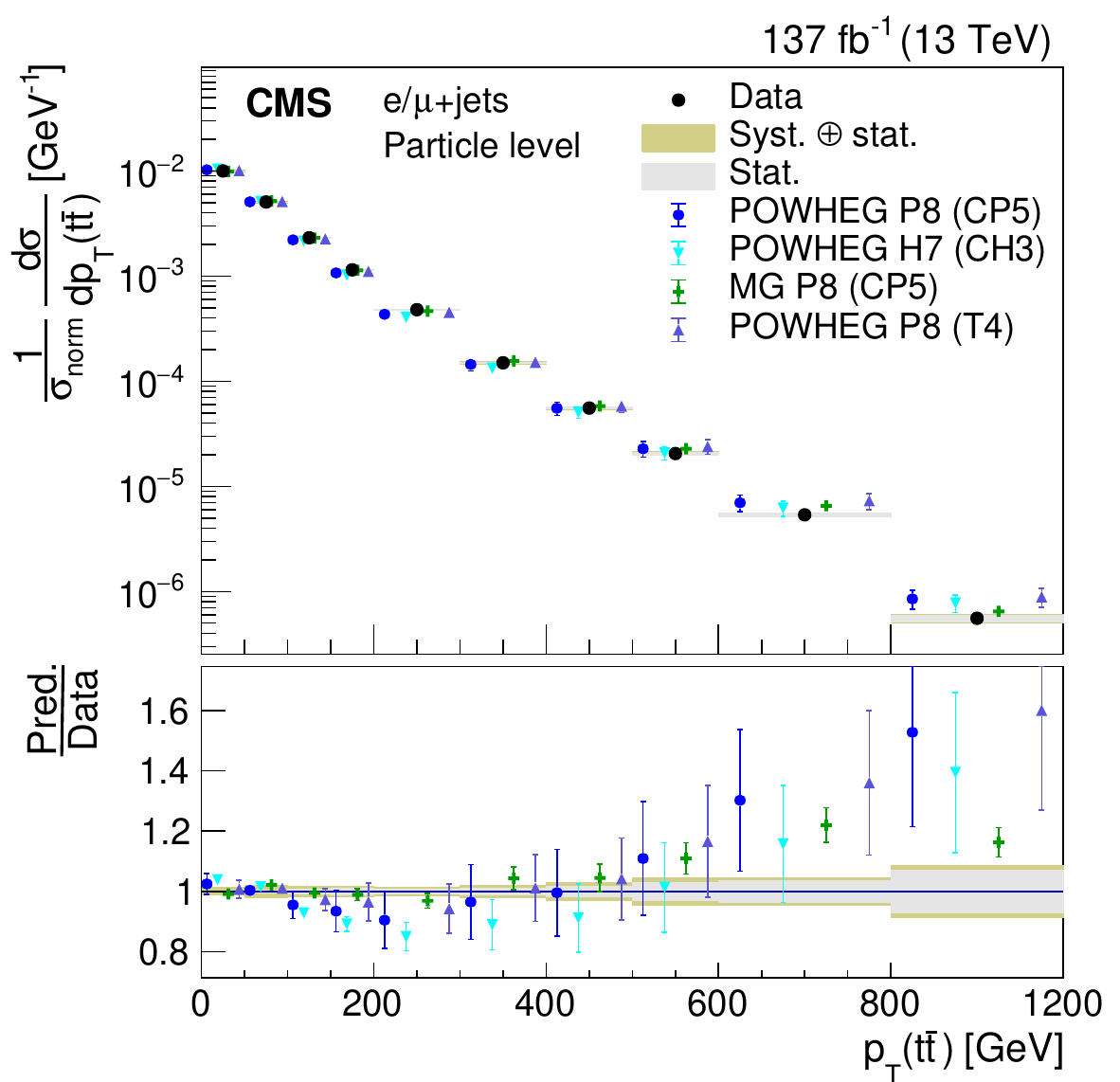}
 \includegraphics[width=0.42\textwidth]{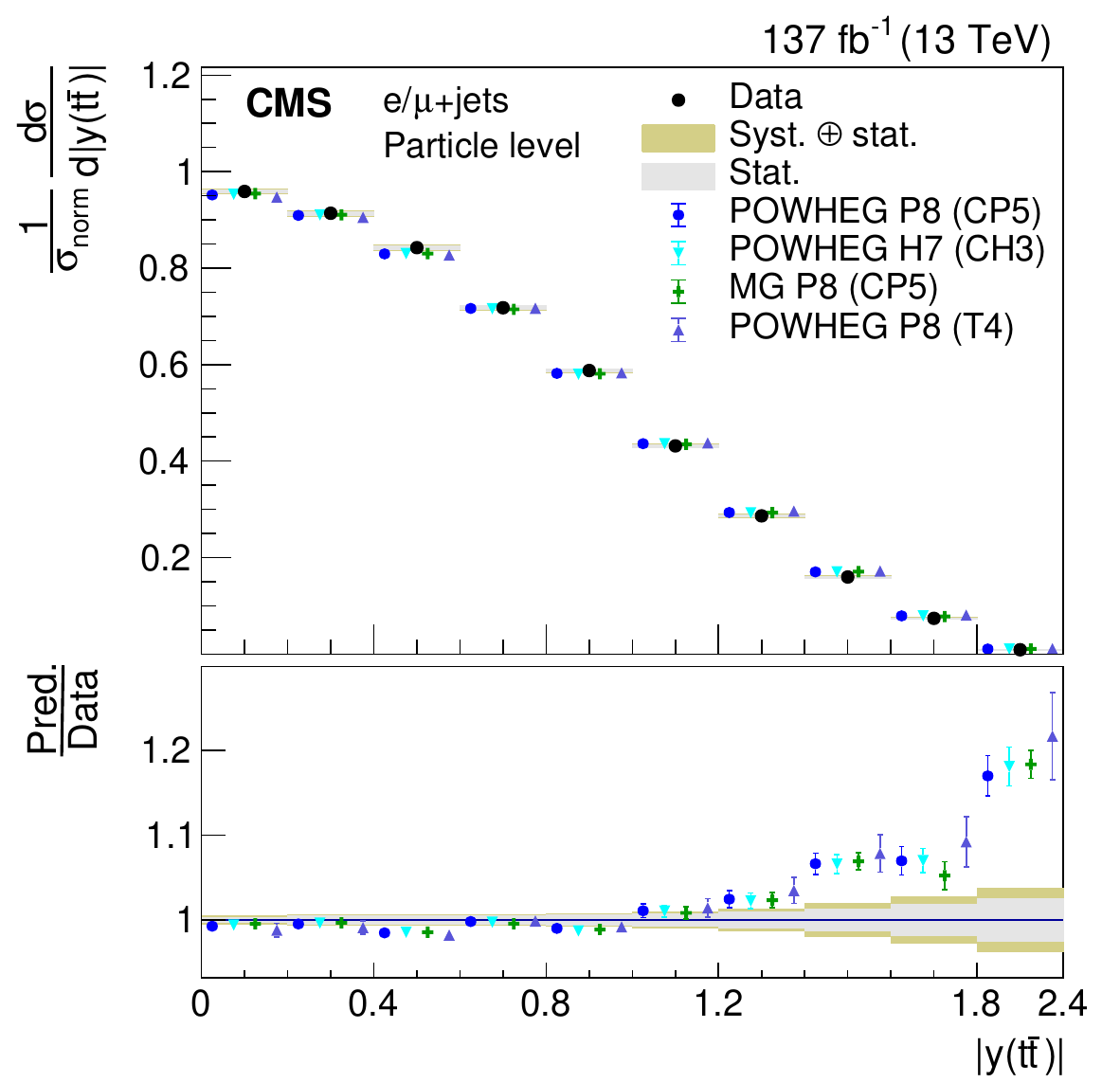}
 \includegraphics[width=0.42\textwidth]{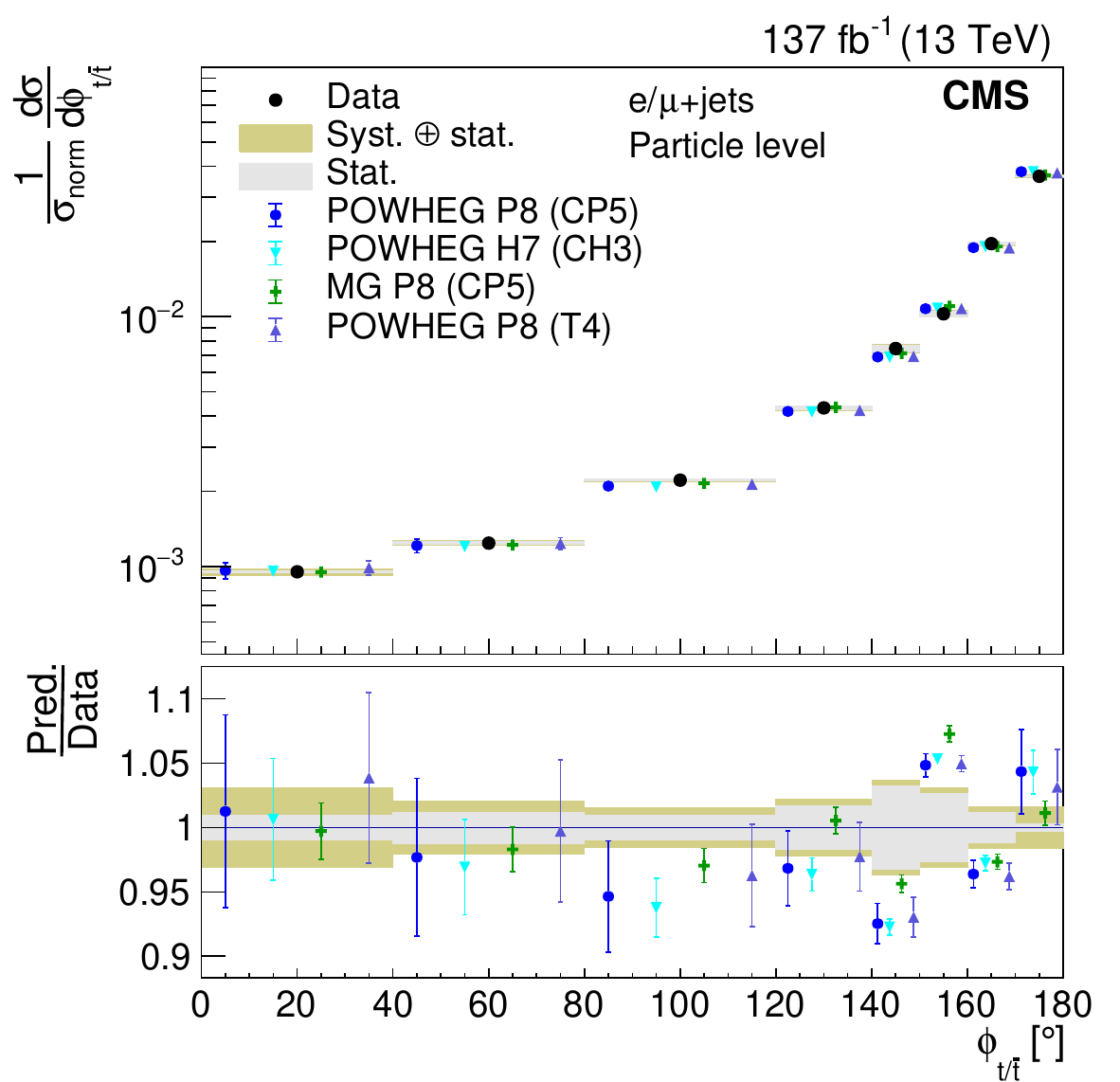}
 \includegraphics[width=0.42\textwidth]{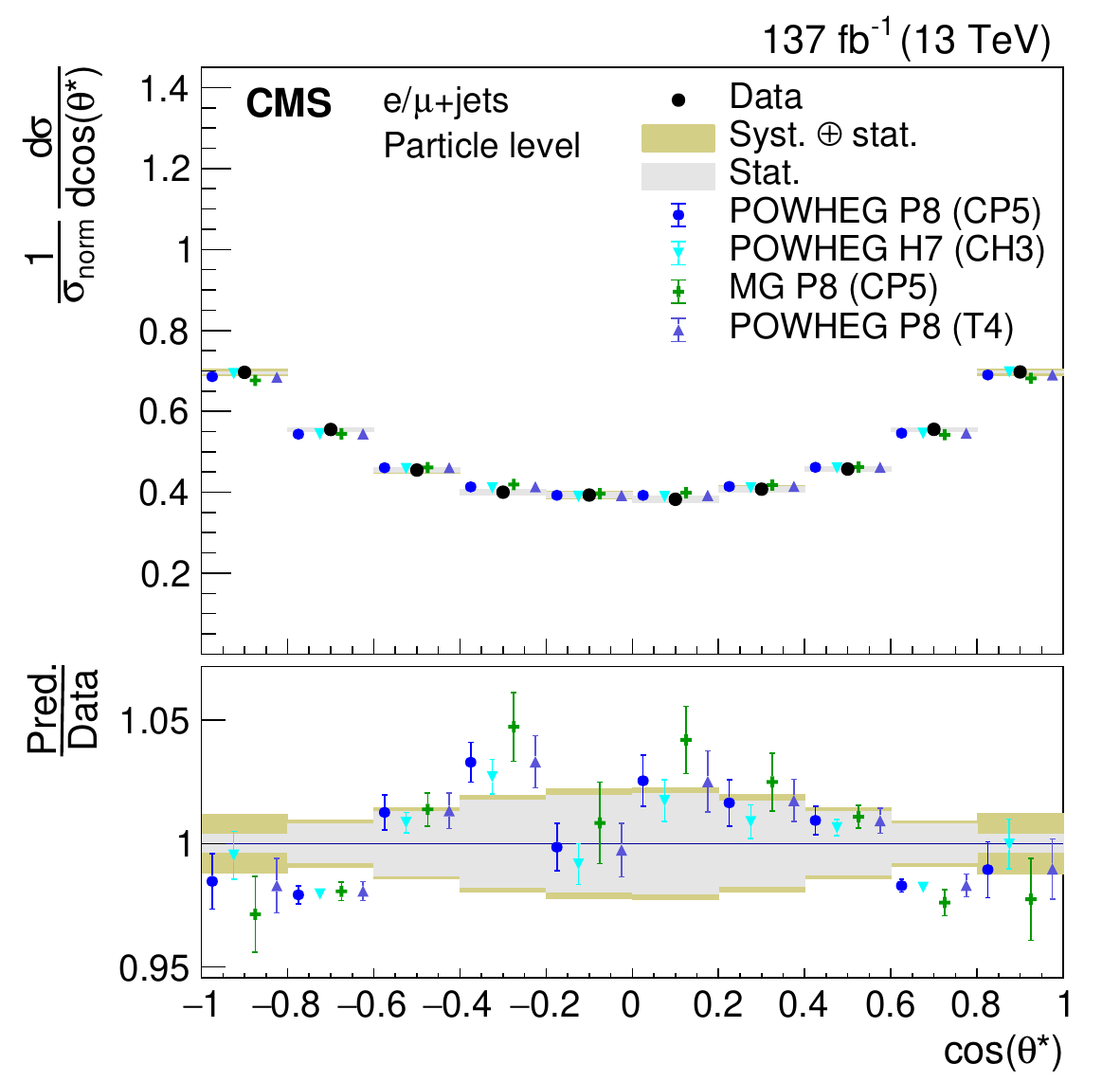}
 \caption{Normalized differential cross sections at the particle level as a function of kinematic variables of the \ttbar system. \XSECCAPPS}
 \label{fig:RESNORMPS3}
\end{figure*}

\begin{figure*}[tbp]
\centering
 \includegraphics[width=0.42\textwidth]{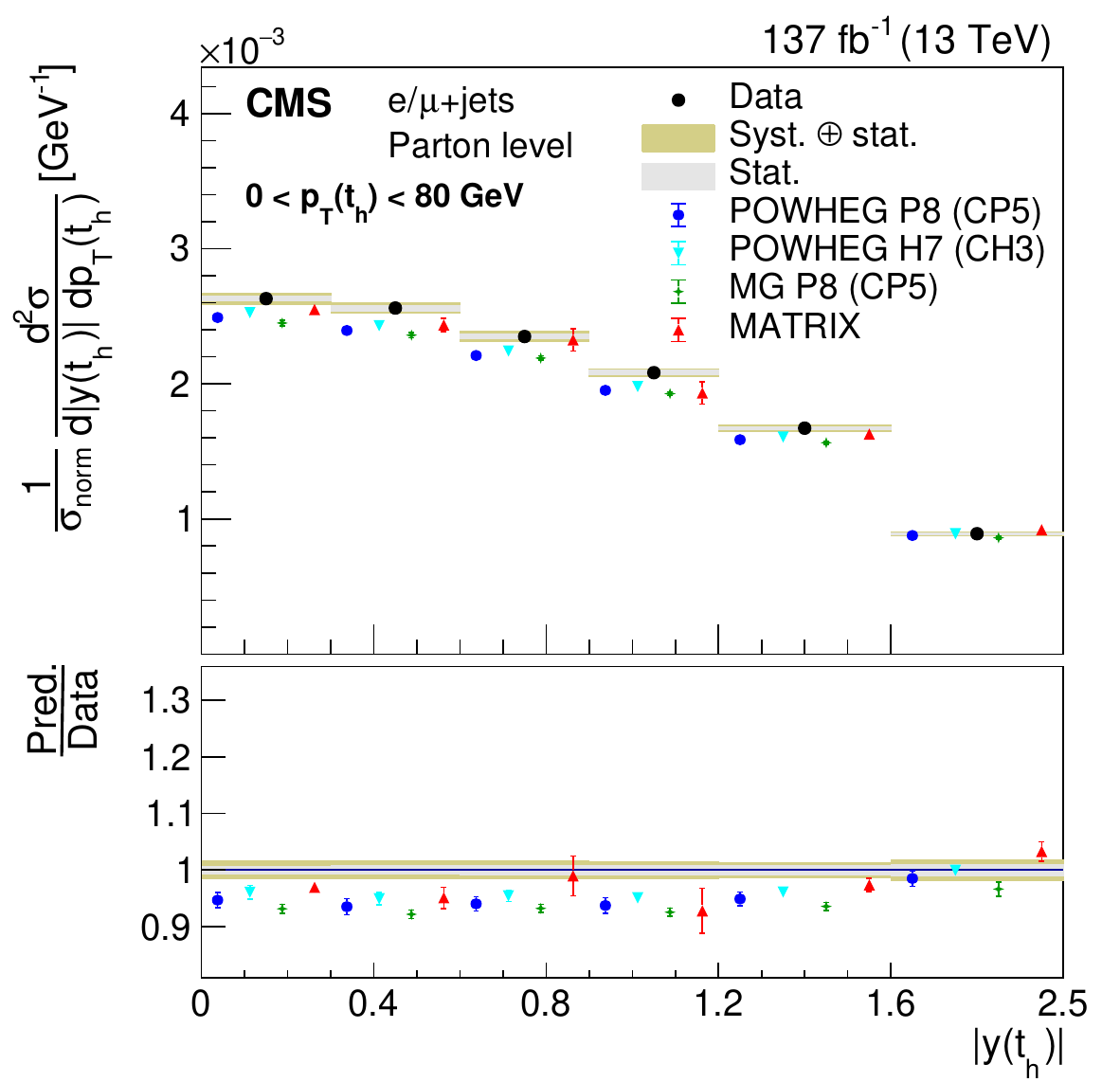}
 \includegraphics[width=0.42\textwidth]{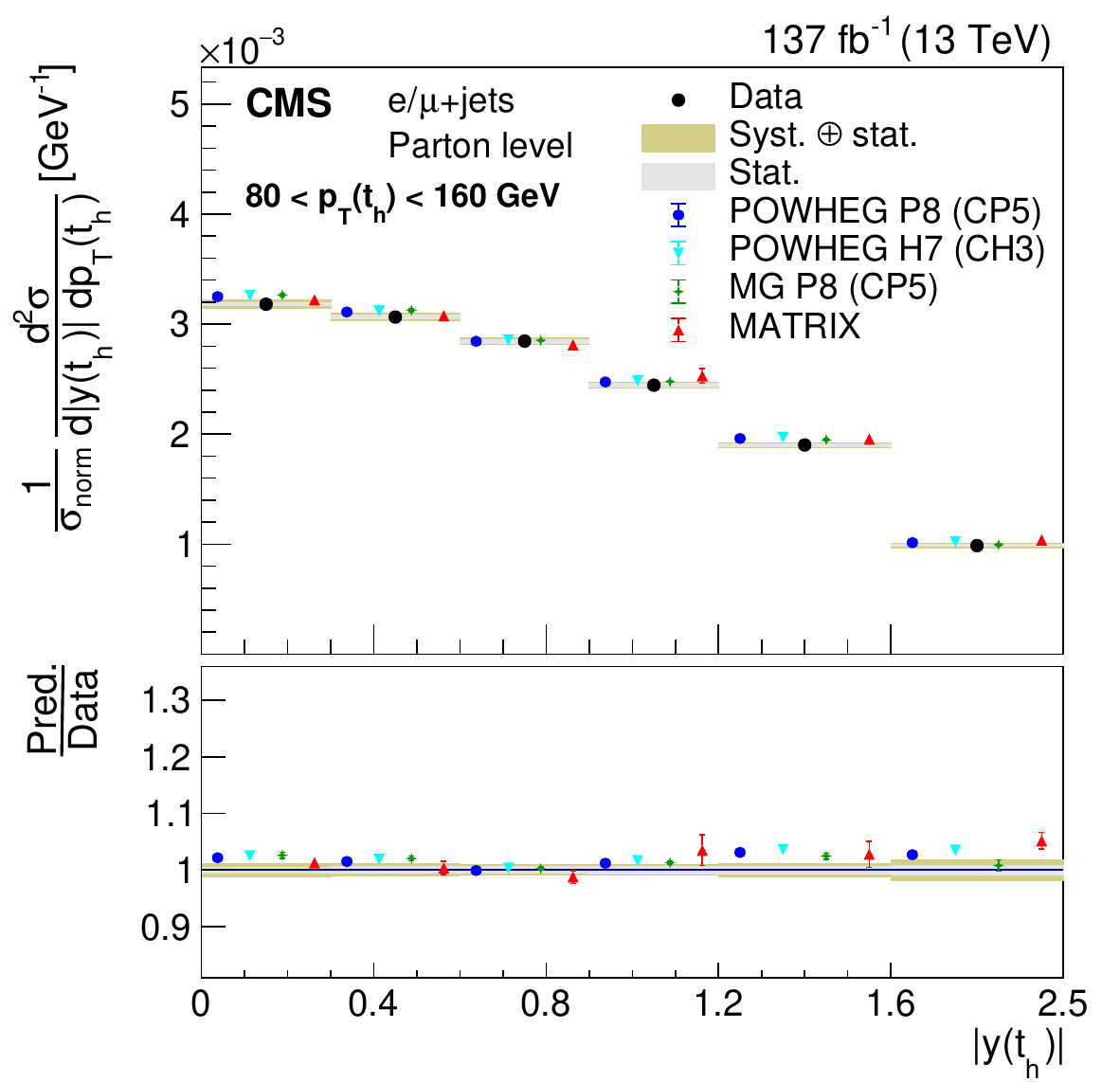}\\
 \includegraphics[width=0.42\textwidth]{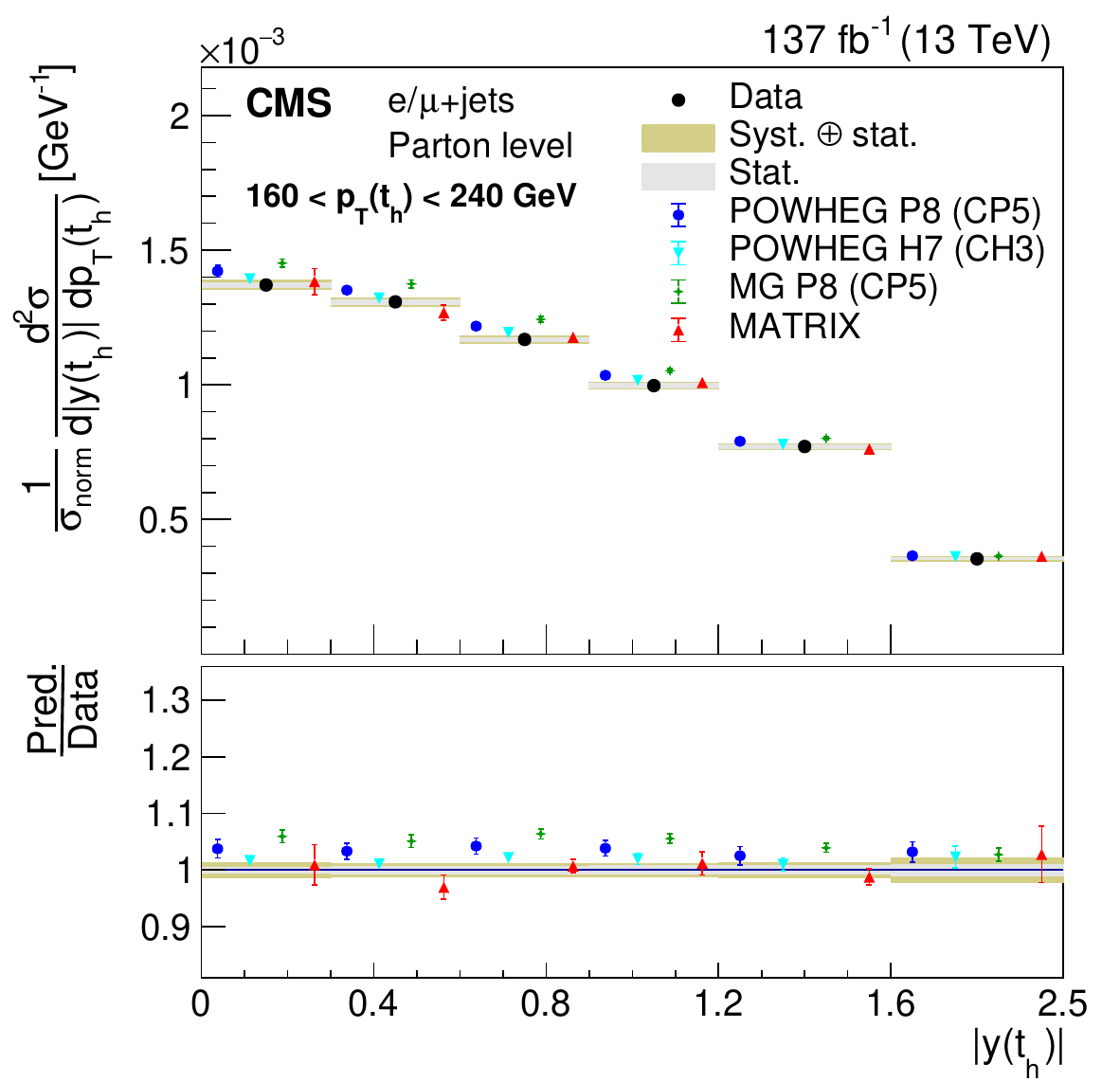}
 \includegraphics[width=0.42\textwidth]{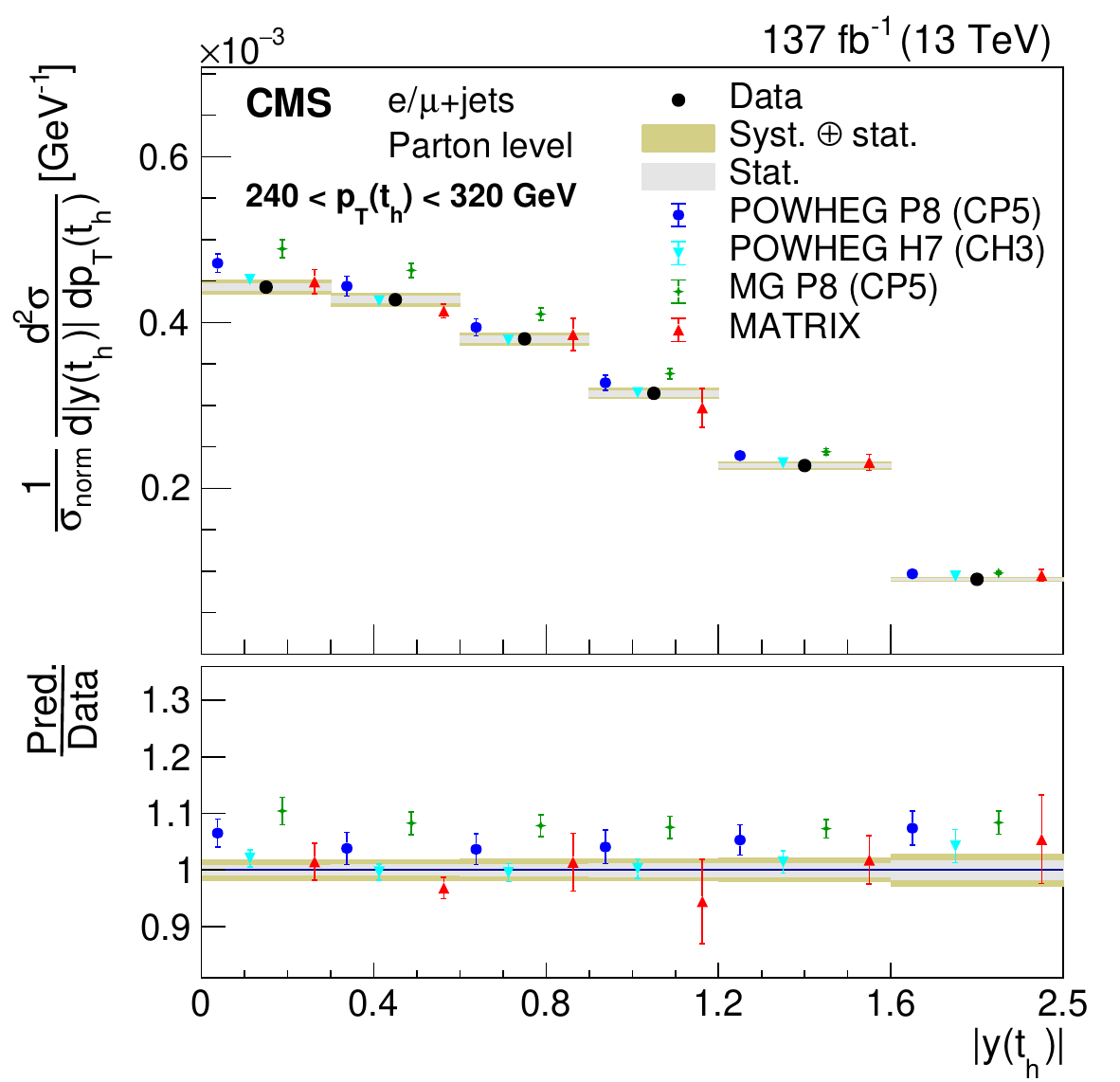}\\
 \includegraphics[width=0.42\textwidth]{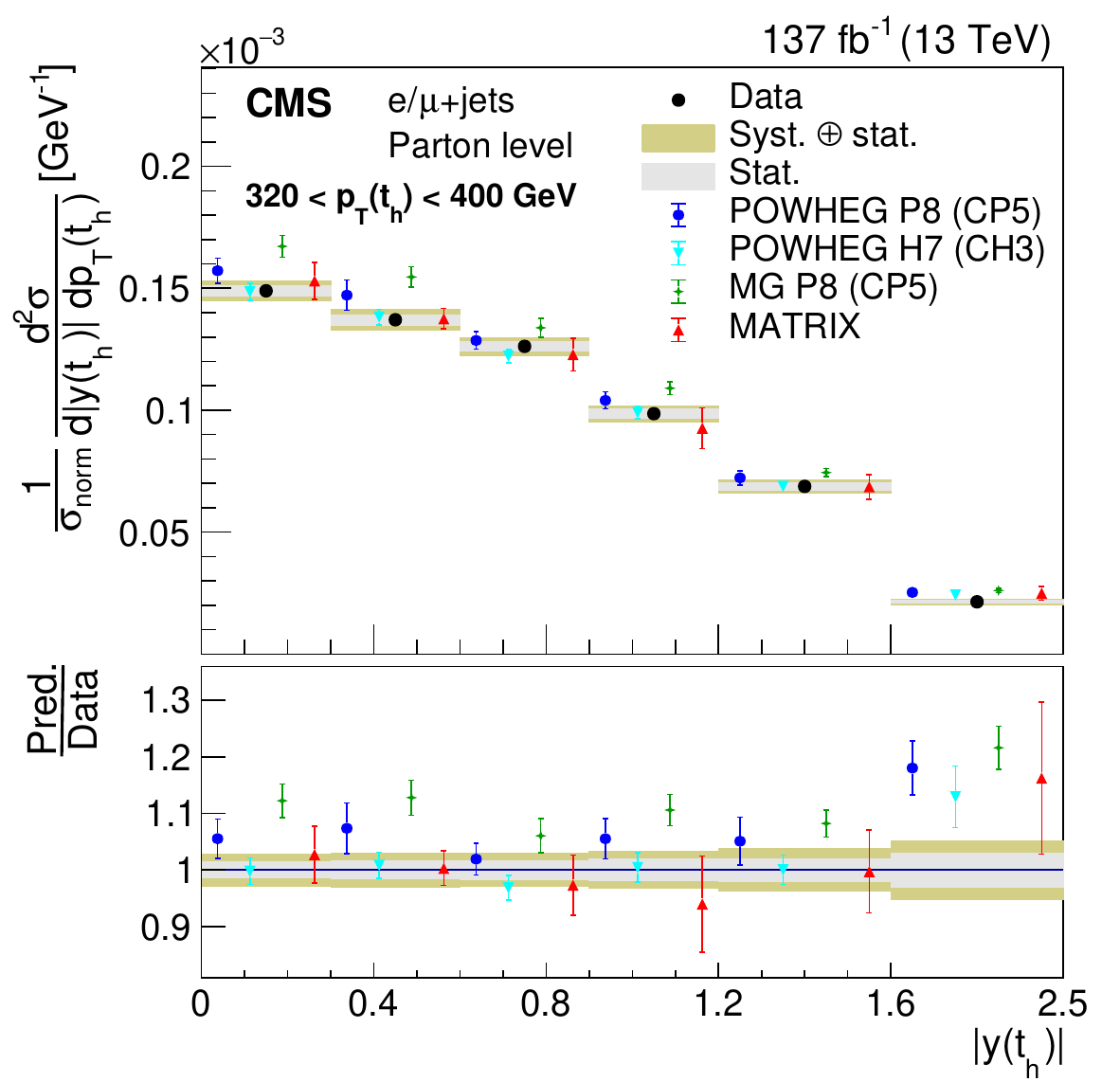}
 \includegraphics[width=0.42\textwidth]{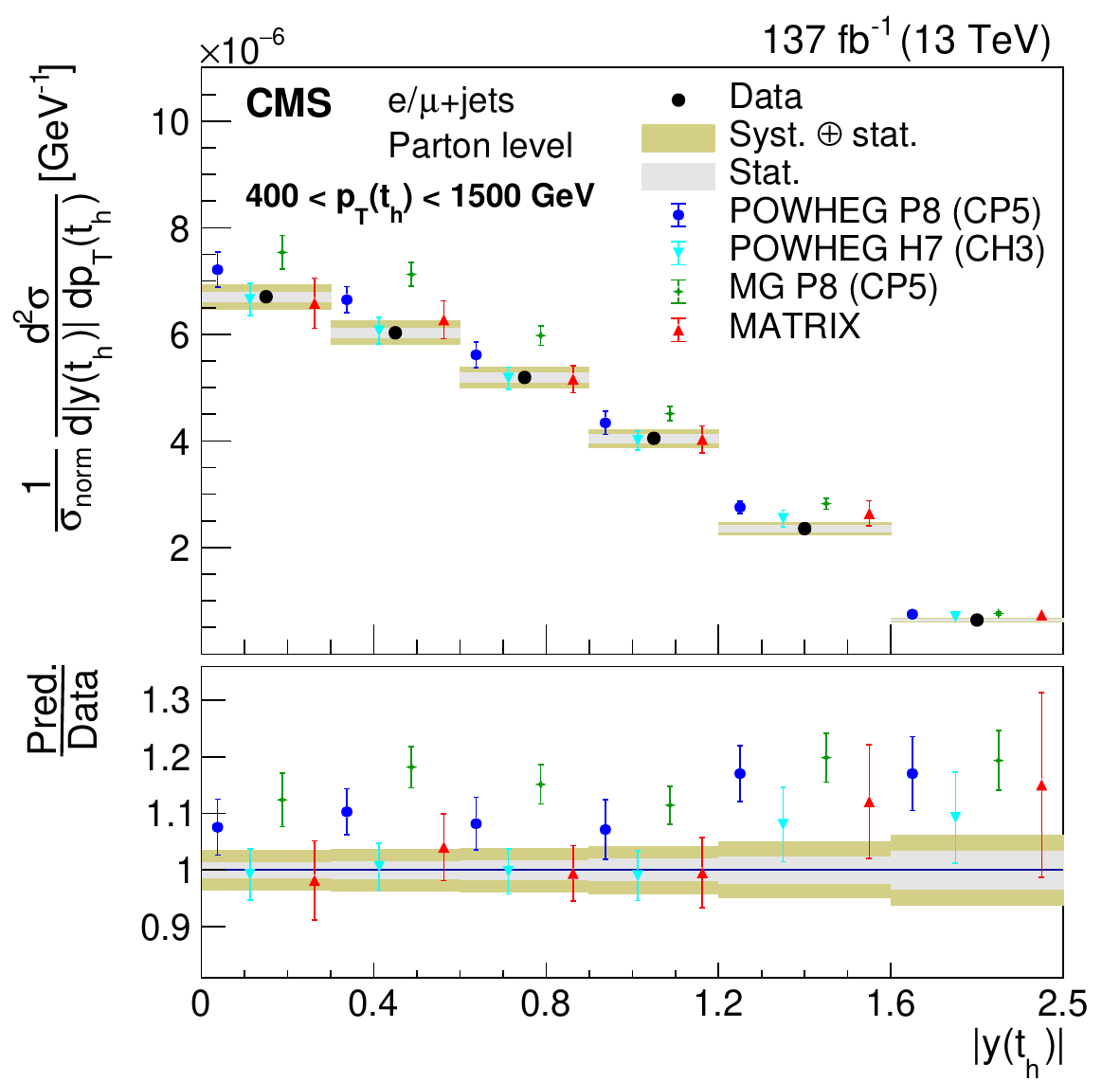}
 \caption{Normalized double-differential cross section at the parton level as a function of \thadptvsthady. \XSECCAPPA}
 \label{fig:RESNORM4}
\end{figure*}

\begin{figure*}[tbp]
\centering
 \includegraphics[width=0.42\textwidth]{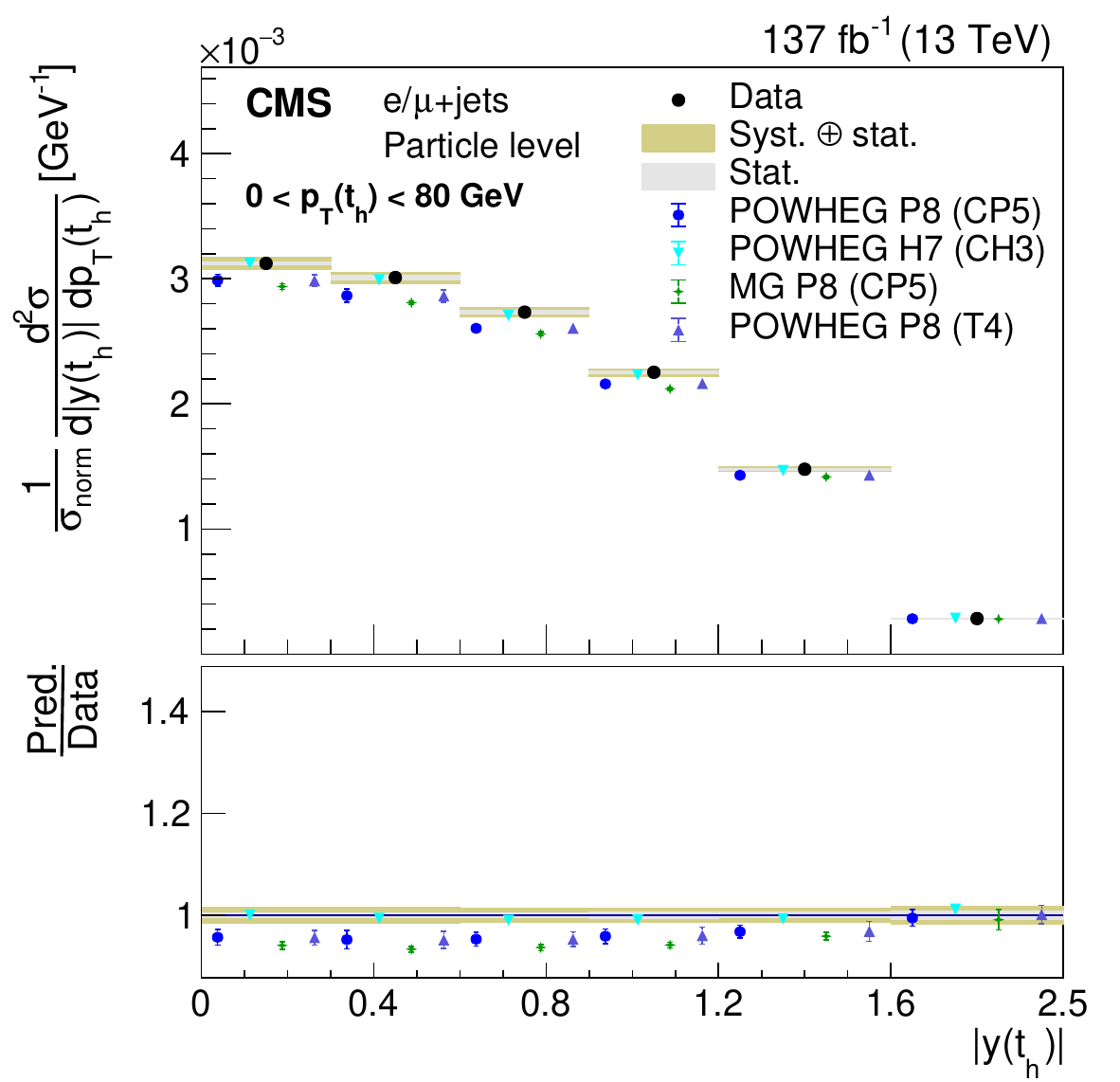}
 \includegraphics[width=0.42\textwidth]{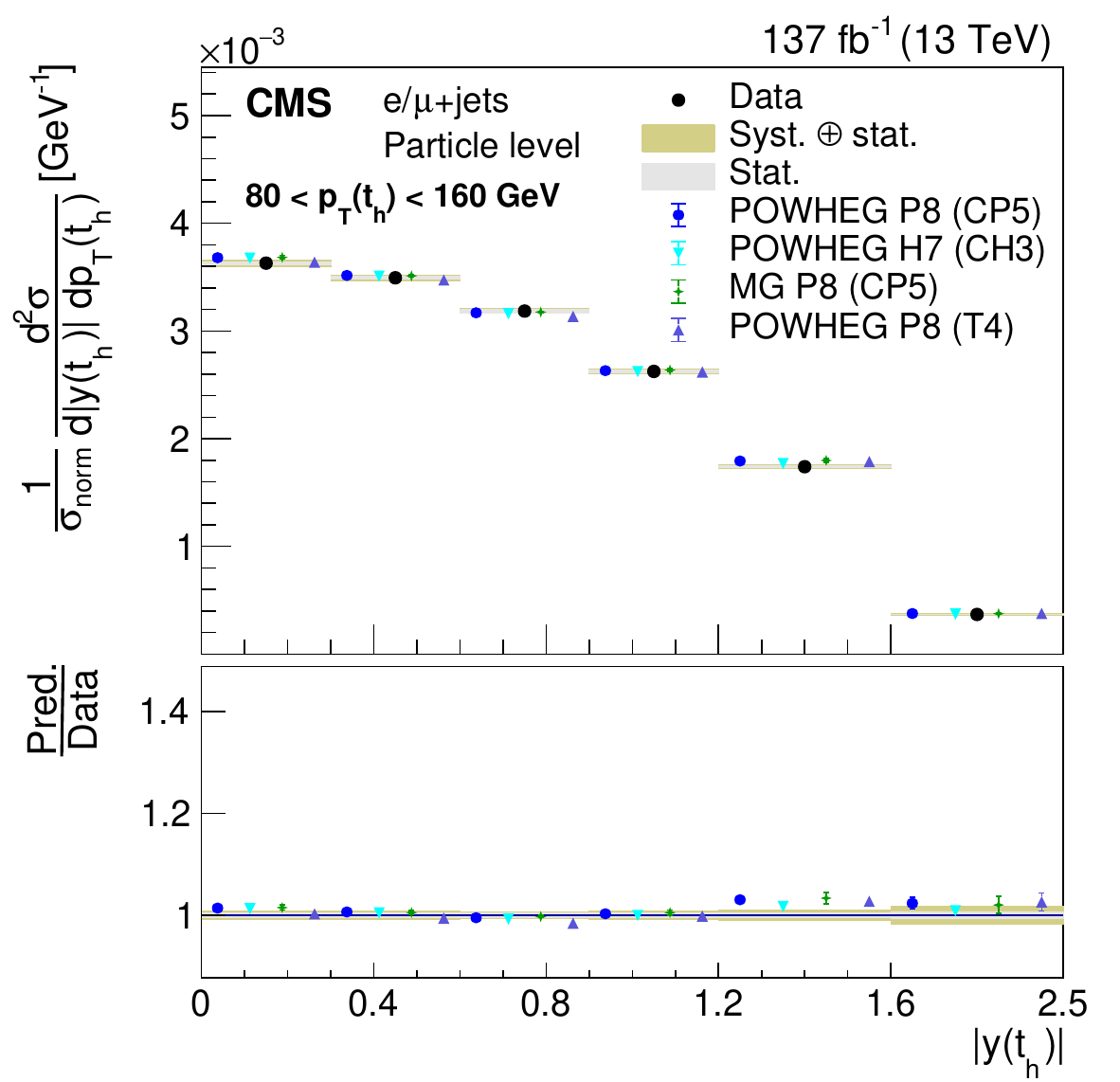}\\
 \includegraphics[width=0.42\textwidth]{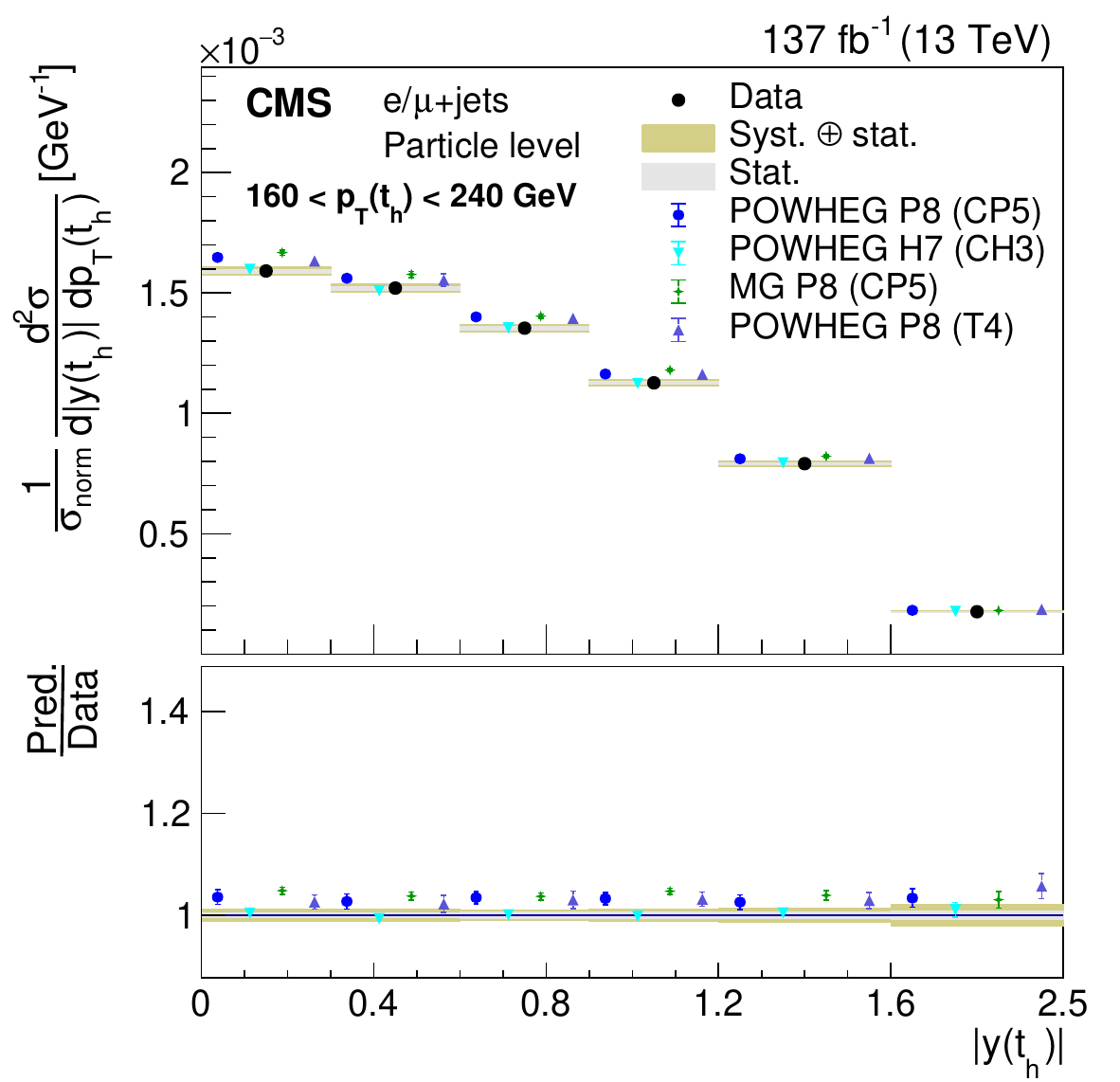}
 \includegraphics[width=0.42\textwidth]{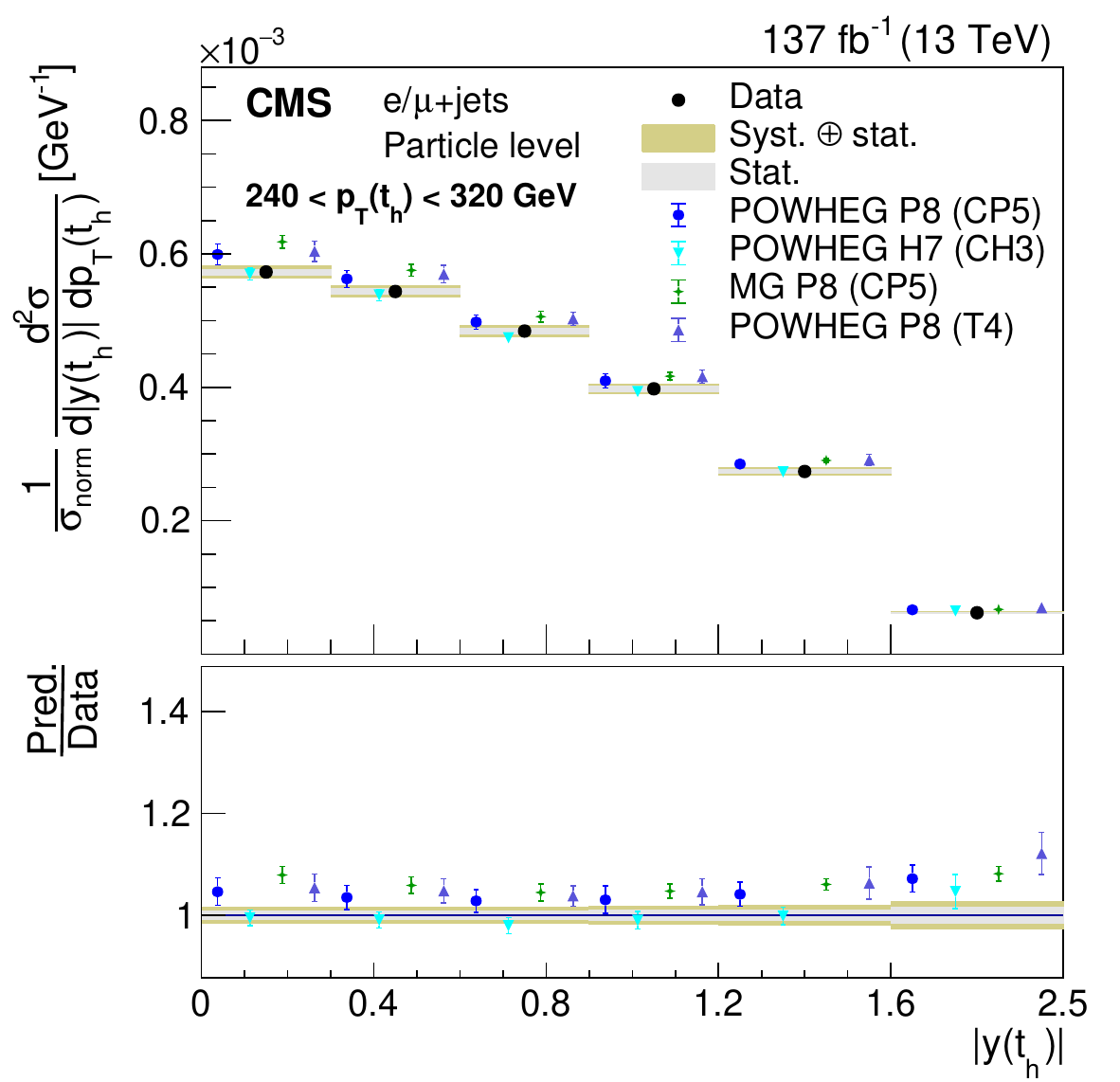}\\
 \includegraphics[width=0.42\textwidth]{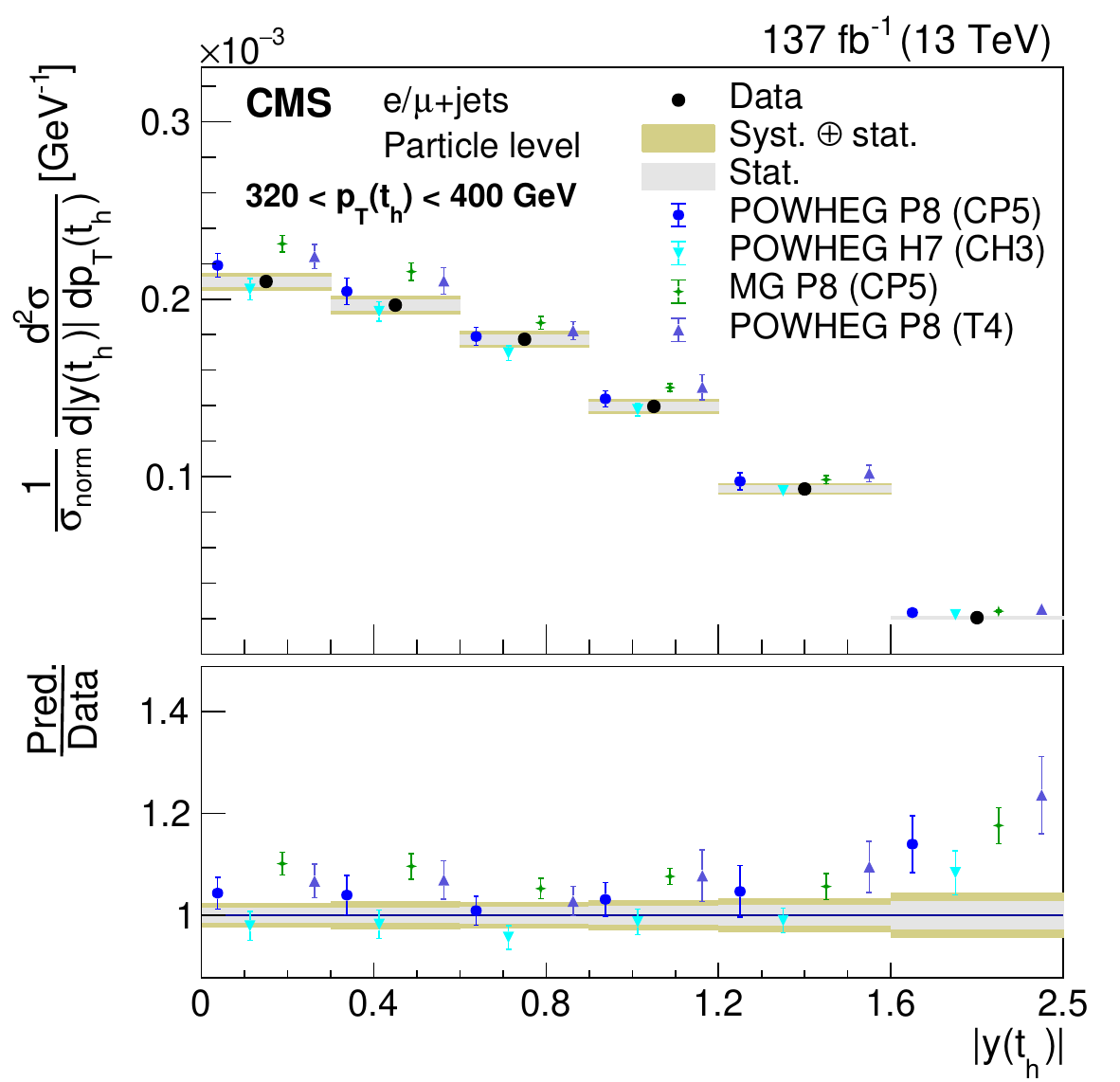}
 \includegraphics[width=0.42\textwidth]{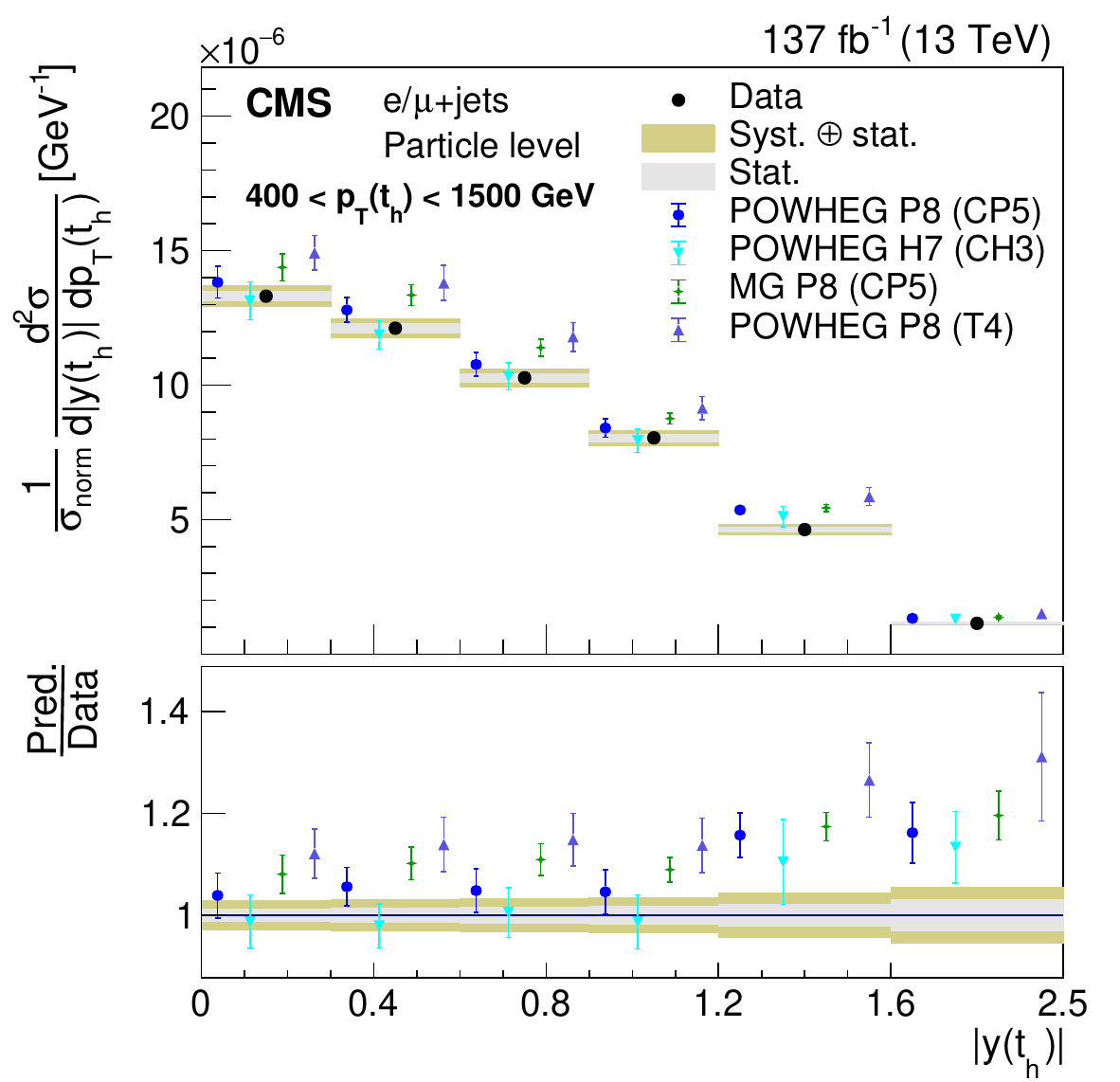}
 \caption{Normalized double-differential cross section at the particle level as a function of \thadptvsthady. \XSECCAPPS}
 \label{fig:RESNORMPS4}
\end{figure*}

\begin{figure*}[tbp]
\centering
 \includegraphics[width=0.42\textwidth]{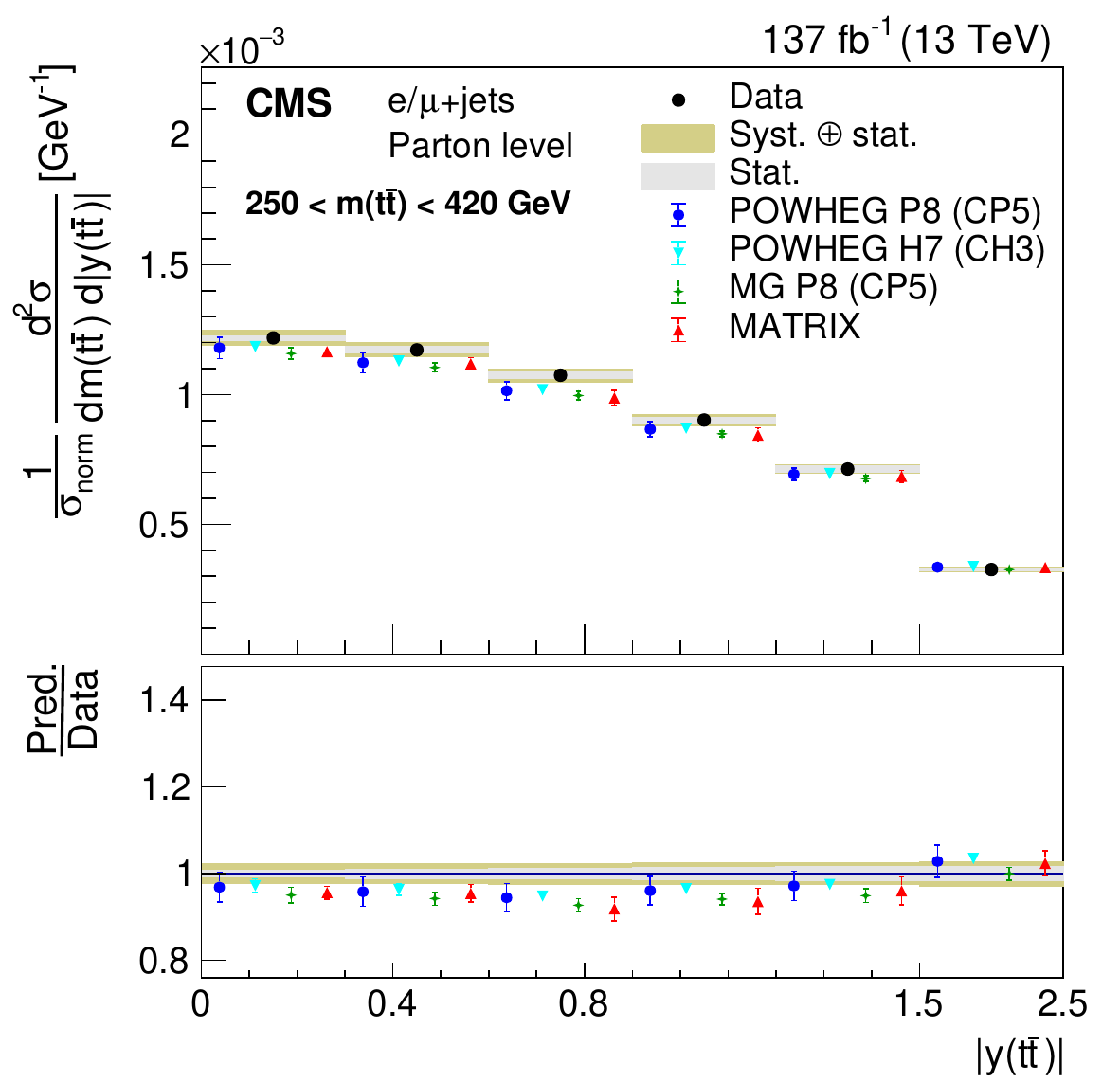}
 \includegraphics[width=0.42\textwidth]{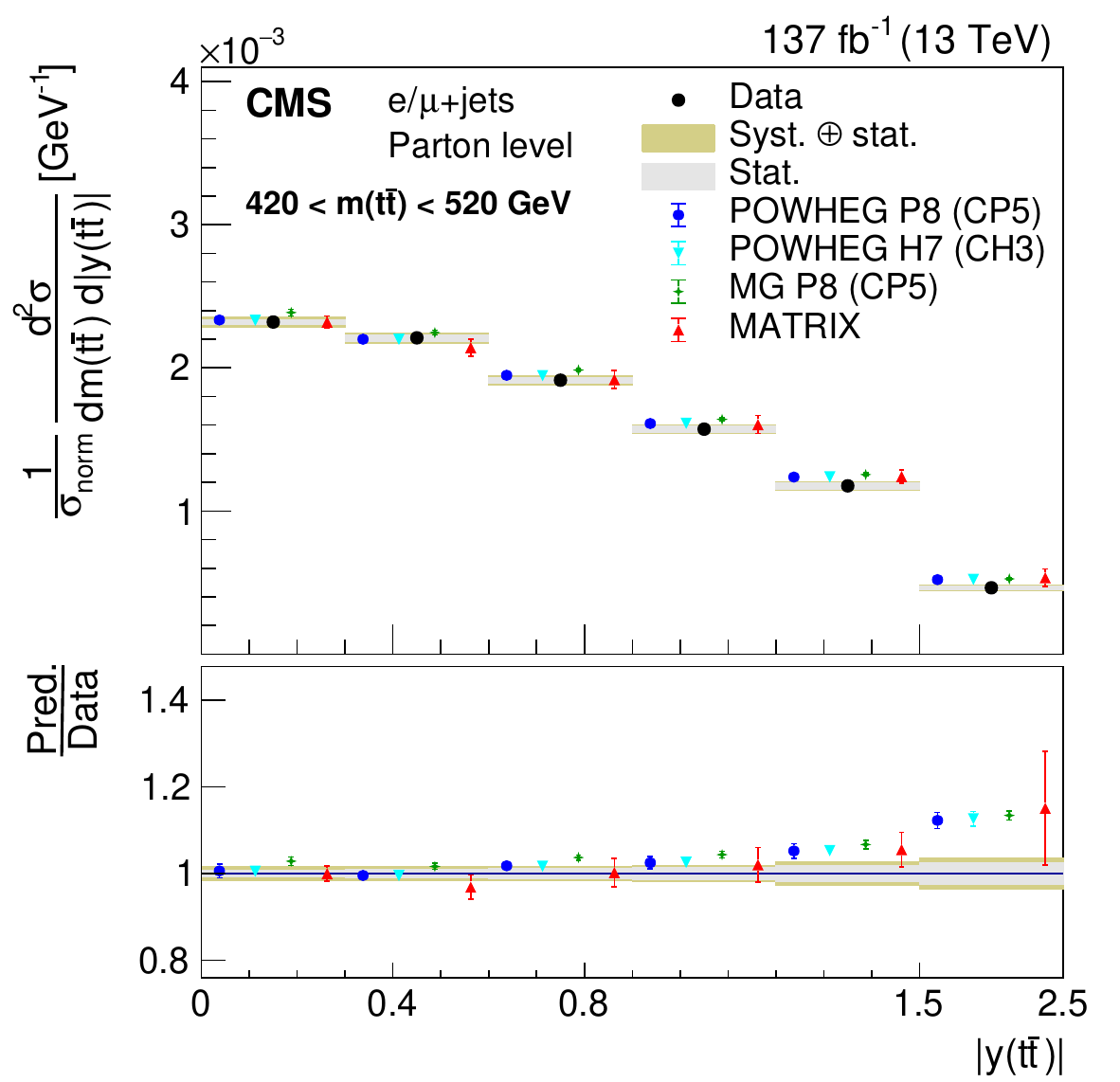}\\
 \includegraphics[width=0.42\textwidth]{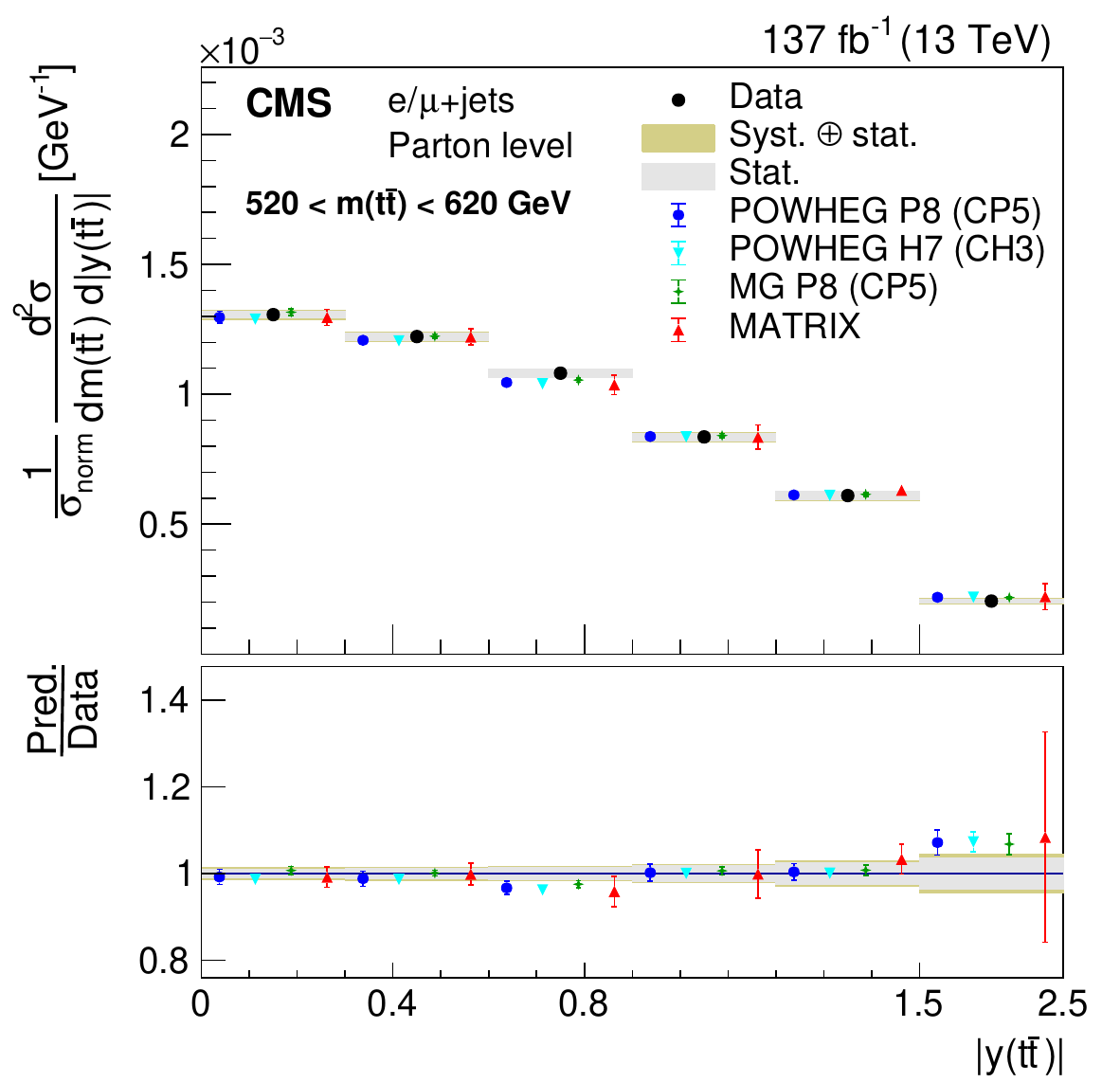}
 \includegraphics[width=0.42\textwidth]{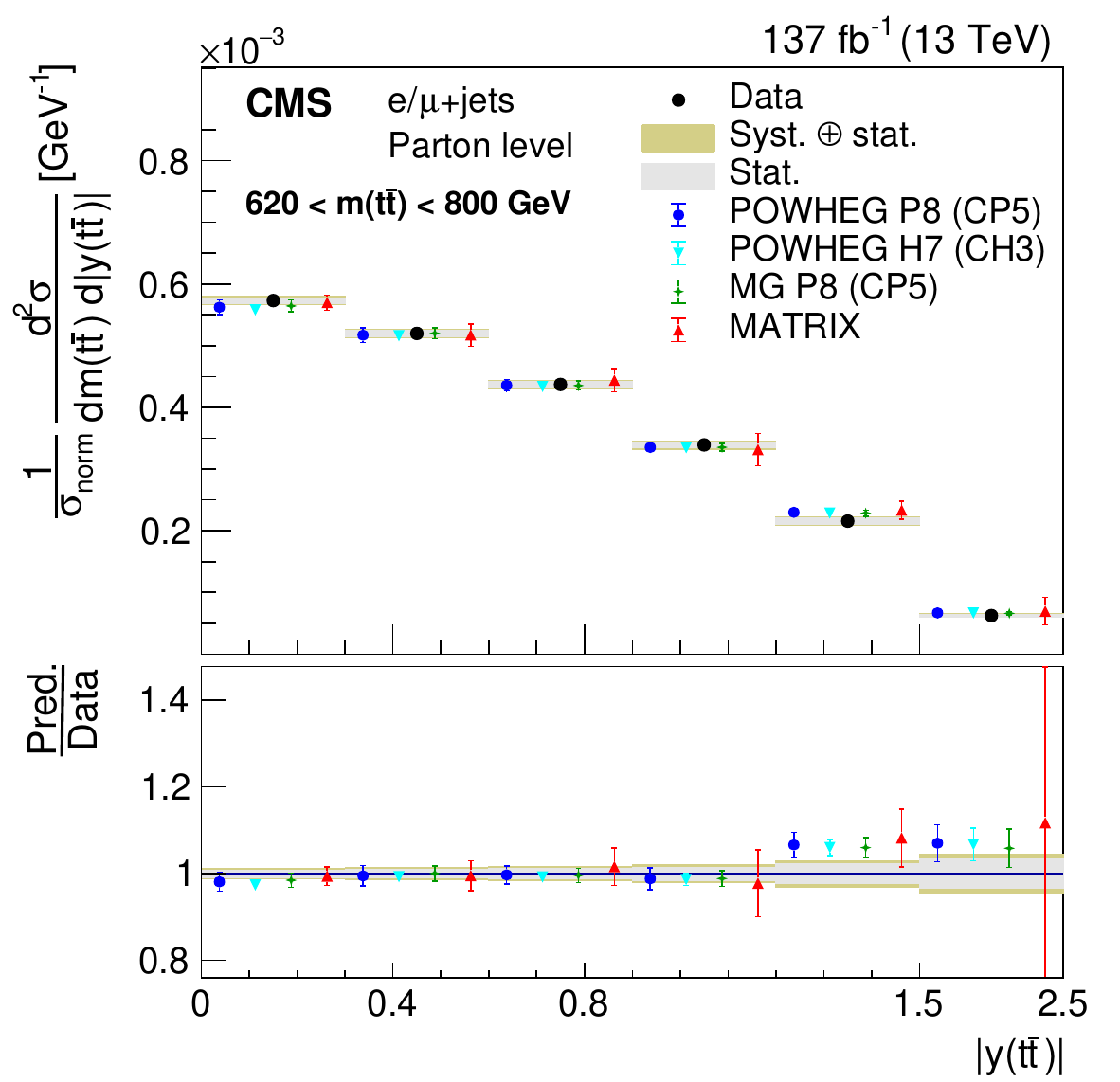}\\
 \includegraphics[width=0.42\textwidth]{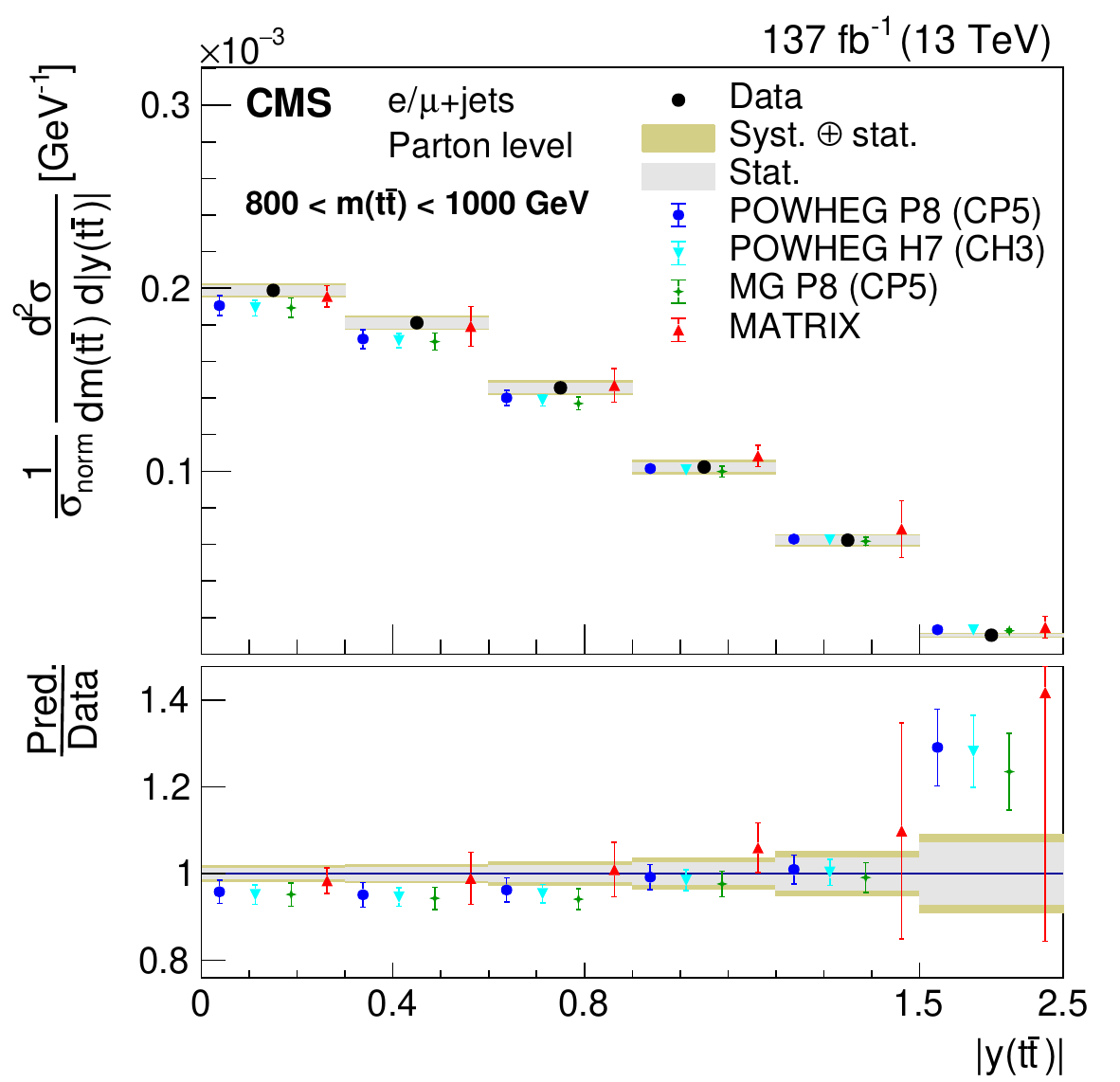}
 \includegraphics[width=0.42\textwidth]{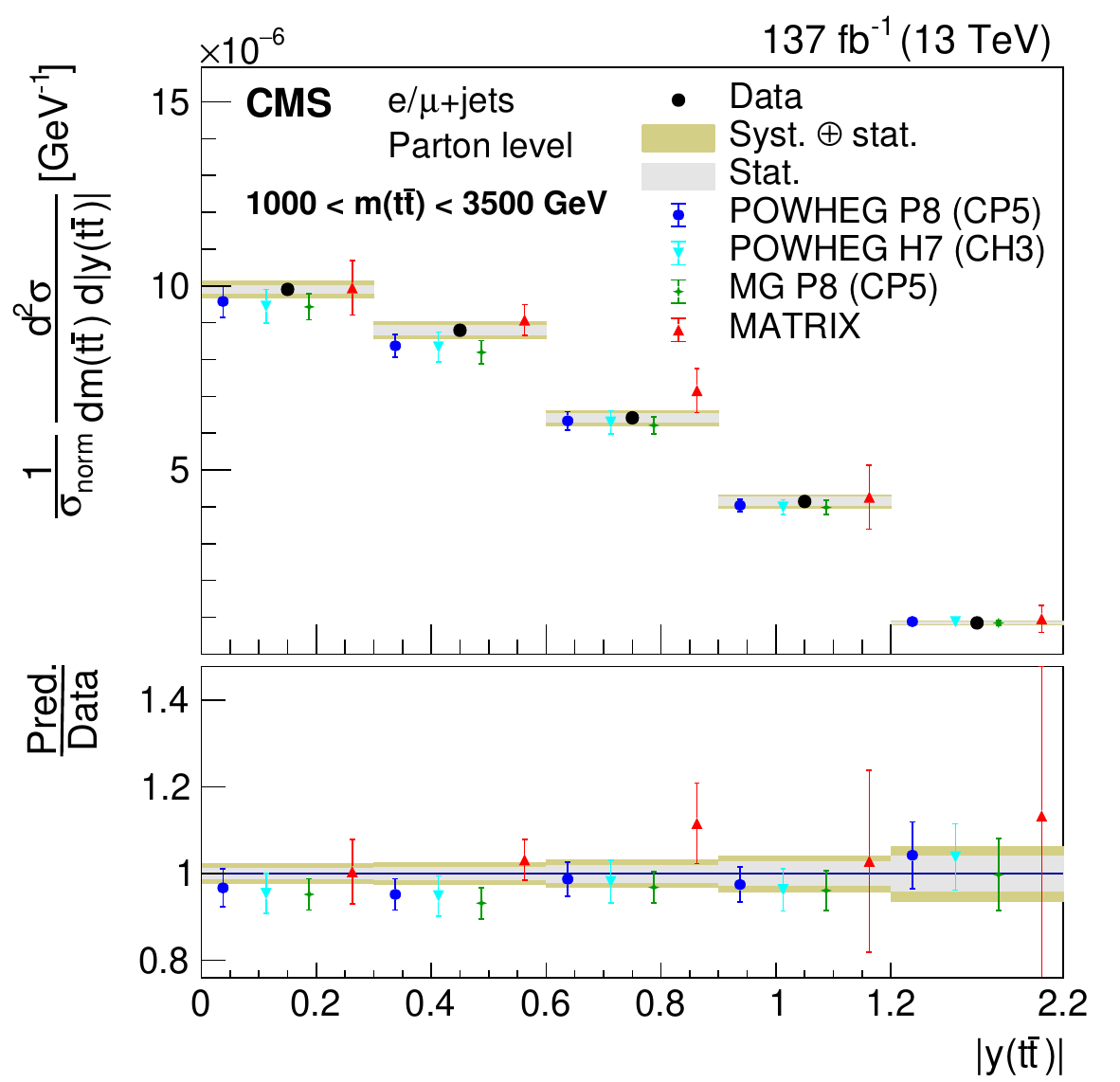}
 \caption{Normalized double-differential cross section at the parton level as a function of \ttmvstty. \XSECCAPPA}
 \label{fig:RESNORM5}
\end{figure*}

\begin{figure*}[tbp]
\centering
 \includegraphics[width=0.42\textwidth]{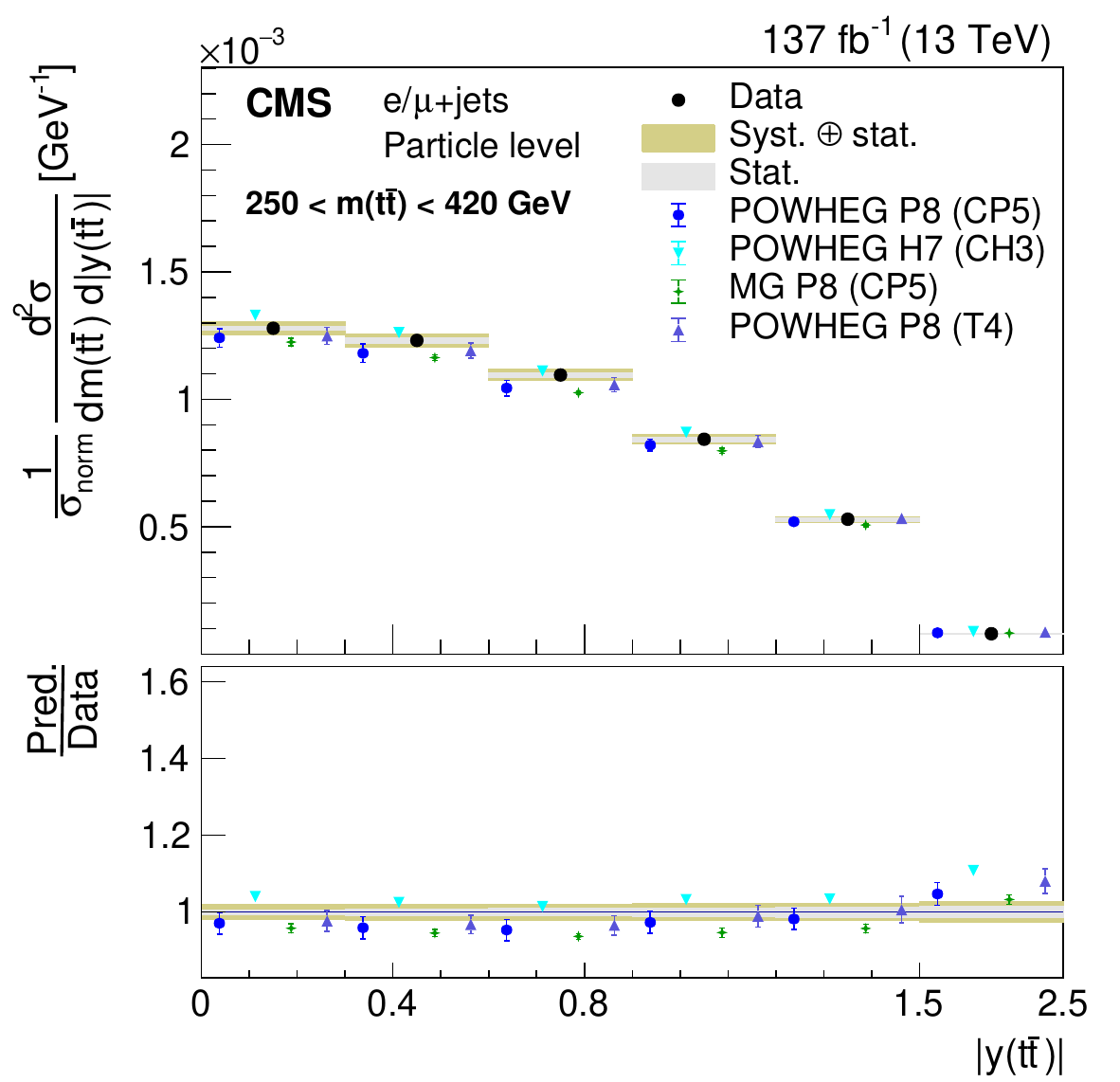}
 \includegraphics[width=0.42\textwidth]{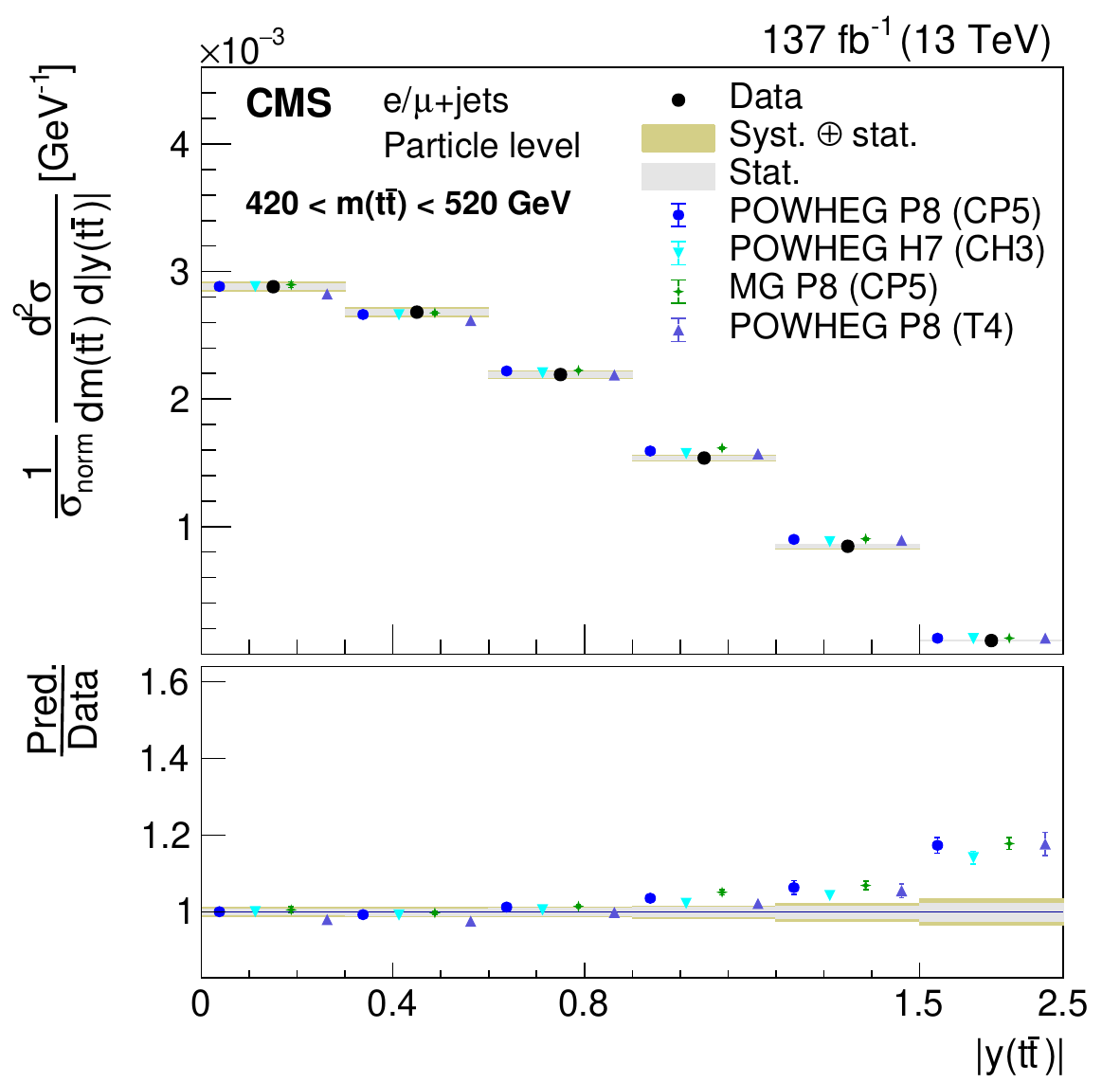}\\
 \includegraphics[width=0.42\textwidth]{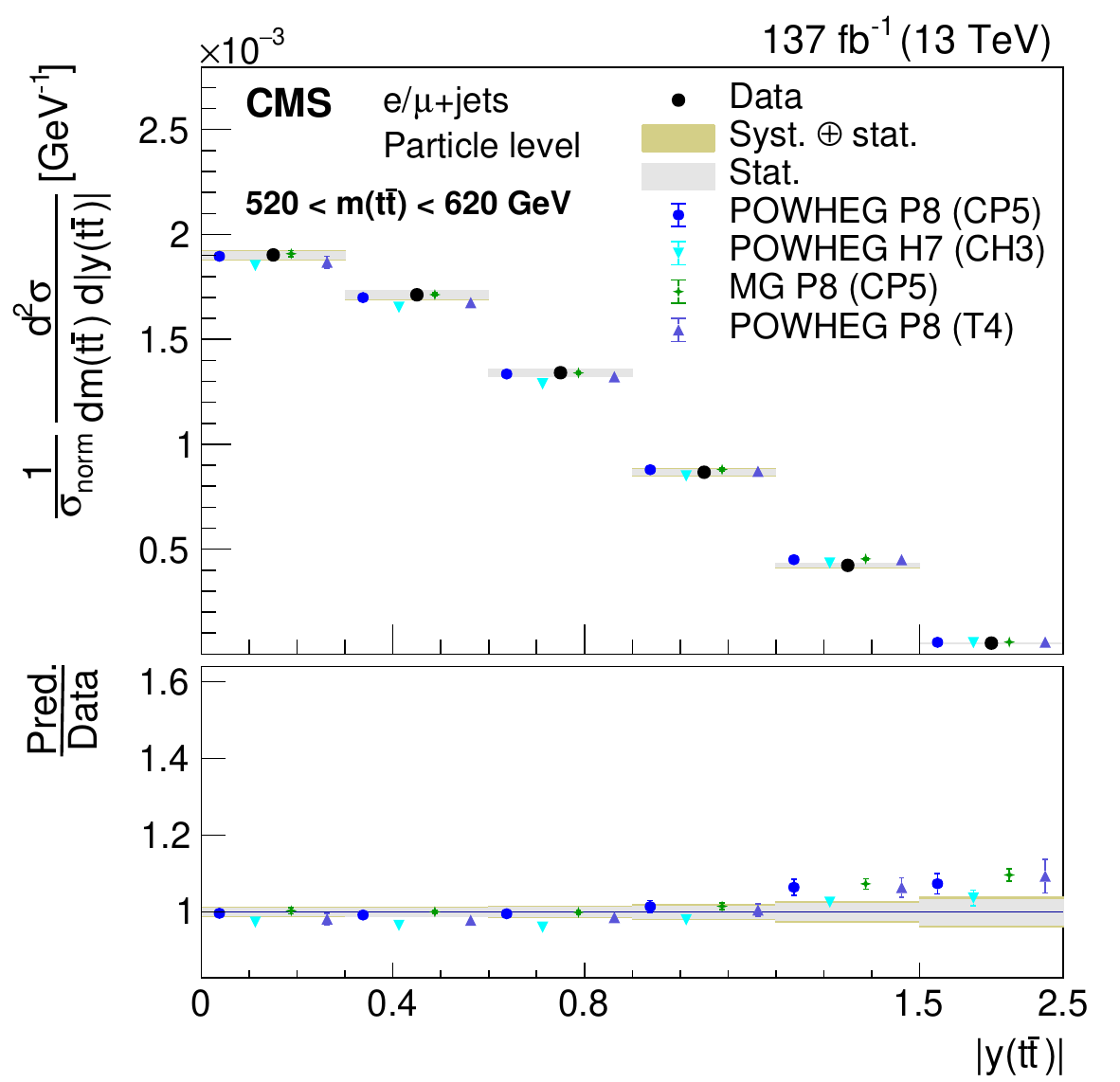}
 \includegraphics[width=0.42\textwidth]{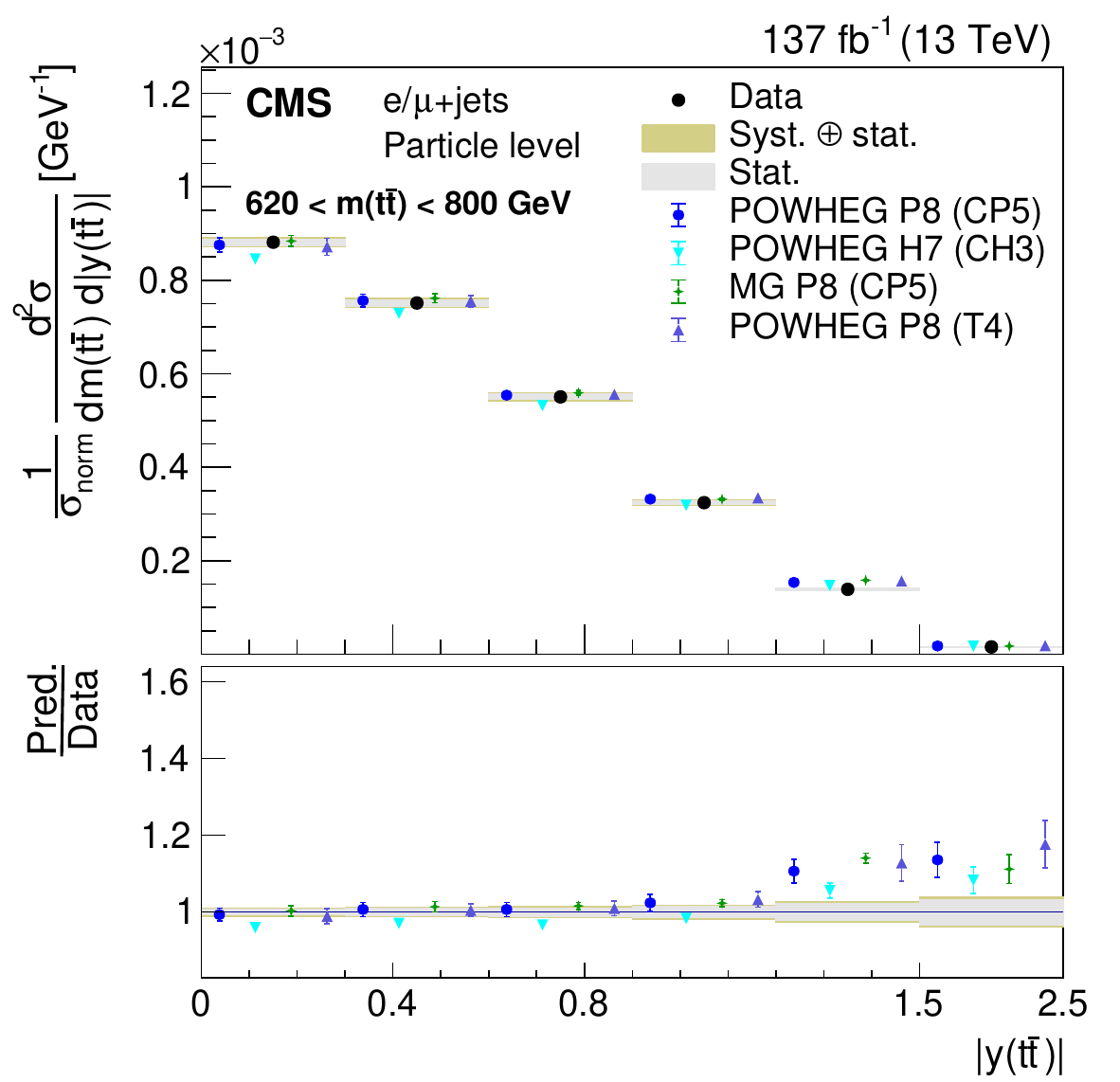}\\
 \includegraphics[width=0.42\textwidth]{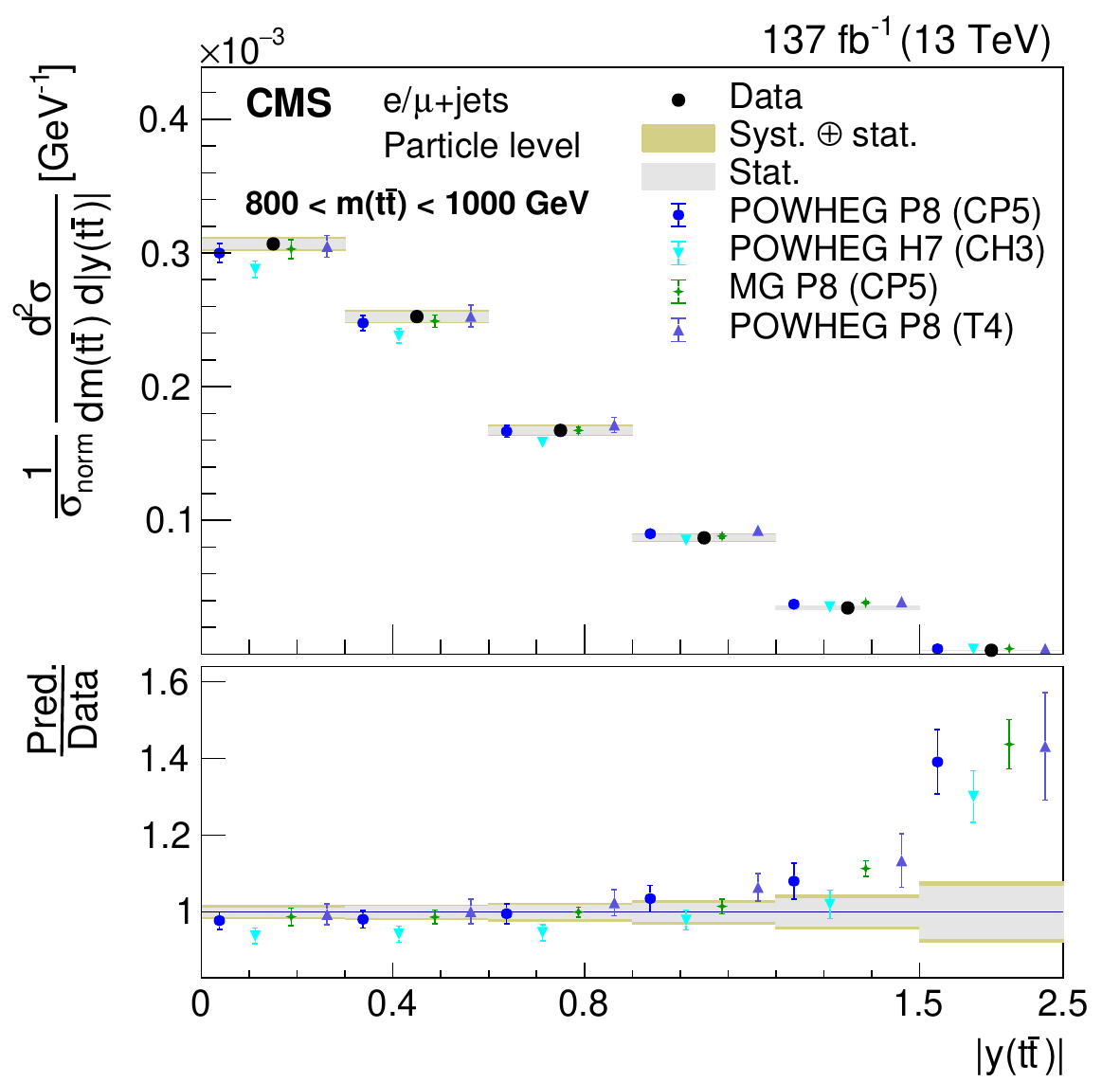}
 \includegraphics[width=0.42\textwidth]{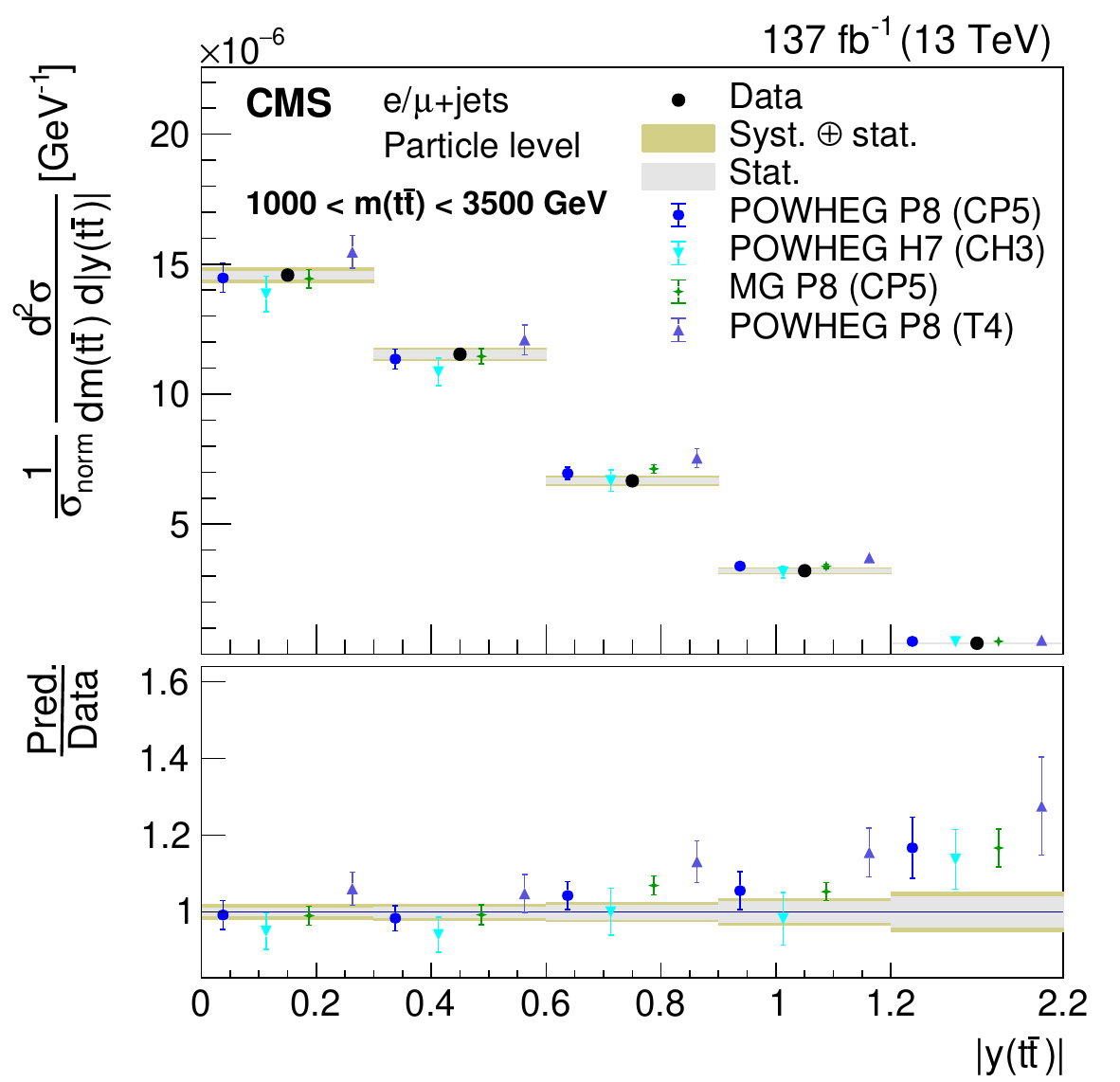}
 \caption{Normalized double-differential cross section at the particle level as a function of \ttmvstty. \XSECCAPPS}
 \label{fig:RESNORMPS5}
\end{figure*}

\begin{figure*}[tbp]
\centering
 \includegraphics[width=0.42\textwidth]{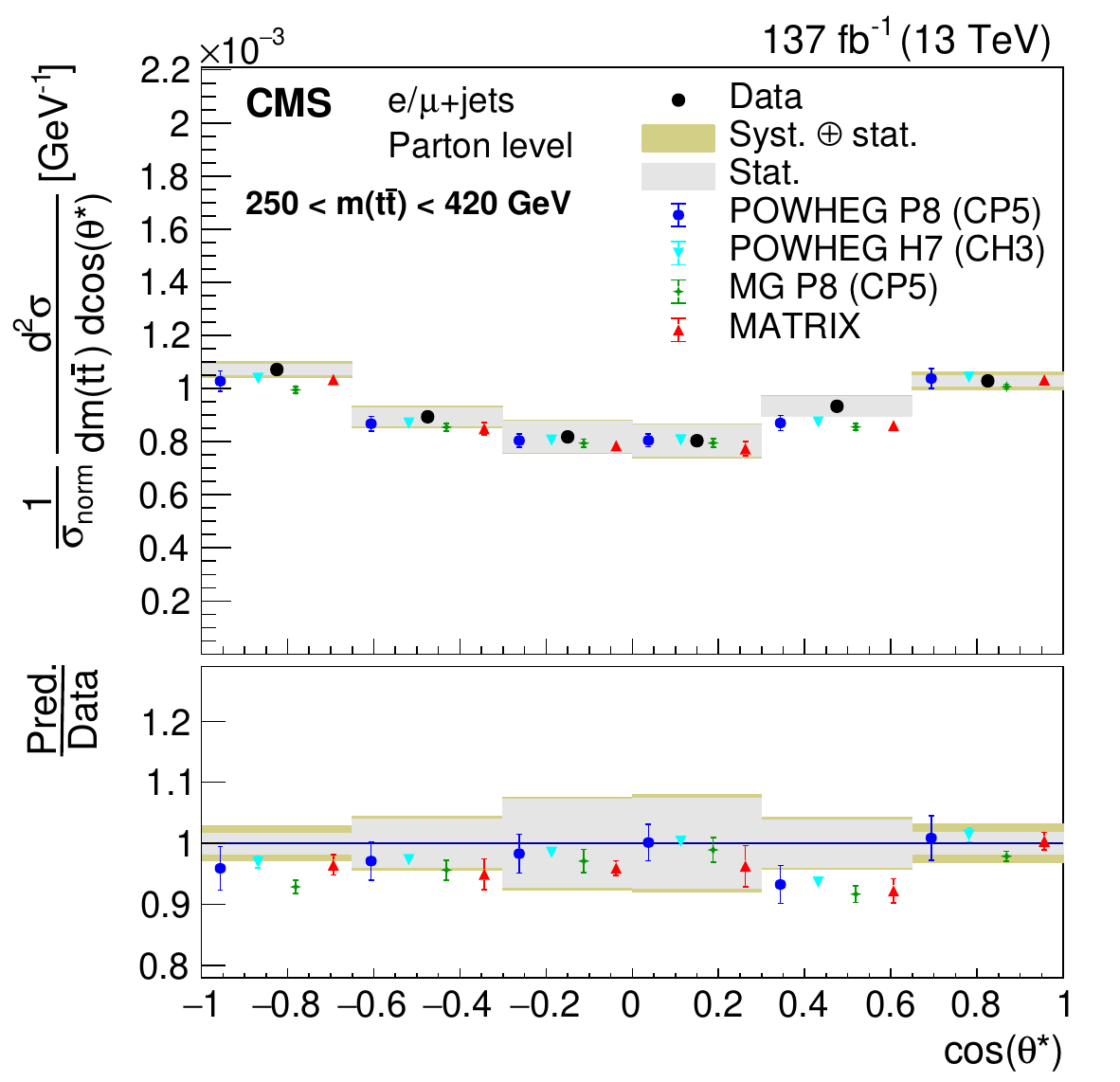}
 \includegraphics[width=0.42\textwidth]{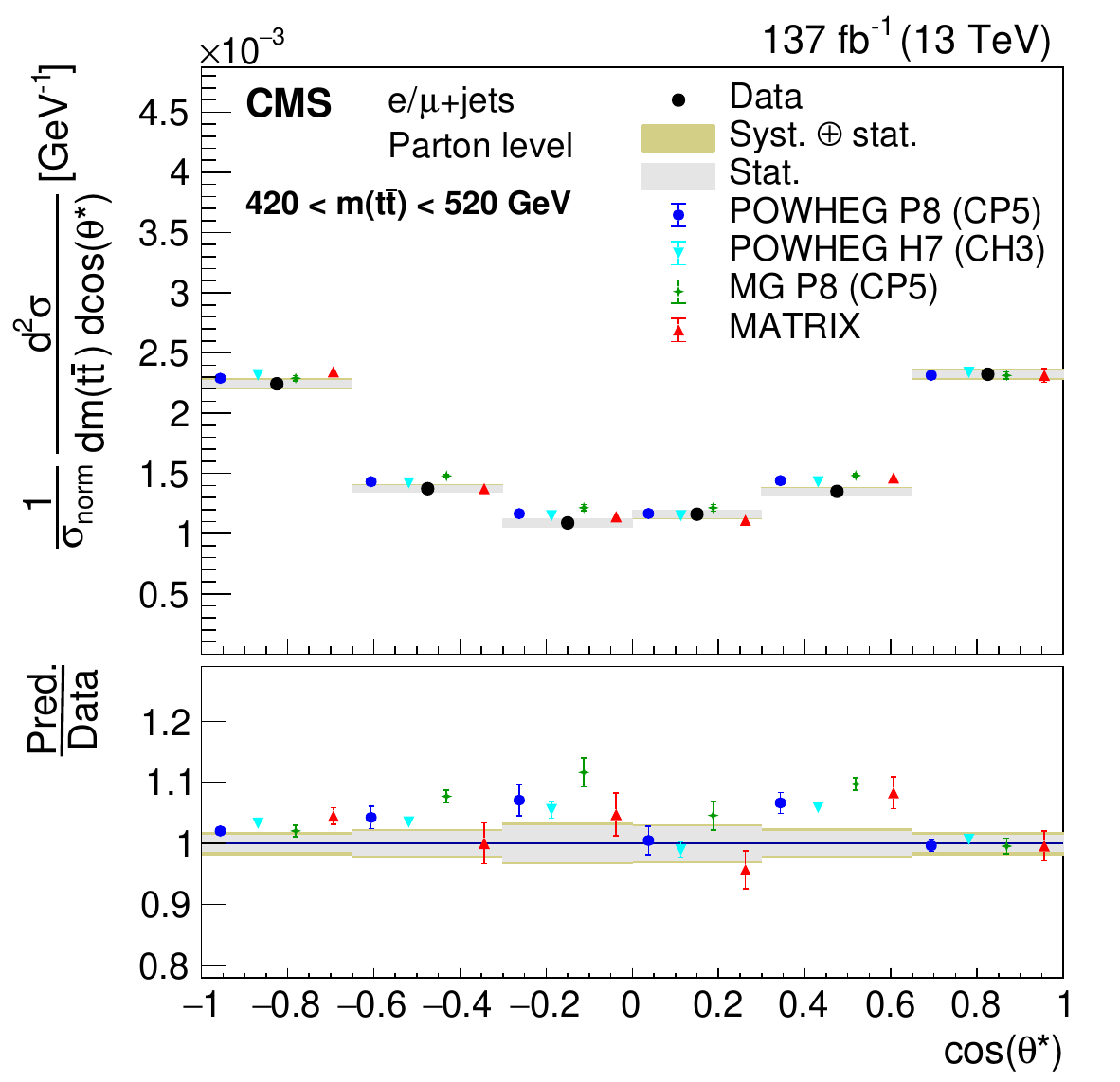}\\
 \includegraphics[width=0.42\textwidth]{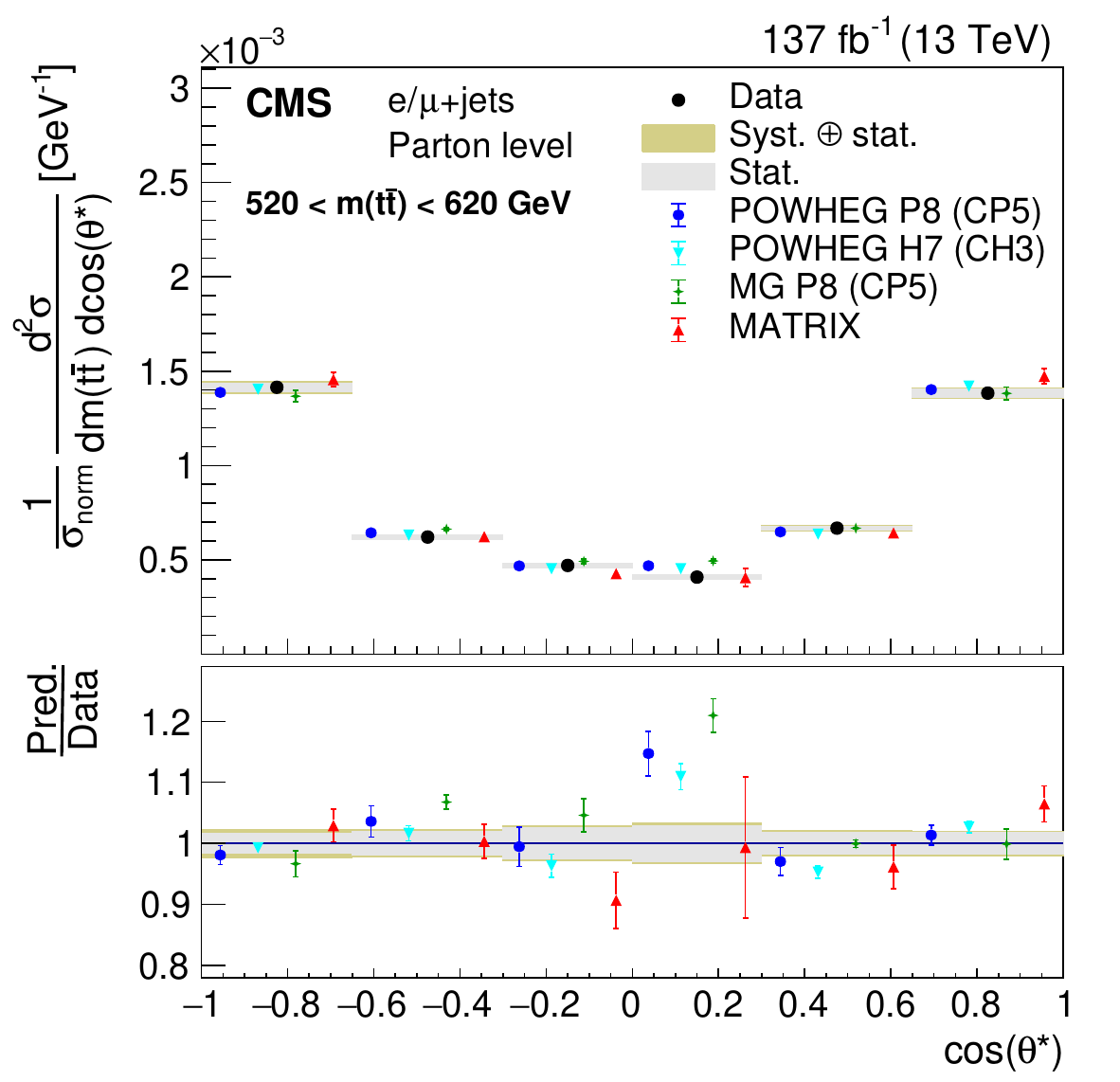}
 \includegraphics[width=0.42\textwidth]{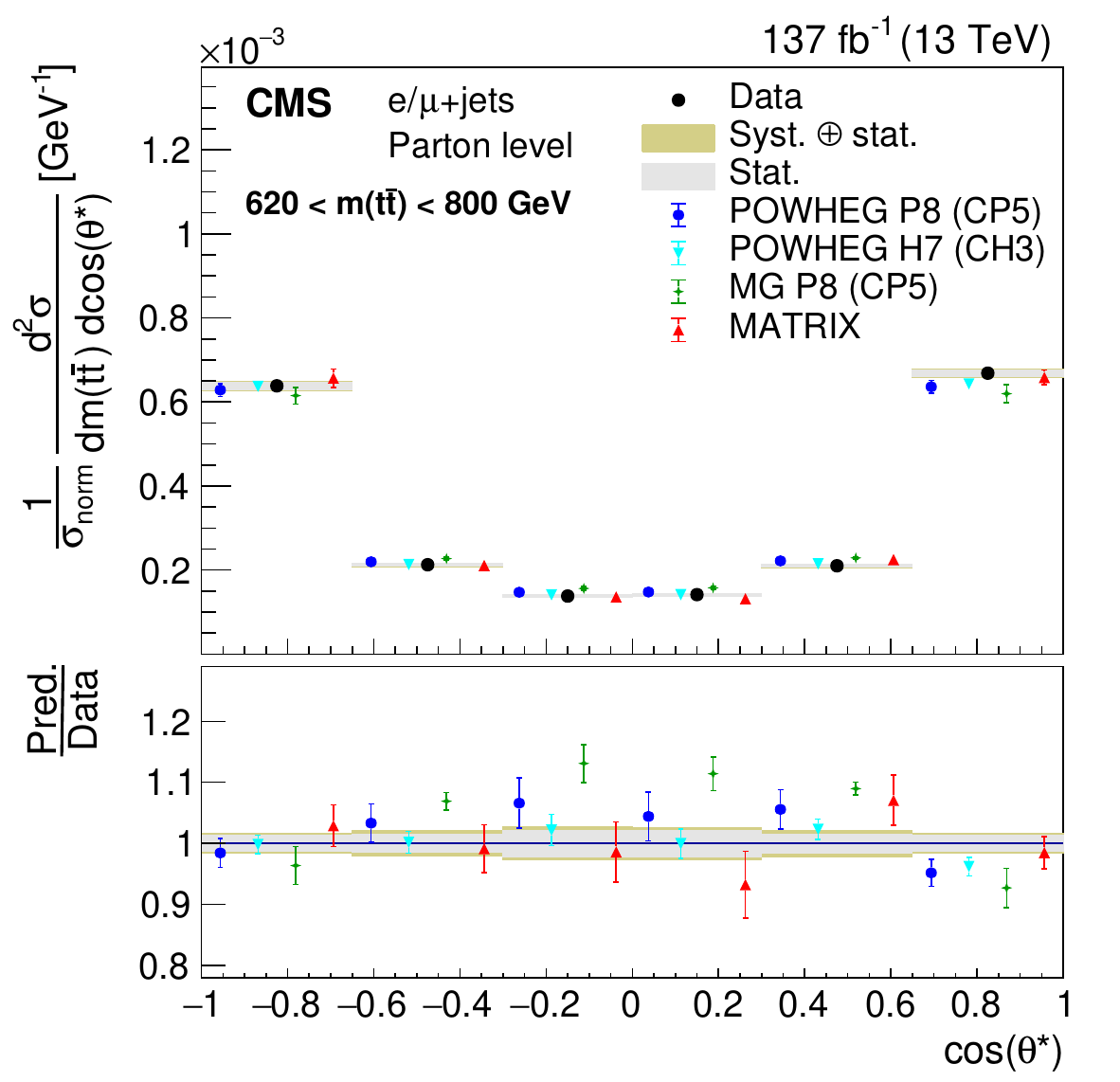}\\
 \includegraphics[width=0.42\textwidth]{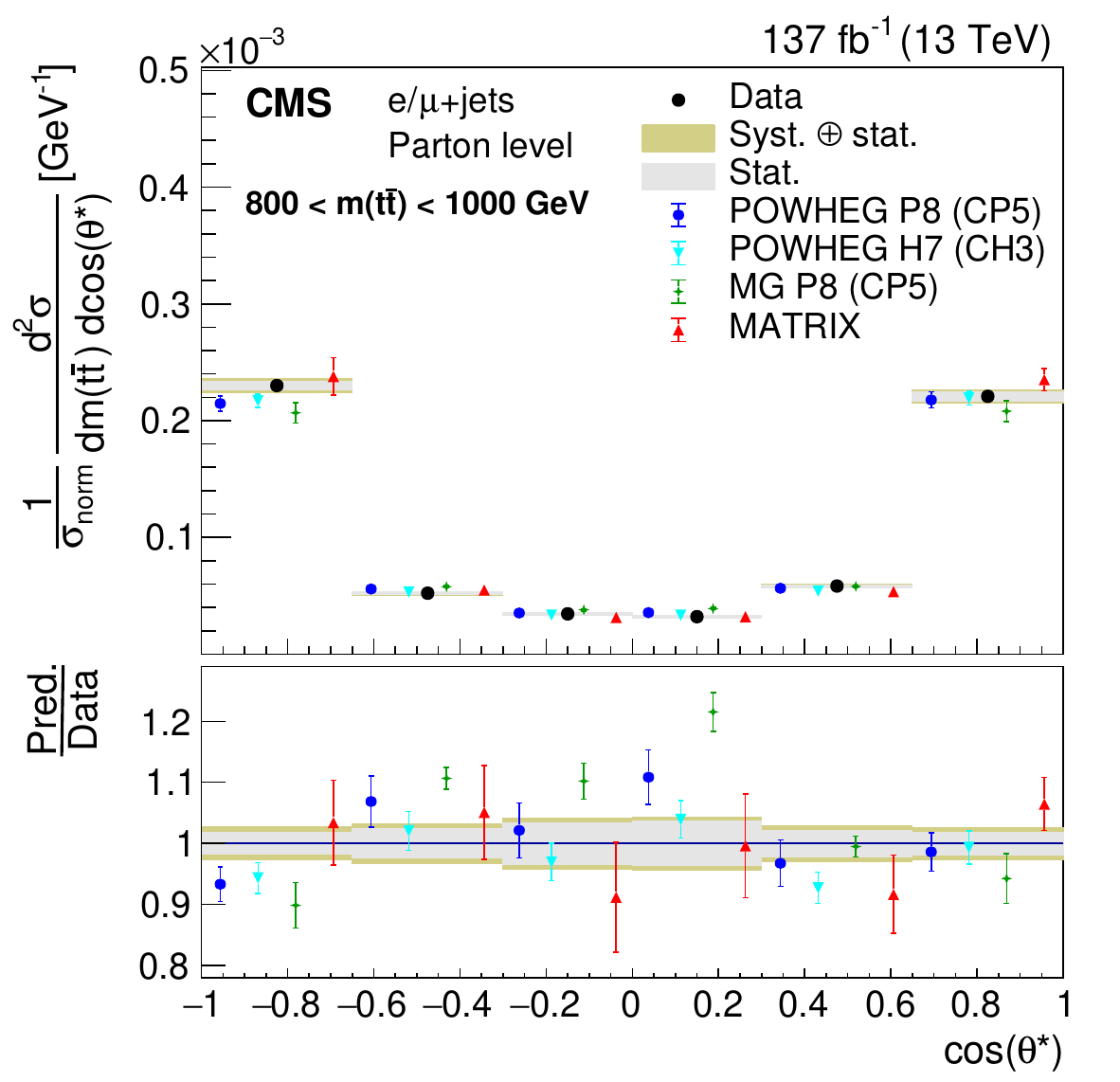}
 \includegraphics[width=0.42\textwidth]{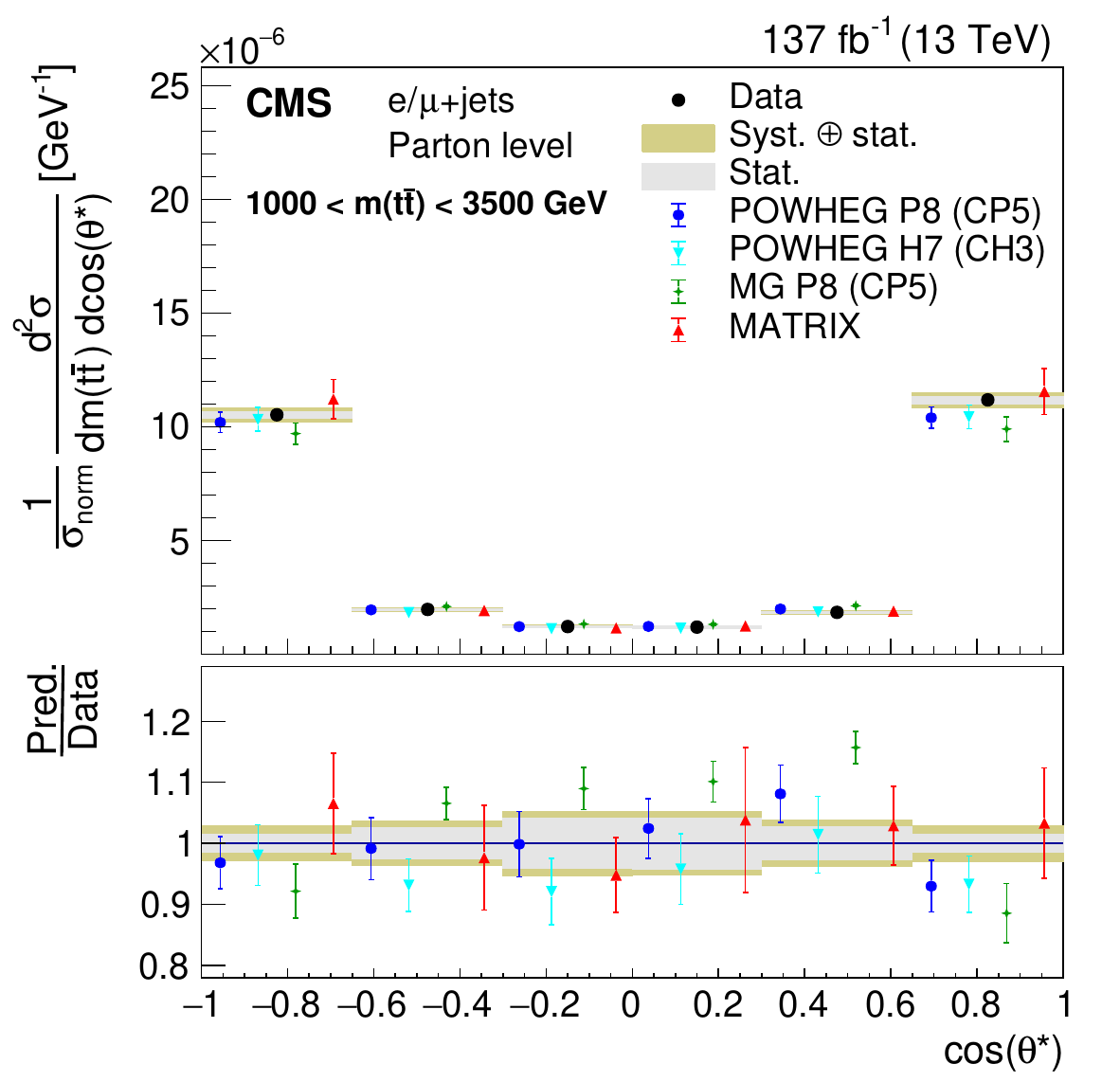}
 \caption{Normalized double-differential cross section at the parton level as a function of \ttmvscts. \XSECCAPPA}
 \label{fig:RESNORM6}
\end{figure*}

\begin{figure*}[tbp]
\centering
 \includegraphics[width=0.42\textwidth]{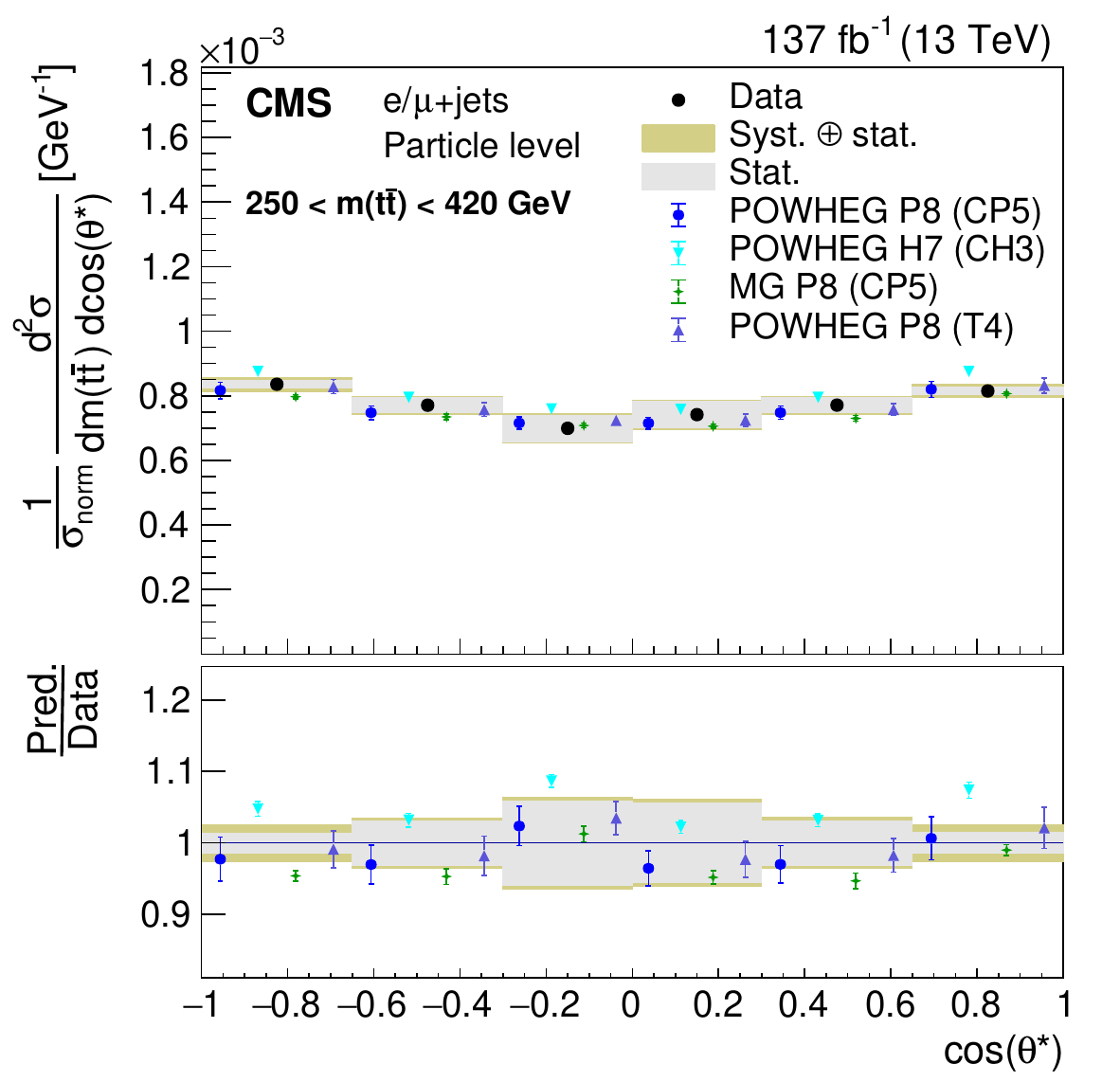}
 \includegraphics[width=0.42\textwidth]{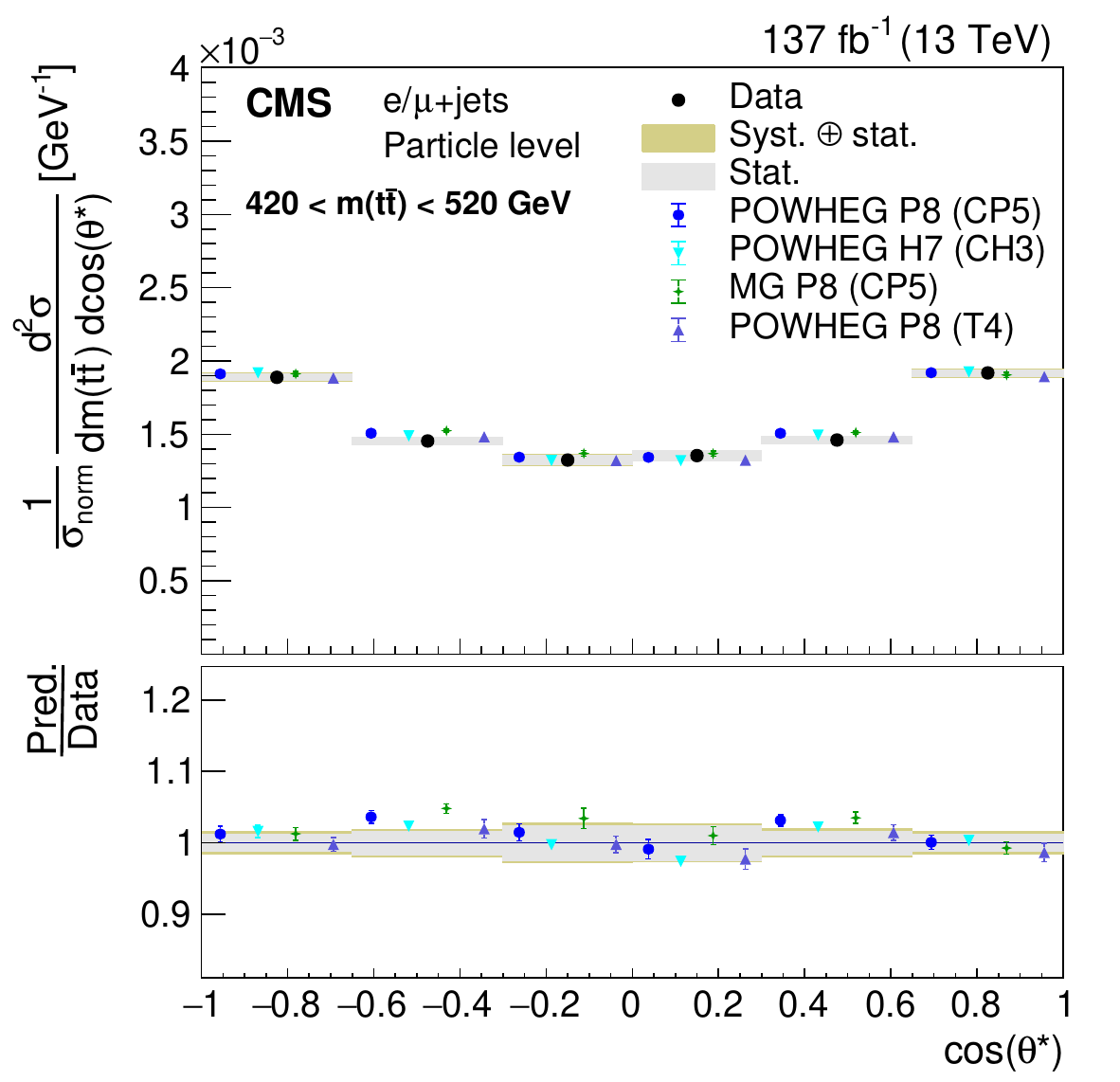}\\
 \includegraphics[width=0.42\textwidth]{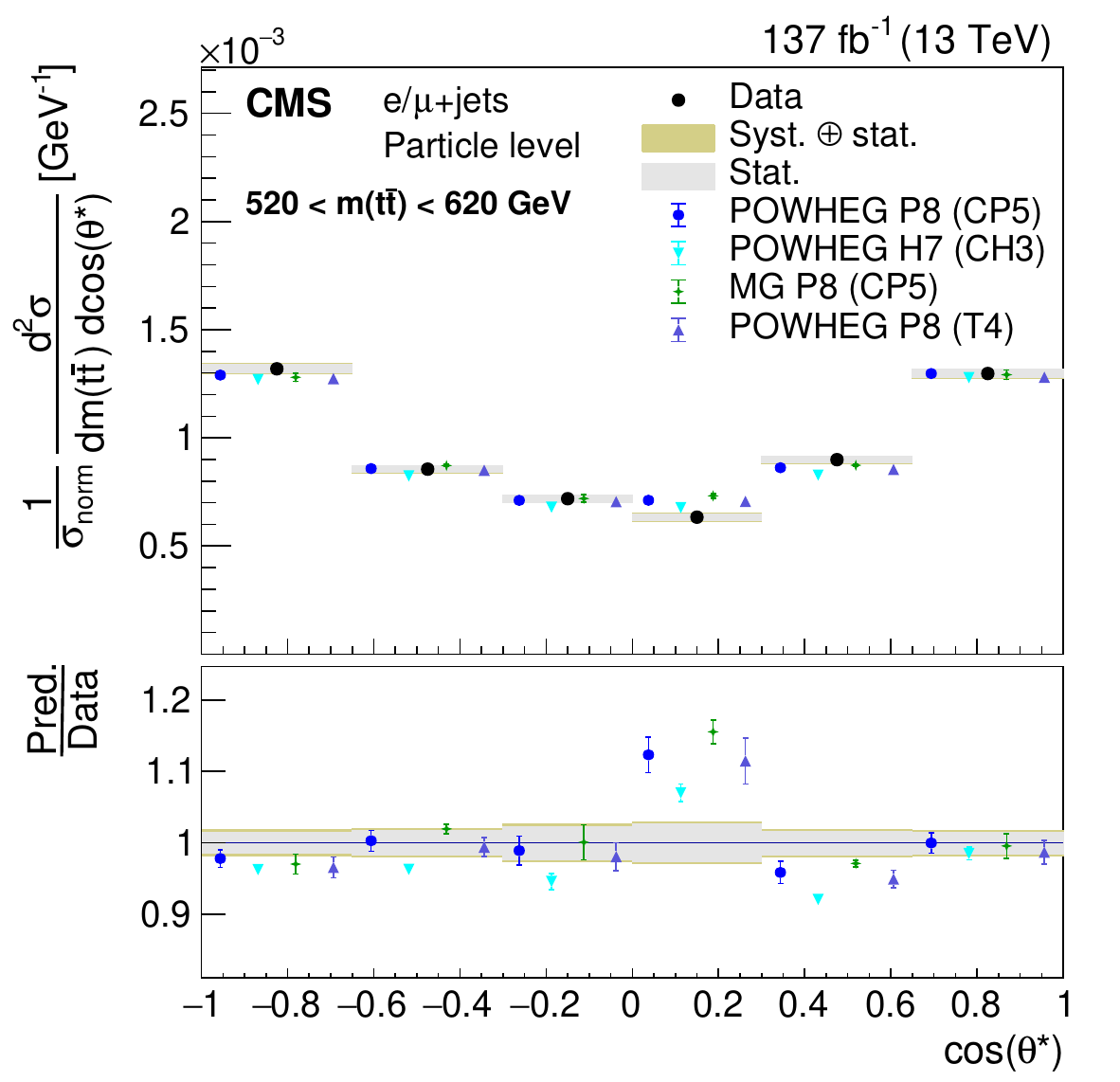}
 \includegraphics[width=0.42\textwidth]{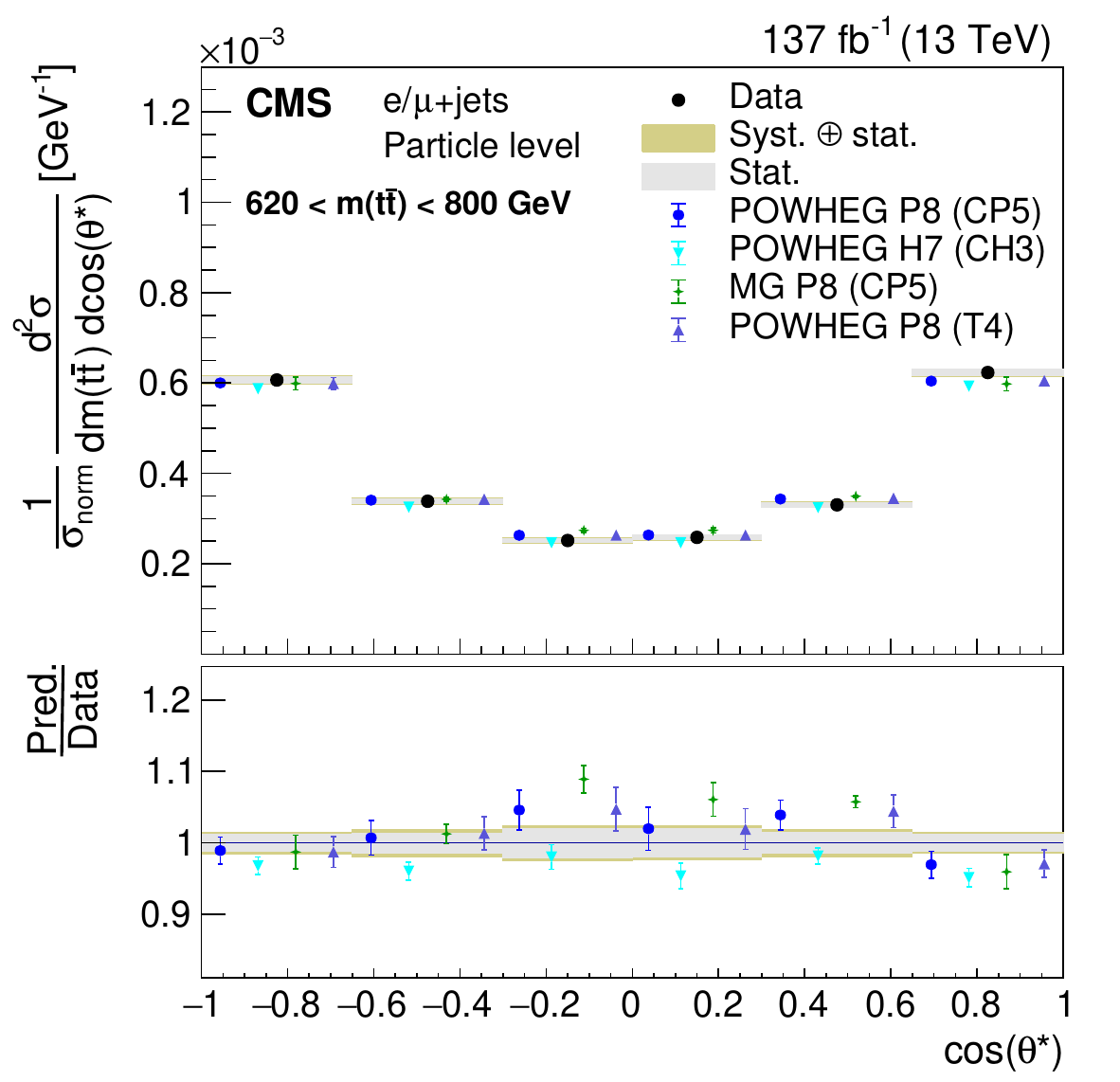}\\
 \includegraphics[width=0.42\textwidth]{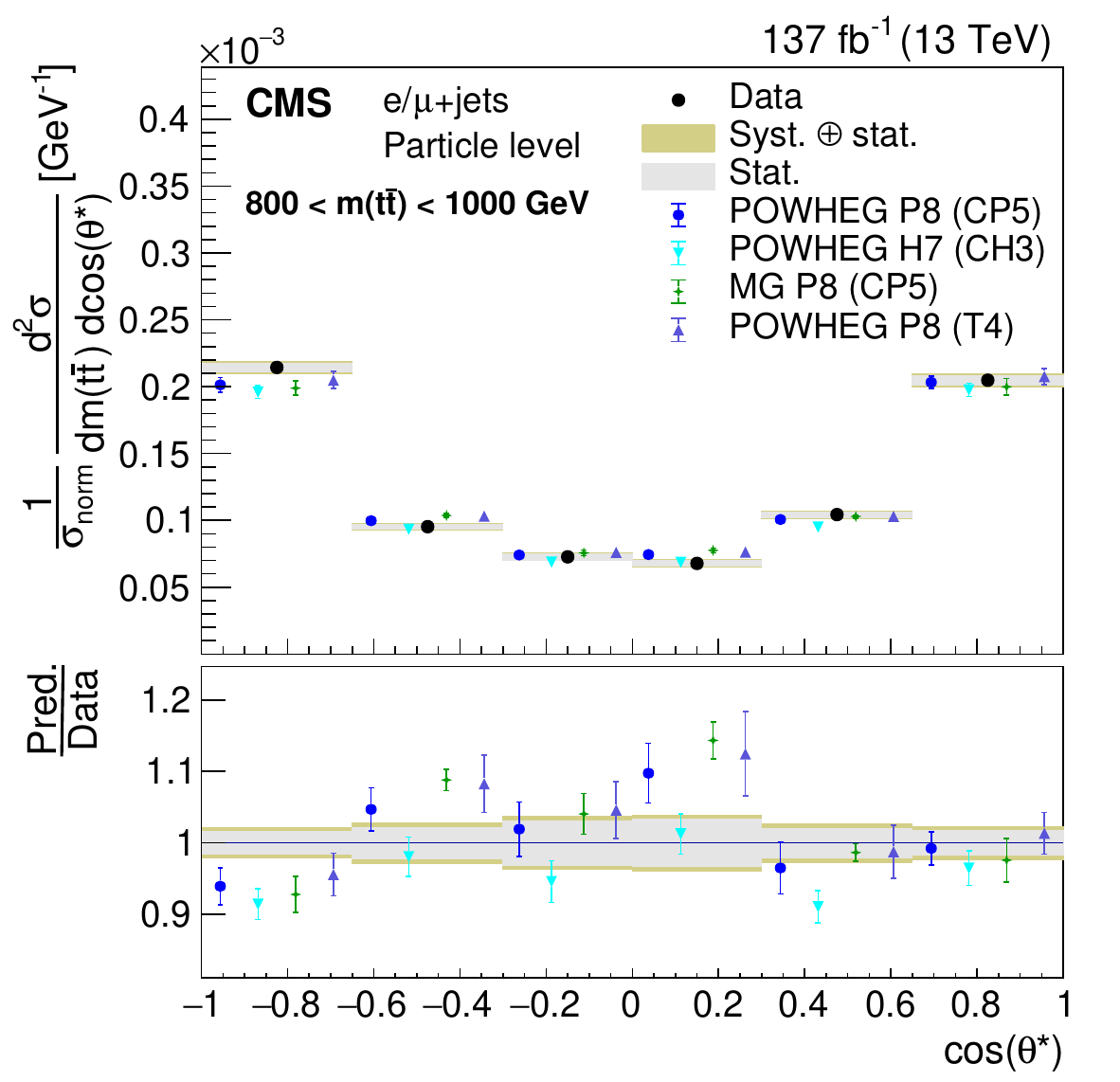}
 \includegraphics[width=0.42\textwidth]{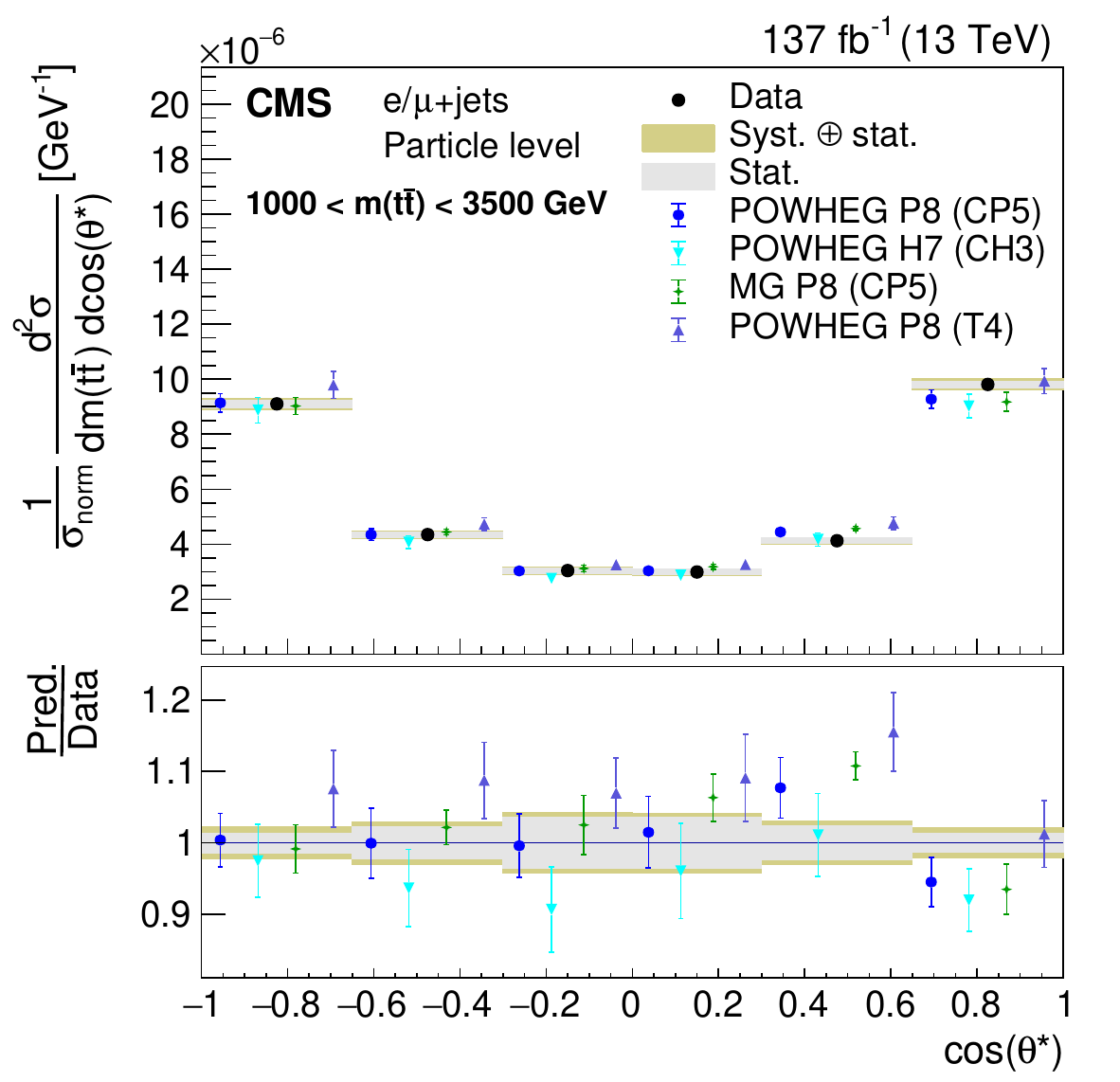}
 \caption{Normalized double-differential cross section at the particle level as a function of \ttmvscts. \XSECCAPPS}
 \label{fig:RESNORMPS6}
\end{figure*}

\begin{figure*}[tbp]
\centering
 \includegraphics[width=0.42\textwidth]{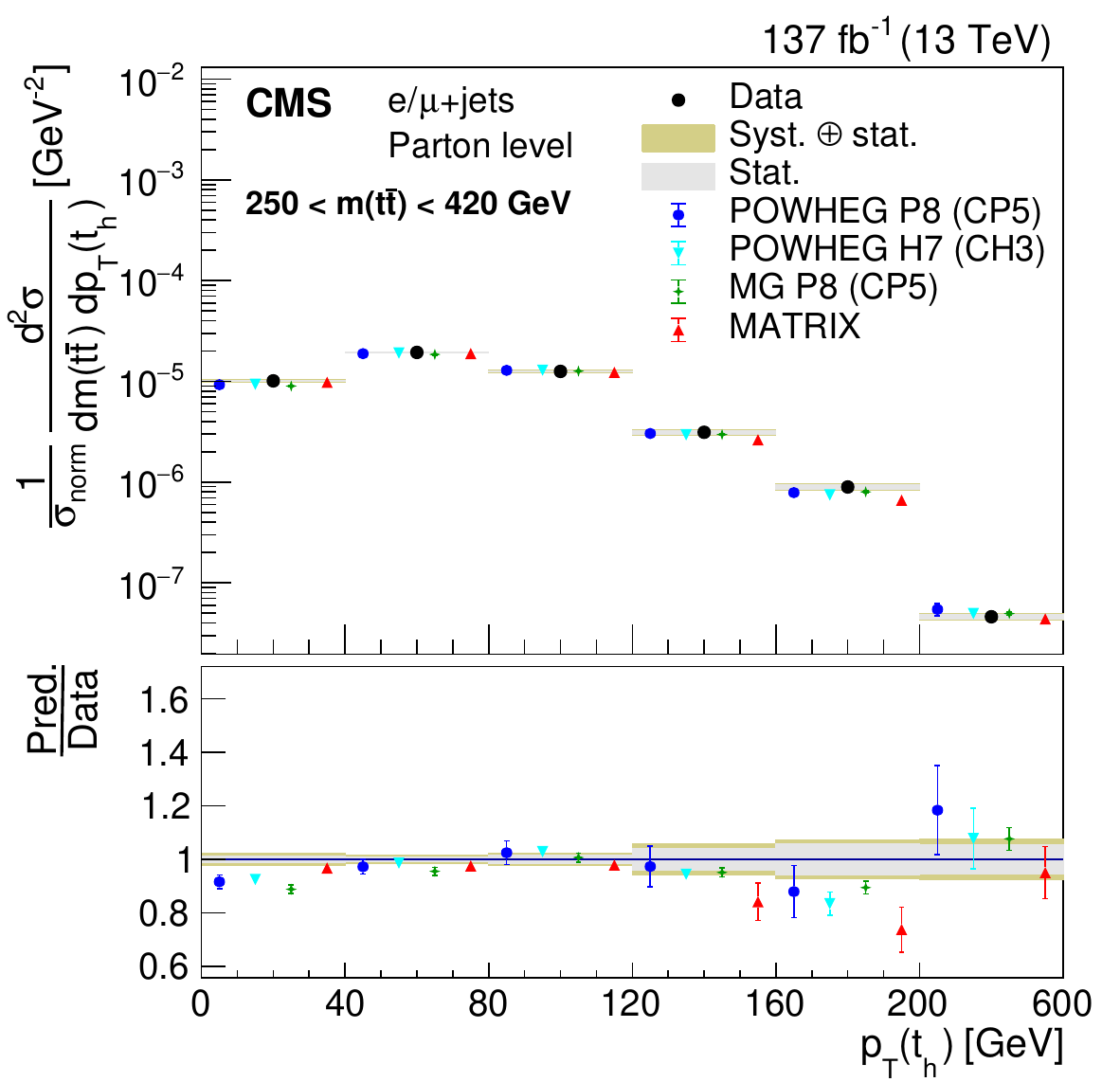}
 \includegraphics[width=0.42\textwidth]{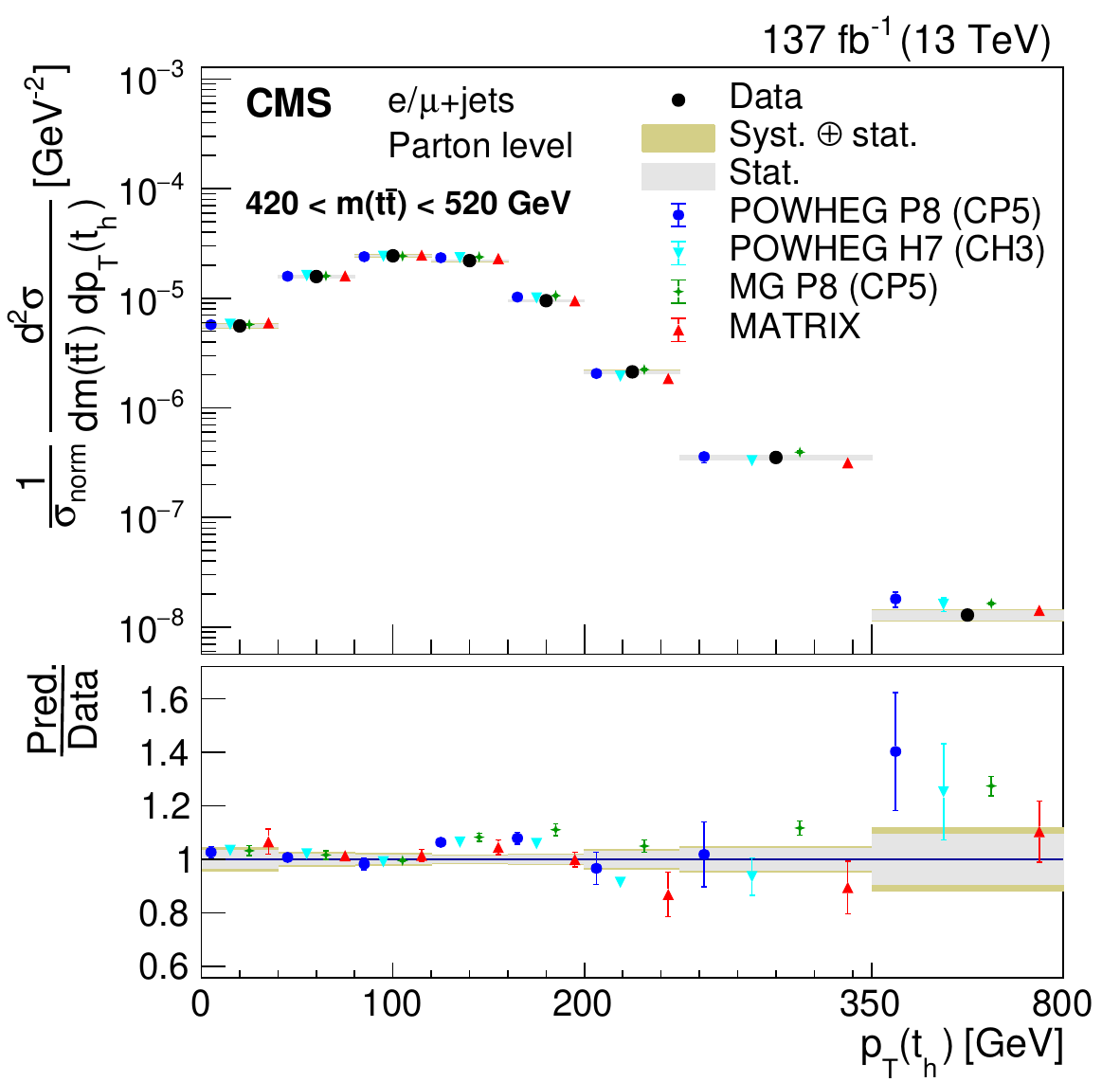}\\
 \includegraphics[width=0.42\textwidth]{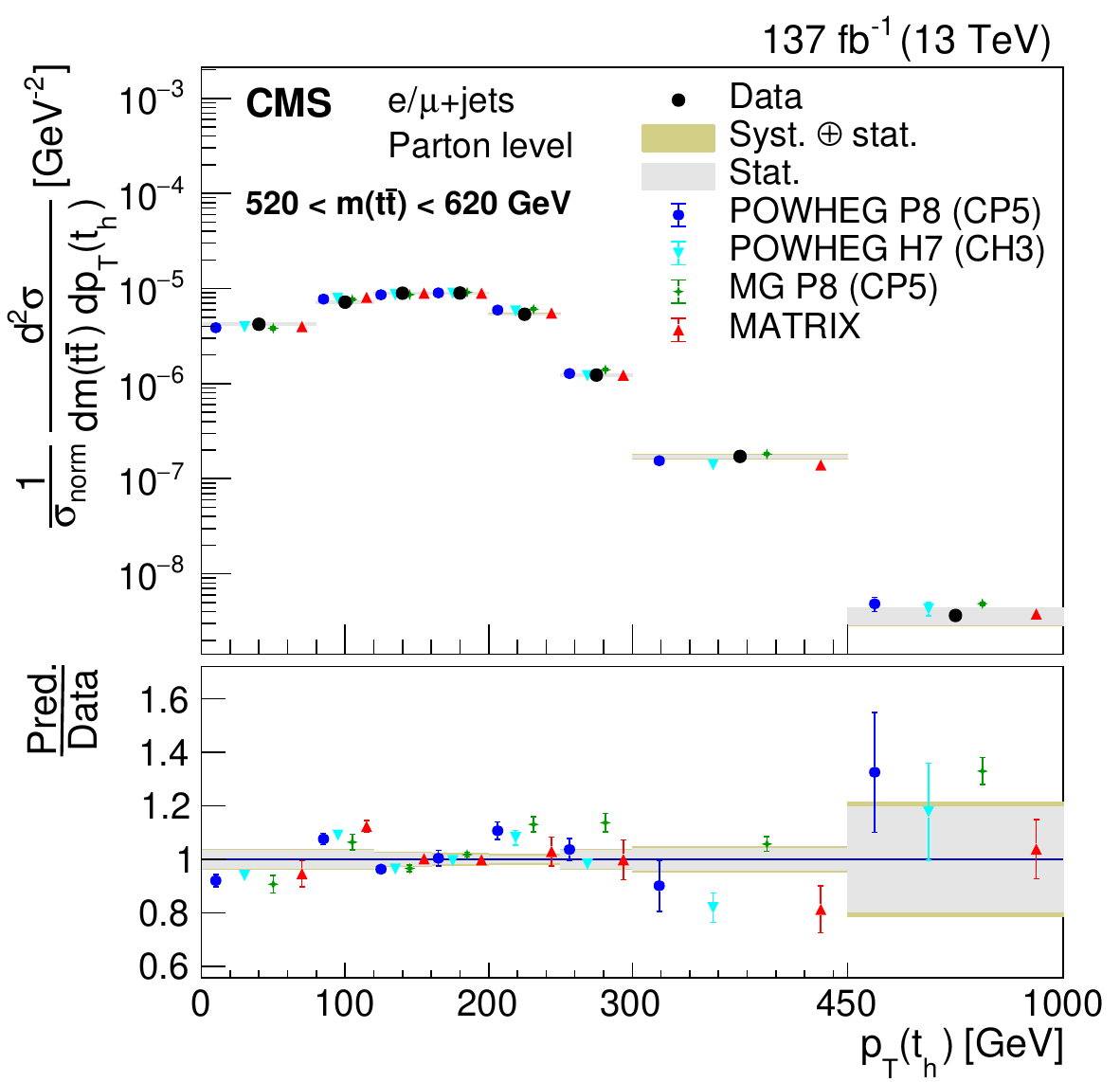}
 \includegraphics[width=0.42\textwidth]{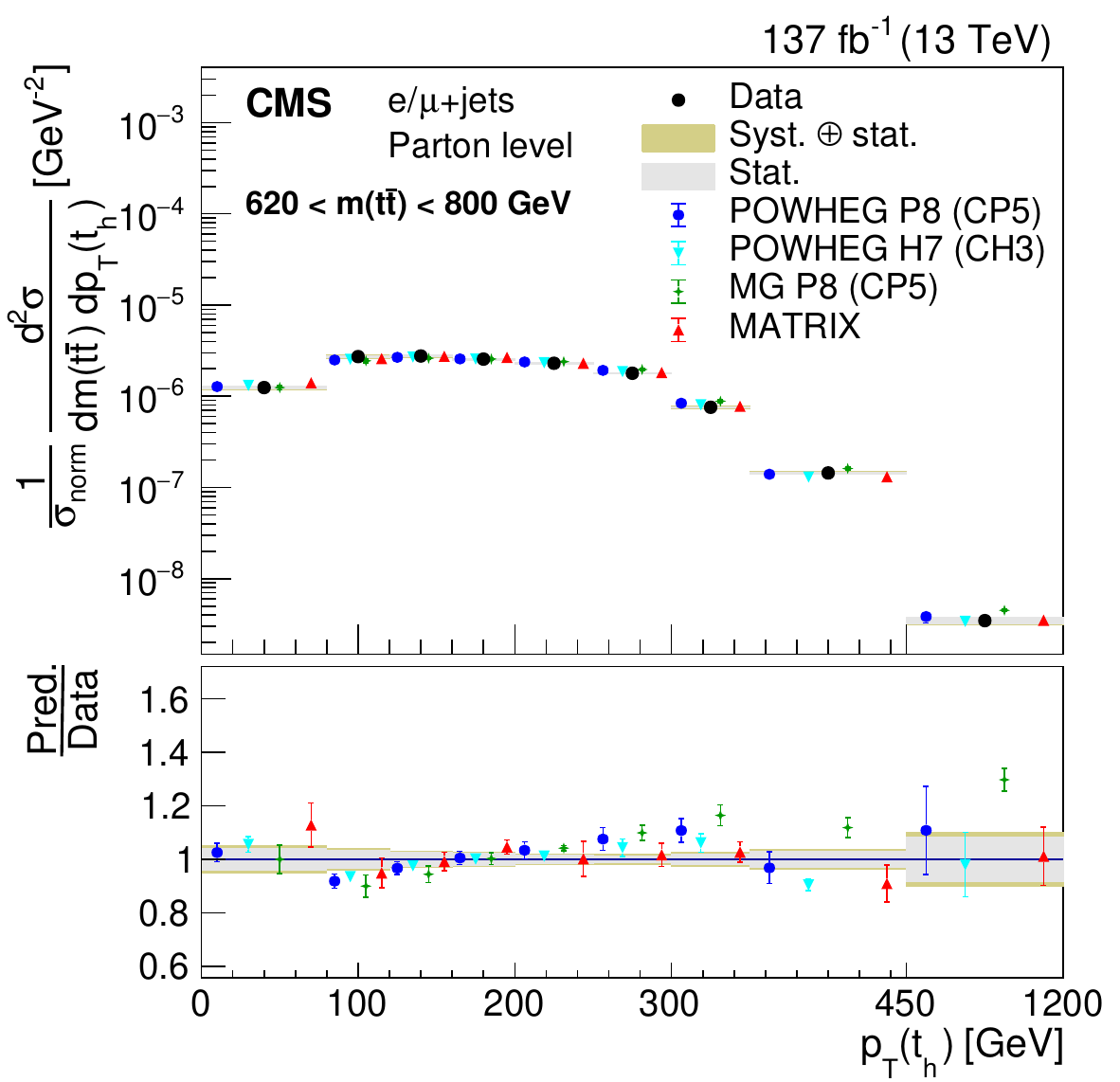}\\
 \includegraphics[width=0.42\textwidth]{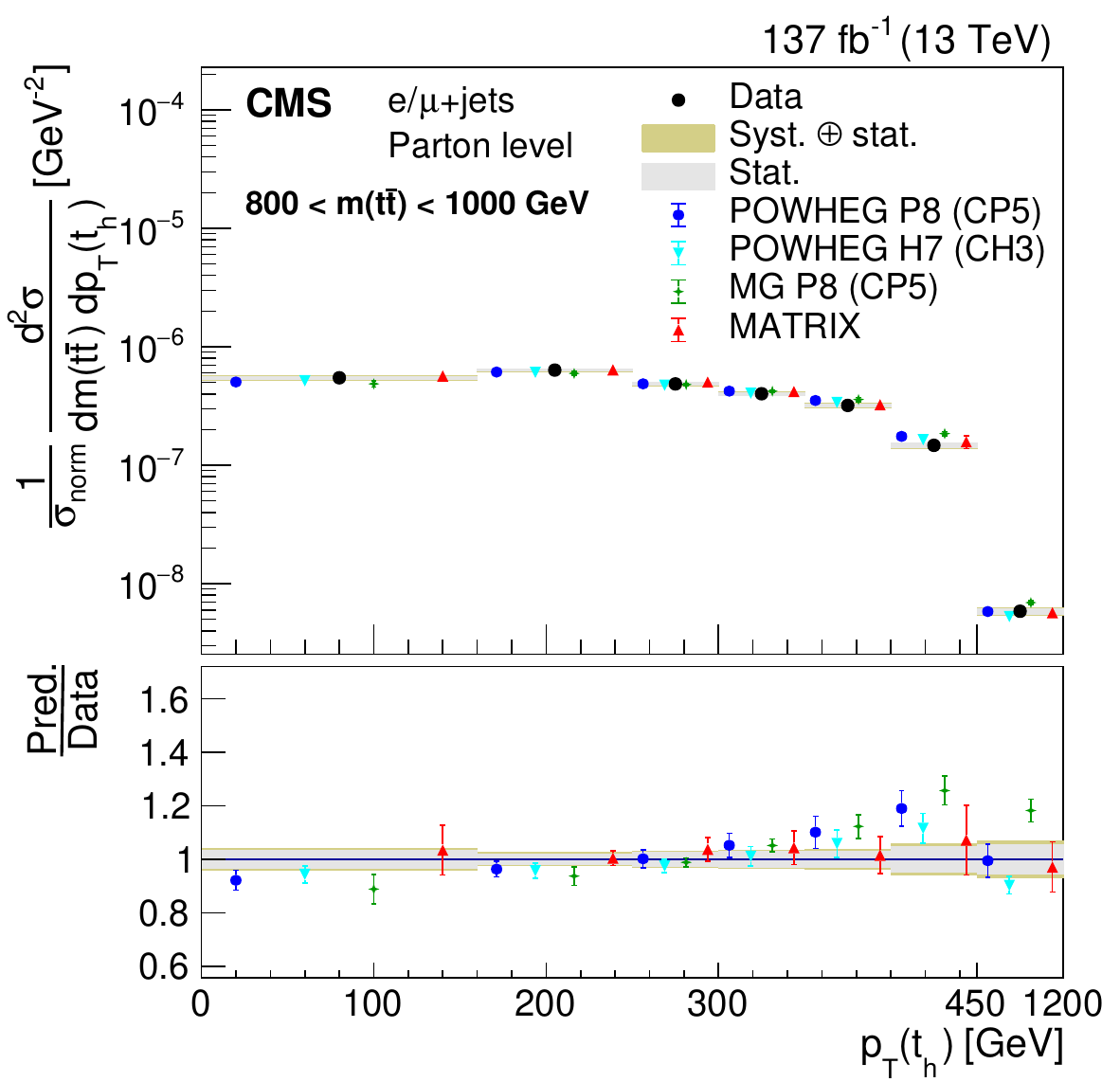}
 \includegraphics[width=0.42\textwidth]{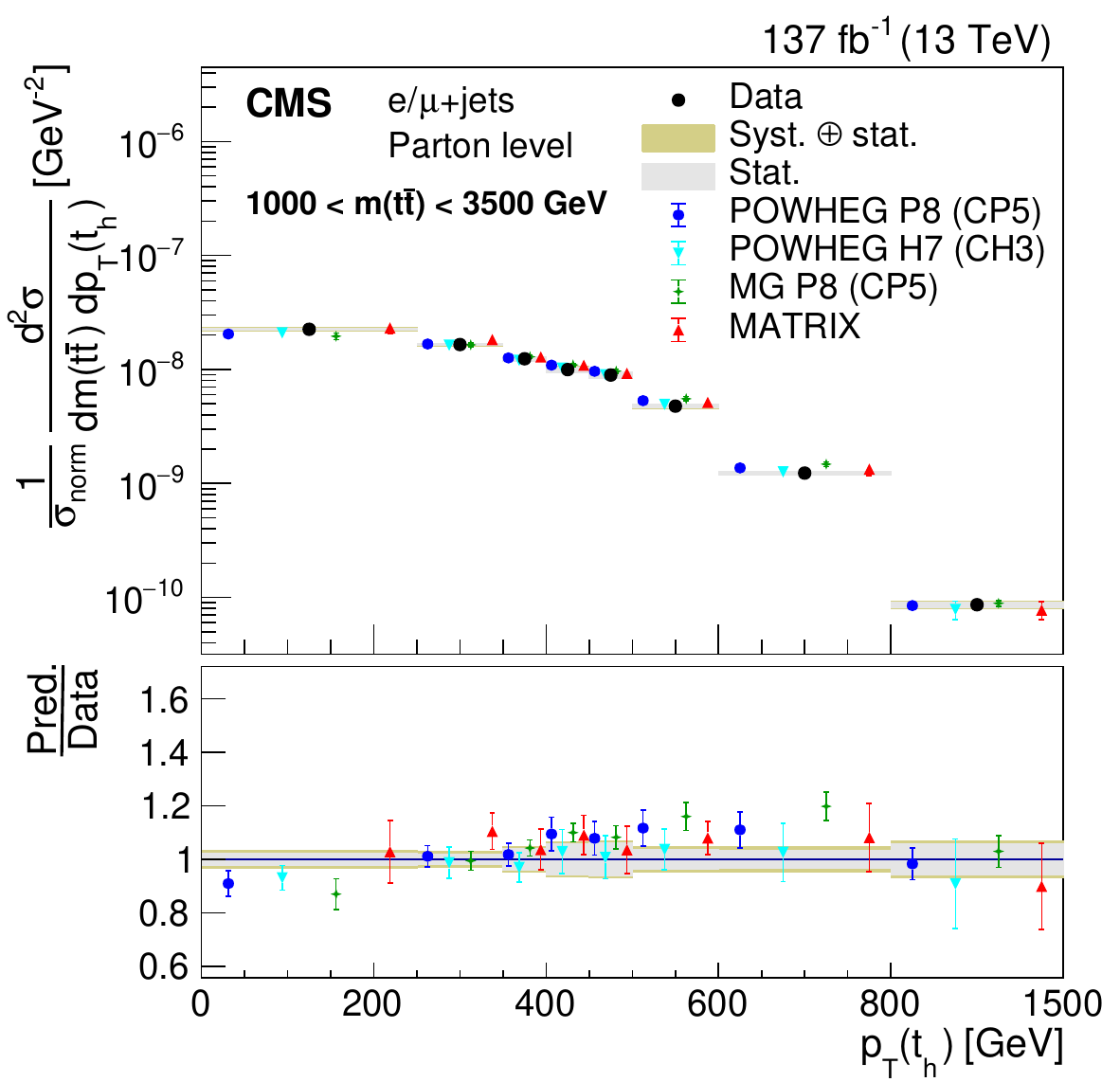}
 \caption{Normalized double-differential cross section at the parton level as a function of \ttmvsthadpt. \XSECCAPPA}
 \label{fig:RESNORM7}
\end{figure*}

\begin{figure*}[tbp]
\centering
 \includegraphics[width=0.42\textwidth]{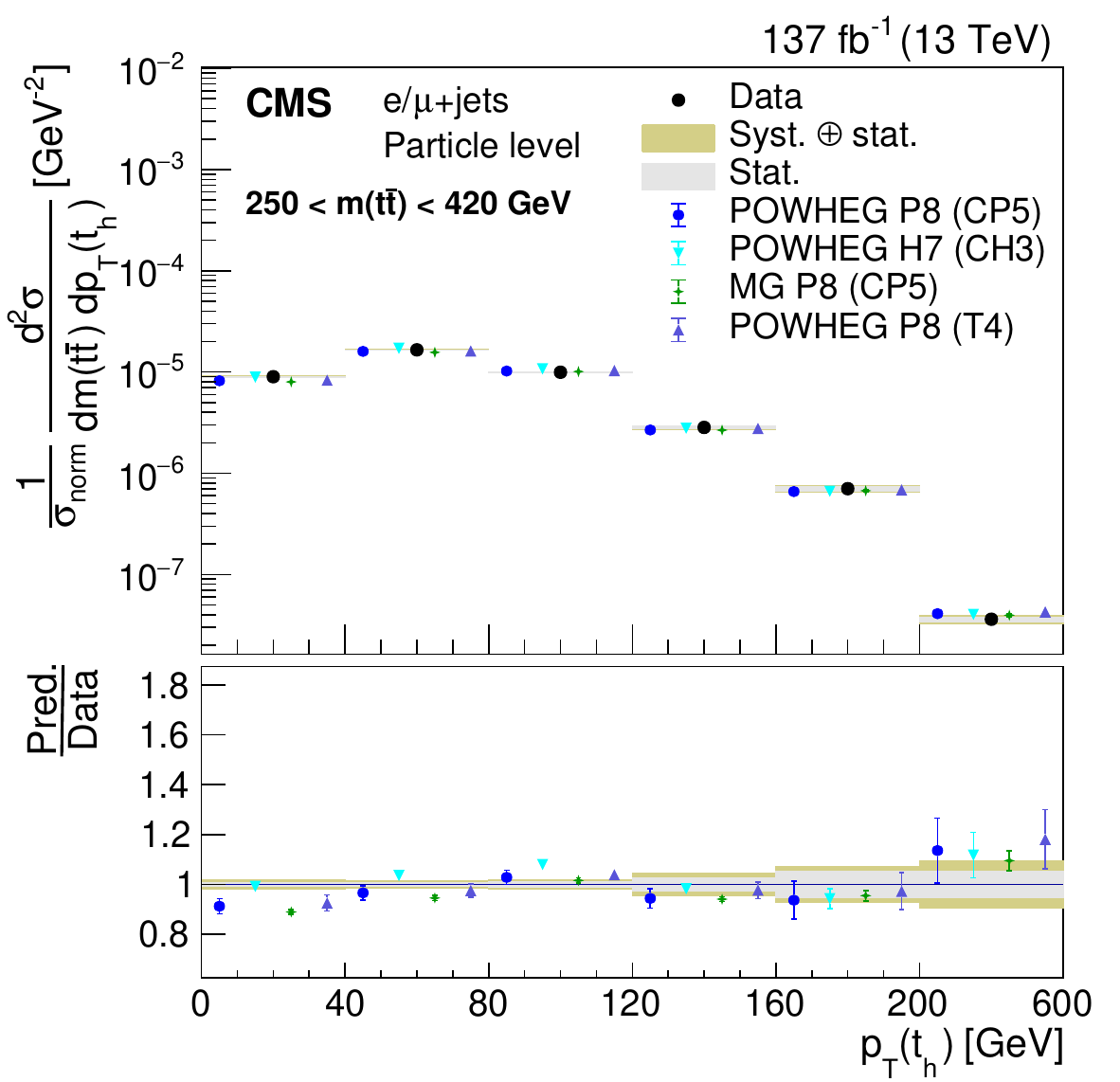}
 \includegraphics[width=0.42\textwidth]{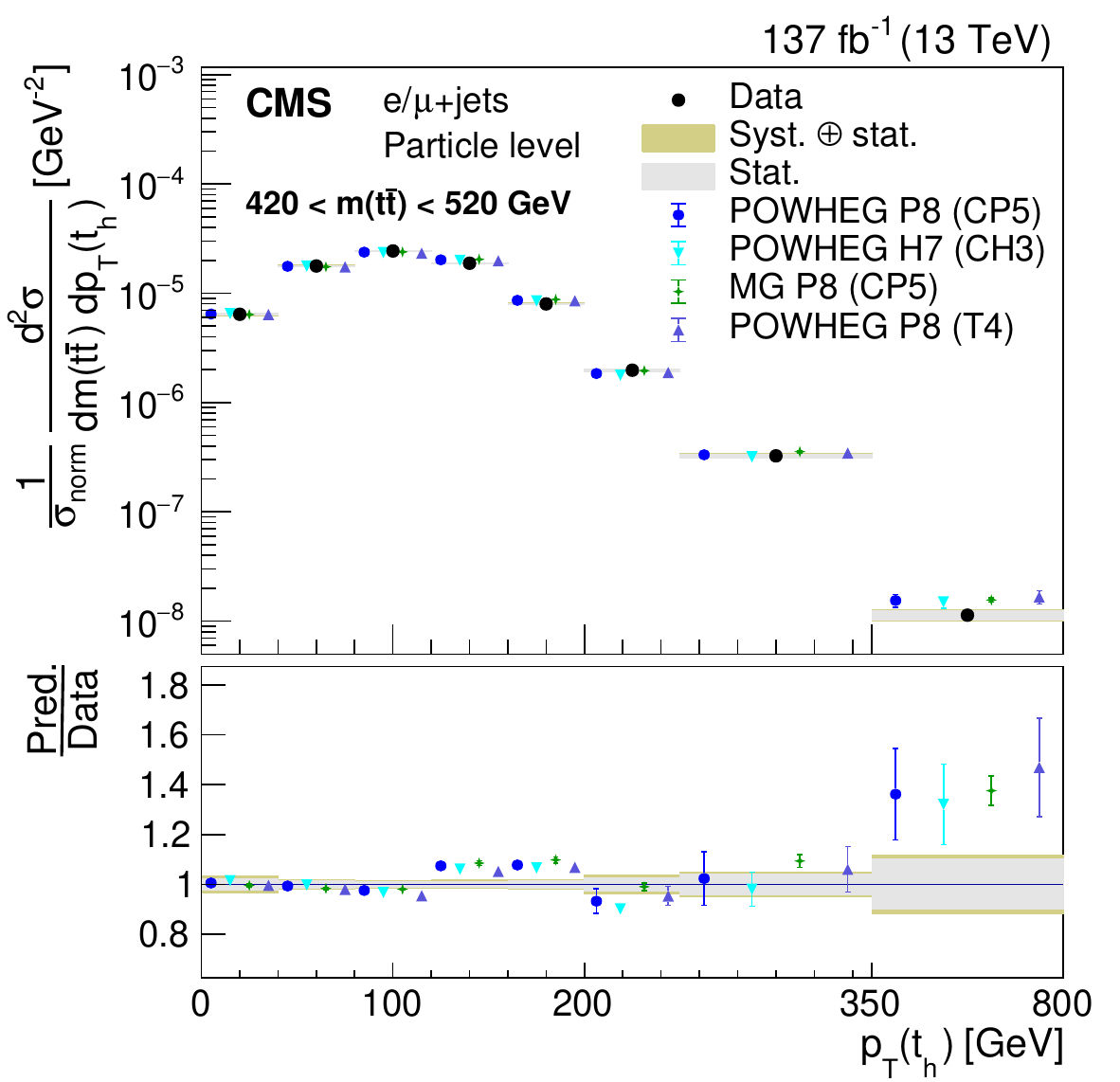}\\
 \includegraphics[width=0.42\textwidth]{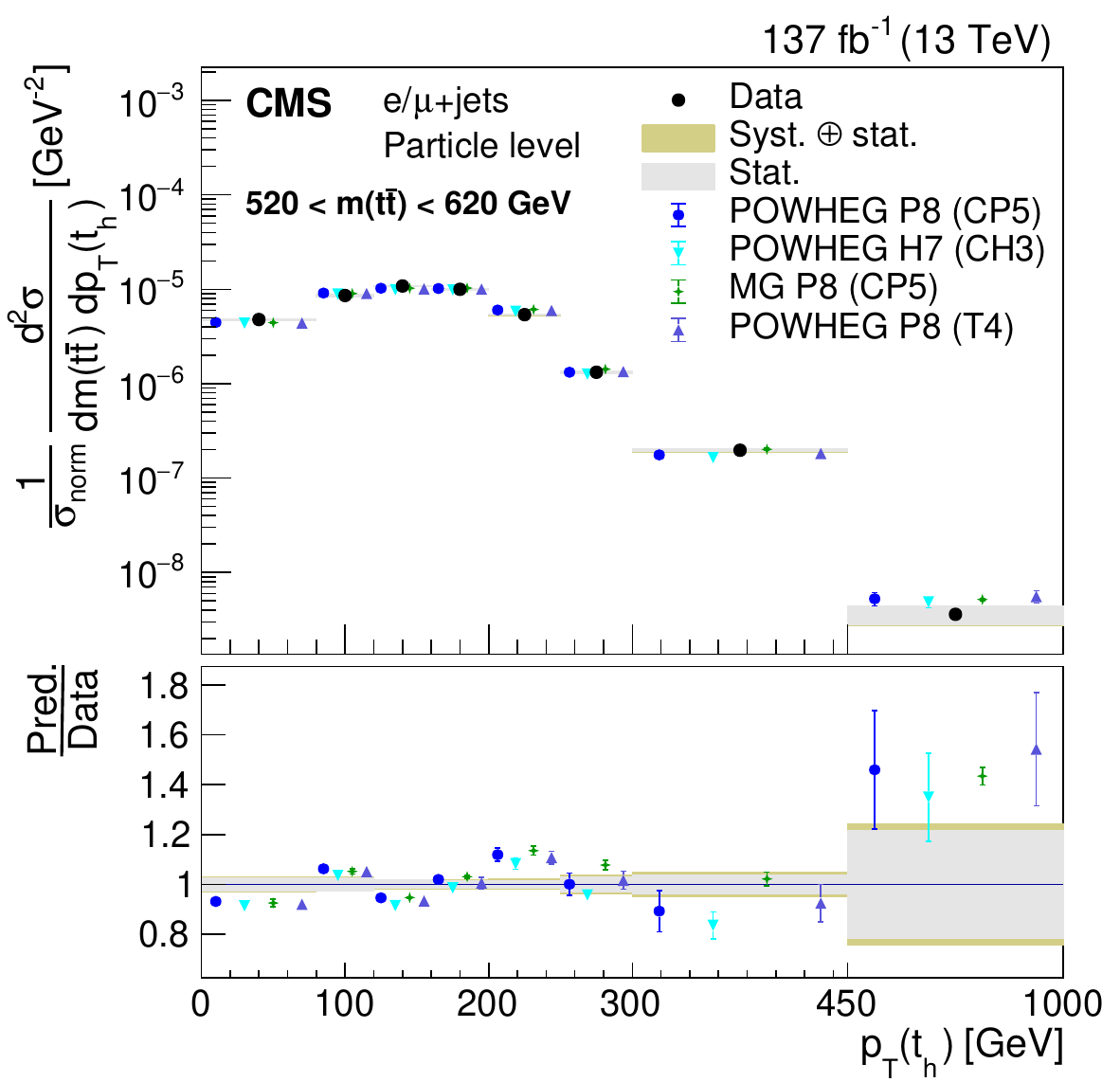}
 \includegraphics[width=0.42\textwidth]{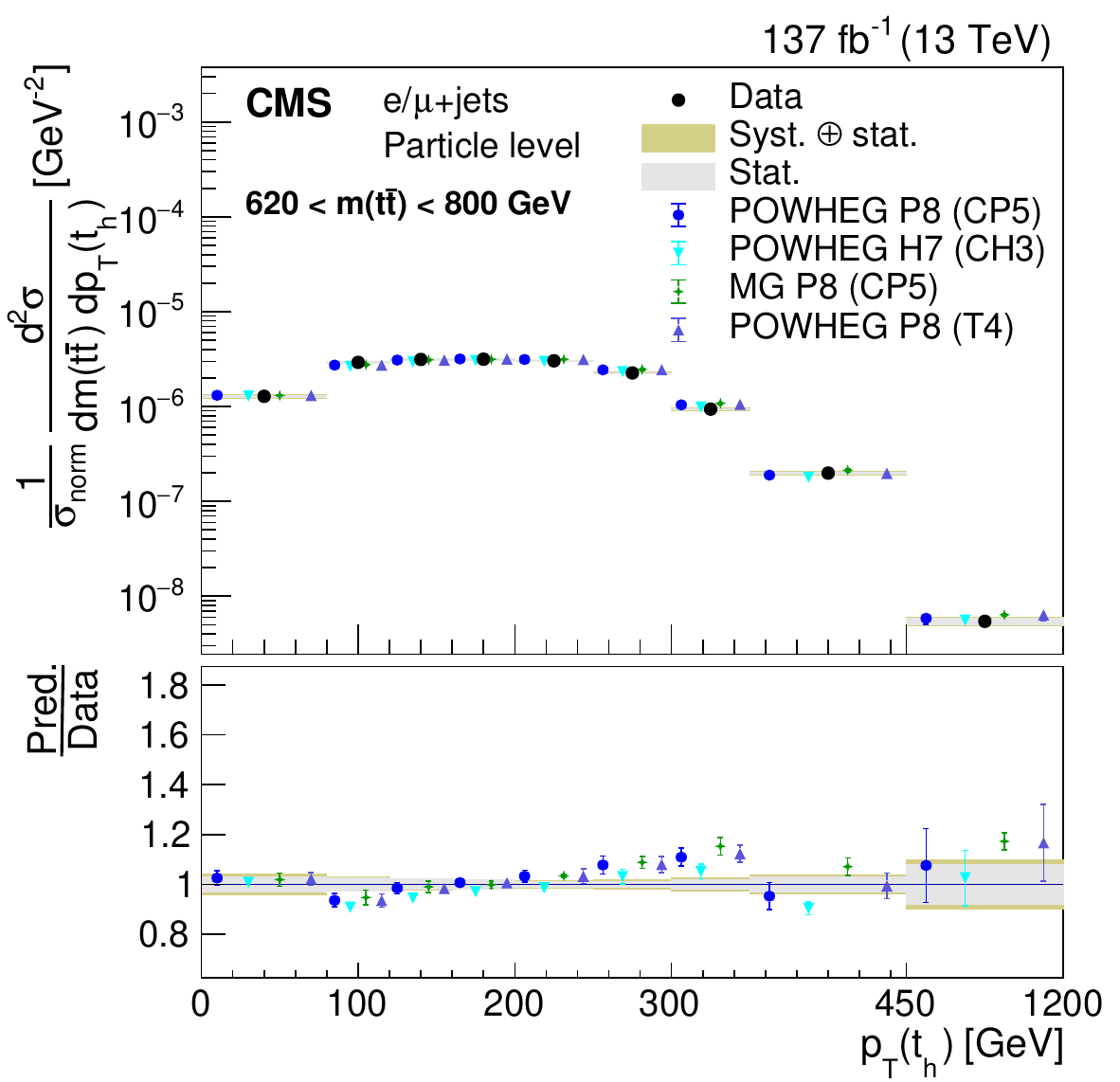}\\
 \includegraphics[width=0.42\textwidth]{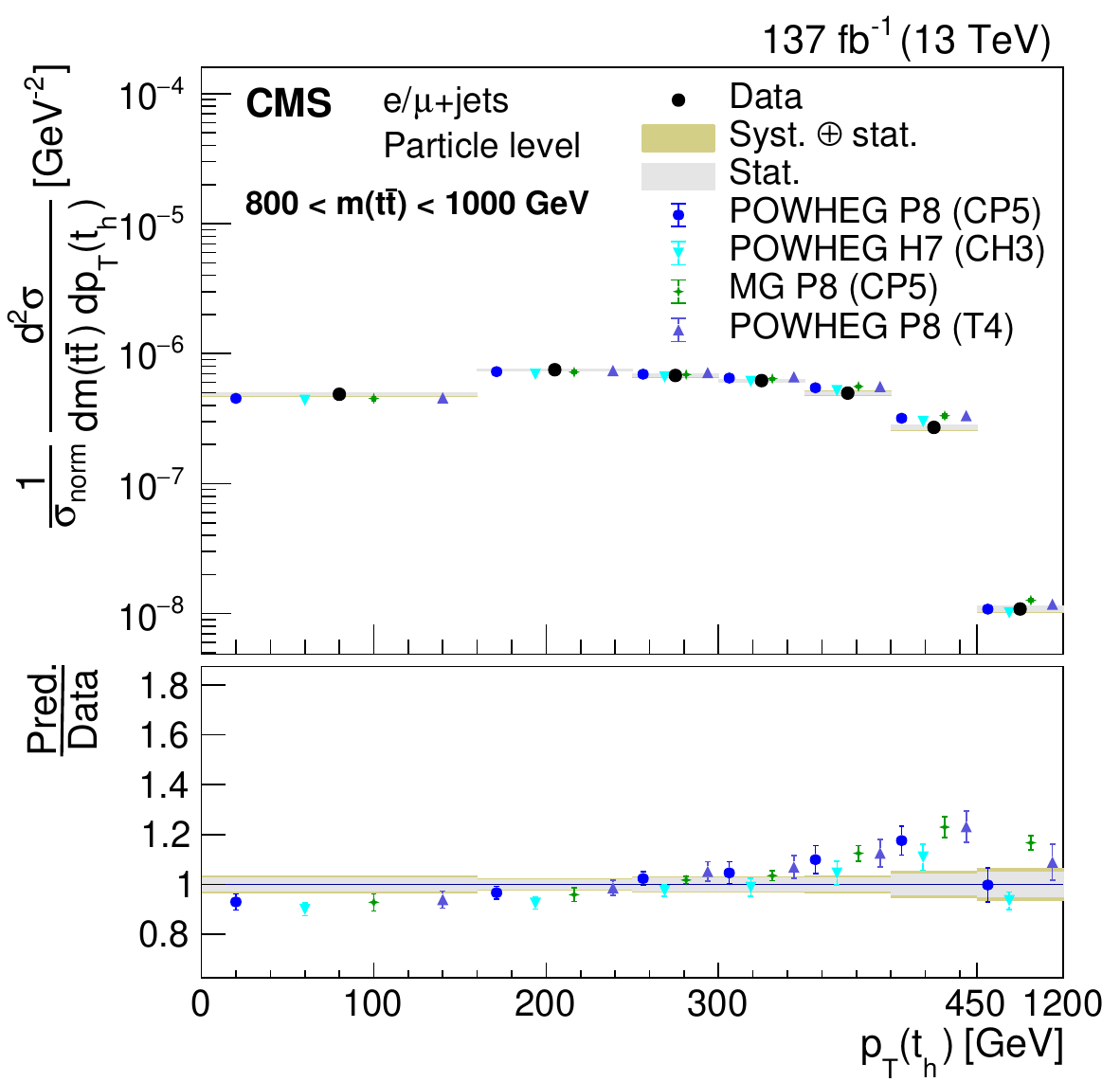}
 \includegraphics[width=0.42\textwidth]{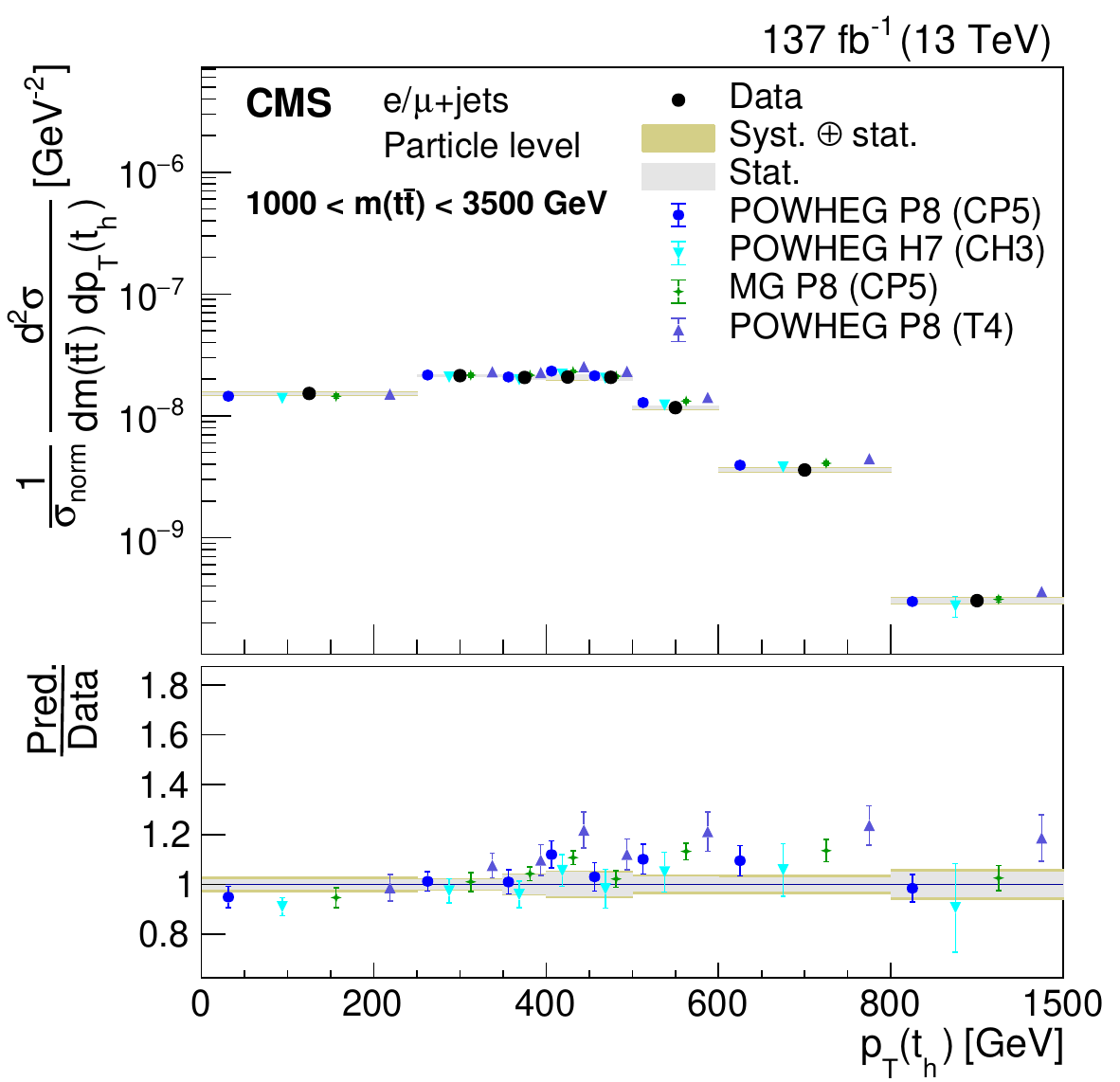}
 \caption{Normalized double-differential cross section at the particle level as a function of \ttmvsthadpt. \XSECCAPPS}
 \label{fig:RESNORMPS7}
\end{figure*}

\begin{figure*}[tbp]
\centering
 \includegraphics[width=0.42\textwidth]{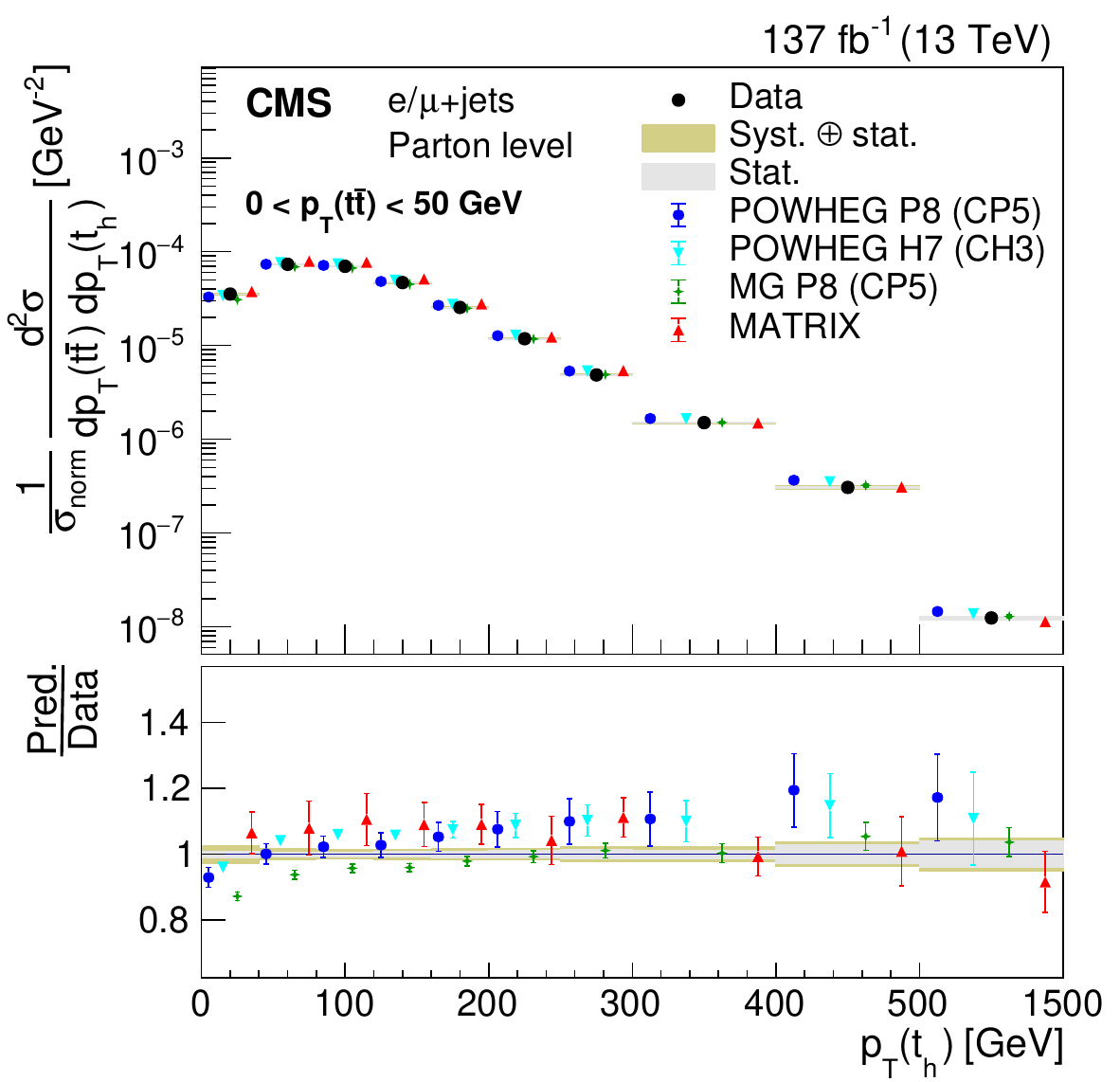}
 \includegraphics[width=0.42\textwidth]{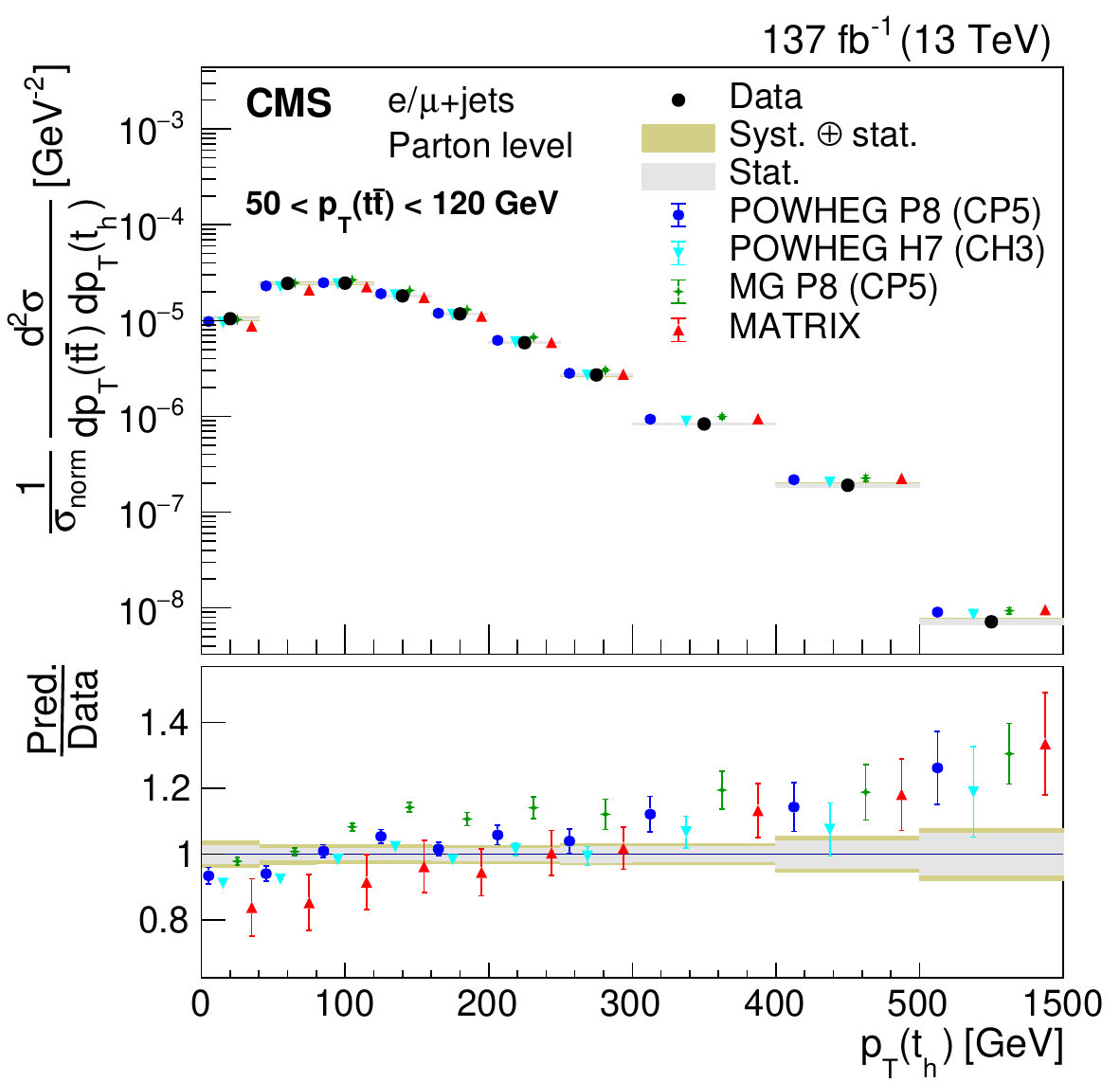}\\
 \includegraphics[width=0.42\textwidth]{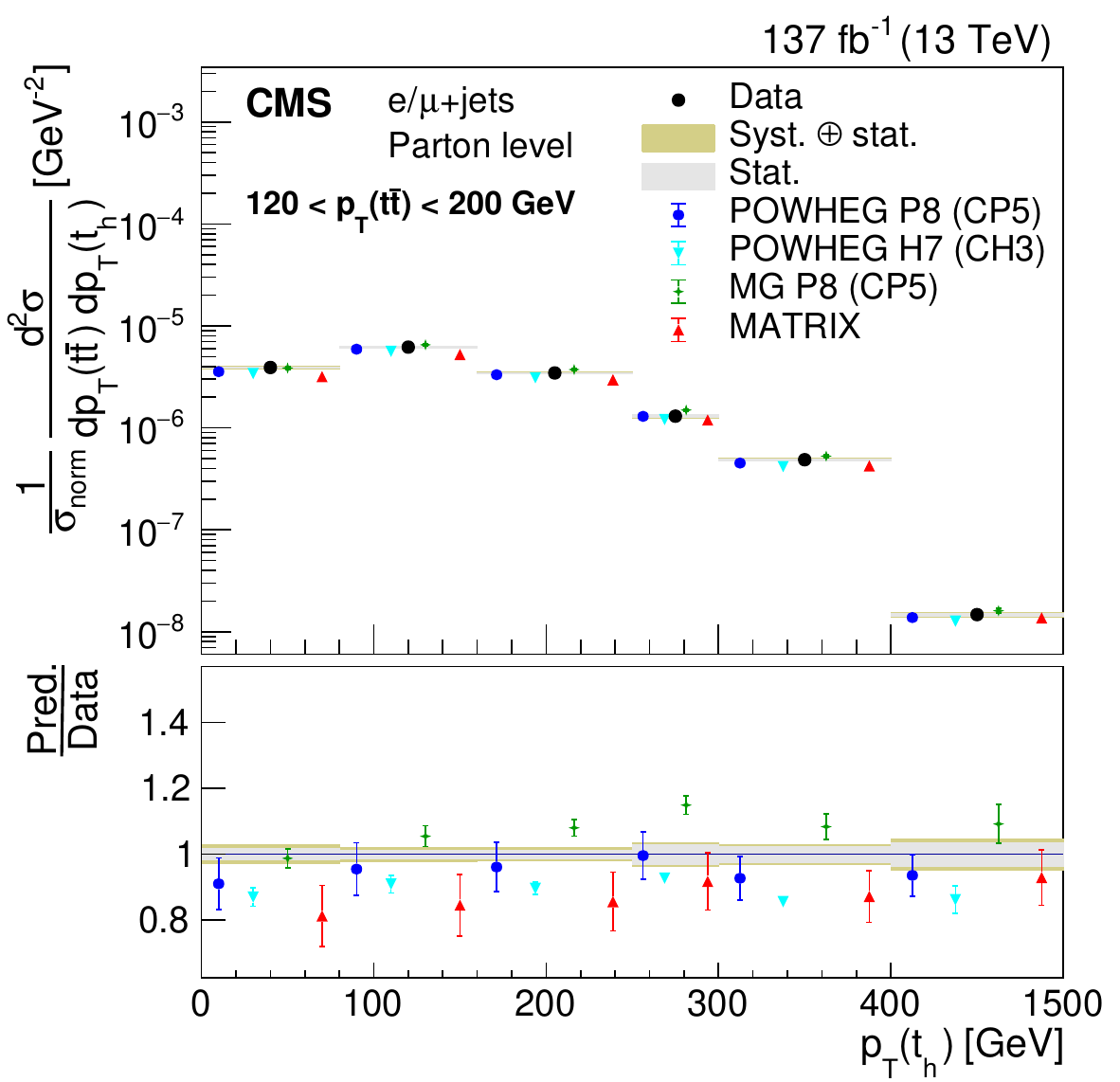}
 \includegraphics[width=0.42\textwidth]{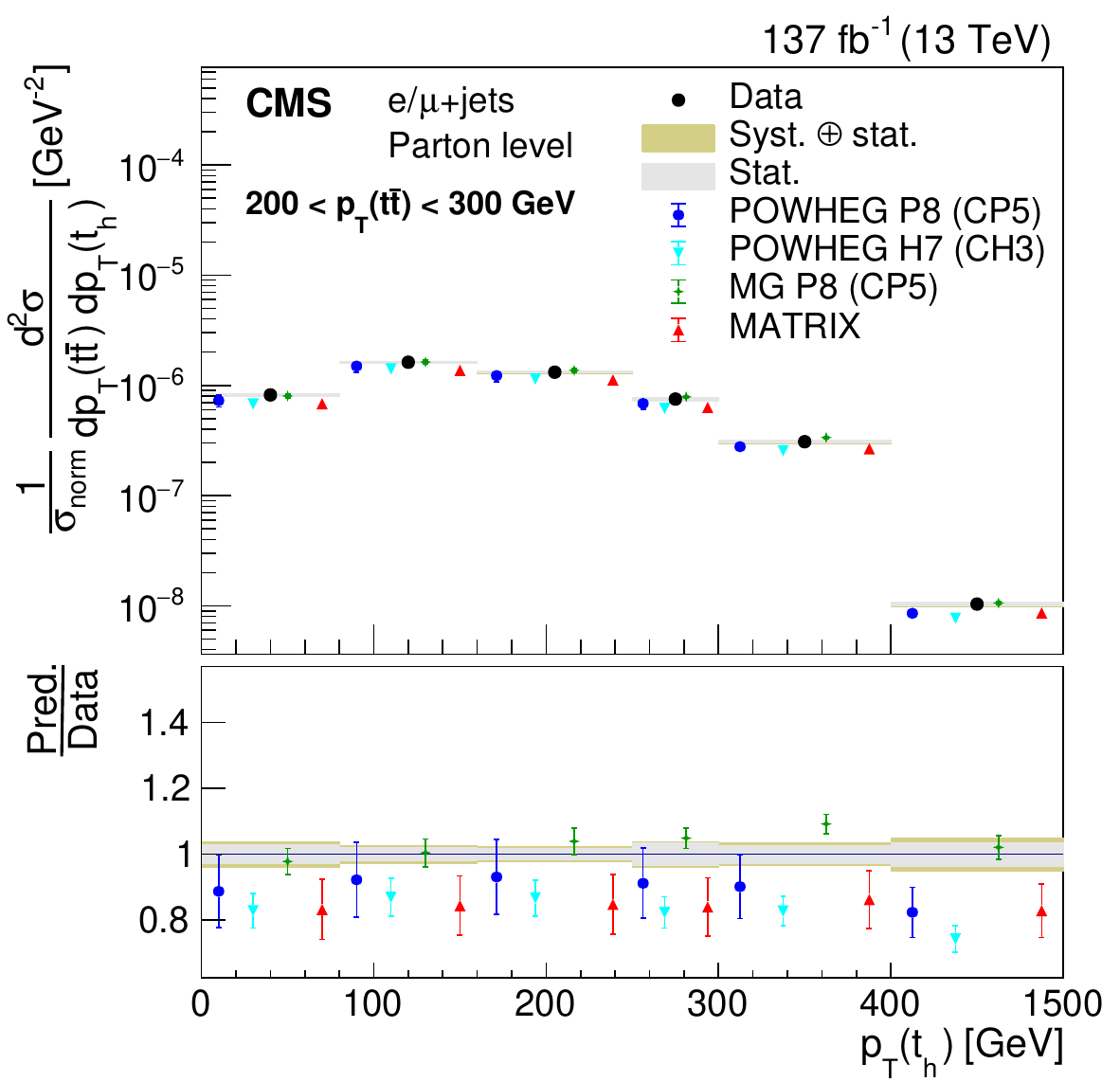}\\
 \includegraphics[width=0.42\textwidth]{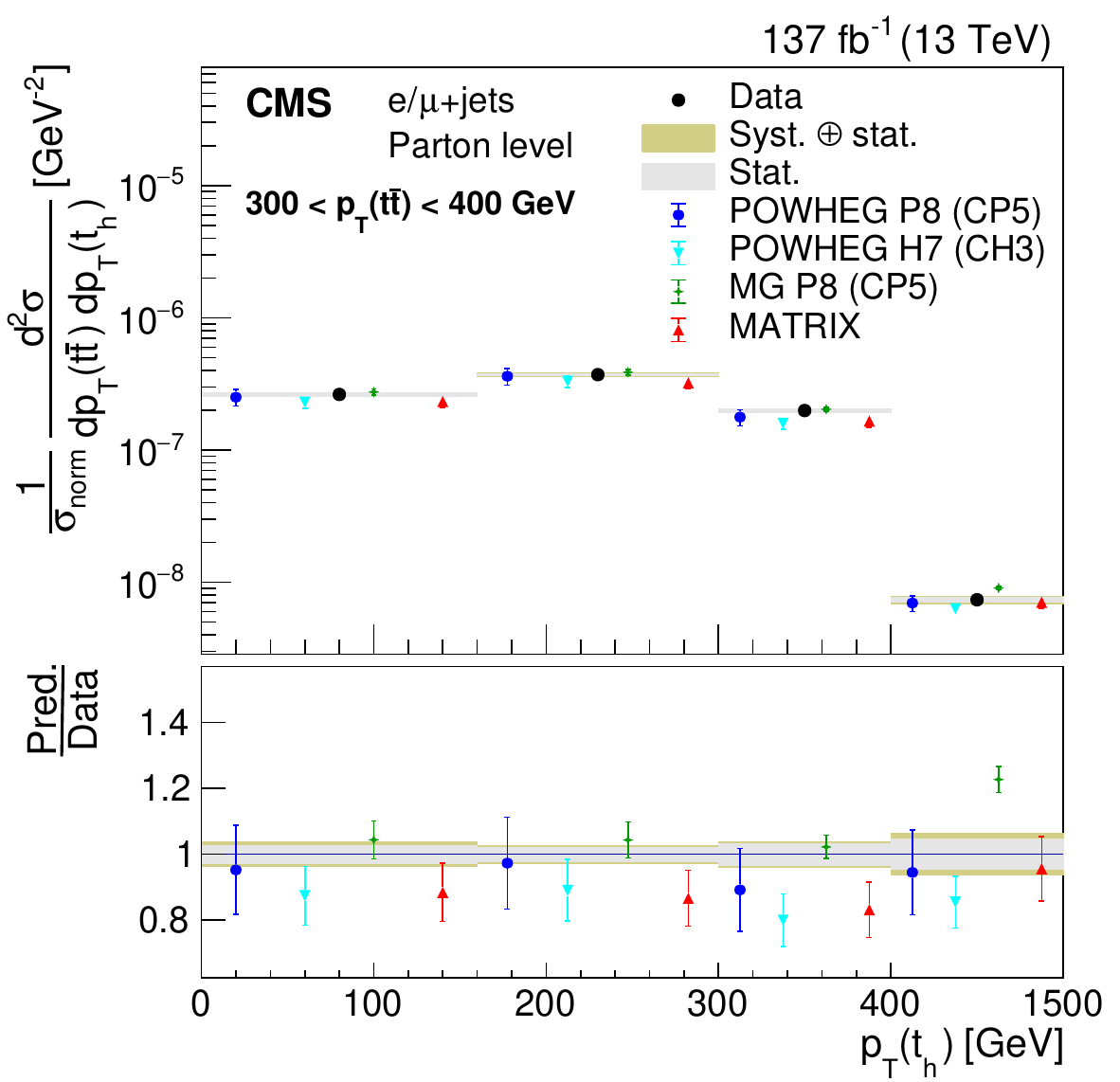}
 \includegraphics[width=0.42\textwidth]{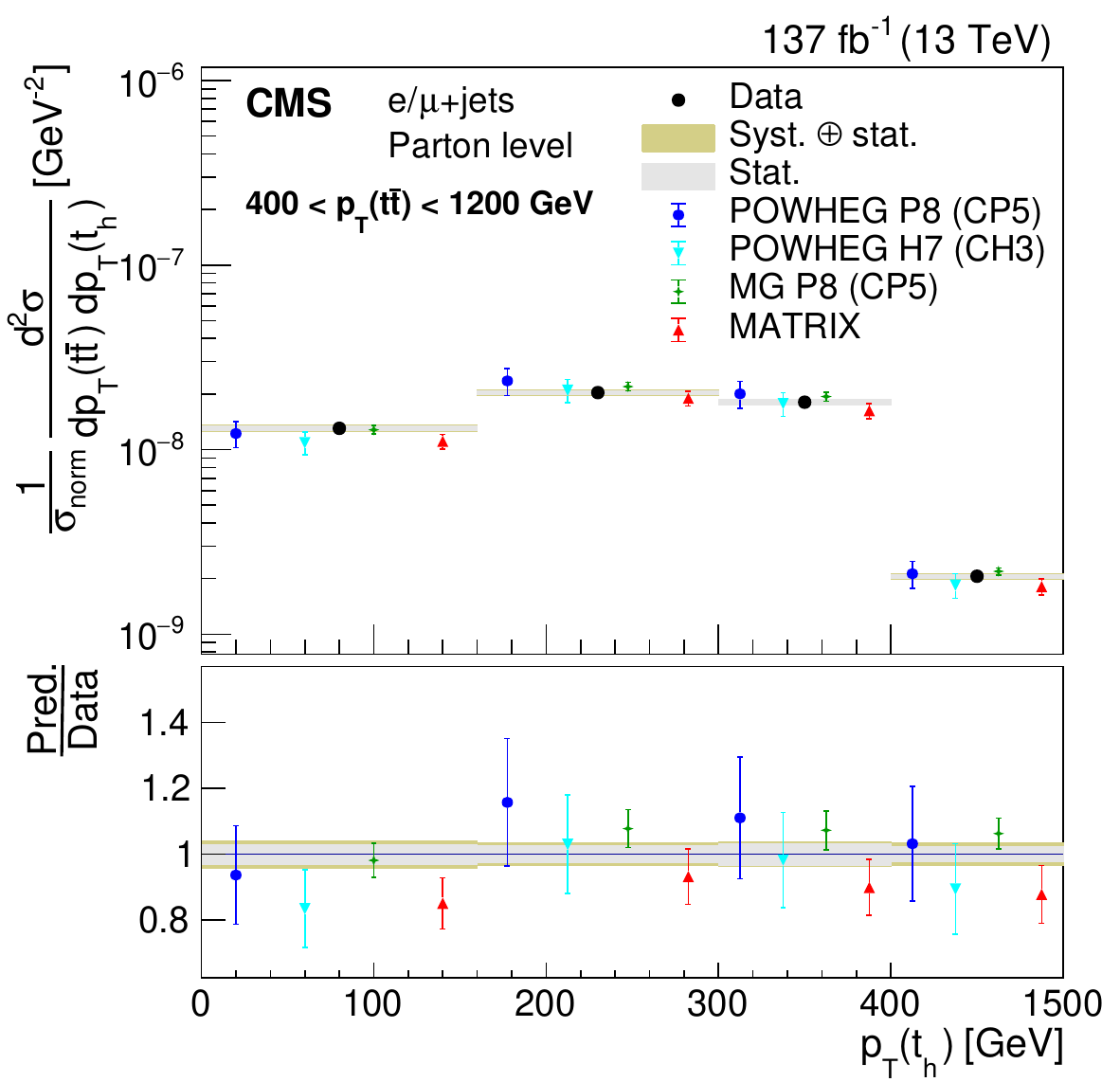}
 \caption{Normalized double-differential cross section at the parton level as a function of \ttptvsthadpt. \XSECCAPPA}
 \label{fig:RESNORM7b}
\end{figure*}

\begin{figure*}[tbp]
\centering
 \includegraphics[width=0.42\textwidth]{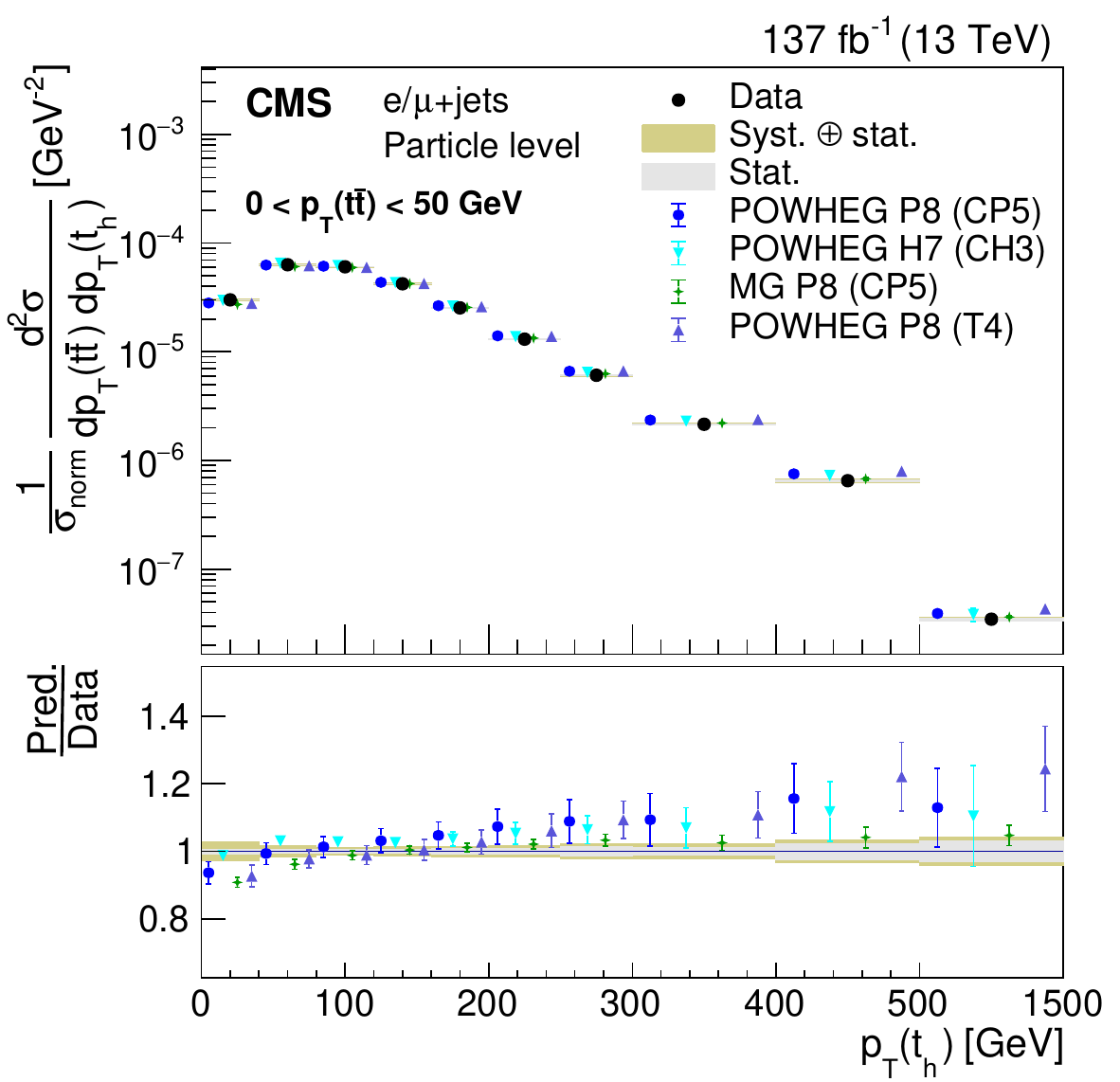}
 \includegraphics[width=0.42\textwidth]{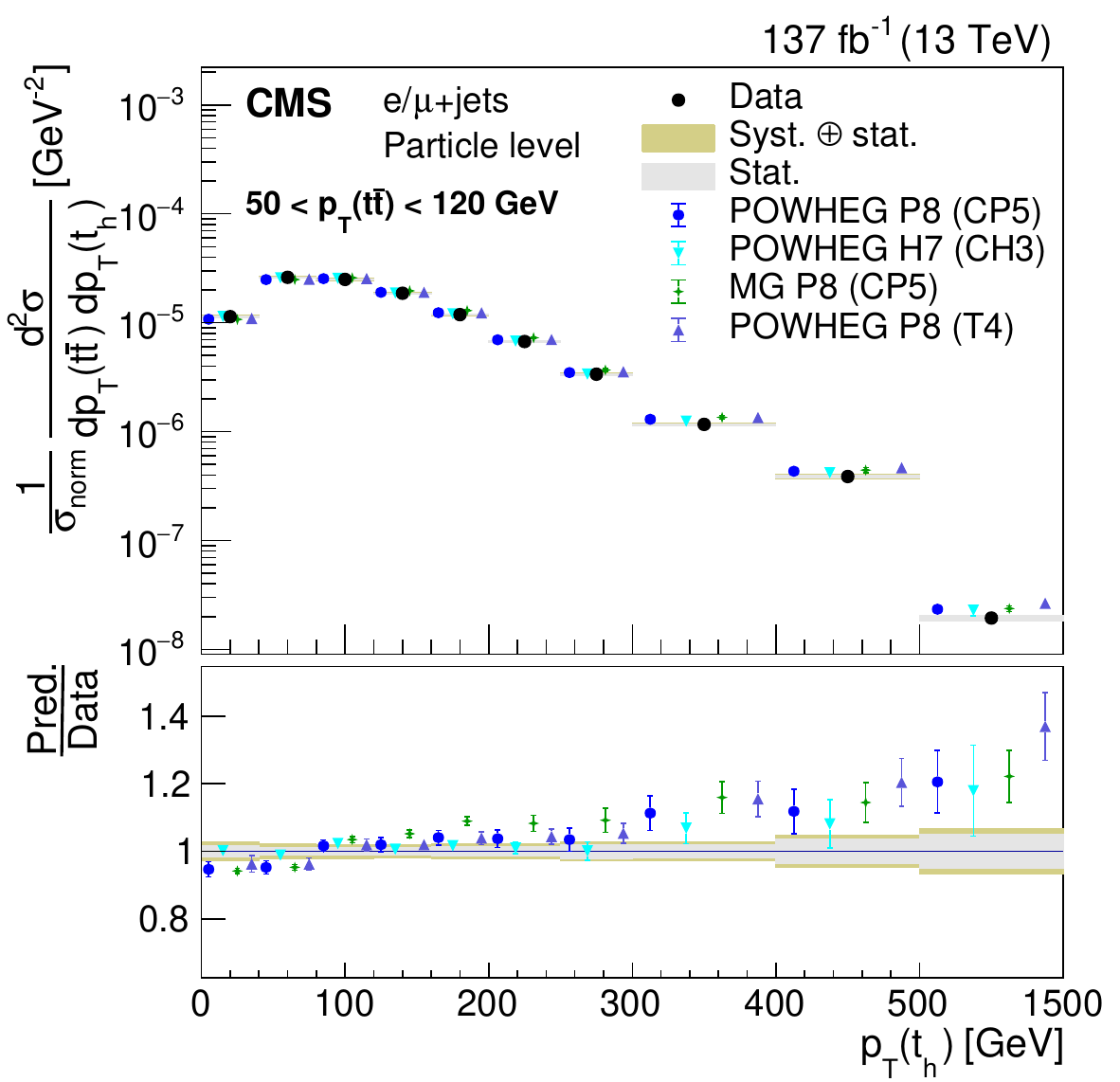}\\
 \includegraphics[width=0.42\textwidth]{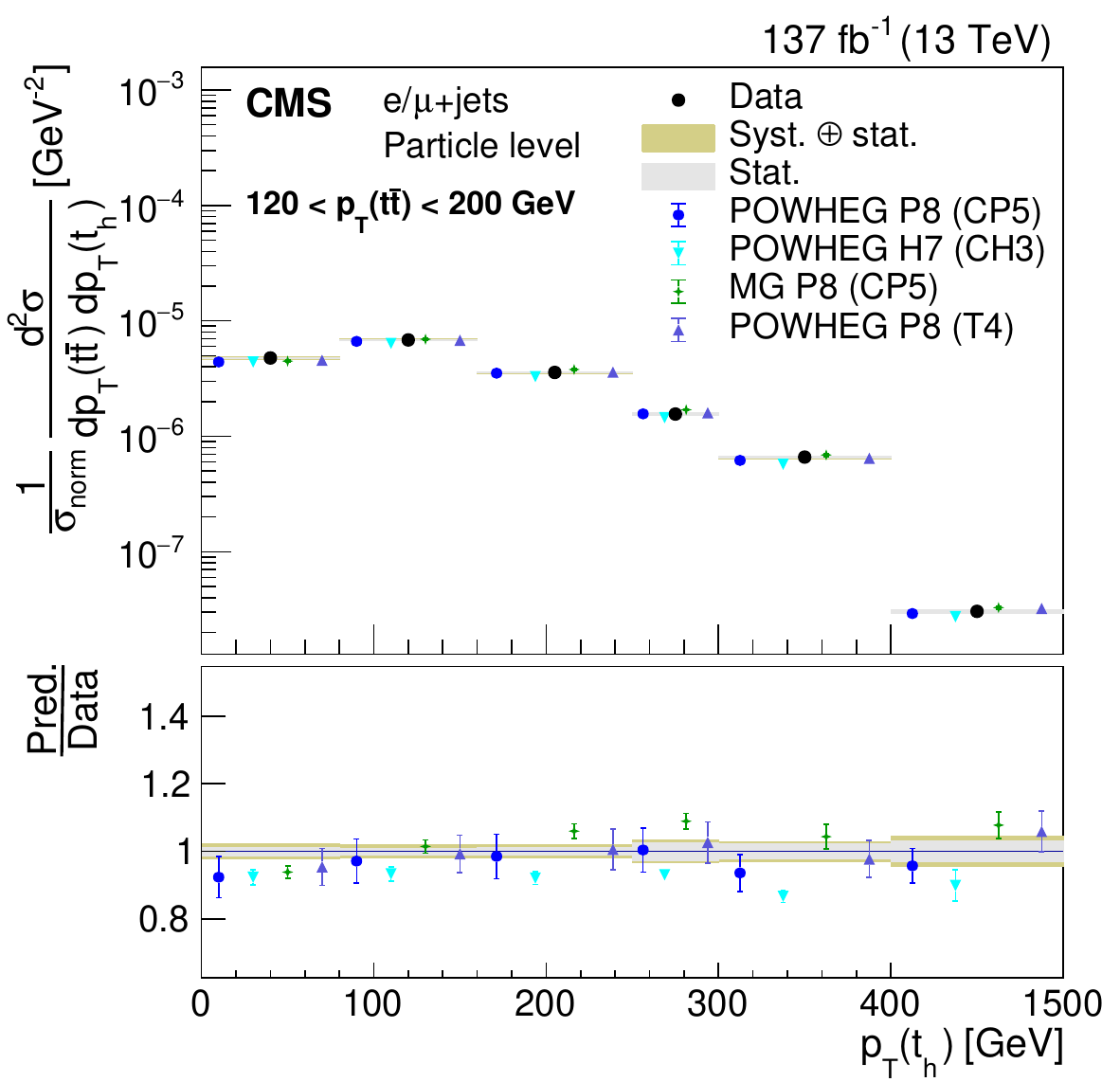}
 \includegraphics[width=0.42\textwidth]{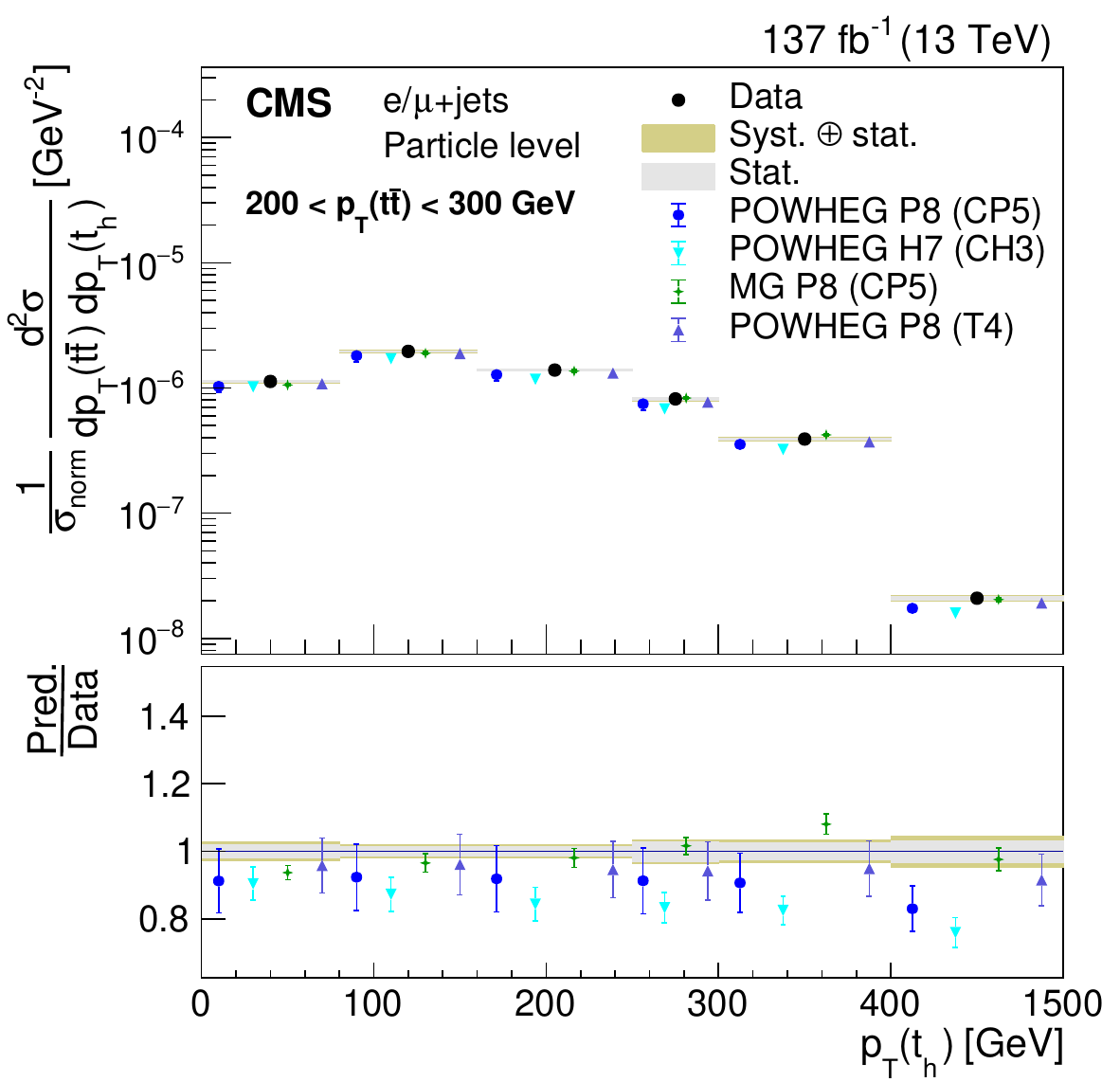}\\
 \includegraphics[width=0.42\textwidth]{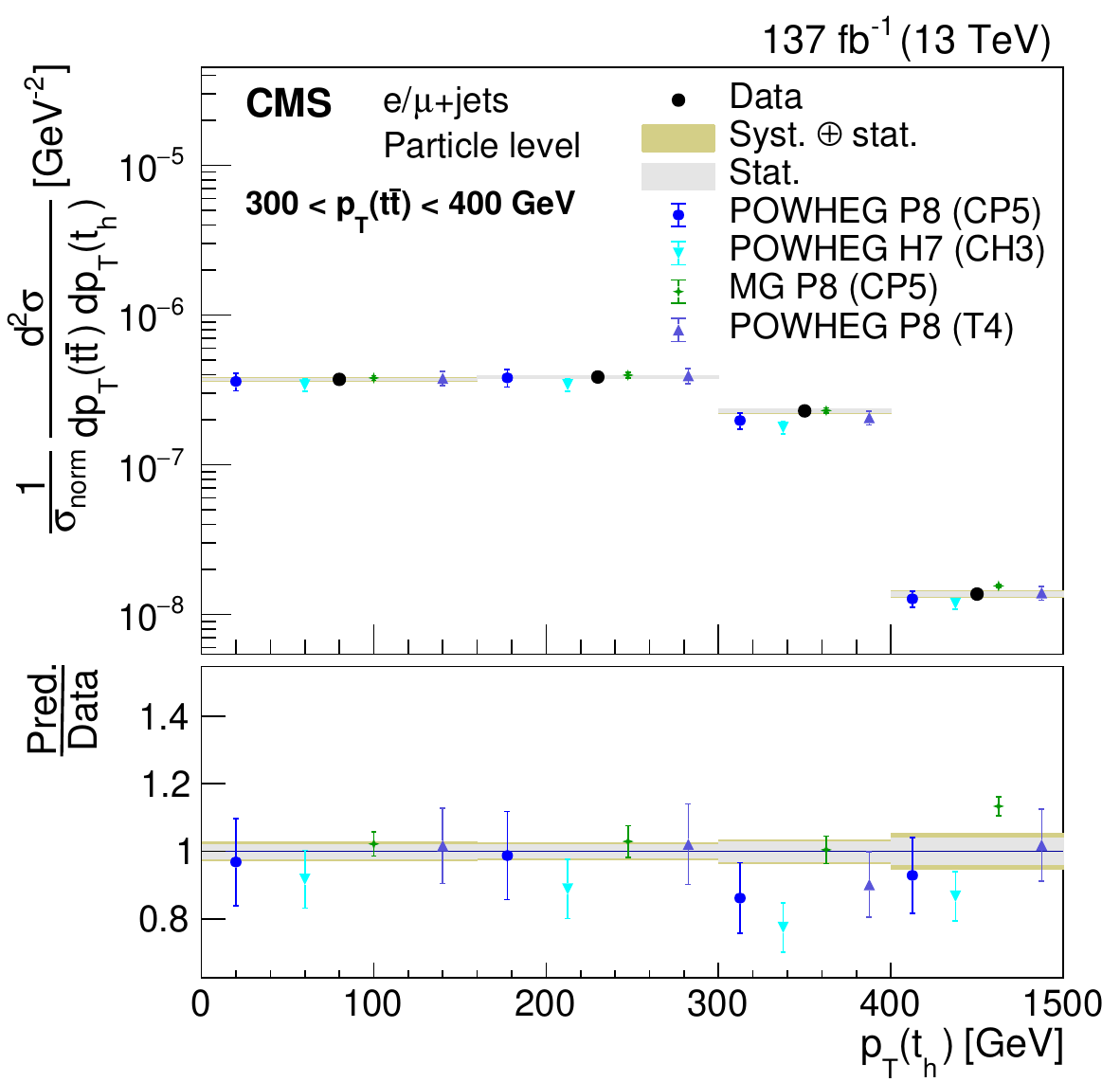}
 \includegraphics[width=0.42\textwidth]{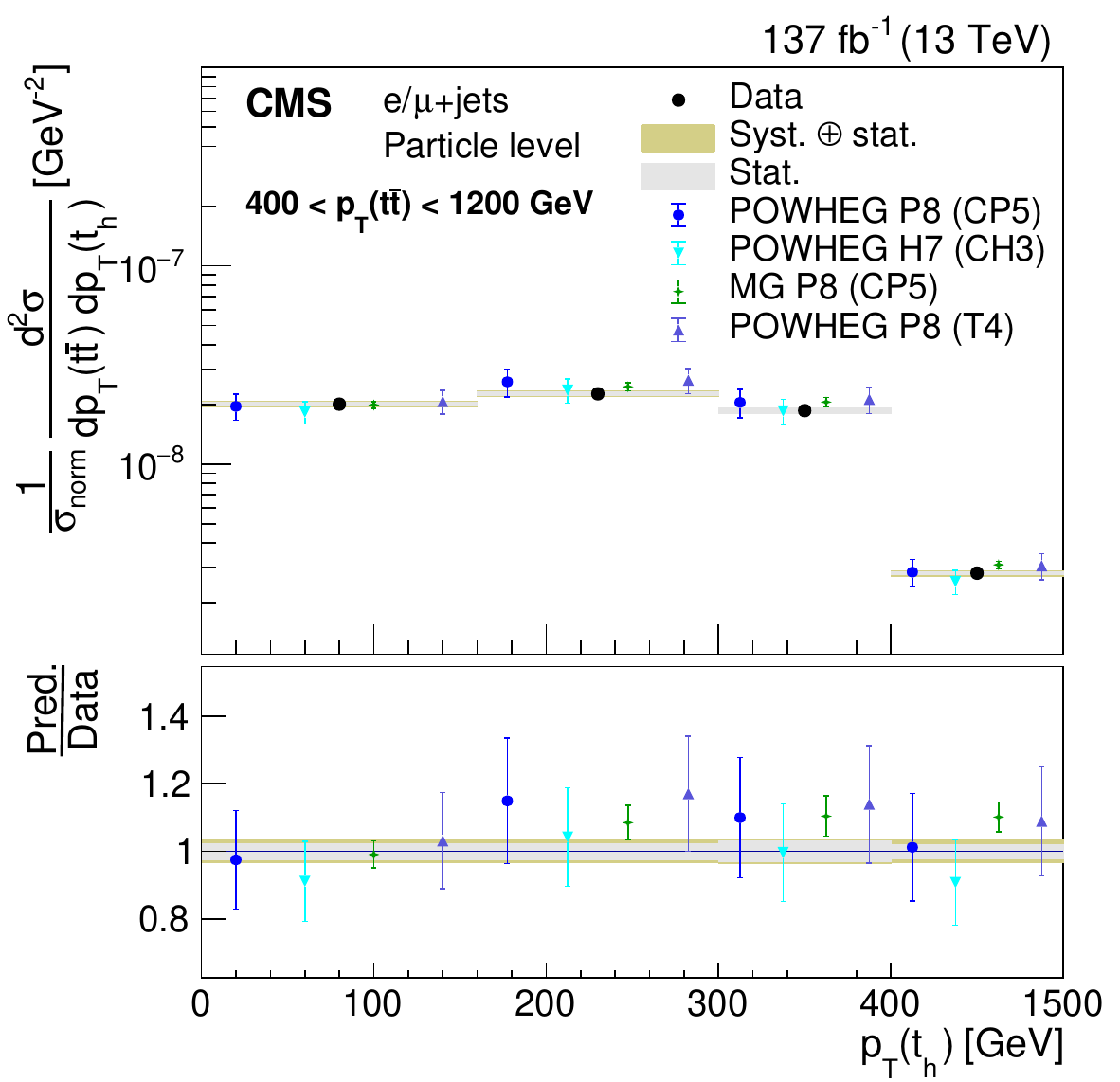}
 \caption{Normalized double-differential cross section at the particle level as a function of \ttptvsthadpt. \XSECCAPPS}
 \label{fig:RESNORMPS7b}
\end{figure*}

\begin{figure*}[tbp]
\centering
 \includegraphics[width=0.42\textwidth]{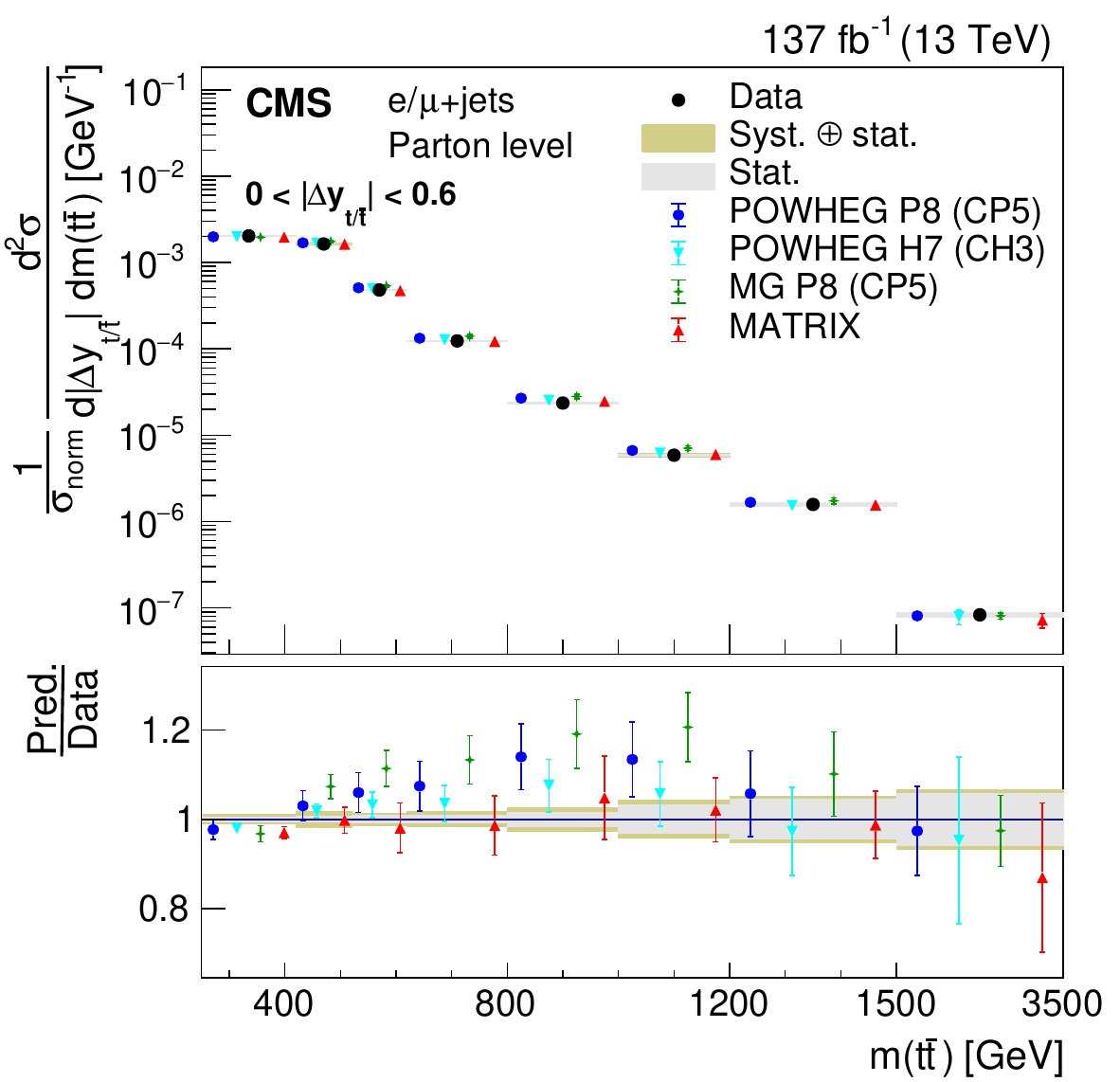}
 \includegraphics[width=0.42\textwidth]{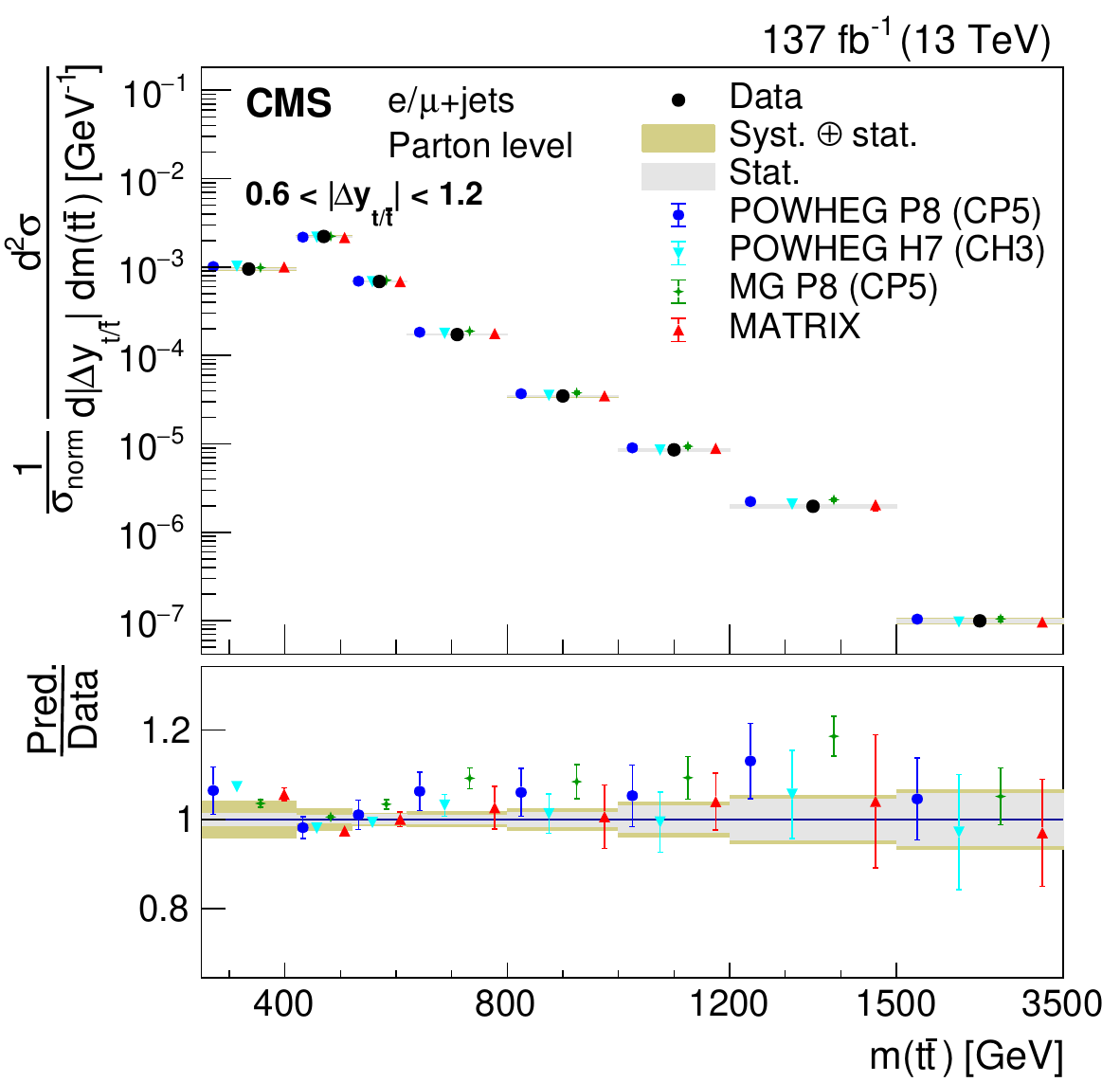}\\
 \includegraphics[width=0.42\textwidth]{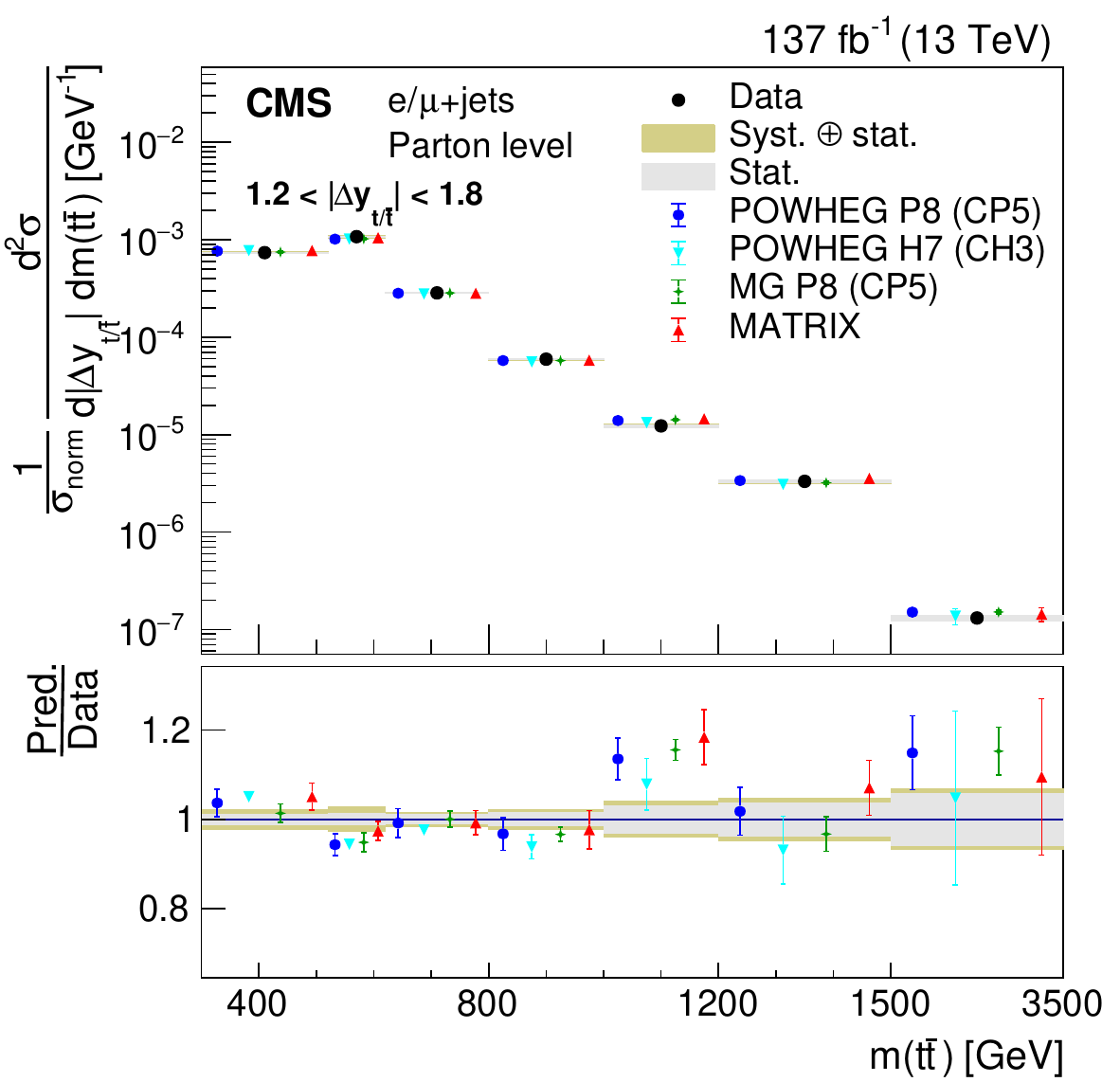}
 \includegraphics[width=0.42\textwidth]{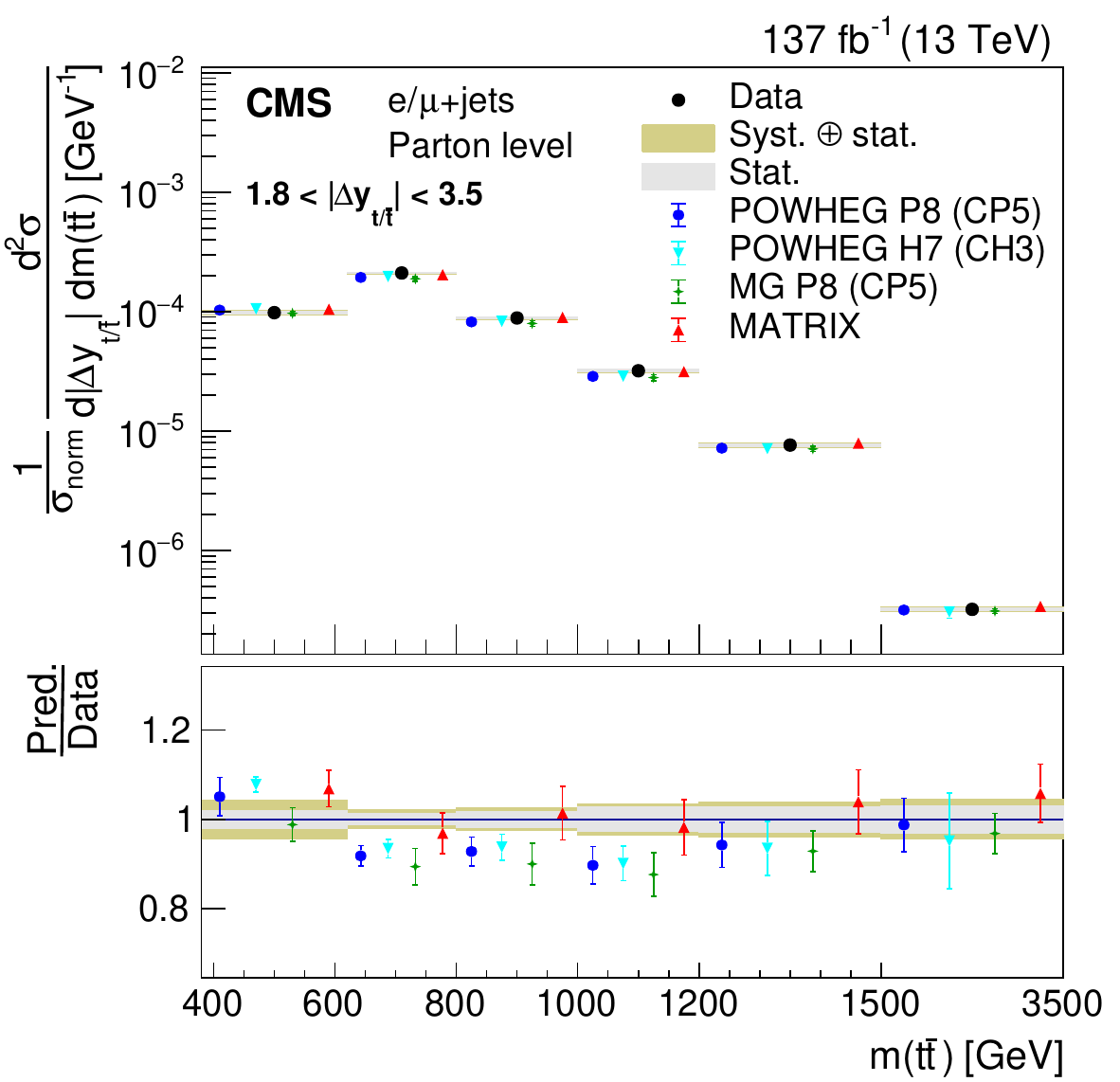}
 \caption{Normalized double-differential cross section at the parton level as a function of \adyvsttm. \XSECCAPPA}
 \label{fig:RESNORM8}
\end{figure*}

\begin{figure*}[tbp]
\centering
 \includegraphics[width=0.42\textwidth]{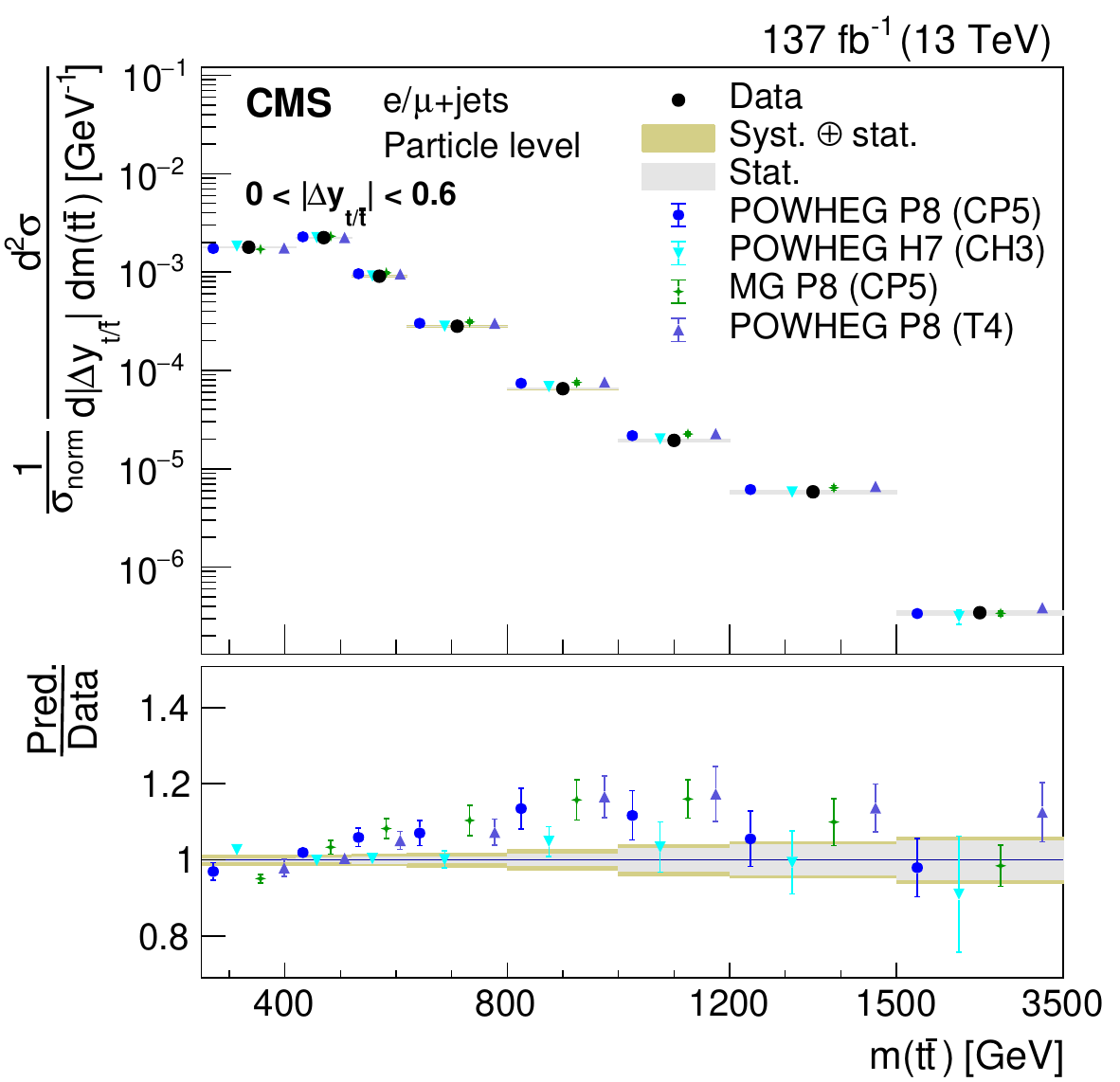}
 \includegraphics[width=0.42\textwidth]{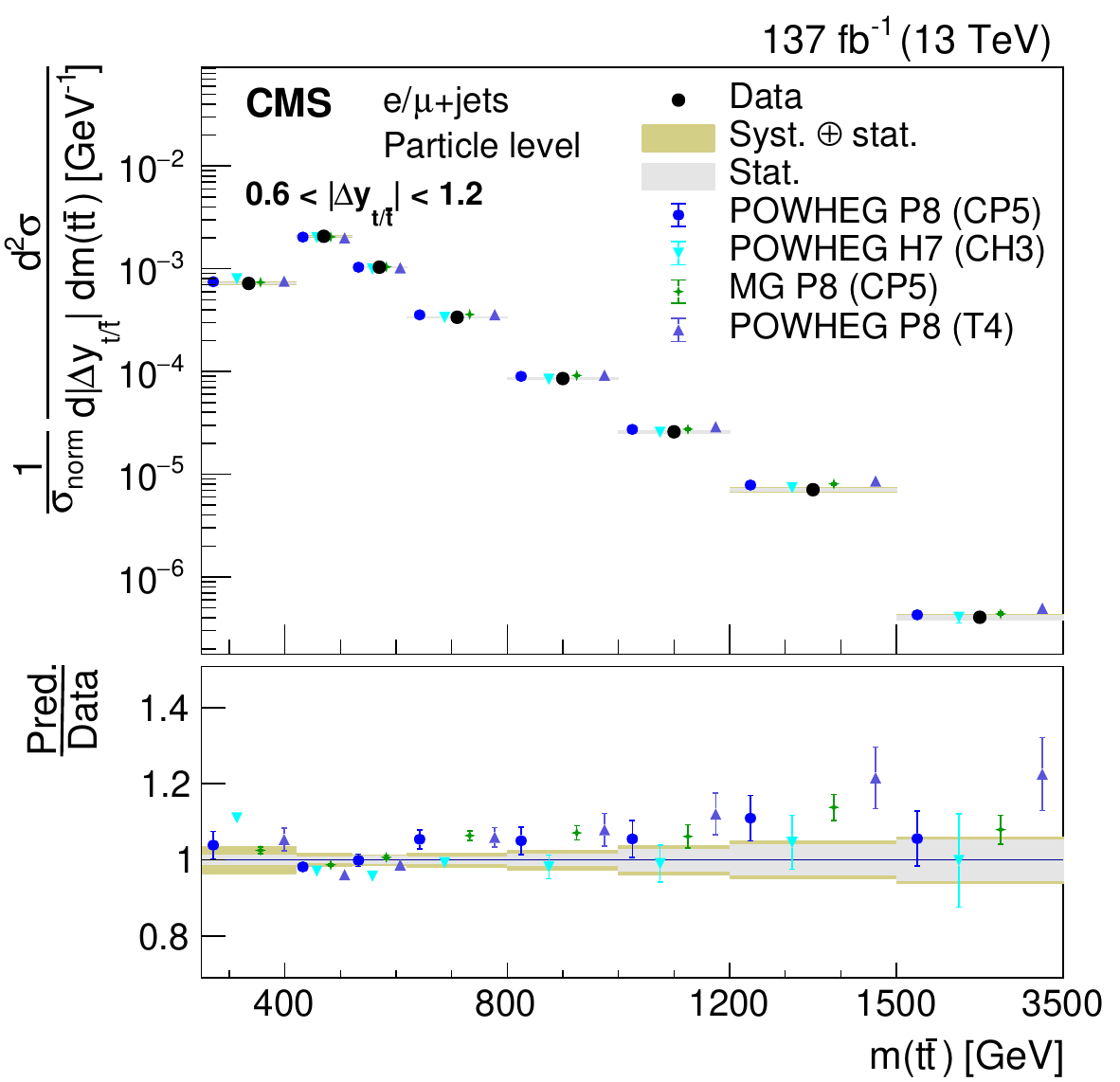}\\
 \includegraphics[width=0.42\textwidth]{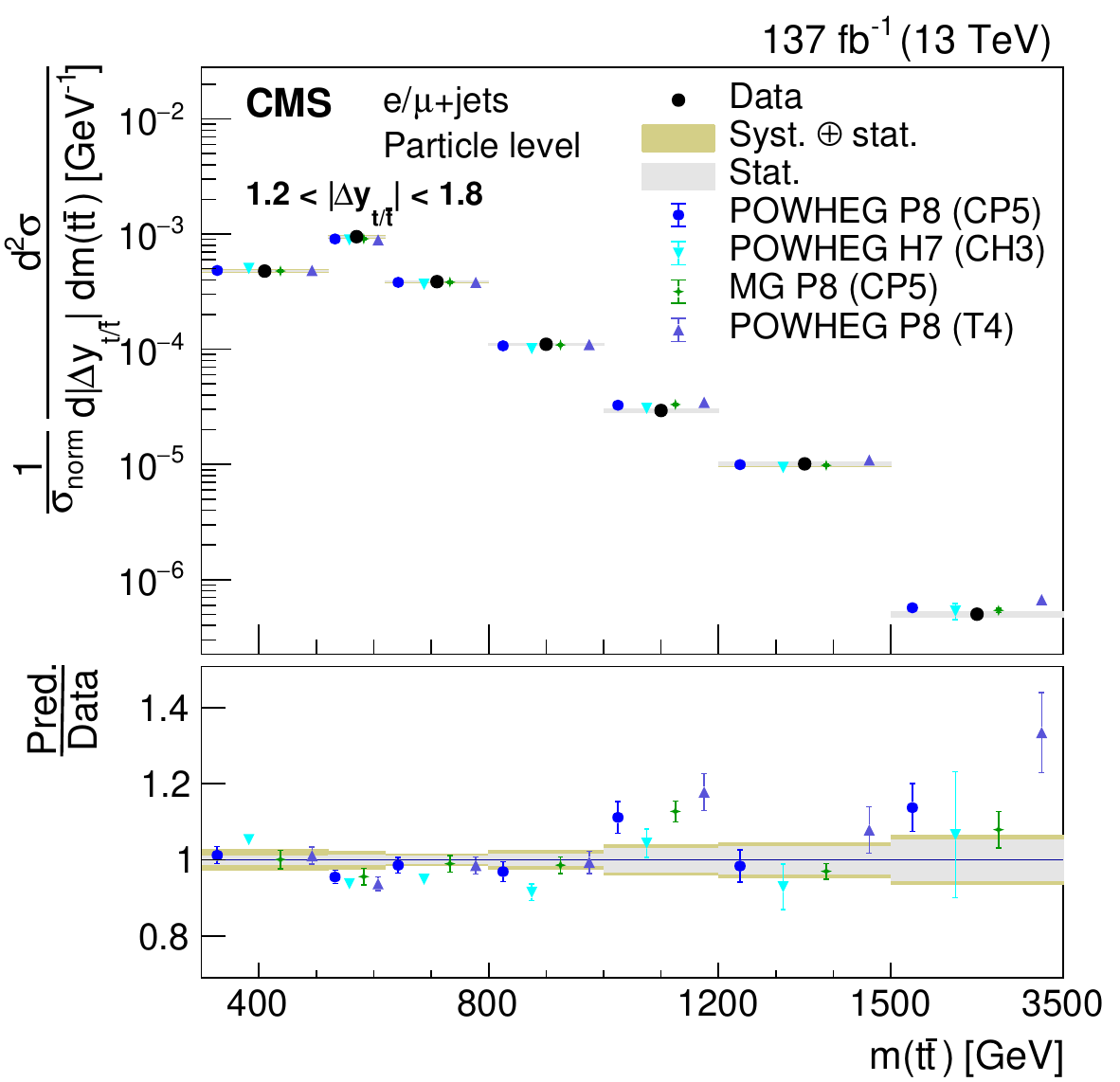}
 \includegraphics[width=0.42\textwidth]{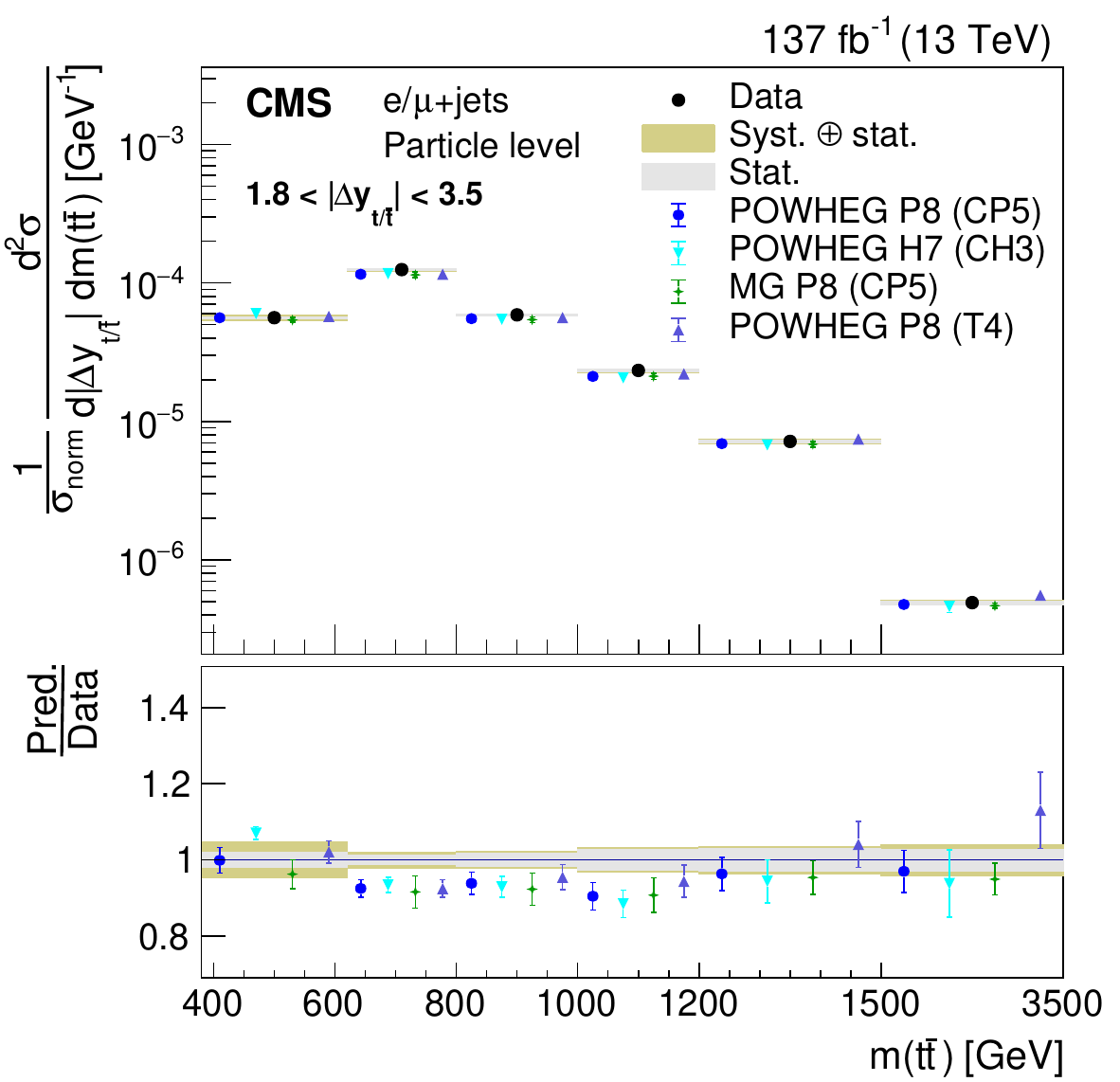}
 \caption{Normalized double-differential cross section at the particle level as a function of \adyvsttm. \XSECCAPPS}
 \label{fig:RESNORMPS8}
\end{figure*}

\begin{figure*}[tbp]
\centering
 \includegraphics[width=0.42\textwidth]{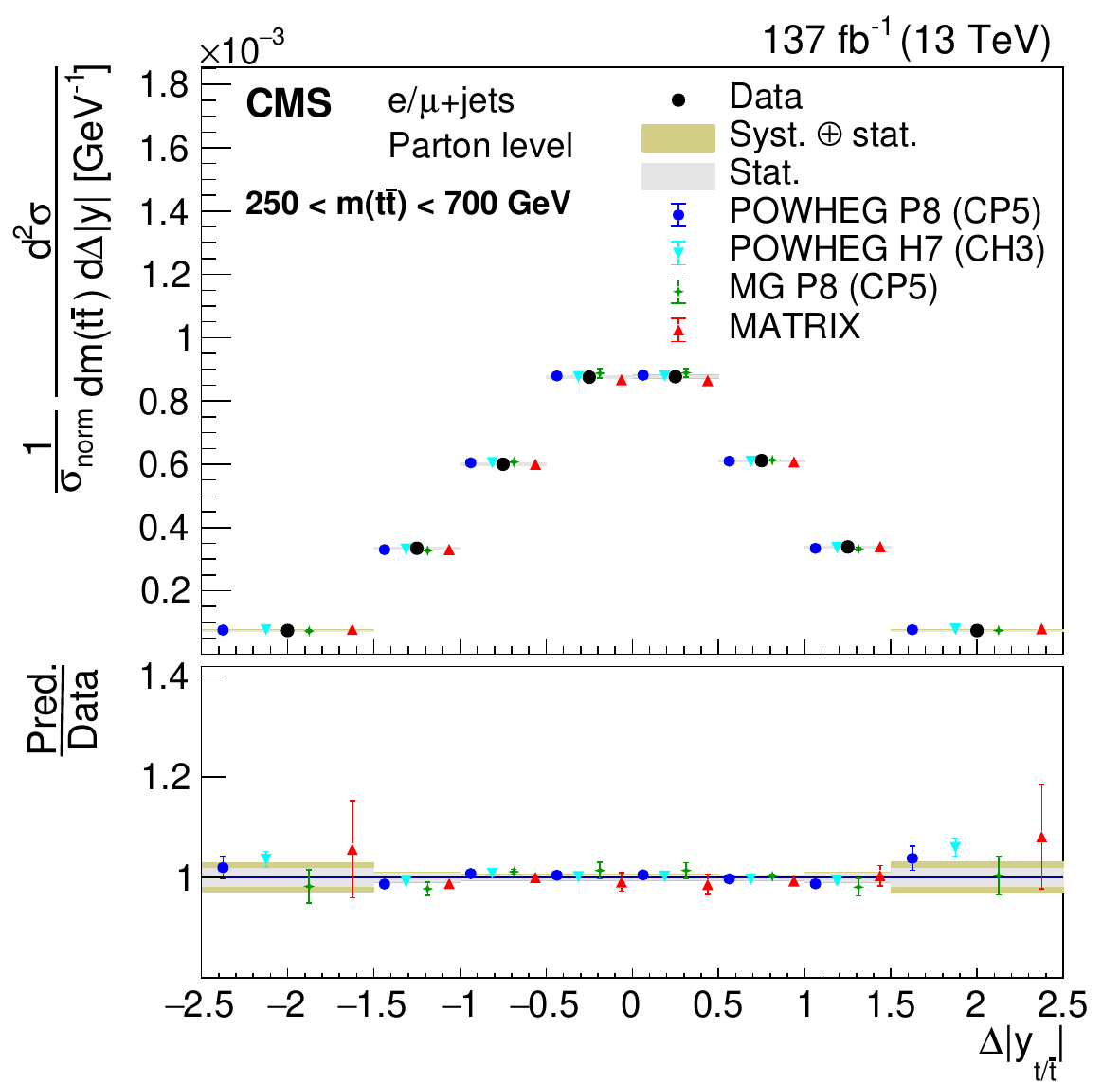}
 \includegraphics[width=0.42\textwidth]{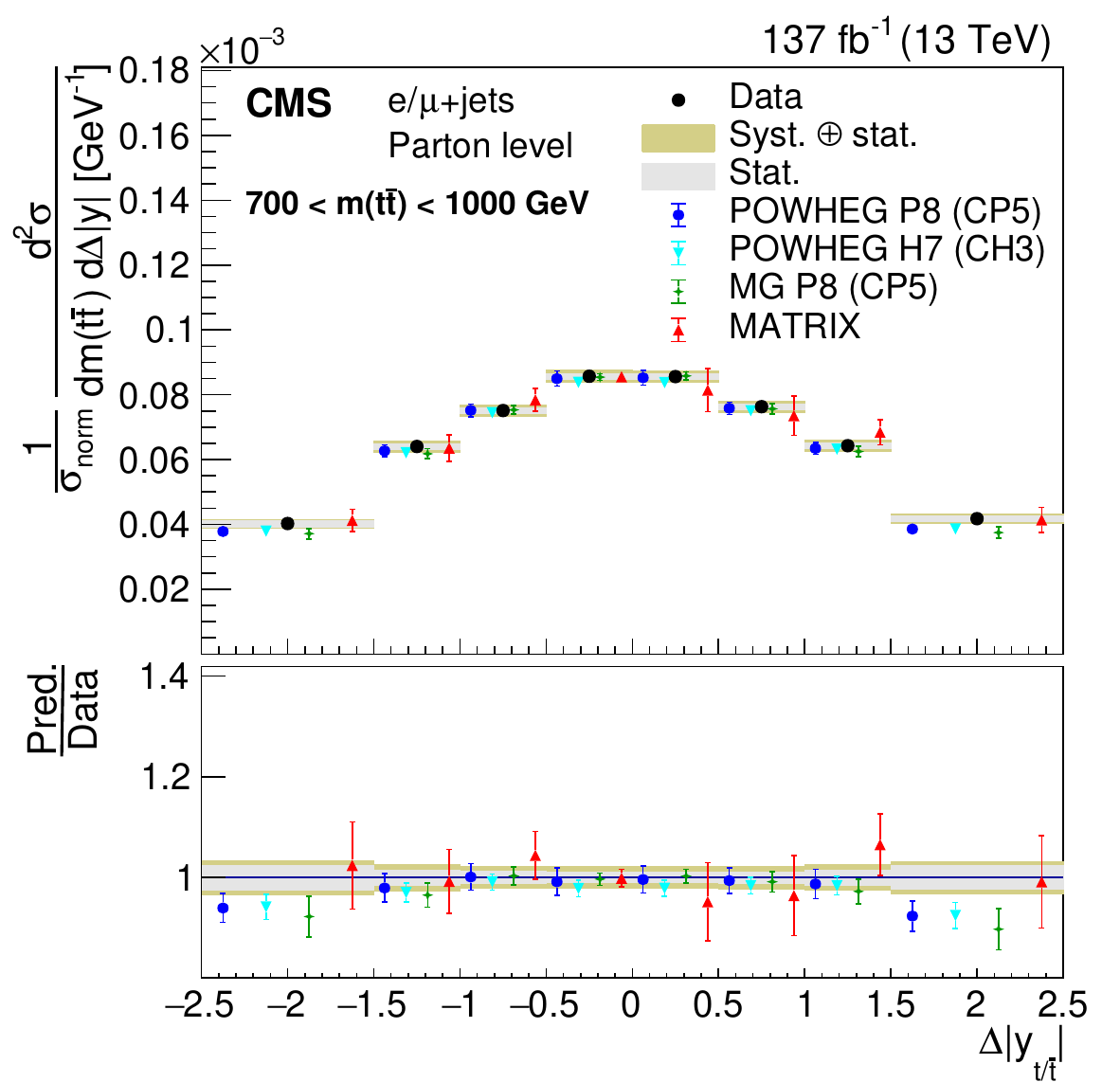}\\
 \includegraphics[width=0.42\textwidth]{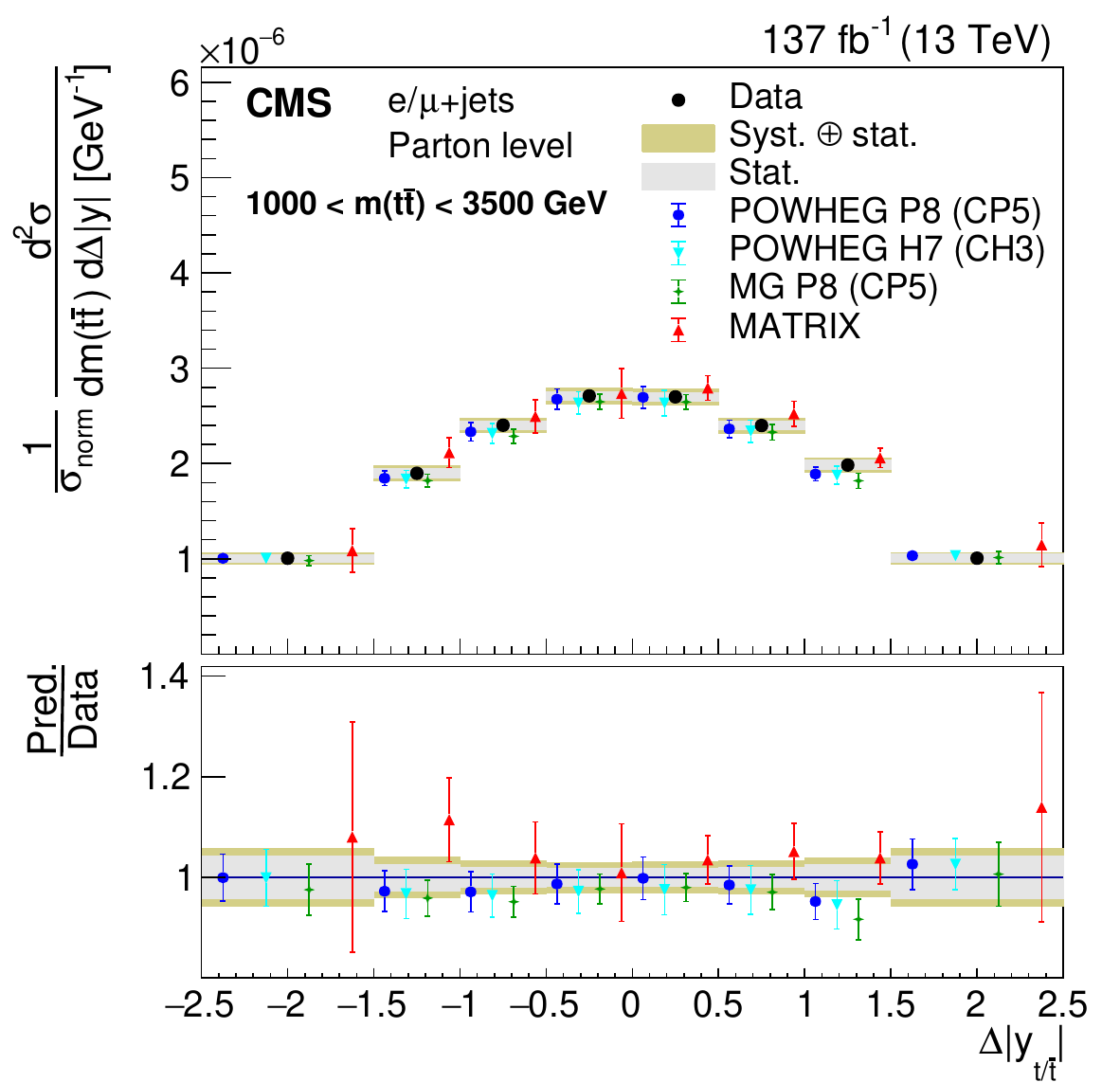}
 \caption{Normalized double-differential cross section at the parton level as a function of \ttmvsdy. \XSECCAPPA}
 \label{fig:RESNORM9}
\end{figure*}

\begin{figure*}[tbp]
\centering
 \includegraphics[width=0.42\textwidth]{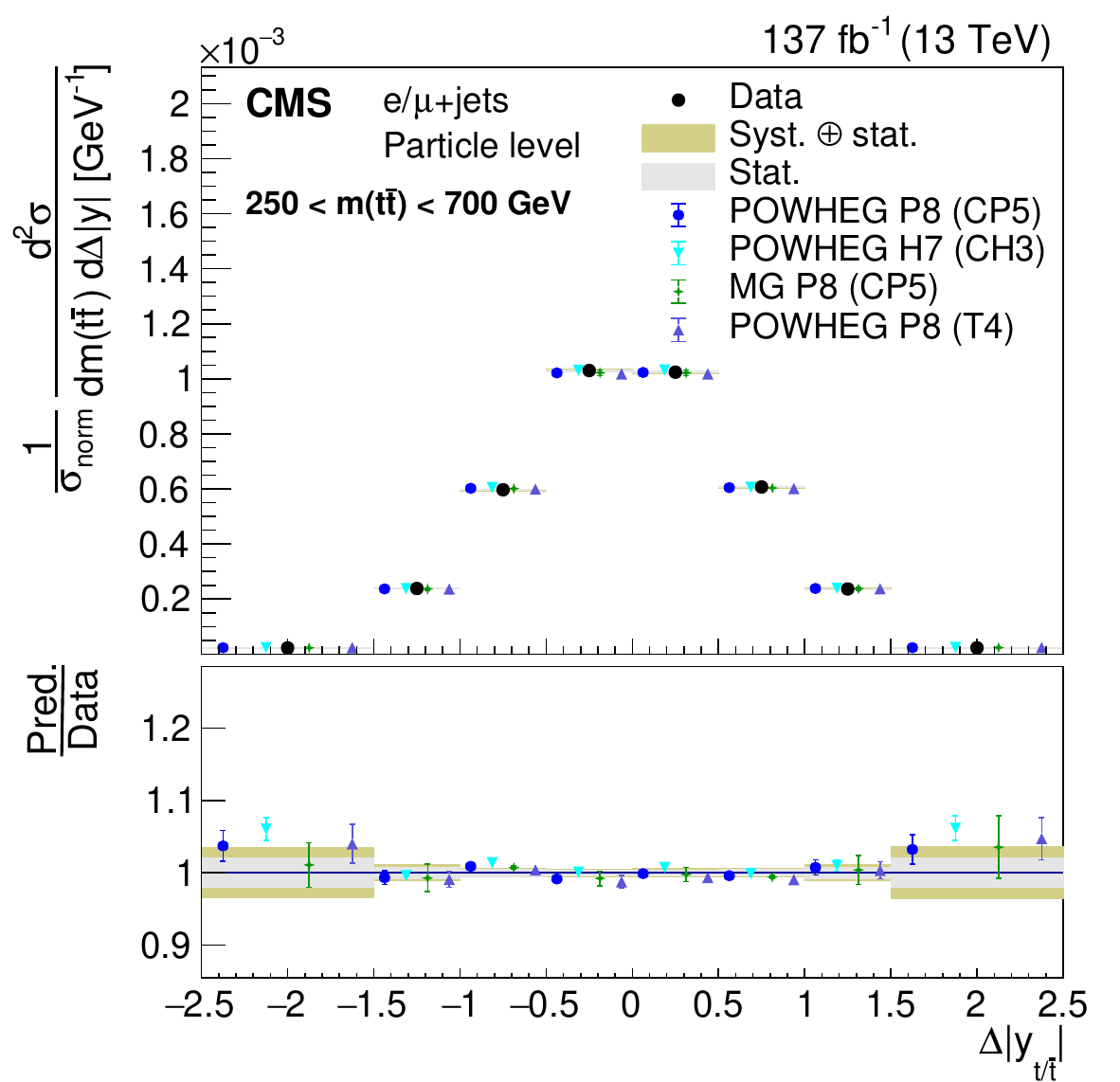}
 \includegraphics[width=0.42\textwidth]{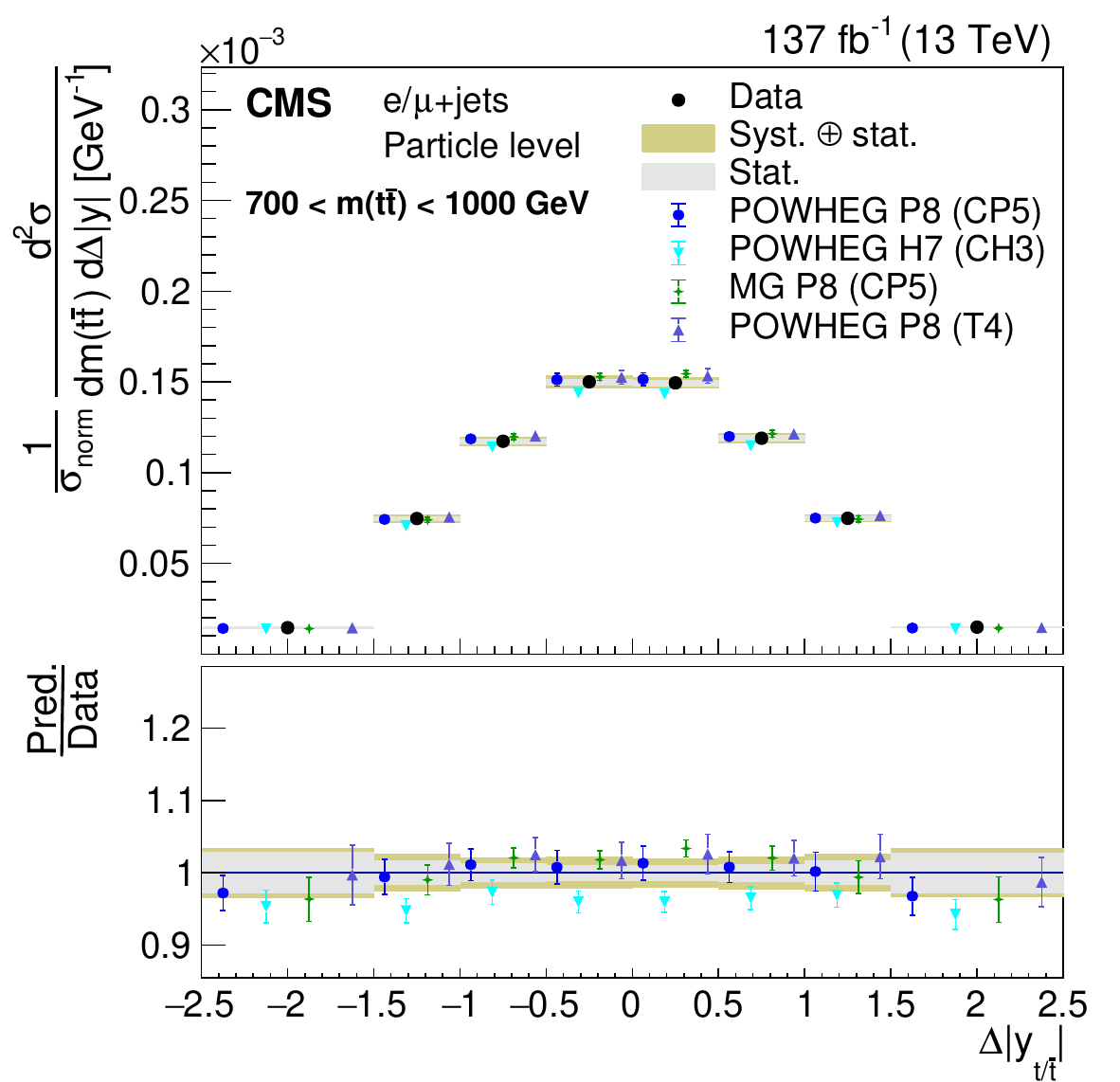}\\
 \includegraphics[width=0.42\textwidth]{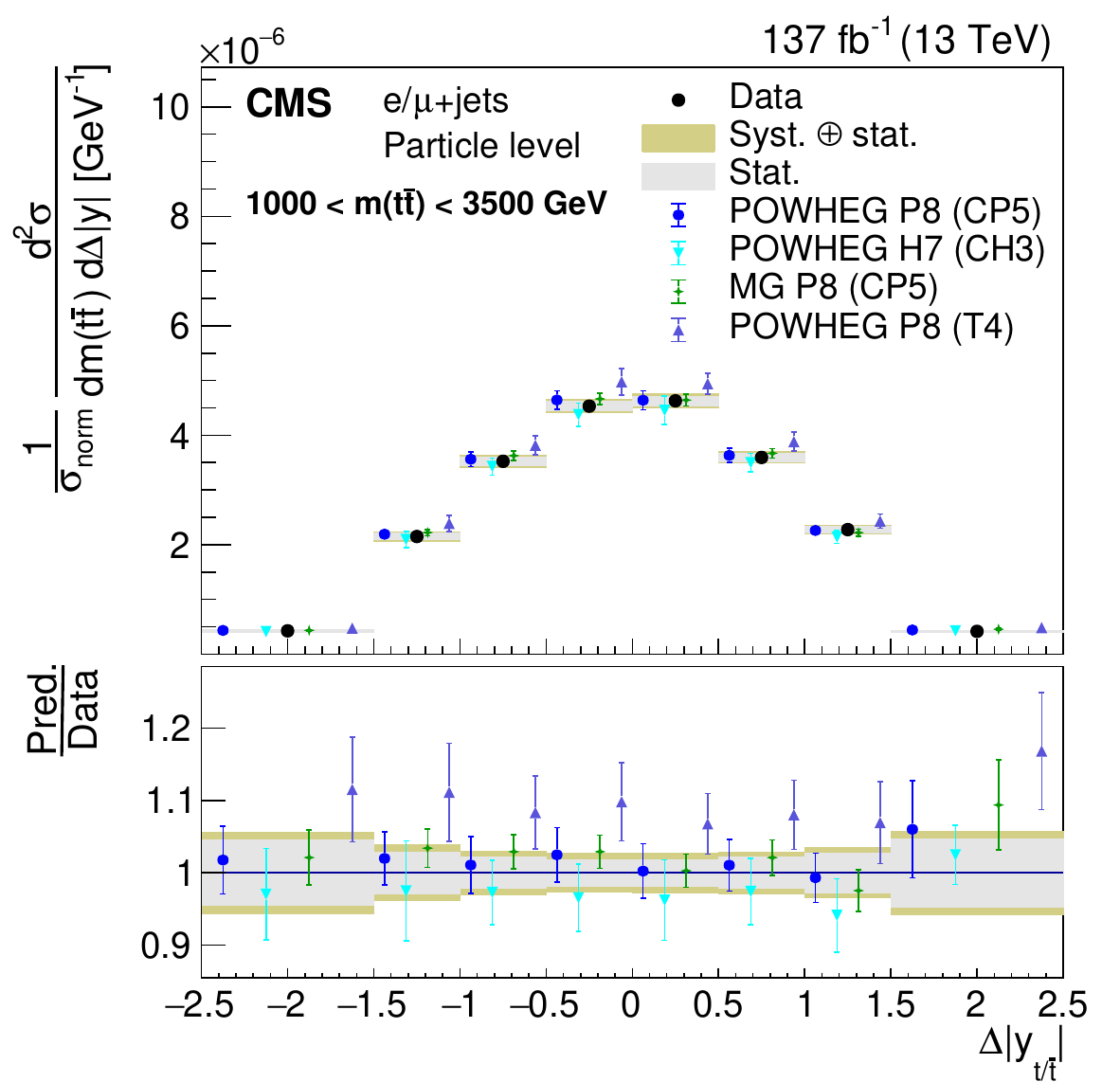}
 \caption{Normalized double-differential cross section at the particle level as a function of \ttmvsdy. \XSECCAPPS}
 \label{fig:RESNORMPS9}
\end{figure*}

\begin{figure*}[tbp]
\centering
 \includegraphics[width=0.42\textwidth]{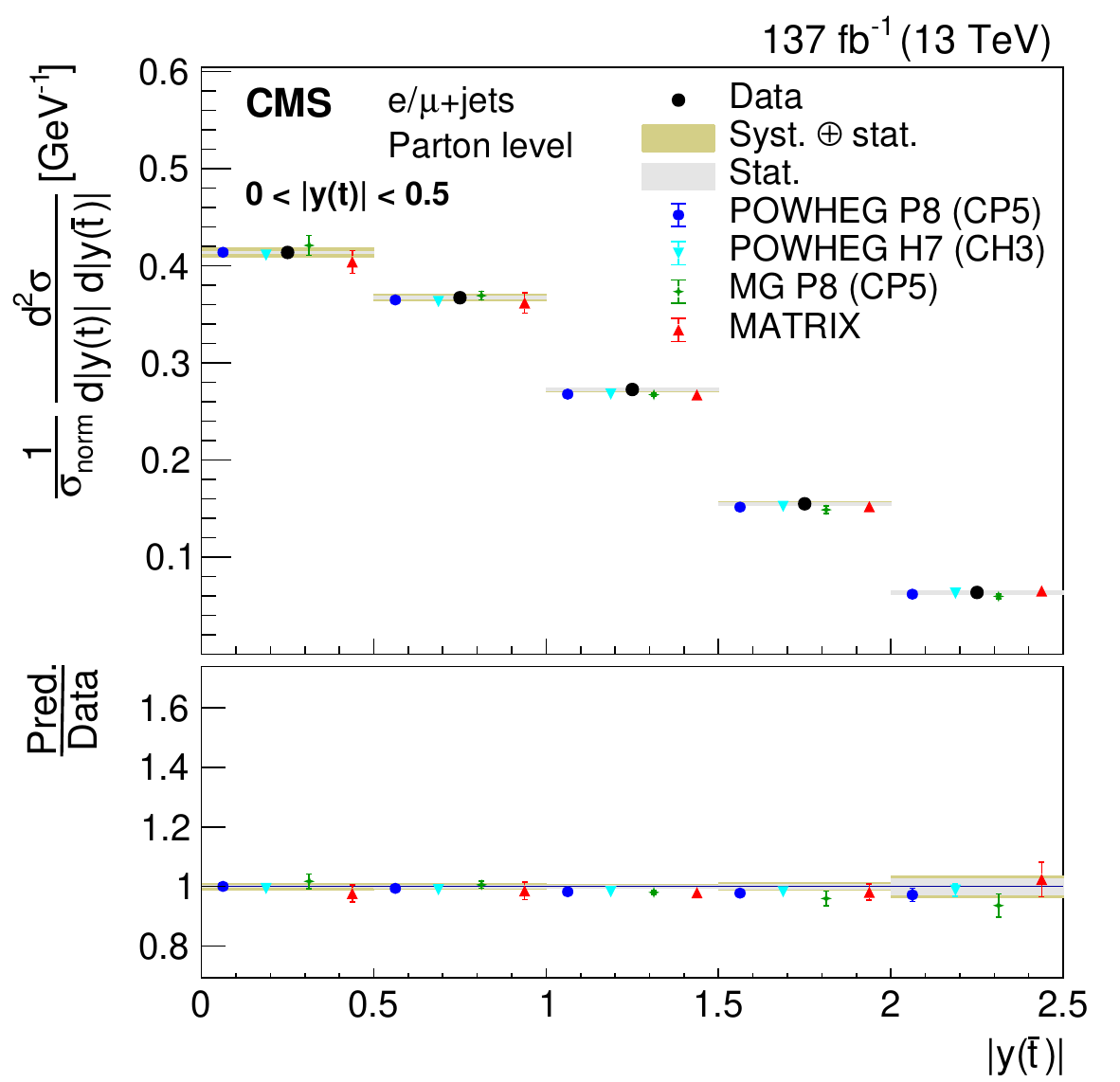}
 \includegraphics[width=0.42\textwidth]{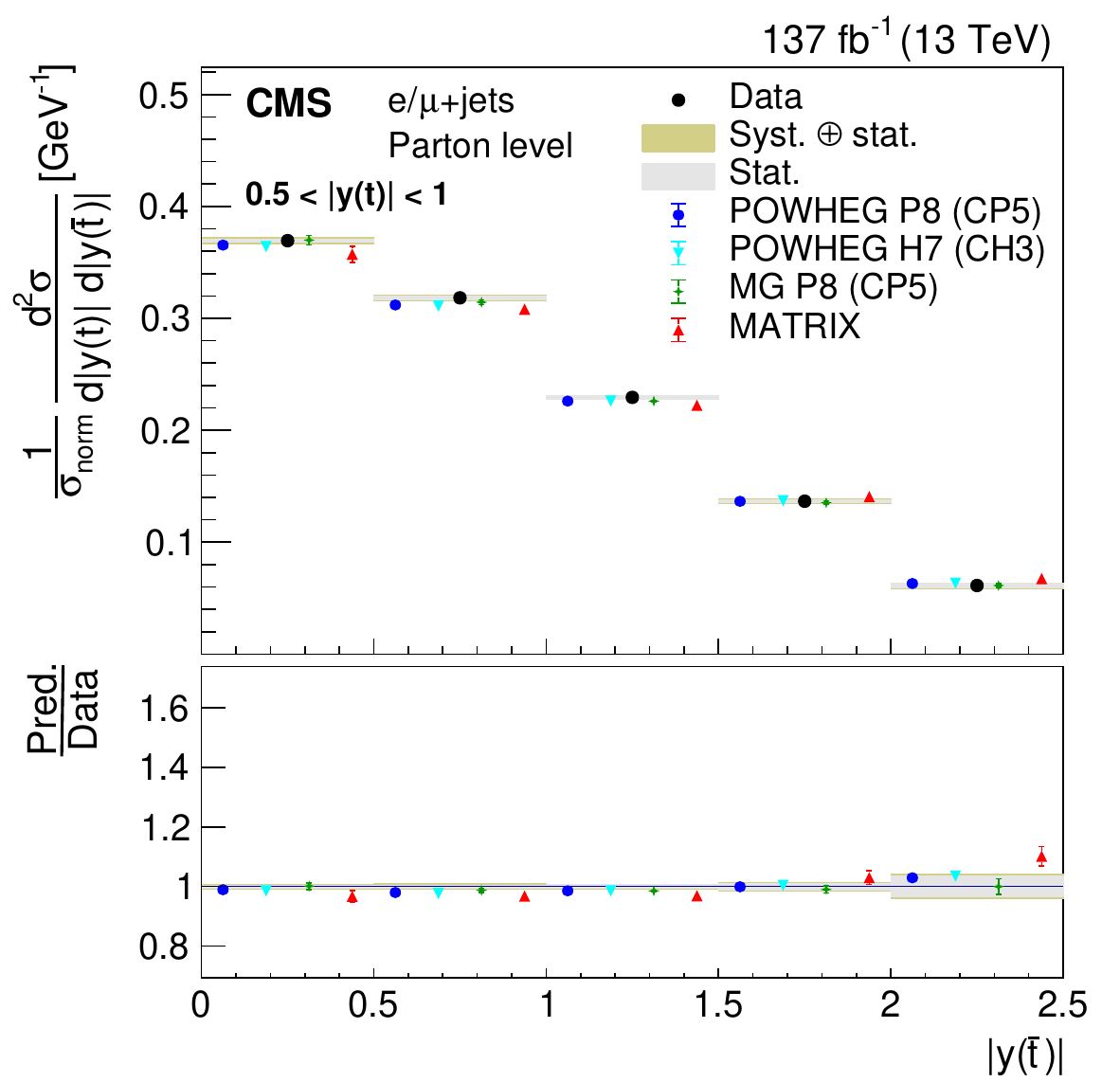}\\
 \includegraphics[width=0.42\textwidth]{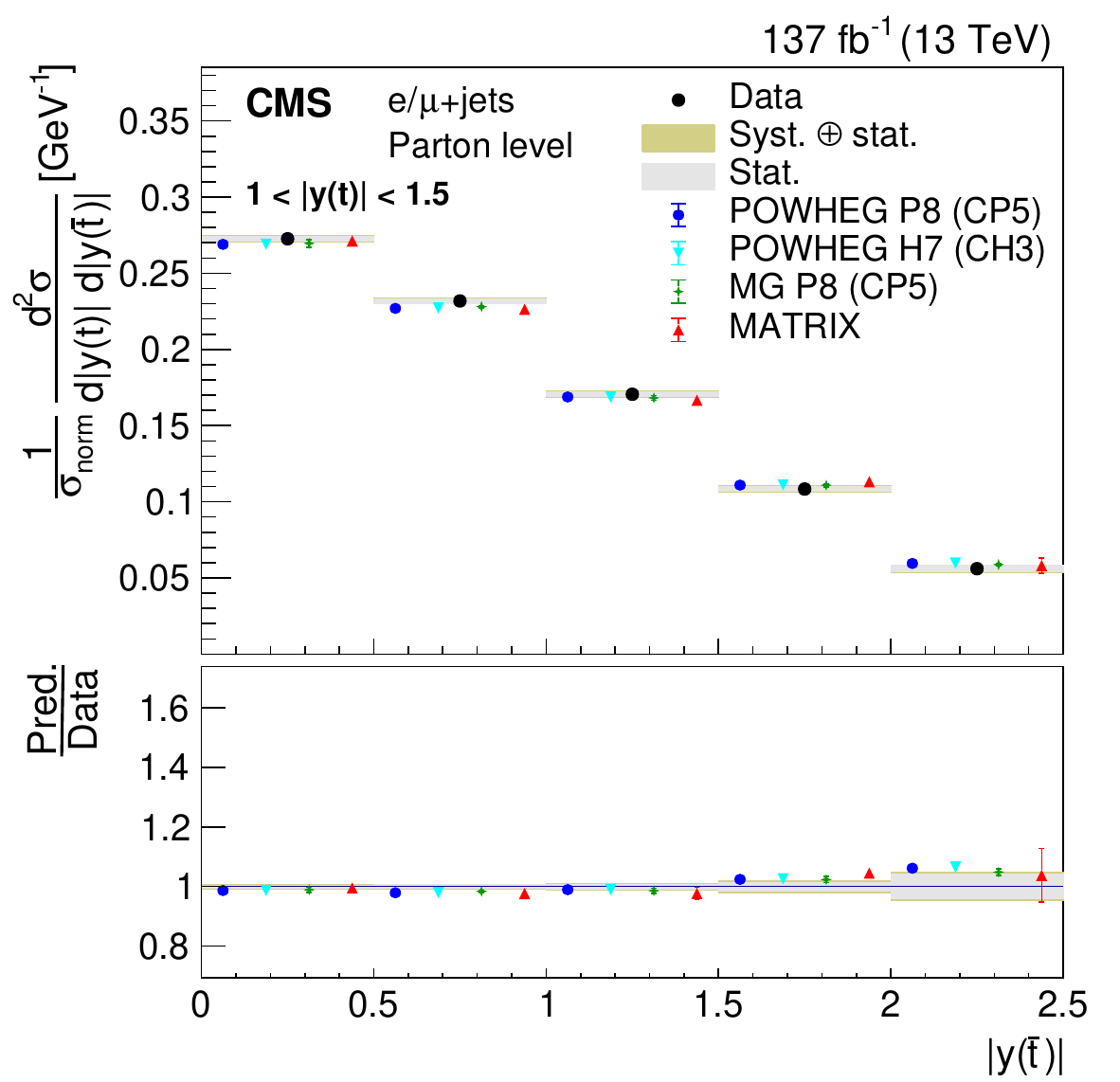}
 \includegraphics[width=0.42\textwidth]{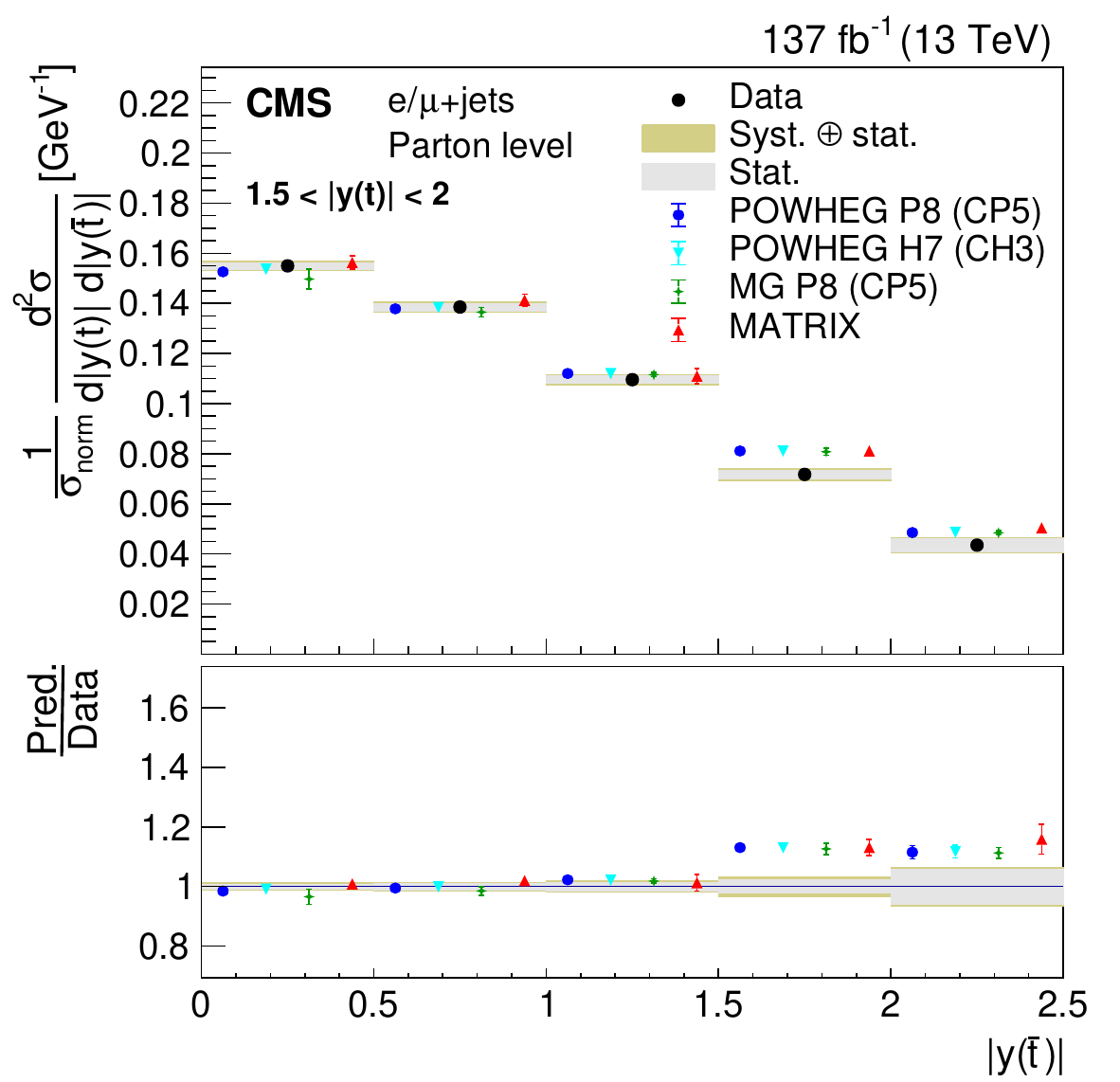}\\
 \includegraphics[width=0.42\textwidth]{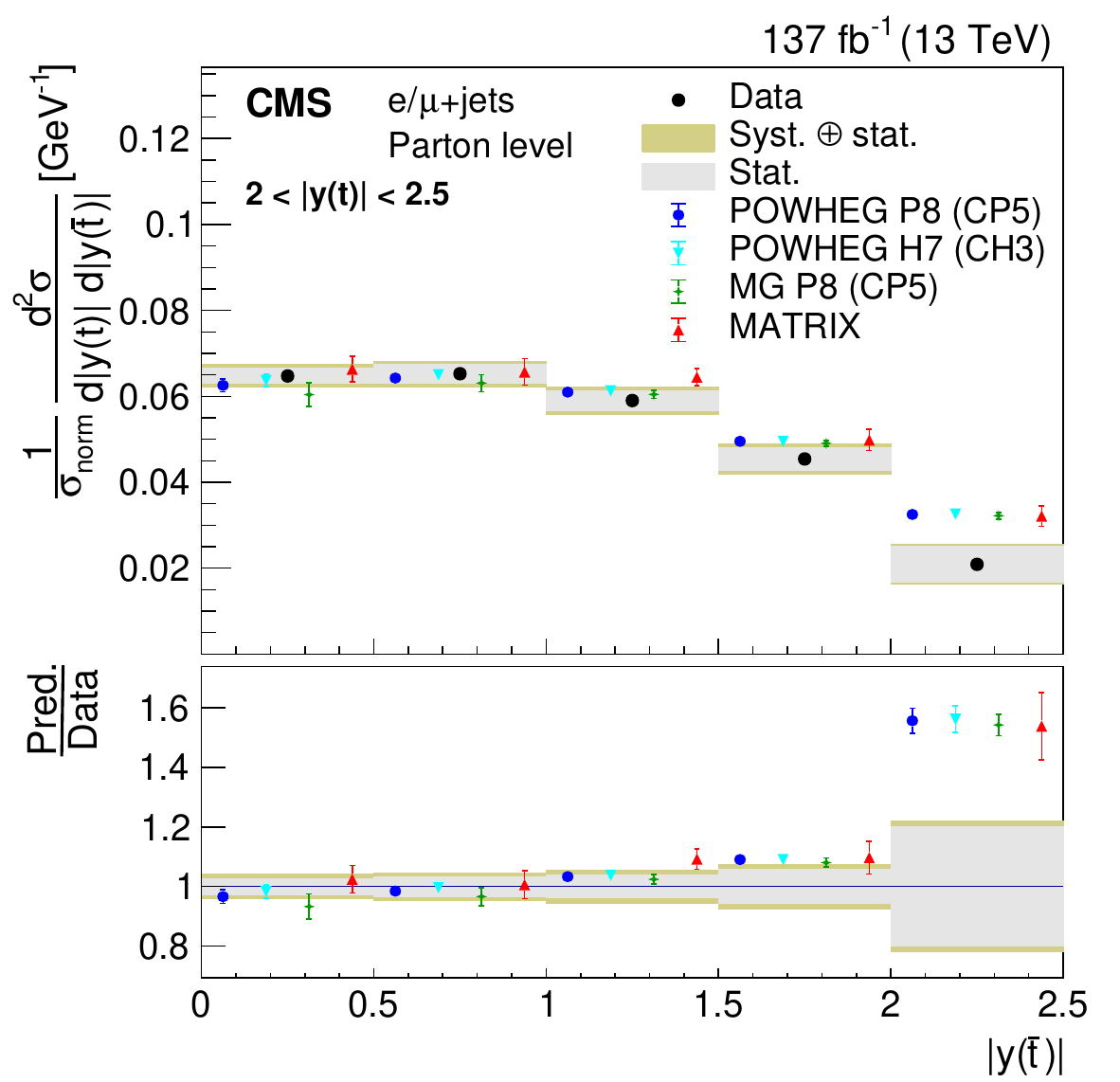}
 \caption{Normalized double-differential cross section at the parton level as a function of \topyvstopbary. \XSECCAPPA}
 \label{fig:RESNORM10}
\end{figure*}

\begin{figure*}[tbp]
\centering
 \includegraphics[width=0.42\textwidth]{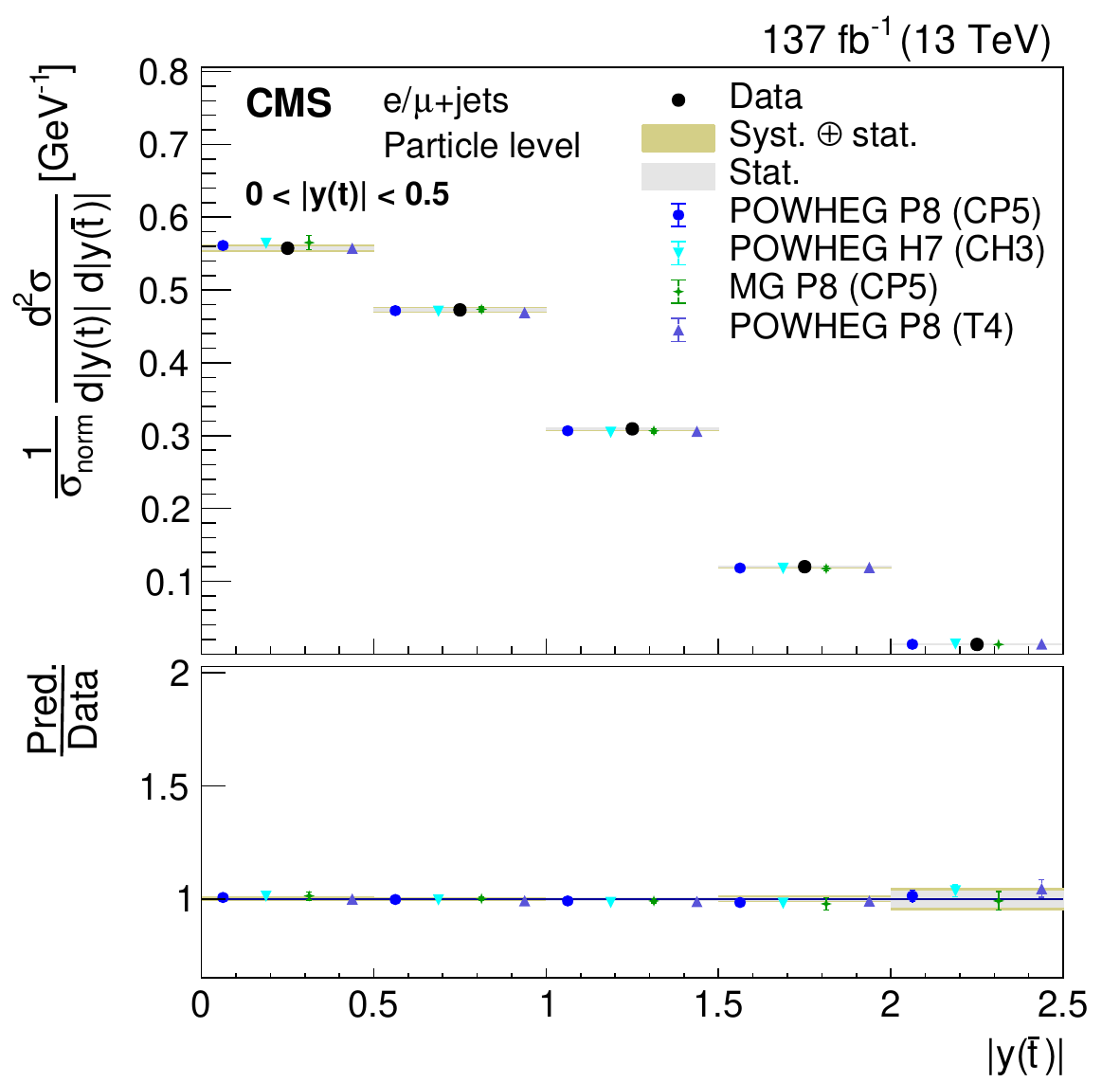}
 \includegraphics[width=0.42\textwidth]{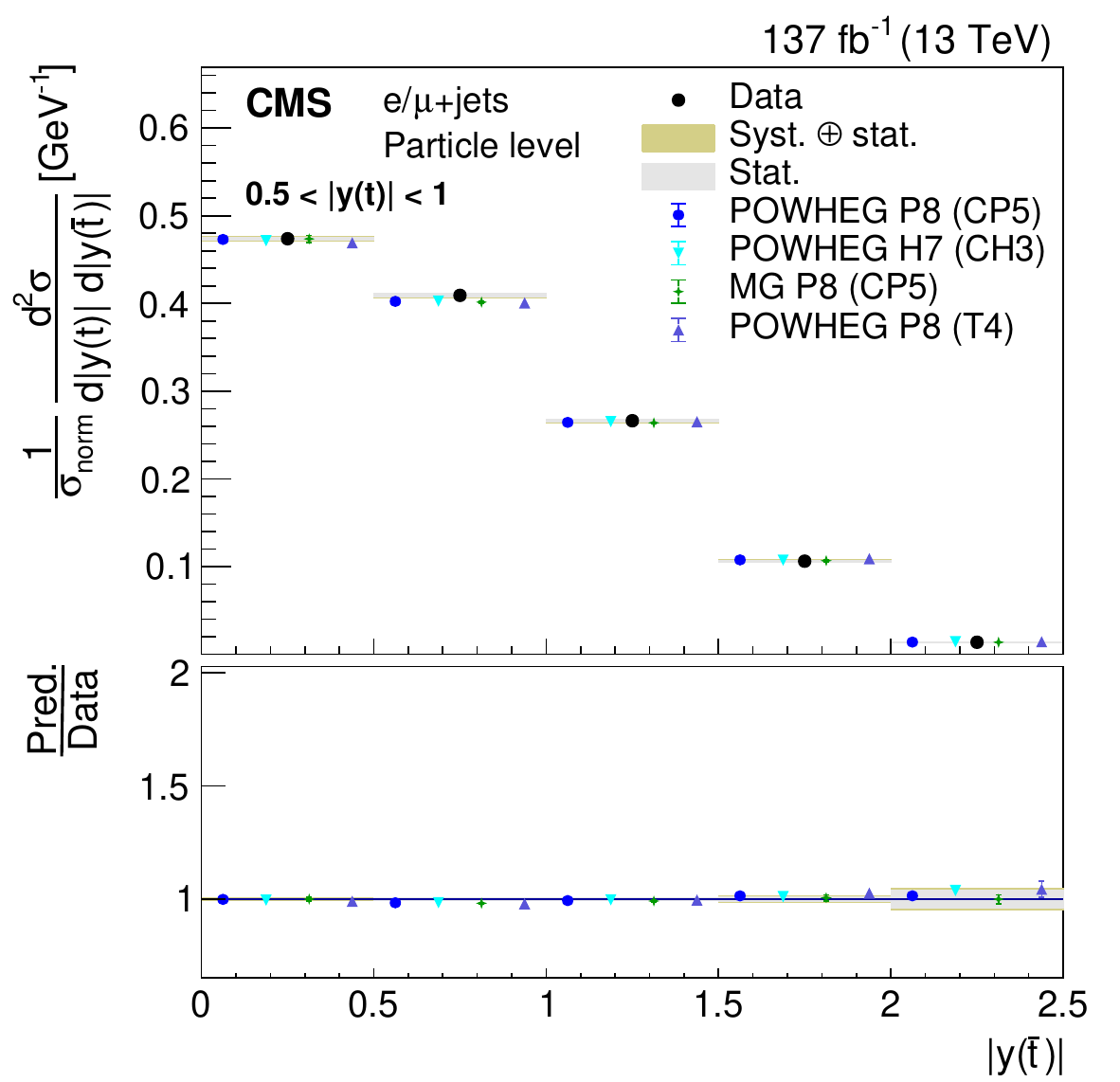}\\
 \includegraphics[width=0.42\textwidth]{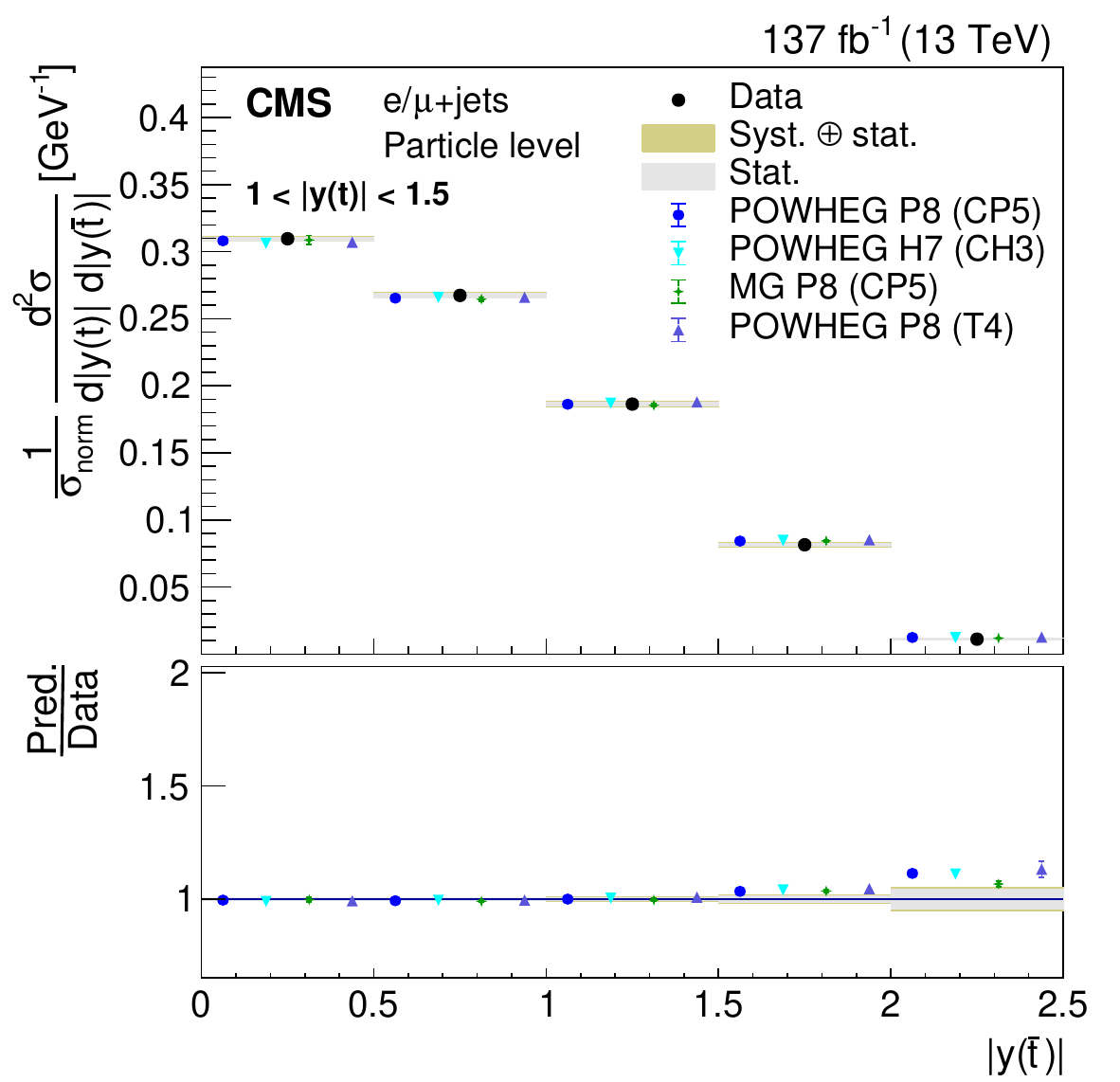}
 \includegraphics[width=0.42\textwidth]{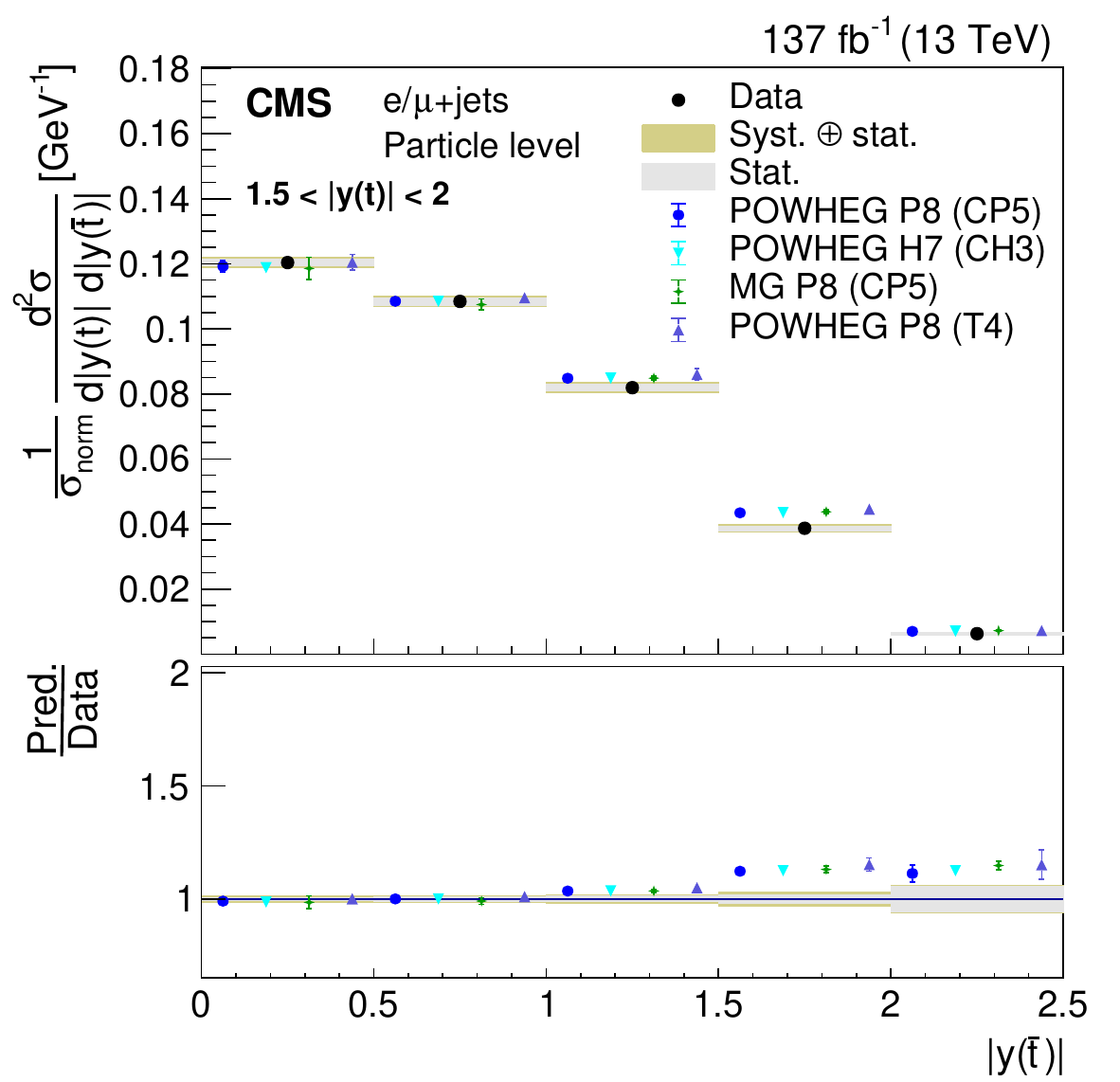}\\
 \includegraphics[width=0.42\textwidth]{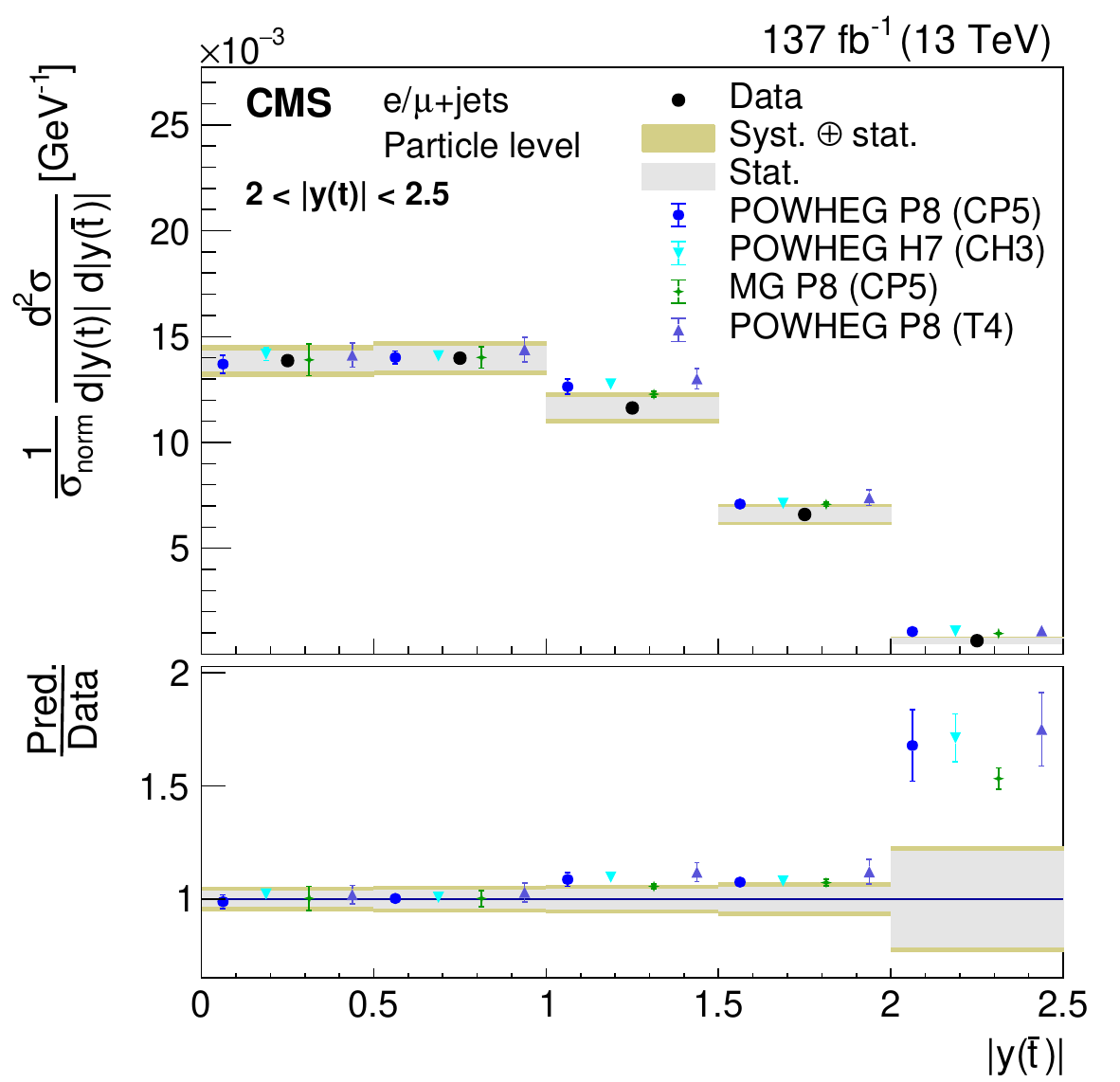}
 \caption{Normalized double-differential cross section at the particle level as a function of \topyvstopbary. \XSECCAPPS}
 \label{fig:RESNORMPS10}
\end{figure*}

\begin{figure*}[tbp]
\centering
 \includegraphics[width=0.42\textwidth]{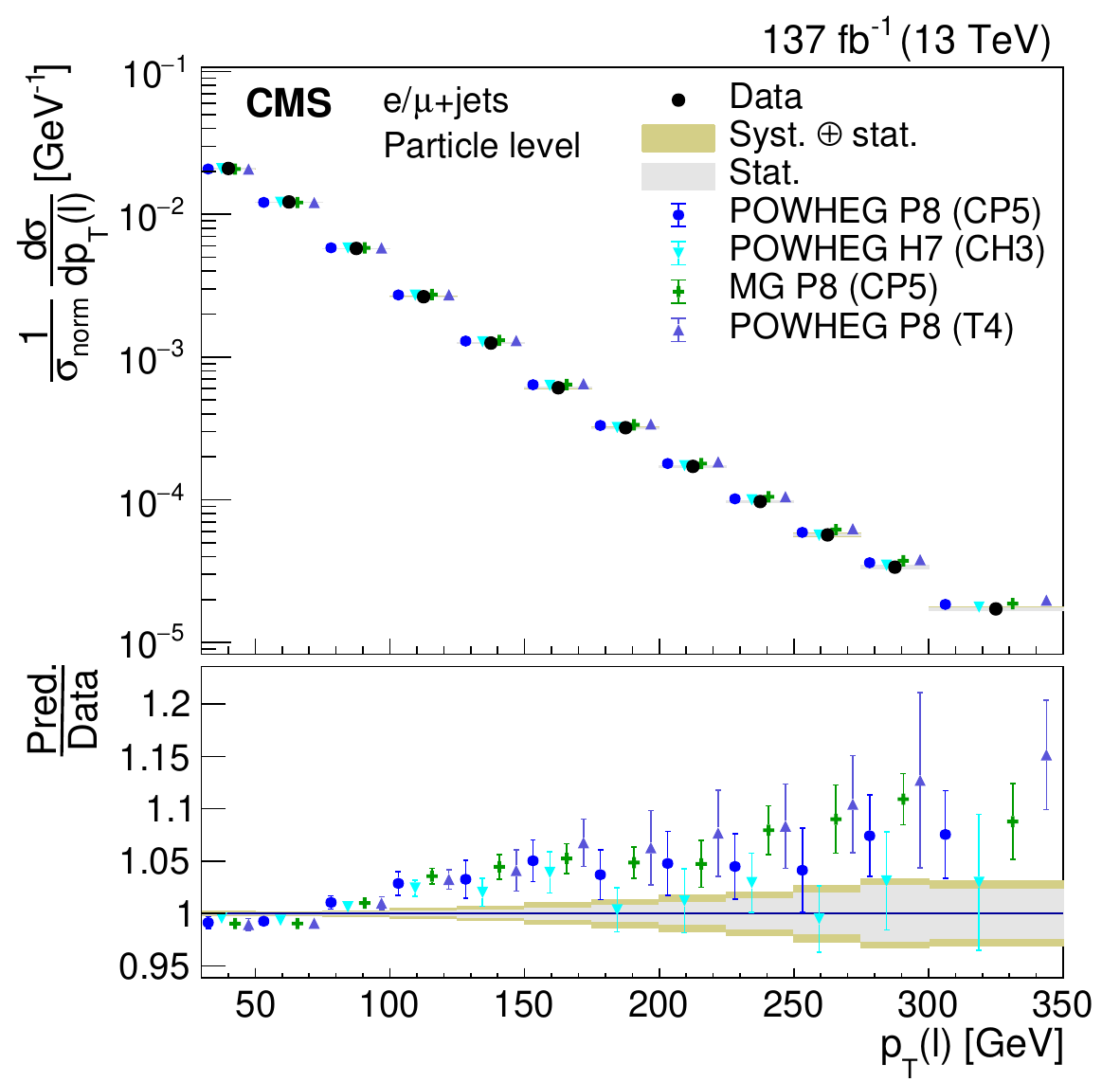}
 \includegraphics[width=0.42\textwidth]{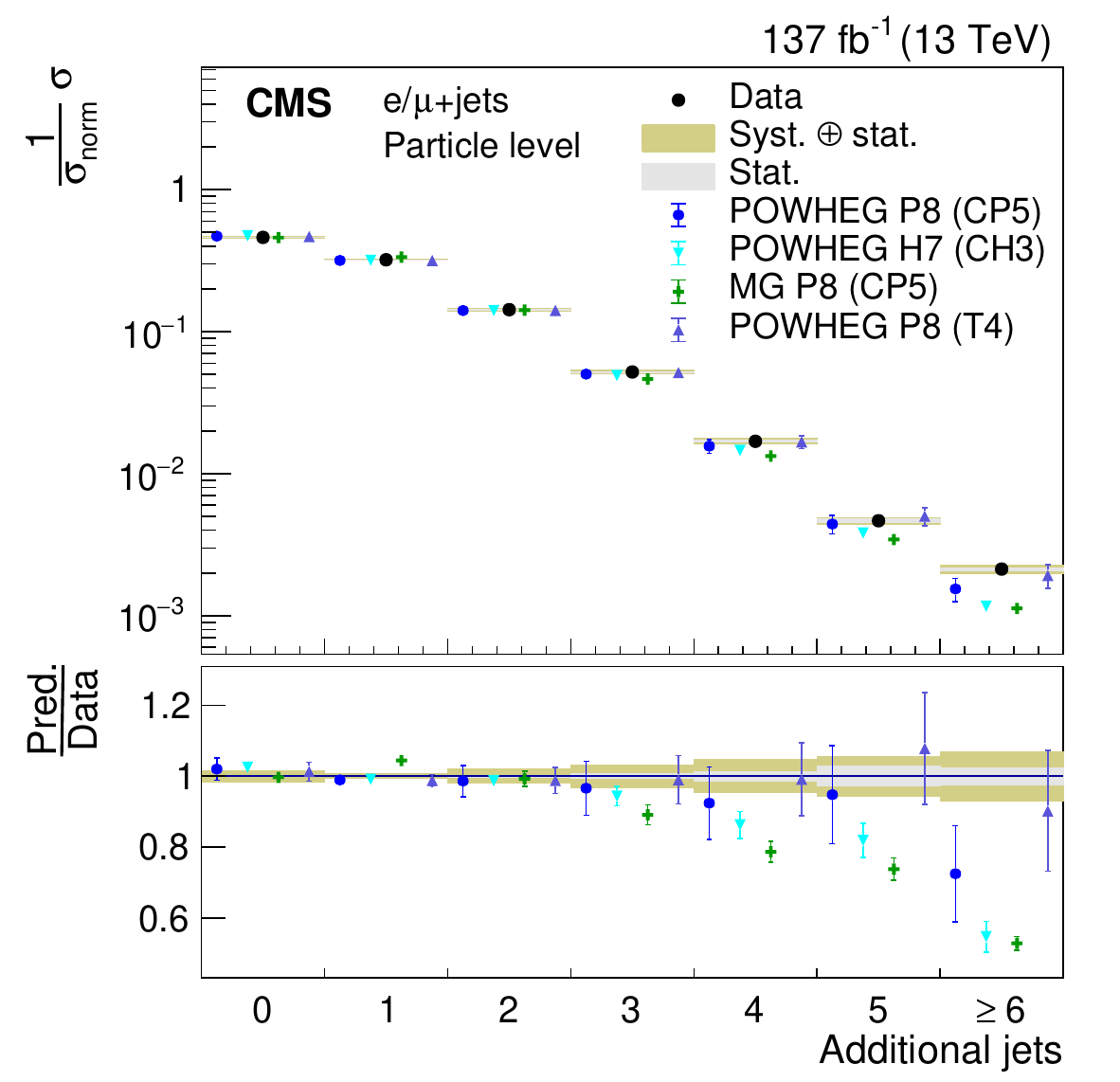}
 \includegraphics[width=0.42\textwidth]{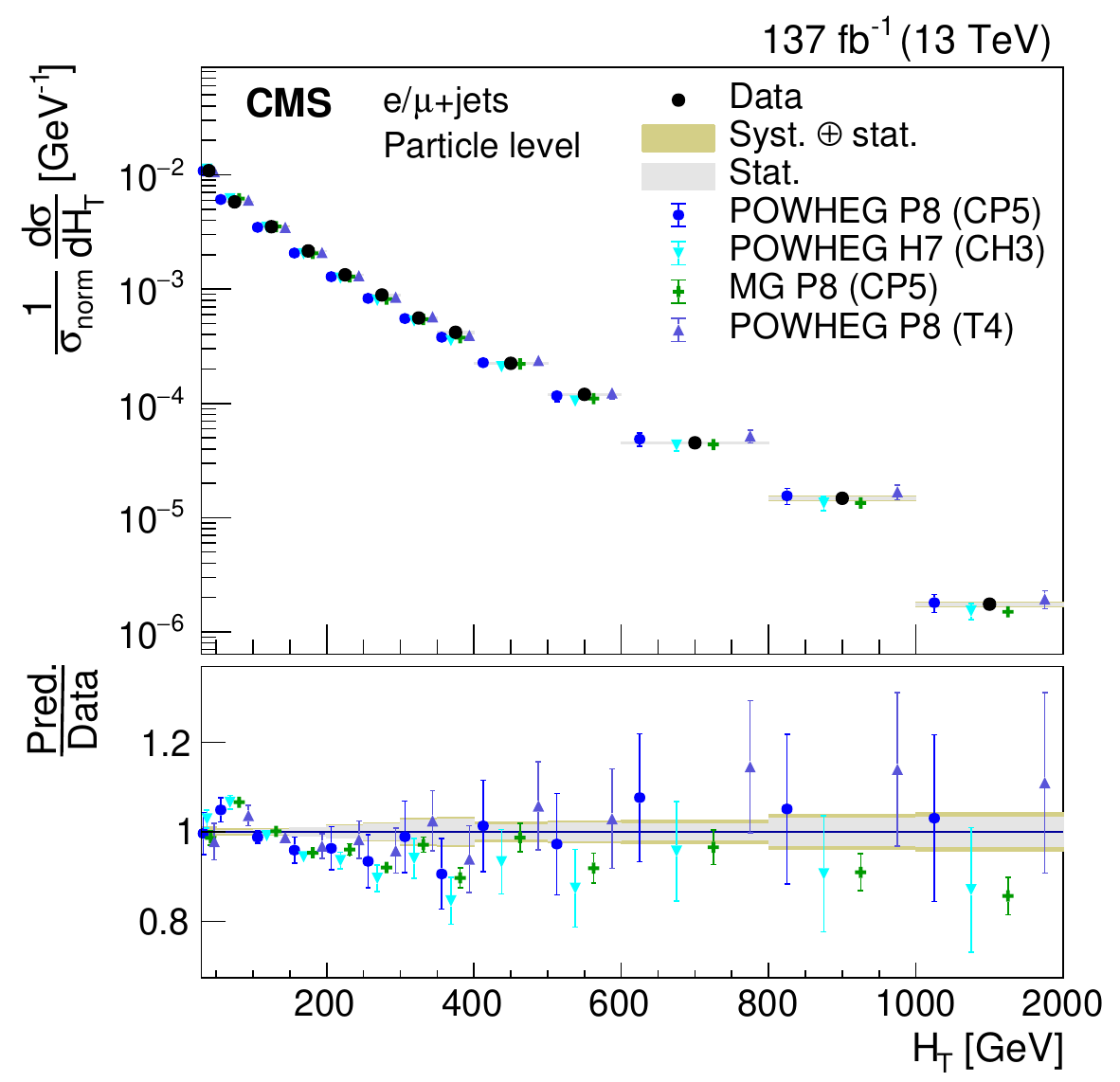}
 \includegraphics[width=0.42\textwidth]{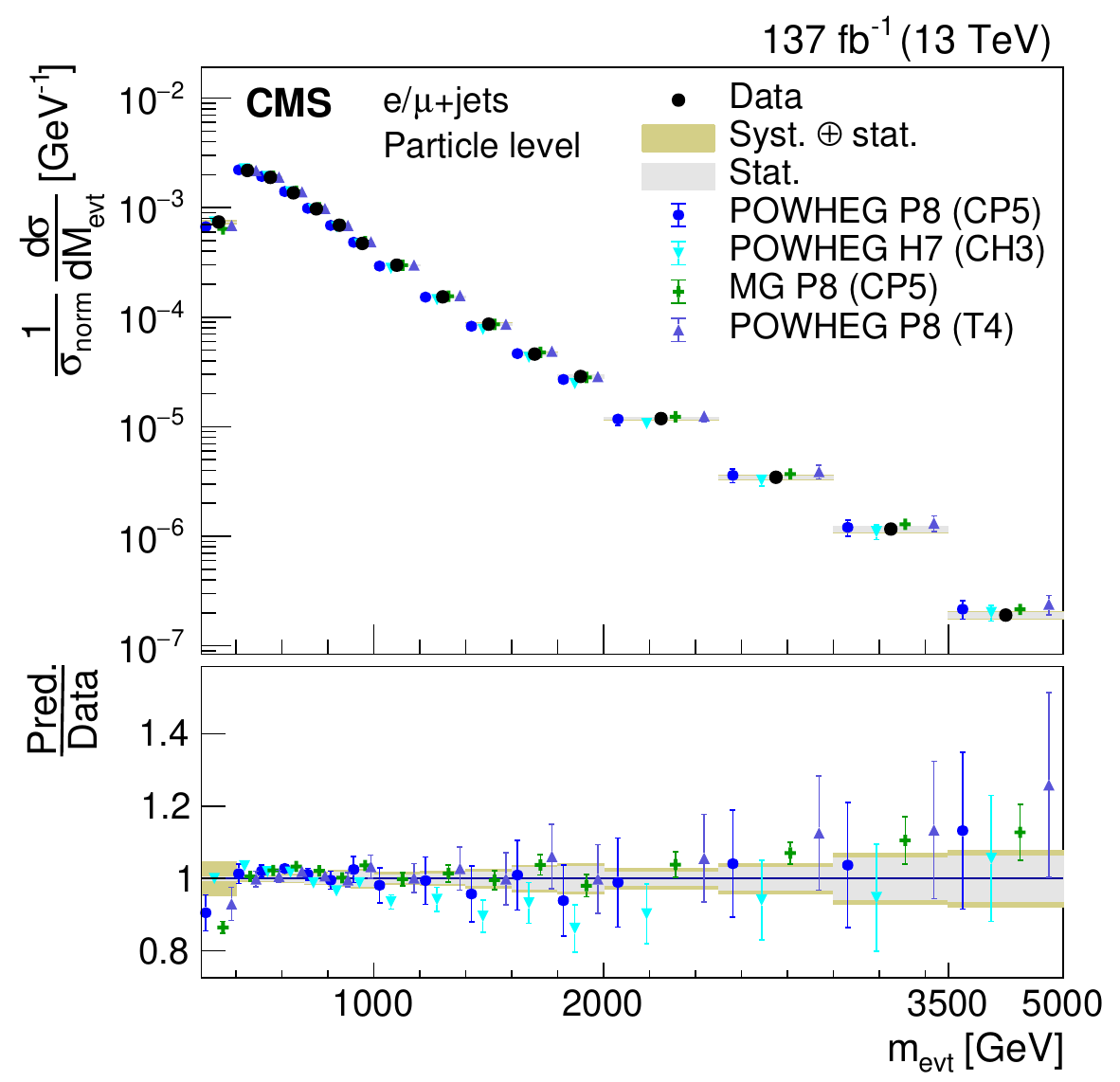}
 \caption{Normalized differential cross sections at the particle level as a function of \ptl, jet multiplicity, \HT, and \mevt. \XSECCAPPS}
 \label{fig:RESNORMPS13}
\end{figure*}

\begin{figure*}[tbp]
\centering
 \includegraphics[width=0.42\textwidth]{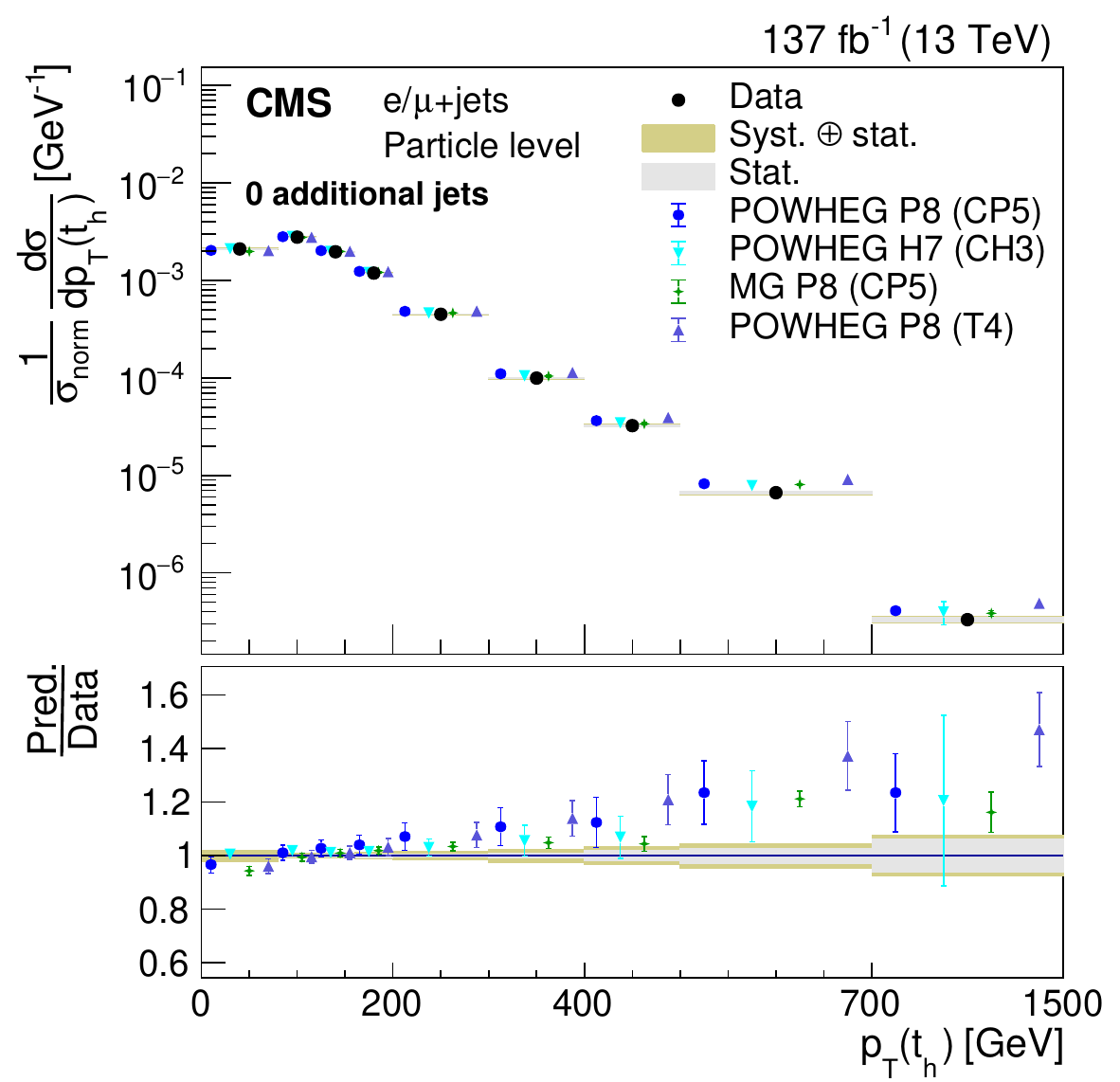}
 \includegraphics[width=0.42\textwidth]{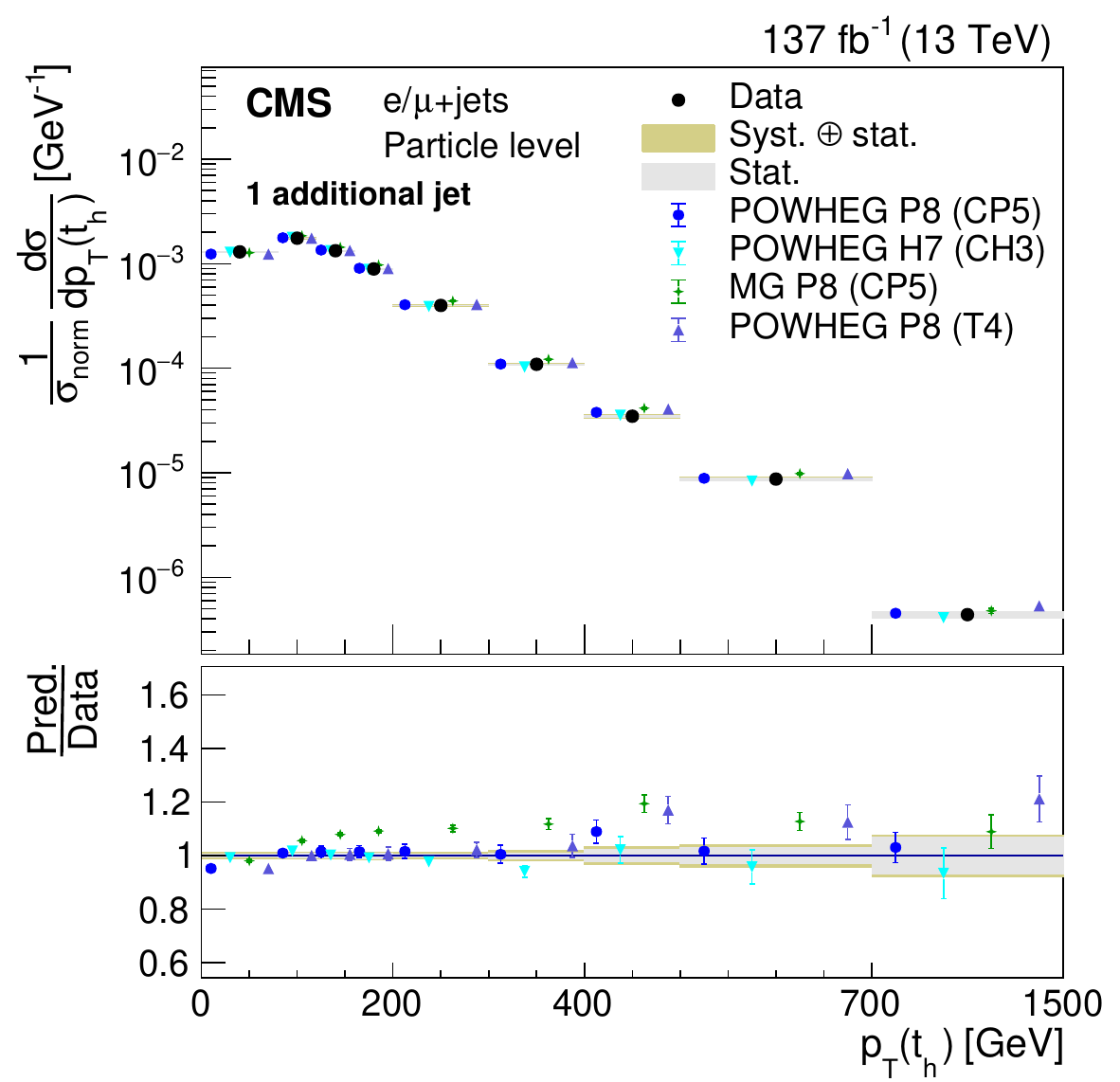}\\
 \includegraphics[width=0.42\textwidth]{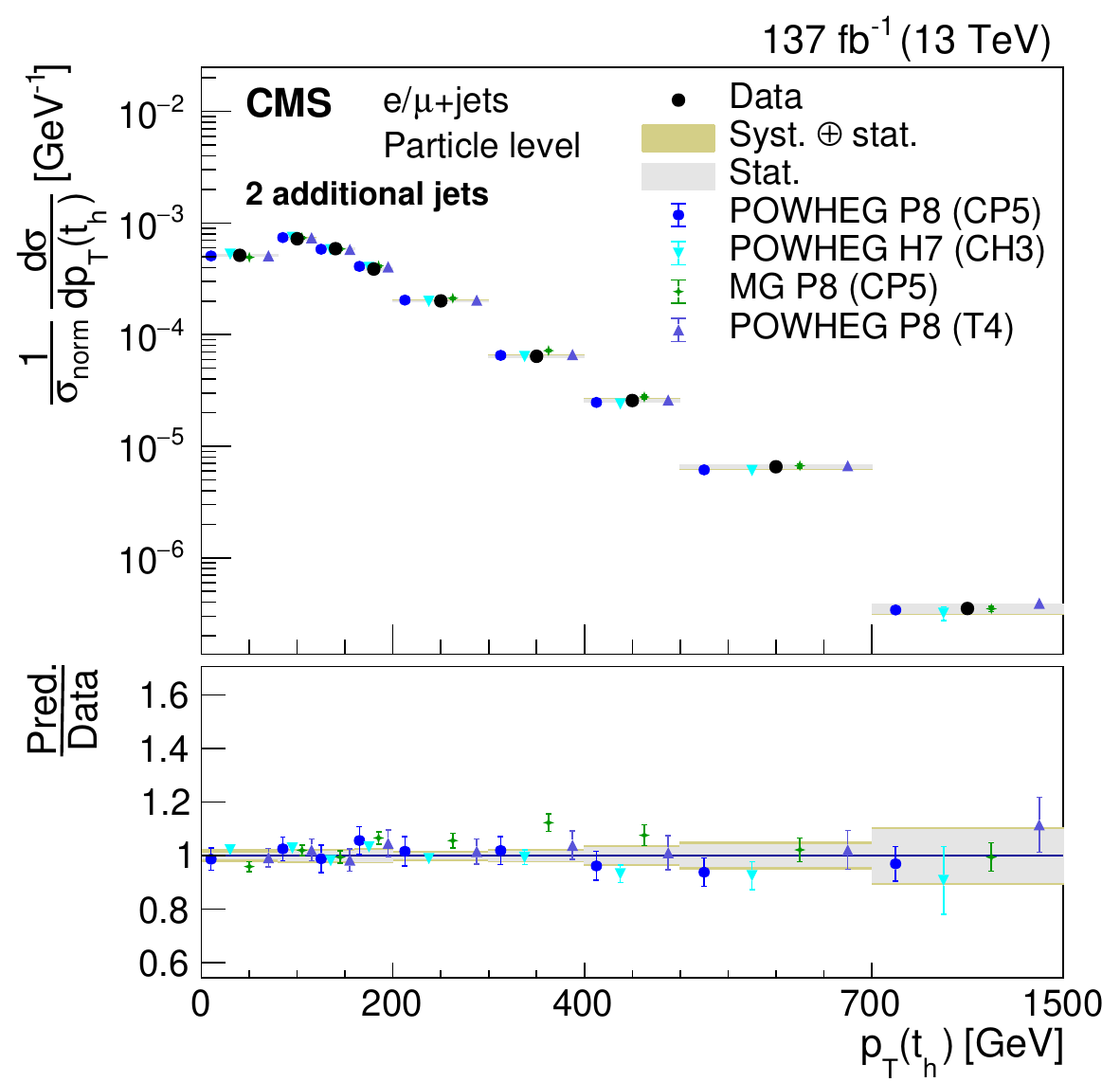}
 \includegraphics[width=0.42\textwidth]{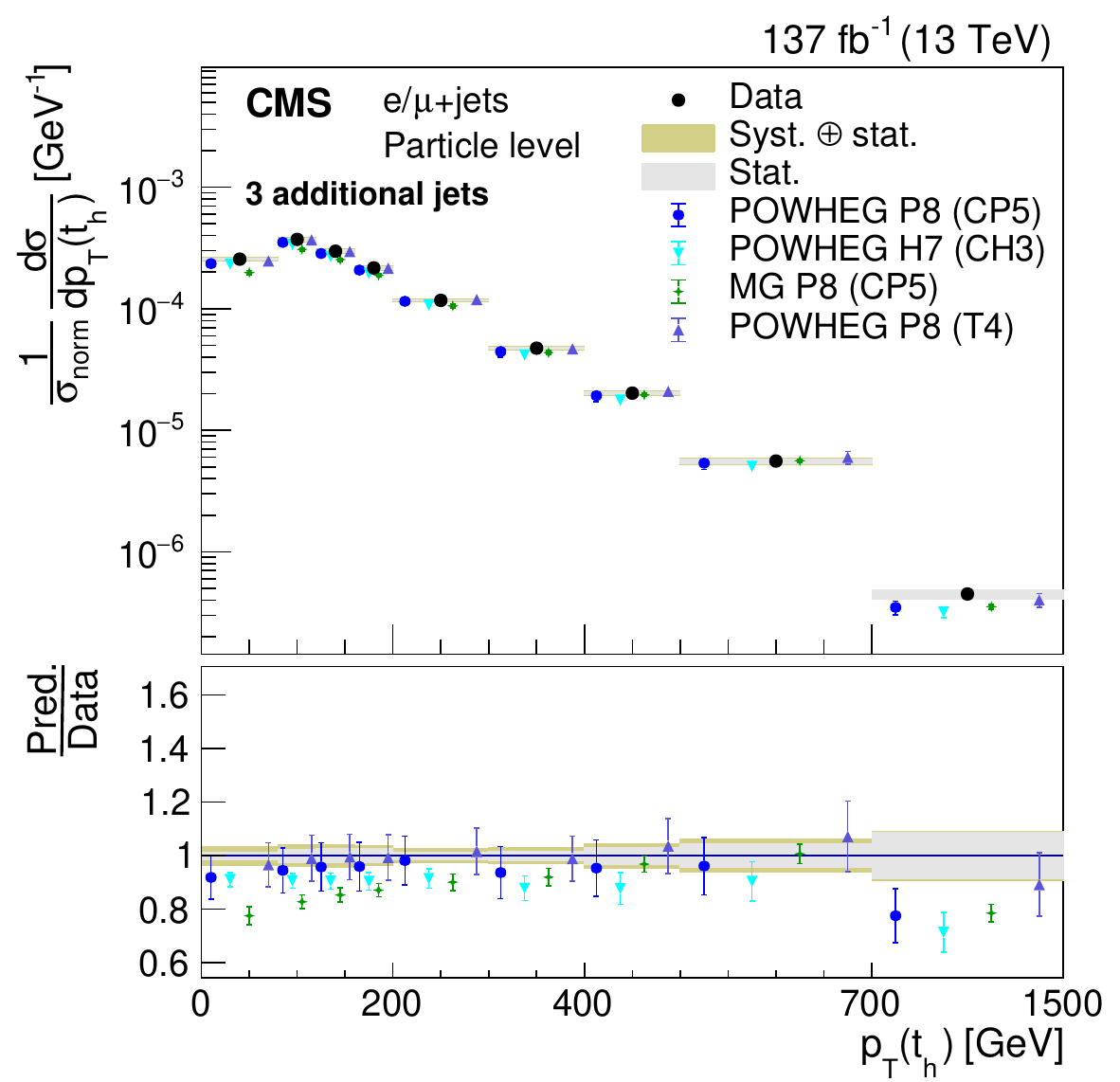}
 \caption{Normalized differential cross section at the particle level as a function of \thadpt in bins of jet multiplicity. \XSECCAPPS}
 \label{fig:RESNORMPS14}
\end{figure*}

\begin{figure*}[tbp]
\centering
 \includegraphics[width=0.42\textwidth]{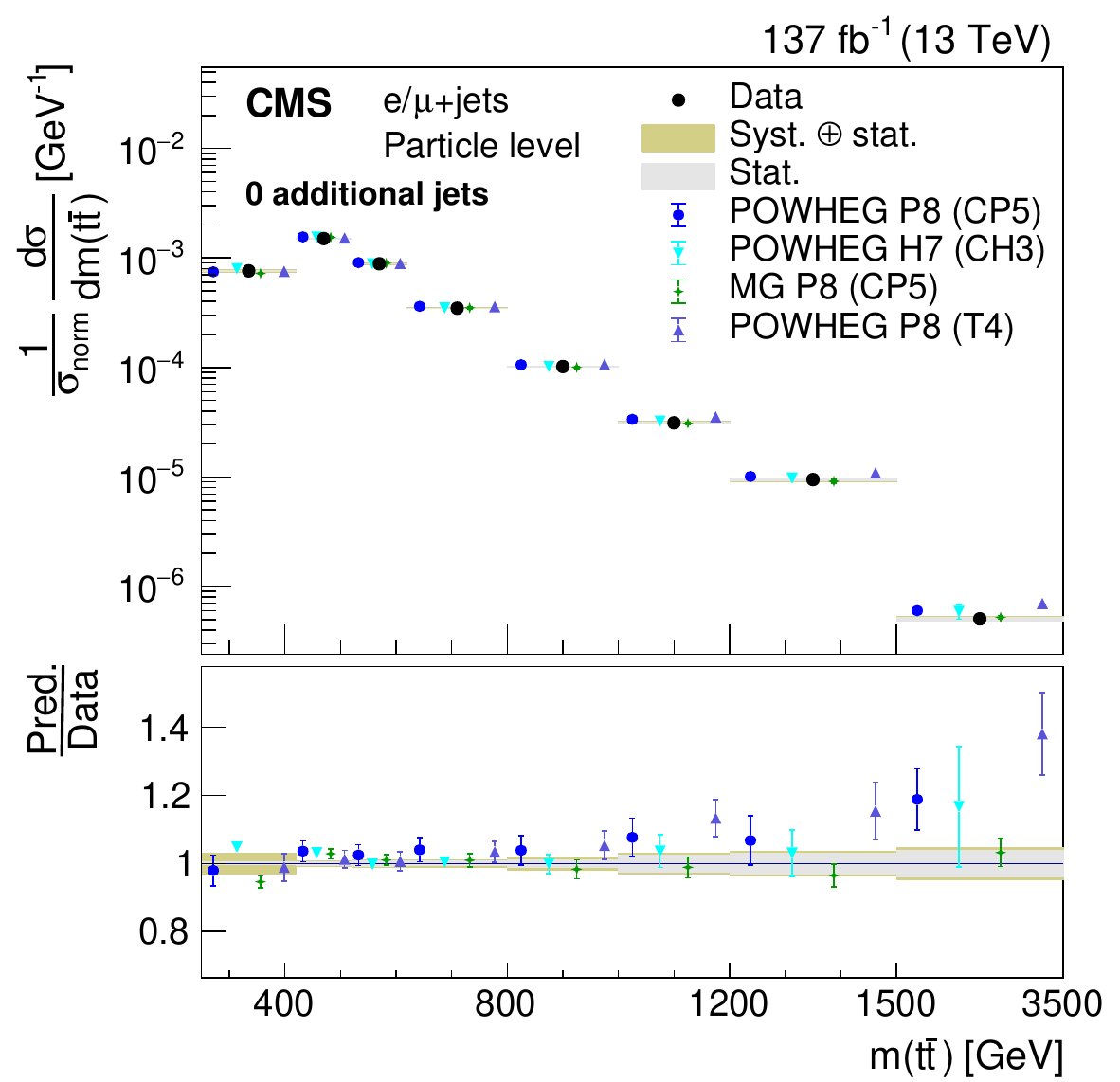}
 \includegraphics[width=0.42\textwidth]{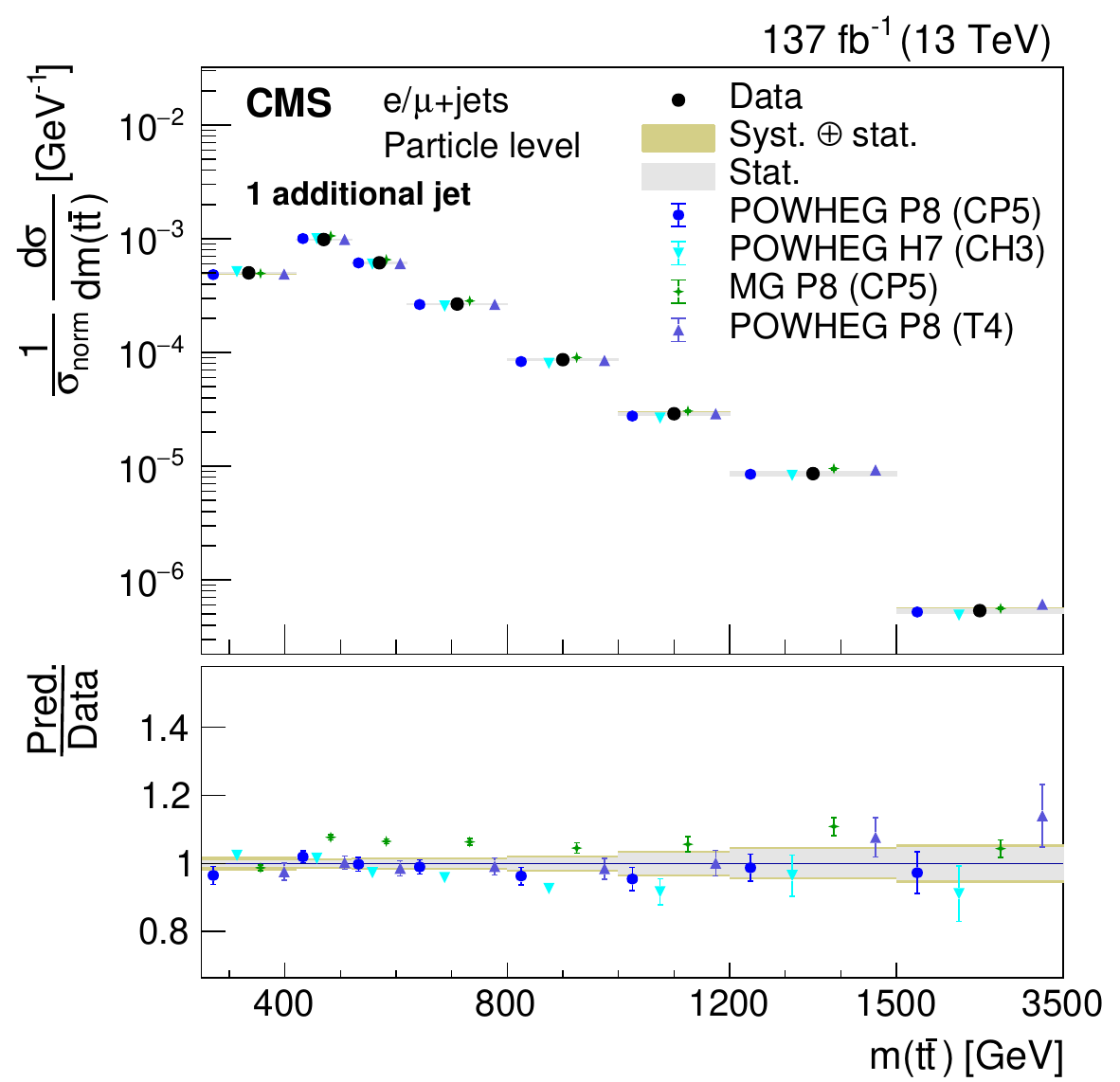}\\
 \includegraphics[width=0.42\textwidth]{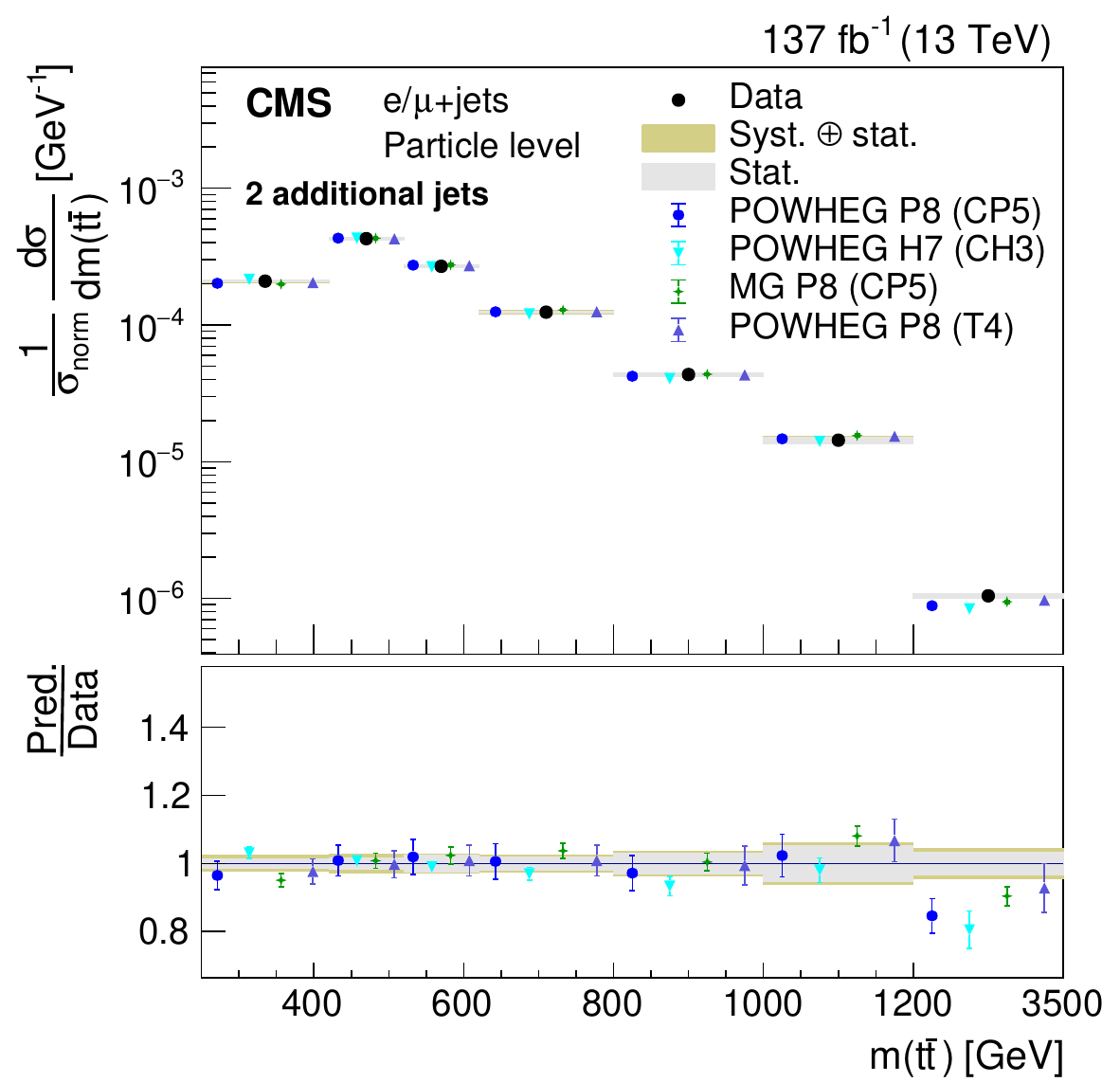}
 \includegraphics[width=0.42\textwidth]{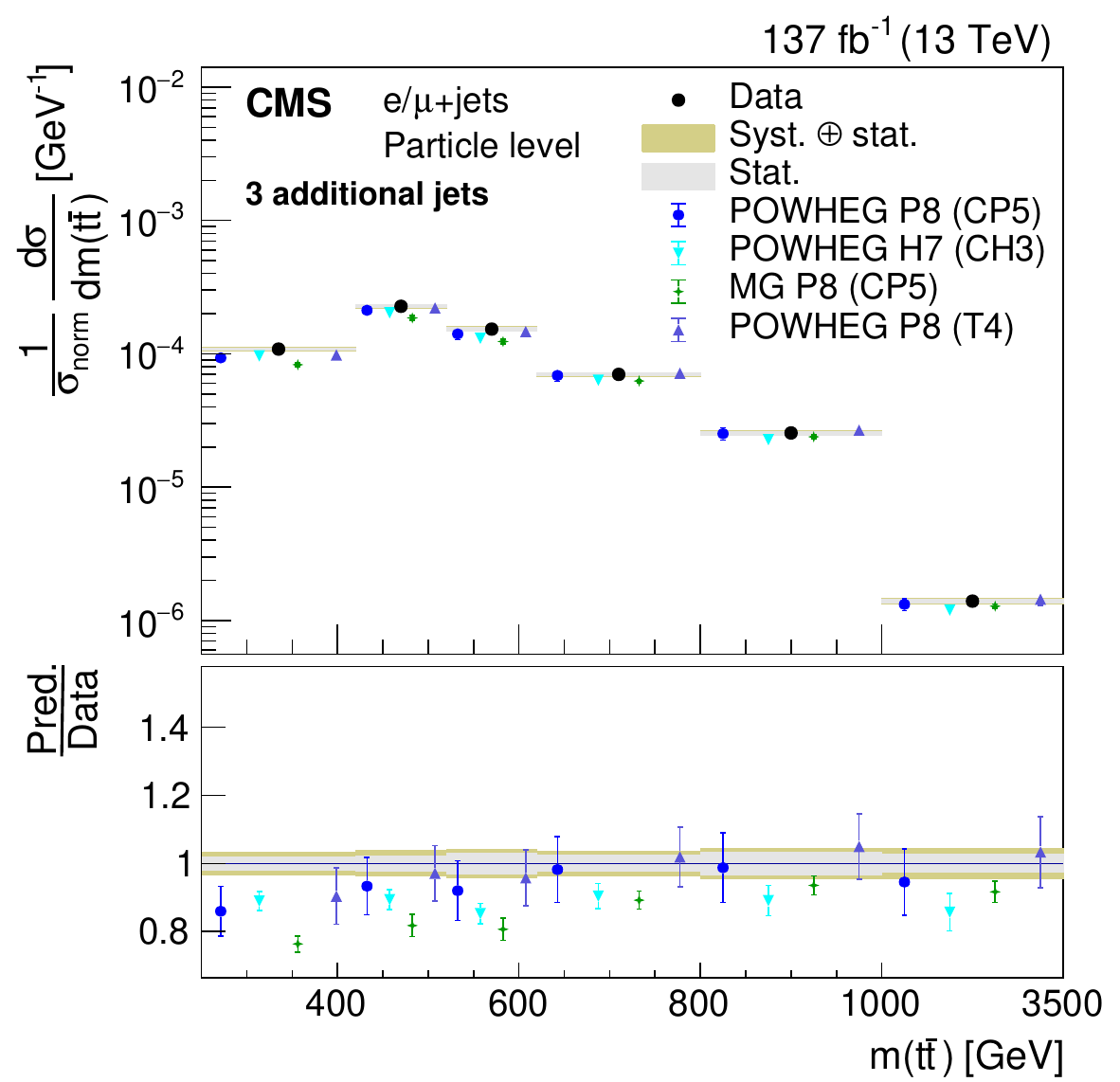}
 \caption{Normalized differential cross section at the particle level as a function of \ttm in bins of jet multiplicity. \XSECCAPPS}
 \label{fig:RESNORMPS15}
\end{figure*}

\begin{figure*}[tbp]
\centering
 \includegraphics[width=0.42\textwidth]{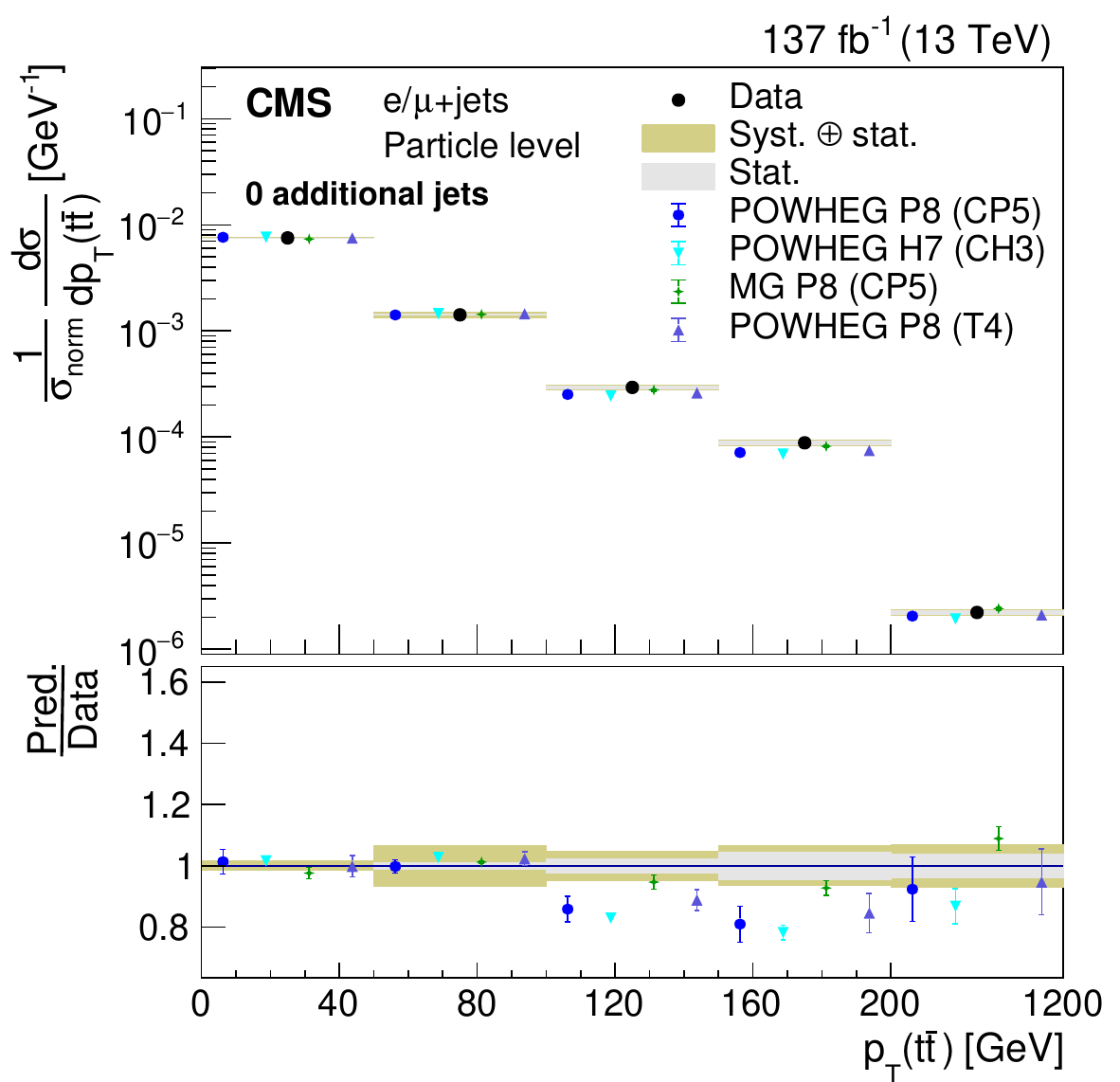}
 \includegraphics[width=0.42\textwidth]{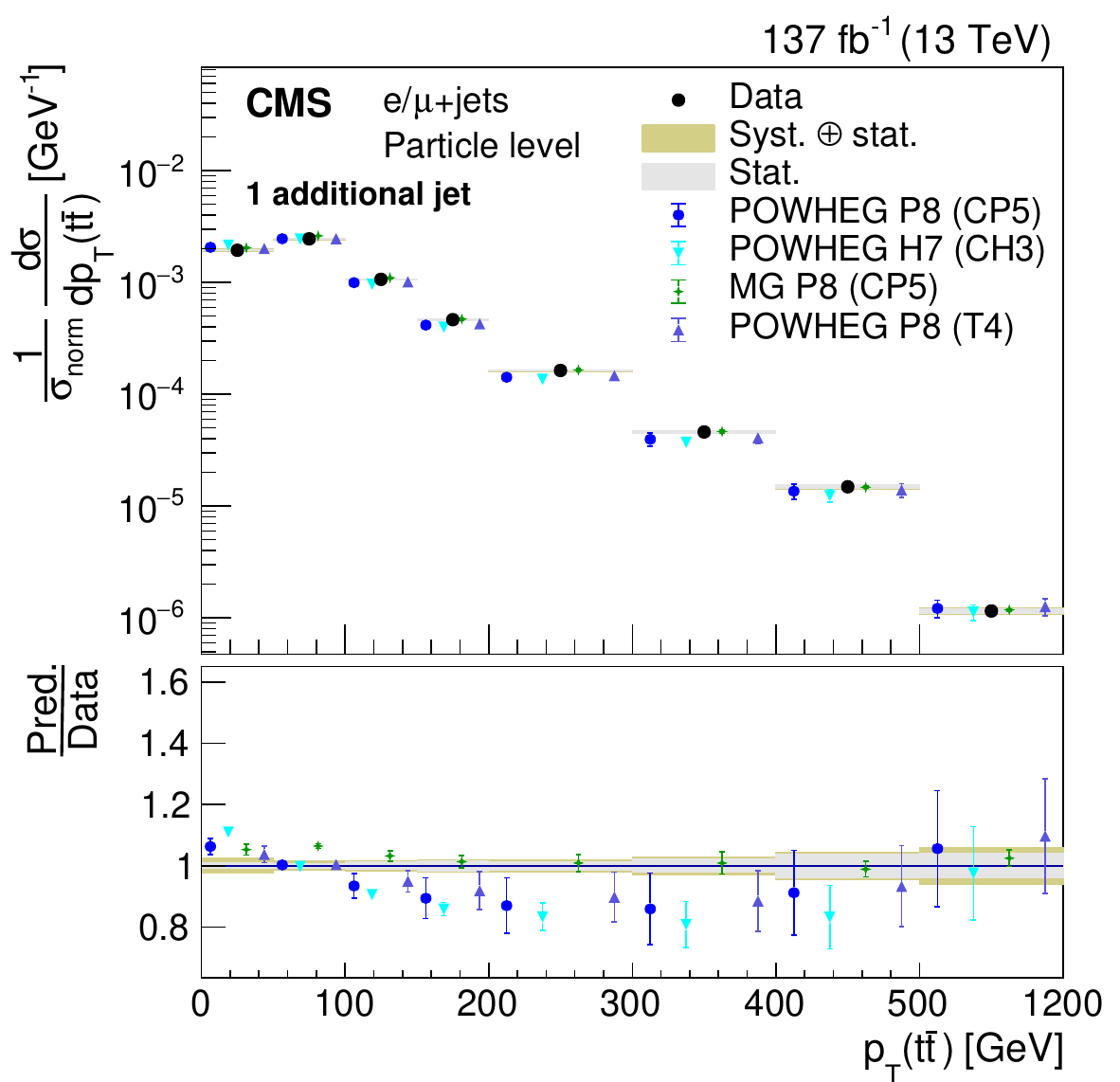}\\
 \includegraphics[width=0.42\textwidth]{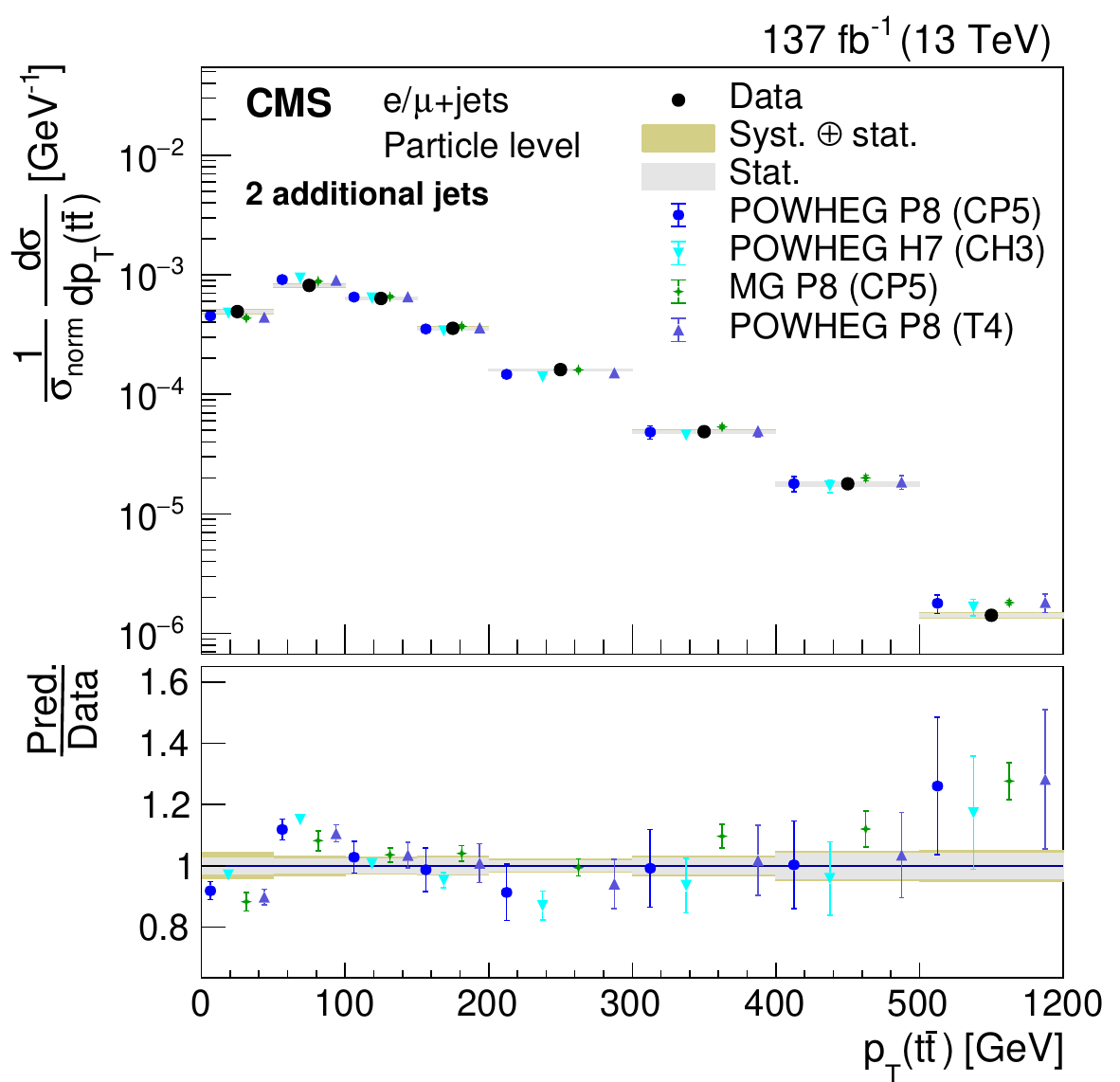}
 \includegraphics[width=0.42\textwidth]{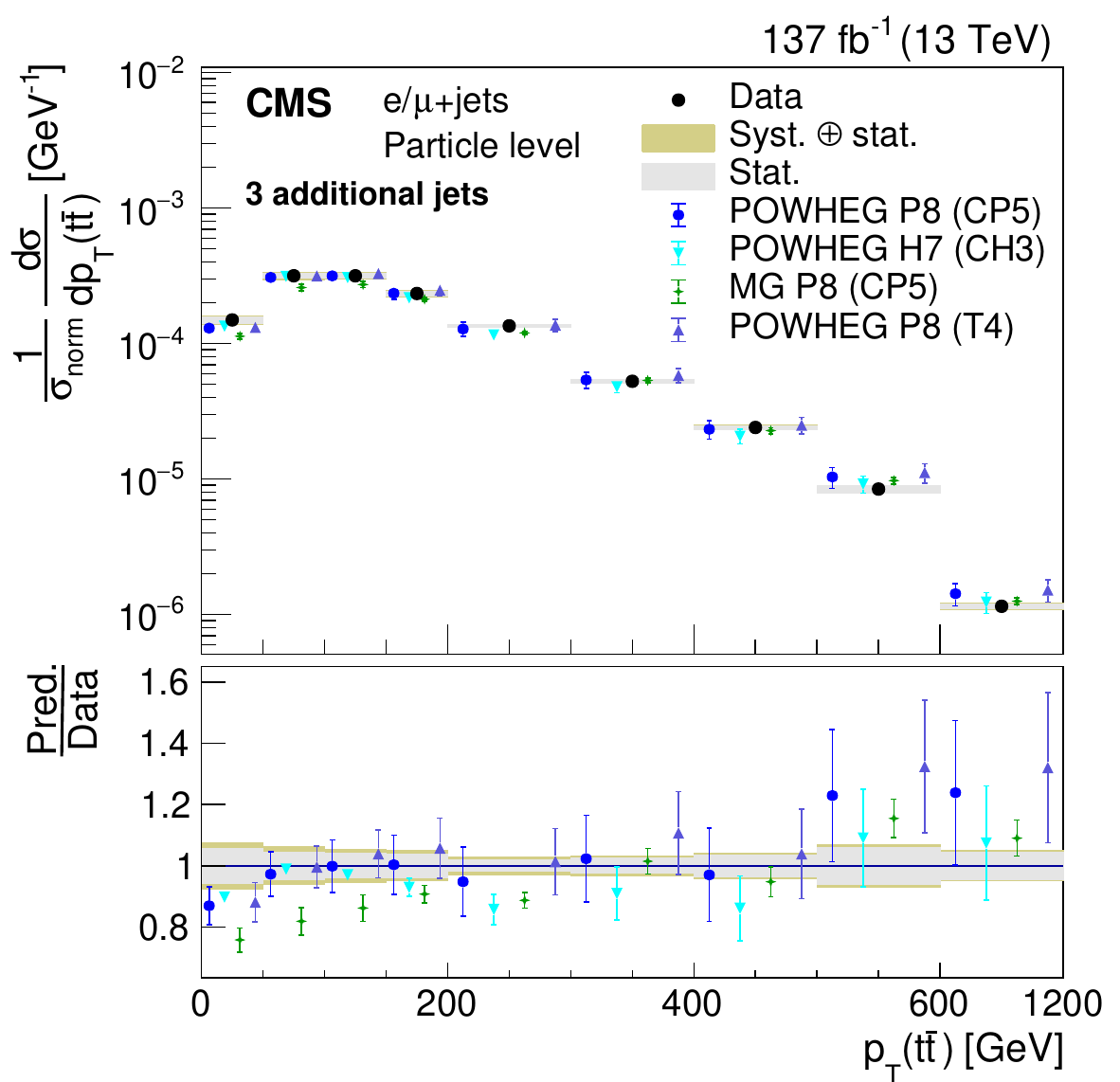}
 \caption{Normalized differential cross section at the particle level as a function of \ttpt in bins of jet multiplicity. \XSECCAPPS}
 \label{fig:RESNORMPS16}
\end{figure*}

\begin{figure*}
\centering
 \includegraphics[width=0.75\textwidth]{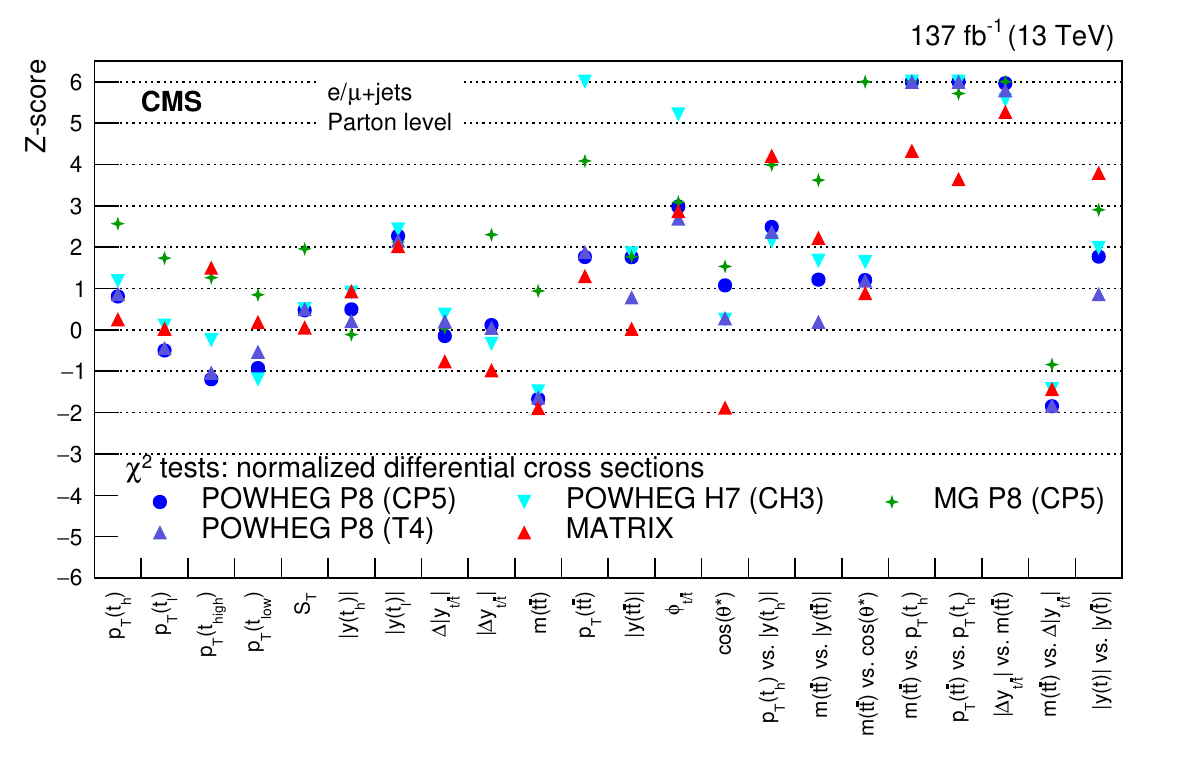}
 \includegraphics[width=0.75\textwidth]{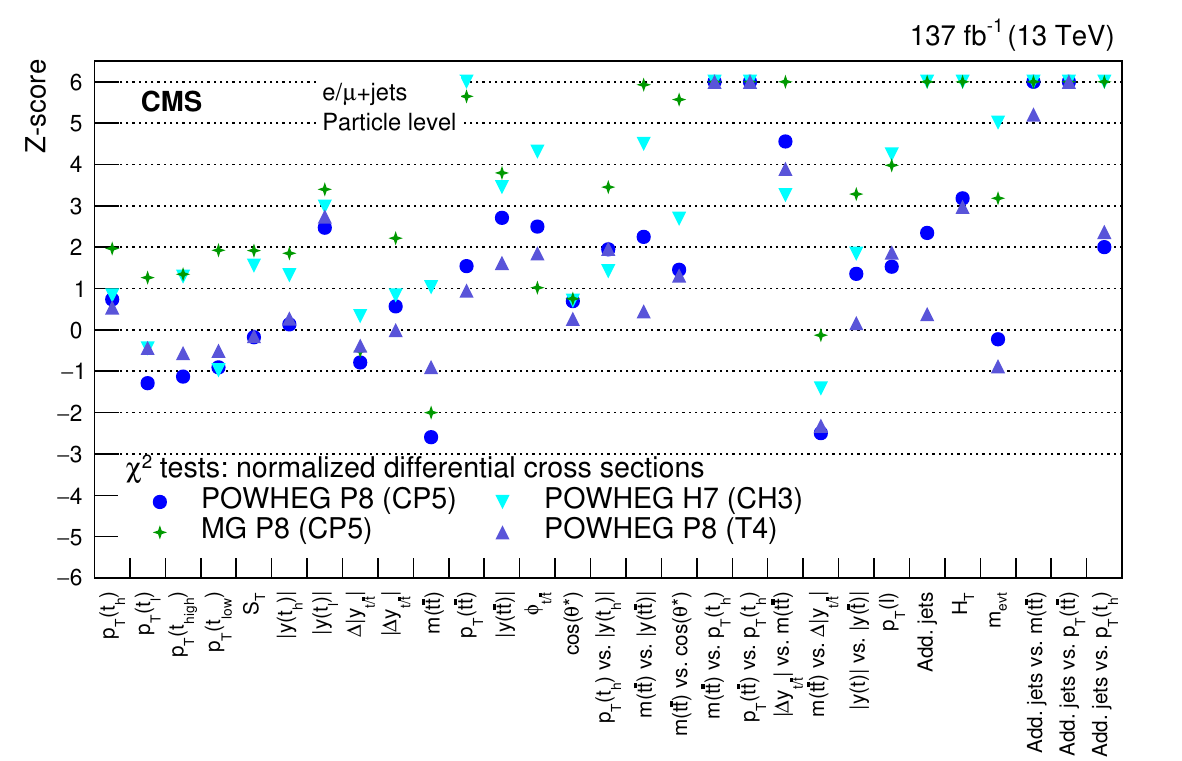}
 \caption{Results of $\chi^2$ tests in $Z$-scores comparing the measured normalized cross sections at the parton (upper) and particle (lower) levels to the predictions of \POWHEG{}+\PYTHIA (P8) for the CP5 and \CUET (T4) tunes, \POWHEG{}+\HERWIG (H7), the multiparton simulation \AMCATNLO(MG)+\PYTHIA FxFx, and the NNLO QCD calculations obtained with \MATRIX. The $Z$-scores are truncated at an upper limit of six. The uncertainties in the measurements and the predictions are taken into account for the calculation of the $\chi^2$.}
 \label{fig:RESNORM17}
\end{figure*}

In Figs.~\ref{fig:RESNORM1}--\ref{fig:RESNORMPS16}, all differential cross sections are presented normalized to unity. This has the advantage of canceling out the systematic uncertainties affecting only the overall normalization, and the differences in the shapes between data and predictions are more apparent. For this purpose the differential cross sections are divided by the \ttbar cross sections $\sigma_\mathrm{norm}$, which are obtained for each measurement as the sum of the cross sections in all bins of the corresponding kinematic observable or observables in the one- or two-dimensional range. The uncertainties in the normalized distributions are evaluated using error propagation and take into account the correlations between uncertainties in the individual measurements and $\sigma_\mathrm{norm}$. The results of the $\chi^2$ tests, comparing the normalized differential cross sections to the various predictions, are shown in Fig.~\ref{fig:RESNORM17}. The $p$-values are very similar to those obtained without normalization. This confirms that the normalizations are well predicted and that the $\chi^2$ tests are sensitive to differences in the shapes.

\cleardoublepage \section{The CMS Collaboration \label{app:collab}}\begin{sloppypar}\hyphenpenalty=5000\widowpenalty=500\clubpenalty=5000\vskip\cmsinstskip
\textbf{Yerevan Physics Institute, Yerevan, Armenia}\\*[0pt]
A.~Tumasyan
\vskip\cmsinstskip
\textbf{Institut f\"{u}r Hochenergiephysik, Vienna, Austria}\\*[0pt]
W.~Adam, J.W.~Andrejkovic, T.~Bergauer, S.~Chatterjee, M.~Dragicevic, A.~Escalante~Del~Valle, R.~Fr\"{u}hwirth\cmsAuthorMark{1}, M.~Jeitler\cmsAuthorMark{1}, N.~Krammer, L.~Lechner, D.~Liko, I.~Mikulec, P.~Paulitsch, F.M.~Pitters, J.~Schieck\cmsAuthorMark{1}, R.~Sch\"{o}fbeck, M.~Spanring, S.~Templ, W.~Waltenberger, C.-E.~Wulz\cmsAuthorMark{1}
% ...
\vskip\cmsinstskip
% ...
1:  Also at TU Wien, Wien, Austria\\
% ...
\end{sloppypar}
\end{document}